\documentclass[b5paper,10pt,times,authoryear,print,index]{Classes/PhDThesisPSnPDF}

\input{Preamble/preamble}

\title{Theory \& observations of the PWN-SNR complex} 

\author{Jonatan Mart\'in Rodr\'iguez}

\dept{Institut d'Estudis Espacials de Catalunya (IEEC-CSIC)}

\university{Universitat Aut\`onoma de Barcelona}
\crest{
\includegraphics[width=0.3\textwidth]{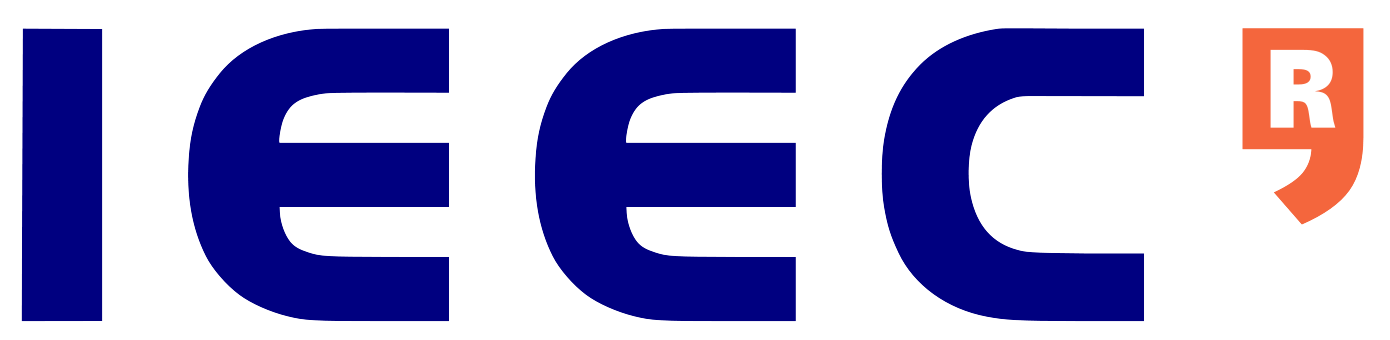}
\raisebox{-1cm}{\includegraphics[width=0.3\textwidth]{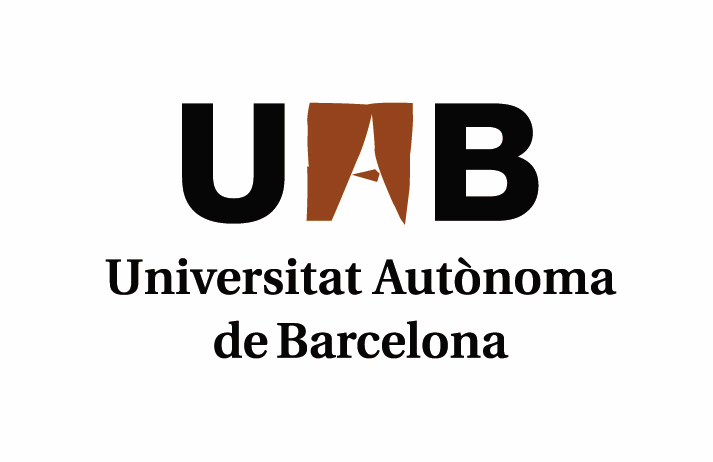}}
\raisebox{-0.35cm}{\includegraphics[width=0.3\textwidth]{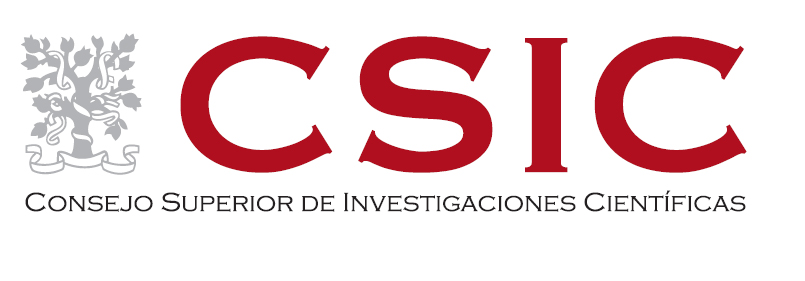}}
}

\degree{Prof. Diego F. Torres\\
Dr. Nanda Rea\\
Tutor: Dr. Llu\'is Font Guiteras}
 
\college{Barcelona (Spain)}

\degreedate{July 2014} 

\subject{LaTeX} \keywords{{LaTeX} {PhD Thesis} {Astrophysics} {Universitat Aut\`onoma de Barcelona}}

\ifdefineAbstract
\fi

\ifdefineChapter
\fi

\begin{document}

\frontmatter

\begin{titlepage}

\maketitle

\end{titlepage}


\begin{dedication} 

{\it Pels que pensen,\\
pels que lluiten,\\
pels que pateixen,\\
pels que somien}

\end{dedication}


\begin{declaration}

I hereby declare that except where specific reference is made to the work of others, the contents of this dissertation are original
and have not been submitted in whole or in part for consideration for any other degree or qualification in this, or any other
University. This dissertation is the result of my own work and includes nothing which is the outcome of work done in collaboration,
except where specifically indicated in the text. This dissertation is based in works already published detailed below:

\begin{itemize}
 \item Mart\'in, J., Torres, D. F. \& Rea, N. 2012, MNRAS, 427,415
 \item Torres, D. F., Cillis, A. N. \& Mart\'in Rodr\'iguez, J. 2013, ApJL, 763, L4
 \item Torres, D. F., Mart\'in, J., de O\~na Wilhelmi, E. \& Cillis, A. N. 2013, MNRAS, 436, 3112
 \item Torres, D. F., Cillis, A. N., Mart\'in, J. \& de O\~na Wilhelmi, E. 2014, JHEAp, 1, 31
 \item Mart\'in, J., Rea, N., Torres, D. F. \& Papitto, A. 2014a, accepted in MNRAS, arXiv: 1409.1027
 \item Mart\'in, J., Torres, D. F., Cillis A., \& de O\~na Wilhelmi, E. 2014b, MNRAS, 443, 138
\end{itemize}


\end{declaration}


\begin{acknowledgements}      

Com resumir una cosa infinita en un espai finit? Aquest \'es el gran repte amb el que em trobo escrivint aquestes l\'inies. Darrere d'aquest
treball, no nom\'es hi ha hagut la feina, l'esfor{\c c} i la t\`ecnica, que s\'on molt importants, per\`o no hagu\'es estat possible de cap
manera si no m'hagu\'es trobat amb totes les persones amb les que he pogut conviure durant aquest temps, que m'han donat tot el suport
necessari i m'han ensenyat all\`o que no es troba als llibres. A tots vosaltres, des del primer p\`aragraf, us vull donar les gr\`acies.

En primer lugar, me gustar\'ia dar las gracias a mis dos directores de tesis, Diego y Nanda, por la confianza que han depositado en mi desde el
primer d\'ia. Vuestro esfuerzo, vuestro inter\'es, vuestra paciencia (que conmigo a veces es necesaria) y vuestro buen trato han hecho de
trabajar y pasar el d\'ia a d\'ia con vosotros una grata experiencia que espero que en el futuro siga dando sus frutos. Tamb\'e vull agra\"ir al
Llu\'is Font que hagi acceptat ser el tutor d'aquesta tesi i pel seu inter\'es.

Tambi\'en quiero agradecer a mis padres (Domingo y Carmen), a mi hermano Alejandro y a toda mi familia el apoyo que me han dado durante todo este
tiempo. Sin su ayuda incondicional, todo esto hubiera sido imposible.

Especial agradecimiento a mis \'angeles que hacen que todo sea m\'as f\'acil y llevadero. Nataly, podr\'ia escribir p\'aginas y p\'aginas y no
acabar\'ia. Muchas gracias por tu compa\~n\'ia y por tu afecto. Por las bromas y por las risas, por las conversaciones m\'as serias, por los paseos
por los pasillos, por molestarte a cruzar al otro lado del cyber cada d\'ia a saludar, y como no, por los azucarillos :-). !`No cambies nunca,
porque eres \'unica! Daniela, vielen Dank f\"ur dein L\"acheln und deine Hilfe. Tu experiencia y simpat\'ia me han ayudado mucho. ?`Que hubiera
hecho yo sin una alemana que supiera bailar salsa? En Barcelona siempre tendr\'as un amigo (o como m\'inimo, un amigo barcelon\'es). Elsa, a
pesar de haber compartido cyber y grupo solo un a\~no, has sido un referente y un gran apoyo en los inicios de esta aventura. A\'un se echan de
menos tus planes mal\'eficos para salir los d\'ias laborables o momentos paranormales en el cyber en que pelotitas antiestr\'es volaban por el
aire. Alicia, gracias por tu alegr\'ia y por tu buen humor. Esos momentos de escapada al IFAE para tomar el caf\'e o tu motivaci\'on para hacer
planes en el tiempo libre no tienen precio. Ahora es el turno de los no-tan-\'angeles ;-). Jacobo, el ma\~no que ha dejado un hueco en el cyber
que no se puede llenar. Por eso, a\'un hay carteles dando una recompensa por encontrarte. Muchas gracias por tu espontaneidad, tu buen humor y
tus locuras ya sean en persona o en la lejan\'ia, por skype o por whatsapp. Adiv, eres un grande. Gracias por tu sencillez y por todos los
momentos vividos ya sea en un bar llamado Kentucky o cuando hemos hecho alguna escapada. Felipe, moltes gr\'acies per tot amic, mai oblidar\'e la
teva simpatia i els bons moments que hem passat junts. Hem de trobar el moment de tornar al ``Magic'', sense que la Daniela s'enteri ;-).

No penseu que m'oblidava de vosaltres: Kike, Antonia, Albert, Carles, Laura, Carmen, Padu i Marina. M'alegro molt d'haver-vos conegut. M'ho he
passat molt b\'e anant amb vosaltres al ``laser-tag'' (quina motivaci\'o!), els sopars, els partits de futbol... Espero poder seguir vient-vos i
que els vostres somnis es compleixen. Josep, t'agraeixo molt\'issim la teva ajuda en el fosc m\'on de la inform\`atica i que, tot i les teves
amena{\c c}es, no m'hagis portat a veure al Pepe. Alina i Aina, molt\'issimes gr\`acies per la vostra ajuda desinteressada. El vostre suport no
t\'e preu. Tambi\'en a la gente del IFAE a pesar que no he podido compartir mucho tiempo con ellos: Rub\'en, Dani Garrido (esas barbacoas!),
Alba, Daniel (argentino) y Quim. Amb el vostre perm\'is, deixo un petit raconet per la N\'uria i l'Arnau. Moltes gr\`acies pels vostres \`anims
durant la meva tesi, els sopars improvitzats i per fer dels matins al cotxe una cosa divertida i distreta. N\'uria, amb qui far\`as ara carreres
de cadires al cyber? ;-). Estic conven{\c c}ut que tot t'anir\`a b\'e. Et deixo l'enc\`arrec d'administrar les ``meves'' taules del cyber, del pa
bimbo i la ``nocilla''. Arnau, vigila que la N\'uria faci b\'e la seva feina i no li robis el pa bimbo. No canvi\"is mai, crack!

I would like to thank insistently all the people that has been in our group during these years: Andrea, Giovanna, Choni, Ana and Eric. Special
thanks to Anal\'ia (gracias por tu inter\'es y tu afecto), Emma (gracias por tu trato y sentido del humor) and Alessandro (Daje, Lazio!) for 
working directly with me in some projects. Jian, thank you for your support. I'm very happy to know someone like you. Daniele, tot i que fa poc
que est\`as a Barcelona, tamb\'e has sigut una gran ajuda quan el vent ha bufat en contra i m'alegro d'haver conegut alg\'u amb la teva
senzillesa i sinceritat. Moltes gr\'acies, amic.

No voldria acabar sense dirigir-me als amics de sempre de l'escola que han estat al peu del can\'o: Albert (el GR-11 nos espera!), Joan
(``au va!''), Marcel (``amigooo''), Angi, Dela, Pau (``tungs-t\`e!''), Cris, Carlos i Coke. Igualment pels del cau: Andr\'es (viva Las Vegas!),
Lluna, Marta, Germ\'an, Irene, Olga i Ignasi, i a tots els membres a l'Agrupament Escolta Champagnat pel seu suport i la seva comprensi\'o.
\\

Finalment, vull donar les gr\`acies a tots els contribuients que amb la seva aportaci\'o, encara que sigui en moments de dificultat, fan possible
les beques que donen l'oportunitat a molts estudiants que, com jo, simplement lluitem pels nostres somnis i una vida digna.

\end{acknowledgements}
\begin{abstract}

In this work, we study theoretical and observational issues about pulsars (PSRs), pulsar wind nebulae (PWNe) and supernova remnants (SNRs). In
particular, the spectral modeling of young PWNe and the X-ray analysis of SNRs with magnetars comparing their characteristics with those remnants
surrounding canonical pulsars.

The spectra of PWNe range from radio to $\gamma$-rays. They are the largest class of identified Galactic sources in $\gamma$-rays increasing the
number from 1 to $\sim$30 during the last years. We have developed a detailed spectral code which reproduces the electromagnetic spectrum of PWNe
in free expansion ($t_{age} \lesssim$10 kyr). We shed light and try to understand issues on time evolution of the spectra, the synchrotron
self-Compton dominance in the Crab Nebula, the particle dominance in PWNe detected at TeV energies and how physical parameters constrain the
detectability of PWNe at TeV. We make a systematic study of all Galactic, TeV-detected, young PWNe which allows to find correlations and trends
between parameters. We also discuss about the spectrum of those PWNe not detected at TeV and if models with low magnetized nebulae can explain
the lack of detection or, on the contrary, high-magnetization models are more favorable.

Regarding the X-ray analysis of SNRs, we use X-ray spectroscopy in SNRs with magnetars to discuss about the formation mechanism of such extremely
magnetized PSRs. The alpha-dynamo mechanism proposed in the 1990's produces an energy release that should have influence in the energy of the SN
explosion. We extend the work done previously done by \cite{vink06} about the energetics of the SN explosion looking for this energy release
and we look for the element ionization and the X-ray luminosity and we compare our results with other SNRs with an associated central source.

\end{abstract}


\tableofcontents

\listoffigures

\listoftables 


\mainmatter


\chapter{Introduction}  
\label{chap1}

\ifpdf
    \graphicspath{{Chapter1/Figs/Raster/}{Chapter1/Figs/PDF/}{Chapter1/Figs/}}
\else
    \graphicspath{{Chapter1/Figs/Vector/}{Chapter1/Figs/}}
\fi

\section[A brief historical view]{A brief historical view} 
\label{sec1.1}

{\it Supernova} (SN) explosions of stars have been observed for centuries. For the first civilizations, these phenomena were known simply as very
bright stars that appeared spontaneously in the sky, and thought to be related to extraordinary events. The Romans called these ``stars'' as
{\it novae}, which in Latin means ``new stars''. There are many historical records of SN observations, but the Chinese were the ones who took the
most detailed record of these events. The reminiscent gas and dust structures of these explosions, the {\it supernova remnants} (SNRs), were
observed for the first time during the 18th century. The first claimed SNR was the Crab Nebula, reported by the english astronomer John Bevis in
1731. In 1757, Charles Messier confused the Crab Nebula with the Halley's comet. When he realized his error, he included it in the Messier's
catalog as a non-comet like object two years later.

\begin{figure}[t!]
\centering
\includegraphics[width=0.5\textwidth]{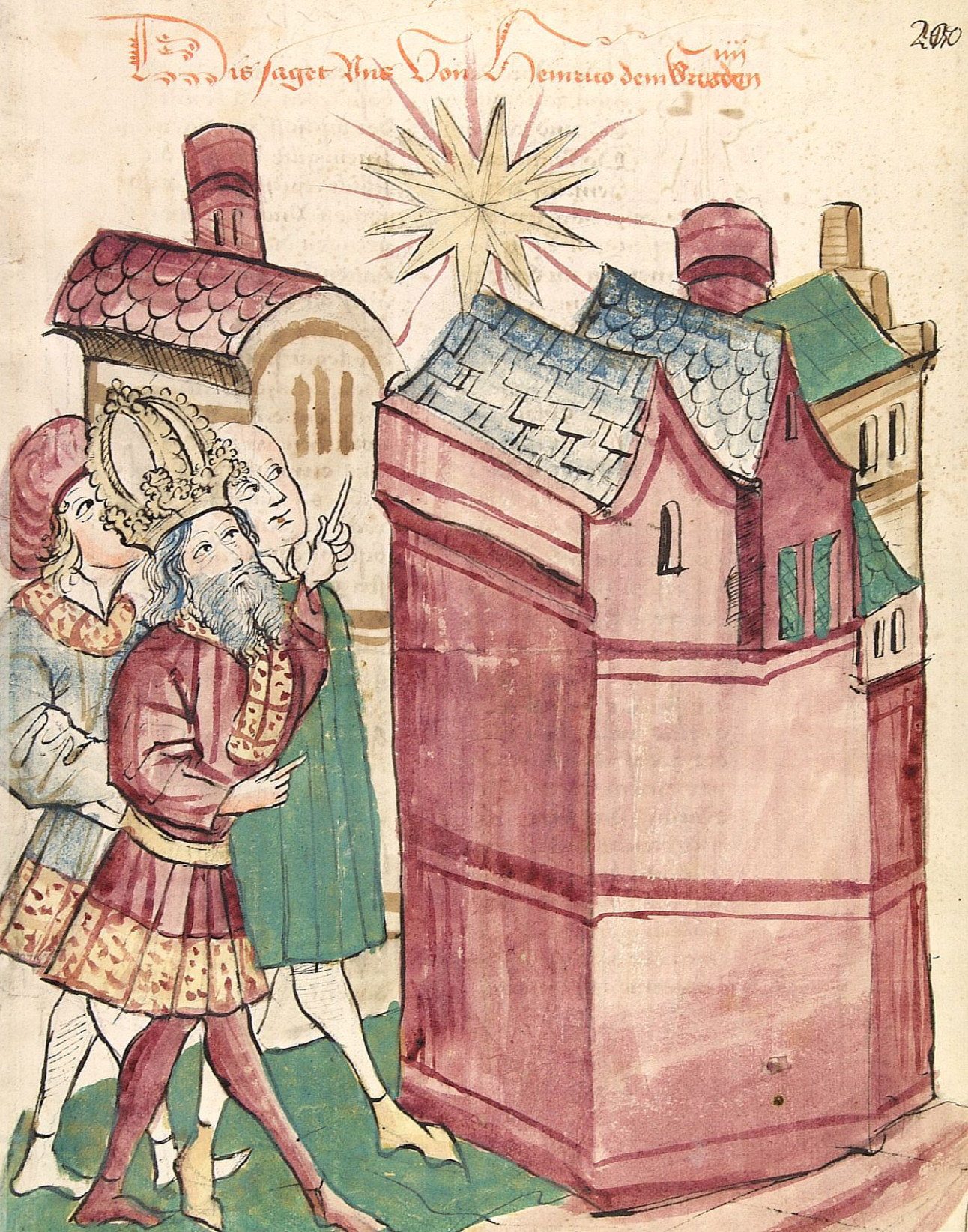}
\caption[Henry III observing a new rised star]{Picture of the Holy Roman Emperor Henry III with two more people observing a new rised star over
the city of Tivoli (Italy).}
\label{european}
\end{figure}

\begin{figure}
\centering
\includegraphics[width=0.75\textwidth]{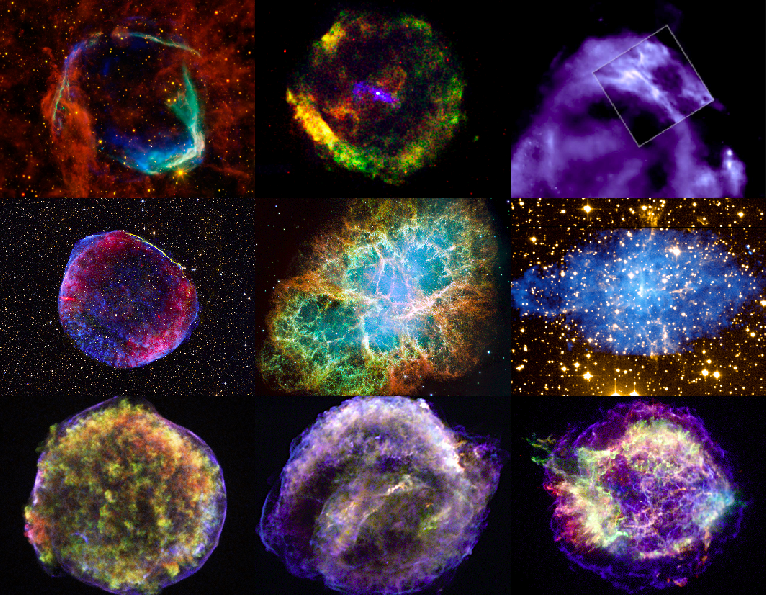}
\caption[The nine historical SNRs]{The nine historical SNRs. From left to right, first row: RCW 86, G11.2-0.3, G347.3-0.5. Second row: SN 1006,
Crab Nebula, 3C 58. Third row: Tycho's SNR, Kepler's SNR, Cassiopeia A.}
\label{snr_collage}
\end{figure}

In the beginning of the 20th century, other objects started to be identified as SNRs. Thanks to records collected by several ancient civilizations, the years of the SN explosions are reported for nine of the brightest SNRs (see \citealt{clark1982}): RCW 86 (possibly observed by the
Chinese in 185 D.C.), G11.2-0.3 (possibly observed by the Chinese in 386 D.C.), G347.3-0.5 (possibly observed by the Chinese in 393 D.C.), SN
1006 (observed by the Chinese, Japanese, Arabic and Europeans in 1006 D.C.), Crab Nebula (observed by the Chinese, Japanese, Arabic and probably
native Americans in 1054 D.C.), 3C 58 (possibly observed by the Chinese and Japanese in 1181 D.C.), Tycho's SNR (observed by the Europeans,
Chinese and Koreans in 1572 D.C.), Kepler's SNR (observed by the Europeans, Chinese and Koreans in 1604 D.C.) and Cassiopeia A (possibly observed by
the Europeans in 1680 D.C.). An image of these remnants is shown in figure \ref{snr_collage}. Depending on the SNR morphology, they are now defined 
as: shell-like and composite SNRs. Historically, composite SNRs are characterized by a central non-thermal
emission. These non-thermal nebulae are now recognized as {\it pulsar wind nebulae} (PWNe) or {\it plerions}, name derived from the ancient greek ``pleres'', which means ``full'' (name coined by \citealt{weiler78}).

In the 1930's, even before the discovery of the neutron \citep{chadwick32}, \citet{landau32} suggested the existence of stars which would look
like giant atomic nuclei. Later, \citet{chandrasekhar35} suggested models of stars with degenerate cores, and \citet{oppenheimer39} proposed the
first equation of state for a neutron degenerate gas. \citet{baade34} already suggested that these {\it neutron stars} (NS) could be formed in a
supernova explosion. NS were also considered as candidates for unidentified sources in X-rays during the sixties (e.g., \citealt{morton64}). In
1967, Jocelyn Bell dicovered the first rapidly rotating neutron star (or {\it pulsar}, which is an abbreviation for pulsating radio star) in the
Mullard Radio Astronomy Observatory while she was analyzing radio data from quasars. The discovery was published one year later \citep{hewish68}.
A few years later, \citet{giacconi71} discovered a 4.8 s pulsation from Cen X-3 in X-rays with the {\it UHURU} telescope, being the first X-ray pulsar discovered.
Using radio observations from the Arecibo antenna, \citet{taylor74} discovered PSR B1913+16, the first binary pulsar (a binary system with two
NSs, but only one is pulsating). \citet{backer82} discovered the first millisecond pulsar ($P \sim 1.5$ ms), one of the fastest rotators known.
\citet{burgay03} discovered the first double neutron star system where both components are detectable as pulsars (PSR J0737-3039). Finally,
the most massive neutrons stars were discovered only recently by \citet{demorest10} (PSR J1614-2230) and \citet{antoniadis13} (PSR J0348+0432),
with (1.97$\pm$0.04)$M_\odot$ and (2.01$\pm$0.04)$M_\odot$, respectively.

\begin{figure}
\centering
\includegraphics[width=0.75\textwidth]{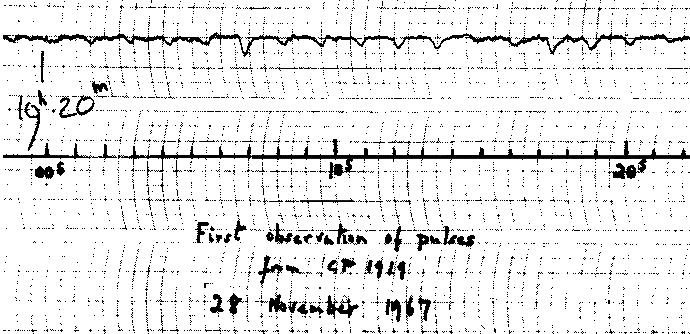}
\caption[Radio pulses of CP 1919]{Radio pulses observed for CP 1919 (PSR J1921+2153), the first pulsar known \protect\citep{hewish68}.}
\label{cp1919}
\end{figure}

In 1979, a new class of neutron stars was discovered through the detection of repeated bursts in hard X-rays and soft $\gamma$-ray energies in the
source currently known as SGR 1900+14 \citep{mazets79a,mazets79b} in the SNR N 49, in the Large Magellanic Cloud (LMC). This kind of objects were
called Soft Gamma Repeaters (SGRs). Two years later, \citet{fahlman81} reported an X-ray pulsar (1E\,2259+586) in SNR CTB 109, with an X-ray luminosity larger than could be explained via its rotational power alone. It was thought that maybe a companion
star was accreting material onto the neutron star surface, but no direct or indirect sign of a binary system was observed. This object and the others following this behavior were called ``Anomalous X-ray Pulsars'' (AXPs). Only in the 1990's, \citet{duncan92,duncan95,thompson96} proposed that
the origin of the emission of SGRs (and later also of AXPs) was their magnetic energy, they were hence labelled as ``magnetars'' (see section
\ref{sec1.2} for details). This picture has been amplified with the discovery of low magnetic field magnetars
\citep{rea10,rea12,rea14}.

\section[Pulsars]{Pulsars}  
\label{sec1.2}

{\it Pulsars} (PSRs) are compact objects left over from the reminiscent core of a star which has exploded as a supernova (SN). Their density is
around $\sim 10^{14}$ g cm$^{-3}$ and have a mass of $\sim1-2M_\odot$. These quantities imply a radius of $\sim 10$ km. PSRs are the fastest rotators
known in the Universe with periods of $\sim 10^{-3}-10$ s and have associated dipole magnetic fields of $10^8-10^{15}$ G, also the highest known.
This make pulsars excellent astrophysical laboratories to study matter, hydrodynamics, electrodynamics, particle acceleration and radiation
processes under extreme conditions.

\subsection{The magnetic dipole model}

Two of the most important parameters of PSRs obtained by observations, generally in radio, X-rays and $\gamma$-rays in some cases, are the rotational period
$P$, and the period derivative $\dot{P}$. Using these parameters, we can deduce some important formulae from the magnetic dipole model. This model
assumes that the pulsar rotates in vacuum with frequency $\Omega$ ($\Omega=2\pi/P$), with a magnetic moment $\vec{m}$, and an angle $\alpha$
between the magnetic moment and the rotation axis. The magnetic moment for a pure rotating magnetic dipole is defined (e.g., \citealt{shapiro04})
\begin{equation}
\lvert \vec{m} \rvert=\frac{B_p R_{PSR}^3}{2},
\end{equation}
where $B_p$ is the dipolar magnetic field and $R_{PSR}$ is the radius of the PSR. The vector $\vec{m}$ is expressed as
\begin{equation}
\vec{m}=\frac{1}{2} B_p R_{PSR}^3({\bf e_\parallel}\cos \alpha+{\bf e_\perp}\sin \alpha \cos \Omega t+{\bf e'_\perp}\sin \alpha \sin \Omega t),
\end{equation}
where ${\bf e_\parallel}$ is the parallel component with respect the rotation axis and ${\bf e_\perp}$ and ${\bf e'_\perp}$ are perpendicular
and mutually orthogonal vectors. As the configuration of the vector changes in time, the energy radiated is given by the Larmor formula
\begin{equation}
\dot{E}=-\frac{2}{3 c^3}\lvert \ddot{\vec{m}} \rvert^2.
\end{equation}
being $c$ the speed of light. Substituting $\vec{m}$ in equation above, we obtain
\begin{equation}
\label{larmor}
\dot{E}=-\frac{B^2_p R^6 \Omega^4 \sin^2 \alpha}{6 c^3}.
\end{equation}

The total rotational energy of the PSR is given by
\begin{equation}
\label{energy}
E=\frac{1}{2}I \Omega^2=\frac{2 \pi^2 I}{P^2},
\end{equation}
where $I$ is the moment of inertia of the PSR. This value is typically assumed as $\sim 10^{45}$ g cm$^2$. The time derivative of equation
(\ref{energy}) is
\begin{equation}
\label{edot}
\dot{E}=I \Omega \dot{\Omega}=-\frac{4 \pi^2 I \dot{P}}{P^3}
\end{equation}
$\dot{E}<0$ as we have seen in equation (\ref{larmor}), so $\dot{\Omega}<0$. Now, we define the {\it characteristic age} of the PSR $\tau_c$ as
\begin{equation}
\tau_c=-\frac{P}{2\dot{P}}=-\frac{1}{2} \left(\frac{\Omega}{\dot{\Omega}} \right)_{now}
\end{equation}
where the subindex $now$ means at the present time. Combining equation (\ref{larmor}) with the latter, we find
\begin{equation}
\tau_c=\frac{3 I c^3}{B_p^2 R_{PSR}^6 \Omega_t^2 \sin^2 \alpha},
\end{equation}
and now we can integrate equations (\ref{larmor}), (\ref{energy}) and (\ref{edot}) to solve the rotation frequency evolution of the magnetic
dipole
\begin{equation}
\Omega(t)=\Omega_0 \left[1+\left(\frac{\Omega_0}{\Omega_{now}} \right)^2 \frac{t}{\tau_c} \right]^{-1/2}.
\end{equation}
In terms of the period,
\begin{equation}
\label{period}
P(t)=P_0 \left[1+\left(\frac{P_{now}}{P_0} \right)^2 \frac{t}{\tau_c} \right]^{1/2}.
\end{equation}
We can compute the age of the pulsar from equation (\ref{period}) just setting $P(t)=P_{now}$. Thus,
\begin{equation}
t_{age}=\tau_c \left[1-\left(\frac{P_0}{P_{now}} \right)^2 \right].
\end{equation}
Note that if $P_0 \ll P_{now}$, then $t_{age} \simeq \tau_c$. Using the solution for the rotation frequency, we can rewrite the spin-down
luminosity in terms of the initial period, period and period derivative
\begin{equation}
\dot{E}=\frac{2 \pi^2 I}{\tau_c P_0^2} \left(\frac{P_{now}}{P_0} \right)^2 \left[1+\frac{t}{\left(\frac{P_0}{P_{now}} \right)^2 \tau_c} \right]^{-2},
\end{equation}
or to simplify,
\begin{equation}
\label{edotevol}
\dot{E}=\dot{E}_0 \left(1+\frac{t}{\tau_0} \right)^{-2},
\end{equation}
where $\tau_0$ is the initial spin-down age and $\dot{E}_0$ is the initial spin-down luminosity.

Note that for a purely rotating magnetic dipole, the angular frequency evolution is ruled by an equation of the kind
$\dot{\Omega} \propto \Omega^3$. For some pulsars, the rotation evolves with a different power index $n$, also called the {\it braking index}
and defined as
\begin{equation}
n=-\frac{P\ddot{P}}{\dot{P}}.
\end{equation}
Under this condition, the period evolution is
\begin{equation}
P(t)=P_0 \left[1+\frac{n-1}{2} \left(\frac{P_{now}}{P_0} \right)^{n-1} \frac{t}{\tau_c} \right]^\frac{1}{n-1},
\end{equation}
the age of the pulsar,
\begin{equation}
\label{tage}
t_{age}=\frac{2 \tau_c}{n-1} \left[1-\left(\frac{P_0}{P_{now}} \right)^{n-1} \right].
\end{equation}
and the spin-down evolution,
\begin{equation}
\label{edotevol2}
\dot{E}=\dot{E}_0 \left(1+\frac{t}{\tau_0} \right)^{-\frac{n+1}{n-1}},
\end{equation}
with
\begin{equation}
\dot{E}_0=\frac{2 \pi^2 I}{\tau_c P_0^2} \left(\frac{P_{now}}{P_0} \right)^{n-1}\qquad \tau_0=\frac{2}{n-1} \left(\frac{P_0}{P_{now}} \right)^{n-1} \tau_c.
\end{equation}
A relation for $\tau_0$ and $t_{age}$ can be derived using equation (\ref{tage}), such that
\begin{equation}
\label{tau0}
\tau_0=\frac{2\tau_c}{n-1}-t_{age}.
\end{equation}

Another important physical property is the polar magnetic field $B_p$. Its expression can be easily obtained from equations (\ref{larmor}) and
(\ref{edot}) giving
\begin{equation}
\label{bdip}
B_p=\sqrt{\frac{3 I c^3}{2 \pi^2 R_{PSR}^6} P \dot{P}} \simeq 6.4 \times 10^{19} \sqrt{P \dot{P}}\ \textrm{G}.
\end{equation}
It is possible to find in the literature that the magnetic field is a factor 2 lower ($B_e \sim 3.2 \times 10^{19} \sqrt{P \dot{P}}$). This
difference depends on where we define the magnetic moment. Here, we are defining the magnetic moment in the pole, but if we do it in the equator,
then we find this difference of a factor 2 ($B_p=2B_e$).

\subsection{Electric potential and $\sigma$-parameter}

According to the current picture, a charge-filled magnetosphere surrounds the PSR and the particle acceleration occurs in charged gaps in outer
regions that extend to the light cylinder (defined as $R_{LC}=c/P$). The first magnetosphere model was proposed by \citet{goldreich69} and they
calculated the maximum electric potential generated by an aligned rotating magnetic field (i.e, magnetic and spin axes co-aligned). The
expression is
\begin{equation}
\label{elecpot}
\Delta V=\frac{B_p \Omega^2 R_{PSR}^3}{2c}.
\end{equation}
The associated particle current is $\dot{N}=(\Omega^2 B_p R_{PSR}^3)/c Z e$, where $Z e$ is the ion charge.

The magnetization parameter ($\sigma$) is the ratio between the Poynting flux and the particle energy flux. This parameter is defined by
\citet{kennel84a}:
\begin{equation}
\label{sigma}
\sigma=\frac{B^2_p}{4 \pi \rho \gamma m c^2},
\end{equation}
where $\rho$ is the number density of particles, and $\gamma m c^2$ is the energy of each particle. This ratio is expected to be dominated by the
Poynting flux term as the wind flows from the light cylinder ($\sigma>10^4$, see \citealt{arons02}), but for the structure of the Crab Nebula, we
need $\sigma \ll 1$ just behind the termination shock in order to meet flow and pressure boundary conditions at the outer edge of the PWN
\citep{rees74,kennel84a}. The particle-dominated wind is also required by the high ratio of the synchrotron luminosity to the total spin-down
power \citep{kennel84b}, and implies $\gamma \sim 10^6$, a value considerably higher than that expected in the freely expanding wind
\citep{arons02}. This change in the conditions of the pulsar wind between the termination shock and the light cylinder is still unclear (see
\citealt{melatos98,arons02}), and is known as the ``{\it $\sigma$ problem}''.

\subsection{What do we observe?}

Pulses of neutron stars are commonly detected at radio frequencies. This pulses are very stable which allow to measure the period of the sources
with very high precision ($\delta P \sim 10^{-10} s$). Despite the number of PSRs detected in radio, the physical mechanism that generates the
coherent radio emission is not well understood. Thanks to this precision, long observations of these objects allow to measure also the
period derivative ($\dot{P}$) and, in some cases, the second derivative of the period ($\ddot{P}$) and with it, the braking index.

\begin{figure}
\centering
\includegraphics[width=0.6\textwidth]{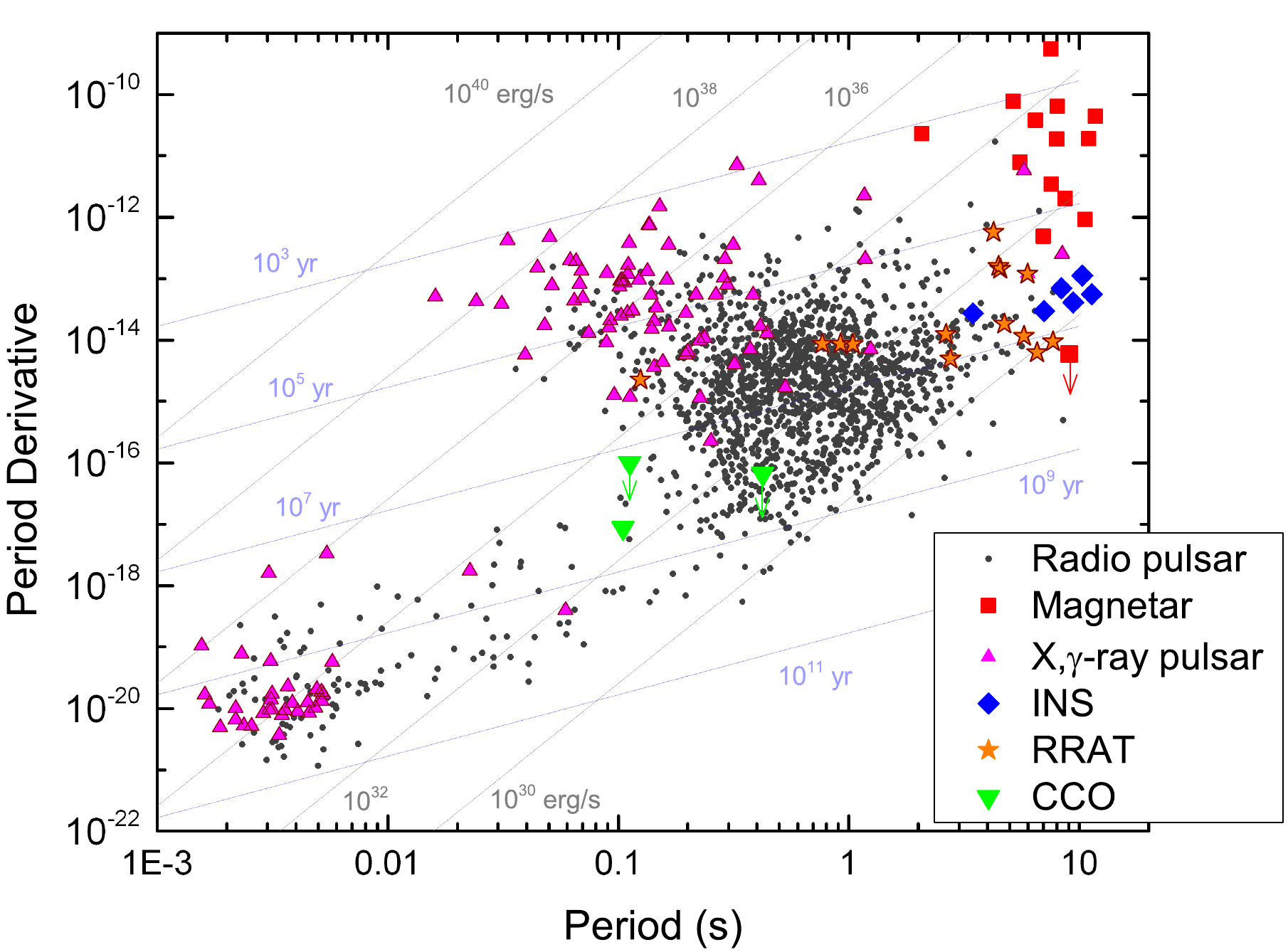}
\caption[$P \dot{P}$-diagram]{$P \dot{P}$-diagram for the known rotation-powered pulsars, isolated neutron stars (INS), central compact objects
(CCO), rotating radio transients (RRAT) and magnetars \protect\citep{harding13}. The lines of constant characteristic age and spin-down
luminosity are superposed.}
\label{ppdot}
\end{figure}

When we plot the location of the PSRs in the period and period derivative phase space, we generate the so-called {\it $P \dot{P}$-diagram} (see
figure \ref{ppdot}). In this plot we can distinguish different populations. In the lower-left part of the diagram, we find the recycled PSRs.
Despite the large characteristic age of these PSRs, they have very low periods ($P \lesssim 0.01$ s). Recycled PSRs are in binary systems where
they have been spun-up due to the presence of an accretion disk which transfers angular momentum to the PSR. The center of the diagram is
dominated by middle-aged pulsars. Regarding the band where we detect the pulses, we find radio, X-ray and $\gamma$-ray PSRs. Finally,
{\it magnetars} are located in the upper-right part. Neutron stars with a thermal spectrum and with no signals of pulsations (or just hints), are
referred as {\it Central Compact Objects} (CCO).

In some PSRs, we detect sudden spin-ups called glitches. It is thought that glitches are produced by superfluid neutron vortices in the crust
\citep{anderson75}.

Regarding other wavelengths as the ultraviolet, optical and infrared, NS are very faint at these energies, but some counterparts has been
identified (e.g., \citealt{mignani12b}. Optical observations are useful to determine the presence of debris disks in isolated neutron stars and
their good resolution allows to measure proper motions and parallaxes to determine distances.

\subsection{Magnetars}

As we have explained in section \ref{sec1.1}, magnetars are a class of pulsars which show high energy transient burst and flaring activity with
luminosities higher than the spin-down luminosity. Nowadays, we know $\sim$24 magnetars \citep{olausen14}. They are characterized by having long
periods ($P > 1$ s) and associated dipolar magnetic fields greater than $B_{crit}=m_e^2 c^3/e \hbar \sim 4.4 \times 10^{13}$ G, which is the
elecron critical magnetic fields at which the cyclotron energy of an electron reaches the electron rest mass energy. The latter characteristic is
now misleading since, in the last years, magnetar-like behaviour has been discovered also in low-magnetic X-ray pulsars \citep{rea10,rea12,rea14}.
Regarding the flaring and bursting activity, it might involve their emission from radio to hard X-rays, with an increase of the soft X-ray flux
of a factor between 10 and 1000 with respect the flux in quiescense. The decay timescales of this flux is very varied raging from weeks to years.
The same happens with the decay of the light curve, which can be characterized by an exponential or a power-law function \citep{rea11}.

The exact mechanism playing a key role in the formation of such strong magnetic fields is currently debated; in particular it is not clear which
are the characteristics of a massive star turning into a magnetar instead of a normal radio pulsar, after its supernova explosion. Preliminary
calculations have shown that the effects of a turbulent dynamo amplification occurring in a newly born neutron stars can indeed result in a
magnetic field of a few $10^{17}$ G. This dynamo effect is expected to operate only in the first $\sim$10 s after the supernova explosion of the
massive progenitor, and if the proto-neutron star is born with sufficiently small rotational periods (of the order of 1-2 ms). The resulting
amplified magnetic fields are expected to have a strong multipolar structure, and toroidal component \citep{duncan92,thompson93,duncan96}.

However, this scenario is encountering more and more difficulties: i) if magnetic torques can indeed remove angular momentum from the core via
the coupling to the atmosphere in a pre-SN phase, then the core soon after the SN might not spin rapidly enough for this convective dynamo
mechanism to take place \citep{heger05}; ii) such a fast spinning proto-neutron stars would require a supernova explosion one order of magnitude
more energetic than normal supernovae, possibly an hypernova, which is yet not clear on whether it can indeed form a neutron star instead of a
black hole. Recent simulations have shown that GRBs and hyper-luminous supernovae can indeed be powered by recently formed millisecond magnetars
\citep{metzger11,bucciantini12}, although no observational evidence of the existence of such fast spinning and strongly magnetized neutron stars
have been collected thus far.

Besides the fast spinning proto-neutron star, a further idea on the origin of these high magnetic fields is that they simply reflect the high
magnetic field of their progenitor stars. Magnetic flux conservation \citep{woltjer64} implies that magnetars must then be the stellar remnants
of stars with internal magnetic fields of $B>1$ kG, whereas normal radio pulsars must be the end products of less magnetic massive stars.

Recent theoretical studies showed that there is a wide spread in white dwarf progenitor magnetic fields \citep{wickramasinghe05}, which, when
extrapolated to the more massive progenitors implies a similar wide spread in neutron stars progenitors \citep{ferrario06}. Hence, apparently it
seems that a fossil magnetic field might be the solution of the origin of such strongly magnetized neutron stars, without the need of invoking
dynamo actions on utterly fast spinning proto-neutron stars.

However, this lead to the problem of the formation of such high $B$ progenitor stars. The most common idea is that the magnetic field in the star
reflects the magnetic field of the cloud from which the star is formed. The best studied very massive stars (around $\sim$40$M_\odot$) with a
directly measured magnetic field are $\theta$ Orion C and HD191612, with dipolar magnetic field of 1.1 kG and 1.5 kG, respectively
\citep{donati02,donati06}. Very interestingly, the magnetic fluxes of both these stars ($1.1 \times 10^{27}$ G cm$^2$ for $\theta$ Orion C and
$7.5 \times 10^{27}$ G cm$^2$ for HD191612) are comparable to the flux of the highest field magnetar SGR 1806$-$20 ($5.7 \times 10^{27}$ G
cm$^2$; \citealt{woods06}). Other high magnetic field stars are reported in \citet{oskinova11}.

Recent observations of the environment of some magnetars revealed strong evidence that these objects are formed from the explosion of very
massive progenitors ($M>30M_\odot$). In particular: i) a shell of HI has been detected around 1E 1048.1-–5937, and interpreted by ISM displaced
by the wind of a progenitor of 30–40$M_\odot$ \citep{gaensler05}; SGR 1806$-$20 and SGR 1900+14 have been claimed to be a member of very
young and massive star clusters, providing a limit on their progenitor mass of $>50M_\odot$ \citep{fuchs99,figer05,davies09} and $>20M_\odot$
\citep{vrba00}. Finally, CXOU 010043$-$7211 it is a member of the massive cluster Westerlund 1 \citep{muno06,ritchie10}, with a progenitor with
mass estimated to be $>40M_\odot$ (see also \citealt{clark14}).

In chapter \ref{chap6}, we study the X-ray spectrum of some SNRs with an associated magnetar to check some of these statements and compare their
spectra with the features found in SNRs with associated normal pulsars. Other important questions and observational properties are discussed in
some reviews \citep{woods06,mereghetti08,kaspi10,rea11,olausen14}.

\section[Pulsar Wind Nebulae]{Pulsar Wind Nebulae}  
\label{sec1.3}

In addition to their electromagnetic emission, PSRs dissipate their rotational energy via relativistic winds of particles. Because the
relativistic bulk velocity of the wind is supersonic with respect to the ambient medium, such a wind produces a termination shock. In turn, the
wind particles, moving trough the magnetic field and the ambient photons, produce radiation that we observe as pulsar wind nebulae. As the
pulsars themselves, the PWN emits at all wavelengths from radio to TeV energies.

\subsection{Morphology}

\begin{figure}
\centering
\includegraphics[width=0.5\textwidth]{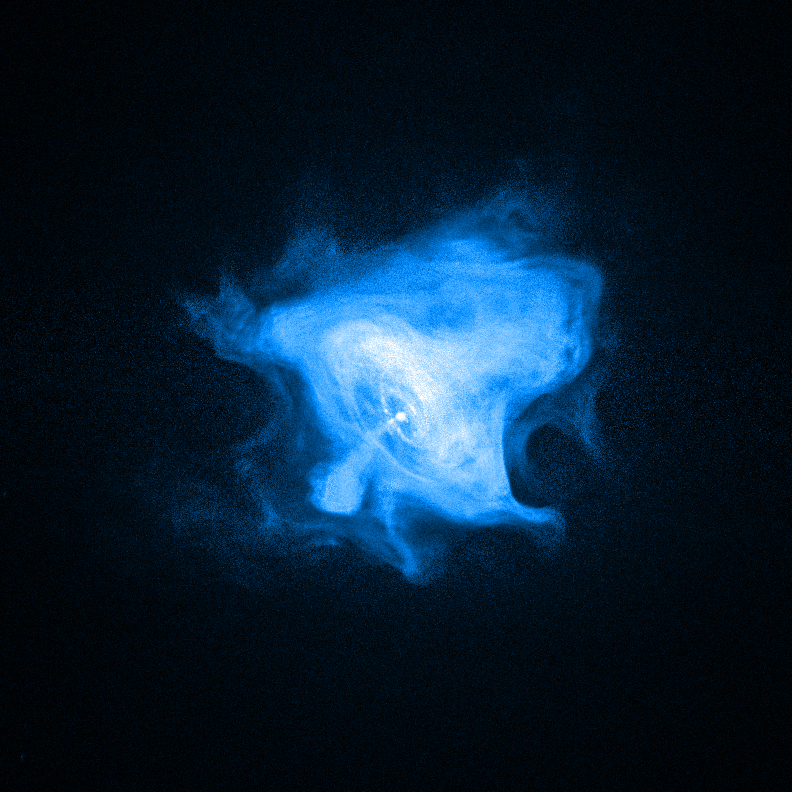}
\caption[Crab Nebula in X-rays]{Crab Nebula in X-rays seen by the Chandra X-ray Observatory.}
\label{crabxray}
\end{figure}

PWNe usually have two main X-ray morphologies, depending on the velocity of the pulsar proper motion and the ambient medium. A classic example is
the Crab Nebula (see figure \ref{crabxray}) For slow pulsars, images taken with the Chandra X-ray Observatory (see e.g., \citealt{kargaltsev08})
show a toroidal shape around the pulsar equator, with two possible jets starting from the pulsars poles. Instead, pulsars moving with high
velocity in the interstellar medium produce on average PWNe with the characteristic bullet-like or bow-shock morphology, with the tail developed
along the pulsar motion. Thus, the study of PWNe can lead to knowledge of pulsar winds, the properties of the ambient medium, and the wind-medium
interaction.

The pulsar wind expands in its wound-up toroidal magnetic fields and it is confined by the expanding shell of the SN ejecta. As the wind
decelerates to match the boundary condition imposed by the slowly-expanding SN material at the nebula radius, a wind termination shock is formed
with radius $R_w$ such that (e.g., \citealt{gelfand09}):
\begin{equation}
R_w=\sqrt{\frac{\dot{E}}{4 \pi \chi c P_{PWN}}}
\end{equation}
where $\chi$ is the equivalent filling factor for an isotropic wind ($\chi=1$ when the wind is isotropic). $P_{PWN}$ is the pressure of the gas in
the PWN interior. High resolution X-ray observations have shown the ring-like emission from the termination shock of the Crab Nebula, but it has
not been detected in other cases as, for example, 3C 58 \citep{slane02b}, G21.5–0.9 \citep{camilo06} and G292.0+1.8 \citep{hughes01}.

In the case of the Crab Nebula (also similar for 3C 58), the X-ray morphology consists in a tilted torus with jets along the toroid axis (see
figure \ref{crabxray}). The jets extend nearly 0.25 pc from the PSR \citep{gaensler06}. A faint counter-jet accompanies the structure and the
X-ray emission is significantly enhanced along one edge of the torus presumably as the result of Doppler beaming of the outflowing material.

Chandra X-ray observations revealed a similar structure for G54.1+0.3 with a point-like central source surrounded by an X-ray ring with an
inclination of 45$^\circ$ \citep{lu02}. In the eastern part of the ring, the X-ray emission is brighter. Jets of shocked material are also
observed aligned with the projected axis of the ring \citep{bogovalov05}. An important difference of this case with the Crab Nebula or 3C 58 is
that the X-ray flux contribution of the torus and the jets is similar, when for the latter two, the central torus is brighter by a large factor.

Some effort has been done in order to understand the formation of these structures. Modeling of the flow conditions across the shock shows that
magnetic collimation produces jet-like flows along the rotation axis \citep{komissarov04,bogovalov05}. The collimation of the jet is highly
dependent on the magnetization of the wind. For $\sigma \gtrsim 0.01$, magnetic hoop stresses are sufficient to divert the toroidal flow back
toward the pulsar spin axis, collimating and accelerating the flow to speeds of $\sim 0.5c$ \citep{delzanna04}. Smaller values of the
magnetization allow to increase the radius at which the flow is diverted. Near the poles, $\sigma$ is large, resulting in a small termination
shock radius and strong collimation, while near the equator, it is much smaller and the termination shock is larger \citep{bogovalov02}.

Note that all these structures and time variability are also observed at other wavelengths (radio \& optical) indicating that the acceleration of
the associated particles must have the same origin as for the X-ray-emitting population \citep{bietenholz04}.

\begin{figure}[t!]
\centering
\includegraphics[width=1.0\textwidth]{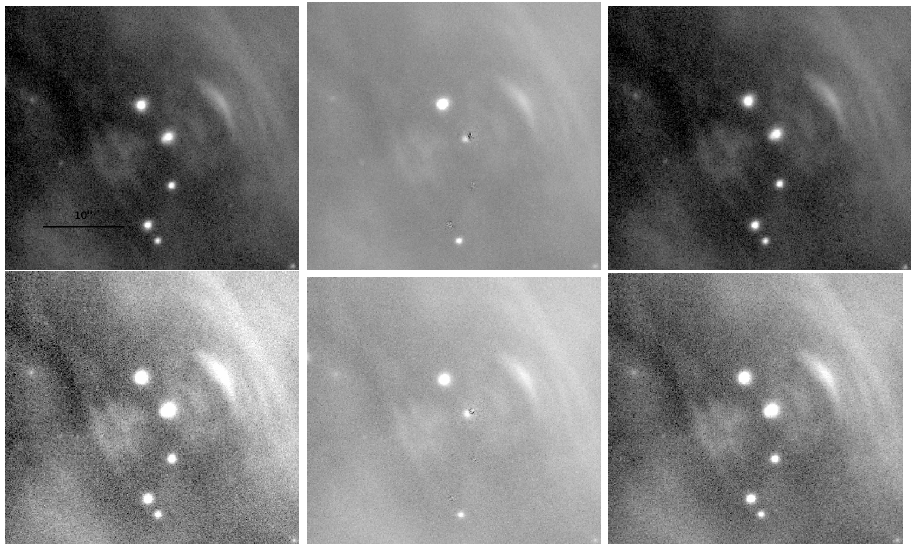}
\caption[Crab Nebula knots in infrared]{Time variability of the knots and wisps near the Crab Nebula PSR observed in the $H$ band (upper panel)
and $K_s$ band (lower panel) by the Nordic Optical Telescope (NOT). The images were taken December 13th 2007 and September 30th 2007
\protect\citep{tziamtzis09}.}
\label{knots}
\end{figure}

The filaments formed by Rayleigh-Taylor instabilities are also observed in the optical for the Crab Nebula \citep{hester96}. MHD simulations
indicate that 60$-$75\% of the swept-up mass can be concentrated in such filaments \citep{bucciantini04b}. These filaments compress the expanding
bubble and increase the magnetic field forming sheets of enhanced synchrotron emission \citep{reynolds88a}, but this emission is not detected
in the Crab Nebula \citep{weisskopf00}.

Loop-like filaments are observed in 3C 58 in X-rays \citep{slane04b} and they are coincident with those observed in radio \citep{reynolds88b}.
Optical fainter filaments are also observed \citep{vandenbergh78} and their origin seems similar as in the Crab Nebula, but they are not coincident
with the X-ray ones, revealing that the mechanisms of formation must be different. \citet{slane04b} proposed that the bulk of the discrete structures
seen in the X-ray and radio images of 3C 58 are magnetic loops torn from the toroidal field by kink instabilities.

Finally, there are time variable structures which appear and disappear on timescales of months as the compact knots near the PSR observed in the
Crab Nebula \citep{tziamtzis09} and others, as for  PSR B1509-58 \citep{gaensler02}. It is believed that they actually correspond to unstable,
quasi-stationary shocks in the region just outside the termination shock, at high latitudes where the shock radius is small due to larger values of
$\sigma$ (e.g., \citealt{komissarov04}).

\subsection{Spectrum characteristics}

Pulsar wind nebulae emit radiation from ratio to TeV energies. The most observed PWN is, by far, the Crab Nebula, for which we find detailed
observations in the whole electromagnetic spectrum. From radio to X-rays, the emission consist in synchrotron radiation coming from the particles
accelerated by the magnetic field of the nebula. Depending on the density of the magnetic field and the lifetime of the particles, we can get an
idea about the frequency where the cooling cut-off is located in the synchrotron spectrum \citep{ginzburg65}:
\begin{equation}
\nu_b=10^{21} \left(\frac{B_{PWN}}{10^{-6}\,\textmd{G}} \right)^{-3} \left(\frac{t}{1\,\textmd{kyr}} \right)^{-2} \textmd{Hz}
\end{equation}

Using a $\delta$-approximation for the synchrotron cross-section \citep{ginzburg79}, it is easy to relate the energy of synchrotron photons
($E_{sync}$) with the energy of the electrons that produce the radiation itself ($E_e$)
\begin{equation}
\label{esync}
E_{sync}=\frac{h \nu_c}{3},
\end{equation}
where $h$ is the Planck constant and $\nu_c$ is the so-called {\it critical frequency} which depends on the magnetic field and the energy of the
electrons
\begin{equation}
\label{critfreq}
\nu_c=\frac{3 e B(t) E_e^2}{4 \pi m_e^3 c^5}.
\end{equation}

Particles radiating photons beyond this frequency decrease rapidly their energy before reaching the outer shell of the PWN. Observationally, we
see a decreasing radius of the nebula as we increase the frequency as we observe in Crab Nebula (e.g., \citealt{hillas98}). At lower magnetic
fields, this effect is less important, because the synchrotron loss time is longer. It is also detected in some PWNe, an infrared and optical
excess and some recombination lines in the spectrum \citep{hester96} coming from thermal radiation produced by the filament structures
surrounding the PWN.

Generally speaking, the PWN radio spectra are characterized by a flat power-law index at radio wavelengths ($\alpha \approx -0.3$). We find an
intrinsic energy break at infrared/optical frequencies and the slope changes for the X-ray emission ($\Gamma \approx 2$). The nature of this
spectral shape is still not understood. Relic breaks in the spectrum can be produced by a rapid decline in the pulsar output over time, and these
breaks propagate to lower frequencies as the PWN ages \citep{pacini73}. In cases where we can observe the spectrum with radial resolution, it is
detected radial steepening in the spectrum \citep{slane00,willingale01,slane04b}. This steepening is less than expected in the outer part of the
nebula \citep{kennel84b,reynolds03}, but diffusion processes can be involved in some mixing of electrons with different ages at each radius.

\begin{figure}
\centering
\includegraphics[width=0.6\textwidth]{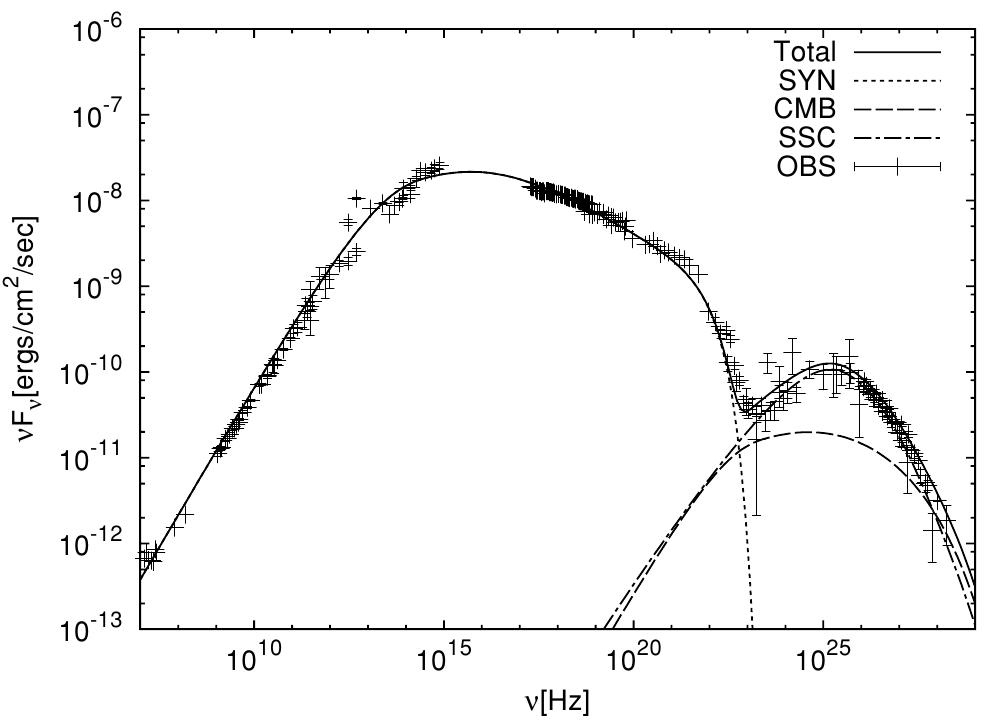}
\caption[An example of PWN spectrum]{The most complete example of PWN spectrum, the Crab PWN. From radio to X-rays, the emission is well
described by synchrotron emission, and from hard X-rays to VHE, the spectrum is described by the IC scattering of the electrons with the
background photon fields. Image taken from \citet{tanaka10}.}
\label{pwn_spectrum}
\end{figure}

From X-rays to VHE, the radiation produced comes basically from inverse Compton (IC) interaction of the high energy pairs with the low energy
photons of the ambient medium. Three main target photon fields are considered in the current spectral models: the cosmic microwave background
(CMB), the far-infrared contribution coming from the galactic ISM (FIR) and the infrared/optical contribution coming from the surrounding stars
(NIR). Using the aproximation done by \citet{ginzburg79} for the IC cross-section in the Thompson limit, the IC photons energy ($E_{IC}$) and
the energy of the electron is related as
\begin{equation}
\label{eic}
E_{IC}=\frac{4}{3}h \nu_i \left(\frac{E_e}{m_e c^2} \right)^2,
\end{equation}
where $\nu_i$ is the energy of the target electrons. We can relate equations (\ref{esync}) and (\ref{eic}) to connect the characteristic energies
of the synchrotron and IC photon produced by the same electrons. Assuming the CMB as the only contributor for the IC emission, we obtain
\citep{aharonian97}
\begin{equation}
E_{sync} \simeq 0.07 \left(\frac{E_{IC}}{1\,\textmd{TeV}} \right) \left(\frac{B}{10\,\mu G} \right)\,\textmd{keV}.
\end{equation}
This relation is useful to obtain a first idea on the value of the magnetic field of the nebula from the spectrum. An example of a
multi-wavelength spectrum of a PWN is shown in figure \ref{pwn_spectrum}.

PWNe constitute the largest class of identified Galactic very high energy (VHE) $\gamma$-ray sources, with the number of TeV detected objects
increasing from 1 to $\sim$30. in the last years. These statistics shine in comparison with the $\sim$30, 10, or 40 PWNe known in radio,
optical/IR, or X-rays, respectively, detected in decades of observations.

Regarding the time evolution of the spectra, as the magnetic field diminishes with age, the synchrotron flux decreases and the IC component
becomes more important. This is because the electrons required to emit in hard X-rays are more energetic than those necessary to emit in TeV by
IC, thus when these very high energy electrons lose energy, they increase the electron population which contributes to the TeV emission via IC.
This is one of the reasons we have detected middle-aged PWNe in $\gamma$-rays that have not been detected in other wavelengths.

\subsection{Current models}

In studying PWNe, there are two distinct theoretical approaches. On one hand, detailed magnetohydrodynamic (MHD) simulations have succeeded in
explaining the morphology of PWNe. On the other hand, spherically symmetric 1D PWNe spectral models, with no energy-dependent morphological
output, have been constructed since decades. We will review some of these models and others in section \ref{sec1.3}.

There has been a great effort trying to reproduce the spectral and morphological features of PWNe. The first model for the Crab Nebula was
proposed by \citet{rees74}. Ten years later, \citet{kennel84a,kennel84b} did a further step introducing an analytical model with a magnetic field
with radial dependence. The diffusion-loss equation was solved analytically by \citet{syrovatskii59} applied to the distribution of relativistic
electrons in the Galaxy (explained in detail in section \ref{sec2.1}). The solution of this equation was calculated considering no
time-dependence on the energy losses or the magnetic fields. \citet{atoyan96} and \citet{aharonian97} applied the same equation to PWNe but
neglecting the diffusion term to study the IC radiation from these sources. In the latter models, the $\gamma$-ray and VHE radiation is produced
by pairs, but some others proposed that there could be an important contribution from ions by pion decay \citep{bednarek03,bednarek05,li10}.

Time-dependent models have been presented lately (e.g., \citealt{bednarek03,bednarek05,busching08,zhang08,fang10a,fang10b,li10,tanaka10,
bucciantini11,tanaka11,vanetten11,martin12,tanaka13,torres13a,torres13b,vorster13c,torres14}). \citet{zhang08} integrate the energy loss
equation considering only the synchrotron and Bohm diffussion lifetimes of the particles to obtain an analytical solution for the electron
population varying the magnetic field with time. \citet{fang10a} extrapolate the electron injection function fitted by \citet{spitkovsky08} from
numerical simulations of collisionless shocks in unmagnetized plasmas. \citet{tanaka10,tanaka11,tanaka13} integrate numerically the energy loss
equation for pairs taking into account the energy losses of the particles, but neglecting escape time term. The magnetic field is not time
parametrized as in other works (e.g., \citealt{venter07}), but it is calculated by magnetic field energy conservation. \citet{vanetten11}
included radial dependence and the diffusion term to calculate the electron population. \citet{vorster13b} also studied in detail the effect of
the diffussion in the electron population in PWNe. In \citet{martin12,torres13a,torres13b,torres14}, the energy loss equation is integrated
considering the energy losses and the escape terms in a time-dependent way and taking into account adiabatic losses for the magnetic field
calculation as in \citet{pacini73}.

Due to the complexity of the problem, the majority of these models are only capable to reproduce the first stage of the evolution (i.e, free
expansion phase) with high precision. Other caveat is the morphology, which is completely neglected. Some models have been more dedicated to
reproduce better the dynamical evolution of the system. For example, \citet{vanderswaluw01} proposed an analytical model to study the PWN
evolution during the Sedov phase of the SNR. \citet{chevalier82,blondin01} studied the interaction of the PWN with the reverse shock of the SNR
using numerical simulations. \citet{gelfand09,fang10b,bucciantini11,vorster13c} combined the works done by \citet{chevalier82,blondin01} with
simple spectral models to study the evolution of the spectrum and applied to some particular cases.

In order to understand better the magnetic configuration of a PWN and its morphology, several works have been focused to reproduce the X-ray
morphology of the Crab Nebula using MHD multidimensional time-dependent models (e.g., \citealt{delzanna04,komissarov04,vanderswaluw04b,
delzanna06,volpi08}).

\section[PWN-SNR complex evolution]{PWN-SNR complex evolution}  
\label{sec1.4}

After the SN explosion, the ejected mass expands throught the interstellar material. The central PSR starts to accelerate particles inside the
termination shock and these particles interact with the material inside the SNR. Due to the explosion, the PSR has a certain kick velocity and
moves outside its proper PWN and SNR. The typical energy that a PSR injects into the PWN during its lifetime is only $\sim$1\% of the SN
explosion ($10^{51}$ erg). Therefore, the presence of an energetic PSR has little effect on the global evolution of the SNR, but the evolution
of the PWN strongly depends on its interaction with the SNR. In this section we will explain briefly the different evolution stages of the
system. Note that this scheme is a simplification when we study particular SNRs, since the SN explosion and the ISM density have important
asymmetries and different parts of the system may be in different phases. In any case, this provides a useful framework to have a first idea on
how these systems evolve.

\subsection{Free expansion phase}

\begin{figure}
\centering
\includegraphics[width=0.75\textwidth]{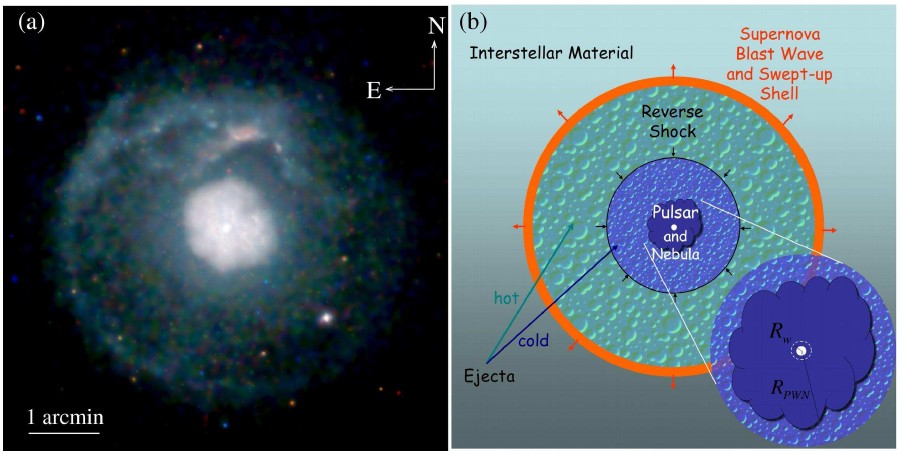}
\caption[Free expansion stage scheme]{Free expansion stage scheme \citep{gaensler06}. Left image: Composite SNR G21.5-0.9 image obtained by
Chandra. This is a canonical example of a free expansion phase system. Right image: Free expansion stage scheme. The SN blast wave expands
through the ISM and sweeps up the material, increasing the mass in the shock. The reverse shock heats the material of the interior of the SNR.
The PWN expands supersonically inside the SNR and the PSR remains in the center due to the low kick velocity in comparison with the expansion
velocity of the nebula.}
\label{free_expansion}
\end{figure}

At the beginning, the mass ejected by the SN explosion ($M_{ej}$) sweeps up the ISM ($M_{sw}$) like a piston with a constant velocity
\citep{taylor46}. The shock wave moves at a speed of $>(5-10) \times 10^3$ km s$^{-1}$, while asymmetry in the SN explosion gives the pulsar a
random space velocity of typical magnitude 400-500 km s$^{-1}$ and stars to move from the center. In this phase, $M_{ej}$ dominates over
$M_{sw}$. As $t \ll \tau_0$, we can consider that the PWN has constant energy input such that $L(t) \simeq L_0$ (we recall $\dot{E}$ to
$L(t)$ for convenience) (see equation \ref{edotevol}). The pulsar wind is highly over-pressured with respect to its environment, and the PWN
thus expands rapidly, moving supersonically and driving a shock into the ejecta. In the spherically symmetric case, the radius of the PWN evolves
as (e.g., \citealt{vanderswaluw01}):
\begin{equation}
\label{pwnrad}
R_{PWN}=C \left(\frac{L_0 t}{E_{SN}} \right)^{1/5} V_0 t,
\end{equation}
where $E_{SN}$ is the energy of the supernova explosion and $V_0$ is the velocity of the SN ejecta at the center of the explosion. $V_0$ is
calculated assuming that the energy $E_{SN}$ is converted into kinetic energy and a uniform density medium. Its expression is then
\begin{equation}
\label{v0}
V_0=\sqrt{\frac{10 E_0}{3 M_{ej}}}.
\end{equation}
The numerical constant $C$ depends on the adiabatic coefficient of the pulsar wind $\gamma_{PWN}$,
\begin{equation}
C=\left(\frac{6}{15(\gamma_{PWN}-1)}+\frac{289}{240} \right)^{-1/5},
\end{equation}
which in this case is $\gamma_{PWN}=4/3$, since the gas is relativistically hot. Because the PWN expansion velocity is steadily increasing, and
the sound speed in the relativistic fluid in the nebular interior is $c/\sqrt{3}$, the PWN remains centered on the pulsar. There are some systems
discovered at this stage, but a good example would be the composite SNR G21.5–0.9 with the PSR J1833-1034 (see figure \ref{free_expansion}).

For the SNR, as the swept-up mass becomes comparable to the ejected mass, two effects become important. The first one is that the pressure
difference between the shocked ISM and the ejecta drives a shock wave into the ejecta due to the low pressure in the ejected material which has
been adiabatically expanding. This is the so-called reverse shock. The reverse shock wave is the beginning of the deceleration of the supernova
ejecta, which leads to the second effect: the Rayleigh-Taylor instability at the interface between the dense shell and the ambient medium. At
this moment, significant deceleration is expected and the SNR evolution follows the self-similar adiabatic blast wave solution for a point
explosion in a uniform medium described by \citet{sedov59}. The PWN expansion will be affected later when the reverse shock collides with the
front shock of the pulsar wind.

\subsection{Sedov phase}

As we explained in the previous section, when $M_{sw}>M_{ej}$, the expansion of the SNR follows the self-similar solution given by
\citet{sedov59}. It assumes that the energy of the SN explosion is injected into the ISM instantaneously with an uniform density $\rho_0$. As in
the free expansion phase, radiative energy losses are neglected. A simple formula for the evolution of the SNR is obtained
\begin{equation}
R_{SNR}(t)=\left(\xi \frac{E_{SN} t^2}{\rho_0} \right)^{1/5},
\end{equation}
with $\xi=2.026$ for a non-relativistic, monatomic gas ($\gamma_{ej}=5/3$). The Sedov-Taylor solution can be generalized to a gas medium with a
power-law density profile $\rho(r) \propto r^{-s}$, $R_s \propto t^\beta$, $V_s=\beta R_s/t$, with the parameter $\beta=2/(5-s)$. A SNR shock
moving through the progenitor's stellar wind corresponds to the case where $s=2$ ($\beta=2/3$). A physical example would be Cas A (e.g.
\citealt{vanveelen09}), where the observed value in x-rays of $\beta$ is $0.63 \pm 0.02$ \citep{vink98,delaney03,patnaude09}.

Regarding the reverse shock, firstly it expands outwards behind the forward shock and separated by the contact discontinuity where the
Rayleigh-Taylor instabilities are produced, but eventually the reverse shock moves inwards. The reverse shock reaches the center of the SNR in a
characteristic timescale assuming that there is neither PWN nor PSR. This timescale is \citep{reynolds84}
\begin{equation}
t_{Sedov} \approx 10 \left(\frac{M_{ej}}{15M_\odot} \right)^{5/6} \left(\frac{E_{SN}}{10^{51}\,\textmd{erg}} \right)^{-1/2} \left(\frac{\rho_0}{1\,\textmd{cm}^{-3}} \right)^{-1/3} \textmd{kyr}
\end{equation}

\begin{figure}
\centering
\includegraphics[width=0.4\textwidth]{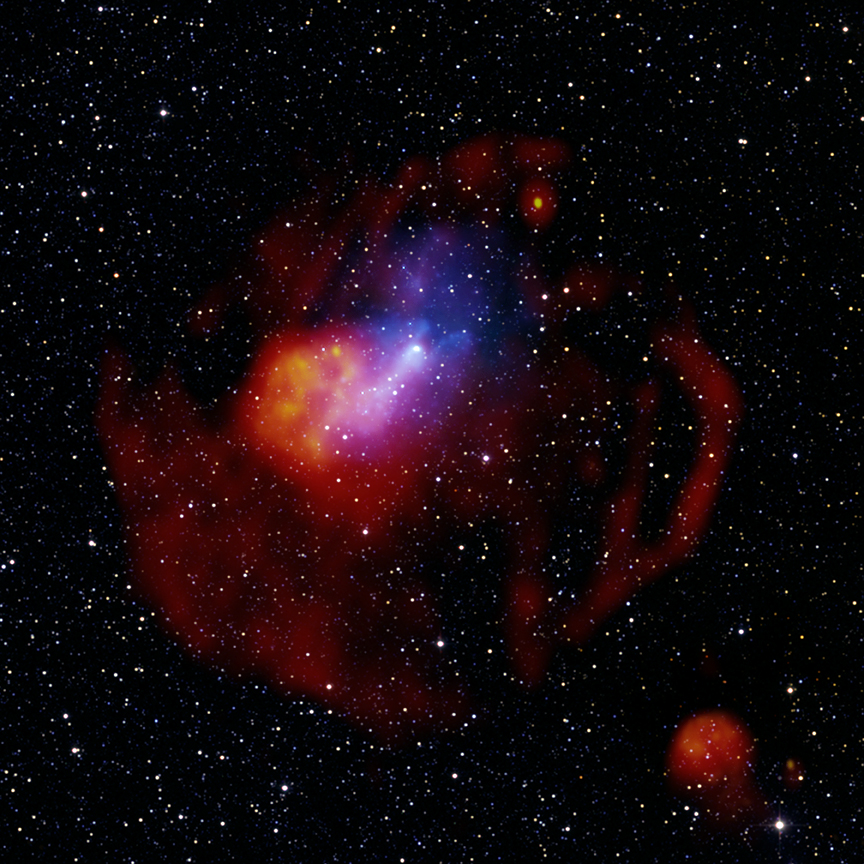}
\caption[Composite image of SNR G327.1-1.1]{Composite image of SNR G327.1-1.1 \citep{temim09}. The blue emission corresponds to the X-ray
emission of the PWN and the red emission is the radio emission obtained by 843 MHz Molonglo Observatory. The peak of the X-ray emission is
displaced from the center coming from the new particles injected by the pulsar. The radio emission corresponds to the original PWN (relic PWN).}
\label{sedov_phase}
\end{figure}

At this point the SNR interior is entirely filled with shock-heated ejecta, and the SNR is in a fully self-similar state that can be completely
described by a small set of simple equations \citep{cox72}. But, of course, this is not the case, and we are considering a young PWN and PSR
inside the remnant. In a few thousand years, the reverse shock arrives at the PWN shell and compresses the PWN by a large factor, increasing the
pressure and producing a bounce of the nebula. The magnetic field also increases and burns off the highest energy electrons
\citep{reynolds84,vanderswaluw01,bucciantini03}. Rayleigh-Taylor instabilities produce filamentary structures and mixes thermal and non-thermal
material within the PWN \citep{chevalier98,blondin01}. At this stage, the central PSR can move outside the PWN and re-enter afterwards due to
the expansion of the PWN. After this reverberation phase, the PSR powers again the PWN (when it re-enters) and there are two solutions depending
on whether $t<\tau_0$ or $t>\tau_0$. In the former case, $L(t) \simeq L_0$ and $R_{PWN} \propto t^{11/15}$ when the braking index $n=3$
\citep{vanderswaluw01} or $R_{PWN} \propto t^{3/10}$ in the latter case \citep{reynolds84}. When the distance traveled by the PSR is large
enough, the PSR escapes from the original PWN and does not power it anymore, leaving a so-called {\it relic PWN}. As the PWN travels through the
SNR material, it creates a new smaller PWN \citep{vanderswaluw04b}. Observationally, this appears as a central, possibly distorted radio PWN,
showing little corresponding X-ray emission. The pulsar is to one side of or outside this region, with a bridge of radio and X-ray emission
linking it to the main body of the nebula. An example is the PWN in the SNR G327.1–1.1 (figure \ref{sedov_phase}).

\begin{figure}
\centering
\includegraphics[width=0.5\textwidth]{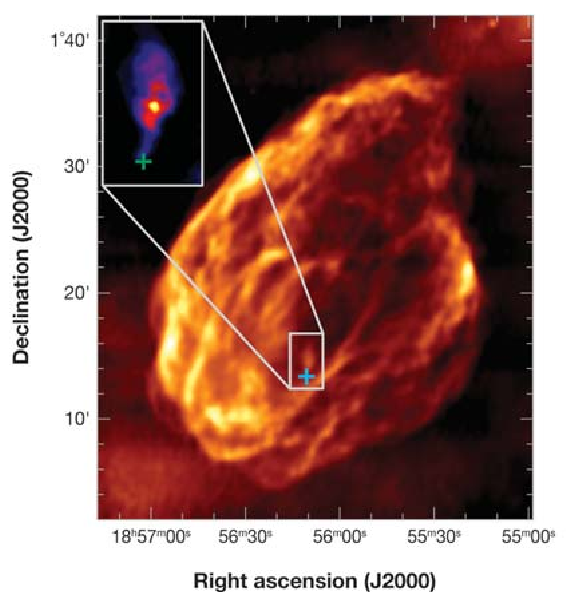}
\caption[1.4 GHz radio image of SNR W44]{1.4 GHz radio image of SNR W44 made by the {\it Very Large Array} telescope (VLA) \citep{giacani97}.
In the upper left small panel, it is shown a 8.4 GHz VLA image of the bow-shock pulsar wind nebula powered by PSR B1853+01 \citep{frail96b}.}
\label{bow_shock}
\end{figure}

When the speed of the  PSR becomes supersonic inside the SNR, it drives a bow shock \citep{chevalier98,vanderswaluw98}. The pressure produced by
the PSR's motion confines the PWN, which is in equilibrium with the material of the SNR (for example, W44, see figure \ref{bow_shock}). For a SNR
in the Sedov phase, the transition to a bow shock takes place when the pulsar has moved 68\% of the distance between the center and the forward
shock of the SNR \citep{vanderswaluw98,vanderswaluw03}. At this moment, the PWN takes a comet-like shape. A pulsar will typically cross its SNR
shell after $\sim$40000 years. If the SNR is still in the Sedov phase, the bow shock has a Mach number $\sim3.1$ \citep{vanderswaluw03}. After
crossing the SNR shell, the PWN maintains this bow-shock, but this time it is propagating through the ISM. It can be detected from radio to
X-rays. The shock driven by the PWN has H$_\alpha$ emission produced by excitation of the ISM. Finally, the spin-down luminosity of the PSR drops
and it is not capable to power an observable synchrotron nebula and the PSR is surrounded by a static or slowly expanding cavity of relativistic
material in equilibrium with the pressure of the ISM.

\subsection{Later stages for SNRs}

As the SNR forward shock accumulates matter form the ISM, it weakens and radiative cooling starts to be an important energy loss contribution.
Generally, radiative losses become important when the post-shock temperature falls below $\sim 5 \times 10^5$ K, in which case oxygen line
emission becomes an important coolant (e.g., \citealt{schure09}). In this moment, the evolution of the shock radius is described using a
momentum conservation law, such that
\begin{equation}
M_{sw} V_{SNR}=\frac{4}{3}\pi R_{SNR}^3 \rho_0 \frac{dR_{SNR}}{dt}=\textmd{const}.
\end{equation}
From this conservation law, one can calculate the moment when radiative losses become important
\begin{equation}
t_{rad}=\frac{4 \pi R_{rad}^3}{3 \rho_0 V_{rad}},
\end{equation}
and finally integrate to obtain an implicit function for the SNR radius (e.g. \citealt{toledoroy09})
\begin{equation}
t=t_{rad}+\frac{R_{rad}}{4V_{rad}}\left[\left(\frac{R(t)}{R_{rad}} \right)^4-1 \right],
\end{equation}
with
\begin{equation}
t_{rad}=44.6 \left(\frac{E_{51}}{\rho_0} \right)^{1/3} \textmd{kyr},
\end{equation}
\begin{equation}
R_{rad}=23 \left(\frac{E_{51}}{\rho_0} \right)^{1/3} \textmd{pc}.
\end{equation}

After this stage, the SNR continues to expand slowly and decreasing its temperature untill it megers completely with the ISM (merging phase).

\section[Supernova Remnants in X-rays]{Supernova Remnants in X-rays}  
\label{sec1.5}

{\it Supernova remnants} (SNRs) are gas and dust structures formed as a consequence of a supernova explosion. Supernovae can be produced in
different ways depending on if the progenitor star is isolated or in a binary system. In the isolated case, at the beginning of its life, the
star sustains the force of gravity using the radiation pressure produced by photons generated in the core which merge hydrogen atoms (H) to form
helium (He) by thermonuclear reactions. This stage, which is the longest during the life of the star, is called {\it main sequence}. At the end
of the main sequence, if the mass of the star is not very large ($M \lesssim 9M_\odot$, $M_\odot=1.9889 \times 10^{33} g$), the density in the
core is so high that the gas becomes degenerate before reaching the temperature to merge He nuclei to carbon and oxygen (C and O). For a
degenerate gas, the pressure depends almost only on its density. This pressure is much higher than for an ideal gas. This makes the core stable
against gravity without thermonuclear burning of C. Regarding the expanding envelope, it finally becomes unstable and the stellar winds strip the core
forming a planetary nebula around it. This stripped core, with a typical mass of $1M_\odot$ and a radius of $\sim$5000 km is called a
{\it white dwarf}. The evolution of this white dwarf is only a thermal cooling, but if it forms a binary system with a companion star which
transfers mass, depending on the transferred mass rate, we can observe sudden explosions due to H burning in the surface of the white dwarf,
called {\it novae}, or if the transfer rate is very high, this could activate the thermonuclear burning of C of the white dwarf and explode
destroying completely the star as a {\it thermonuclear supernovae} (or Type Ia SNe).

Stars with $M \gtrsim 9M_\odot$ pass through all the thermonuclear burning phases acquiring a burning layer structure, where lighter materials
as H and He are found in the envelope surface and silicon and iron (Si and Fe) in the core. The energy to merge two atoms of Fe is higher than
the nuclear potential energy released in the reaction, thus when the core burns almost all the Si, the contraction of the core becomes
unavoidable. The bounce of the core creates a shock wave propagating outwards which accelerates the thermonuclear reactions of the outer layers
releasing a huge amount of energy of $10^{51}$ erg. The star explodes as a {\it core collapse supernovae} (Type II and Type I SNe, with the
exception of Type Ia SNe). If the mass of the surviving core has a mass of 1-2$M_\odot$, then the gas degenerates and becomes stable forming a
neutron star. In this conditions, almost all the electrons has fallen into the nuclei of the atoms and formed neutrons (giving the name of
neutron star). If the mass of the core is higher than $\sim$2-3$M_\odot$ (Oppenheimer-Volkov limit), then the collapse is completely unavoidable
and becomes a time-space singularity called {\it black hole}.

The elements created during the life of the star determine the chemical composition of the SNR. The shock wave arisen from the explosion sweeps
up the star envelope and the surrounding interstellar medium (ISM) and creates a thermal shell which can be visible at different wavelengths (from
radio to X-rays, generally).

SNRs are objects of interest for many applications in astrophysics. They are important in the study of the local population of SNe (2 or 3 per
century in a spiral galaxy like ours). In addition, they can give more information about how the explosion mechanism of the SN revealing
different asymmetries and velocity distribution of the material. SNR shocks provide the best laboratories to study high Mach number,
collisionless shocks. It is thought that cosmic-rays are accelerated in these shocks. This is supported by the detection of SNRs of synchrotron
emission from radio to X-rays and $\gamma$-ray emission from pion decay. X-ray observations and spectroscopy of these objects are essential to
know more about the abundances and nucleosynthesis of the elements generated in the SN explosion and the state of the plasma. We will see some
results obtained using these techniques in chapter \ref{chap6}.

More than 100 SNRs have been observed in X-rays (e.g., Chandra SNR catalog\footnote{\url{http://hea-www.cfa.harvard.edu/ChandraSNR/}}) and
many more if we take into account the rest of the electromagnetic spectrum (e.g, the Green SNR catalog\footnote{\citet{green09}} or the University of
Manitoba SNR catalog\footnote{\citet{ferrand12}}). Observations in X-rays are very important in many aspects of SNRs, and particularly, the X-ray
spectroscopy. Using X-ray spectroscopy we can study the abundances of the elements created during the life of the progenitor star, the so-called
{\it $\alpha$-elements} (C, O, Ne, Mg, Si, S, Ar, Ca, Fe) and other elements that can be produced during the SN explosion. The emission lines
between 0.5-10 keV are very prominent for plasma temperatures between 0.2-5 keV, which are the temperatures that we usually find in the SNR
shocks. The plasma in SNR is optically thin at this energy range, which makes the measurement of the abundances quite confident \citep{vink12}.
Analyzing the X-ray spectra is also possible to see the existence of non-thermal emission coming from synchrotron radiation produced by
cosmic-rays and inferred the magnetic field that accelerates particles.

The last generation of X-ray telescopes as XMM-Newton and Chandra and the spectrometers installed in these observatories allowed as to do
imaging spectroscopy, which is very useful to differentiate the regions in the SNR where thermal and non-thermal radiation is produced and to
produce temperature maps of the plasma to understand better the dynamics of the shocks.

In this section, we will explain the main features of the X-ray emission of SNRs (see reviews of \citealt{mewe99,kaastra08} for an detailed
explanation) and which physical processes are involved, which will be useful to understand better the work done in chapter \ref{chap5}.

\subsection{Thermal emission}

Most of the spectral characteristics of the thermal emission, i.e. continuum shape and emission line ratios, are determined by the electron
temperature. Note that this temperature is not necessarily the same as the ion temperature. SNR plasmas are optically thin in X-rays, which
makes X-ray spectroscopy a very interesting tool for measuring element abundances. In the case of old SNR, this is also useful to study the
abundances in the ISM (e.g. \citealt{hughes98}).

Thermal X-ray spectra has different components: continuum emission by Bremsstrahlung (free-free emission), recombination continuum (free-bound
emission), which arises when an electron is captured into one of the atomic shells, and two-photon emission caused by a radiative electron
transition from a metastable quantum level.

The total emissivity for a Maxwellian energy distribution of the electrons is \citep{vink12}
\begin{equation}
\label{ff}
\varepsilon_{ff}=\frac{32 \pi e^6}{3 m_e c^3} \sqrt{\frac{2 \pi}{3 k m_e}}g_{ff}(T_e)T_e^{-1/2}\exp{\left(-\frac{h \nu}{k T_e} \right)} n_e \sum_i n_i Z_i^2\quad\textmd{(erg\,s$^{-1}$\,cm$^{-3}$\,Hz$^{-1}$)}
\end{equation}
with $g_{ff} \approx 1$, the gaunt factor. The subscript $i$ denotes the ion species with charge $e Z_i$. The emissivity at a given temperature
is determined by the factor $n_e \sum_i n_i Z_i^2$. Collisions with H and He nuclei dominate and this is why we usually simply this factor by
taking $n_e n_H$ or $n_e^2$. The normalization factor fitted by spectral analysis tools (e.g., {\it xspec}) is
$\int n_e n_H \mathrm{d}V/(4 \pi d^2)$ (also called {\it emission measure}), where we integrate the emissivity by the observed volume observed
and divide by $4 \pi d^2$ factor to obtain the measured flux. This simplification of the factor $n_e \sum_i n_i Z_i^2$ may be not valid for
shocked SN ejecta electrons where collisions with heavy ions can also be an important contribution. Neglecting these contributions can derive
erroneous density and mass estimates from the Bremsstrahlung emissivities (e.g. \citealt{vink96}).

The other mentioned components (recombination and two-photon emission) are normally neglected, but in some situations, they can also be
important, in particular for metal-rich plasmas in young SNRs \citep{kaastra08}.

SNR plasmas are often out of ionization equilibrium or {\it non-equilibrium ionization} plasmas (NEI). Plasmas of cool stars and clusters of
galaxies are referred to as {\it collisional ionization equilibrium} (CIE). SNR plasmas are in NEI because for the low densities involved, not
enough time has passed since the plasma was shocked, and few ionizing collisions have occurred for any given atom \citep{itoh77}. The number
fraction of atoms in a given ionization state $F_i$ is governed by the following differential equation:
\begin{equation}
\label{fracat}
\frac{1}{n_e} \frac{dF_i}{dt}=\alpha_{i-1}(T)F_{i-1}-\left[\alpha_i(T)+R_{i-1}(T) \right]F_i+R_i(T)F_{i+1}
\end{equation}
with $\alpha_i(T)$ being the ionization rate for a given temperature and $R_i$ the recombination rate. For NEI plasmas, $dF_i/dt \neq 0$ and the
ionization fractions have to be solved using equation (\ref{fracat}) as a function of $n_e t$, the {\it ionization age}. To solve this system is
CPU expensive, but fast approach were proposed by \citet{hughes85,kaastra93,smith10}. The main effect of NEI in young SNRs is that the ionization
states at a given temperature are lower than in CIE.

\subsection{Non-thermal emission}

X-ray synchrotron radiation has been traditionally related with composite SNRs (SNR with a PWN), but recently this kind of radiation has been
also detected in young SNR shells \citep{koyama97}. X-ray synchrotron spectra of young SNRs have rather steep indices ($\Gamma$=2-3.5)
indicating a rather steep underlying electron energy distribution. Electrons which produce X-ray synchrotron radiation are close to the maximum
energy of the distribution. When this maximum is defined where the acceleration gains are comparable to the radioactive losses, we say that we are
in the {\it loss-limited case}. When the shock acceleration process has not had enough time to accelerate particles, then we say that we are in the
{\it age-limited case} \citep{reynolds98}. The energy cut-off in these two situations is defined differently: in the age-limited case, the energy
cut-off is $\alpha \exp(-E/E_{max})$, whereas in the loss-limited case the cut-off is super-exponential $\alpha \exp(-E/E_{max})^2$
\citep{zirakashvili07}. In the loss-limited case, the cut-off photon energy is independent of the magnetic field \citep{aharonian99}.
Values of the shock velocity obtained through this energy cut-off are $\sim$2000 km s$^{-1}$ only encountered in young SNRs.

Another source of non-thermal X-ray emission comes from Bremsstrahlung and IC scattering. IC scattering is for SNRs important in the GeV-TeV band
\citep{hinton09}, but for the magnetic fields inside SNRs, $B \approx 5-500 \mu$G, it is generally not expected to be important in the soft X-ray
band. Bremsstrahlung would be caused by the non-thermal electron distribution. This contribution has been considered in some works (e.g.,
\citealt{asvarov90,vink97,bleeker01,laming01}). The electrons involved in the production of X-ray Bremsstrahlung emission have non-relativistic
energies. This means that identifying non-thermal Bremsstrahlung would be useful to obtain information about the low energy end of the electron
cosmic-ray distribution. However, it is unlikely that non-thermal Bremsstrahlung contributes enough to identify it with the current generation of
hard X-ray telescopes.

\subsection{Line emission}

\begin{figure}
\centering
\includegraphics[width=0.5\textwidth]{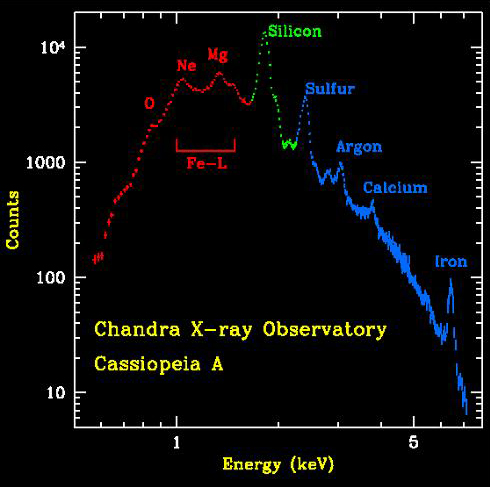}
\caption[Tycho's SNR X-ray spectrum]{X-ray spectrum of Tycho's SNR obtained with the {\it Chandra} X-ray Observatory
(\url{http://chandra.harvard.edu}).}
\label{snr_spectrum}
\end{figure}

Line emission in SNR results from excitation or recombination of electrons in ions. Since density is very low, ions can be assumed to be in the
ground state, and thus, collisional de-excitation or further excitation or ionization can be neglected. This also means that the ionization
balance can be treated independently of the line emission properties (see \citealt{mewe99}, for a full treatment). The spectrum of Tycho's SNR by
the {\it Chandra} X-ray observatory is shown in figure \ref{snr_spectrum} as an example. Tycho's SNR is the reminiscent of a Type Ia SN (e.g,
\citealt{lopez09}). The most important lines in the spectrum from 0.3 to 1 keV are the oxygen (O), iron (Fe) and neon (Ne) lines. Magnesium (Mg),
sulfur (S) and silicon (Si) lines dominate from 1 to 3 keV followed by the lines of argon (Ar) at 3.1 keV, calcium (Ca) at 3.8 keV and the Fe
line at 6.4 keV.

\begin{figure}
\centering
\includegraphics[width=0.6\textwidth]{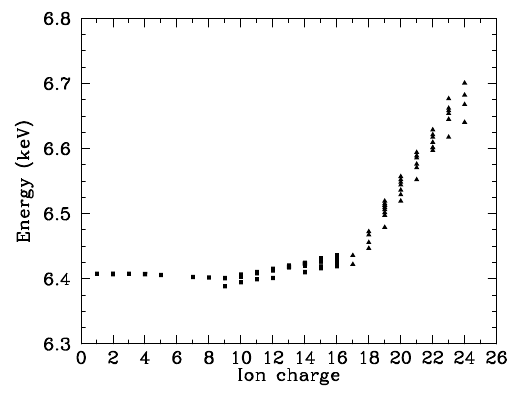}
\caption[Fe-K shell line as a function of the ionization state]{Energy of the Fe-K shell line as a function of the ionization state
\protect\citep{vink12}. The data is given by \protect\citet{beiersdorfer93,palmeri03,mendoza04}. For ionization states from Fe I to Fe XVII the
average Fe-K shell line is $\sim$6.4 keV.}
\label{feion}
\end{figure}

Fe line emission is an useful tool to characterize the state of the plasma. Fe-K shell ($n=1$) emission can be observed for all ionization states
of Fe, because of its high fluorescence yield and high abundance, provided that the electron temperature is high enough ($k T_e \gtrsim 2$ keV).
The average line energy of the Fe-K shell emission provides information about the dominant ionization state. Figure \ref{feion} shows how the
energy of the lines depends on the ionization state of the atom. For ionization states from Fe I to Fe XVII the average Fe-K shell line is
$\sim$6.4 keV.

Regarding the Fe-L shell ($n=2$), there are prominent transitions between 0.7 and 1.2 keV. Fe-L shell line emission occurs for lower temperatures
and ionization ages than the Fe-K lines ($k T_e \gtrsim$0.15 keV). The ionization state of the plasma can be also accurately determined by
combining the observations of the Fe-K and Fe-L shell lines. This is especially useful in high resolution X-ray spectroscopy, where you can
resolve the individual lines produced by the Fe-L shell. Fe-K emission around 6.4 keV could be caused by Fe XVII-XIX, or by lower ionization
states, but in the latter case, Fe-L lines should not be detected. In higher ionization states, as Fe XXV and Fe XXVI, displace this line at 6.7
keV and 6.96 keV (see figure \ref{feion}). 

Additional lines caused by radioactivity could be detected. During the first year after the SN explosion, the most important radioactive element
is $^{56}$Ni, which decays in 8.8 days into $^{56}$Co, which subsequently decays into $^{56}$Fe. $^{56}$Fe is the most abundant isotope in the
Universe. Type Ia supernovae (thermonuclear SNe) produce typically 0.6$M_\odot$ per explosion, which makes the larger part of the production of
this isotope. Other important element is $^{44}$Ti. Its production is about 10$^{-5}$--10$^{-4}$M$_\odot$ per explosion (e.g.
\citealt{prantzos11}). The longer decay time makes it interesting for studying SNR in hard X-rays (85 yr, \citealt{ahmad06}). The decay chain of
$^{44}$Ti results in line emission at 67.9 keV and 78.4 keV, which are caused by the nuclear de-excitation of $^{44}$Sc. The emission lines of
this element are sensitive to the expansion speed of the inner layers of the ejecta. In addition, $^{44}$Ti is sensitive to the boundary between
the accreted material onto the proto-neutron star and the ejected material. Also is useful to identify and study explosion asymmetries
\citep{nagataki98}.

\section{This thesis} 
\label{sec1.6}

There are still many unanswered questions about how pulsars interact with the ambient interstellar medium and how these interactions affect their
evolution. During their life,  PSRs accelerate particles in the termination shock creating what we know as a pulsar wind nebula (PWN). Here, we
explore models of spectra and magnetic field. These magnetized nebulae show spectral features which are still difficult to reproduce.

We have developed a new code to reproduce the spectra of PWNe, which we call TIDE-PWN ({\it TIme DEpendent-Pulsar Wind Nebulae}). This code
solves the electron diffusion-loss equation as a function of time for the pairs accelerated and injected to the ambient medium from the
termination shock of the PSR considering synchrotron, inverse Compton (IC), adiabatic and Bremmstrahlung energy losses. The resulting electron
population is integrated in order to obtain the synchrotron, IC and Bremmstrahlung spectra of the PWN. The expansion of the nebula is considered
during the free expansion. The model is described in detail in chapter \ref{chap2}.

We use this code to study different approximations made on the diffussion-loss equation and how they affect the spectra and their evolution. We
have also performed a parameter space exploration with $\sim$100 simulations covering a wide range of ages, magnetic fractions and spin-down
luminosities, in order to understand better the behavior of the spectra of Crab-like PWNe and shed some light on general issues as the dominance
of the synchrotron self-Compton component at VHE for the Crab Nebula, the low magnetic fraction deduced from multi-wavelength observations and
some general constrains on the detectability of young PWNe at high energies. Other project in this field has been the systematic parameterization
of the already detected young PWNe. We analyzed the spectra of 10 young PWNe and made a consistent comparison between the parameters
obtained in each of them and looked for correlations (see chapter \ref{chap3}). All this work is explained in detail in chapter \ref{chap4}. In some
cases, despite of the high spin-down power of the central pulsar, the associated PWN is not detected in TeV energies. We discuss about their
detectability at TeV and magnetization state in chapter \ref{chap5}.

The formation mechanisms of magnetars and how it could influence the surrounding medium, i.e. the SNR is discussed in chapter \ref{chap6}. There
are also unsolved questions about how magnetars are created after the supernova explosion. Two main models are still under debate to generate
their huge magnetic fields: increase of the magnetic field of the pulsar through magnetic field conservation of the progenitor star or the
alpha-dynamo process through vigorous convection of the core during the first few seconds after the supernova event. In this second process, it
is expected to observe an excess of rotational energy generated during the process, but previous works done on this did not find clear evidences.
Using the X-ray data available in the XMM-Newton and Chandra telescopes archive, we want to extend these works done before and look for features
not only in the spectral lines, but also in the photometry and other parameters in comparison with other well studied SNRs. Finally, the
conclusions of this dissertation and future projects are described in chapter \ref{chap7}.


\chapter[Time-dependent spectra of PWNe]{Time-dependent spectra of pulsar wind nebulae}
\label{chap2}

\ifpdf
    \graphicspath{{Chapter2/Figs/Raster/}{Chapter2/Figs/PDF/}{Chapter2/Figs/}}
\else
    \graphicspath{{Chapter2/Figs/Vector/}{Chapter2/Figs/}}
\fi

In this chapter, we describe in detail the main characteristics of the 1D spectral model for PWNe that we have developed and its technical
structure. We apply this model and fit the most known and complete PWN spectrum, the Crab Nebula. We also discuss about the improvements and
caveats of our model and how the parameters of the Crab Nebula have changed with new implementations of physics in each version of the code.

This chapter is based on the work done in \citet{martin12}.

\section{Description of the code}
\label{sec2.1}

\subsection{The difussion-loss equation}

The difussion-loss equation describes the evolution of the distribution of particles per unit energy and per unit volume in a certain time. We
represent this function by $N_i(\gamma,\vec{r},t)$, where the subscript $i$ represents the particle species, $\gamma$ the energy Lorentz factor,
$\vec{r}$ the position vector where we consider the distribution and $t$ the current time. The most general form of this equation is (e.g.,
\citealt{ginzburg64})

\begin{multline}
\label{gdle}
\frac{\partial N_i(\gamma,\vec{r},t)}{\partial t}=\vec{\nabla} \cdot \left[D_i(\gamma,\vec{r},t),t \right]-\frac{\partial}{\partial E}\left[\dot{\gamma}_i(\gamma,\vec{r},t) N_i(\gamma,\vec{r},t) \right]+\frac{1}{2}\frac{\partial^2}{\partial E^2}\left[d_i(\gamma,\vec{r},t)N_i(\gamma,\vec{r},t) \right]\\
+Q(\gamma,\vec{r},t)-\frac{N_i(\gamma,\vec{r},t)}{\tau_i(\gamma,\vec{r},t)}+\sum_k \int P_i^k(\gamma',\gamma)N_k(\gamma,\vec{r},t)\mathrm{d}\gamma.
\end{multline}

The term on the left-hand side of the equation is the variation of the distribution in time. The first term on the right-hand side of the
equation describes the spatial diffusion of the particles and $D_i(\gamma,\vec{r},t)$ is the diffusion coefficient. The space and time-dependence
of the diffusion coefficient is due to changes in time in the structure and composition of the PWN (expansion and interaction with the ISM) and
also changes in the magnetic field structure and density. Note that the diffusion coefficient also depends on the particle species, because
the motion of particles in the magnetic field depends on their charge. The second term leads continuous change in energy of the particles due to
acceleration mechanisms or energy losses in collisions. The function $\dot{\gamma}_i(\gamma,\vec{r},t)$ is the summation of the energy losses due
to all the mechanisms or collisions. The third term takes into account the fluctuations of this continuous variation of the energy of the
particles. The coefficient $d_i(\gamma,\vec{r},t)$ is the variation in time of the mean square increment of energy of each kind of particle
\begin{equation}
d_i(\gamma,\vec{r},t)=\frac{d}{dt}\bar{(\Delta E)^2}.
\end{equation}

The function $Q_i(\gamma,\vec{r},t)$ represents the injection of particles from the termination shock per unit energy and unit volume in a
certain time. The fifth term allows for the disappearance of particles due to escape from the distribution. $\tau_i(\gamma,\vec{r},t)$ is the
characteristic escape time of each particle. When energy losses are very important, we can consider them as an escape term defining
\begin{equation}
\tau_i(\gamma,\vec{r},t)=\frac{\gamma}{\dot{\gamma}_i(\gamma,\vec{r},t)}.
\end{equation}

Finally, the last term of equation (\ref{gdle}) takes into account all the collisions which allow creation and annihilation of particles.
$P_i^k(\gamma',\gamma)$ is the probability per unit time and per unit energy of the appearance of a particle of kind $i$ with an energy $\gamma$
produced by a collision of a particle of kind $k$ with an energy $\gamma'$. Note that the fluctuations in the energy variations due to the
creation or annihilation of particles are not included. If the particles are atoms, $P_i^k(\gamma',\gamma)$ gives also the probability of
fragmentation of the nuclei.

The solution of equation (\ref{gdle}) considering all the terms is an extraordinary difficult task and it is useful to make approximations in
some terms which are not important in our problem. First of all, in PWNe the spectral emission can be explain just considering electrons-positron
pairs. We do not consider creation or annihilation of other particles species. Fluctuations in the variation of the energy are also neglected
as we will use the mean value of the energy losses per unit energy, which typically are known and they have an analytical expressions for pairs.
We assume an isotropic injection in the whole nebula and no morphology in the magnetic field is taken into account, thus we do not consider
diffusion effects. Applying these approximations, typically found in the literature, equation (\ref{gdle}) yields
\begin{equation}
\label{dle}
\frac{\partial N(\gamma,t)}{\partial t}=-\frac{\partial}{\partial \gamma}\left[\dot{\gamma}(\gamma)N(\gamma,t) \right]-\frac{N(\gamma,t)}{\tau(\gamma,t)}+Q(\gamma,t).
\end{equation}
This equation can be solved using a Green function (see e.g., \citealt{aharonian97}). The solution is
\begin{equation}
\label{nana}
N(\gamma,t)=\frac{1}{\dot{\gamma}(\gamma,t)} \int_{-\infty}^t \dot{\gamma}(\gamma_0,t_0)Q(\gamma_0,t_0) \exp \left(-\int_{t_0}^t \frac{\mathrm{d}x}{\tau(\gamma_x,t_x)} \right) \mathrm{d}t_0.
\end{equation}
being $\gamma_0$ the initial energy of the electron at time $t_0$. The initial and final energy are related by
\begin{equation}
t-t_0=\int_\gamma^{\gamma_0} \frac{\mathrm{d}\gamma'}{\dot{\gamma}(\gamma',t')}.
\end{equation}

To calculate these equations has a high computational cost. Because of this, we preferred to do a numerical approach using a first order
approximation implicit scheme. Equation (\ref{dle}) is then
\begin{multline}
\frac{N(\gamma,t+\Delta t)-N(\gamma,t)}{\Delta t}=-\frac{\dot{\gamma}(\gamma+\Delta \gamma,t+\Delta t)N(\gamma+\Delta \gamma,t+\Delta t)-\dot{\gamma}(\gamma,t+\Delta t)N(\gamma,t+\Delta t)}{\Delta \gamma}\\
-\frac{N(\gamma,t+\Delta t)}{\tau(\gamma,t+\Delta t)}+Q(\gamma,t),
\end{multline}
where $\Delta t$ and $\Delta \gamma$ are the increments in time and energy. To abbreviate the notation, we will write a superscript $t$ or
$t+1$ all those parameters which depend on the current time $t$ or on the next time step $\Delta t$. We do the same with the energy using the
subscripts $g$ and $g+1$. Reordering the elements of the equation, we obtain the implicit scheme
\begin{equation}
\label{impscheme}
\left(1-\frac{\Delta t}{\Delta \gamma}\dot{\gamma}_g^{t+1}+\frac{\Delta t}{\tau_g^{t+1}} \right)N_g^{t+1}+\frac{\Delta t}{\Delta \gamma}\dot{\gamma}_{g+1}^{t+1}N_{g+1}^{t+1}=N_g^t+Q_g^t \Delta t.
\end{equation}

Equation (\ref{impscheme}) can be written in matrix representation, such that
\begin{equation}
\begin{pmatrix}
\alpha_1 & \beta_2 & \ldots & \ldots & 0\\
0 & \alpha_2 & \beta_3 & \ldots & \vdots\\
\vdots & \ddots & \ddots & \vdots & \vdots \\
0 & \ldots & \ldots & \alpha_{M-1} & \beta_M
\end{pmatrix}
\cdot
\begin{pmatrix}
N_1^{t+1}\\
N_2^{t+1}\\
\vdots \\
N_M^{t+1}
\end{pmatrix}
=
\begin{pmatrix}
N_1^t\\
N_2^t\\
\vdots \\
N_M^t
\end{pmatrix}
+\Delta t
\begin{pmatrix}
Q_1^t\\
Q_2^t\\
\vdots \\
Q_M^t
\end{pmatrix}
\end{equation}
where $\alpha_g=\left(1-\frac{\Delta t}{\Delta \gamma}\dot{\gamma}_g^{t+1}+\frac{\Delta t}{\tau_g^{t+1}} \right)$ and
$\beta_{g+1}=\frac{\Delta t}{\Delta \gamma}\dot{\gamma}_{g+1}^{t+1}$ and $M$ is the number of points where we compute the spectrum. Solving
this system of equations in each time step is needed to obtain the evolution of the pair population.

The injection $Q(\gamma,t)$ acts as a source term. It is usually supplied by the user. The most typical form of injection for PWNe is a
broken power law,
\begin{equation}
\label{inj}
Q(\gamma,t)=Q_0(t)\left \{
\begin{array}{rl}
\left(\frac{\gamma}{\gamma_b} \right)^{-\alpha_1} & \text{for }\gamma_{min} \le \gamma \le \gamma_b,\\
\left(\frac{\gamma}{\gamma_b} \right)^{-\alpha_2} & \text{for }\gamma_b<\gamma \le \gamma_{max}.
\end{array} \right.
\end{equation}
where $\gamma_{min}$ and $\gamma_{max}$ are the minimum and maximum energy of particles. The energy break is represented by $\gamma_b$ and
$Q_0$ is the normalization factor of the injection. In our model, the spin-down losses of the PSR accelerate particles ($\dot{E}_p$) and
maintain the magnetic field of the nebula ($\dot{E}_B$), thus $\dot{E}=\dot{E}_p+\dot{E}_B$. We define the magnetic fraction $\eta$ as (e.g., 
\citealt{tanaka10})
\begin{equation}
\eta=\frac{\dot{E}_B}{\dot{E}}=\frac{\dot{E}_B}{\dot{E}_B+\dot{E}_p}.
\end{equation}
Note that this definition is different from the magnetization factor $\sigma$ defined in equation (\ref{sigma}), where we calculate the ratio
between the energy that goes to the magnetic field and the energy that goes to particles ($\sigma=\dot{E}_B/\dot{E}_p$). This implies that the
relation between both factors is
\begin{equation}
\eta=\frac{\sigma}{1+\sigma}.
\end{equation}

We assume that $\eta$ is constant during the life of the PWN. We use the magnetic fraction to compute the injection normalization factor $Q_0$,
such that
\begin{equation}
\label{norminj}
(1-\eta)L(t)=\int_{\gamma_{min}}^{\gamma_{max}} \gamma m_e c^2 Q(\gamma,t) \mathrm{d}\gamma.
\end{equation}
\\

In our model, $\gamma_{min}$ is a free parameter fixed by the user, but $\gamma_{max}$ is calculated demanding particle confinement into the
acceleration zone, i.e. the Larmor radius of particles $R_L$ must be smaller than the termination shock radius $R_w$
\begin{equation}
R_L=\varepsilon R_w,
\end{equation}
where $\varepsilon<1$ is the containment factor. The Larmor radius is given by $R_L=\gamma_{max} m_e c^2/e B_w$, thus $\gamma_{max}$ can be
written as
\begin{equation}
\label{gmax1}
\gamma_{max}=\frac{e \varepsilon B_w R_w}{m_e c^2}.
\end{equation}
The magnetic field in the termination shock $B_w$, in terms of the magnetic fraction $\eta$, is \citep{kennel84b}
\begin{equation}
\label{bw}
B_w=\kappa \sqrt{\eta \frac{\dot{E}}{c}} \frac{1}{R_w},
\end{equation}
where $\kappa$ is the magnetic compression ratio which lies between 1, for Vela-like shocks, and 3, for strong shocks ($\sigma \ll 1$). Combining
equations (\ref{gmax1}) and (\ref{bw}), we obtain
\begin{equation}
\label{gmax2}
\gamma_{max}=\frac{e \varepsilon \kappa}{m_e c^2} \sqrt{\eta \frac{\dot{E}}{c}}.
\end{equation}

The containment factor depends on the coherent length of the magnetic field ($\varepsilon=\sqrt{l_c/3 R_L}$), but as our model does not take into
account the magnetic field morphology, we consider $\varepsilon$ as a free parameter.

Extrapolations to the PWN case of the simulations done by \citet{spitkovsky08} for collisionless shock (e.g., \citealt{fang10a,holler12}). In
this case, the injection function consists in a Maxwellian distribution for pairs at low energies plus a power law component only important at
high energy
\begin{equation}
Q(\gamma,t)=\left \{
\begin{array}{lr}
Q_{0,1}(t) \gamma \exp \left(-\frac{\gamma}{\Delta \gamma_1} \right) & \text{for }\gamma<\gamma_b,\\
Q_{0,1}(t) \gamma \exp \left(-\frac{\gamma}{\Delta \gamma_1} \right)+Q_{0,2}(t) \gamma^{-\alpha} & \text{for }\gamma_b \le \gamma<\gamma_{cut},\\
Q_{0,1}(t) \gamma \exp \left(-\frac{\gamma}{\Delta \gamma_1} \right)+Q_{0,2}(t) \gamma^{-\alpha} \exp \left(-\frac{\gamma-\gamma_{cut}}{\Delta \gamma_{cut}} \right) & \text{for }\gamma \ge \gamma_{cut},
\end{array} \right.
\end{equation}
where $\Delta \gamma_1$ is the width of the Maxwellian distribution and $\gamma_{cut}$ and $\Delta \gamma_{cut}$, the energy and the energy width
of the cut-off. The index $\alpha$ should be between 2.3--2.5 and there are ratios established between some parameters, i.e.
$\gamma_b/\Delta \gamma_1 \approx 7$,  $\gamma_{cut}/\gamma_b \approx 7.5$ and $\gamma_{cut}/\Delta \gamma_{cut} \approx 3$.

\subsection{Magnetic field evolution}

The magnetic field evolution is balanced solving the differential equation (e.g., \citealt{pacini73})
\begin{equation}
\frac{dW_B(t)}{dt}=\eta \dot{E}(t)-\frac{W_B(t)}{R_{PWN}(t)} \frac{dR_{PWN}(t)}{dt},
\end{equation}
where $W_B=B^2 R_{PWN}^3/6$ is the total magnetic energy. Note that the variation of the magnetic energy depends on the spin-down luminosity
magnetic fraction and the adiabatic energy losses due to expansion of the nebula. Multiplying both sides of the equation and reordering, we find
that
\begin{equation}
R_{PWN}(t)\frac{dW_B(t)}{dt}+\frac{dR_{PWN}(t)}{dt} W_B(t)=\frac{d}{dt}[W_B(t)R_{PWN}(t)]=\eta \dot{E}(t)R_{PWN}(t).
\end{equation}
Integrating, the equation yields
\begin{equation}
W_B(t)=\frac{\eta}{R_{PWN}(t)} \int_0^t L(t')R_{PWN}(t') \mathrm{d}t',
\end{equation}

and the magnetic field is then
\begin{equation}
\label{bfield}
B(t)=\frac{1}{R_{PWN}^2(t)} \sqrt{6 \eta \int_0^t L(t')R_{PWN}(t') \mathrm{d}t'}.
\end{equation}

\subsection{Energy losses and escape}
\label{sec2.1.3}

The non-thermal emission received from the PWN comes from the energy losses of pairs due to different radiative (or mechanical) processes
described in many papers and books (e.g, \citealt{blumenthal70,blumenthal71,ginzburg64,ginzburg65,haug04,longair94,rybicki79}). We describe
briefly the main energy losses taken into account in the diffusion-loss equation for leptonic models: synchrotron, inverse Compton, adiabatic and
Bremsstrahlung energy losses. For more detail in the formulae derivation, see appendix \ref{appa}. Some particles escape from the distribution
due to diffusion or catastrophic energy losses (mechanisms in which particles lose practically all the energy). In our model, we consider the
escaping particles due to Bohm diffusion, which is the diffusion effect due to the presence of a magnetic field (e.g., \citealt{vorster13c})
\begin{equation}
\tau_{Bohm}=\frac{e B(t) R_{PWN}^2(t)}{2 \gamma m_e c^3},
\end{equation}
where $e$ is the electron charge. Note that the effect of this diffusion grows linearly with the energy $\gamma$ as the timescale diminishes.

When particles are accelerated by a magnetic field, synchrotron radiation is emitted. The synchrotron energy losses suffered by a relativistic
electron passing through a magnetic field are given by
\begin{equation}
\label{synclosses}
\dot{\gamma}_{sync}(\gamma,t)=-\frac{4}{3}\frac{\sigma_T}{m_e c}U_B(t)\gamma^2 ,
\end{equation}
where $\sigma_T=(8\pi/3)r^2_0$ is the Thomson cross section for electrons, $r_0$ is the electron classical radius, and $U_B(t)=B^2(t)/8\pi$ is
the energy density of the magnetic field. The dependence of equation (\ref{synclosses}) with the magnetic field and the energy is quadratic and
typically dominates in the high energy range of the pair distribution for young PWNe.

Inverse Compton interaction (IC) consist on collisions of high energy electrons with soft photons of the environment (e.g., CMB) that lend part
of their energy to photons, upgrading them to $\gamma$-rays. For low photon energies ($h \nu \ll m_e c^2$), the scattering of radiation from free
charges reduces to the classical case of Thomson scattering. In the Thomson limit, the IC energy losses have the form \citep{blumenthal70}
\begin{equation}
\label{icthomson}
\dot{\gamma}_{IC}(\gamma,t)=-\frac{4}{3}\frac{\sigma_T}{m_e c}U_{\gamma} \gamma^2,
\end{equation}
where $U_{\gamma}$ is the total energy density of the photon background in the PWN. Note that equation~\ref{icthomson} has a similar form with
equation~\ref{synclosses}. This means that in the Thomson limit, the synchrotron and the IC energy losses domain depending on the energy density
of the magnetic field and the target photon field density rate.

Thomson approximation fails when $h \nu \gg m_e c^2$. In order to take into account both regimes, we use the IC energy losses calculated using
the exact Klein-Nishina cross section \citep{klein29}. The IC energy losses in this case yield
\begin{equation}
\dot{\gamma}_{IC}(\gamma)=-\frac{3}{4}\frac{\sigma_T h}{m_e c}\frac{1}{\gamma^2}\int_0^\infty \nu_f \mathrm{d}\nu_f \int_0^\infty \frac{n(\nu_i)}{\nu_i} f(q,\Gamma_{\varepsilon}) H(1-q) H \left(q-\frac{1}{4\gamma^2} \right) \mathrm{d}\nu_i,
\end{equation}
being $H$ the Heaviside step function ($H(x)=\int_{-\infty}^x \delta(y) \mathrm{d}y$) and $n(\nu)$ the photon background distribution. The
subscripts $i$ and $f$ refer to frequencies of the photons before and after scattering, respectively. The other terms are defined as
\begin{equation}
\label{fqg}
f(q,\Gamma_{\varepsilon})=2q\ln q+(1+2q)(1-q)+\frac{1}{2}(1-q)\frac{(\Gamma_{\varepsilon}q)^2}{1+\Gamma_{\varepsilon}q},
\end{equation}
\begin{equation}
\label{geps}
\Gamma_{\varepsilon}=\frac{4 \gamma h \nu_i}{m_e c^2},
\end{equation}
\begin{equation}
\label{qkin}
q=\frac{h \nu_f}{\Gamma_{\varepsilon}(\gamma m_e c^2-h \nu_f)}.
\end{equation}

Bremsstrahlung radiation is caused by the deceleration of pairs due to the presence of electric fields that modifies the original trajectory. We
consider two contributions: the electron-ion Bremsstrahlung and the electron-electron Bremsstrahlung. The electron-ion Bremsstrahlung is due to
the interaction of the electron with the electromagnetic field produced by the ionized nuclei of the ISM. The electron-atom bremsstrahlung energy
losses have the form \citep{haug04}
\begin{equation}
\dot{\gamma}_{Brems}^{e-a}=-\frac{3}{\pi} \alpha \sigma_T c S \frac{\gamma^2}{\gamma^2+p^2} \left [\gamma \ln(\gamma+p)-\frac{p}{3}+\frac{p^3}{\gamma^6} \left(\frac{2}{9}\gamma^2-\frac{19}{675}\gamma p^2-0.06 \frac{p^4}{\gamma} \right) \right],
\end{equation}
with
\begin{equation}
\label{s}
S=\sum_Z Z^2 N_Z=N_H \left[1+\sum_{Z \ge 2} \left(\frac{N_Z}{N_H} \right)Z^2 \right].
\end{equation}

The parameter $p=\sqrt{\gamma^2-1}$ is the linear moment of the electron. $N_H$ is the number density of ISM hydrogen and $N_Z$, the number
density of the elements with atomic number $Z$ and $\alpha \simeq 1/137$ is the fine-structure constant. 

For the second contribution, the electron-electron Bremsstrahlung, the energy losses are given by
\begin{equation}
\dot{\gamma}_{Brems}^{e-e}=-c \left(\sum_Z Z N_Z \right) \frac{p}{\gamma} (\gamma-1) \Phi_{rad}^{ee}(\gamma).
\end{equation}
It is not possible to obtain approximated formulae for the function $\Phi_{rad}^{ee}(\gamma)$, but we use function coming from fits of numerical
computations \citep{haug04}

\begin{equation}
\label{bremsfit}
\Phi_{rad}^{e-e} \approx \frac{3\alpha}{8\pi} \sigma_T \left \{
\begin{array}{lr}
\frac{\gamma+1}{\gamma} \left(0.6664+43.935 p-2.272 p^2-3.055 p^3 \right) \left(1-e^{2\pi \alpha \gamma/p} \right), & \gamma \le 1.02,\\
\sigma_T \frac{\gamma+1}{\gamma} \left(0.5754+1.14492 \gamma-0.0665 \gamma^2 \right), & 1.02<\gamma \le 3,\\
(\gamma+1) \frac{4.181 \gamma \ln (\gamma+p)-2.676 \gamma-2.256}{\gamma^2+1.022 \gamma-3.871}, & 3 < \gamma \le 980,\\
4 \left[\ln(2\gamma)-\frac{1}{3} \right], & \gamma>980,
\end{array} \right.
\end{equation}
\\
\\
Note that the last term of equation (\ref{bremsfit}) coincides with the formula given by \citet{blumenthal70}.

Finally, we take into account a non-radiative energy loss term, i.e the adiabatic losses. It is referred to the loss of internal energy of
particles due to work applied over the ISM in the expansion of the PWN. For a relativistic gas, this term yields
\begin{equation}
\dot{\gamma}_{ad}=-\frac{1}{3} \left(\vec{\nabla} \cdot \vec{v} \right) \gamma.
\end{equation}
In our model, we consider the PWN as an uniform expanding sphere, thus, we get the simple expression
\begin{equation}
\dot{\gamma}_{ad}(\gamma,t)=-\frac{v_{PWN}(t)}{R_{PWN}(t)}\gamma.
\end{equation}
where $v_{PWN}(t)=dR_{PWN}(t)/dt$. Usually, adiabatic energy losses dominate in the low energy range of the pair distribution and is the most
important contribution together with synchrotron losses.

\subsection{Photon luminosity}
\label{sec2.1.4}

\begin{figure}
\centering
\includegraphics[width=0.6\textwidth]{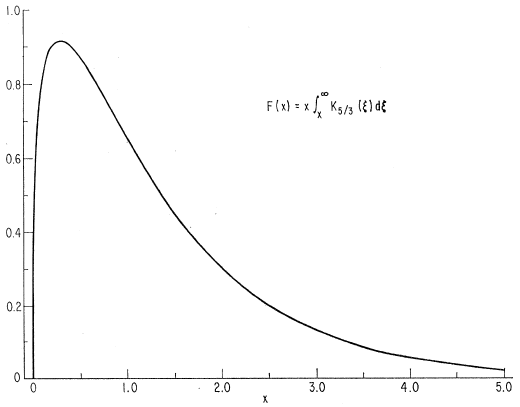}
\caption[The $F$ function]{Plot of the $F$ function \protect\citep{blumenthal70}. The peak of the function is near $x=$0.29.}
\label{ffunc}
\end{figure}

Once we integrate the diffusion-loss equation, we obtain the energy distribution of pairs $N(\gamma,t)$. To compute the spectrum luminosity, we
need to multiply the pair population by the power emission of each contribution. A more detailed derivation of the formulae is given in appendix
\ref{appb}. For synchrotron luminosity, the power emitted by each electron is given by (e.g., \citealt{ginzburg65,blumenthal70,rybicki79})
\begin{equation}
\label{psync}
P_{syn}(\nu,\gamma,B(t))=\frac{\sqrt{3}e^3 B(t)}{m_e c^2}F \left(\frac{\nu}{\nu_c(\gamma,B(t))} \right),
\end{equation}
where $\nu_c$ is the critical function defined in equation (\ref{critfreq}) and the dimensionless function $F$ is defined as
\begin{equation}
\label{ffunction}
F(x)=x \int_x^\infty K_{5/3}(y) \mathrm{d}y,
\end{equation}
where $K_{5/3}$ is the modified Bessel function of order $5/3$. The function $F$ peaks near 0.29$\nu_c$ as shown in figure \ref{ffunc}. In the
literature we find approximations to the synchrotron emission, where all the radiation at a given energy comes is concentrated at this frequency.
In our code, we do not consider this monochromatic approximation and we compute numerically the function $F$. Some useful asymptotic expressions
of $F$ are given by \citet{ginzburg65}
\begin{equation}
\begin{array}{lr}
F(x)=\frac{4 \pi}{\sqrt{3} \Gamma \left(\frac{1}{3} \right)} \left(\frac{x}{2} \right)^{1/3} \left[1-\frac{\Gamma \left(\frac{1}{3} \right)}{2} \left (\frac{x}{2} \right)^{2/3}+\frac{3}{4} \left(\frac{x}{2} \right)^2-\frac{9}{40} \frac{\Gamma \left(\frac{1}{3} \right)}{\Gamma \left(\frac{5}{3} \right)} \left(\frac{x}{2} \right)^{10/3}+\cdots \right] & \text{for } x \ll 1,\\
F(x)=\frac{2 \pi}{\sqrt{3} \Gamma \left(\frac{1}{3} \right)} \left(\frac{x}{2} \right)^{1/3} \left[1-\frac{\Gamma \left(\frac{1}{3} \right)}{\Gamma \left(\frac{5}{3} \right)} \left(\frac{x}{2} \right)^{4/3}+3 \left(\frac{x}{2} \right)^2-\frac{3}{5} \frac{\Gamma \left(\frac{1}{3} \right)}{\Gamma \left(\frac{5}{3} \right)} \left(\frac{x}{2} \right)^{10/3}+\cdots \right] & \text{for }x \gg 1.
\end{array}
\end{equation}

Multiplying by the pair distribution and integrating, the synchrotron luminosity gives (in erg cm$^{-2}$ s$^{-1}$ Hz$^{-1}$)
\begin{equation}
\label{synclum}
L_{syn}(\nu,t)=\int_{\gamma_{min}}^{\gamma_{max}} N(\gamma,t)P_{syn}(\nu,\gamma,B(t)) \mathrm{d}\gamma.
\end{equation}

Regarding the IC radiation, the power emitted by each electron is given by \citep{blumenthal70}
\begin{equation}
\label{pic}
P_{IC}(\gamma,\nu,t)=\frac{3}{4} \frac{\sigma_T c h \nu}{\gamma^2} \int_0^\infty \frac{n(\nu_i)}{\nu_i}f(q,\Gamma_{\varepsilon}) H(1-q) H \left(q-\frac{1}{4\gamma^2} \right) \mathrm{d}\nu_i.
\end{equation}
Proceeding as for the synchrotron case, we obtain
\begin{equation}
\label{iclum}
L_{IC}(\nu,t)=\frac{3}{4}\sigma_T c h \nu \int_{\gamma_{min}}^{\gamma_{max}} \frac{N(\gamma,t)}{\gamma^2} \mathrm{d}\gamma \int_0^\infty \frac{n(\nu_i)}{\nu_i}f(q,\Gamma_{\varepsilon}) H(1-q) H \left(q-\frac{1}{4\gamma^2} \right) \mathrm{d}\nu_i.
\end{equation}

In the code, the target photon field $n(\nu)$ can be defined in two ways. One possibility is using a renormalized black body (grey body) with
energy density $w$ and temperature $T$ for each one of the target photon fields considered, such that $n(\nu)=\sum_j n_j(\nu),$ with
\begin{equation}
n_j(\nu)=\frac{15 w_j h^3}{\pi k T_j^4} \frac{\nu^2}{\exp \left(\frac{h \nu}{k T_j} \right)-1}.
\end{equation}

The other possibility allowed by the code is to introduce a synthesized target photon distribution coming from other codes (e.g.,
GALPROP\footnote{\citet{porter06}}). The code provides tools to transform the GALPROP outputs into a proper format (see section \ref{sec2.1.7}).

The synchrotron spectrum generated by pairs is also a target photon field for IC interaction. This process is called synchrotron self-Compton
interaction (SSC). The photon distribution coming from the synchrotron spectrum is calculated assuming an homogeneous distribution of sources in
a spherical volume $V_{sync}=(4/3)\pi R_{sync}^3$, such that $R_{sync} \le R_{PWN}$ \citep{atoyan96},
\begin{equation}
n_{SSC}(\nu,R_{sync}(t),t)=\frac{L_{sync}(\nu,t)}{4\pi R_{sync}^2(t) c} \frac{\bar{U}}{h \nu}.
\end{equation}

The volume $V_{sync}$ corresponds to the equivalent volume where the synchrotron radiation is contained. It is observed in many PWNe that the
radius of the synchrotron volume varies with the frequency. For instance, the radius of the Crab PWN in radio is 2 pc, while in X-rays is only
0.6 pc. The parameter $\bar{U}$ is $\sim$2.24 and corresponds to the spherical mean of the function $U(x)$ given by
\citep{atoyan96}
\begin{equation}
U(x)=\frac{3}{2} \int_0^\frac{R_{sync}(t)}{R_{PWN}(t)} \frac{y}{x} \ln \frac{x+y}{|x-y|} \mathrm{d}y.
\end{equation}

\begin{figure}
\centering
\includegraphics[width=0.5\textwidth]{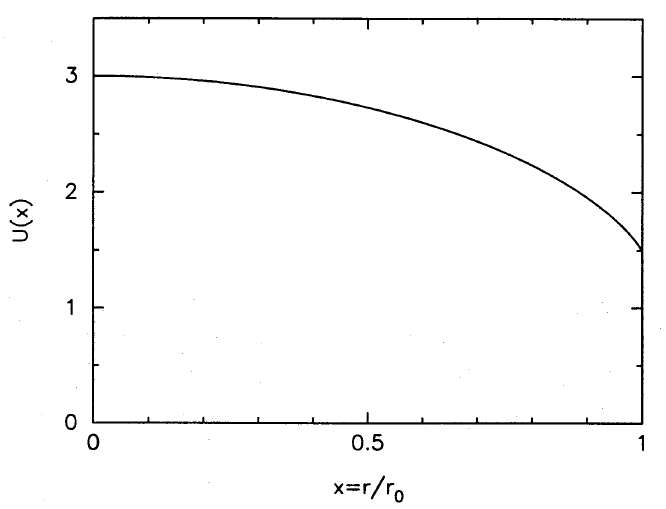}
\caption[The $U$ function]{Plot of the $U$ function inside the spherical source, i.e. the PWN \protect\citep{atoyan96}.}
\label{ux}
\end{figure}

In figure \ref{ux}, we see that the value of $U(x)$ at $x=0$ is $3$ and decreases until $1.5$ when $x=1$. If we would consider a distribution
greater than the radius of the PWN, then $U(x) \simeq 1/x^2$ and $n_{SSC}(\nu,R_{sync},t)=L(\nu,t)/(4 \pi R_{PWN}^2 c)$, as we should expect.

Finally, Bremsstrahlung luminosity per electron has the form \citep{blumenthal70}
\begin{equation}
P_{Brems}(\gamma_i,\nu)=\frac{3}{2 \pi} \frac{\alpha \sigma_T h c S}{\gamma_i^2} \left(\gamma_i^2+\gamma_f^2-\frac{2}{3} \gamma_i \gamma_f \right) \left(\ln \frac{2 \gamma_i \gamma_f m c^2}{h \nu}-\frac{1}{2} \right),
\end{equation}
where $S$ as defined in equation (\ref{s}). $\gamma_i$ and $\gamma_f$ are the energy of the electron before and after interaction and are related
by the kinematic condition
\begin{equation}
\gamma_i-\gamma_f=\frac{h \nu}{m_e c^2}.
\end{equation}
For a given frequency, there is a minimum initial energy for the electron to produce the scattered photon. This minimum energy is
\begin{equation}
\gamma_i^{min}(\nu)=\frac{1}{2}\left[\frac{h \nu}{m_e c^2}+\sqrt{\left(\frac{h \nu}{m_e c^2} \right)^2+\frac{2h \nu}{m_e c^2} \exp \left(\frac{1}{2} \right)} \right].
\end{equation}
With all these conditions, the total Bremsstrahlung luminosity is
\begin{equation}
\label{bremslum}
L_{Brems}(\nu,t)=\frac{3}{2 \pi} \alpha \sigma_T h c S \int_{\gamma_i^{min}(\nu)}^{\gamma_{max}} \frac{N(\gamma_i)}{\gamma_i^2} \left(\gamma_i^2+\gamma_f^2-\frac{2}{3} \gamma_i \gamma_f \right) \left(\ln \frac{2 \gamma_i \gamma_f m c^2}{h \nu}-\frac{1}{2} \right) \mathrm{d} \gamma_i.
\end{equation}

\subsection{PWN expansion}

The radius of the nebula is necessary not only for the calculation of the magnetic field, but also for the adiabatic losses of the nebula. The
evolution is described only for the free expansion phase. For $\dot{E} \simeq \dot{E}_0$, the expression for the radius is described by equation
(\ref{pwnrad}) and used in the first versions of the code. Currently, we do a more detailed calculus. We use the first law of the thermodynamics
\begin{equation}
\label{dq}
\mathrm{d}q=\dot{E}(t)\mathrm{d}t-P_{PWN}(t)\mathrm{d}V_{PWN},
\end{equation}
where $q$ is the thermal energy of the nebula, $P_{PWN}$ the internal pressure and $V_{PWN}$ the volume of the nebula. The spin-down luminosity
is given by equation (\ref{edotevol2}). Equation (\ref{dq}) can be written in a more explicit way
\begin{equation}
\label{pwnexpan}
\frac{d}{dt}\left[\frac{4 \pi P_{PWN}(t) R_{PWN}^3(t)}{3(\gamma_{PWN}-1)} \right]=\eta \dot{E}_0\left(1+\frac{t}{\tau_0} \right)^{-\frac{n+1}{n-1}}-\frac{4}{3}\pi R_{PWN}^2(t)P_{PWN}(t)\left(\frac{dR_{PWN}(t)}{dt} \right).
\end{equation}

\begin{figure}
\centering
\includegraphics[width=0.6\textwidth]{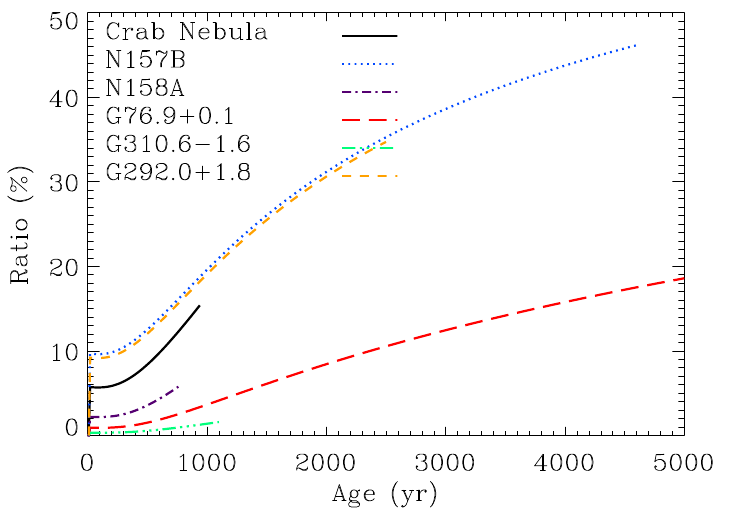}
\caption[Ratio of the PWN radii from analytical and numerical models]{Ratio of the PWN radii resulting from the analytical and numerical models
as commented in the text for the nebulae studied in \protect\citet{martin14b}.}
\label{pwnexpanfig}
\end{figure}

The internal pressure is given by \citep{vanderswaluw01}
\begin{equation}
P_{PWN}(t)=\frac{3}{25}\rho_{ej}(t)\left(\frac{R_{PWN}(t)}{t} \right)^2,
\end{equation}
being $\rho_{ej}(t)=3M_{ej}/(4 \pi R_{ej})$ the density of the ejecta. $R_{ej}=V_0 t$ is the radius of the SNR forward shock with $V_0$ defined
as in equation (\ref{v0}). Equation (\ref{pwnexpan}) is integrated numerically using an first order explicit scheme, which is stable enough in
this case. In figure \ref{pwnexpanfig}, we show the difference as a function of time of using equation (\ref{pwnrad}) and the numerical solution
of equation (\ref{pwnexpan}) applied to the PWNe modeled in \citet{martin14b}. Note that the difference in some cases could reach more than
40\%.

\subsection{Code design}

TIDE-PWN is written in FORTRAN language. It is divided in five independent executables, which compute separately the pair distribution
({\it nspectrum}) and the synchrotron ({\it synclum}), IC ({\it iclum}), SSC ({\it ssclum}), and Bremsstrahlung spectra ({\it bremslum}). There
are also useful tools incorporated to sum up the components of the spectrum, integrate the spectra, etc. These tools are describe in section
\ref{sec2.1.7}.

\begin{figure}
\centering
\includegraphics[width=1.0\textwidth]{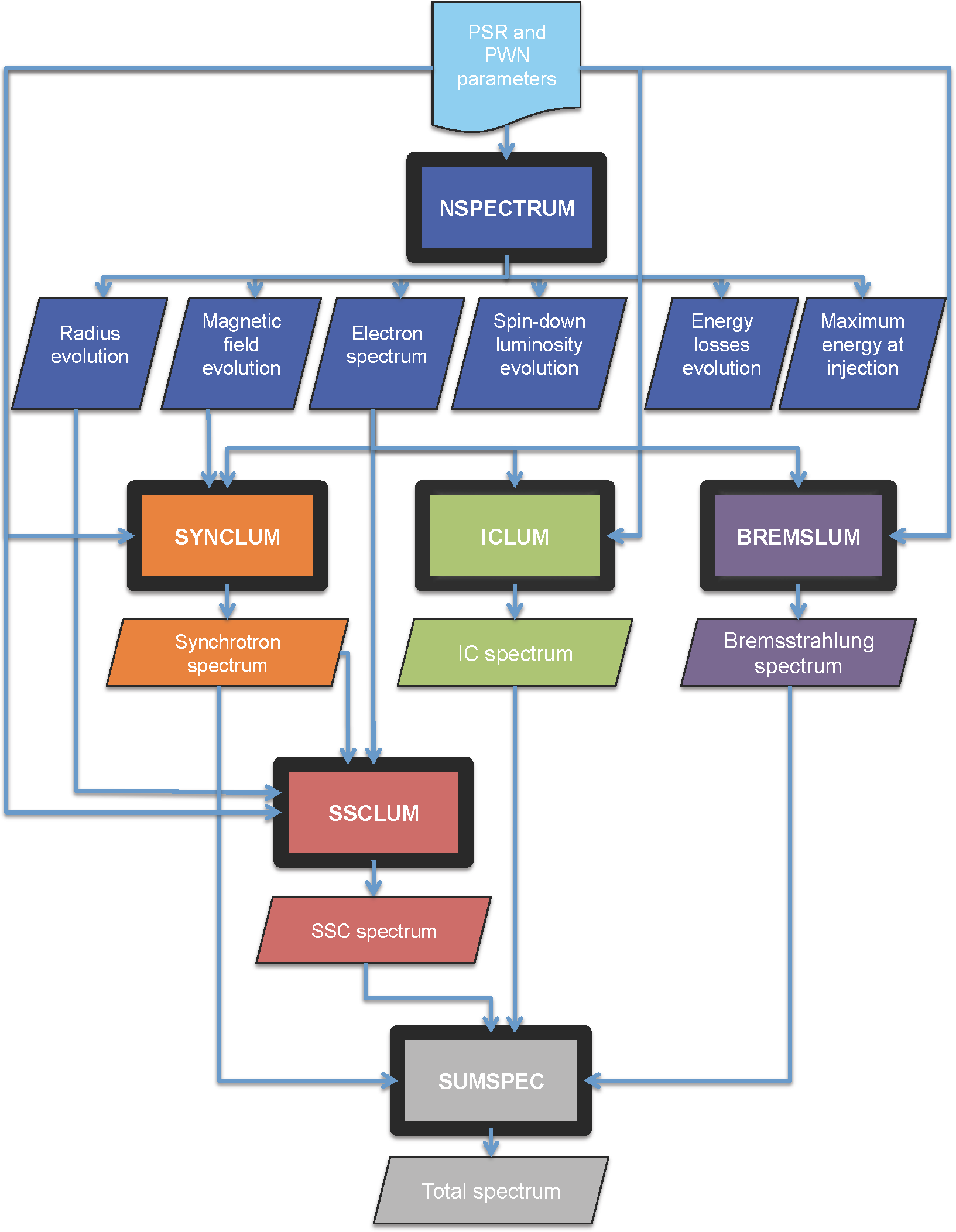}
\caption[Flowchart for a complete run of TIDE-PWN.]{Flowchart for a complete run of TIDE-PWN. The parameters, modules and files involved in each
program are explained in the text.}
\label{tidefc}
\end{figure}

In a complete normal run of the code (see figure \ref{tidefc}), the parameters of the problem are edited in the external file
\texttt{parameters.txt}. The list of physical parameters is: age of the PSR (yr), braking index, distance (kpc), energy of the SN explosion
(erg), ejected mass (in $M_\odot$), minimum energy of injection (Lorentz factor units), injection low and high energy indices, containment
factor, magnetic fraction, temperature (K) and energy density (erg cm$^{-3}$) of the target photon field (up to five different target fields),
hydrogen ISM density (cm$^{-3}$) and helium/hydrogen density ratio, the time step in the integration of the diffusion-loss equation (yr) and the
number of points in energy where we compute the pair distribution. Other important parameters in the file, but less usual to modify, are those
which define the frequency range where we integrate the spectra components, the name of the output files and the path where the code has to save
the outputs and read them when it is needed. Additional outputs as the magnetic field, spin-down luminosity, radius and energy losses evolution
can be demanded or avoided. As we explained in section \ref{sec2.1.4}, the target photon field can be also provided by an external code. The
TIDE-tool {\it gtotide} converts the target fields generated by GALPROP to TIDE-format files called \texttt{cmb.txt}, \texttt{fir.txt} and
\texttt{nir.txt}, which are inputs of our simulation. As we observe in figure \ref{tidefc}, {\it nspectrum} computes the pair distribution
function at the given age of the PSR (file named by default \texttt{electron\_spectrum.txt}). The file \texttt{electron\_spectrum.txt} provides
the pair distribution where we integrate the PWN spectrum in the rest of the codes. All the photon spectra are summed using the tool
{\it sumspec}, which requires the additional input file \texttt{spectrum\_list.txt} where we enter the list of the spectra that we want to join.

TIDE-PWN is written to have an easy interaction with the user and to allow the implementation of user-supplied routines to extend its
applicability to other kind of physical problems as radiative processes in SNRs, cosmic-rays, etc. In order to get this, the code has been
written using a modular structure, where the different ingredients are located in independent files. The executables work independently allows to
compute only those component in which we are interested or use pair distributions given by other codes or given by analytical formulae. A brief
description of each executable and their modules is explained below.

\subsubsection{nspectrum}

{\it nspectrum} is the program that we use when we want to obtain the pair distribution function solving the difussion-loss equation. The code is
written in six different archives which include functions and subroutines to compute the evolution of the parameters of the diffusion-loss
equation. In a complete run, {\it nspectrum} generates the outputs not only of the pair distribution, but also the magnetic field, energy losses,
spin-down luminosity and PWN radius evolution. The code \texttt{nspectrum.f} includes the general algorithm to read the parameters, call the
subroutines needed during the execution of the program and generates the output files. \texttt{injection.f} computes the particles injected into
the nebula in each time step taking into account the evolution of the spin-down luminosity. The evolution of the latter is also written in the
same file. The evolution of the magnetic field is performed by \texttt{magfield.f} and the PWN expansion is driven by \texttt{pwnexp.f}. The
escape terms are calculated in \texttt{escape.f} and the energy losses formulae are written in \texttt{losses.f}. The main output is a
three-column file called \texttt{electron\_spectrum.txt} (name by default), where we keep the pair distribution as a function of the energy. The
information description by columns is the following: energy (Lorentz factor units), $N(\gamma,t)$ (Lorentz factor units$^{-1}$) and
$\gamma^2 m_e c^2 N(\gamma,t)$ (erg).

Apart from the file \texttt{electron\_spectrum.txt}, the other generated files are: \texttt{magfield.txt} (magnetic field evolution),
\texttt{linj.txt} (spin-down luminosity evolution), \texttt{rpwn.txt} (PWN expansion) and \texttt{losses(1-10).txt} (energy losses evolution).
The current age (yr), magnetic field (G), spin-down luminosity (erg s$^{-1}$), radius of the PWN (pc) and maximum energy of the particles at
injection (Lorentz factor units) is shown in the screen at the end of the run.

\subsubsection{synclum}

The synchrotron spectrum of a given pair distribution is computed by {\it synclum}. Apart from the parameters given in \texttt{parameters.txt},
there are two additional inputs. The first one is the pair distribution, which can be provided by {\it nspectrum} or an external code in a list
format, or we can modify the analytical function in the module \texttt{analytic.f}. The second additional input is the current magnetic field of
the nebula. We introduce it by hand in the screen. The module \texttt{synclum.f} contains the integration routine to compute equation
(\ref{synclum}). {\it synclum} generates a five-column output file called \texttt{sync\_spectrum.txt} (name by default). The description of
columns is: frequency (Hz), differential luminosity $L(\nu)$ (erg s$^{-1}$ Hz$^{-1}$), luminosity $\nu L(\nu)$ (erg s$^{-1}$), differential flux
$F(\nu)$ (erg s$^{-1}$ cm$^{-2}$ Hz$^{-1}$) and flux $\nu F(\nu)$ (erg s$^{-1}$ cm$^{-2}$).

\subsubsection{iclum}

The program {\it iclum} computes the IC spectra coming from the target photon fields given in the file \texttt{parameters.txt} or the fields
provided in a list format, i.e. \texttt{cmb.txt}, \texttt{fir.txt} and \texttt{nir.txt}. As in {\it synclum}, the pair distribution function is
given in a list format or using the analytical defined in the module \texttt{analytic.f}. The integration routine is written in \texttt{iclum.f}.
{\it iclum} can generate up to five output files depending on the number of the target photon fields considered for interaction (one file for
each field). The format of the output files is the same as in {\it synclum}.

\subsubsection{ssclum}

The integration algorithm of {\it ssclum} is very similar as in {\it iclum}, but both routines has been written separately because the target
photon spectrum (in this case the, synchrotron spectrum) can be treated with a greater level of complexity and include some morphology. The pair
distribution input works as in the other programs already explained, but we need to introduce the current radius of the PWN (or generally, the
radius of the volume where pairs are contained) as an additional input by screen. The synchrotron radiation produced by pairs is also demanded
(only in list format), thus it is necessary to run first {\it synclum} before start {\it ssclum}. The output file is called
\texttt{ssc\_spectrum.txt} by default and has the same format as in {\it synclum}.

\subsubsection{bremslum}

{\it bremslum} computes the Bremsstrahlung radiation produced by the interaction with the electric field generated by pairs and the ions of the
ejecta. Apart from the \texttt{parameters.txt} file, we only need the pair distribution function, which can be provided in a list format or by a
function in the module \texttt{analytic.f}. The integration routine is written in \texttt{bremslum.f}. The output file is
\texttt{brems\_spectrum.txt} by default and has the same format as in {\it synclum}.

\subsection{Other tools}
\label{sec2.1.7}

TIDE-PWN provides other tools useful for the analysis of the spectra obtained. In this section we make a brief description of these supplementary
codes.

\begin{itemize}
\item {\bf sumspec}: it is used to sum up all the contributions of the PWN spectrum (see figure \ref{tidefc}). The input file is
\texttt{spectrum\_list.txt}, which contains a list with the names of the files containing the spectrum information. {\it sumspec} generates the
output file \texttt{photon\_spectrum.txt} (name by default) with the same format described for {\it synclum}. Up to eight components can be
summed at the same time.
\item {\bf lsdlum}: this program computes the total energy injected by the central pulsar during its life via spin-down luminosity. It solves the
equation
\begin{equation}
E_{total}=\int_0^t \dot{E}_0 \left(1+\frac{t'}{\tau_0} \right)^{-\frac{n+1}{n-1}} \mathrm{d}t'.
\end{equation}
The input parameters are demanded by screen: initial luminosity (erg s$^{-1}$), initial spin-down age (yr), braking index and age of the PSR
(yr). The output is given by screen in erg.
\item {\bf nelec}: it computes the total energy contained in the electron spectrum $E_p$, such that
\begin{equation}
E_p=\int_{\gamma_{min}}^{\gamma_{max}} \gamma m_e c^2 N(\gamma,t) \mathrm{d}\gamma.
\end{equation}
The only input is the electron spectrum in a list format, i.e. \texttt{electron\_spectrum.txt}. The result is given by screen in erg.
\item {\bf luminteg}: this program integrates the photon spectrum in a given frequency range. It solves the integral
\begin{equation}
L=\int_{\nu_l}^{\nu_h} L(\nu) \mathrm{d}\nu,
\end{equation}
where $\nu_l$ and $\nu_h$ are the low and high limits of the desired frequency range. These limits are introduced by screen. The photon spectrum
has to be written in a list format. The output is shown by screen and gives the integrated flux (erg s$^{-1}$ cm$^{-2}$) and its decimal
logarithm, and the integrated luminosity (erg s$^{-1}$) and its decimal logarithm.
\item {\bf radiationFields}: the program {\it radiationFields} is written in IDL language and obtains the target photon fields calculated by
GALPROP. {\it radiationFields} reads a fits file incorporated in the TIDE-PWN package which contains the simulation of the CMB, FIR and NIR
photon fields. The input files introduced by screen are: galactic latitude and longitude (degrees), and distance from the Earth (kpc). It
generates two output files called \texttt{table1.dat} and \texttt{table2.dat}. \texttt{table1.dat} is a one-column file with the wavelengths
where the photon fields are computed in $\mu$m, and \texttt{table2.dat} contains the energy density distribution of the CMB, FIR and NIR fields
in $\mu$m erg cm$^{-3}$ $\mu$m$^{-1}$.
\item {\bf integ\_ph}: it computes the energy densities of the CMB, FIR and NIR fields provided by GALPROP. There input files are
\texttt{table1.dat} and \texttt{table2.dat}. In this case, there is no need to used {\it gtotide} previously. The energy densities are shown in
the screen in eV cm$^{-3}$.
\item {\bf gtotide}: This program reads the files provided by GALPROP and change its format creating three new files in a tide-friendly format.
The inputs are \texttt{table1.dat} and \texttt{table2.dat}. The outputs files are called \texttt{cmb.txt}, \texttt{fir.txt} and \texttt{nir.txt}.
In each file, we find two columns: the first one is frequency (Hz) and the second one is the photon density per unit frequency (cm$^{-3}$
Hz$^{-1}$).
\end{itemize}

\section{The Crab Nebula}
\label{sec2.2}

We apply the model to the fit the spectrum of the Crab Nebula. The Crab Nebula is the most studied PWN in the whole electromagnetic spectrum. The
distance to the Crab nebula is only 2 kpc \citep{manchester05}. The period of the PSR and its derivative are obtained from \citep{taylor93b}.
Assuming that the moment of inertia of the Crab PSR is $I=10^{45}$ g cm$^2$, and using equation (\ref{edot}), we obtain the spin-down luminosity
power today ($4.5 \times 10^{38}$ erg s$^{-1}$). The expansion of the PWN is considered using the free expansion approximation given by
\citet{vanderswaluw01} (equation \ref{pwnrad}). We consider a characteristic energy for the SN explosion of $10^{51}$ erg and an ejected mass of
9.5$M_\odot$ \citep{bucciantini11}. All the parameters used for the Crab nebula are summarized in table \ref{crabconstrains}, including those
determined by the code.

\begin{table}[t!]
\centering
\scriptsize
\caption[Crab Nebula fit parameters]{Summary of the physical magnitudes used or obtained for the Crab Nebula fit at the current age. A few
parameters are fixed based on prior input or hypothesis.}
\vspace{0.2cm}
\label{crabconstrains}
\begin{tabular}{llll}
\hline
Magnitude & Symbol & Value & Origin or Result\\
\hline
Age (yr) & $t_{age}$ & 940 & fixed\\
Period (ms) & $P(t_{age})$ & 33.4033474094 & from \citet{taylor93b}\\
Period derivative (s  s$^{-1}$) & $\dot{P}(t_{age})$ & 4.209599 $\times 10^{-13}$ & from \citet{taylor93b}\\
Spin-down luminosity now (erg s$^{-1}$) & $L(t_{age})$ & $4.53 \times 10^{38}$ & equation (\ref{edot})\\
Moment of inertia (g cm$^2$) & $I$ & $10^{45}$ & fixed\\
Breaking index & $n$ & 2.509 & from \citet{lyne88}\\
Distance (kpc) & $d$ & 2 & from \citet{manchester05}\\
Ejected mass ($M_\odot$) & $M_{ej}$ & 9.5 & from \citet{bucciantini11}\\
SN explosion energy (erg) & $E_0$ & $10^{51}$ & from \citet{bucciantini11}\\
\hline
Minimum energy at injection & $\gamma_{min}$ & $1$ & assumed\\
Maximum energy at injection at $t_{age}$ &  $\gamma_{max}(t_{age})$ & $7.9 \times 10^9$ & result of the fit\\
Break energy & $\gamma_b$ & $7 \times 10^5$ & result of the fit\\
Low energy index & $\alpha_1$ & 1.5 & result of the fit\\
High energy index & $\alpha_2$ & 2.5 & result of the fit\\
Containment factor & $\varepsilon$ & 0.3 & result of the fit\\
\hline
Initial spin-down luminosity (erg s$^{-1}$) & $L_0$ & $3.1 \times 10^{39}$ & result of the fit\\
Initial spin-down age (yr) & $\tau_0$ & 730 & equation (\ref{tau0})\\
Magnetic field ($\mu$G) & $B(t_{age})$ & 97 & result of the fit\\
Magnetic fraction & $\eta$ & 0.012 & result of the fit\\
PWN radius today (pc) & $R_{PWN}(t_{age})$ & 2.1 & equation (\ref{pwnrad})\\
\hline
CMB temperature (K) & $T_{CMB}$ & 2.73 & fixed\\
CMB energy density (eV/cm$^3$) & $w_{CMB}$ & 0.25 & fixed \\
FIR temperature (K) & $T_{FIR}$ & 70 & as in \citet{marsden84}\\
FIR energy density (eV/cm$^3$) & $w_{FIR}$ & 0.5 & as in \citet{marsden84}\\
NIR temperature (K) & $T_{NIR}$ & 5000 & as in \citet{aharonian97}\\
NIR energy density (eV/cm$^3$) & $w_{NIR}$ & 1 & as in \citet{aharonian97}\\
Hydrogen density (cm$^{-3}$) & $n_H$ & 1 & assumed\\
\hline
\hline
\end{tabular}
\end{table}

At the date of the pulsar period ephemerides (year 1994), the age of the pulsar was 940 yr. This is consistent with equation (\ref{edot}) and
helps to minimize the bias produced by the non-simultaneity of the multi-wavelength data points used, obtained from $\sim$1970 (radio) to 2008
(VHE). We checked that changing the ephemeris to the latest one (e.g. the one used by the Fermi–LAT
Collaboration\footnote{\url{http://fermi.gsfc.nasa.gov/ssc/data/access/lat/ephems/}}) introduces no visible change in the results.

In order to compute the IC energy losses and spectrum, we consider the CMB, FIR and NIR as a target photon fields. Each one of the latter two is
considered as a diluted blackbody \citep{schlickeiser02}. The temperature of the FIR (NIR) is considered as 70 (5000) K. The CMB is a blackbody
of temperature 2.73 K.

Regarding the magnetic field, in this fit we have assumed magnetic energy conservation as in \citet{tanaka10},
\begin{equation}
\int_0^t \eta L(t') \mathrm{d}t'=\frac{4 \pi}{3}R_{PWN}^3(t) \frac{B^2(t)}{8 \pi},
\end{equation}
thus, using equation (\ref{edot}) and solving for the field, we obtain
\begin{equation}
\label{bfield2}
B(t)=\sqrt{\frac{3(n-1)\eta L_0 \tau_0}{R^3_{PWN}(t)}\left[1-\left(1+\frac{t}{\tau_0} \right)^{-\frac{2}{n-1}} \right]}.
\end{equation}
A summary of how the parameters has changed with the new versions is done in the next section.

For the injection we use equation (\ref{inj}), where $Q_0(t)$ is calculated using equation (\ref{norminj}). The final luminosity power is given
by equation (\ref{edot}) and the initial spin-down power is determined using equation (\ref{edotevol2}), since we know the luminosity power
nowadays and the age of the PSR. For the ISM density in the Crab nebula, we take a fiducial value of 1 cm$^{-3}$. Thus, our free parameters in
order to fit the spectrum are the magnetic fraction $\eta$ and the containment factor $\varepsilon$. For the final fit, we get $\eta=0.012$ and
$\varepsilon=0.3$. The magnetic field value we get today is 97 $\mu$G, which is close to the 100 $\mu$G calculated by the MHD simulations done by
\citet{volpi08}. Table \ref{crabconstrains} clarifies which parameters come from observations or assumptions, and which parameters are used to
fit the data.

\begin{figure}
\centering
\includegraphics[width=1.0\textwidth]{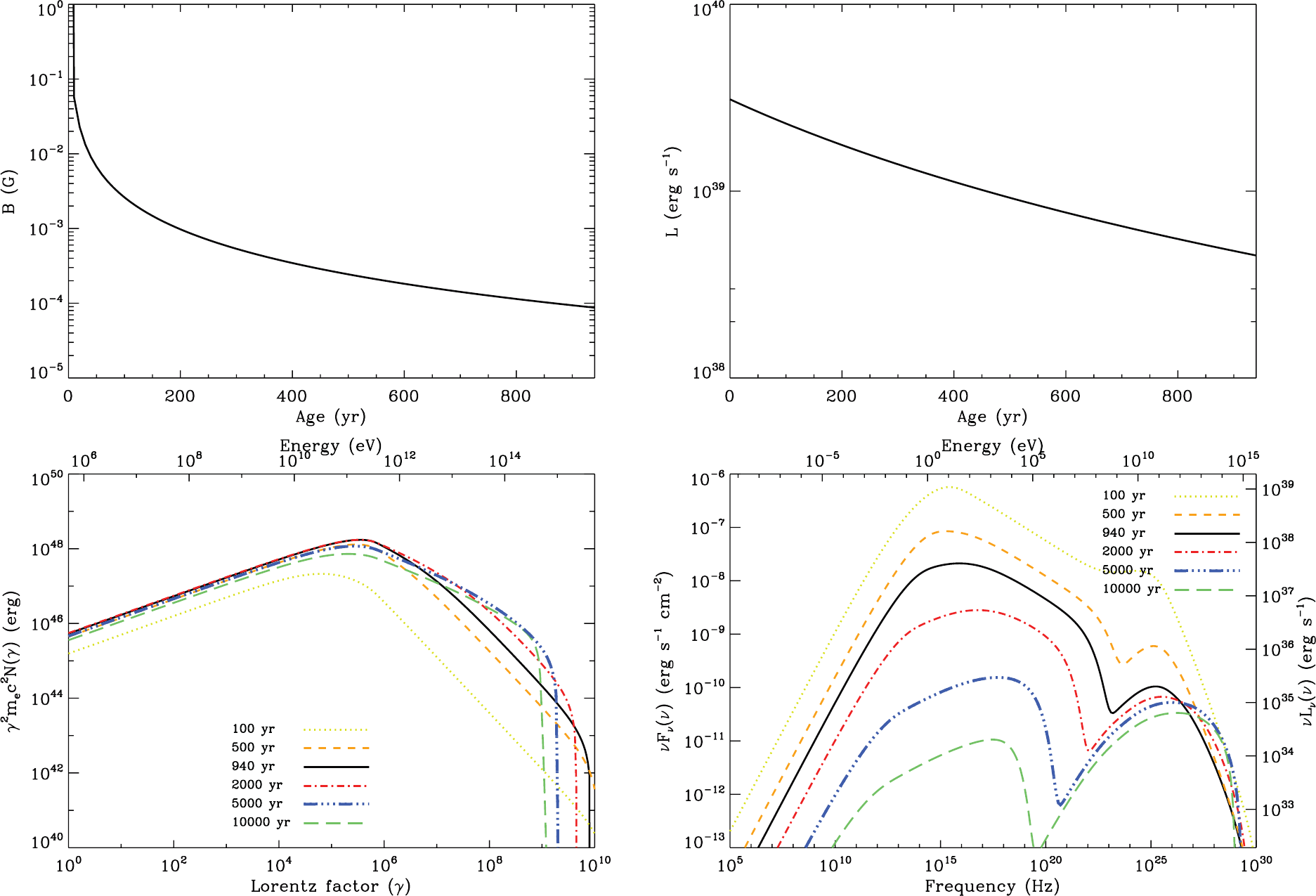}
\caption[Crab Nebula evolution]{Crab Nebula parameters and spectrum evolution. From top to bottom, left to right: Magnetic field, spin-down
power, lepton population, and spectral energy distribution of the Crab Nebula as a function of time.}
\label{crabevol}
\end{figure}

\begin{figure}
\centering
\includegraphics[width=0.7\textwidth]{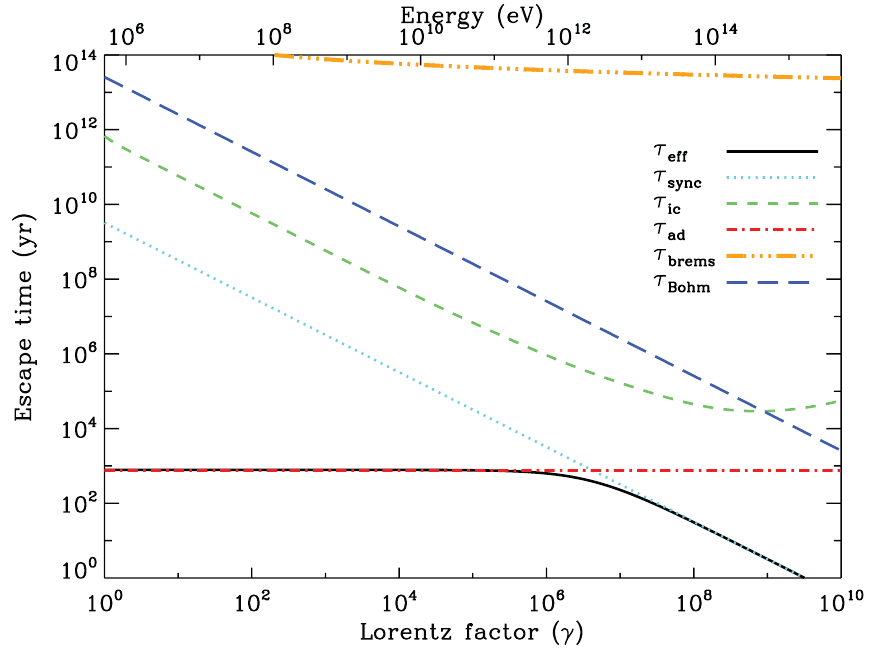}
\caption[Crab Nebula cooling times]{Cooling times for the Crab Nebula at $t_{age}=940$ yr. At low energies, the adiabatic losses are dominant
because their cooling time is of the same order of the pulsar age. At high energies, synchrotron losses become the most important.}
\label{crabtimes}
\end{figure}

\begin{figure}
\centering
\includegraphics[width=0.7\textwidth]{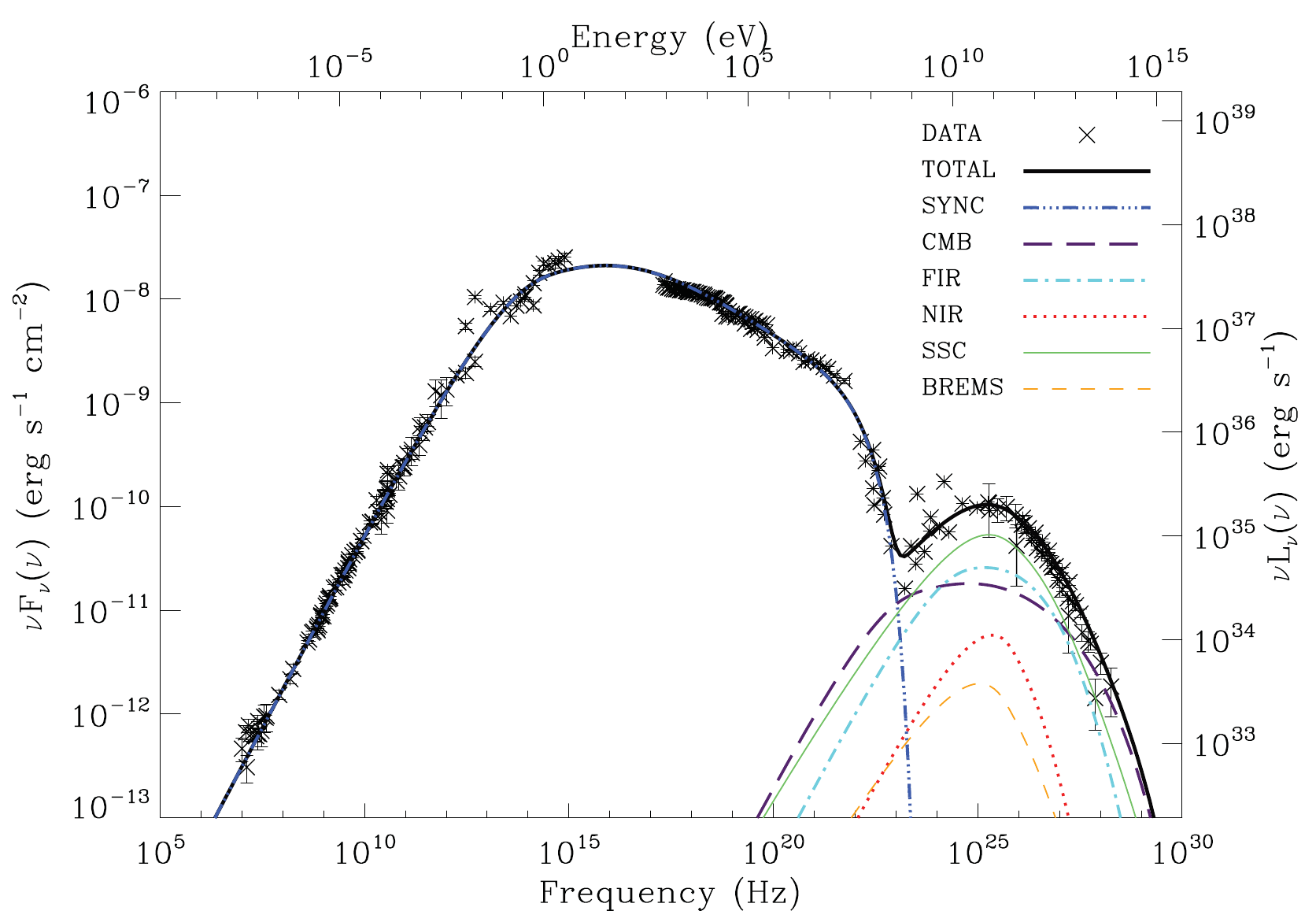}
\caption[Crab Nebula fitted spectrum]{Spectrum of the Crab Nebula fitted by our model. The data points are obtained from
\protect\citet{baldwin71,maciasperez10} for the radio band; \protect\citet{ney68,grasdalen79,green04,temim06} for the infrared;
\protect\citet{veroncetty93} for the optical; \protect\citet{hennessy92} for the ultraviolet; \protect\citet{kuiper01} for the X-rays and soft
$\gamma$-rays; and \protect\citet{abdo10a,aharonian04,aharonian06a,albert08a} for $\gamma$-rays.}
\label{crabfit}
\end{figure}

Figure \ref{crabevol} shows the magnetic field, spin-down power, lepton population and spectral energy distribution of the Crab nebula as a
function of time, resulting from our code after normalization to current measurements. The current cooling times for the different processes
considered are shown in Fig. \ref{crabtimes}, whereas the current spectrum is shown in figure \ref{crabfit}.

The SSC flux is the strongest contributor to the high-energy spectra, followed by IC with the CMB and the FIR. The Bremsstrahlung contribution is
not very important, but as it is similar to the NIR radiation, we do not neglect it in favour of the other contributions. Most of the radiative
considerations of \citet{tanaka10} are similarly obtained in our model, since they are driven by SSC domination. Our resulting value of the
magnetic field today is lower than that used by \citet{atoyan96} in their time-independent approach, who in turn adopted it from the
\citet{kennel84a} model, followed by an adjustment on the relativistic particle density to enable the data fitting. This value of magnetic field
is unrealistically high for our time-dependent spectral model, and a lower value is preferred also by MHD simulations.

\begin{figure}
\centering
\includegraphics[width=0.6\textwidth]{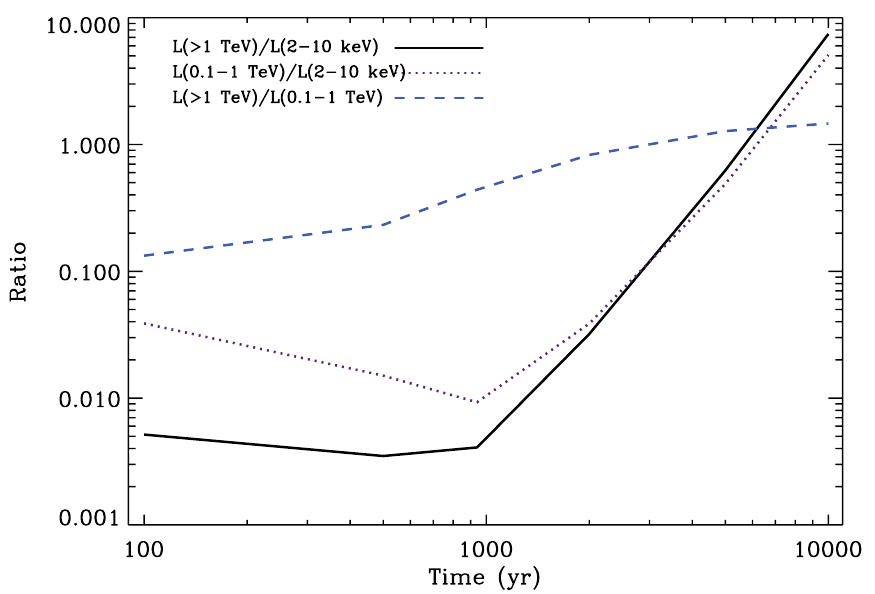}
\caption[Luminosity ratios for the Crab nebula]{Luminosity ratios for the Crab nebula evolution: $L(>$1 TeV)/$L$(2--10 KeV),
$L$(0.1--1 TeV)/$L$(2--10 KeV) and $L(>$1 TeV)/$L$(0.1--1 KeV).}
\label{crabratio}
\end{figure}

Regarding the time evolution presented in figure \ref{crabevol}, it is interesting to note how the peak of the electron distribution moves from
lower Lorentz factors to the energy break in the injection. This displacement of the peak is due to the high energy losses for energies lower
than the break at early ages. The maximum energy of the injection is decreasing with time and the maximum energy of the electron population is
decreasing also through energy losses, but at a slower rate due to the presence of high-energy electrons that were previously injected. The slope
of the distribution at VHE becomes flattened with time also due to evolution in time of the dominant cooling process, increasing the power of the
IC radiation. As the magnetic field falls, the synchrotron radiation diminishes with respect the IC radiation and at later ages (e.g. towards 10
kyr), the IC radiation contains most of the emitted flux. This is in agreement with the idea of older PWNe being still detectable at high
energies but being devoid of lower-energy counterparts \citep{dejager09b}. This is shown in figure \ref{crabratio}. We see that the flux at
energies $>$1 TeV and the gamma-ray flux are equal for an age of $\sim$5 kyr.

The radio and optical flux evolution of the Crab nebula show a decreasing-with-time behaviour. Measurements of the radio flux decrease were done
by \citet{vinyaikin07}, using data from 1977 to 2000 at 86, 151.5, 927 and 8000 MHz. The mean flux-decrease rate averaged obtained was (-0.17
$\pm$ 0.02)\% yr$^{-1}$. Using data obtained from our code at the same frequencies for the same time interval we obtained an averaged rate of
-0.2\% yr$^{-1}$. In optical frequencies, the continuum flux decreases 0.5$\pm$0.2\% yr$^{-1}$ at 5000 {\AA} \citep{smith03}. In this case, we
obtained directly from the model a flux decrease of 0.3\% yr$^{-1}$. The evolution of both luminosities as extracted from our model is in
agreement with observations.

\section{Improvements and caveats}
\label{sec2.3}

Since the first version of the code (TIDE-PWN v0.0), where the diffusion-loss equation was solved using the time-independent analytical solution
given by \citet{aharonian97} (equation \ref{nana}), there has been a lot of improvements in terms of the physical assumptions, design and
interaction with the user.

The time dependent code was introduced in the version v.1.0, where the spin-down luminosity varied as in equation \ref{edotevol2} and the
magnetic field as the prescription given by \citet{venter07}
\begin{equation}
B(t)=\frac{B_0}{\left(1+\frac{t}{\tau_B} \right)^\alpha},
\end{equation}
where $B_0$ is the initial magnetic field, $\tau_B$ a characteristic timescale and $\alpha$ is the index which regulates how fast the magnetic
field decays. The PWN expansion was considered in a ballistic approximation ($R_{PWN}(t)=v_{PWN}t$).

The CPU time cost to solve equation (\ref{nana}) is high and it is not flexible for future versions of the code or for taking into account other
similar physical problems. TIDE-PWN v1.1 solves the diffusion-loss equation in the numerical approach and the modular code struture described in
section \ref{sec2.1}. The magnetic evolution prescription given by \citep{tanaka10} using the magnetic field energy conservation assumption was
implemented in TIDE-PWN v1.2 (equation \ref{bfield2}). In this version, we also implemented a better approximation in the expansion of the PWN
and we introduced the free expansion model for constant spin-down luminosity deduced by \citet{vanderswaluw01} (equation \ref{pwnrad}). The model
for the Crab Nebula in the previous section was done using this version of the code \citep{martin12}.

In TIDE-PWN v1.3, we improved the estimate of the nebular magnetic field implementing the solution given in equation (\ref{bfield}) proposed in
\citet{pacini73}. Another important implementation were the creation of the TIDE-tools. In particular, {\it integ\_ph}, {\it lsdlum},
{\it nelec}, {\it radiationFields} and {\it sumspec}.

The rest of implementations as the tools {\it luminteg} and {\it gtotide}, the possibility to use the target photon fields generated by GALPROP,
the module \texttt{analytic.f} which allows an analytic function for the pair population, the free expansion model with varying $\dot{E}$ and the
current flowchart were introduce in the current version of the code presented in this chapter (v1.4).

In table \ref{crabcomp}, we show a summary of the parameters obtained for the Crab Nebula in different versions of the code. As we should expect,
the parameters that changes are those related with the magnetic field and the expansion of the nebula. The IC target photon fields change from
version 1.3 since we use the values provided by GALPROP.

\begin{table}
\centering
\scriptsize
\caption[Crab Nebula parameters for different versions of TIDE-PWN]{Physical magnitudes obtained for the Crab Nebula in the different versions of TIDE-PWN. The age considered
in all cases is 940 yr.}
\vspace{0.2cm}
\label{crabcomp}
\begin{tabular}{llll}
\hline
Magnitude & TIDE-PWN v1.2 & TIDE-PWN v1.3 & TIDE-PWN v1.4\\
\hline
$M_{ej}$ ($M_\odot$) & 9.5 & 9.5 & 8.5\\
$E_0$ (erg) & $10^{51}$ & $10^{51}$ & $10^{51}$\\
\hline
$\gamma_{min}$ & 1 & 1 & 1\\
$\gamma_{max}(t_{age})$ & $7.9 \times 10^9$ & $9.3 \times 10^9$ & $8.0 \times 10^9$\\
$\gamma_b$ & $7 \times 10^5$ & $7 \times 10^5$ & $7 \times 10^5$\\
$\alpha_1$ & 1.5 & 1.5 & 1.5\\
$\alpha_2$ & 2.5 & 2.5 & 2.5\\
$\varepsilon$ & 0.3 & 0.25 & 0.25\\
\hline
$L_0$ (erg s$^{-1}$) & $3.1 \times 10^{39}$ & $3.1 \times 10^{39}$ & $3.1 \times 10^{39}$\\
$\tau_0$ (yr) & 730 & 730 & 730\\
$B(t_{age})$ ($\mu$G) & 97 & 84 & 80\\
$\eta$ & 0.012 & 0.03 & 0.022\\
$R_{PWN}(t_{age})$ (pc) & 2.1 & 2.1 & 2.1\\
\hline
$T_{CMB}$ (K) & 2.73 & 2.73 & 2.73\\
$w_{CMB}$ (eV/cm$^3$) & 0.25 & 0.25 & 0.25\\
$T_{FIR}$ & 70 & 70 & 70\\
$w_{FIR}$ & 0.5 & 0.23 & 0.23\\
$T_{NIR}$ & 5000 & 5000 & 5000\\
$w_{NIR}$ & 1 & 0.56 & 0.56\\
$n_H$ (cm$^{-3}$) & 1 & 1 & 1\\
\hline
\hline
\end{tabular}
\end{table}

Despite the improvements, there is still a lot of work to do in order to reproduce the PWN spectrum in the most complete and realistic way as
possible. The first caveat of our model is that we only can reproduce the spectrum of young PWNe, i.e. PWNe which are still in its free expansion
phase. We are developing a new module to reproduce the evolution of the SNR and the trajectory of the reverse shock to take into account its
interaction with the PWN shell. When the reverse shock and the PWN shell collide, there is a bouncing process where the radius of the nebula is
reduced and the magnetic field increases and reduces the energy of particles through synchrotron losses, increasing the emitted synchrotron
radiation. After the bounce, the PWN reaches its Sedov phase, where the expansion behaves differently, as we already explaned in chapter
\ref{chap1}. The proper motion of the PSR may be not important during the free expansion phase, but after the interaction with the reverse shock,
the PSR can leave the PWN and powers it anymore (relic PWN) or re-enter in the nebula after the re-expansion. This possible discontinuity in the
injection is nowadays neglected. Our code is, as all the other radiative codes we know, a 1D spectral code, thus other effects related with
morphology are also neglected as injection and radiation anisotropies, magnetic field morphology, formation of the torus and jets, etc. All this
phenomenology is only observed for a few relatively young PWNe, but observations with the next generation observatories (e.g., CTA) will
increase, not only the number, but also the resolution of the images and morphology will be a key factor to understand the radiation and
acceleration mechanisms of these objects and their evolution.

\chapter[Effects in time-dependent modelling of PWNe]{Impact of approximations and effects of parameter variation in time-dependent modelling of
PWNe}
\label{chap3}

\ifpdf
    \graphicspath{{Chapter3/Figs/Raster/}{Chapter3/Figs/PDF/}{Chapter3/Figs/}}
\else
    \graphicspath{{Chapter3/Figs/Vector/}{Chapter3/Figs/}}
\fi

PWNe detected at TeV energies are associated with young ($\tau_c<10^5$ years, here, only the very young are considered) and energetic PSRs
($\dot{E}>10^{33}$ erg s$^{-1}$), and usually display extended emission up to a few tens of parsecs \citep{rieger13}. The majority of PWNe were
observed by the H.E.S.S. experiment during the Survey of the Galactic plane performed since 2004 (see \citealt{gast11} for the current status).
Up to that time, only the Crab Nebula has been detected having a steady $\gamma$-ray flux about 1 TeV \citep{weekes89}. Due to the lack of a
previous systematic study of the known young PWNe, some basic questions remain: How do approximations done in many radiative PWNe models affect
to the spectra in a time-dependent context? Why is Crab the only PWN that is self-synchrotron (SSC) dominated? Why are the PWNe that we see at
TeV energies particle dominated? Is there any observational bias behind this fact? At which sensitivity do we expect to map the whole phase space
between particle and magnetic dominated nebula? What defines TeV observability of PWNe?

In this chapter, we deal with these questions based on the works done in \citet{martin12,torres13b}.

\section[Impact of approximations]{Impact of approximations}
\label{sec3.1}

Apart from the approximations we focus below, one can also find many radiative approximations too in PWNe models: using a priori guesses for
which field is dominant in each environment, using mono-chromatic assumptions for synchrotron and IC, or using Thompson cross-section instead of
Klein–Nishina. These assumptions certainly simplify the treatment, but at the expense of assuming approximations for which their impact is
usually not checked. We have not adopted any of them here.

Regarding the diffusion-loss equation, the most usual approximation is to neglect the escape term (see e.g. \citealt{tanaka10,tanaka11}) to
obtain an advective differential equation (to abbreviate, we call it ADE). Using just this approximation in our complete model would lead to very
similar values for the magnetic field and magnetic fraction (needed to obtain a good fit for today's Crab nebula, when imposing a correct
contribution of the SSC such that it fits the high energy data today). This is because the Bohm time-scale is larger than the age of the Crab
Nebula and is not affecting strongly the particles' evolution (see figure \ref{crabtimes}). Another common (and additional) approximation is
neglecting the treatment of energy losses and instead replace it by the particle's escape time (see e.g. \citealt{zhang08,qiao09}). In this case,
equation (\ref{dle}) has the form
\begin{equation}
\label{tde}
\frac{\partial N(\gamma,t)}{\partial t}=-\frac{N(\gamma,t)}{\tau(\gamma,t)}+Q(\gamma,t).
\end{equation}
In equation (\ref{tde}), particles are not losing energy, but they are rather removed from the distribution after a certain time. This makes
equation (\ref{tde}) a partial differential equation in time (from now, TDE).

We use the fit of Crab Nebula to do the comparison between models. We include the fits in the ADE and TDE cases with the complementary
approximations done by \citet{tanaka10} (hereafter ADE-T) and \citet{zhang08} (hereafter TDE-Z). In ADE-T, the Bremsstrahlung energy losses and
its spectrum, and the FIR and NIR contributions into the IC energy losses and their spectrum, are ignored. Also, the maximum energy at injection
is fixed and the expansion of the PWN is modelled in a ballistic approximation ($R_{PWM}=v_{PWN} t$). All these approximations are not done in
the full treatment presented before, against which we compare. In the TDE-Z, only the synchrotron escape time is considered (thus ignoring all
other time-scales) and Bohm diffusion is used.

Table \ref{approxcomp} shows the parameters for each of the models needed to obtain a good fit of the Crab nebula data at the current age. The
column labelled value gives the parameters of the complete model in section \ref{sec2.2}.

\begin{table}
\centering
\scriptsize
\caption[Comparison of the values obtained in the different fits of the Crab Nebula today]{Comparison of the values used or obtained in the
different fits of the Crab Nebula today. We use dots for those parameters which have the same values as in the complete model. The dots appear
when no change is needed from those values.}
\label{approxcomp}
\vspace{0.2cm}
\begin{tabular}{llll}
\hline
Symbol & Value & {ADE-T} & {TDE-Z}\\
\hline
$L(t_{age})$ & $4.5 \times 10^{38}$  & \ldots & $2.5 \times 10^{38}$\\
\hline
$\gamma_{min}(t)$ & 1 & $10^2$ & \ldots\\
$\gamma_{max}(t)$ & $7.9 \times 10^9$ & $7 \times 10^9$ (fixed) & $6.5 \times 10^9$\\
$\gamma_b$ & $7 \times 10^5$  & $7 \times 10^5$ & $9 \times 10^5$\\
$\alpha_1$ & 1.5 & \ldots & \ldots\\
$\alpha_2$ & 2.5 & \ldots & \ldots\\
$\varepsilon$ & 0.3  & \ldots & \ldots\\
\hline
$L_0$ & $3.1 \times 10^{39}$ & \ldots & $1.7 \times 10^{39}$ \\
$B(t_{age})$ & 97 & \ldots & 93\\
$\eta$ & 0.012 & 0.006 & 0.015\\
$R_{PWN}$ & 2.1 & 1.7 & 1.9\\
\hline
$T_{CMB}$ & 2.73 & \ldots & \ldots\\
$w_{FIR}$ & 0.25 & \ldots & \ldots\\
$T_{FIR}$ & 70 & 0 & \ldots\\
$w_{FIR}$ & 0.5 & 0 & \ldots\\
$T_{NIR}$ & 5000 & 0 & \ldots\\
$w_{NIR}$ & 1 & 0 & \dots\\
$n_H$ & 1 & 0 & \ldots\\
\hline
\hline
\end{tabular}
\end{table}

Given that the observational parameters such as the age, the braking index, the period and the period derivative are fixed, they continue to
determine $\tau_0$ and $\tau_c$ in all models. For the TDE model, the break energy and the containment factor (and, in consequence, the maximum
energy at injection) have decreased. A smaller shock radius diminishes the number of VHE electrons, which is necessary due to the lack of energy
losses affecting the population, and the smaller energy break corrects the lack of radio flux. The initial spin-down luminosity is smaller
because the lack of losses makes electrons' disappearance slower. The magnetic fraction is larger to power the SSC contribution, and the energy
break increases to compensate the lack of escaping particles at low energies and correct the radio flux. In the ADE-T case, we take the expansion
velocity given by \citet{tanaka10} of 1800 km s$^{-1}$, which gives a radius for the PWN of 1.7 pc. This means that the synchrotron radiation is
confined in a smaller volume, so the synchrotron photon density is larger and the magnetic energy fraction needed to obtain the correct SSC
contribution is smaller than in our case. Note that the minimum and maximum energy at injection are also fixed in time according to the values
used in \citet{tanaka10}.

\begin{figure}
\centering
\includegraphics[width=1.0\textwidth]{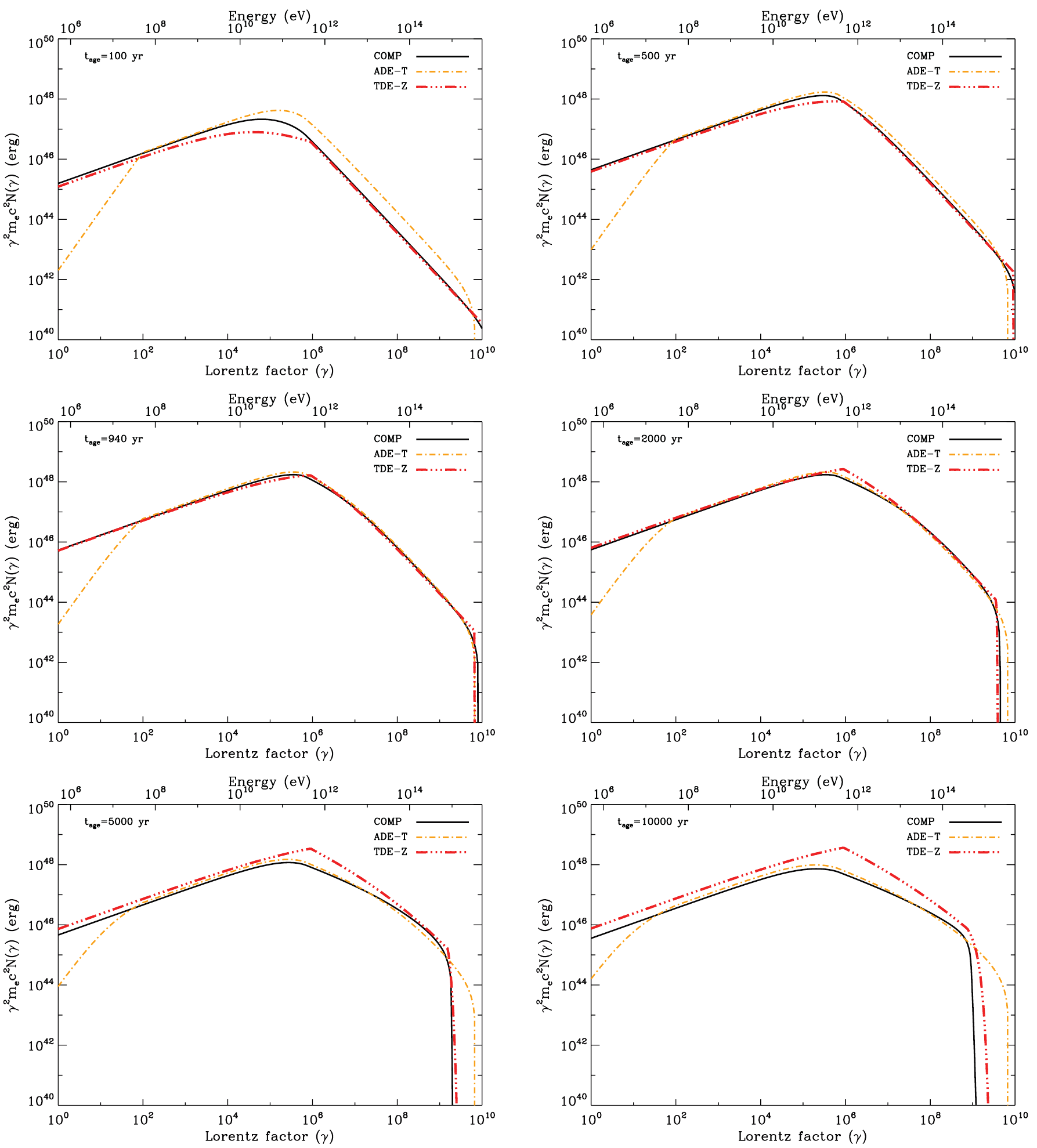}
\caption[Crab Nebula electron evolution in comparison with other models]{Electron distribution of the Crab Nebula computed for different ages
using the complete model together with the obtained results under the ADE-T and TDE-Z approximations.}
\label{elec_approx}
\end{figure}

\begin{figure}
\centering
\includegraphics[width=1.0\textwidth]{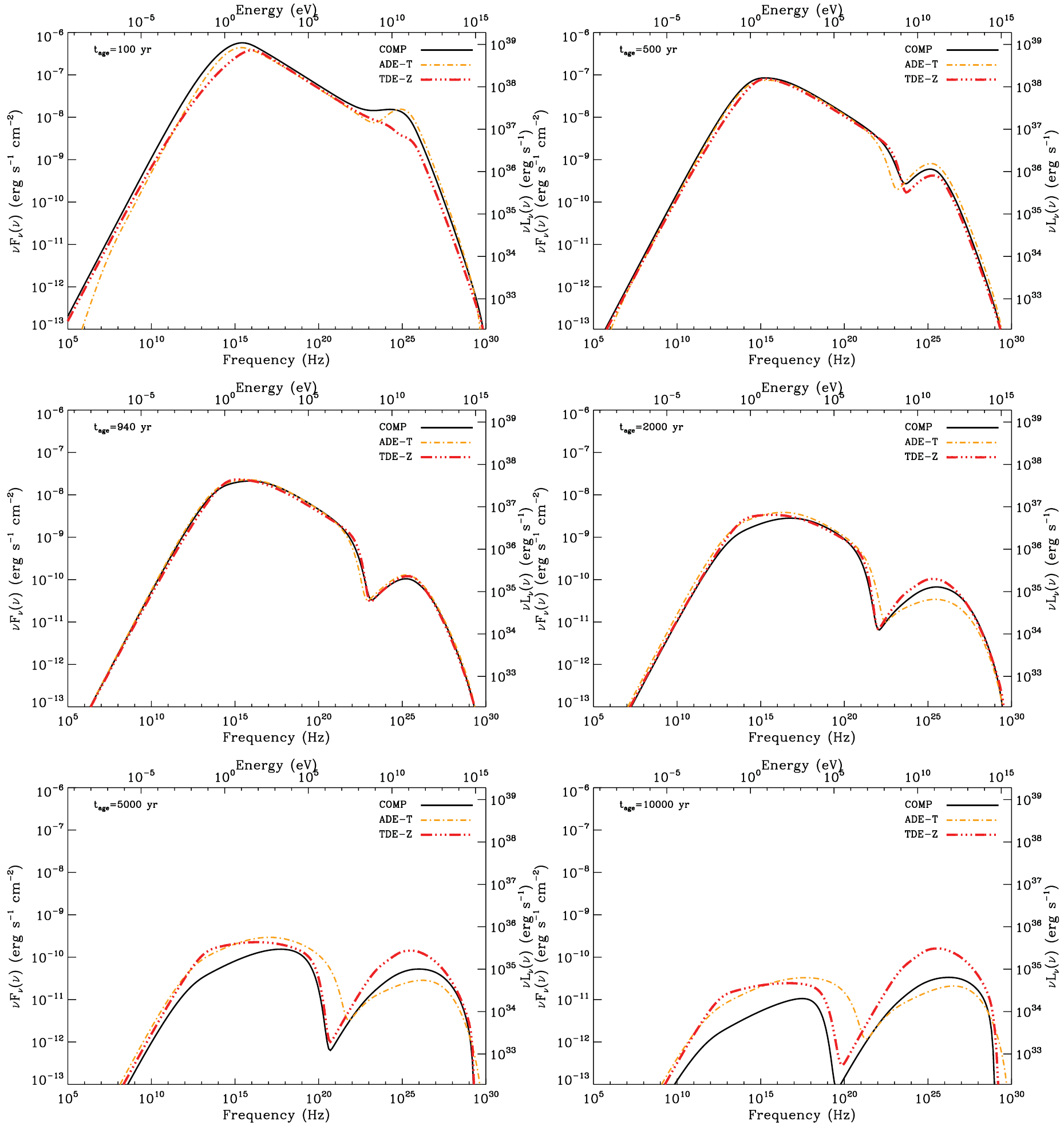}
\caption[Crab Nebula spectrum evolution in comparison with other models]{Photon spectrum of the Crab Nebula computed for different ages using the
complete model, together with the obtained results under the ADE-T and TDE-Z approximations.}
\label{spec_approx}
\end{figure}

\begin{figure}
\centering
\includegraphics[width=1.0\textwidth]{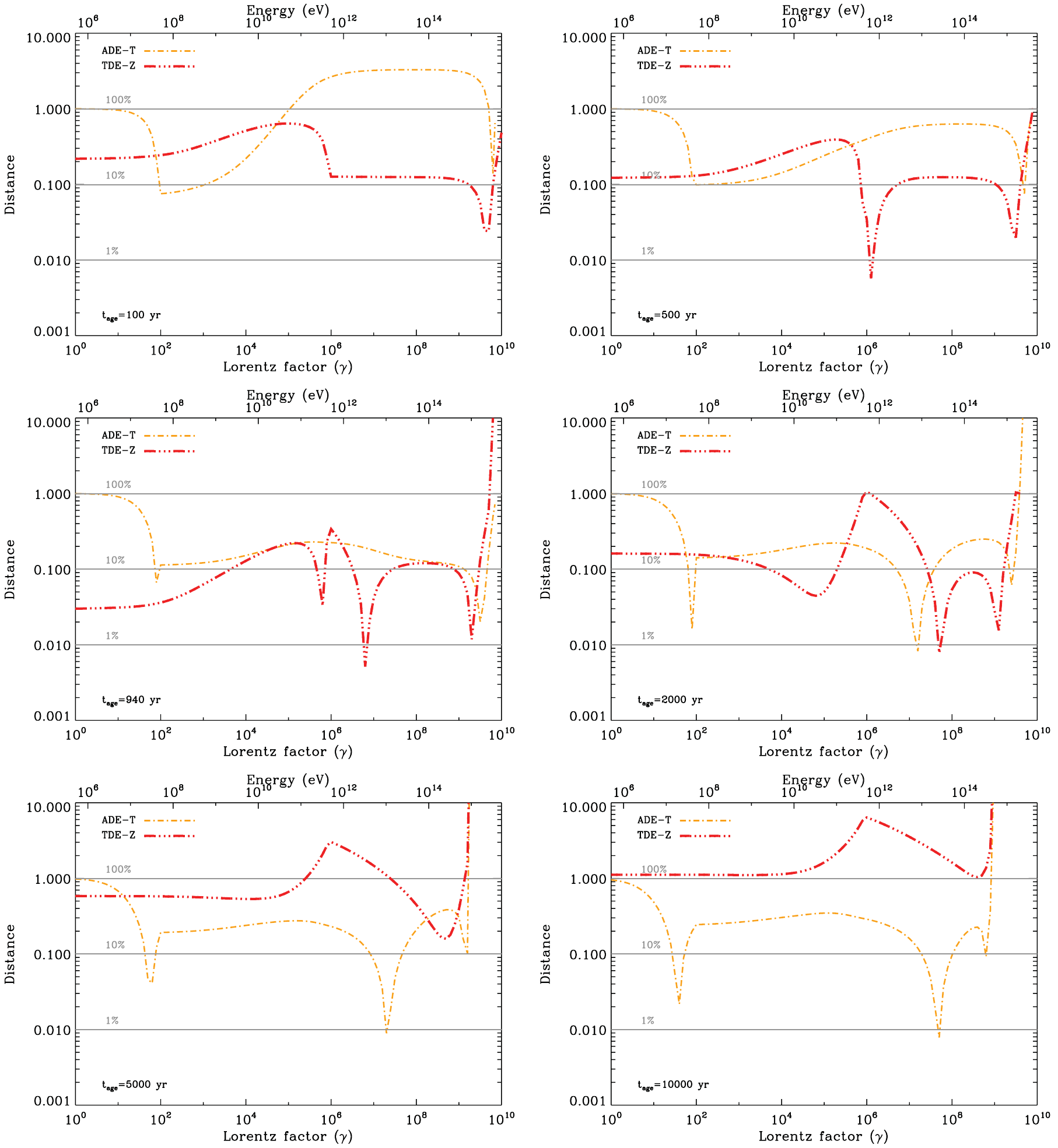}
\caption[Relative distance of the Crab Nebula electron distribution with other models]{Relative distance of the results for the electron
distribution between the complete model, the ADE-T and TDE-Z approximations for the Crab Nebula at different ages.}
\label{elec_dif}
\end{figure}

\begin{figure}
\centering
\includegraphics[width=1.0\textwidth]{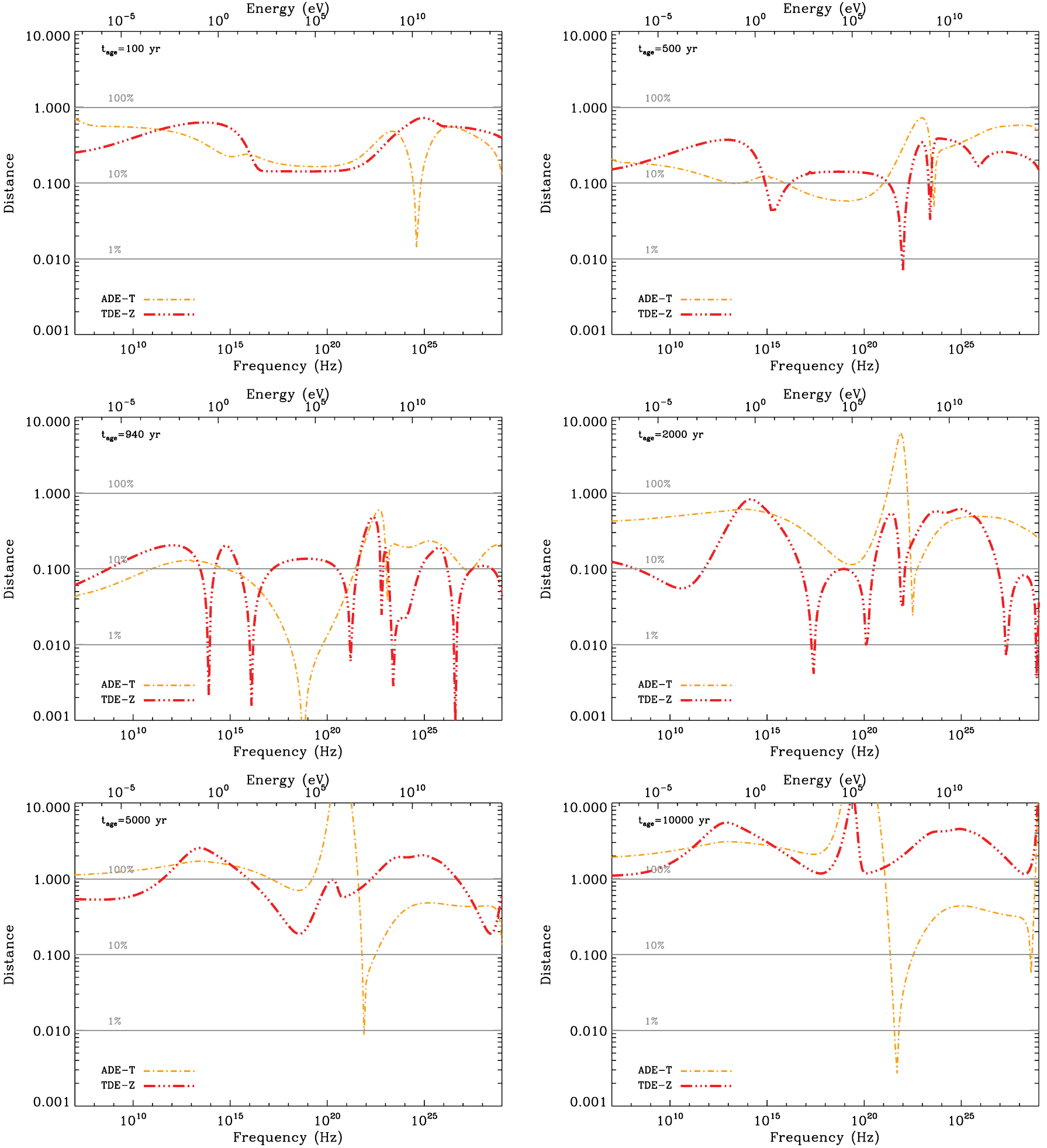}
\caption[Relative distance of the Crab Nebula spectrum with other models]{Relative distance of the results for the photon spectrum between the
complete model, the ADE-T and TDE-Z approximations for the Crab Nebula in different ages.}
\label{spec_dif}
\end{figure}

It is clear that at the current age, and particularly due to the fact of the strong SSC domination of the Crab nebula, one can find acceptable
sets of parameters in both approximated models that fit the data well. However, this does not mean that the time evolution of these models would
be similarly close to the complete analysis. Figures \ref{elec_approx} and \ref{spec_approx} compare the evolution of the results of the complete
model and the approximate ones for the electron population and photon spectra, respectively. Differences increase with the time elapsed off the
normalization age (the current Crab nebula), and are clear at a few hundred and a few thousand years. To have a better idea on how the spectra
are changing, we use our model as reference data and calculate the relative theoretical distance of the ADE-T and TDE-Z models with respect to it
as a function of frequency, both for the electron population and spectrum. We thus compute the theoretical distance $D$ as
$D=|F_{comp}-F_{approx}|/F_{comp}$, so that distance times 100 per cent is the percentile value of the deviation. $F_{comp}$ and $F_{approx}$
are the fluxes obtained by the complete and approximated model, respectively. These results are given in figures \ref{elec_dif} and
\ref{spec_dif}. The dips visible in these figures correspond to the crossing of the curves (lepton Lorentz factors or photon energies where both
the approximate and complete models, as plotted in figures \ref{elec_approx} and \ref{spec_approx}, coincide). The recovering of the curves after
these dips correspond to the use of the absolute value in the Distance definition. Visually removing these dips gives an idea of the average
deviation between the approximate and complete models across all energies. Note that the peaks in the relative distance evolution are broader,
and represent bona fide increases in the deviation of the approximate results.

Regarding the underlying electron population, we see that differences among models range from 10 to 100 per cent and beyond. The TDE-Z models
have a more deviating behaviour than the ADE-T models at later ages. Regarding the photon spectrum deviation, we find that for ages close to
$t_{age}=940$ yr and lower, the relative distance of all models with respect to the complete one is below 40 per cent with the exception of the
frequency range between 10$^{22}$ and 10$^{23}$ Hz, where there is a transition between the synchrotron and IC dominated radiation. At larger
times, deviations can be larger than 100 per cent. From 2 to 10 kyr, the relative distance in the optical range and in gamma-rays increases with
age. As soon as the Crab nebula is let to evolve beyond a few thousand years, and consistently with the results found for the electron
population, the relative distance between the spectra of the complete and the approximate models goes up to a factor of a few (i.e. percentile
distance is a factor of a few 100 per cent) over large portions of the electromagnetic spectrum. These changes in the nebula evolution are only
the by-product of the approximations used in the models and do not represent the expected behaviour of the source.

In conclusion, the time-evolution of the electron population and the photon spectrum deviation from the complete analysis are larger than 100 per
cent, when they evolve in time off the normalization age. This puts in evidence the risks of considering approximations when studying time
evolution, as well as, equivalently, when members of a population observed at different ages are analysed with the intention of extracting
statistical conclusions.

\section[Effects of $B$, age and $\dot{E}$ on Crab-like PWNe]{Effects of magnetic field, age and intrinsic luminosity on Crab-like PWNe}
\label{sec3.2}

To investigate the behavior of a Crab-like PWNe with different parameters, we generated a set of fake (i.e. synthetic) PWNe-models using the
Crab Nebula as starting scaling. The Crab Nebula model we adopt is the one obtained with version 1.3 of TIDE-PWN (see table \ref{crabcomp}).

\begin{figure}
\centering
\includegraphics[width=0.6\textwidth]{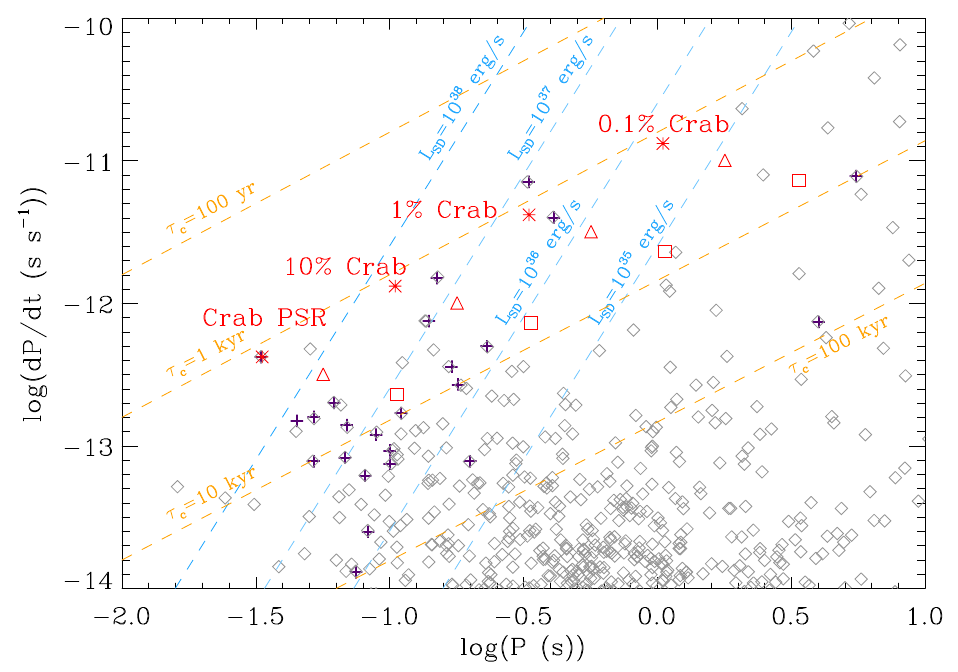}
\caption[$P \dot{P}$-diagram of ATNF PSRs and 4 fake PSRs]{$P \dot{P}$-diagram of ATNF PSRs (in grey), together with the TeV detected PWNe (in
violet), and the 4 fake PSRs adopted for this study (in red). The latter are shown at three different ages. See the text for a discussion.}
\label{fakeplot}
\end{figure}

We consider 4 different intrinsic luminosities with respect to the Crab one ($\dot{E}_0$=$\{$1, 0.1, 0.01, 0.001$\} \times \dot{E}_{0,{\rm Crab}}$).
Additionally, we require $\tau_0$ and $\tau_c$ to be the same as those of Crab, as well as we take the same moment of inertia and braking index.
All pulsars which have a braking index measurement show an n-value smaller than 3 \citep{espinoza11}, like Crab. These requirements sets the
properties of the fake pulsars, as well as define that all of them are young. We use equation (\ref{tau0}) and the definition of $\tau_c$ to
derive, e.g., $\dot{P}$ as a function of $P$, and equation (\ref{edotevol2}) to derive P as a function of the chosen $\dot{E}_0$. The definition
of $n$ can then be used to define $\ddot{P}$. Using this approach, and an initial spin-down power equaling that of Crab, 10\%, 1\%, and 0.1\% of
the latter, we have defined the properties of 4 fake pulsars, which we show in table \ref{fakepwn} at an age of 940 years (the first row in that
table is Crab's observational data). Their position in the $P \dot{P}$-diagram is shown in figure \ref{fakeplot}. These 4 simulated PSRs cover a
wide range of young systems putatively powering a nebula, from the powerful Crab, to a magnetar-like case with 0.1\% of its power (i.e. like PSR
J1550-5418). The two intermediate cases, with luminosities of 10\% and 1\% of Crab are similar to, e.g., PSR J1124-5916 or J1930+1852, and
J1119-6127, respectively.

\begin{table}
\centering
\caption[Properties of the fake PSRs considered at 940 years]{Properties of the fake PSRs considered for the study, at an age of 940 years.}
\vspace{0.2cm}
\begin{tabular}{lll}
\hline
$L_0$ & P & $\dot P$\\
($L_{0,Crab}$) & (s) & (s s$^{-1}$)\\
\hline
1& 0.0334  & $4.2 \times 10^{-13}$\\
$ 0.1$& 0.1048  &$1.3 \times 10^{-12}$\\
$ 0.01$ & 0.3314  & $4.2\times 10^{-12}$\\
$ 0.001$ & 1.0479  & $1.3\times 10^{-11}$\\
\hline
\hline
\label{fakepwn}
\end{tabular}
\end{table}

The spectral energy distribution (SED) of the nebulae (consisting of synchrotron and inverse Compton (IC) radiation) is determined by several
factors, including the magnetic field strength, the age of the system, and the background photon fields. To account for the PWNe phase space, we
considered 8 values of magnetic fraction $\eta$ (0.001, 0.01, 0.03, 0.1, 0.5, 0.9, 0.99, and 0.999, from fully particle dominated to fully
magnetically dominated nebulae), and 3 distinctive ages: 940, 3000, and 9000 years, in addition of the 4 values of $L_0$. Therefore, the explored
phase space of PWNe models is constructed by $8 \times 4 \times 3=96$ cases. The supernova (SN) explosion energy is fixed in our models, as is
the ejected mass, the injection parameters, and the environmental variables. It should be noted that these assumptions (particularly, to assume
the same injection or environmental parameters than those in the Crab Nebula) will not necessarily reflect the reality of a particular PWN (below
we present a discussion on how the photon field and injection spectrum would affect the results). Here, we are not looking for fits of the
multiwavelength emission of a particular source, but rather searching for common trends in the phase-space of PWN models. For the study we are
doing, maintaining these parameters fixed is essential to shed light on the behavior of the generated luminosities and SEDs as a function of the
initial spin-down power and the magnetic fraction.

For instance, the contribution of the IC yield on the far infra-red (FIR) background (T$\sim$70 K) would increase with respect to the cosmic
microwave background (CMB) if we consider a steeper spectrum of electrons than the one in the Crab Nebula. In such a case, more electrons (with
energies of a few TeV) are able to generate TeV photons via interacting with the FIR background, increasing its contributions relatively to the
one from the CMB \citep{aharonian97}. On the contrary, the optical / near infra-red (NIR) background (T$\sim$5000 K) hardly plays any role.
Assuming the Thompson limit, the IC emissivity $q(E) \propto w T^{(\alpha-3)/2} E^{-(\alpha+1)/2}$, where $w$ is the energy density and $T$ is
the temperature of the photon background, and $\alpha$ is the slope of the electron distribution \citep{blumenthal70}. It is possible to compare
the two contributions by estimating the ratio $q_1/q_2=(w_1/w_2)(T_1/T_2)^{(\alpha-3)/2}$. Supposing $\alpha \sim$1.5, then the ratio between the
IC contribution of NIR (T$\sim$5000 K) to FIR (T$\sim$70 K) is about 0.045 (for equal energy densities; and as per the quoted formula). In
addition, the Klein-Nishina effect, operative for VHE production with NIR photons, would reduce the IC-NIR yield significantly. If we use the
ratio above to compare the IC-FIR with the IC-CMB yield, the contribution of dust with T$\sim$70 K interacting with electrons distributed with a
slope of $\alpha=$2.5 will be similar to that of the CMB when $w_{FIR} \sim 2.2 w_{CMB} \sim 0.5$ eV cm$^{-3}$. The fixed values of the photon
backgrounds on our Crab-like models are close to the Galactic averages (see \citealt{porter06}) and should be enough to gather general trends,
which is the aim of this exercise. Finally, the distance to Crab is taken as fiducial. Given the relatively short distance to the Crab Nebula
($D=2$ kpc), the conclusions reached on the lack of detectable TeV emission for some configurations will hold for pulsars located farther away.
In any case, we provide both, luminosities and fluxes, when showing SEDs.

\begin{table}
\centering
\scriptsize
\caption[Physical magnitudes for the fake PWNe sets]{Physical magnitudes for the fake PWNe sets. The symbol ``\ldots" stands for the same value
shown in the column to the left.}
\vspace{0.2cm}
\begin{tabular}{l@{$\quad$}l@{$\quad$}l@{$\quad$}l@{$\quad$}l@{$\quad$}l@{$\quad$}l@{$\quad$}l}
\hline
\hline
{\textcolor{red}{$\dot{E}_0=\dot{E}_{0,Crab}$=$3.1 \times 10^{39}$ erg s$^{-1}$}} \\
\hline
\hline
Magnitude & Symbol & {\textcolor{blue}{$\eta$=0.001}} & {\textcolor{blue}{$\eta$=0.03}} & {\textcolor{blue}{$\eta$=0.1}} & {\textcolor{blue}{$\eta$=0.5}} & {\textcolor{blue}{$\eta$=0.9}} & {\textcolor{blue}{$\eta$=0.999}}  \\
\hline
Age (yr) & $t_{age}$ & 940\\
Spin-down luminosity at $t_{age}$ (erg/s) & $\dot{E}(t_{age})$ & $4.5 \times 10^{38}$ & \ldots & \ldots & \ldots & \ldots & \ldots\\
Maximum energy at injection at $t_{age}$ &  $\gamma_{max}(t_{age})$ & $2.3 \times 10^9$ & $1.2 \times 10^{10}$ & $2.3 \times 10^{10}$ & $5.1 \times 10^{10}$ & $6.8 \times 10^{10}$ & $7.2 \times 10^{10}$\\
Magnetic field ($\mu$G) & $B(t_{age})$ & 15.4 & 84.2 & 153.8 & 343.8 & 461.3 & 486.0\\
PWN radius today (pc) & $R_{PWN}(t_{age})$ & 2.1 & \ldots & \ldots & \ldots & \ldots & \ldots\\
\hline
Age (yr) & $t_{age}$ & 3000\\
Spin-down luminosity at $t_{age}$ (erg/s) & $\dot{E}(t_{age})$ & $7.0 \times 10^{37}$ & \ldots & \ldots & \ldots & \ldots & \ldots\\
Maximum energy at injection at $t_{age}$ &  $\gamma_{max}(t_{age})$ & $9.0 \times 10^8$ & $4.9 \times 10^9$ & $9.0 \times 10^9$ & $2.0 \times 10^{10}$ & $2.7 \times 10^{10}$ & $2.8 \times 10^{10}$\\
Magnetic field ($\mu$G) & $B(t_{age})$ & 1.7 & 9.3 & 17.0 & 38.0 & 50.9 & 53.7\\
PWN radius today (pc) & $R_{PWN}(t_{age})$ & 8.5 & \ldots & \ldots & \ldots & \ldots & \ldots\\
\hline
Age (yr) & $t_{age}$ & 9000\\
Spin-down luminosity at $t_{age}$ (erg/s) & $\dot{E}(t_{age})$ & $7.5 \times 10^{36}$ & \ldots & \ldots & \ldots & \ldots & \ldots\\
Maximum energy at injection at $t_{age}$ &  $\gamma_{max}(t_{age})$ & $2.9 \times 10^8$ & $1.6 \times 10^9$ & $2.9 \times 10^9$ & $6.6 \times 10^9$ & $8.8 \times 10^9$ & $9.3 \times 10^9$\\
Magnetic field ($\mu$G) & $B(t_{age})$ & 0.2 & 0.9 & 1.7 & 3.8 & 5.1 & 5.4\\
PWN radius today (pc) & $R_{PWN}(t_{age})$ & 31.6 & \ldots & \ldots & \ldots & \ldots & \ldots\\
\hline
\hline
{\textcolor{red}{$\dot{E}_0= 0.1 \, \dot{E}_{0,Crab}$=$3.1 \times 10^{38}$ erg s$^{-1}$ }} \\
\hline
\hline
Age (yr) & $t_{age}$ & 940\\
Spin-down luminosity at $t_{age}$ (erg/s) & $\dot{E}(t_{age})$ & $4.5 \times 10^{37}$ & \ldots & \ldots & \ldots & \ldots & \ldots\\
Maximum energy at injection at $t_{age}$ &  $\gamma_{max}(t_{age})$ & $7.2 \times 10^8$ & $3.9 \times 10^9$ & $7.2 \times 10^9$ & $1.6 \times 10^{10}$ & $2.2 \times 10^{10}$ & $2.3 \times 10^{10}$\\
Magnetic field ($\mu$G) & $B(t_{age})$ & 9.7 & 53.1 & 97.0 & 216.9 & 291.0 & 306.6\\
PWN radius today (pc) & $R_{PWN}(t_{age})$ & 1.3 & \ldots & \ldots & \ldots & \ldots & \ldots\\
\hline
Age (yr) & $t_{age}$ & 3000\\
Spin-down luminosity at $t_{age}$ (erg/s) & $\dot{E}(t_{age})$ & $7.0 \times 10^{36}$ & \ldots & \ldots & \ldots & \ldots & \ldots\\
Maximum energy at injection at $t_{age}$ &  $\gamma_{max}(t_{age})$ & $2.8 \times 10^8$ & $1.6 \times 10^9$ & $2.8 \times 10^9$ & $6.3 \times 10^9$ & $8.5 \times 10^9$ & $9.0 \times 10^9$\\
Magnetic field ($\mu$G) & $B(t_{age})$ & 1.1 & 5.9 & 10.7 & 23.9 & 32.1 & 33.9\\
PWN radius today (pc) & $R_{PWN}(t_{age})$ & 5.3 & \ldots & \ldots & \ldots & \ldots & \ldots\\
\hline
Age (yr) & $t_{age}$ & 9000\\
Spin-down luminosity at $t_{age}$ (erg/s) & $\dot{E}(t_{age})$ & $7.5 \times 10^{35}$ & \ldots & \ldots & \ldots & \ldots & \ldots\\
Maximum energy at injection at $t_{age}$ &  $\gamma_{max}(t_{age})$ & $9.3 \times 10^7$ & $5.1 \times 10^8$ & $9.3 \times 10^8$ & $2.1 \times 10^9$ & $2.8 \times 10^9$ & $2.9 \times 10^9$\\
Magnetic field ($\mu$G) & $B(t_{age})$ & 0.1 & 0.6 & 1.1 & 2.4 & 3.2 & 3.4\\
PWN radius today (pc) & $R_{PWN}(t_{age})$ & 19.9 & \ldots & \ldots & \ldots & \ldots & \ldots\\
\hline
\hline
{\textcolor{red}{$\dot{E}_0= 0.01 \, \dot{E}_{0,Crab}$=$3.1 \times 10^{37}$ erg s$^{-1}$}} \\
\hline
\hline
Age (yr) & $t_{age}$ & 940\\
Spin-down luminosity at $t_{age}$ (erg/s) & $\dot{E}(t_{age})$ & $4.5 \times 10^{36}$ & \ldots & \ldots & \ldots & \ldots & \ldots\\
Maximum energy at injection at $t_{age}$ &  $\gamma_{max}(t_{age})$ & $2.3 \times 10^8$ & $1.2 \times 10^9$ & $2.3 \times 10^9$ & $5.1 \times 10^9$ & $6.8 \times 10^9$ & $7.2 \times 10^9$\\
Magnetic field ($\mu$G) & $B(t_{age})$ & 6.1 & 33.5 & 61.2 & 136.9 & 183.6 & 193.5\\
PWN radius today (pc) & $R_{PWN}(t_{age})$ & 0.8 & \ldots & \ldots & \ldots\\
\hline
Age (yr) & $t_{age}$ & 3000\\
Spin-down luminosity at $t_{age}$ (erg/s) & $\dot{E}(t_{age})$ & $7.0 \times 10^{35}$ & \ldots & \ldots & \ldots & \ldots & \ldots\\
Maximum energy at injection at $t_{age}$ &  $\gamma_{max}(t_{age})$ & $9.0 \times 10^7$ & $4.9 \times 10^8$ & $9.0 \times 10^8$ & $2.0 \times 10^9$ & $2.7 \times 10^9$ & $2.8 \times 10^9$\\
Magnetic field ($\mu$G) & $B(t_{age})$ & 0.7 & 3.7 & 6.8 & 15.1 & 20.3 & 21.4\\
PWN radius today (pc) & $R_{PWN}(t_{age})$ & 3.4 & \ldots & \ldots & \ldots & \ldots & \ldots\\
\hline
Age (yr) & $t_{age}$ & 9000\\
Spin-down luminosity at $t_{age}$ (erg/s) & $\dot{E}(t_{age})$ & $7.5 \times 10^{34}$ & \ldots & \ldots & \ldots & \ldots & \ldots\\
Maximum energy at injection at $t_{age}$ &  $\gamma_{max}(t_{age})$ & $2.9 \times 10^7$ & $1.6 \times 10^8$ & $2.9 \times 10^8$ & $6.7 \times 10^8$ & $8.8 \times 10^8$ & $9.3 \times 10^8$\\
Magnetic field ($\mu$G) & $B(t_{age})$ & 0.07 & 0.4 & 0.7 & 1.5 & 2.0 & 2.1\\
PWN radius today (pc) & $R_{PWN}(t_{age})$ & 12.6 & \ldots & \ldots & \ldots & \ldots & \ldots\\
\hline
\hline
{\textcolor{red}{$\dot{E}_0= 0.001 \, \dot{E}_{0,Crab}$=$3.1 \times 10^{36}$ erg s$^{-1}$}} \\
\hline
\hline
Age (yr) & $t_{age}$ & 940\\
Spin-down luminosity at $t_{age}$ (erg/s) & $\dot{E}(t_{age})$ & $4.5 \times 10^{35}$ & \ldots & \ldots & \ldots & \ldots & \ldots\\
Maximum energy at injection at $t_{age}$ &  $\gamma_{max}(t_{age})$ & $7.2 \times 10^7$ & $3.9 \times 10^8$ & $7.2 \times 10^8$ & $1.6 \times 10^9$ & $2.2 \times 10^9$ & $2.3 \times 10^9$\\
Magnetic field ($\mu$G) & $B(t_{age})$ & 3.8 & 21.2 & 38.6 & 86.4 & 115.9 & 122.1\\
PWN radius today (pc) & $R_{PWN}(t_{age})$ & 0.5 & \ldots & \ldots & \ldots & \ldots & \ldots\\
\hline
Age (yr) & $t_{age}$ & 3000\\
Spin-down luminosity at $t_{age}$ (erg/s) & $\dot{E}(t_{age})$ & $7.0 \times 10^{34}$ & \ldots & \ldots & \ldots & \ldots & \ldots\\
Maximum energy at injection at $t_{age}$ &  $\gamma_{max}(t_{age})$ & $2.8 \times 10^7$ & $1.6 \times 10^8$ & $2.8 \times 10^8$ & $6.3 \times 10^8$ & $8.5 \times 10^8$ & $9.0 \times 10^8$\\
Magnetic field ($\mu$G) & $B(t_{age})$ & 0.4 & 2.3 & 4.3 & 9.5 & 12.8 & 13.5\\
PWN radius today (pc) & $R_{PWN}(t_{age})$ & 2.1 & \ldots & \ldots & \ldots & \ldots & \ldots\\
\hline
Age (yr) & $t_{age}$ & 9000\\
Spin-down luminosity at $t_{age}$ (erg/s) & $\dot{E}(t_{age})$ & $7.5 \times 10^{33}$ & \ldots & \ldots & \ldots & \ldots & \ldots\\
Maximum energy at injection at $t_{age}$ &  $\gamma_{max}(t_{age})$ & $9.3 \times 10^6$ & $5.1 \times 10^7$ & $9.3 \times 10^7$ & $2.0 \times 10^8$ & $2.8 \times 10^8$ & $2.9 \times 10^8$\\
Magnetic field ($\mu$G) & $B(t_{age})$ & 0.04 & 0.2 & 0.4 & 1.0 & 1.3 & 1.4\\
PWN radius today (pc) & $R_{PWN}(t_{age})$ & 7.9 & \ldots & \ldots & \ldots & \ldots & \ldots\\
\hline
\hline
\end{tabular}
\label{fakepar}
\end{table}

Table \ref{fakepar} shows the results for the scaled models for the different parameters considered. The results varying the total luminosity
from the largest to the smallest are listed from top to bottom, each one considering three evolutionary ages (940, 3000 and 9000 years), and
different magnetic fractions, increasing from left to right. Table \ref{fakepar} also quotes the intrinsic sizes of the simulated nebulae,
magnetic fields, and maximum energy at the selected age, for the different models. A systematic comparison among the different results will be
done in the following.

\subsection[IC contributions for different age and $\dot{E}$]{IC contributions for different age and PSR spin-down power}

\begin{landscape}
\begin{figure}
\centering
\includegraphics[width=1.3\textwidth]{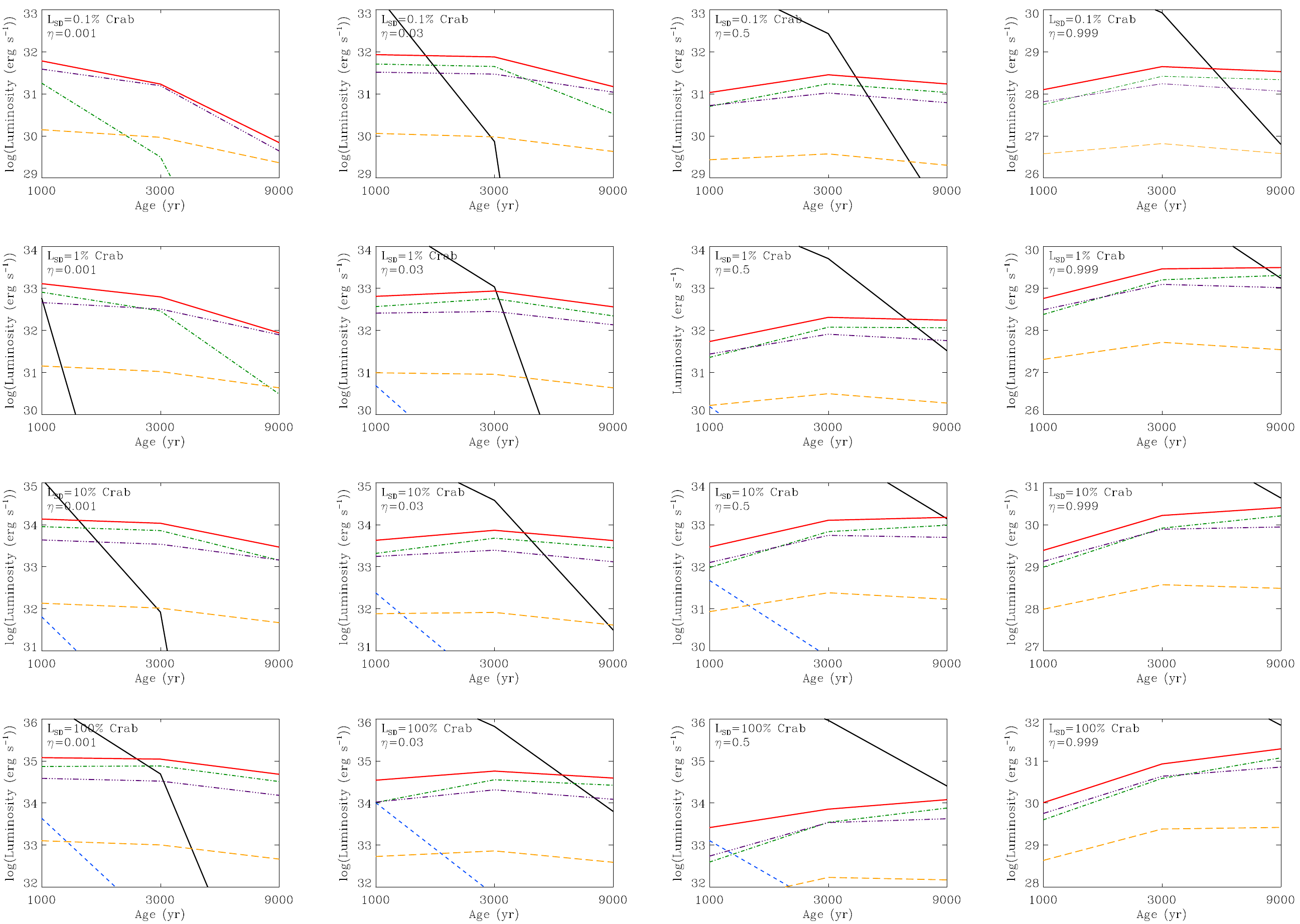}
\caption[Luminosities between 1 and 10 TeV of the IC contributions of the spectrum as a function of age]{Luminosities between 1 and 10 TeV of the
IC contributions of the spectrum as a function of age. We fix $\dot{E}=$0.1, 1, 10, and 100\% of the Crab Nebula (from top to bottom) and a
magnetic fraction of 0.001, 0.03, 0.5 \& 0.999 (from left to right). The black solid line is the synchrotron luminosity calculated between 1 and
10 keV. The other components are: total IC (red solid line), IC-CMB (green dot-dashed line), IC-FIR (purple triple dotted-dashed line), IC-NIR
(orange dashed line) and SSC (blue-dashed line).}
\label{tev_age}
\end{figure}
\end{landscape}

\begin{landscape}
\begin{figure}
\centering
\includegraphics[width=1.45\textwidth]{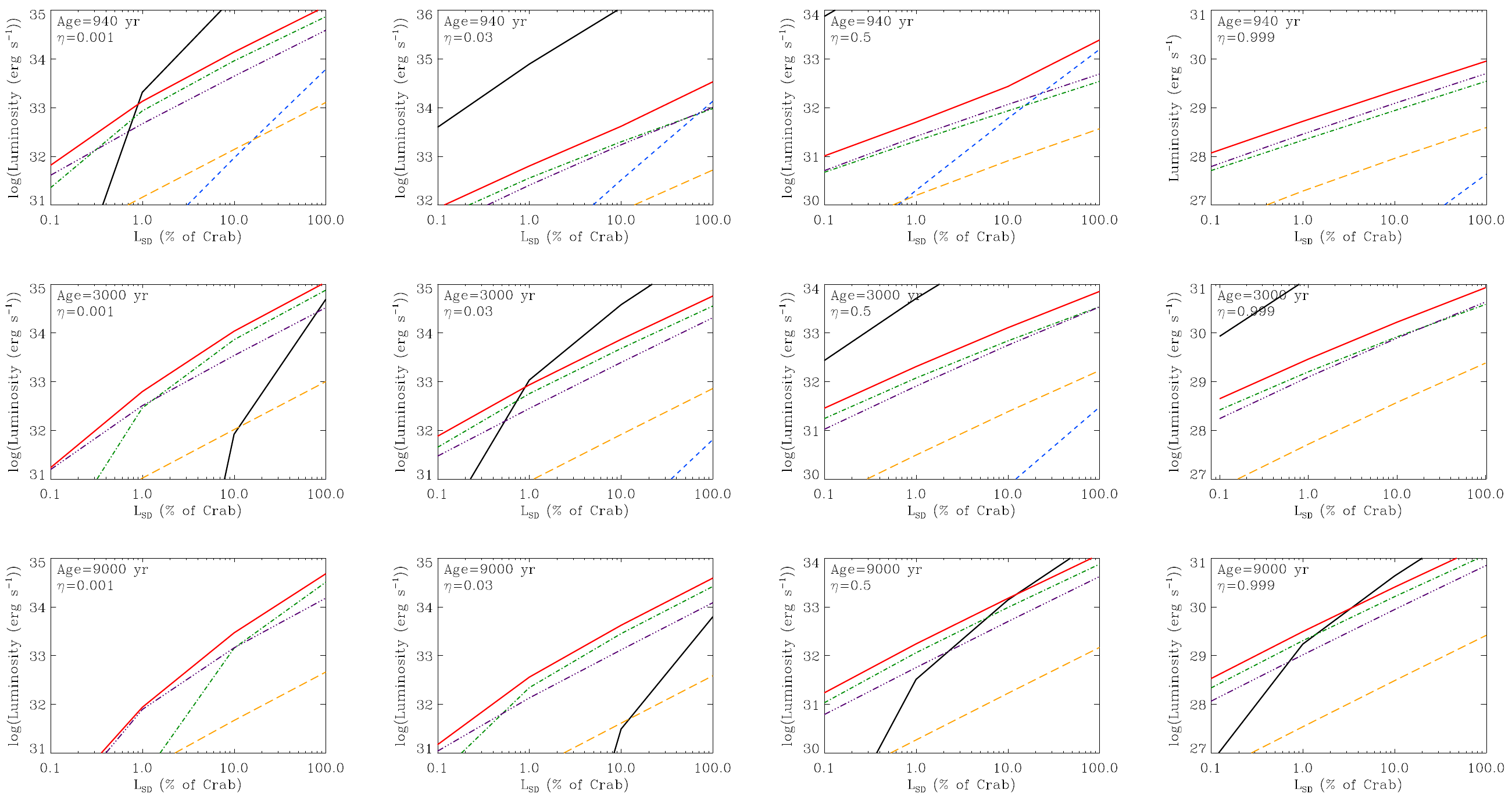}
\caption[Luminosities between 1 and 10 TeV of the contributions of the spectrum as a function of $\dot{E}$]{Luminosities between 1 and 10 TeV of
the contributions of the spectrum as a function of the spin-down luminosity. We fix an age of 940, 3000, and 9000 years (from top to bottom) and
a magnetic fraction of 0.001, 0.03, 0.5 \& 0.999 (from left to right). The color coding is as in figure \ref{tev_age}}
\label{tev_lsd}
\end{figure}
\end{landscape}

To compare the TeV luminosities, we integrated the simulated gamma-ray emission between 1 and 10 TeV. For comparison, we have also computed the
synchrotron luminosity integrated between 1 and 10 keV. We compare the contributions of different photon backgrounds, namely SSC, FIR, NIR, and
CMB, to the total IC yield of each of the nebulae. The results for the luminosity as a function of age are shown in figure \ref{tev_age} (for
fixed LSD=0.1, 1, 10, and 100\% of the Crab, from top to bottom, and a magnetic fraction of 0.001, 0.03, 0.5 \& 0.999, from left to right). The
results for the luminosity as a function of spin-down power are shown in figure \ref{tev_lsd} (for fixed increasing age, from top to bottom, and
a magnetic fraction of 0.001, 0.03, 0.5 \& 0.999, from left to right).

The IC components have a very similar behavior one to another, with the exception of the SSC, which has a similar slope as the synchrotron
contribution. This slope similarity between the synchrotron and the SSC luminosity is seen for most of the plots in this section. There are some
particular cases in which this is not the case, though. In the top-left panel of figure \ref{tev_age}, the CMB contribution decays with age much
more steeply than the FIR contribution to the total yield. This is the result of cutting the energy range in a small band, from 1 to 10 TeV,
where, in this case, the IC contribution off the CMB is falling. The latter dominates the FIR contribution at 1 TeV in this case, where it starts
to fall steeply; due to the value of $\gamma_{max}$ (see table \ref{fakepar}), there are not enough electrons to generate higher energy photons
interacting with the CMB background.

If we consider the SSC contribution, depicted by the blue-dashed line, we note it is only visible in the y-axis scale of the different panels of
figure \ref{tev_age} in only a few occasions. It is irrelevant for $\dot{E}=$0.1, 1, and 10\% of the Crab power, disregarding the age and the
magnetic fraction of the nebulae. On the contrary, it only becomes relevant for highly energetic (Crab-like) particle dominated nebulae at low
ages (of less than a few thousand years). The Crab Nebula today corresponds to the bottom row, second column plot of figure \ref{tev_age} when
the age (in the x-axis) is taken as 940 years. It is seen there how uncommon the SSC domination is: A lower or higher magnetic fraction (left or
right panels), or a higher age (movement along the x-axis), and the SSC contribution would quickly be sub-dominant to the IC-FIR or even to the
IC-CMB components.

Figure \ref{tev_lsd} shows the IC contributions of the spectrum as a function of spin-down. As we increase the spin-down power, all the IC
contributions increase their luminosity due to the presence of additional high-energy electrons, but the SSC depends also on the power of the
synchrotron emission, which is increasing too due to the higher magnetic field. This effect makes the SSC a steeper function of $\dot{E}$
compared to the other contributions. Consistently with the results of figure \ref{tev_age}, the SSC contribution requires a young age and
$\sim$70\% of the Crab's power to become relevant. Fully magnetized nebulae ($\eta=0.999$), if they exist, are never SSC-dominated no matter the
age or pulsar spin-down power. This is partly also a result of the increased synchrotron losses produced by the very high magnetic field, which
diminishes the relative importance of all IC components. Finally, we note that –mimicking the SSC behaviour– for lower $\dot{E}$ and
older ages than that of the Crab Nebula, the synchrotron luminosity falls down very quickly. This is partly because the energy range where we are
integrating the luminosity is in the synchrotron cutoff regime produced by the electron population cut at high energies. The former results
clarify why the Crab Nebula, and only it, is SSC dominated: There are no other PSRs we know, young and powerful enough so that SSC could
play any role against the comptonization of FIR, or CMB photons.

\begin{figure}
\centering
\includegraphics[width=1.0\textwidth]{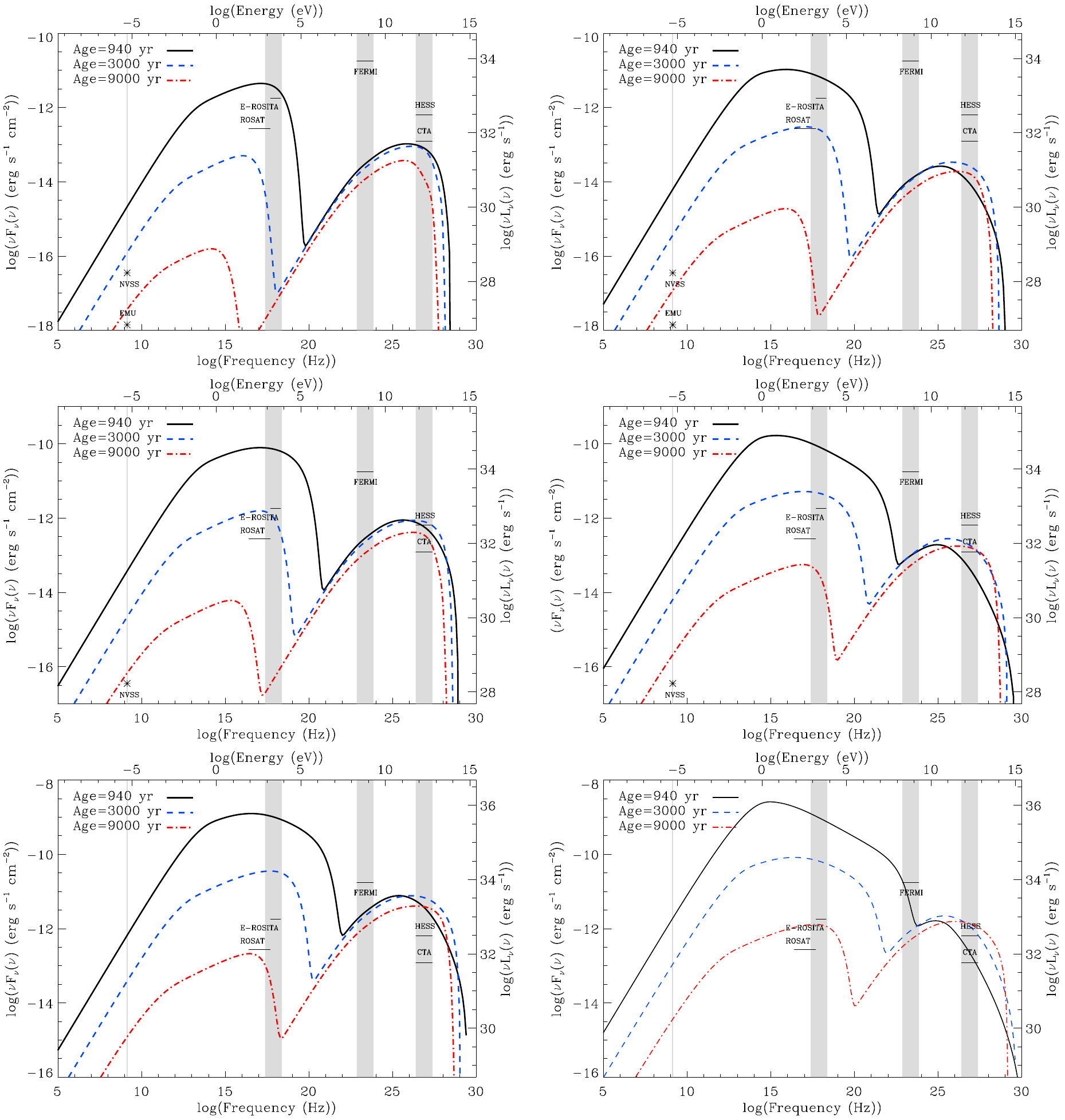}
\caption[SEDs for $\dot{E}$=0.1, 1, and 10\% of the Crab as a function of the age]{Comparison of the SEDs for $\dot{E}$=0.1, 1, and 10\% of the
Crab (from top to bottom) as a function of the age. The magnetic fraction is fixed at 0.03 (left) and 0.5 (right). The shadowed columns
correspond to the frequency intervals in radio, X-rays, the GeV, and the TeV bands where we have integrated the luminosity. The sensitivity of
some surveys and telescopes in these energy ranges are shown by thin black lines.}
\label{sed_age}
\end{figure}

\begin{figure}
\centering
\includegraphics[width=1.0\textwidth]{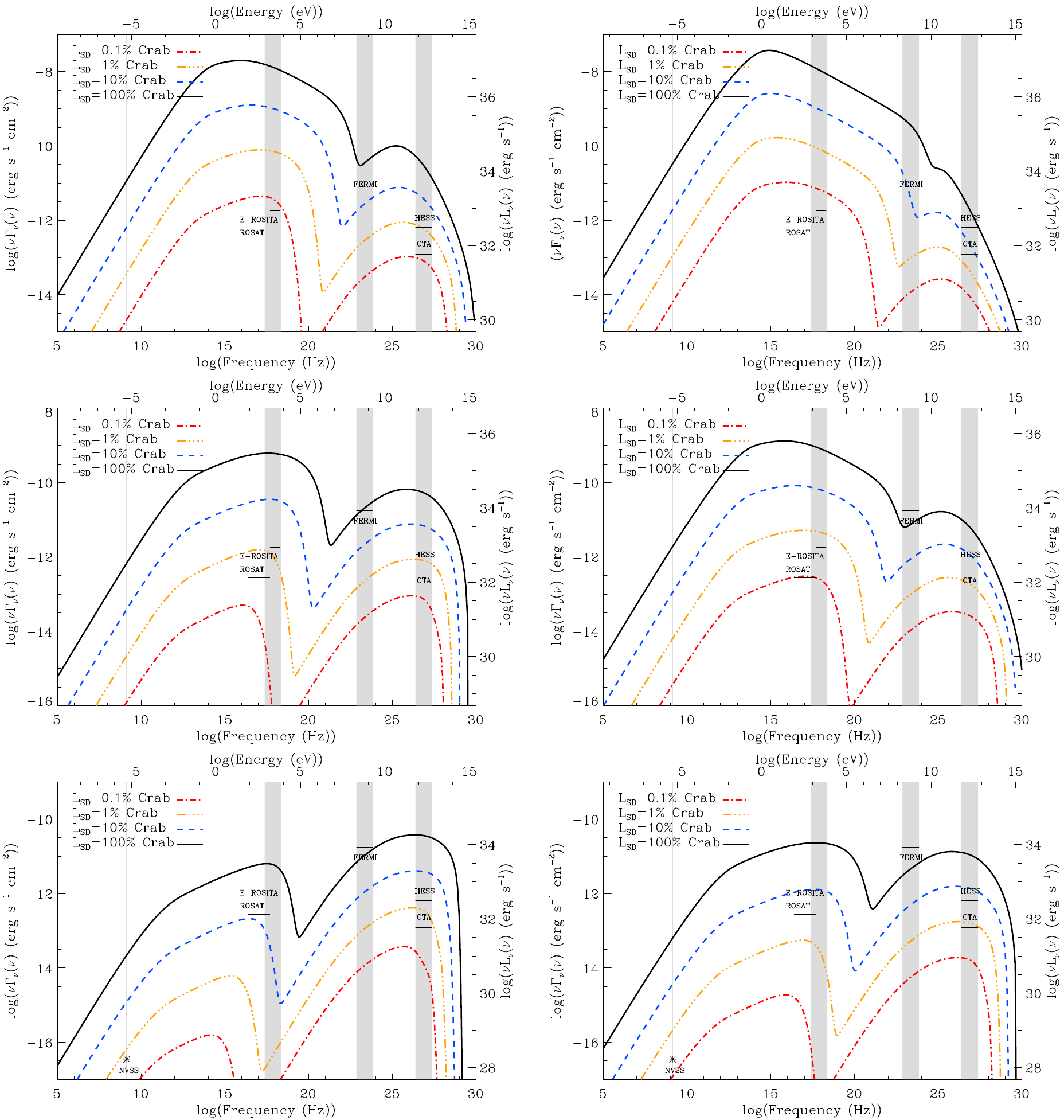}
\caption[SEDs for an age of 940, 3000, and 9000 years (from top to bottom) as a function of $\dot{E}$]{Comparison of the SEDs for an age of 940,
3000, and 9000 years (from top to bottom) as a function of the spin-down luminosity. The magnetic fraction is fixed at 0.03 (left) and 0.5
(right).}
\label{sed_lsd}
\end{figure}

Similar considerations can be done by inspecting the SED as a function of age and spin-down power (figure \ref{sed_age} and \ref{sed_lsd}). In
each Figure, the SED showed in the left panel is calculated for a particle-dominated nebula ($\eta=0.03$) whereas the right one is computed for a
nebula in equipartition ($\eta=0.5$). The shadowed areas correspond to the frequency intervals in radio, X-rays, GeV, and TeV bands where we
integrate the luminosity to compare their ratios (see below). Several instrument sensitivities (in survey mode) are also shown, corresponding to
the NVSS and EMU in radio\footnote{See for instance \url {http://askap.pbworks.com/w/page/14049306/RadioSurveys}}, e-Rosita and ROSAT in
X-rays\footnote{See \url {http://www.mpe.mpg.de/455799/instrument}}, Fermi (3-yr Galactic) in the GeV band\footnote{From the Fermi-LAT
performance \url{http://www.slac.stanford.edu/exp/glast/groups/canda/lat_Performance.htm}}, and the current (H.E.S.S.) and future (CTA)
experiments in the TeV band (for 50 hours and 5$\sigma$ detection) (e.g., \citealt{gast11,actis11}).

It is interesting to note that, for the considered sensitivities, no young PWN at TeV energies (for any age or magnetic fraction, top row in
figure \ref{sed_age}) would be detectable if the pulsar’s spin-down power is 0.1\% Crab or lower (and under the caveats of the assumptions
discussed in the previous section, e.g., assuming the same spectral slope in the injection than those in the Crab Nebula). This conclusion is
particularly stable for H.E.S.S.-like telescopes; the youngest of the pulsar's considered is more than one order of magnitude below the
sensitivity considered. The effect of using different injection or FIR energy density to this conclusion is discussed below.

For more energetic pulsars (1\% of Crab, middle row) distinctions in age and magnetic fraction appears to reflect strongly on the TeV flux and
therefore on the detectability of the nebulae. For instance, only low magnetic fraction, i.e., particle dominated nebulae, can be detected by
H.E.S.S.-like telescopes if young enough (a few thousand years). If the same nebulae were in equipartition, the TeV luminosity would be very much
suppressed and the detection even with CTA would require a deep observation. Which exact ages of the nebulae will CTA detect in these conditions
will ultimately depend on the injection and environmental parameters. For the ones we have assumed, larger ages are preferred, when enough
electrons are available for interaction. On the contrary, nebulae powered by pulsars with spin-down of 10\% Crab or more are all detectable by
H.E.S.S.-like telescopes if they are particle dominated, no matter the age (bottom-left panel of figure \ref{sed_age}). The possibility of
detection is less clear in case of larger magnetic fractions (bottom-right panel, same figure). We recall that 2 kpc is assumed for the distance
in the scale of the left-axis of figures \ref{sed_age} and \ref{sed_lsd}, but the general trend of these conclusions should scale with it,
worsening the chances of detection the farther the nebula is located. This already points to an interesting observational bias, which we discuss
further below: if there are magnetically dominated PWNe, similar to the ones simulated here, it would be hard to detect them with the current
generation of TeV telescopes.

To illustrate the effect of the initial spin-down power injected, the SED is shown in figure \ref{sed_lsd} for each of the 4 $\dot{E}_0$
proposed. Similarly to figure \ref{sed_age} the figures on the left panels are calculated for a particle dominated nebula ($\eta=0.03$) and the
right ones for equipartition ($\eta=0.5$). The trends noted above are more clearly shown here. In particular, for relatively old PWNe, at ages of
9000 years (bottom row), the increase of the X-ray nebula to detectable levels with the current instruments has a strong dependence on the
magnetic fraction considered. For instance, a relatively bright pulsar with a spin-down energy of 10\% of the Crab, at 9000 years would be
detectable by ROSAT if in equipartition, but not for smaller magnetic fields.

\subsection[The effect of $\eta$ on X-ray and TeV luminosity]{The effect of magnetic fraction on the X-ray and TeV luminosity}

\begin{landscape}
\begin{figure}
\centering
\includegraphics[width=1.45\textwidth]{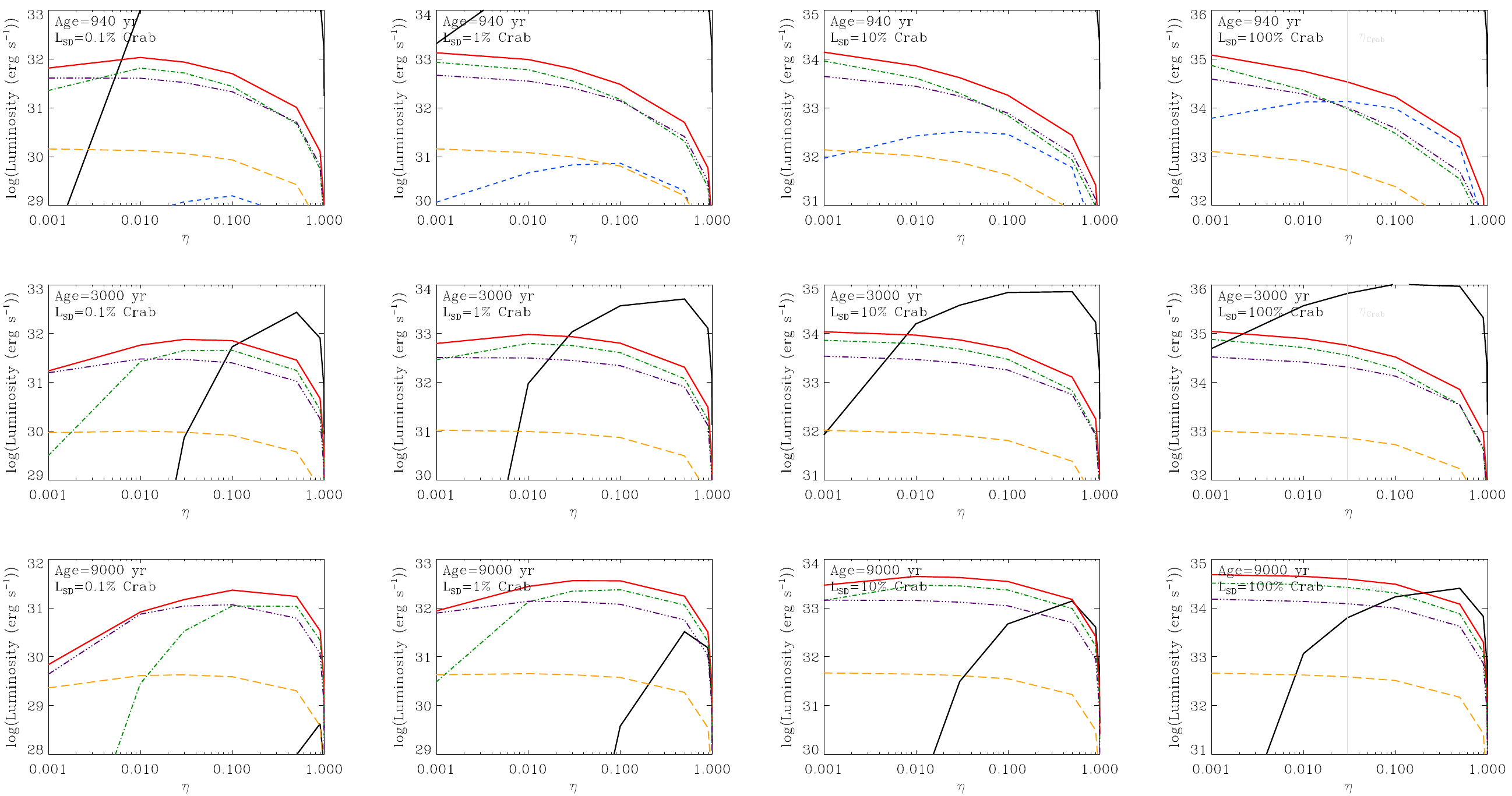}
\caption[X-ray and TeV luminosities as a function of $\eta$]{Luminosities between 1 and 10 TeV of the IC, and between 1 and 10 keV of the
synchrotron contributions to the spectrum as a function of the magnetic fraction. We fix an age of 940, 3000, and 9000 years (from top to bottom)
and vary the spin-down luminosity (0.1\%, 1\%, 10\% \& 100\% of Crab, from left to right). The color coding is as in figure \ref{tev_age}.}
\label{tev_eta}
\end{figure}
\end{landscape}

Generally, magnetic equipartition is assumed when discussing X-ray nebulae, but recent TeV observations have shown that many (if not all) of
these PWNe are particle dominated. To analyze in more detail the impact of the magnetic fraction parameter on the detectability of the nebulae
(or on their flux level), we represent the IC contribution to the spectral flux between 1 and 10 TeV as a function of the magnetic fraction
(figure \ref{tev_eta}). As discussed before, large spin-down and very young ages (top right panels) are required to observe a relevant
contribution of SSC. The rightmost top panel corresponds to a pulsar such as Crab, having its age but different magnetic fraction. The
contribution of SSC dominates for $\eta>0.02$ whereas for lower $\eta$-values the total luminosity would be dominated by FIR even for a 940 years
pulsar.

Figure \ref{tev_eta} shows the corresponding SEDs for the three ages under consideration, and three spin-down powers (1, 10, and 100\% of
Crab's). The impact of the magnetic fraction on the final SED is large, generating orders of magnitude variations in the luminosity even when
keeping all other system's parameters fixed (same spin-down, $P$, $\dot{P}$, injection, and environment). At TeV energies, the simulations show
(for Crab-like photon field background and injection parameters) that H.E.S.S.-like telescopes would not be sensitive enough to fully explore the
$\eta>0.5$ regime, independently from the pulsar age (when lower than $10^4$ years) or spin-down power. Even for CTA, a complete coverage of the
phase space of young nebulae (assuming that strongly magnetic field-dominated nebulae exist, of course) can only be partially achieved for up to
$\eta<0.9$ and $\dot{E}_0>$10\% Crab, for near PWNe. Below, we give details on the impact that a different injection or a different FIR
background energy density have on this conclusion.

Finally, we calculate the total bolometric power integrating the total luminosity $L(\nu)$, corresponding to the spectra in figure \ref{tev_eta}.
The results for 1\% and 10\% of the Crab luminosity are shown in table \ref{bol_ratio}. The total radiated power is in all cases less than the
injected spin-down (see table \ref{fakepar}) at the age considered, amounting a few percent for young PWNe (with $\eta \sim$0.01 - 0.1).
Equipartition naturally produces the maximum of the radiated power in all cases. The integrated-in-time spin-down power ranges from
$\sim 4 \times 10^{47}$ erg, for 1\% of Crab, to $\sim 5 \times 10^{48}$ erg, for 10\% of Crab.

\begin{table}
\centering
\scriptsize
\caption[Ratio of the bolometric radiated power of the spectra divided by $\dot{E}$ at the given age]{Ratio of the bolometric radiated power (erg
s$^{-1}$) of the spectra in figure \ref{sed_eta} divided by the spin-down power (erg s$^{-1}$) at the given age. Two examples are shown for 1\%
and 10\% of Crab's spin-down.}
\vspace{0.2cm}
\begin{tabular}{llllllll}
\hline
	$\eta$	&	940 yr	&	3000 yr	&	9000 yr \\
\hline
1\% Crab \\
\hline	
	0.001	&	0.00580	&	0.00471	&	0.01480 \\
	0.01	&	0.04356	&	0.00794	&	0.01880 \\
	0.03	&	0.09022	&	0.01350	&	0.02013 \\
	0.1	&	0.16356	&	0.02757	&	0.02000 \\
	0.5	&	0.18889	&	0.04229	&	0.01480 \\
	0.9	&	0.04733	&	0.01169	&	0.00292 \\
	0.99	&	0.00491	&	0.00123	&	0.00030 \\
	0.999	&	0.00049	&	0.00012	&	0.00003 \\
\hline
10\% Crab \\
\hline
	0.001	&	0.01689	&	0.00640	&	0.01960 \\
	0.01	&	0.08822	&	0.01457	&	0.02280 \\
	0.03	&	0.15733	&	0.02800	&	0.02320 \\
	0.1	&	0.25778	&	0.05443	&	0.02307 \\
	0.5	&	0.26667	&	0.07157	&	0.01747 \\
	0.9	&	0.06467	&	0.01900	&	0.00427 \\
	0.99	&	0.00667	&	0.00199	&	0.00044 \\
	0.999	&	0.00067	&	0.00020	&	0.00004 \\
\hline
\hline
\label{bol_ratio}
\end{tabular}
\end{table}

\begin{landscape}
\begin{figure}
\centering
\includegraphics[width=1.45\textwidth]{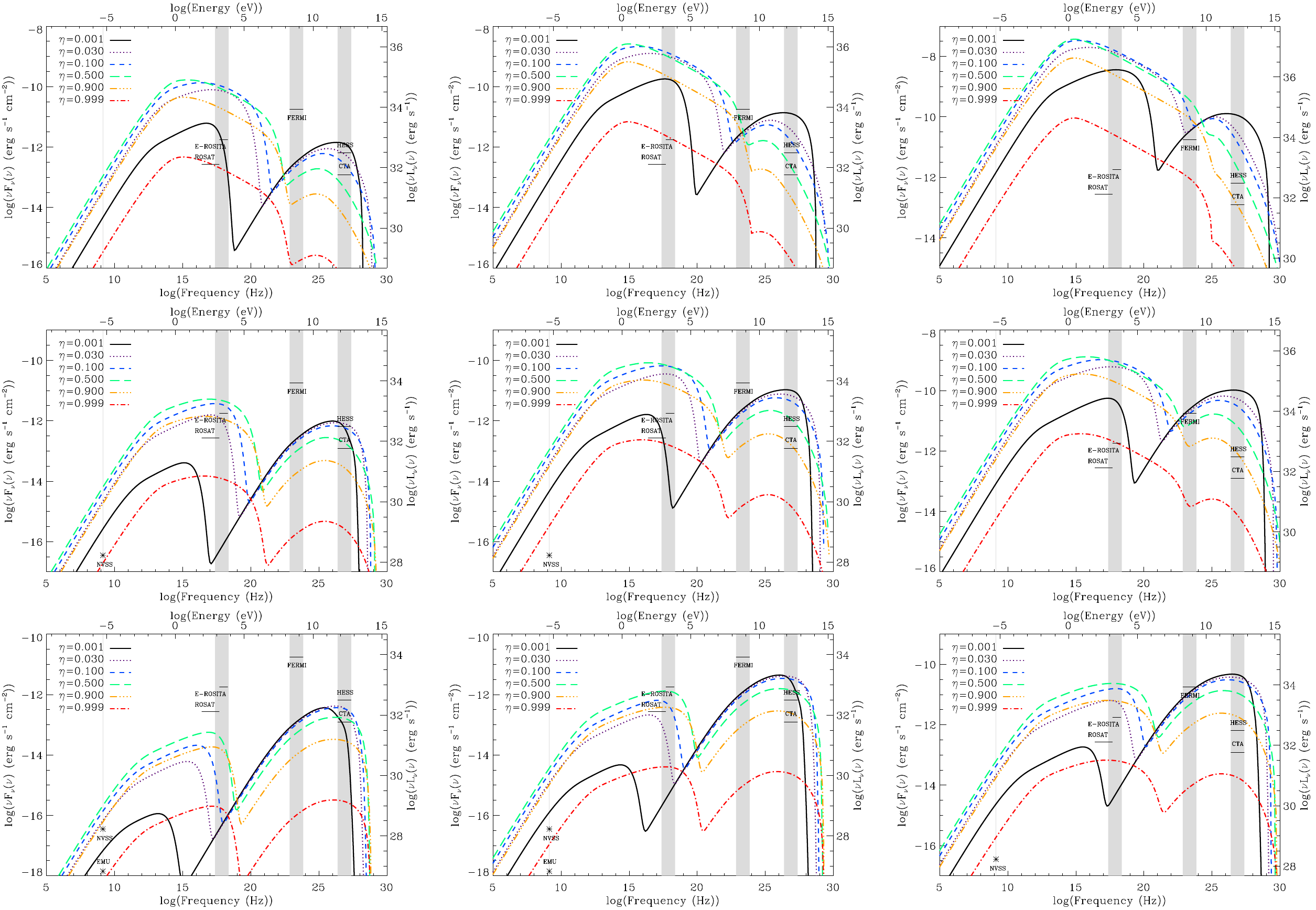}
\caption[SEDs for an age of 940, 3000, and 9000 years as a function of $\eta$]{Comparison of the SEDs for an age of 940, 3000, and 9000 years
(from top to bottom) as a function of the magnetic fraction. The spin-down luminosity is fixed at 1\% 10\%, and 100\% (panels from left to right)
of Crab.}
\label{sed_eta}
\end{figure}
\end{landscape}

\subsection{Luminosity ratios for different wavelengths}

\begin{landscape}
\begin{figure}
\centering
\includegraphics[width=1.45\textwidth]{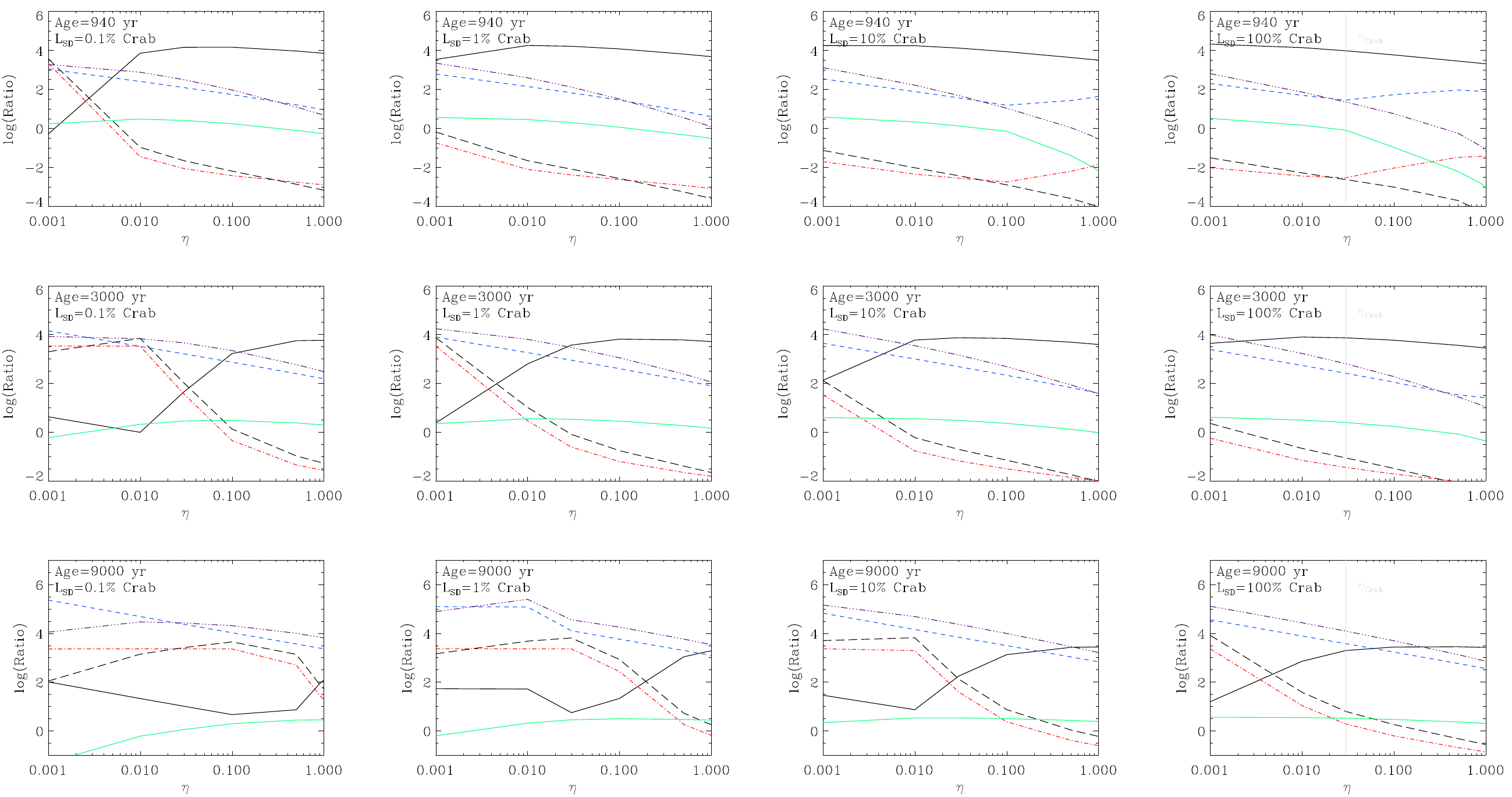}
\caption[Luminosity ratios at 940, 3000, and 9000 years as a function of $\eta$]{Luminosity ratios at 940, 3000, and 9000 years (from top to
bottom) as a function of the magnetic fraction. We show the cases for $\dot{E}$=0.1, 1, 10, and 100\% of Crab (from left to right) and depict the
following ratios: X-ray/radio: black solid line, $\gamma$-ray/radio: blue dashed line, VHE/radio: purple triple-dot-dashed line,
$\gamma$-ray/X-ray: red dot-dashed line, VHE/X-ray: black dashed line, VHE/$\gamma$-ray: green solid line. The bands used for defining the ratios
corresponds to the shaded regions in previous figures. For further details, see text.}
\label{ratio_eta}
\end{figure}
\end{landscape}

Figure \ref{ratio_eta} represents the distance-independent luminosity ratios at 940, 3000, and 9000 years (from top to bottom) as a function of
the magnetic fraction; for different spin-down powers. All the ratios can, of course, be directly measured if such pulsars exist. As an example,
the vertical line in the rightmost panels shows the ratios from the spectrum of the Crab Nebula along time (the right-top panel corresponds to
the values of the ratios as measured today). They correspond to one and the same magnetic fraction (in the framework of the model assumptions),
at $\eta=0.03$. Note that the ratios are, mostly, monotonic functions of $\eta$, and thus, a measurement or upper limits on the luminosities of a
PWN can be used to estimate a value of the magnetic fraction. Exceptions to the monotonic character of the ratios happen. Examples of those are
the VHE/X-ray ratio (represented by the black dashed line) of pulsars having 1 to 10\% of the Crab’s spin down and ages of 9000 years (the two
bottom-middle panels). If we are to measure a VHE/X-ray ratio only, there could be two magnetic fractions corresponding to it and its value is
then degenerate. However, not all ratios are, and measurements of other luminosity ratios would break the degeneracy and inform on a plausible
value of $\eta$.

One can also consider that the ratios between luminosities can be ideally measured even if we do not know the $P$ and $\dot{P}$ of the
corresponding pulsar, say, after a blind discovery of a PWN in the foreseen CTA Galactic Plane survey. Having several luminosity ratios, if they
correlate with a single value of $\eta$ would inform of a plausible value not only of the magnetic fraction, but also of the age and power
(always under the assumption of Crab-like injection and environmental variables, which we challenge below). The efficiencies of the radiative
power at each of the bands play a similar role to figure \ref{ratio_eta} when $P$ and $\dot{P}$ are known quantities.

The caveat is that for particular PWNe, neither the injection parameters, nor the densities of the background photons will be exactly as assumed
here. Thus, there is no escape from individual modeling; figure \ref{ratio_eta} can only be taken as an approximation if we are to compare with
directly measurable quantities. However, we can imagine having a set of figures \ref{ratio_eta} and/or the efficiencies at each of the bands,
spanning different assumptions for the injection or the environmental parameters. Using such expanded phase space, an automatic procedure of
interpolation could inform on plausible values of $\eta$, age, and luminosity starting only from observational data, like the ratios of
luminosities or efficiencies. This is to be considered at CTA times, when hundreds of PWN are expected to be discovered blindly.

\subsection{Discussion and conclusions}

After considering a phase space of $\sim 100$ Crab-like PWNe of different magnetization, spin-down power, and age we concluded that:
\begin{itemize}
\item The SSC contribution to the total IC yield is irrelevant for $\dot{E}$=0.1, 1, and 10\% of the Crab power, disregarding the age and the
magnetic fraction of the nebulae. It only becomes relevant for highly energetic ($\sim 70\%$ of the Crab) particle dominated nebulae at low ages
(of less than a few thousand years). 
\item No young (rotationally powered) PWN would be detectable at TeV energies if the pulsar spin-down power is 0.1\% Crab or lower. For 1\% of
the Crab spin-down, only particle dominated nebulae can be detected by H.E.S.S.-like telescopes if young enough (with the detail of the
detectability analysis depending on the precise injection and environmental parameters). Above 10\% of the Crab's power, all PWNe are detectable
by H.E.S.S.-like telescopes if they are particle dominated, no matter the age.
\item The magnetic fraction is an important order parameter in the TeV observability of nebulae, and induces orders of magnitude variations in
the luminosity output for systems that are otherwise the same (same spin-down, $P$, $\dot P$, injection, and environment). For Crab-like photon
field background and injection parameters, H.E.S.S.-like telescopes would not be sensitive enough to fully explore the $\eta>$0.5 regime,
independently from the pulsar age (when lower than $10^4$ years) or spin-down power. 
\end{itemize}

\begin{figure}
\centering
\includegraphics[width=1.0\textwidth]{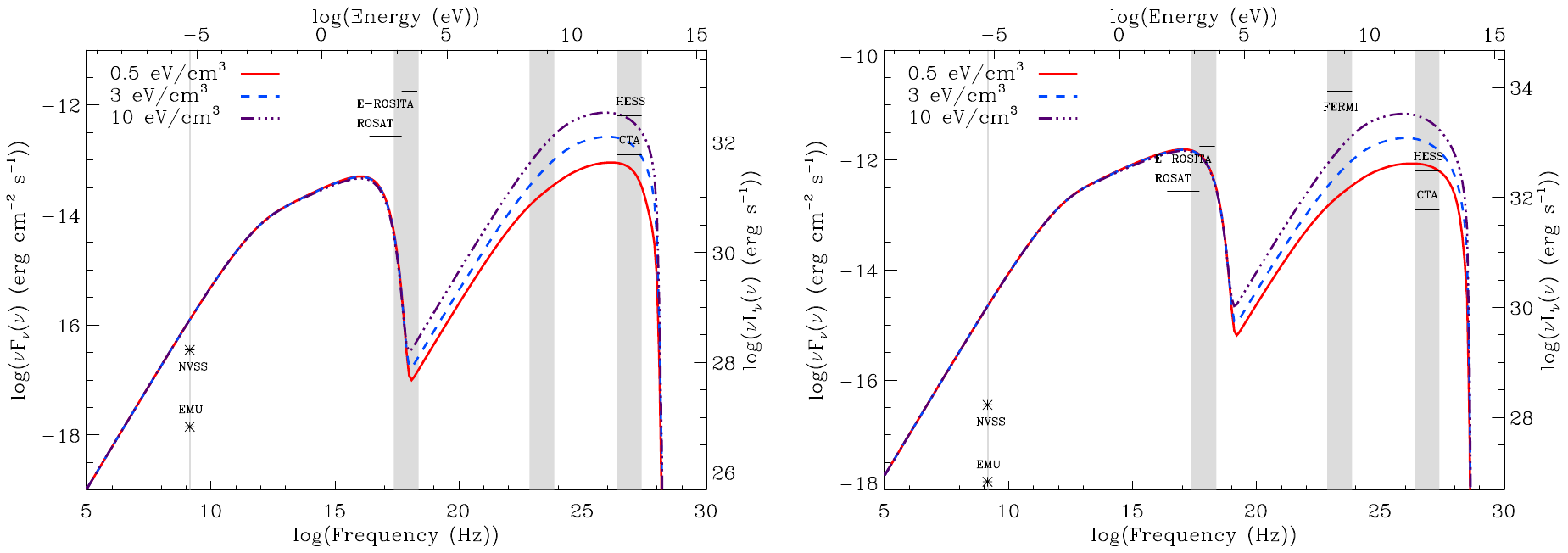}
\caption[SEDs for Crab's injection parameters and different FIR photon density]{SEDs for Crab's injection parameters and different FIR photon
density, from 0.5 to 10 eV cm$^{-3}$. The left panel shows the results for 0.1\% of Crab's energetics, at an age of 3000 years. The right panel
shows the same analysis for 1\% of Crab's energetics. A low magnetization of 0.03 is assumed.}
\label{sed_fir}
\end{figure}

Based on the above results, we pose that if extreme differences between the environmental or injection variables do not occur (in comparison with
the Crab Nebula's) it is the magnetic fraction what decides detectability of TeV nebulae. It is thus important to analyze the impact that a
different injection or environmental parameters have on our conclusions. Here we specify on the stability of the results against changes on
$\alpha_1$, $\alpha_2$, and $w_{FIR}$. The IR energy density in specific regions of the Galaxy can well exceed the 0.5 eV cm$^{-3}$ considered,
for instance close to star formation sites.

\begin{figure}[t!]
\centering
\includegraphics[width=1.0\textwidth]{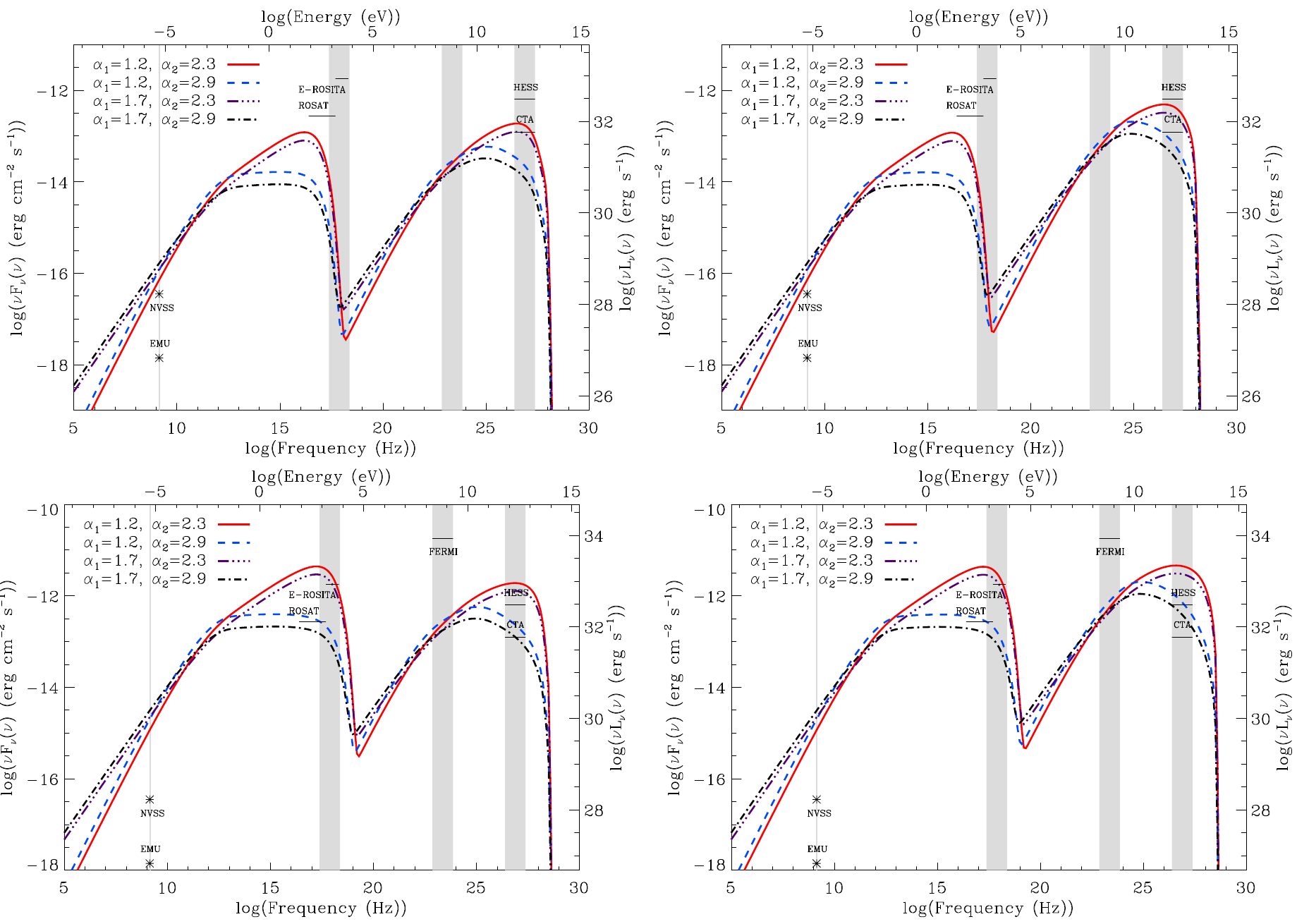}
\caption[SEDs for different injection parameters and different FIR photon density]{SEDs for different injection parameters (as detailed in the
legend) and different FIR photon density (0.5 eV cm$^{-3}$ in the left panel, and 3 eV cm$^{-3}$ in the right one) for a pulsar with 0.1\% (top
row) and 1\% (bottom row) of Crab's energetics. A low magnetization of 0.03 is assumed.}
\label{sed_inj}
\end{figure}

Figure \ref{sed_fir} (left) compares the results for 0.1\% of Crab's energetic, at an age of 3000 years. It can be seen that even in the extreme
case of 10 eV cm$^{-3}$; a factor of 20 in excess of Crab’s energy density, a H.E.S.S.-like telescope will not detect this PWN. The conclusion is
thus stable for the current generation of telescopes. CTA detectability, instead, will depend on the FIR density. We recall that the distance
assumed for the flux-sensitivity comparison is that of Crab, and thus, that for farther PWNe, the ability of the telescope for detection will be
diminished further. The right panel of figure \ref{sed_fir} shows the same analysis but for 1\% of Crab's energetics. As stated above, the FIR
energy density (and distance, of course) is the decisive parameter concerning the detectability of PWNe in H.E.S.S.-like telescopes in this case.
Note that the extreme case of 10 eV cm$^{-3}$ , which at 2 kpc produces about one order of magnitude in excess of a H.E.S.S.-like telescope
sensitivity, would be invisible when the pulsar that drives the nebula is located instead at 5 kpc or beyond. Thus, if having the same injection
parameters, most of of the PWNe with 1\% of the Crab's energetics would not be seen by the current generation of instruments.

The detectability of PWNe with 0.1\% of Crab’s energetics in H.E.S.S.-like telescopes is not affected either by changes in the injection
parameters. The top row of figure \ref{sed_inj} shows such changes together with an increased FIR background. Four different pairs of injection
slopes are assumed and results are shown for an age of 3000 years. The differences produced by the the changes in injection are indeed large, as
expected, but still, a low FIR background would preclude most of these PWNe having 0.1\% of Crab’s energetics to be detected even by CTA. The
bottom row of \ref{sed_inj} shows the same results for the case of 1\% of Crab. Note that for an average value (0.5 eV cm$^{-3}$ in the left
example) or even a significantly increased value (3 eV cm$^{-3}$ in the right example) of FIR photon density, only hard spectra will lead to a
clear detection in the current generation of instruments. None of these pulsars featuring 1\% Crab's energetics, if located at 5 kpc instead of 2
kpc, would be detected by H.E.S.S.-like instruments.

\begin{figure}[t!]
\centering
\includegraphics[width=1.0\textwidth]{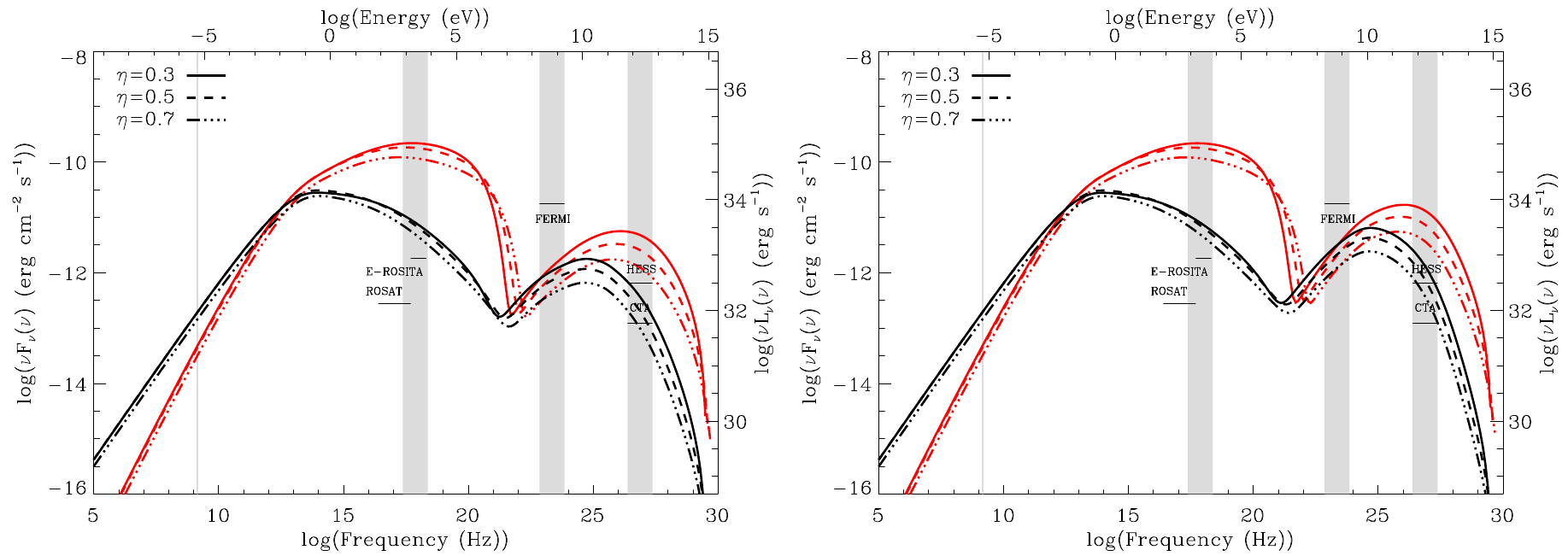}
\caption[SEDs for different magnetic fraction and different FIR photon density]{SEDs for different magnetic fraction (as detailed in the legend)
and different FIR photon density (0.5 eV cm$^{-3}$ in the left panel, and 3 eV cm$^{-3}$ in the right one) for a pulsar with 10\% of Crab's
energetics. In red, a hard spectrum of particles with $\alpha_1=1.2$, $\alpha_2=2.3$ is assumed, whereas the black curves stand for a steep case
with $\alpha_1=1.7$, $\alpha_2=2.9$.}
\label{sed_etafir}
\end{figure}

We also consider how stable is the assertion that at 10\% of Crab's spin down, a H.E.S.S.-like instrument needs $\eta<0.5$ for detecting PWNe,
against variations of injection or FIR background. To do that we consider a PWN of a pulsar with 10\% of Crab's energetics, subject to two energy
densities, 0.5 and 3 eV cm$^{-3}$, with electron distributions having slopes of $\alpha_1=1.2$, $\alpha_2=2.3$ and $\alpha_1=1.7$, $\alpha_2=2.9$
represented in red and black in figure \ref{sed_etafir}, respectively. Obviously, the most favorable cases for detection are given by the harder
slopes of injection and the largest FIR densities. In those particular cases with $w_{FIR}$=3 eV cm$^{-3}$ (0.5 eV cm$^{-3}$) H.E.S.S.-like
instruments could detect the PWNe up to a magnetic fraction of 0.7 if located closer than $\sim$5 kpc ($\sim$3.5 kpc). Results for pulsars with
10\% of Crab's energetic uniformly produce detectable PWNe for magnetization parameters lower than 0.5. The fact that most of the PWNe detected
have strong particle dominance is thus not affected by observational biases when their spin-down exceeds 10\% of the Crab.

\chapter{Systematic modeling of young PWNe}
\label{chap4}

\ifpdf
    \graphicspath{{Chapter4/Figs/Raster/}{Chapter4/Figs/PDF/}{Chapter4/Figs/}}
\else
    \graphicspath{{Chapter4/Figs/Vector/}{Chapter4/Figs/}}
\fi

PWNe constitute the largest class of identified Galactic sources at VHE. Since the observation of the first unidentified TeV $\gamma$-ray source
\citep{aharonian02,albert08b}, more than 20 PWNe have been identified at very-high-energies (VHE; $E>100$ GeV) by the current generation of
Cherenkov telescopes. With the forthcoming generation of Cherenkov telescopes as CTA, the number of these objects will increase to hundreds
\citep{deonawilhelmi13}, providing an unprecedented database to study the fraction of the pulsar energy that is transferred to the particles, or
the magnetic field in the nebula, or what rules the injection power in the surroundings of the pulsar.

In this chapter, we present a systematic study of the Galactic TeV-detected young PWNe. There are previous systematic studies with other models
(see e.g., \citealt{bucciantini11,tanaka11,tanaka13}), but they are partial. We have modelled the spectra of all the known objects of this kind
consistently and search correlations between parameters, extending the work done previously by \citet{mattana09}. We also compare the results
with the predictions extracted form the parameter space exploration done in \citet{torres13b}.

This chapter is based on the work done in \citet{torres14}.

\section{Characteristics of the sample}
\label{sec4.1}

A compilation of PSRs with known rotational parameters and characteristic age of $\tau_c<10^4$ years is presented in table \ref{datapsr},
which is obtained from the updated ATNF catalog \citep{manchester05} and includes the recently detected magnetar close to the Galactic Center
\citep{mori13,rea13}. The value of the period $P$, period derivative $\dot{P}$, distance $D$, characteristic age $\tau_c$, dipolar field $B_d$,
spin-down power $\dot{E}$, and $\dot{E}/D^2$ is listed. Their definitions are given below. These values are obtained directly from the catalog,
neglecting some better estimations on the distances, such as those of e.g., G0.9+0.1 or pulsars at the LMC, in favor of uniformity when compiling
the table. According to their position in the sky, we added the label H, M or V (for H.E.S.S., MAGIC or VERITAS respectively) to indicate the
visibility from different Cherenkov telescopes. The names of the TeV putative PWNe (or at least co-located TeV sources even if the TeV source is
likely not associated to the pulsar in some cases) are also included. The majority of these pulsars, located in the inner part of the Galaxy,
were in the reach of the H.E.S.S. Galactic Plane Survey (GPS), which attains a roughly uniform sensitivity of 20 mCrab \citep{gast11}. Some of
the pulsars in the northern sky have been observed by either MAGIC or VERITAS, with comparable sensitivity.

To compare the pulsar sample in table \ref{datapsr} we consider their characteristic ages. Even if this is not the pulsar real age, which is
usually uncertain, it can be considered a good approximation when the pulsar braking index is $n \sim 3$ and the initial pulsar spin-down period
is much shorter than the current one. In order to give an idea of relative strength, the $\dot{E}$ of any pulsar is compared to that of the
Crab extrapolated to the corresponding characteristic age. The last three columns in table \ref{datapsr} represent, respectively, the age of Crab
(assuming no change in braking index) at which it would have the same characteristic age as the corresponding pulsar ($T_{\tau_c}^{Crab}$), the
Crab's spin-down power at that age ($\dot{E}^{Crab}(T_{\tau_c}^{Crab})$), and the $\dot{E}$ of the pulsar in terms of percentage of
$\dot{E}^{Crab}(T_{\tau_c}^{Crab})$, which we refer to as CFP (or Crab fractional power). When looked in this way, the Crab PSR is no longer
special.

Considering the characteristic ages provides the possibility of assessing the total power input into the nebula. Take as an example PSR
J1617-5055 and J1513-5908, and assume for the sake of the argument that both generate TeV emission via a PWN. Both pulsars have essentially the
same, and relatively high spin-down power, $1.7 \times 10^{37}$ erg s$^{-1}$. However, one has likely been injecting this power for a much longer
time, since the characteristic age of PSR J1617-5055 is a factor of 5 larger. The electrons that populate the nebulae will sustain energy losses
and live, in most conditions, for more than 10$^4$ years. Thus, it is reasonable to suppose that there will be more high-energy electrons with
which generate TeV radiation in

\begin{landscape}
\begin{table}
\centering
\scriptsize
\caption[Pulsars in the ATNF catalog with less than $\tau_c$=10 kyr.]{Pulsars in the ATNF catalog with known period $P$, and period derivative
$\dot P$, and less than 10 kyr of characteristic age $(\tau_c)$. The first few columns are taken from ATNF data. The column ``TeV Obs.?" answers
whether the pulsar has been observed in TeV range, and, if so, by which telescope (noting H for H.E.S.S., M for MAGIC, and V for VERITAS). The
column ``TeV PWN?" indicates whether there has been a detection of a PWN or in general a TeV source spatially co-located with the pulsar. This
information comes from published literature. The last  three columns represent, respectively, the age of Crab (assuming today's braking index) at
which it would have the same characteristic age than the corresponding pulsar ($T^{Crab}_{\tau}$), the Crab's spin-down power at that age
($\dot{E}^{Crab}(T^{Crab}_{\tau})$), and the spin-down power of the pulsar in terms of percentage of $\dot{E}^{Crab}(T^{Crab}_{\tau})$, which we
refer to as CFP (Crab fractional power). Sources maked with $\dag$ are magnetars, which low rotational power is not expected to contribute
significantly to the corresponding TeV sources (marked in red). Names of the TeV sources shown in blue are the ones studied in this work.}
\vspace{0.2cm}
\begin{tabular}{l@{\quad}c@{\ }c@{\ }c@{\quad }c@{\quad}c@{\quad}c@{\quad}c@{\ \ }c@{\ }c@{\ }c@{\ \ }c@{\quad}c}
\hline
Name & $P$ (s) & $\dot P$ (s s$^{-1}$) & $D$ (kpc) & $\tau_c$ (yr) & $B_d$ (G) & $\dot{E}$ (erg s$^{-1}$) & $\dot{E}/ D^2$ (erg s$^{-1}$ kpc$^{-2}$) & TeV Obs.? & TeV source & $T^{Crab}_{\tau_c}$ (yr) &$ \dot{E}^{Crab}(T^{Crab}_{\tau_c})$ (erg s$^{-1}$) & CFP (\%)\\
\hline            
J1808$-$2024  $\dag$  &     7.5559                  &     $5.49\times 10^{-10}$            &     13.0 &  218  &    2.0 $\times 10^{15}$ &  5.0 $\times 10^{34}$ &  3.0 $\times 10^{32}$ & H & {\textcolor{red}{J1809-194/G11.0+0.08}} & \ldots & \ldots  & \ldots   \\
J1846$-$0258    &     0.3265     &     $7.10\times 10^{-12}$   &      5.8  & 728   &   4.9 $\times 10^{13}$ &  8.1 $\times 10^{36}$ & 2.4 $\times 10^{35}$ & H & {\textcolor{blue}{Kes 75}} & 238 & 1.6 $\times 10^{39}$ & 0.5\\
J1907+0919 $\dag$   &     5.1983                &     $9.20\times 10^{-11}$               &      \ldots     &  895    &  7.0 $\times 10^{14}$   &  2.6 $\times 10^{34}$ & \ldots & H & {\textcolor{red}{J1908+063/G40.1-0.89} }& 459 & 1.0 $\times 10^{39}$ & 0.003  \\
J1714$-$3810 $\dag$    &     3.8249                &     $5.88\times 10^{-11}$             &      \ldots     &  1030  & 4.8 $\times 10^{14}$ &  4.1 $\times 10^{34}$ &  \ldots & H & {\textcolor{red}{J1718--385/CTB37A}} & 638 & 7.2 $\times 10^{38}$  & 0.006\\
J0534+2200   &     0.0334     &      $4.21\times 10^{-13}$         &   2.0  &  1258   &  3.8 $\times 10^{12}$ & 4.5 $\times 10^{38}$ & 1.2 $\times 10^{38}$ & HMV & {\textcolor{blue}{Crab nebula}} & 940 & 4.5 $\times 10^{38}$ & 100\\
J1550$-$5418    &     2.0698              &     $2.32\times 10^{-11}$            &     9.7  & 1410   & 2.2 $\times 10^{14}$ & 1.0 $\times 10^{35}$ & 1.1 $\times 10^{33}$ & H & \ldots & 1141 & 3.5 $\times 10^{38}$  & 0.03\\
J1513$-$5908    &     0.1512   &     $  1.53\times 10^{-12}$   &       4.4  &  1560  & 1.5 $\times 10^{13}$ & 1.7 $\times 10^{37}$ & 9.0 $\times 10^{35}$ & H & {\textcolor{blue}{J1514--281/MSH 15--52}} & 1340 & 2.8 $\times 10^{38}$ &6 \\
J1119$-$6127    &     0.4079            &     $ 4.02\times 10^{-12}$       &      8.4  &  1610  & 4.1 $\times 10^{13}$  & 2.3 $\times 10^{36}$ &  3.3 $\times 10^{34}$ & H & {\textcolor{blue}{J1119-6127/G292.1--0.54}} & 1406 & 2.6 $\times 10^{38}$ & 0.9  \\
J0540$-$6919    &     0.0504        &    $ 4.79\times 10^{-13}$      &      53.7 &  1670  & 5.0 $\times 10^{12}$ & 1.5 $\times 10^{38}$ & 5.1 $\times 10^{34}$ & H & \ldots & 1486 & 2.4 $\times 10^{38}$  & 63\\ 
J0525$-$6607    &     8.0470                   &    $ 6.50\times 10^{-11}$              &   \ldots     & 1960  & 7.3 $\times 10^{14}$ & 4.9 $\times 10^{33}$ & \ldots & \ldots & \ldots & 1871 & 1.6 $\times 10^{38}$  & 0.003\\
J1048$-$5937    &     6.4520               &   $  3.81\times 10^{-11}$         &     9.0 &  2680  & 5.0 $\times 10^{14}$ & 5.6 $\times 10^{33}$ & 6.9 $\times 10^{31}$ & H & \ldots & 2825 & 7.8 $\times 10^{37}$ & 0.007\\
J1124$-$5916    &     0.1354         &  $   7.52\times 10^{-13}$    &     5.0  & 2850  & 1.0 $\times 10^{13}$ & 1.2 $\times 10^{37}$ & 4.8 $\times 10^{35}$ & H & \ldots & 3050 & 6.8 $\times 10^{37}$ &18 \\
J1930+1852   &     0.1368         &  $   7.50\times 10^{-13}$       &     7.0  & 2890  & 1.0 $\times 10^{13}$ & 1.2 $\times 10^{37}$ & 2.4 $\times 10^{35}$ & V & {\textcolor{blue}{J1930+188/G54.1+0.3}} & 3103 & 6.6 $\times 10^{37}$ & 18 \\
J1622$-$4950    &     4.3261                   &      $1.70\times 10^{-11}$              &     9.1 &  4030  & 2.7 $\times 10^{14}$ &  8.3 $\times 10^{33}$ & 9.9 $\times 10^{31}$ & H & \ldots & 4614 & 3.0 $\times 10^{37}$  & 0.03\\
J1841$-$0456    &     11.7789             &   $ 4.47\times 10^{-11}$           &   9.6 &  4180  & 7.3 $\times 10^{14}$ & 1.1 $\times 10^{33}$ & 1.2 $\times 10^{31}$ & H & \ldots & 4813 & 2.8 $\times 10^{37}$ & 0.004\\
J1023$-$5746    &     0.1115        &  $   3.84\times 10^{-13}$        &     {8.0}    &  4600  &  6.6 $\times 10^{12}$ &  1.1 $\times 10^{37}$ & 1.7 $\times 10^{35}$ &H & J1023+575 & 5370 & 2.2 $\times 10^{37}$   & 50\\
J1833$-$1034    &     0.0618         &  $   2.02\times 10^{-13}$          &     4.10 &  4850  & 3.6 $\times 10^{12}$ &  3.4 $\times 10^{37}$ &  2.0 $\times 10^{36}$ &H & {\textcolor{blue}{J1833--105/G21.5--0.9}} & 5701 & 2.0 $\times 10^{37}$ & 170 \\
J1838$-$0537    &     0.1457             &   $ 4.72\times 10^{-13}$           &    \ldots    &  4890  & 8.4 $\times 10^{12}$ &  6.0 $\times 10^{36}$ &\ldots &H & J1841--055 & 5754 & 1.9 $\times 10^{37}$ & 32\\
J0537$-$6910    &     0.0161         &  $   5.18\times 10^{-14}$   &     53.7 & 4930  & 9.2 $\times 10^{11}$ &  4.9 $\times 10^{38}$ &  1.7 $\times 10^{35}$ &H & N157B (in the LMC) & 5807 & 1.9 $\times 10^{37}$ & 2579 \\
J1834$-$0845 $\dag$    &     2.4823                &    $ 7.96\times 10^{-12}$          &    \ldots    &  4940  & 1.4 $\times 10^{14}$ & 2.1 $\times 10^{34}$ & \ldots &H & {\textcolor{red}{J1834--087/W41}} & 5820 & 1.9 $\times 10^{37}$ & 0.1\\
J1747$-$2809    &     0.0521              &    $ 1.55\times 10^{-13}$      &     17.5   & 5310  & 2.9 $\times 10^{12}$ & 4.3 $\times 10^{37}$ & 1.4 $\times 10^{35}$ &H & {\textcolor{blue}{J1747--281/G0.9+0.1}} & 6311 & 1.6 $\times 10^{37}$ & 269 \\
J0205+6449   &     0.0657     &    $ 1.94\times 10^{-13}$  &     3.2  & 5370  & 3.6 $\times 10^{12}$ & 2.7 $\times 10^{37}$ & 2.6 $\times 10^{36}$ &MV & \ldots & 6390 & 1.6 $\times 10^{37}$  & 169 \\
J1813$-$1749    &     0.0446             &    $  1.26\times 10^{-13}$           &      \ldots   &   5600  & 2.4 $\times 10^{12}$ & 5.6 $\times 10^{37}$  & \ldots &H & {\textcolor{blue}{J1813--178/G12.8--0.02}} & 6695 & 1.4 $\times 10^{37}$ &400  \\
J0100$-$7211    &     8.0203                 &     $1.88\times 10^{-11}$            &     62.4 & 6760  & 3.9 $\times 10^{14}$ & 1.4 $\times 10^{33}$ & 3.7 $\times 10^{29}$ & \ldots & \ldots & 8233 & 9.1 $\times 10^{36}$ & 0.02\\
J1357$-$6429    &  0.1661         &     $3.60\times 10^{-13}$    &     4.1 &  7310  &  7.8 $\times 10^{12}$ &  3.1 $\times 10^{36}$ & 1.9 $\times 10^{35}$ &H & {\textcolor{blue}{J1356--645/G309.9--2.51}} & 8962 & 7.6 $\times 10^{36}$ &41 \\
J1614$-$5048   &   0.2316            &   $4.94\times 10^{-13}$        &  7.2  & 7420 & 1.1 $\times 10^{13}$ & 1.6 $\times 10^{36}$ & 3.0 $\times 10^{34}$ &H &  \ldots & 9107 & 7.3 $\times 10^{36}$ & 22\\
J1734$-$3333   &     1.1693             & $2.28\times 10^{-12 }    $     &     7.4  & 8130   & 5.2 $\times 10^{13}$ & 5.6 $\times 10^{34}$ & 1.0 $\times 10^{33}$ &H & \ldots & 10048 & 5.9 $\times 10^{36}$ & 0.9\\
J1617$-$5055    &     0.0693              & $    1.35\times 10^{-13 }  $       &     6.4  &  8130 & 3.1 $\times 10^{12}$ &  1.6 $\times 10^{37}$ &  3.8 $\times 10^{35}$ &H & J1616-508 & 10048 & 5.9 $\times 10^{36}$ & 271 \\
J2022+3842    &     0.0242         &$4.32\times 10^{-14 }$      &     10.0 & 8910 & 1.0 $\times 10^{12} $& 1.2 $\times 10^{38}$ & 1.2 $\times 10^{36}$ &\ldots & \ldots & 11082 & 4.8 $\times 10^{36}$ & 2500 \\
J1708$-$4009 $\dag$    &     11.0013              &     $1.93\times 10^{-11}      $     &     3.8 &  9010 & 4.7 $\times 10^{14}$ &  5.7 $\times 10^{32}$ & 4.0 $\times 10^{31}$ &H &  {\textcolor{red}{J1708-443/G343.1--2.69}} & 11215 & 4.7 $\times 10^{36}$ & 0.01\\
J1745$-$2900 $\dag$ & 3.76356 & $6.5 \times 10^{-12}$ & 8.0 & 9170 & 1.6 $\times 10^{14}$ &  $4.8 \times 10^{33}$ & 7.5 $\times 10^{31}$ & HM & {\textcolor{red}{(in the Galactic Center)}} & 11427 & 4.4 $\times 10^{36}$ & 0.99\\
\\
\hline
\end{tabular}
\label{datapsr}
\end{table}
\end{landscape}

\hspace{-0.5cm}the older pulsar than in the younger one. The differences between PSR J1617-5055 and J1513-5908 are reflected in
the comparison with Crab at the moment when its characteristic age is correspondingly the same to the pulsar in question. PSR J1617-5055 is
approximately three times as luminous than Crab will be at the same $\tau_c$. Instead PSR J1513-5908 spin-down corresponds to only a few percent
of the Crab will have at its $\tau_c$. Thus, even when both have the same $\dot{E}$ we are speaking of very different nebulae.

\begin{figure}
\centering
\includegraphics[width=0.6\textwidth]{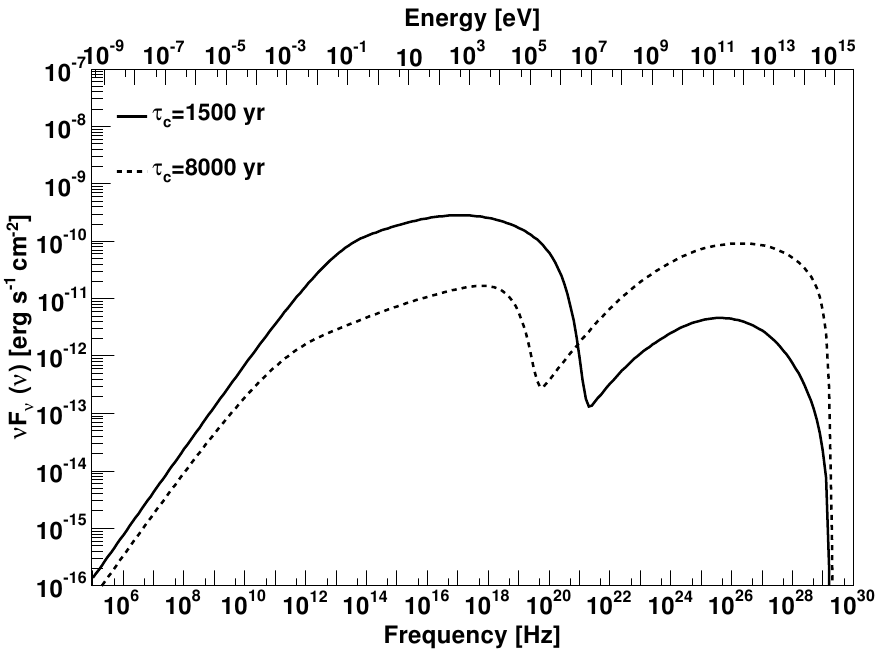}
\caption[PWNe models that differ only in age]{Comparing two PWNe models that differ only in age. Parameters of these models are as those used for
the Crab nebula, and both have the same spin-down power.}
\label{difage}
\end{figure}

To exemplify further this point, consider two mock pulsars having the same spin-down evolution, magnetic fraction, injection spectrum, and photon
background parameters than Crab (see below for precise definition of all these quantities) and both having also the same spin-down power,
$1.7 \times 10^{37}$ erg s$^{-1}$, but two different characteristic ages of 1500 and 8000 years, respectively. The modeled PWNe (details of the
model itself are discussed below) when every parameter is the same but just the $\tau_c$ and the corresponding real age vary turn out to be
different: For instance, the resulting magnetic field varies from 1 to 30 $\mu$G. The SEDs shown in figure \ref{difage} show that the spin-down
power $\dot{E}$, or the parameter $\dot{E}/D^2$ (which is the same for the SEDs in the figure), unless of course when $\dot{E}$ is extremely low,
cannot by themselves blindly define dectectability of PWNe, and further considerations about the PWNe age, injection, and environment have to be
taken into account. This conclusion is emphasized when the photon background, the injection, and the magnetic fraction, among other key
parameters, may vary from one pulsar to the next.

Table \ref{datapsr} shows that most of the young pulsars we know of were indeed surveyed for TeV emission. This has motivated developing detailed
radiative models to tackle the complexities in each of the PWNe. However, whereas some of these models are time-dependent, which is essential for
a proper accounting of the nebula evolution and electron losses as per the discussion above, they are different to one another, and are
constructed under different approximations and assumptions (see section \ref{sec1.6} for a summary). The magnitude of spectral results introduced
by different underlying assumptions has been quantified only in some cases (e.g., see section \ref{sec3.1} or \citealt{martin12})). Having a
clear conversion of results from one model to another, in order to generate a uniform theoretical setting where PWNe fittings can be compared, is
simply impossible. In addition, apart from the obvious mismatches in the models per se, the nebulae that have been studied with each of them are
scarce. Table \ref{models} gives some examples using a certainly incomplete span of the literature. Our interpretation of observations is based
on uncommon modeling, undermining our conclusions.

\begin{table}
\scriptsize
\centering
\caption[Examples of radiative time-dependent models used to fit observations of young PWNe]{Examples of radiative time-dependent models used to
fit observations of young PWNe.}
\begin{tabular}{lllllllllll}
\hline
 & \rotatebox{90}{Tanaka  \& Takahara 2011}  & \rotatebox{90}{Zhang et al. 2008}  & \rotatebox{90}{Bucciantini et al. 2011}&  \rotatebox{90}{Fang \& Zhang 2010} & 
\rotatebox{90}{Qiao, Zhang, \& Fang 2009} & \rotatebox{90}{Li, Chen, \& Zhang 2010}&  \rotatebox{90}{Venter \& de Jager 2007} & \rotatebox{90}{{ {This work}}}\\  
 \hline   
Crab nebula & X  & X & X & --  & -- & -- & -- & X \\
G54.1+0.3 & X & -- & --  & -- & -- &  X & -- & X\\
G0.9+0.1 & X & -- & X & X & X & -- & X & X \\
G21.5--0.9 & X & -- & -- & -- & -- & -- &-- & X \\
MSH 15--52 & -- & X & X &   X & -- & -- &-- & X\\
G292.2--0.5 & -- &-- &-- & -- & -- & -- &-- & X \\
Kes 75 & X & X & X &-- & -- & -- &-- & X \\
HESS J1356--645 & -- &-- &-- & -- & -- & -- &-- &X \\ 
CTA-1 & -- &-- &-- & -- & -- & -- &-- & X \\
HESS J1813--178    & -- &-- &-- & -- & -- & -- &-- &X \\ 
\hline
\hline
\end{tabular}
\label{models}
\end{table}

In this work, we put at least a partial remedy to this situation, and provide a study of several young, TeV detected PWN. In order to do that we
have used version 1.3 of TIDE-PWN. Our sample is formed by 10 TeV detected, possibly Galactic PWNe, taken from table \ref{datapsr} plus the
recently detected CTA-1, which has a characteristic age slightly larger than 10$^4$ years. We comment also on why we do not consider in our study
the cases of HESS J1023–575, J1616–508, J1834–087/W41, and J1841–055 (in most cases, the information gathered on them imply that the TeV emission
is not univocally associated with a PWN) as well as Boomerang and HESS J1640–465. We find that not all of the 10 cases studied are best
interpreted with a PWN. In particular, we conclude that the case of HESS J1813–178 is most likely related to the SNR rather than to the PWN.

\section{Individual modelling of young PWNe}
\label{sec4.2}

Table \ref{param} presents all the fit parameters and assumed physical magnitudes for all the models fitted, including the Crab nebula. We
introduce below each of the TeV detections in our sample, proppse a PWN model, and discussiong the complexities of each case, surfacing caveats
of our model when appropiate.

\begin{table}
\centering
\scriptsize
\caption[Physical magnitudes obtained for the young PWNe in our study]{Physical magnitudes obtained for the young PWNe in our study. The dot
symbols are used to represent the same value of the corresponding left column.}


  \begin{tabular}{  l l  ll ll l l ll l   }
  \hline
                               &{\bf \textcolor{red}{Crab nebula}} & {\bf \textcolor{red}{G54.1+0.3}}           &  {\bf \textcolor{red}{G0.9+0.1}} & \ldots  &  {\bf \textcolor{red}{G21.5$-$0.9}}  &  {\bf \textcolor{red}{MSH 15--52}} 
                                &  \ldots   \\
                               &                       &         & Model 1     & Model 2   &    &    Model 1  & Model 2 \\
 \hline
{\textcolor{blue}{Pulsar \& Ejecta}}  \\
 \hline
$P(t_{age})$ (ms) & 33.40 & 136  & 52.2 & \ldots & 61.86 & 150 &  \ldots \\

$\dot{P} (t_{age})$ (s s$^{-1}$)  	& 4.2$\times 10^{-13}$ &  7.5$\times 10^{-13}$ & 1.5$\times 10^{-13}$ & \ldots & 2.0$\times 10^{-13}$ & 1.5$\times 10^{-12}$  &   \ldots \\

$\tau_{c}$   (yr)  & 1296   &  2871 & 5305 &\ldots &  4860  & 1600  &  \ldots \\

$\dot{E}(t_{age})$  (erg s$^{-1}$) 	& 4.5 $\times 10^{38}$  &  1.2$\times 10^{37}$  & 4.3$\times 10^{37}$ & \ldots & 3.4$\times 10^{37}$ &  1.8 $\times 10^{37}$  &    \ldots  \\

$n$ &  2.509 &   3  & 3 &\ldots & 3 & 2.839 &  \ldots\\

$t_{age}$  (yr) & 940  &  1700 & 2000 & 3000 & 870 &  1500 &  \ldots \\        

$D$  (kpc)  & 2.0   &  6  & 8.5 & 13   & 4.7 & 5.2  &    \ldots \\             

$\tau_{0}$ (yr) & 730  & 1171   & 3305 & 2305 & 3985 &   224    &  \ldots \\          

$\dot{E}_{0}$   (erg s$^{-1}$)     	& 3.1 $\times 10^{39}$  & 7.2$\times 10^{37}$  & 1.1$\times 10^{38}$ & 2.3$\times 10^{38}$ &5.0$\times 10^{37}$ & 1.3$\times 10^{39}$  
&  \ldots \\ 


$M_{ej}$ ($M_{\odot}$)                    	&  9.5  &    20 &  11 & 17 & 8  & 10  &  \ldots  \\

$R_{PWN}(t_{age})$ (pc)             & $2.1$      		&1.4      & 2.5 & 3.8  & 0.9 & 3  &  \ldots \\
\hline
 {\textcolor{blue}{Environment}}\\
\hline
$T_{FIR}$ (K)                                  	& 70  	 &       20       & 30 & \ldots & 35 & 20 &  20 \\

$w_{FIR}$ (eV/cm$^{3}$)       	&   0.5      &    2.0       & 2.5 & 3.8 & 1.4  & 5 &   4 \\ 

$T_{NIR}$ (K)     	&   5000     &     3000      & 3000 & \dots  & 3500 &  3000 &  400 \\

$w_{NIR}$   (eV/cm$^{3}$)	&  1.0       	 &       1.1     & 25 & \dots & 5.0 & 1.4 & 20 \\ 

$n_{H}$    	&   1.0         &    10  & 1.0 &\ldots & 0.1 &  0.4 &   \ldots \\
 \hline
 {\textcolor{blue}{Particles and field}} \\
 \hline
$\gamma_{max} (t_{age})$ & $7.9\times10^9$  & $7.5\times10^8$   & $1.3 \times10^9$ & $1.9 \times10^9$ & $2.4 \times10^9$ & $1.9 \times10^9$ 
&   $2.3 \times10^9$ 
&   \\

$\gamma_{b}$                      & $7\times 10^5$    & $5\times 10^5$    & $1.0 \times10^5$ & $0.5 \times10^5$ & $1.0 \times10^5$ &  $5.0 \times10^5$ 
&  \ldots \\

$\alpha_{1}$              & $1.5$      	 	& 1.20          & 1.4 & 1.2 & 1.0 & 1.5  &  \ldots  \\

$\alpha_{2}$              & $2.5$      		& 2.8          & 2.7 & 2.5 & 2.5 & 2.4 &   \ldots \\

$\epsilon$                 & $0.2$      		& 0.3         & 0.2 & \ldots & 0.2 & 0.2 &  \ldots \\

$B(t_{age})$    ($\mu$G)          & $84$       		& 14   & 14 & 15 & 71 & 21 & 25    \\

$\eta$                         & $0.03$    		& 0.005       &0.01 & 0.02 & 0.04 & 0.05 & 0.07  \\
\hline
\end{tabular}
\label{param}
\end{table}

\begin{table}
\centering
\scriptsize
  Continued.


  \begin{tabular}{  l l  lll ll l l ll l   }
  \hline
                                & {\bf \textcolor{red}{G292.2--0.5}}  & {\bf \textcolor{red}{ Kes 75}}  & \ldots    &  {\bf \textcolor{red}{HESS J1356--645  }}  & \dots  & {\bf \textcolor{red}{CTA~1}} \\
                                &   & Model 1      & Model 2   &   Model 1 & Model 2  & \\
 \hline
{\textcolor{blue}{Pulsar \& Ejecta}}  \\
 \hline
$P(t_{age})$ (ms) & 408 &  324 & \ldots   & 166 & \ldots  & 316.86 & \\

$\dot{P} (t_{age})$ (s s$^{-1}$)  	&   4.0$\times 10^{-12}$ & 7.1$\times 10^{-12}$ & \ldots &  3.6$\times 10^{-13}$ & \ldots  & 3.6$\times 10^{-13}$ \\

$\tau_{c}$   (yr)  &   1610& 724 & \ldots  & 7300 & \ldots & 13900 &  \\

$\dot{E}(t_{age})$  (erg s$^{-1}$) 	&  2.3 $\times 10^{36}$ & 8.2$\times 10^{36}$ & \ldots &  3.1$\times 10^{36}$ & \ldots    &  4.5$\times 10^{35}$ &  \\

$n$  & 1.72 &  2.16  & \ldots & 3 & 2 & 3 &  \\

$t_{age}$  (yr) & 4200 &  700 & 800 & 6000 & 8000 & 9000 & \\        

$D$  (kpc)     &  8.4 & 6  & 10.6  & 2.4 & \ldots & 1.4 &  \\             

$\tau_{0}$ (yr)   & 270 & 547   & 447 & 1311 & 6622 & 4901 &  \\          

$\dot{E}_{0}$   (erg s$^{-1}$)   &  9.3 $\times 10^{40}$ & 7.7$\times 10^{37}$ & 1.3$\times 10^{38}$ &  9.6$\times 10^{37}$ & 3.3$\times 10^{37}$ & 3.6$\times 10^{36}$ &  \\ 


$M_{ej}$ ($M_{\odot}$)                & 35   	  &   6 & 7.5 & 10 & 12 & 10 &  \\

$R_{PWN}(t_{age})$ (pc)            &13		& 0.9     & 1.0 &  9.5 & 9.9 & 8.0 &  \\
\hline
 {\textcolor{blue}{Environment}}\\
\hline
$T_{FIR}$ (K)                                 & 70 	  	 &       25      & \ldots  & 25 & \ldots & 70 & \\

$w_{FIR}$ (eV/cm$^{3}$)       & 3.8	      &    2.5       & 5.0  & 0.4 & \ldots & 0.1 & \\ 

$T_{NIR}$ (K)    &4000 	 &     5000    & \ldots   & 5000 & \ldots & 5000 &  \\

$w_{NIR}$   (eV/cm$^{3}$)	&1.4       	 &       1.4    & 1.4 & 0.5 & \ldots & 0.1 &   \\ 

$n_{H}$    	&0.02 &      1.0 & \ldots  & 1.0 & \ldots & 0.07 &  \\
 \hline
{\textcolor{blue}{Particles and field}}\\
 \hline
$\gamma_{max} (t_{age})$ & $8.0\times10^8$   & $5.2\times10^8$   & $4.9 \times10^8$  & $8.8\times10^8$ & $1.5\times 10^{9}$ & $8.6\times 10^{8}$ &  \\

$\gamma_{b}$          &                  $5.0\times 10^6$    & $2.0\times 10^5$    & $1.0 \times 10^5$ & $3.0 \times 10^5$ & -- & $0.8\times 10^{5}$ & \\

$\alpha_{1}$                   	& 1.5 	& 1.4          & 1.6  & 1.2 & -- & 1.5 & \\

$\alpha_{2}$                   &4.1  		& 2.3          & 2.1 & 2.52 & 2.6 & 2.2 &  \\

$\epsilon$                    &  0.3 		& 0.2          & 0.1  & 0.2 & 0.3 & 0.2 &  \\

$B(t_{age})$    ($\mu$G) & 4                		& 19    & 33  & 3.1 & 3.5 & 4.1 & \\

$\eta$                         &0.03   		& 0.008       &0.03  & 0.06 & 0.08 & 0.4 &  \\
\hline
\end{tabular}
\end{table}

\subsubsection{VER J1930+188 (G54.1+0.3)}

The central pulsar in G54.1+0.3 (PSR J1930+1852) is observed in radio and X-rays to have a period of 136 ms, and a period time derivative of
$7.51 \times 10^{-13}$ s s$^{-1}$, implying a characteristic age of $\tau_c \sim$2.9 kyr \citep{camilo02}. The braking index is unknown, we
assume it to be 3. Considering a possible range of braking indices and initial spin periods, \citet{camilo02} estimated the age of G54.1+0.3 to
be between 1500 and 6000 yr.

The PWN was first discovered by \citet{reich85} in radio wavelengths. The later observation by \citet{lu01,lu02} reveals the X-ray non-thermal
spectrum and the ring and bipolar jet morphology confirmed the source as a PWN. From the equations describing the PWN evolution in the model by
\citet{chevalier05}, \citet{camilo02} calculated an age of 1500 yr and an initial spin period of 100 ms. Based on HI line emission and absorption
measurements, the distance to G54.1+0.3 was reported to be in the 5–9 kpc range \citep{weisberg08,leahy08a}, while the pulsar dispersion measure
implied a distance less than or equal to 8 kpc \citep{camilo02,cordes03}. \citet{leahy08a} suggested a morphological association between the
nebula and a CO molecular cloud at a distance of 6.2 kpc. However, the absence of X-ray thermal emission and the lack of evidence for an
interaction of the SNR with the cloud are caveats in this interpretation. According to \citet{temim10}, who also assumes a distance of 6 kpc, the
size of the PWN is $2 \times 1.3$ arcmin. Extrapolating these magnitudes to the spherical case by matching the projected area of the nebula to
that of a circle, the radius for the nebula assumed in our model is $\sim$1.4 pc at 6 kpc. We also assume \citet{temim10} estimation of the mass
of the ejecta ($\sim$20M$_\odot$). Since the SNR shell has not been detected, the particle density in the nebula is more uncertain.
\citet{temim10} have derived a density of 30 cm$^{-3}$ at one IR knot that appears to be interacting with one of the jets of the PWN. To be
conservative (see the discussion on the influence of the bremsstrahlung component in the SED below) we will adopt a lower, average density of 10
cm$^{-3}$.

The observations against which we fit the theoretical model are collected from different works. Radio observations are obtained from
\citet{altenhoff79,reich84,reich85,caswell87,velusamy88,condon89,griffith90,hurleywalker09}. X ray data come from \citet{temim10}, where we have
considered the fluxes given in their table 2 except the one corresponding to the central object. For the spectral slope span, we have adopted the
limiting cases of $-$1.8 and $-$2.2, also from \citet{temim10}. We note that the X-ray observations of \citet{lu01,lu02} (used for instance in
\citealt{lang10,li10,tanaka11}) also took into account the central source (region 1 of \citealt{temim10}; leading to a higher flux, and did not
account for pileup effects (see \citealt{temim10} for a discussion). Use of these X-ray flux values are thus disfavored for modeling the PWN.
Finally, TeV observations represent the results of the VERITAS array \citep{acciari10}. Fermi-LAT did not detect G54.1+0.3 \citep{acero13}.

For the ISRF, the region around G54.1+0.3 has been observed in the infrared by \citet{koo08}, and \citet{temim10}. These observations suggest
that the ISRF around G54.1+0.3 is larger than that resulting from Galactic averages as obtained, for instance, from CR propagation models. We
concur (see table \ref{param}). Considering further additional components in the ISRF, as for instance \citet{li10} did with the optical/UV
contribution from nearby YSOs, does not yield to any significant changes in the fit.

\begin{figure}[t!]
\centering
\includegraphics[width=1.0\textwidth]{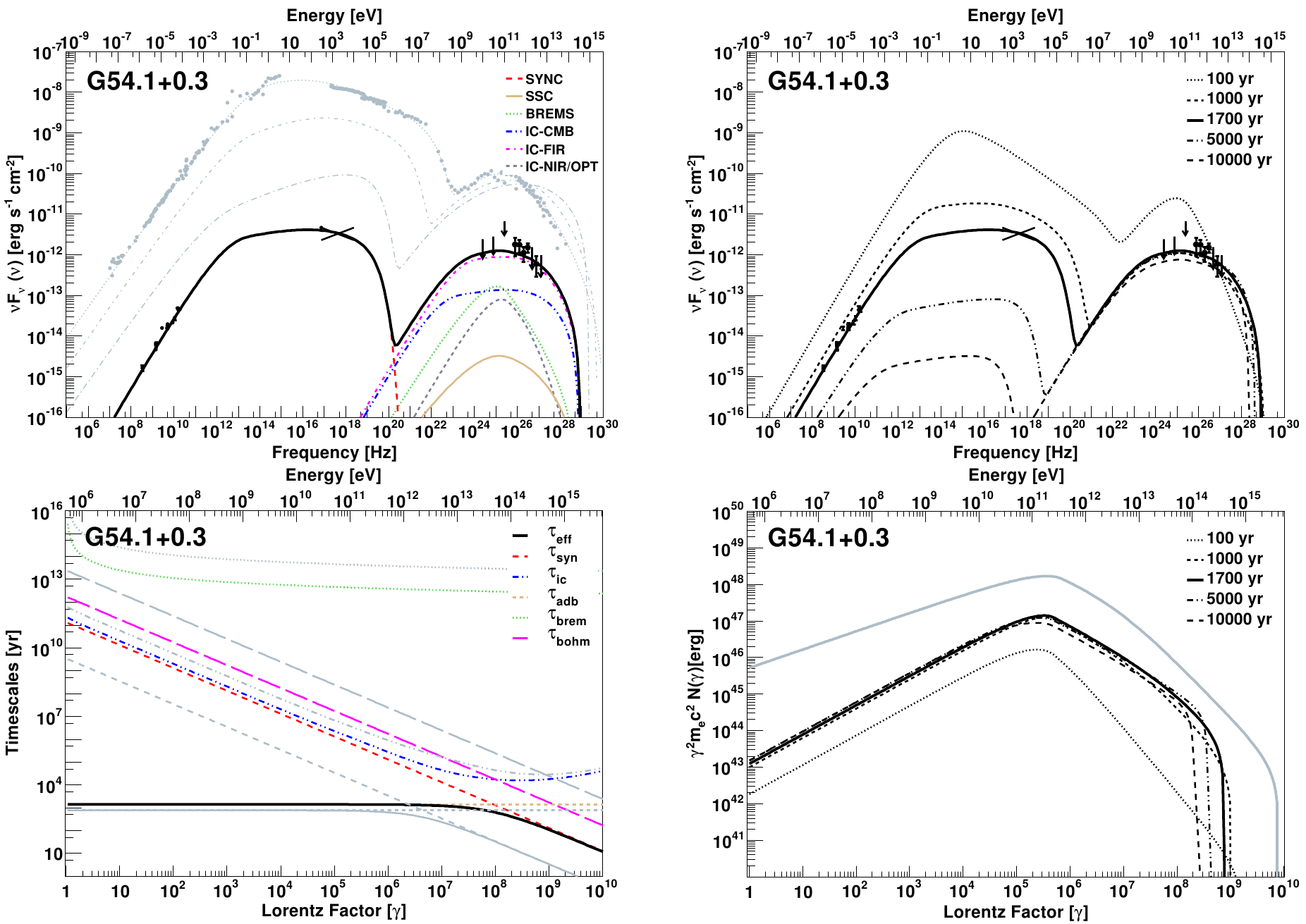}
\caption[SED of G54.1+0.3 as fitted by our model]{Details of the SED (black bold line) of G54.1+0.3 as fitted by our model. The top left panel
shows the SED at the adopted age (i.e., today), whereas the top right panel does it along the time evolution. The bottom panels represent the
timescales for the different losses today (the effective timescale for the losses is represented with bolder curves, both for G54.1+0.3 and the
Crab nebula) and the evolution of the electron spectra in time. Here and in the figures that follow, we use the results of the Crab nebula model
as a comparison. In the top-left panel, we plot (in grey, from top to bottom) three curves corresponding to the Crab nebula's SED at 940, 2000,
and 5000 years. In the bottom left panel we compare the losses of G54.1+0.3 to each of the processes with those of Crab (in grey). In the case of
the electron distribution we compare with the electron population resulting from the Crab nebula model at its current age. For details regarding
the observational data and a discussion of the fit, see the text.}
\label{g54}
\end{figure}

YSOs, does not yield to any significant changes in the fit. Table \ref{param} and figure \ref{g54} present the fitting result of our model of
G54.1+0.3. Radio and X-ray data can be fitted very well with a synchrotron component driven by a low magnetic field of only 14 $\mu$G. We found a
very small parameter dependence for differences in the value of the containment factor; for instance for values of $\varepsilon$=0.5, 0.3, 0.2,
other parameters are only slightly changed. The magnetic fraction in our model is 0.005 (half of a percent). This turns out to be a factor of 6
smaller than that of Crab nebula. Clearly, G54.1+0.3 is a particle dominated nebula.

At high energies, the influence of the SSC, and the NIR-IC contribution is negligible, with the FIR-IC contribution clearly dominating and the
CMB-IC and Bremsstrahlung contributing at the same level at $\sim$100 GeV (albeit both do so at one order of magnitude lower than the dominant
component). The Bremsstrahlung contribution is linear with the uncertain particle density. Then, the selection of 10 cm$^{-3}$ as the average
particle density against which we compute the bremsstrahlung contribution may be subject to further discussion. We note that it is a factor of 3
lower than that measured in the IR knots (see, e.g., \citealt{temim10}). However, the average density of the medium is probably lower than that
found in such IR enhancements, and in addition, relativistic electrons may not be able to fully penetrate into the knots. Other authors, e.g.,
\citet{lu01}, used the IR-knot measured 30 cm$^{-3}$ as average particle density, but did not compute the bremsstrahlung luminosity in his
leptonic models. For such densities, the bremsstrahlung would overcome the IC-CMB contribution to the SED in a narrow range of energies. In
agreement with observations, G54.1+0.3 should not be seen by Fermi-LAT in the framework of this model.

One interesting difference with the results of the work by \citet{tanaka11} is the value of the high-energy index ($\alpha_2$). In our model, it
results in 2.8 where it is 2.55 for \citet{tanaka11}. Contributing to this difference is likely the fact that in the latter model the maximum
energy of electrons is fixed all along the evolution of the nebula, whereas in ours it evolves in time in agreement with the rest of the physical
magnitudes. Having a fixed maximal electron energy hardens the needed slope to fit the data.

\citet{li10} have argued for a lepto-hadronic origin of the TeV radiation from G54.1+0.3. The main reason argued for this case is that a
leptonic-only model would produce a low magnetic field, as indeed we find. This would result, these authors claim, very low in comparison with
estimates of an equipartition magnetic field of 38 $\mu$G, obtained from the radio luminosity of the PWN or a magnetic field of 80–200 $\mu$G
from the lifetime of X-ray emitting particles as discussed by \citet{lang10}. But there is no indication that the PWN is in equipartition (in
fact, models such as ours, including a proper calculation of losses) show that it is not necessary to include any significant relativistic hadron
contribution to fit the SED.

Finally, we have also considered uncertainties in parameters that lead to degeneracies in the fit quality. One such is the age. Indeed,
considering ages around 1700 years would still make possible to produce a good fit to the spectral data if changes to the photon backgrounds are
allowed. For instance, the FIR energy density would need to shift from 2 to 3 eV cm$^{-3}$ in order to have a good fit when the age is 1500 yrs.
Another aspect of note is the degeneracy in $\gamma_b$, which, within a factor of a few, can lead to equal-quality fits requiring a smaller
magnetic field (and magnetic fraction) or small changes in the FIR density.

\subsubsection{HESS J1747-281 (G0.9+0.1)}

The PWN G0.9+0.1 was first identified in radio emission \citep{helfand87}, and then detected in X-rays \citep{mereghetti98,sidoli00}. Its central
pulsar, PSR J1747-2809, was detected years later \citep{camilo09}. The period of this pulsar is 52.2 ms, with a period derivative of
$1.56 \times 10^{-13}$ s s$^{-1}$, leading to a characteristic age $\tau_c$=5300 kyr, and a spin-down luminosity of $4.3 \times 10^{37}$ erg
s$^{-1}$ \citep{camilo09}, one of the largest among Galactic pulsars. The braking index of PSR J1747-2809 is unknown, and we assume $n=3$. The
actual age of G0.9+01 is also unknown. \citet{camilo09} estimated an age between 2000 and 3000 yr, which is compatible with the properties of the
composite SNR in radio and in X-rays \citep{sidoli00}. The average radius of the PWN in radio is $\sim$1 arcmin \citep{porquet03}. G0.9+01 is
close to the Galactic Center. Because of that a distance of 8.5 kpc is usually adopted \citep{aharonian05a,dubner08}. \citet{camilo09} estimated a
distance of 13 kpc according to the dispersion measure and the NE2001 electron model \citep{cordes03}, but this estimation can be especially
faulty towards the inner Galactic regions, and only a range between 8 and 16 kpc can be reliably suggested.

The observational data against which we fit the theoretical models come from different sources. We use new high-resolution radio images from
observations at 4.8 GHz and at 8.4 GHz carried out with the Australia Telescope Compact Array, and from reprocessed archival VLA data at 1.4 GHz
\citep{dubner08}. The X-rays observations we use were done by XMM \citep{porquet03}, and have an unabsorbed flux in the range 2–10 keV of
$5.78 \times 10^{-12}$ erg s$^{-1}$ cm$^{-2}$, with a power-law index 1.99$\pm$0.19. This corresponds to an X-ray luminosity of
$\sim 5 \times 10^{34}$ erg s$^{-1}$, if located at 8.5 kpc. The lack of non-thermal X-ray emission from the shell of G0.9+0.1 argue against the
TeV radiation being leptonically originated there. TeV observations are as in figure 3 of \citet{aharonian05a}.

The values needed of FIR and NIR energy densities for the nebula to be detected in the TeV range, which we found by fitting -see table
\ref{param}-, are higher than what is found in the model by GALPROP \citep{porter06}. This discrepancy is not surprising at the central Galactic
region.

It is interesting to note that different authors have used alternative set of observations for their fits. \citet{aharonian05a} used the XMM
data. \citet{porquet03} like us, but for the radio data they used the work by \citet{helfand87} since their paper is prior to that of
\citet{dubner08}. The latter authors argue for an overestimation of the radio flux of the PWN given by \citet{helfand87}. On the other hand,
\citet{tanaka11} used the data by \citet{dubner08} for radio, but Chandra observations for X-ray data \citep{gaensler01a}, a choice making the
X-ray spectrum higher in the SED, see the discussion in \citep{porquet03}. These differences in the assumed multi-wavelength spectra of the PWN
reflect in the fits, and have to be taken care of when analyzing results.

\begin{figure}[t!]
\centering
\includegraphics[width=1.0\textwidth]{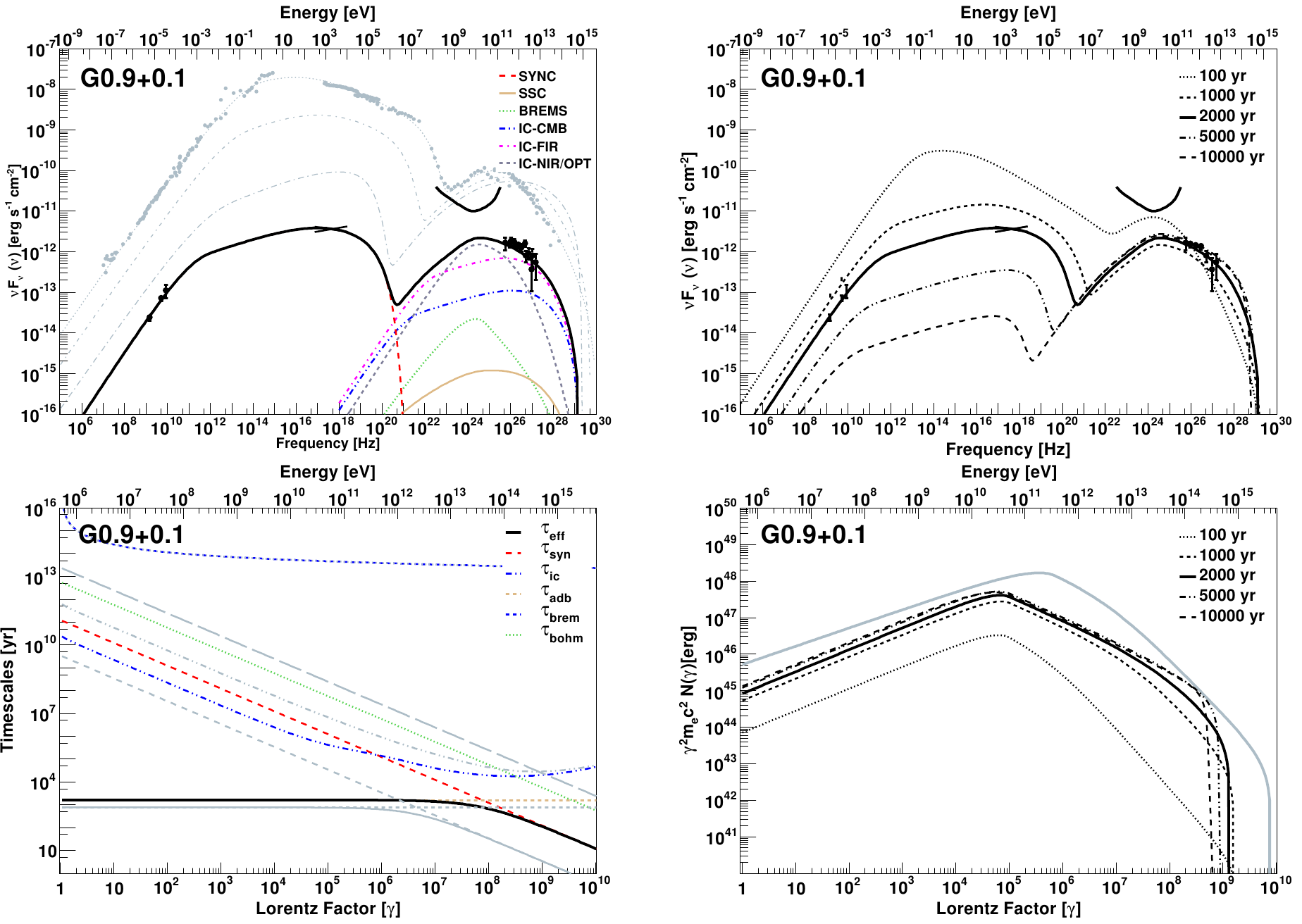}
\caption[SED of G0.9+0.1 as fitted by our model]{Details of the SED of G0.9+0.1 as fitted by our model. The panels are as in figure \ref{g54}.
For details regarding the observational data and a discussion of the fit, see the text.}
\label{g09}
\end{figure}

Due to the uncertainties in the distance, age, and ejected mass, we consider two cases in our fit: In Model 1 (to which the plots in figure
\ref{g09} correspond) we assume a distance of 8.5 kpc, and an age of 2000 yrs. We consider that the PWN is a sphere with a physical radius of 2.5
pc. In Model 2 we assume a larger distance of 13 kpc, and an age of 3000 yrs, leading to a physical radius of 3.8 pc. We assume a value of
11M$_\odot$ (Model 1) and 17M$_\odot$ (Model 2) for the ejected mass. In both models we assume a density of 1 cm$^{-3}$. There are no significant
differences (beyond the defining values for the dynamics and location) between these two models. The magnetic field obtained from our fits is low
$\sim$15 $\mu$G, and the magnetic fraction is in the order of 1–2\%. The spectral break in the electron distribution is equal to 10$^5$ for Model
1 and $0.5 \times 10^5$ for Model 2. The spectral indices for the two cases are given in table \ref{param} and they are very similar for the two
models as well. This similarity gives an idea of the importance of knowing the age and distance of the PWN in fixing model parameters.

We have also analyzed the case in which the injected spectrum is a single power-law; but in practice, this required increasing the minimum energy
of the electrons in the nebula up to the break energy. The values obtained for the energy densities in FIR and NIR in order to fit the data
change accordingly. The SED distribution of all of these models (Models 1 and 2, both described in table \ref{param}, and their analogous with a
single power-law) is essentially exactly the same as the one plotted in figure \ref{g09}, implying that the degeneracy will be hard to break
without precise measurements or modeling of the ISRF backgrounds.

In order to reduce the FIR and NIR densities the only solution is of course to have more high-energy electrons in the nebula. This can be
achieved for instance assuming an injection of electrons in the form of a single power-law with a fixed maximum and minimum energy, as in the
case of \citet{tanaka11}. However, there are no particular reasons to choose given values for the latter parameters. Other differences with the
assumptions in the \citet{tanaka11} model is that their nebula is 4500 years-old (instead of 2000–3000 yrs) and located slightly closer, at 8 kpc
(instead of 8.5 kpc). At this adopted age/distance, which seems not particularly preferred by any observation, the total power would be $\sim$1
order of magnitude larger than that in our Model 1; what explains the lesser need of target photon backgrounds to achieve the same TeV fluxes.
This set of assumptions for the injection and age does not appear preferable or particularly justifiable when confronted with the possibility of
having larger local background in the Galactic Center environment. \citet{fang10a} also studied the spectral evolution of G0.9+0.1; but under the
assumption that the particle distribution at injection is given by a relativistic Maxwellian distribution plus a single power-law distribution.
The latter produces a distinctive feature in the SED at about 10$^{-9}$ MeV for which there is no observational need yet. Even when different
assumptions and modeling techniques are used, a low magnetic field is also singled out by their study.

In agreement with our prediction in all the models analyzed, Fermi-LAT did not detect this PWN, and because of the Galactic Center location, it
has been impossible to impose useful upper limits either \citep{acero13}. The SED fit in figure \ref{g09} shows only a guiding-curve for the
3-years Fermi-LAT sensitivity.

\subsubsection{HESS J1833-105 (G21.5-0.9)}

G21.5-0.9 is a plerionic SNR with an approximately circular shape having a radius of $\sim$40'' in radio, infrared and X-ray. The pulsar is at
its center. The central pulsar of G21.5-0.9, PSR J1833-1034, was observed in radio having a period of 61.8 ms, and a period derivative of
$2.02 \times 10^{-13}$ s s$^{-1}$, yielding a characteristic age $\tau_c$=4860 yr \citep{camilo06}. It was not possible to measure the braking
index, and we take $n=3$. PSR J1833-1034 was also observed pulsating in GeV by Fermi-LAT \citep{abdo10b}, but not in X-rays (see for example,
\citealt{camilo06}).

The pulsar is one of the youngest in the galaxy. A recent age estimate based on measuring the PWN expansion rate in the radio band gives an age
of 870 yr \citet{bietenholz08}. In case of decelerated expansion, this real age could be even lower. However, \citet{wang86} suggested that
G21.5-0.9 might be the historical supernova of 48 BC. Uncertainty remains in this point. We assume the 870 years of age in our model. The
distance to the system was estimated, based on HI and CO measurements, to be 4.7$\pm$0.4 kpc \citep{camilo06}. The same value (within errors) was
obtained by other authors \citep{tian08a}. We approximate the nebula as an sphere of radius $\sim$1 pc. We assumed a mass of 8M$_\odot$ for the
ejected mass. \citet{matheson05} derived an upper limit for the upstream density of $\sim$0.1--0.4 cm$^{-3}$. For our fitting procedure, then, we
assumed that the PWN expands in a low density media with a value of 0.1 cm$^{-3}$.

G21.5-0.9 has been observed at different frequencies. In our analysis we have used the radio data obtained in the works by
\citet{salter89,morsi87,wilson76,becker76}. We have also used the infrared observations performed by \citet{gallant98,gallant99}. There are
additional X-ray and IR data that we are not using in the fit \citep{zajczyk12} and corresponding to the compact nebula only, a region of 2
arcsec surrounding the central pulsar.

\begin{figure}[t!]
\centering
\includegraphics[width=1.0\textwidth]{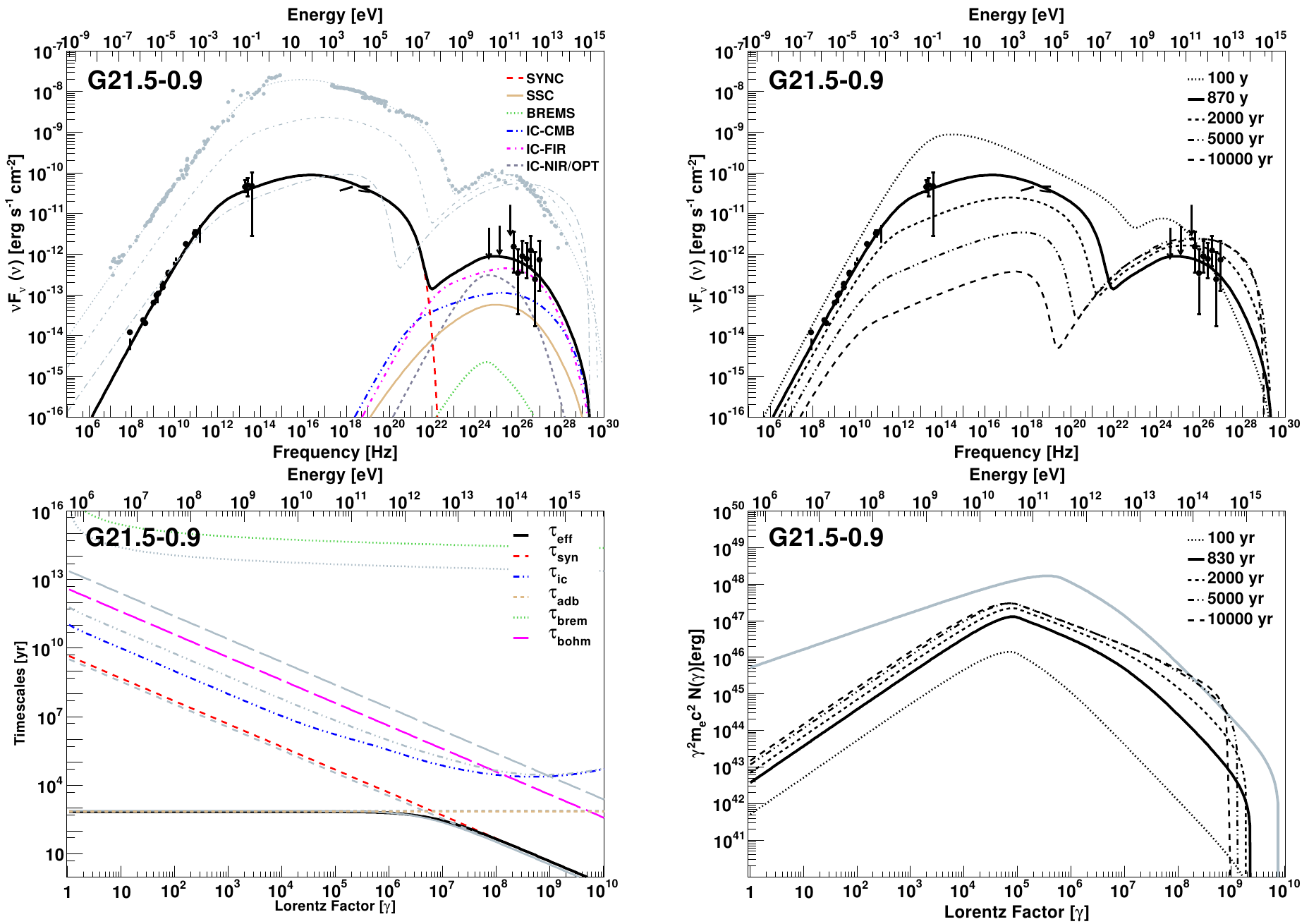}
\caption[SED model of G21.5-0.9]{Details of the SED model of G21.5-0.9. The panels are as in figure \ref{g54}. For details regarding the
observational data and a discussion of the fit, see the text.}
\label{g21}
\end{figure}

G21.5-0.9 is usually taken as a calibration source for X-ray satellites, see for example the works by
\citet{slane00,warwick01,safiharb01,matheson05,matheson10,derosa09}. We have used the joint calibration of Chandra, INTEGRAL, RXTE, Suzaku,
Swift, and XMM-Newton done by \citet{tsujimoto11} when considering the X-ray spectrum. The latter shows an spectral softening with radius
\citep{slane00,warwick01}. Chandra data showed for the first time evidence for variability in the nebula, a similar behavior that occurs in Crab
and Vela \citep{matheson10}. Fermi-LAT data come from \citet{acero13}. Finally, at TeV energies, the data comes from H.E.S.S. observations, which
detected the PWN as the source HESS 1833-105 \citep{gallant08,djannatiatai07}.

G21.5-0.9 was the first PWN discovered to be surrounded by a low-surface brightness X-ray halo that was suggested to be associated with the SNR
shell; its spectrum being non-thermal \citep{slane00}. The halo was not observed in radio wavelengths. \citet{slane00} argued that the halo may
be the evidence of the expanding ejecta and the blast wave formed in the initial explosion. \citet{warwick01} posed that the halo may be an
extension of the central synchrotron nebula. But deep Chandra observations revealed limb-brightening in the eastern portion of the X-ray halo and
wisp-like structures, with the photon index being constant across the halo \citep{matheson05}. Another interpretation of the origin of the halo
is that it could be composed by diffuse extended emission due to the dust scattering of X-ray from the plerion \citep{bocchino05}. Spectroscopy
analysis done by \citet{matheson10} with Chandra data revealed a partial shell on the eastern side of the SNR. \citet{safiharb01} could not find
evidence for line emission in any part of the remnant.

Table \ref{param} summarizes the values of the parameters and the result of the fit. The latter is shown in figure \ref{g21}, which has the same
panels as in the previously analyzed PWNe. It is particularly interesting to note that the electron losses in our model (see bottom left panel of
figure \ref{g21}) are almost exactly the same as those of Crab, and has $\sim$10\% of its spin-down power. Table \ref{param} gives further
account of this similarity as regards of age and energy densities of the photon backgrounds. G21.5-0.9 is a particle dominated nebula, with a
magnetic fraction of 0.03--0.04. This value is higher than the one obtained by \citet{tanaka11}, in correspondence with the different equation
used for the definition of magnetic field. Otherwise, the resulting model parameters are very similar, which is probably due to a significant
domination of the FIR component, almost one order of magnitude above the CMB contribution to the inverse Compton yield at 1 TeV. We fixed the
temperature of FIR and NIR photon distributions at the same values obtained from GALPROP. In order to be detected in the TeV range as has been,
the value for the energy density in the FIR is $\sim$1.4 eV cm$^{-3}$. The Comptonization of these photons dominates the spectrum at the highest
energies. There is some degeneracy in the precise determination of the FIR and NIR densities and temperatures. For instance, we have checked that
our fits would be very similar with temperature of 70 and 5000 K, and densities of 2 eV cm$^{-3}$ in the FIR and NIR, respectively. We have
analyzed the impact of having a smaller braking index (e.g., 2.5), and a different containment factor (from 0.1 to 0.3), but did not find any
significant differences in our fits due to the change in these parameters.

\subsubsection{HESS 1514–591 (MSH 15-52)}

The composite SNR G320.4-1.2/MSH 15-52 \citep{caswell81} is associated with the radio pulsar PSR B1509-58. This pulsar is one of the youngest and
most energetic known, with a 150 ms rotation period. It was discovery by the Einstein satellite \citep{seward82}, and was also detected at radio
frequencies by \citet{manchester82}. It has a period derivative of $1.5 \times 10^{-12}$ s s$^{-1}$, and a characteristic age of $\sim$1600 yr,
leading to a spin-down power of $1.8 \times 10^{37}$ erg s$^{-1}$. It is one of the pulsars with measured braking index
\citep{kaspi94,livingstone05a}; and we adopt for it the value of 2.839. The pulsar was detected also in gamma-rays using Fermi-LAT
\citep{abdo10b}. The central non-thermal source of the system has been interpreted as a PWN powered by the pulsar \citep{seward84a,trussoni96}.
The distance to the system was estimated using HI absorption measurements \citep{gaensler99} to be 5.2$\pm$1.4 kpc, which is consistent with the
value obtained by \citet{cordes03} from dispersion measure estimates, 4.2$\pm$0.6 kpc.

The dimension of the PWN as observed by ROSAT \citep{trussoni96} and H.E.S.S. \citep{aharonian05b} are $10 \times 6$, and $6.4 \times 2.3$ arcmin
respectively. The dimensions obtained in the TeV data, corresponds to a radius of a circle of $\sim$3 pc, at a distance of 5.2 kpc.

The measured braking index of the pulsar implies a young age, lower than $\sim$1700 yr. According to the standard parameters of the ISM, the age
of the system was estimated to be in the range 6-20 kyr, an order of the magnitude larger than the age estimated by the pulsar parameters. A
plausible explanation for this discrepancy is that the SNR has expanded rapidly into a low-density cavity, what can also explain the unusual SNR
morphology, the offset of the pulsar from the apparent center of the SNR, and the faintness of the PWN at radio wavelengths
\citep{gaensler99,dubner02}. The south-southeastern half of the SNR seems to have expanded across a lower density environment of $\sim$0.4
cm$^{-3}$. And the north-northwestern radio limb has instead encountered a dense HI filament. In our models we adopt a density of 0.4 cm$^{-3}$.
However, the morphology of MSH 15–52 is complex and not taken into account in our model (similarly to other analysis alike e.g.,
\citealt{tanaka11,abdo10b,zhang08,nakamori08}).

To perform our multi-wavelength fit we acquired the observational data as follows: Radio observations were obtained from
\citet{gaensler99,gaensler02}. Observations of the nebula in the hard X-rays come from Beppo-SAX \citep{mineo01}, and INTEGRAL-IBIS telescopes
\citep{forot06}. COMPTEL and EGRET measurements \citep{kuiper99} combine the pulsar and the PWN measurement, so we did not consider them in our
fit. The PWN was detected and its spectral distribution in GeV energies was obtained by Fermi-LAT during the first year of operation of this
instrument \citep{abdo10b}. Fermi-LAT observations used in our work come from subsequent work by \citet{acero13}. At even higher energies,
Cangaroo III observations are in agreement with the previous H.E.S.S. observations. Both data sets were used below. In the models presented here
an ejected mass of 10M$_\odot$ is assumed.

\begin{figure}[t!]
\centering
\includegraphics[width=1.0\textwidth]{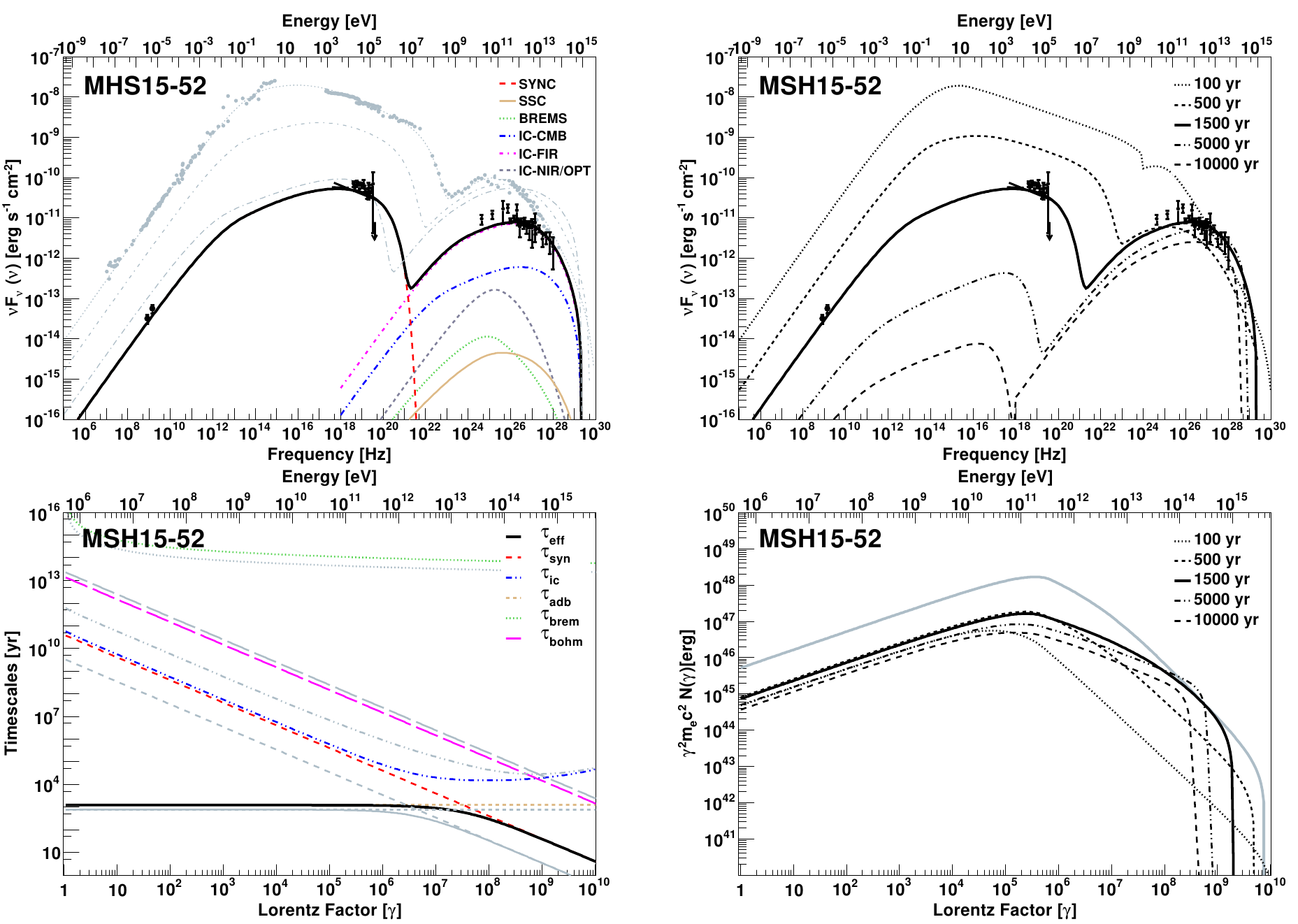}
\caption[SED model of MSH15–52]{Details of the SED model of MSH15–52. The panels are as in figure \ref{g54}. For details regarding the
observational data and a discussion of the fit, see the text.}
\label{msh15}
\end{figure}

We consider different scenarios to fit the multiwavelength data. In the model presented in figure \ref{msh15} we assume that the age of the
system is 1500 yrs, close to the characteristic age of the pulsar. We also assume a broken power-law injection. In order to fit the measured GeV
and TeV data we use a FIR photon field of 5 eV cm$^{-3}$, at a temperature of 20 K. This component is dominating the IC yield, while the
contribution of the optical photon field is much lower in comparison (see table \ref{param}). The other parameters resulting from the fit are
$\alpha_1$=1.5, $\alpha_2$=2.4, a break Lorentz Factor of $5 \times 10^5$, a maximum Lorentz Factor of $1.9 \times 10^9$, a nebula magnetic field
of 21 $\mu$G, and a magnetic fraction of 0.05. It would seem that the Fermi-LAT data is not perfectly well reproduced. This can be cured by
choosing higher densities and temperatures of the photon backgrounds, but we have not been able to find a perfect match in these conditions.

It was already proposed that the local photon background for this PWN could be higher than the average Galactic value, in particular in the FIR
\citep{aharonian05b}. \citet{nakamori08,duplessis95} suggested that the SNR itself could be the origin of the excess of the IR photon field. As
in the work of \citet{bucciantini11}, we have also investigated the possibility of performing our fit assuming a contribution of a local IR
photon field with a temperature of $\sim$400 K. This possibility is presented in our Model 2. Indeed, we have found that we could fit the
observational data with a temperature (energy density) of the IR photon field of 20 K (4 eV cm$^{-3}$), and local IR photon field with a
temperature (energy density) of 400 K (20 eV cm$^{-3}$). The quality and final SED corresponding to these assumptions (leaving all other
parameters unscathed) is better matching also to the Fermi-LAT data, and both M1 and M2 models are compared in figure \ref{msh15}. As the result
of the M2 fit we obtained $\alpha_1$=1.5, $\alpha_2$=2.4, a break Lorentz Factor of $5 \times 10^5$, a maximum Lorentz Factor of
$2.3 \times 10^9$, a nebula magnetic field of 25 $\mu$G, and a magnetic fraction of 0.07.

\begin{figure}
\centering
\includegraphics[width=0.6\textwidth]{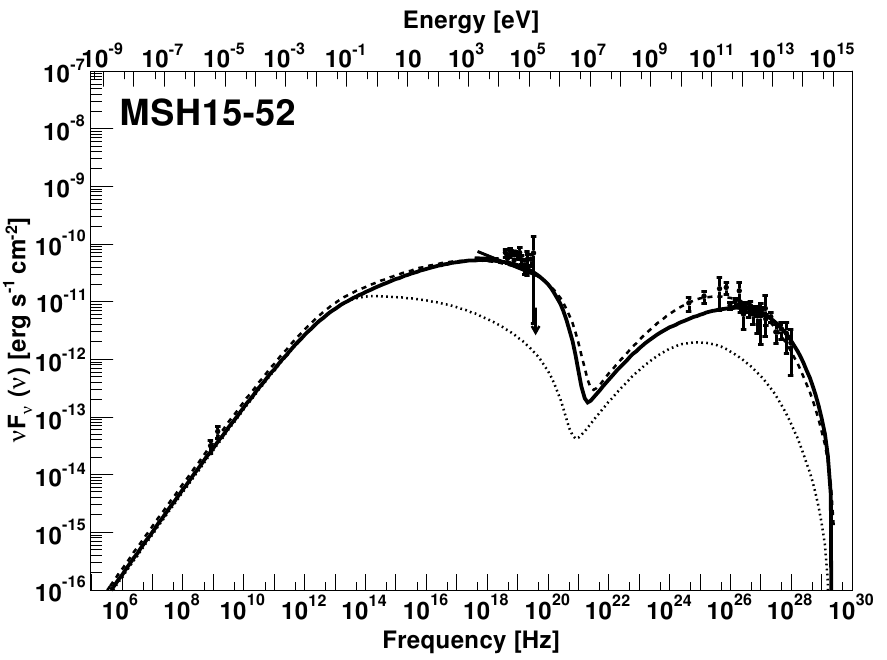}
\caption[SED of MSH 15–52 fitted with different assumptions]{SED of MSH 15–52 fitted with the parameters of the model described in table
\ref{param} (solid line), together with a comparison with the resulting fit using as photon temperatures and corresponding energy densities (20 K
and 4 eV cm$^{-3}$ for the FIR, and 400 K and 20 eV cm$^{-3}$ for the NIR; leaving all other parameters the same, dashed line). We also compare
with the current SED results if the parameters of \protect\citet{abdo10b} are assumed (dotted line).}
\label{msh15_comp}
\end{figure}

Previous to Fermi-LAT observations, \citet{aharonian05b} presented a fit of the X ray and VHE data using a static IC model \citep{khelifi02}.
Using this model they reproduced the VHE spectrum of the whole nebula assuming a power-law energy spectrum for the population of the accelerated
electrons with an spectral index of 2.9. The energy density of the dust component is more than a factor of 2 higher than the nominal value given
by GALPROP, similar to ours. \citet{abdo10b} also performed a fit of the observational data, including radio, X-ray, Fermi-LAT, and TeV
observations using the one-zone, static model described by \citet{nakamori08}. According to their model the gamma-ray emission is dominated by
the IC of the FIR photons from the interstellar dust grains with a radiation density fixed at 1.4 eV cm$^{-3}$ which actually is the nominal
value of GALPROP at the position of MSH 15-52. The energy densities in the model by \citet{aharonian05b} are similar to those assumed by
\citet{abdo10b} when presenting Fermi-LAT results. In these works, no time evolution is considered in any of the quantities. We tried performing
a fit with the same parameters used in \citet{abdo10b}; i.e., assuming their spectral indeces, break in the spectrum of the injected particles,
magnetic field, and energy densities of the photon fields (see table 4 of the mentioned paper). We compare the results of the fits of Model 1 and
2 with the resulting model having the same parameters of \citet{abdo10b} in figure \ref{msh15_comp}. The main difference between \citet{abdo10b}
model and ours reside, apart that the latter is static, is the assumed lower photon field densities and the steeper high-energy slope of the
injected electrons. These changes make for a significant underestimation of the TeV emission. The nebula magnetic field obtained in our model (of
the order of 20–25 $\mu$G) is however similar to the one obtained by \citet{aharonian05b,abdo10b} (17 $\mu$G). Previous estimations
\citet{gaensler02} gave a lower limit of the field (8 $\mu$G), which is also compatible.

\subsubsection{HESS J1119-614 (G292.2-0.5)}

G292.2-0.5 is a SNR associated with the high-magnetic field radio pulsar J1119-6127, which was discovered in the Parkes multibeam pulsar survey
\citep{camilo00}. The pulsar was also detected in X-rays \citep{gonzalez05} and $\gamma$-rays \citep{parent11}. It has a rotational period of 408
ms, and a period derivative of $4 \times 10^{-12}$ s s$^{-1}$, leading to a characteristic age of $\sim$1600 yr, and a spin-down luminosity of
$2.3 \times 10^{36}$ erg s$^{-1}$. The braking index was measured for the first time by \citet{camilo00}, but this value was recently refined
using more than 12 years of radio timing data to 2.684$\pm$0.002 \citep{weltevrede11}. The high value of the pulsar magnetic field,
$\sim 4.1 \times 10^{13}$ G places PSR J1119-6127 between typical radio pulsars and usual magnetars.

A faint PWN surrounding the pulsar was detected in X-rays \citep{gonzalez03,safiharb08}. The X-ray unabsorbed flux between 0.5 and 7 keV was
measured to be $1.9 \times 10^{-14}$ erg s$^{-1}$ cm$^{-2}$ for the compact nebula, and $2.5 \times 10^{-14}$ erg s$^{-1}$ cm$^{-2}$ for the
associated jet, with spectral indices of 1.1$^{+0.9}_{-0.7}$ and 1.4$^{+0.8}_{-0.9}$ respectively. These are extremely low values in comparison
to other PWNe, G292-0.5 is a very faint PWN in X-rays, which remains so even in the case of adding the southern jet flux. The PWN was also
detected at high energies by Fermi-LAT \citep{acero13} and at very high energies by H.E.S.S. \citep{mayer10,djannatiatai09}\footnote{We remark
that these are not official claims of the H.E.S.S. collaboration; they are not confirmed, but not ruled out either. We entertain the possibility
that the final TeV data may differ from the current available spectrum.}. TeV measurements have shown a flux of 4\% of the Crab nebula and a
steeper spectrum (with slope larger than 2.2) compared with other young PWNe. The luminosity in TeV gamma-rays (at 8.4 kpc, see below) is
$3.5 \times 10^{34}$ erg s$^{-1}$ which makes for an efficiency of 1.5\% in comparison of the current pulsar spin-down. Thus, the ratio of
$L_X/L_\gamma$ is $\sim 10^{-3}$, which would imply a low magnetic field.

The mass of the progenitor of the SN explosion is large \citep{kumar12}; these authors inferred that the expansion occurred in a very low-density
medium. We assumed in our calculations that the ejected mass had a value between 30 and 35M$_\odot$, and that the density of the medium was 0.02
cm$^{-3}$. The kinematic distance to the system was suggested to be 8.4$\pm$0.4 kpc based on HI absorption measurements \citep{caswell04}.
According to \citet{safiharb08}, the size of the compact PWN in X-rays is $6 \times 15$ arcsec, with the jet corresponding to a faint structure
of $6 \times 20$ arcsec. For a distance of 8.4 kpc, this size corresponds to $\sim$0.5 pc. In the TeV range, the source is extended and the size
is larger, its diameter is of the order of $\sim$30 pc \citep{kargaltsev10,djannatiatai09}.

\citet{kumar12} estimated the age of the SN in a range between 4200 yr (for a free expansion phase, assuming an expansion velocity of 5000 km
s$^{-1}$ ) and 7100 yr (for a Sedov phase). This estimation is larger than the one obtained using the pulsar parameters, of 1900 yr. In our model
we propose a fit of the data assuming an age of 4200 yr (and $n=1.7$), and compare it with the results of assuming an age of 1900 yrs (and
$n=2.7$) in alternative fittings.

To compute the fit we then consider the H.E.S.S. measurements \citep{kargaltsev10,djannatiatai09}; together with the X-ray flux quoted above
\citet{safiharb08}. These are both crucial assumptions, which, as we shall see, reflect in a very steep injection at high energies. We comment
more on them below. ATCA deep measurements revealed only a 15 arcmin SNR shell \citep{crawford01}, but no radio emission from the PWN. The latter
authors interpreted the absence of a radio PWN as being the result of the pulsar's high magnetic field; which would lead to a short time of high
energy electron injection (due to a large spin-down). What they see is a limb brightening elliptical shell (in fact designated thereafter as
G292-0.5) of dimensions $14 \times 16$ arcmin with a 1.4 GHz flux density of 5.6$\pm$0.3 Jy. At 2.5 GHz, the measured flux density of G292.2-0.5
is 1.6$\pm$0.1 Jy (but this should likely be taken as a lower limit since the shell is larger than the largest scale to which the interferometer
used is sensible). We shall take this SNR flux measurement at 1.4 GHz as a safe upper limit for the PWN radio emission.

\begin{figure}[t!]
\centering
\includegraphics[width=1.0\textwidth]{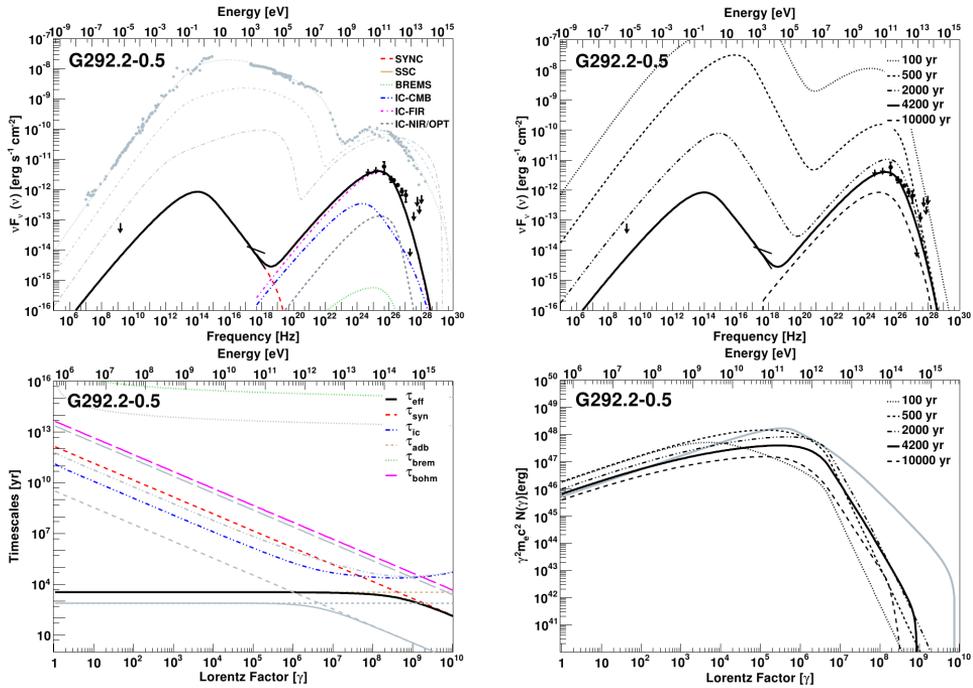}
\caption[SED model of G292.2-0.5]{Details of the SED model of G292.2-0.5. The panels are as in figure \ref{g54}. For details regarding the
observational data and a discussion of the fit, see the text.}
\label{g292}
\end{figure}

We consider first an age of 4200 years, as derived by \citet{kumar12} based on SNR properties. To reconcile the pulsar age with the supernova,
\citet{kumar12} suggested that the braking index has to be smaller than 2 for most of the pulsar lifetime. We assume it to be 1.7. With this age,
a fit can be obtained with the FIR dominating the IC yield, with relatively low energy densities. However, the injected electron spectrum at high
energy needs to be steep (4.1) to achieve good agreement with observational data. This is an interesting result, since it is by far steepest
$\alpha_2$ we shall see in the whole sample, and it is quite constrained by the observations of both GeV and TeV emission from this source.
Another interesting difference in this case is that the spectral break of the injected electron is higher, in the three models, that the one
obtained for other PWNe. However, the extra degree of freedom given by the lack of a detection of the synchrotron at low frequencies peak is a
caveat. The resulting model parameters under this age assumption are given in table \ref{param} and figure \ref{g292}.

\begin{figure}
\centering
\includegraphics[width=0.6\textwidth]{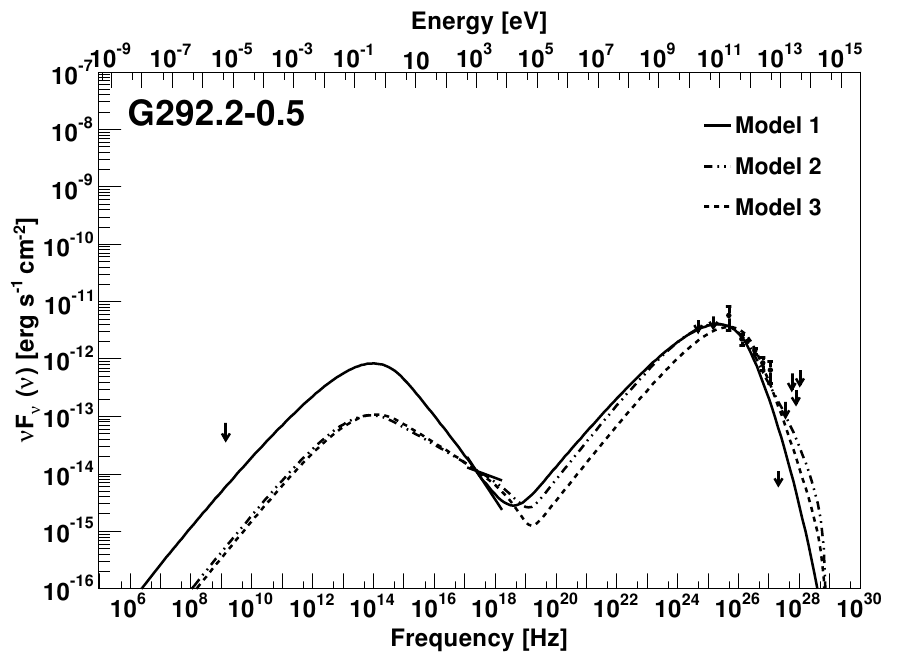}
\caption[Spectrum of the three different models for G292.2–0.5]{Spectrum of the three different models for G292.2–0.5. See the text for a
discussion of the differences and caveats underneath each of these models.}
\label{g292_comp}
\end{figure}

We have also explored models in which the age of the PWN is lower, as resulting from the estimate of the pulsar period, period derivative, and
braking index \citep{weltevrede11}. We have found that it is especially difficult to find models that could consistently fit the whole set of
observations, with the more constraining range being the GeV gamma-rays. In order to fit the MW observational data for lower pulsar ages, either
we assume that the energy densities of the FIR and NIR components are significantly larger (10 and 130 eV cm$^{-3}$, respectively), or we assume
that there is a contribution of a local IR field at 400 K, similar to the alternative model considered above for MSH 15-52; which, in any case,
would need a large energy density (33 eV cm$^{-3}$). These values of NIR densities would make the corresponding IC component to significantly
contribute, or overcome the FIR-IC yield. Both of these models take the measured value of $n \sim 2.7$, show a radius of about 6 pc, and similar
magnetic field, magnetization, injection slopes, and break energies that the corresponding ones shown in table \ref{param}, but are less
satisfying due to the large energy densities involved without a clear a priori justification. In any case, degeneracies in modeling can be broken
at radio and optical frequencies (see figure \ref{g292_comp}).

Interestingly, the three models show a very low magnetic field for the nebula, which is consistent with the expectations coming from the
extremely low value of the ratio of X-ray and gamma-ray luminosities, and about one order of magnitude lower than the one estimated earlier by
\citet{mayer10}, of 32 $\mu$G. A lower ejected mass or a higher energy explosion (that can make the size of the nebula larger) will make the
magnetic field even lower than the ones obtained in the models presented here.

Another point of discussion in this case is the size of the nebula. Whereas the different sizes could be explained due to the larger losses of
X-ray generating electrons, this PWN has one of the largest mismatches. Electrons generating keV photons have, for the resulting B field, a very
high energy, in excess of 70 TeV, much larger than the energy of electrons generating TeV photons. In the model of table \ref{param}, we obtain a
radius of $\sim$13 pc and use it for all frequencies. However, the X-ray and TeV emission regions are probably not the same, and a more detailed
model could be needed for a more proper accounting.

\subsubsection{HESS J1846-029 (Kes 75)}

Kes 75 (also known as G29.7-0.3) is a shell-type supernova remnant with a central core whose observed properties suggest an association with a
PWN. The pulsar associated with this system, PSR J1846-0258, was discovered in a timing analysis of the X-ray data from RXTE and ASCA
\citep{gotthelf00b}. The pulsar has not been detected in the radio band, perhaps due to beaming. Fermi-LAT did not detect the pulsar at high
energies either. PSR J1846-0258 has a spin period of $\sim$324 ms, and a spin-down age of $7.1 \times 10^{-12}$ s s$^{-1}$, implying a large
spin-down luminosity of $8.2 \times 10^{36}$ erg s$^{-1}$, a high surface magnetic field of $\sim 5 \times 10^{13}$ G, and a small characteristic
age $\sim$720 yr \citep{kuiper09}. This pulsar exhibited a magnetar-like outburst with a large glitch in 2006
\citep{gavriil08,kumar08,livingstone11a}. The pulsar's braking index was measured using RXTE observations \citep{livingstone06}. The latter
authors found a value of 2.65$\pm$0.01, which implies a spin-down age of 884 years, placing this pulsar among the youngest in the Galaxy. During
the magnetar-like outburst and the large glitch of 2006, the pulsar presented 5 very short X-ray bursts, changes in the spectra, timing noise,
increase in the flux (6 times larger than in the quiescent state), and softening of the spectral index \citep{ng08,gavriil08,kumar08}. After that
episode the braking index decreased, and has now a value of 2.16$\pm$0.13 and the pulsar and the PWN came back to the previous flux and spectral
index \citep{livingstone11a}. It was proposed that the PWN variability observed in 2006 is most likely unrelated to the outburst and is probably
similar in origin to the variation of small-scale features seen in other PWNe \citep{livingstone11a}. Detailed studies of the variability of the
PWN using deep Chandra observations were also presented by \citet{ng08}. While fitting the multiwavelength emission from Kes 75, we have assumed
a value of 2.16 for the braking index, and analyzed the differences in the predictions entailed by changing the value of $n$ to that valid before
the outburst.

The morphology of the nebula in X-rays is similar to the one observed in radio wavelengths. It is highly structured and it has a dimension,
according to high-resolution Chandra images, of $26 \times 20$ arcsec. A detail of the complex morphology of the nebula according to Chandra
observations is presented by \citet{ng08}. The first estimation of the distance to the system based on neutral hydrogen absorption measurements
was 19 kpc \citep{becker84}. More recently \citet{leahy08b} estimated a new distance between 5.1 and 7.5 kpc from HI and $^{13}$CO maps. However,
\citet{su09} also estimated a new distance to the system of 10.6 kpc based on the association between the remnant and the molecular shells. There
is then a significant uncertainty in the distance to this PWN, and thus we have assumed two different models; with a distance of 6 kpc in our
Model 1 and a distance of 10.6 kpc in our Model 2.

To perform the multiwavelength fit presented below, we took radio observations \citep{salter89,bock05}, and infrared upper limits
\citep{morton07}. The X-ray spectra, resulting from Chandra observations, was taken from \citet{helfand03}. Fermi-LAT upper limits in the photon
flux corresponding to three energy bands are presented in \citet{acero13}. In all of these energy bins, the significance (TS value) is very low
(5 in the range 10–31 GeV, and 0 in the ranges of 31–100 GeV and 100–316 GeV). To obtain the upper limits in energy we multiplied the photon flux
in each bin by the energy of the center of the bin. At very high energies the nebula was detected by H.E.S.S. \citep{djannatiatai07} with an
intrinsic extension compatible with a point-like source and a position in good agreement with the pulsar associated to the nebula.

We present the results of our fit to the multiwavelength observations of Kes 75 assuming that the age and distance to the system are 700 yr and 6
kpc for Model 1, and 800 yr and 10.6 kpc for Model 2. In both models, we have assumed a braking index of 2.16 \citep{livingstone11a} and a
density of the medium of 1 cm$^{-3}$ \citep{safiharb13}. The ejected mass for Model 1 was assumed to be 6M$_\odot$ and 7.5M$_\odot$ for Model 2.
These models span the range of the uncertainties in distance.

\begin{figure}[t!]
\centering
\includegraphics[width=1.0\textwidth]{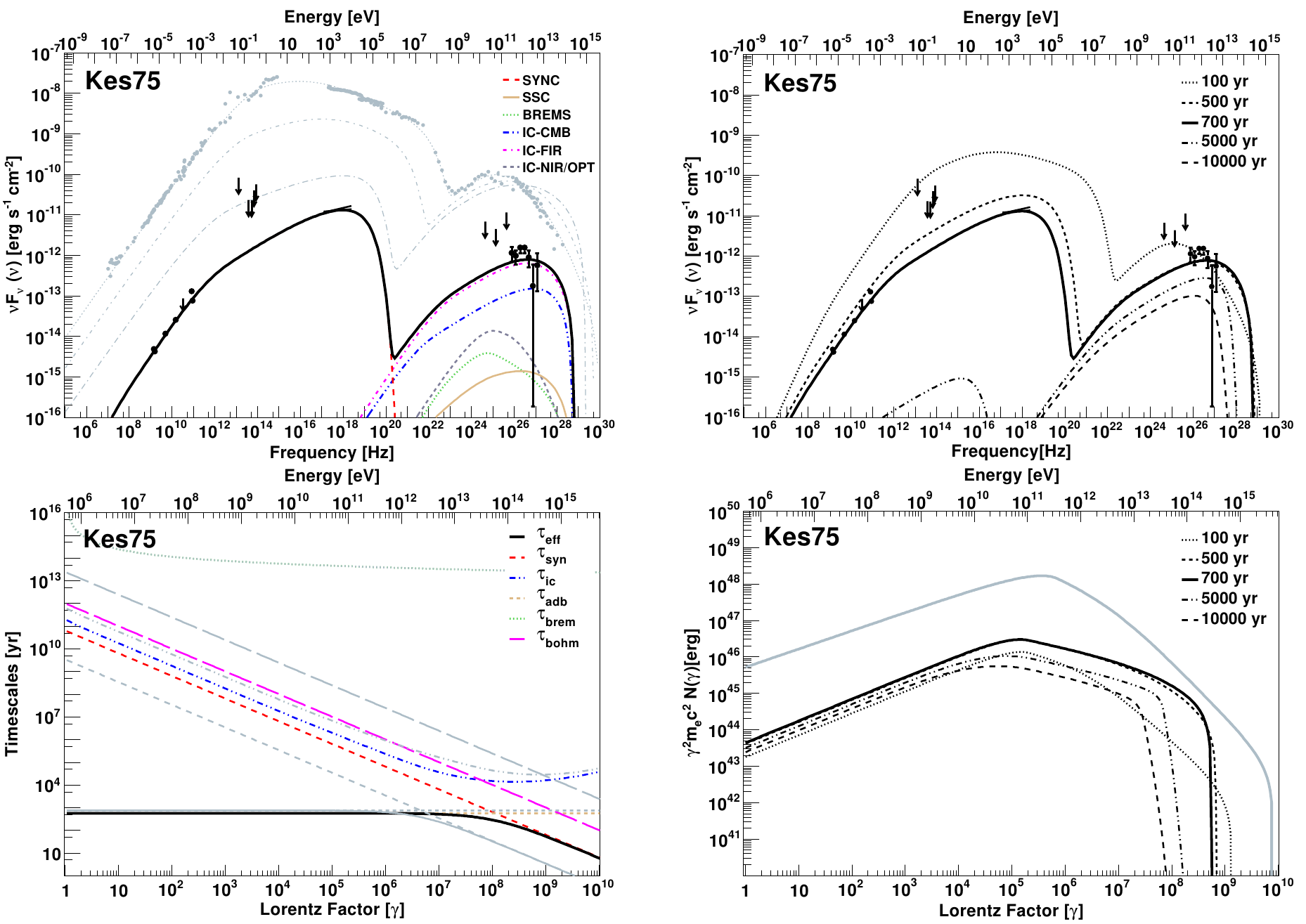}
\caption[SED of Kes 75 as fitted by our model]{Details of the SED of Kes 75 as fitted by our model. The panels are as in figure \ref{g54}. For
details regarding the observational data and a discussion of the fit, see the text.}
\label{kes75}
\end{figure}

To fit the TeV data we assume a temperature (energy density) of 25 K (2.5 eV cm$^{-3}$) for the FIR and 5000 K (1.4 eV cm$^{-3}$) for the NIR
photon field in Model 1. In Model 2 (corresponding to the slightly larger age and farther distance) we need to double the energy density in the
FIR to fit the observational data. We comment more on this below. In both of these models, the IC with the FIR photon field is the most important
component, being the IC with CMB the second contributor to the total yield. The full set of assumed and fitted parameters are shown in table
\ref{param}, whereas the results for Model 1 are presented in figure \ref{kes75}.

The Spitzer upper limits do not constrain the parameters of the models in any significant way. The break in the spectrum between the radio and
X-ray bands appears at $\sim$100 GeV for Model 1 and $\sim$50 GeV for Model 2 in our fit. These low breaks are in agreement with the results
presented by \citet{bock05}. The average magnetic field obtained for the nebula was 19 $\mu$G in Model 1 and 33 $\mu$G in Model 2. In both cases
the magnetic fraction is low and comparable to other PWNe. The average magnetic field obtained are similar to the ones obtained by
\citet{tanaka11}. \citet{djannatiatai07} also suggested a low magnetic field for this nebula of the order of $\sim$10 $\mu$G. The first spectral
index, $\alpha_1$, of the injected spectrum are both also in agreement with the ones obtained by \citet{tanaka11}, but as in other cases, our
second spectral index, $\alpha_2$ are lower than the ones obtained in their fits; which may result from a different treatment of the radiative
losses. The final SED results for Models 1 and 2 are quite similar, showing a problematic degeneracy which cannot be broken by the data now at
hand. In fact, other degeneracies resulting from the uncertainty in age can be accommodated by modifying the high energy slope of the injected
power law, or the magnetic field. Changes are not severe, though, and do not affect the main conclusions.

We could also fit the observational data assuming a braking index of 2.65 (with an age of 700 yrs). For instance, for an ejected mass of
6M$_\odot$, at a distance of 10.6 kpc, a nebula magnetic field of 40 $\mu$G with a magnetic fraction of 0.055, and spectral indices of 1.4 and
2.2 for the injected particle spectrum with a break Lorentz factor at $2 \times 10^5$ would fit the spectrum equally well, for energy densities
and temperatures of photon backgrounds similar to those assumed in Models 1 and 2 presented in table \ref{param}.

All in all, Kes 75 is a difficult case to model in detail: in particular, we find difficult to provide an overall (along all frequencies)
significantly better fit than the one we show in figure \ref{kes75}, which we see a bit dissatisfying at the largest energies. There, the fall
out of the TeV emission is plausibly steeper than in the model we show, what should be studied with future datasets. The VHE energy data seems to
peak around 1 TeV. However, since this is not clear within the reach of the present dataset, we have not tried to model a peak. We have
considered models with larger break energies, different photon background and injection parameters, but they do not provide significant
improvements. We explored increasing the NIR density, i.e., increasing the IC contribution at energies of 10$^{11}$ eV so that the curve at the
highest energies flattens. With $w_{FIR}$=2 eV cm$^{-3}$ at a central temperature of 100 K and $w_{NIR}$=20-25 eV cm$^{-3}$ at 3000 K the
contribution of IC-NIR becomes comparable to that of IC-FIR but peaking at lower energies, thus flattening or even steppening the high-energy
yield.

\subsubsection{HESS J1356-645 (G309.9-2.5)}

HESS J1356-645 is localized at $\sim$5 pc from the pulsar PSR J1357-6429, if at the same distance, and has an intrinsic Gaussian width of
(0.2$\pm$0.02) deg \citep{abramowski11b}. PSR J1357-6429 is a young pulsar with a $\tau_c$=7.3 kyr, a spin-down luminosity of
$3.1 \times 10^{36}$ erg s$^{-1}$, and a period of 166 ms. It was discovered during the Parkes multibeam survey of the Galactic Plane
\citep{camilo04}. \citet{lemoinegoumard11} detected pulsations using data from Fermi-LAT and XMM-Newton observations. A possible optical
counterpart was also reported \citep{danilenko12}. Several authors pointed out the similarities of this pulsar with Vela
\citep{esposito07,abramowski11b,acero13}. Particularly, they both have a low X-ray efficiency, presence of thermal X-ray photons, and a similar
ratio of the compact to diffuse sizes of the nebula. The distance to the pulsar was estimated, based on its dispersion measure, to be 2.4 kpc
\citep{camilo04}.

The first upper limit of the X-rays emission of the PWN of this pulsar was established by \citet{esposito07}. Later, the H.E.S.S. collaboration
studied ROSAT and XMM-Newton images and reported the X-ray spectra of the nebula \citep{abramowski11b}. Radio and X-ray data, although faint, are
coincident in extension with the VHE emission, which provides arguments for the association between the HESS source and the nebula
\citep{abramowski11b}. The morphology of the PWN was also recently studied in detail by \citep{chang12}, who also arrived to the same conclusion
about the possible association of the nebula with the very high energy source. Fermi-LAT detected a faint counterpart to the nebula after 45
months of observations (Acero et al., 2013). The spatial and spectral coincidences between Fermi-LAT and HESS emission also suggests that they
are coming from the same source.

\begin{figure}[t!]
\centering
\includegraphics[width=1.0\textwidth]{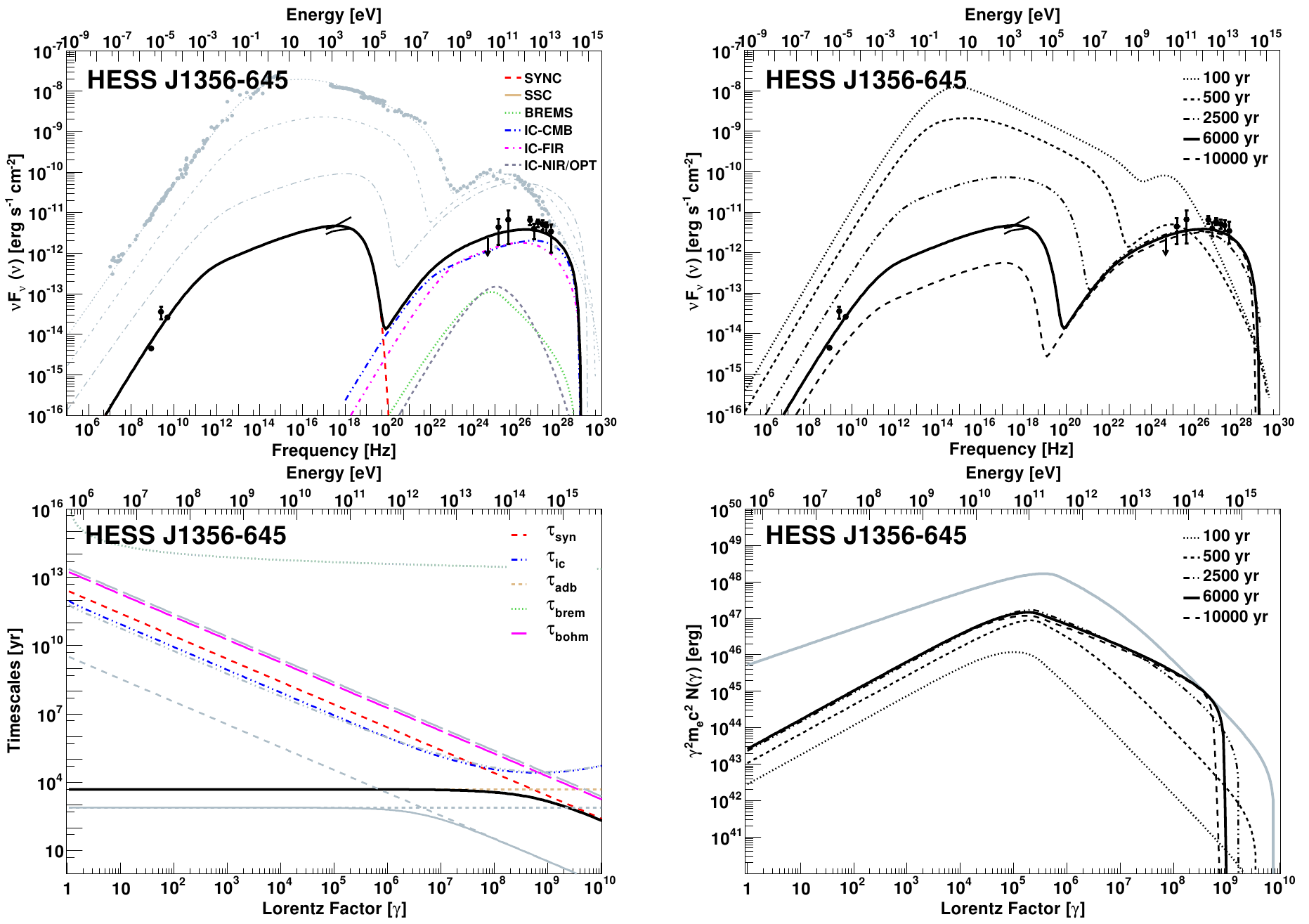}
\caption[SED model of HESS J1356-645]{Details of the SED model of HESS J1356-645. The panels are as in figure \ref{g54}. For details regarding
the observational data and a discussion of the fit, see the text.}
\label{g309}
\end{figure}

To perform our fit we then take the radio, X-ray, and TeV data as quoted in the discovery paper by H.E.S.S. \citep{abramowski11b}: Radio data
comes from the Molonglo Galactic Plane Survey at 843 MHz, Parkes 2.4 GHz, and Parkes-MIT-NRAO (PMN) at 4.85 GHz. The X-ray spectral shape comes
from XMM-Newton observations. Fermi-LAT observations were taken from \citet{acero13}.

To fit the observational data, we have assumed an age of 6000 years, a braking index of 3, an ejected mass of 10M$_\odot$, and a distance of 2.4
kpc (see table \ref{param}). We could fit the data with a broken power-law injection having a hard low-energy spectral index $\alpha_1$=1.2, and
a high-energy slope of $\alpha_2$=2.52. We found no need of adding a constraint on $\gamma_{min}$ in this model. The break in the spectrum
happens at a Lorentz factor of $3 \times 10^5$. We found HESS J1356-645 to be a particle dominated nebulae too, with a magnetic fraction of 0.06.
The FIR and NIR photon fields of the model have temperatures of 25 K and 5000 K, and energy densities of 0.4 and 0.5 eV cm$^{-3}$, respectively.
These values are quite low in comparison with other PWNe we have studied, and near the estimations obtained from GALPROP (see below). The average
magnetic field we obtain is also very low $\sim$3.1 $\mu$G. A magnetic field higher than $\sim$4 $\mu$G would make it impossible to fit the data,
even varying other parameters. The SED today, its evolution over time, the electron population, and the losses are plotted in figure \ref{g309}.
At high and very high energies, the most important contributions are coming from the IC with the CMB and FIR, almost in an equal extent, being
the contributions to the IC coming from the NIR photons, as well as from Bremsstrahlung, negligible in comparison. For comparison, the HESS
Collaboration \citep{abramowski11b} have modeled the source assuming a static one-zone leptonic scenario, with an electron population injected
with an exponential cutoff power-law of index 2.5 and cutoff energy of 350 TeV. They also assumed photon fields with temperatures of $\sim$35 K
and 350 K and optical photon field of temperature of $\sim$4600 K. We do not find the need of incorporating an additional component to the IR
distribution at 350 K in order to fit the data.

\begin{figure}
\centering
\includegraphics[width=0.6\textwidth]{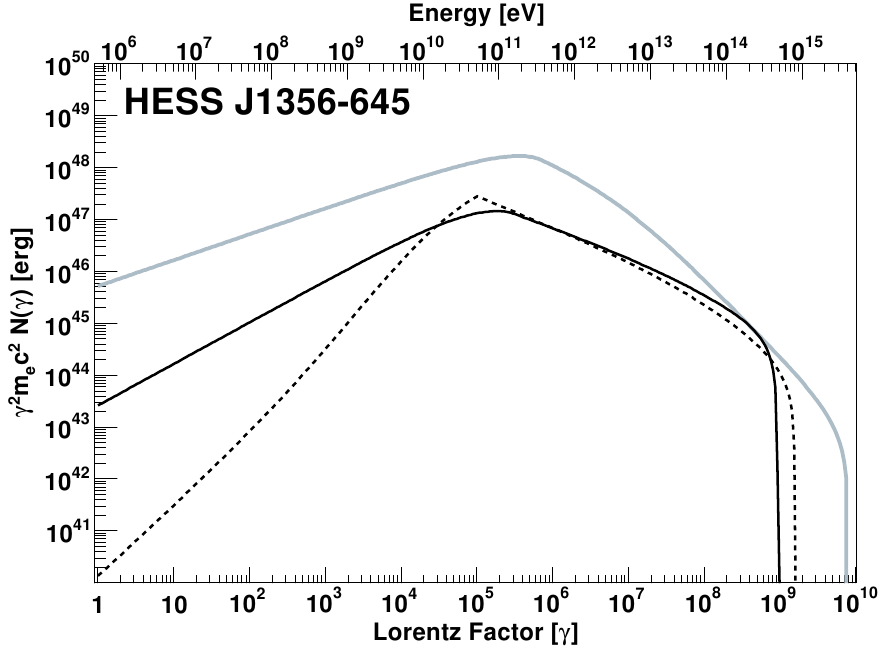}
\caption[Comparison of the electron distributions for the two models considered for HESS J1356–645]{Comparison of the electron distributions for
the two models considered for HESS J1356–645. The solid (dashed) line corresponds to Model 1 (2), with the parameters given in table \ref{param}.
Recall that the age in these two models is different. The grey solid line is the Crab nebula electron distribution today.}
\label{g309_comp}
\end{figure}

We have found that it is also possible to have a good fit to the data with a single power law in the spectrum of injected electrons (with slope
2.6), if electrons are energetic enough. To allow for this possibility the braking index is reduced to 2, so that the initial spin-down age is
increased by about a factor of $\sim$5 (up to 6622 years). With such an spin-down age, the pulsar is injecting more electrons along most of its
lifetime. An slightly larger age (assumed to be 8000 years) and magnetic fraction (0.08) would allow for an equally good SED fit. Finally, the
$\gamma_{min}$ value is here constrained to be larger than 105. In practice, electrons injected are assumed to be above the break energy of the
prior model, and losses populate lower levels in electron energy. These parameters are summarized in table \ref{param}, quoted as Model 2. Figure
\ref{g309_comp} compares the two resulting electron distribution at the corresponding current age. By compensating with a longer injection age
and more energetic electrons, the electron distribution can be made similar in both models, leading to equally acceptable SEDs. This degeneracy
still remains, although preference for model 1 can be argued: the alternative model 2 referred above requires more contrived assumptions to work
and would make the nebula an outlier in comparison with others.

\subsubsection{VER J0006+727 (CTA 1)}

The extended radio source CTA 1 (G119.5+10.2) was first proposed as a SNR by \citet{harris60}. The SNR was first detected in X-rays by ROSAT by
\citet{seward95}. The authors also reported the presence of a faint compact source, RXJ 0007.0+7302, located within the central region.
\citet{slane97} confirmed the non-thermal nature of the central emission using ASCA data. These early detections were indicative of the presence
of a synchrotron nebula powered by an active neutron star, for which the most plausible candidate was the source RX J0007.0+7302. Further studies
performed with the XMM-Newton and ASCA satellites towards RX J0007.0+7302 have resolved the X-ray emission into a point-like source and a diffuse
nebula of 18 arcmin in size \citep{slane04a}. Using the Chandra observatory, \citet{halpern04} have found a point source, RX J0007.0+7302,
embedded in a compact nebula of 3'' in radius, and a jet like extension. At high energies, \citet{mattox96} proposed that the EGRET source 3EG
J0010+7309 (which lies in spatial coincidence with RX J0007.0+7302), was a potential candidate for a radio-quiet gamma-ray pulsar.
\citet{brazier98} also pointed out that this source was pulsar-like, but a search for $\gamma$-ray pulsation using EGRET data failed
\citep{ziegler08}. During the commissioning phase of the Fermi satellite, a radio-quiet pulsar in CTA 1 was finally discovered \citep{abdo08}.
X-rays pulsations from this source were finally detected by XMM-Newton \citep{lin10,caraveo10}. The pulsar in CTA 1 has a period of $\sim$316 ms
and a spin-down power of $\sim 4.5 \times 10^{35}$ erg s$^{-1}$ counterpart. No radio to RX J0007.0+7302 was identified, most likely due to
beaming. No optical counterpart is known either \citep{mignani13}.

\citet{abdo12} reported the detection of an extended source in the off-pulse emission at $\sim 6\sigma$ level using 2 years of Fermi/LAT data.
\citet{acero13} improved on this result (which we use for modeling). The VERITAS Collaboration also detected an extended source of
$0.3 \times 0.24$ deg at 5 min from the Fermi gamma ray pulsar PSR J0007+7303 \citep{aliu13}.

\begin{figure}[t!]
\centering
\includegraphics[width=1.0\textwidth]{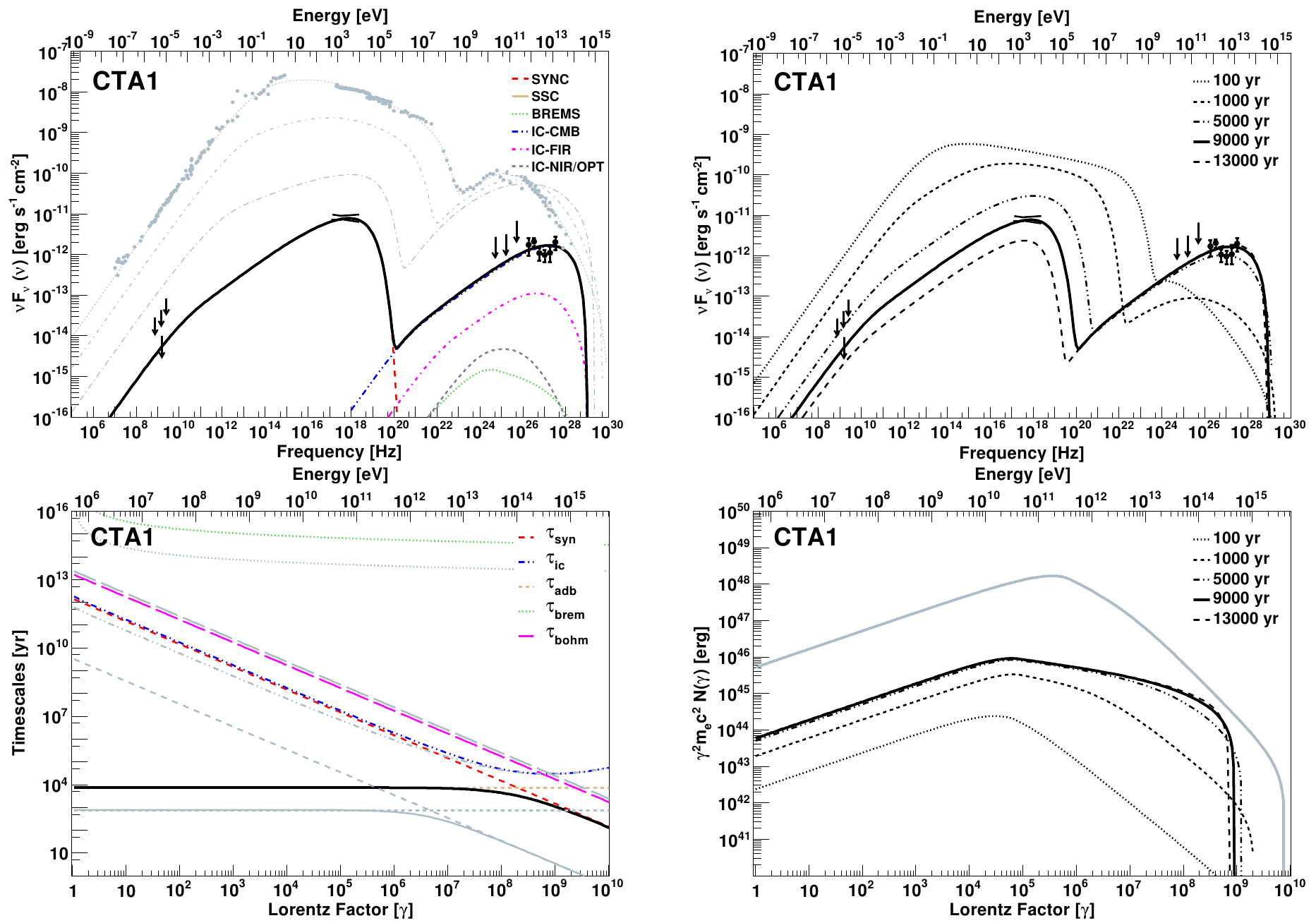}
\caption[SED model of CTA 1]{Details of the SED model of CTA 1. The panels are as in figure \ref{g54}. For details regarding the observational
data and a discussion of the fit, see the text.}
\label{cta1}
\end{figure}

CTA 1 characteristics in radio and X-rays suggest an age between 5000 and 15000 yrs \citep{pineault93,slane97,slane04a} for the SNR, which is in
agreement with the spin-down age of the pulsar ($\sim$14000 yr). \citet{pineault93} derived a kinematic distance of 1.4$\pm$0.3 kpc based on
associating an HI shell found north-western part of the SNR. In order to perform our fit we take the radio upper limits from \citet{aliu13},
where the authors have used a 1.4 GHz image to estimate the flux upper limit within 20 arcmin radius around the pulsar and extrapolated this
upper limit to lower and higher frequencies assuming respectively a radio spectral index of 0.3 and 0. The other UL we use, at 1.5 GHz, was
obtained from a new VLA image \citep{giacani13} considering a size for the nebula of 20 arcmin in radius.

We performed our fit considering a distance to the system of 1.4 kpc, an ejected mass between 6 and 10M$_\odot$, a braking index equal to 3, and
a density of the media of 0.07 similar to the one proposed by the VERITAS Collaboration \citep{aliu13}. We explored the possibility of different
ages for the nebula, between 9000 and 12000 yrs. The best fit of the data was obtained with an age of 9000 yr and 10M$_\odot$ of ejected mass.
The injected spectrum was assumed to follow a power-law with slopes $\alpha_1$=1.5 and $\alpha_2$=2.2. The magnetic field obtained for the model
presented in table \ref{param} was of 4.1 $\mu$G, with an extension of the nebula of 8 pc in radius. For this nebula the main contribution to the
flux at high and very high energies comes from the IC with the CMB, being the IC with the FIR and NIR components almost negligible. Compared to
the other PWNe analyzed in this work, the magnetic fraction of this nebula is much higher, $\eta=0.4$. A low $\eta$ value, like the one obtained
with our model for Crab nebula ($\eta=0.03$), overestimates the flux values at TeV energies compared to the observations of VERITAS.

Previous to Veritas observations, \citet{zhang09} over-predicted the value of the flux at high energies. To model the radio upper limits these
authors assumed that all the emission obtained from the images of \citet{pineault97} was coming from the PWN, which caused also an
over-estimation of the radio flux. In the model presented in figure \ref{cta1}, Fermi upper limits are higher (by about a factor of 8) than the
predictions of our model at those energies.

\section{Other cases}
\label{sec4.3}

\subsubsection{HESS J1813-178 (G12.8-0.0)}

HESS J1813-178 is a TeV source discovered at high energies in the inner galaxy survey done by H.E.S.S. \citep{aharonian06b}. It was also observed
by MAGIC \citep{albert06a}, obtaining its differential $\gamma$-ray spectrum as $(3.3 \pm 0.5) \times 10^{-12}$($E$/TeV)$^{2.1 \pm 0.2}$ s$^{-1}$
cm$^{-2}$ TeV$^{-1}$. The angular extension of the source is 2.2'. With a distance of 4.8 kpc \citep{halpern12}, this gives 3.1 pc of diameter.
The associated central source is the pulsar PSR J1813-1749, which has a period of 44.6 ms \citep{gotthelf09b} and a period derivative of
$1.26 \times 10^{-13}$ s s$^{-1}$ \citep{halpern12}. The spin-down power nowadays is $5.59 \times 10^{37}$ erg s$^{-1}$, and its characteristic
age is 5600 yr.

\citet{brogan05} discovered a radio shell (SNR G12.8-0.0) coincident with the position of HESS J1813-178, having an angular diameter of
$\sim$2.5'. The flux density spectrum was fitted with a power law with an index of 0.48 between 3 cm to 90 cm wavelength. In X-rays, ASCA
detected the source AX J1813-178 also coincident with the position of the SNR and the H.E.S.S. source, but the pointing uncertainty was too large
to distinguish if the origin of the emission is the center of the remnant or from the shell. \citet{helfand07} resolved the X-ray central source
and the PWN using observations from Chandra. The flux of the PWN was fitted with a power law with an index of 1.3 and an absorbed flux of
$5.6 \times 10^{-12}$ erg s$^{-1}$ cm$^{-2}$ between 2 and 10 keV. A distance of 4.5 kpc was assumed and they inferred a luminosity for the PWN
of $1.4 \times 10^{34}$ erg s$^{-1}$. The pulsations of the central source in X-rays were discovered two years later using data from XMM-Newton
\citep{gotthelf09b}. Concerning the age of the system, if the SNR shell were expanding freely, the dynamic age of the system would be about 285 yr
whereas in a Sedov expansion, the age increases until 2520 yr \citep{brogan05}. We adopt an intermediate case of $\sim$1500 yr here, similarly to
other analysis. XMM-Newton also observed this source and could resolve the PWN with an spectral index of 1.8 and a flux between 2 and 10 keV of
$7 \times 10^{-12}$ erg s$^{-1}$ cm$^{-2}$ \citep{funk07a}, which is similar to the one obtained by \citet{helfand07}, but softer.
\citet{ubertini05} observed a soft gamma source with INTEGRAL with an spectral index between 20 and 100 keV of 1.8, as in the XMM-Newton data.
They inferred a luminosity of $5.7 \times 10^{34}$ erg s$^{-1}$ assuming a distance of 4 kpc.

The origin of the emission in the TeV energy range is not clear and we shall use our model to assess the possibility that a PWN produces it.
Other authors have considered this problem before. For instance, \citet{funk07a} considered two scenarios, one in which the VHE and the X-ray
emission are produced leptonically, by electrons in a PWN; and another, in which the VHE and the radio emission are generated in the SNR shell.
They considered two alternatives for the leptonic scenario producing both the X-ray and the VHE photons: a normal FIR and NIR background with a
single power law (with slope 2.4) electron spectrum (model 1); and a significant excess of NIR photons (a factor of 1000 beyond the expected from
GALPROP) subject to an injection spectrum described by a hard, single power law (model 2). In both of these alternatives one is forced to require
that the maximal energy of the electrons is beyond 1.5 PeV, that the minimal energy is also high ($\gamma_{min}$ of the order of $5 \times 10^4$)
and that the magnetic fields are low (a few $\mu$G). The high value needed for $\gamma_{min}$ would convert this PWN in an outlier with respect to the rest
of the population. In any case, these models are both unsatisfying. Model 1 is barely a good fit to the TeV data, significantly overproducing the
measurements at the highest energies. Model 2 has an extremely high photon background, even considering the contribution of the nearby star
forming region W 33 \citep{funk07a}. We have built similar models, and whereas the results cannot be directly compared due to the different
treatments, we essentially find the same trends in the case $\gamma_{max}$ is indeed allowed to reach high values. PWN are capable of
accelerating electrons to PeV energies (see table \ref{param}). However, in the framework of our model (and in a real physical situation), the
maximum Lorentz factor that electrons can achieve is not a free parameter. Here it is set by requesting that the Larmor radius be smaller than
the termination shock (equation \ref{gmax2}). Even assuming that the containment factor is 1, we would attain lower values than 1 PeV, leading
–leaving all other parameters the same– to a bad fitting in both alternatives presented by \citet{funk07a}. For our analogous to their model 1,
the redistribution of the power to lower electron energies would not allow for a good fit to the X-ray peak and the radio emission will increase,
being close or beyond the upper limits. For our analogous to their model 2, we would significantly overproduce the spectral points at all
energies. We need a much lower NIR density of about 55 eV cm$^{-3}$, nevertheless very high, to match the spectrum better. However, particularly
at high TeV energies, it would become impossible to comply with all observational constraints in the case $\gamma_{max}$ is allow to reach a high
value and the slope of the injection power-law is 2, so as to provide a good fit to the X-ray part: the electrons interacting with the CMB would
already overproduce the highest energy data. \citet{fang10b} also studied models for HESS J1813–178, and although the injection is different from
a simple power-law, the general trend is maintained: they cannot attain a good fit to the VHE and X-ray part of the SED with a PWN model either.

Taking into account all of the former, it seems more natural to suppose that HESS J1813–178 VHE emission is generated at the shock of a SNR, or
in the interaction of accelerated protons with the environment (as in \citealt{gabici09,torres10}). We shall not consider this source further in
our sample.

\subsubsection{HESS J1023-575}

HESS J1023-575 was discovered by H.E.S.S. \citep{reimer08}. Its spectrum is fitted by a power law of the form
$dN/dE=4.5 \times 10^{-12}$($E$/TeV)$^{-2.53}$ s$^{-1}$ cm$^{-2}$ TeV$^{-1}$, which implies an integrated flux above 380 GeV of
$1.3 \times 10^{-11}$ s$^{-1}$ cm$^{-2}$. The closest central source is PSR J1022-5746, but the association of these two objects is uncertain due
to the large distance between them, 0.28 degrees, assuming 8 kpc, and the proximity to Westerlund 2, which provides other candidates for the
origin the radiation \citep{abramowski11a}. As far as we are aware there is no synchrotron PWN detected for PSR J1022-5746, leaving any possible
fit of the TeV emission quite unconstrained.

\subsubsection{HESS J1616-508}

HESS J1616-508 is one of the brightest sources in the HESS catalog \citep{aharonian06b}. It is located near RCW 103 (SNR G332.4-0.4) and Kes 32
(G332.4+0.1) and has an extension of 16 arcmin. Its spectrum is fitted by a power-law with an index of 2.35$\pm$0.06 and its flux between 1 and
30 TeV is $2.1 \times 10^{-11}$ erg s$^{-1}$ cm$^{-2}$. PSR J1617-5055 was discovered as a radio pulsar by \citet{kaspi98}. This pulsar was also
detected with INTEGRAL \citep{landi07}, and it was argued that PSR J1617-5055 was the power engine of HESS J1616-508 (e.g., \citealt{mattana09}).
However, there is still some controversy due to the lack of detection in other wavelengths and the position of the PSR in later observations with
Chandra \citep{kargaltsev09}. The latter authors discovered an X-ray PWN surrounding PSR J1617-5055, with a total luminosity between 0.5 and 8
keV of $3.2 \times 10^{33}$ erg s$^{-1}$ assuming a distance of 6.5 kpc. The X-ray efficiency is very low for a young PWN
($L_{PWN}/\dot{E} \sim 2 \times 10^{-4}d^2_{6.5\ \textmd{kpc}}$) as is also for the ratio between luminosities ($L_{PWN}/L_{PSR} \sim 0.18$).
When compared with the TeV source, the size of the putative X-ray nebulae and the TeV emission has one of the largest mismatches. Due to the
controversy in the connection with HESS J1616-508 and the lack of data in the multiwavelength spectrum for the X-ray underluminous PWN, we do not
include this source in our study.

\subsubsection{HESS J1640-465}

HESS J1640-465 is one of the sources discovered by H.E.S.S. during its Galactic Plane survey \citep{aharonian06b}. The source is extended with a
width of 2.7$\pm$0.5 arcmin. Its spectrum is well fitted with a power-law with an index of $\sim$2.4 and a total integral flux above 200 GeV of
$2.2 \times 10^{-11}$ erg s$^{-1}$ cm$^{-2}$. The source is partially coincident with the known radio SNR G338.3-0.0 \citep{whiteoak96}.
XMM-Newton observations \citep{funk07b} showed a hard-spectrum X-ray emitting object at the center of the HESS source, within the shell of the
SNR, most likely a PWN associated with G338.3-0.0 and the counterpart of HESS J1640-465. Chandra observations \citep{lemiere09} constraint the
distance and age of the system between 8 and 13 kpc and 10 and 30 kyr, respectively. For a distance of 10 kpc, the luminosity of the pulsar and
PWN in the range 2-10 keV were estimated as $L_{PSR} \sim 1.3 \times 10^{33}d^2_{10}$ erg s$^{-1}$ and
$L_{PWN} \sim 3.9 \times 10^{33}d^2_{10}$ erg s$^{-1}$ ($d_{10}=d$/10 kpc), respectively. The region of HESS J1640-465 was also detected in Fermi
data \citep{slane10}. No pulsations were found in the Chandra data of this system. Multifrequency radio continuum observations toward SNR
G338.3-0.0 were not able to detect pulsed emission up to a continuum flux density of 2.0 and 1.0 mJy at 610 and 1280 MHZ, respectively; no PWN
was detected in the region of the X-ray PWN was detected \citep{castelletti11}. The lack of the observational data of the period and period
derivative of the pulsar that could be associated with the PWN makes not possible to perform the fit in our model in the same setting as the
others PWNe considered, and thus we do not consider this source in our analysis.

\subsubsection{HESS J1834-087}

The pulsar we quote being positionally correlated in table \ref{datapsr} is a magnetar and unlikely related to the TeV emission unless having an
unusually high spin-down power conversion into TeV photons, of the order of 10\% (orders of magnitude larger than typical values we found in
Table \ref{modprop}). HESS J1834-087 is spatially coincident with the supernova remnant (SNR) G23.3-0.3 (W41) and was detected in the Galactic
Plane survey \citep{aharonian06b}. The MAGIC telescope also observed the source, confirming these results \citep{albert06b}. The TeV emission
seems to have two components, a central source and an extended region surrounding it (see \citealt{mehault11,castro13}). The latter authors have
also reported the GeV detection of this region, with a comparable intrinsic extension and a hard SED between 1 and 100 GeV, of 2.1$\pm$0.1,
somewhat atypical for a PWN spectrum, which smoothly join with the TeV detection. Only a single component is found at GeV energies; the compact
TeV emission is not separately seen by Fermi-LAT. The TeV emission region correlates with a local enhancement of molecular material of about
10$^5$ M$_\odot$ (see \citealt{albert06a,tian07}), what makes possible that TeV emission is in fact hadronically produced in this cloud,
similarly to the models explored in \citet{gabici09} or \citet{torres10}. However, details of the comparison between the CO intensity tracing the
mass and the TeV morphology are not perfectly matching. A new pulsar candidate has been identified by \citet{misanovic11}, CXOU J183434.9–084443,
but its $P$ and $\dot{P}$, if indeed a pulsar, are unknown. These uncertainties suggest that we could not consider this source on a par with the
others in our sample.

\subsubsection{HESS J1841-055}

This source is one of the largest and most complex detected by H.E.S.S., with an extension of approximately 1 degree \citep{aharonian08}. It
would appear that there are several emission peaks within the detection, and thus it is likely that HESS J1841–055 could have multiple origins.
In particular, SNR Kes 73, the pulsar within Kes 73, 1E 1841-45, and also the High Mass X-Ray Binary AX 184100.4–0536 could all plausibly play in
a role in partially generating the TeV emission (see e.g., \citealt{sguera09}). In addition, the pulsar we have proposed in table \ref{datapsr}
as a plausible connection to HESS J1841-055. PSR J1838-0537, was discovered by Fermi \citep{pletsch12}, and can also play a role in producing the
TeV source, particularly when a PWN was detected in GeV gamma-rays \citep{acero13}. However, the plethora of possible origins of the TeV
emission, the difficulty in separating the possible contributors if more than one, and the lack of multiwavelength detections of the PSR
J1838-0537 nebula at lower frequencies preclude us to consider it further in our analysis.

\subsubsection{Boomerang}

The Boomerang PWN (G106.6+2.9) is associated with the pulsar PSR J2229+6114. This pulsar is surrounded by an incomplete radio shell
\citep{halpern02} and it is unique due to its extremely flat spectrum in radio ($\alpha=0.0$). Its distance is not clear, and estimates range
from 3, e.g. see \citet{pineault00} or \citet{abdo09b}, to only 0.8 kpc, see e.g., \citet{kothes06}. The period of the central source is 51.6 ms
and the period derivative is $7.8 \times 10^{-14}$ s s$^{-1}$ \citep{halpern01}. The inferred characteristic age is thus 10460 yr, and the
spin-down luminosity is $2.2 \times 10^{37}$ erg s${-1}$. The PWN seems to have been displaced by the reverse shock of the SNR already.
\citet{kothes01} observed that the forward shock of the SNR has been expanding to the north-east where there is a dense HI medium. As a result of
the interaction of the forward shock with the dense medium, a strong reverse shock was created and crushed with the PWN. After the passage of the
reverse shock, the pulsar created another PWN with less luminosity than the first one, explaining the low radio flux of the nebula considering
the spin-down power of the pulsar. The south-west area is almost empty and the PWN is expanding freely. \citet{kothes06} have also studied the
nature of the break in the spectrum at radio frequencies and inferred an age of 3900 yr since the crush with the reverse shock and a magnetic
field of 2.6 mG from the lifetime of the electrons. Due to the interaction with the reverse shock, we do not consider this PWN in our analysis.

\section{Discussion}
\label{sec4.4}

\subsubsection{SED component dominance}

Table \ref{modprop} shows which components dominate the SED at TeV energies (the first and second contributors are given in the first two
columns). It also provides the ratio (integrating our models in the range 1-10 TeV) between the two largest contributions to the SED at very high
energies (third column). The radio (at 1.4 GHz), X-ray (1-10 keV), and gamma-ray luminosities (1-10 TeV), and their corresponding efficiencies
(when compared with the pulsar spin-down), $f_r$, $f_X$, and $f_\gamma$, are also shown in table \ref{modprop}. To obtain the luminosities we use
the distances to each nebulae according to table \ref{param}, and obtained them from an integration on our fits. This allows to uniformize the
energy range, introducing no change in the conclusions given that all fits are reasonably good descriptions of the observational data when such
exist.

We first see that for all the sources studied, only the Crab nebula is SSC dominated. Given the age, power, and photon backgrounds of the PWNe
studied, this is an expected result \citep{torres13b}. It is interesting to see that in the setting of a leptonic model, all the remaining PWNe
except for HESS J1356-645 and CTA 1 are IC-FIR dominated. The dominance of the FIR contribution to IC is always large in these cases, and the
ratio with the second contributor to the SED at 1 to 10 TeV energies spans from 1.3 to $\sim$10, with the outlying PWN G292-0.5, for which the
ratio is 31. The efficiencies of emission are consistently grouped as follows: $\sim 10^{-6 \div 7}$ in radio, $\sim 10^{-2 \div 3}$ in X-rays,
and $\sim 10^{-3 \div 4}$ in gamma-rays, except for G292.2-0.5, which shows a very low X-ray efficiency in comparison with the others.

\begin{table}
\centering
\scriptsize
\caption[Properties of the fitted models]{Properties of the fitted models. For an explanation of all the columns, see the text.}
  \begin{tabular}{l@{\ }l@{\ }l@{\ }c@{\ }c@{\quad}c@{\quad}c@{\quad}c@{\quad}c@{\quad}c@{\quad}c}
  \hline
    
  PWN  & 1$^{st}$ & 2$^{nd}$ & ratio & $L_r$ & $L_X$ & $L_\gamma$ & $f_r$ & $f_X$ & $f_\gamma$  \\
    & cont. & cont. & (1--10 TeV)  & (1.4 GHz) & (1--10 keV) & (1--10 TeV) & \\
      
    \hline
    
     Crab nebula       &  SSC   & IC-FIR & 1.3 & $1.3\times 10^{33}$  & $1.4\times 10^{37}$& $3.4\times 10^{34}$  & $2.8\times 10^{-6}$ & $3.2\times 10^{-2}$ & $7.5\times 10^{-5}$    \\
     
     G54.1+0.3         & IC-FIR & IC-CMB & 5.3 & $5.0\times 10^{30}$  & $3.0\times 10^{34}$& $6.4\times 10^{33}$ & $4.2\times 10^{-7}$ &  $2.5\times 10^{-3}$ &  $5.3\times 10^{-4}$  \\
     
     G0.9+0.1 (M1)& IC-FIR & IC-NIR & 4.1 & $5.0\times 10^{31}$  & $6.9\times 10^{34}$& $1.4\times 10^{34}$ & $1.2\times 10^{-6}$ & $ 1.6\times 10^{-3}$ & $3.2  \times 10^{-4}$  \\ 
     
     G0.9+0.1 (M2)& IC-FIR & IC-CMB & 6.6 & $1.2\times 10^{32}$  & $1.6\times 10^{35}$& $3.0\times 10^{34}$ & $2.9\times 10^{-6} $& $3.7 \times 10^{-3}$   & $7.1 \times 10^{-4}$ \\
     
     G21.5--0.9        & IC-FIR & IC-CMB & 3.6 & $5.1\times 10^{31}$  & $3.9\times 10^{35}$& $2.0\times 10^{33}$ & $1.5\times 10^{-6} $& 
     $1.2\times 10^{-2}$ & $5.8\times 10^{-5}$  \\
     
     MSH 15--52  (M1)      & IC-FIR & IC-CMB & 10.1 & $2.8\times 10^{31}$  & $3.9\times 10^{35}$& $5.0\times 10^{34}$& $1.5\times 10^{-6}$ & $2.2 \times 10^{-2}$ & $2.7\times 10^{-3}$    \\
     MSH 15--52    (M2)    & IC-FIR & IC-NIR & 1.3 & $3.4\times 10^{31}$  & $3.8\times 10^{35}$& $5.2\times 10^{34}$& $1.9\times 10^{-6}$ & $2.1 \times 10^{-2}$ & $2.9 \times 10^{-3}$    \\

     G292.2--0.5       & IC-FIR & IC-NIR & 31.1 & $1.1\times 10^{31}$ & $1.1\times 10^{32}$& $8.4\times 10^{33}$& $5.0\times 10^{-6}$ & $4.8\times 10^{-5}$ & $3.7 \times 10^{-3}$  \\
     
     Kes 75 (M1)  & IC-FIR & IC-CMB & 4.1 & $4.2\times 10^{30}$ & $1.3\times 10^{35}$& $7.4\times 10^{33}$& $5.1\times 10^{-7} $& 
     $1.5\times 10^{-2}$ & $9.0\times 10^{-4}$  \\
     
     Kes 75 (M2)  & IC-FIR & IC-CMB & 8.5 & $1.3\times 10^{31}$ & $3.7\times 10^{35}$& $1.5\times 10^{34}$& $1.5\times 10^{-6} $& $4.5 \times 10^{-2}$& $1.8\times 10^{-3}$ \\		
     
     HESS J1356--645 (M1)   & IC-CMB & IC-FIR & 1.3 & $1.6\times 10^{30}$ & $7.1\times 10^{33}$& $5.7\times 10^{33}$& $5.0\times 10^{-7} $& $2.3\times 10^{-3}$& $1.8\times 10^{-3}$   \\
     
     HESS J1356--645 (M2)   & IC-CMB & IC-FIR & 1.3 & $1.6\times 10^{30}$ & $6.0\times 10^{33}$& $4.0\times 10^{33}$& $5.2\times 10^{-7} $& $1.9\times 10^{-3}$& $1.3 \times 10^{-3}$  \\
     
     CTA~1             & IC-CMB & IC-FIR & 14.2 & $2.7\times 10^{29}$ & $4.1\times 10^{33}$& $8.6\times 10^{32}$& $6.1\times 10^{-7}$& 
     $9.1 \times 10^{-3}$& $1.9 \times 10^{-3}$ \\                      
 \hline
 \hline
\end{tabular}
\label{modprop}
\end{table}

\subsubsection{Slopes of injection \& electron population}

We have considered a broken or a single power law for the injection distribution of electrons. Other injections can be tried. However, if we use,
e.g., the injection model based on the particle in cell (PIC) simulations done by \citet{spitkovsky08}, we would have several additional –and
observationally unconstrained– parameters. This kind of injection is not devoid of significant extrapolations when considered in a PWN setting
(e.g., the maximal PIC simulated Lorentz factor is far from the maximal electron energies considered in the PWNe). Thus a priori it would seem
that the power-law distributions are a more reasonable choice for the time being, due to their simplicity. Their ability to produce good fits in
all cases give a posteriori support.

We have found that the energy distribution of the electron population is well described almost in all cases by a broken power law. The high
energy slope is found to be in the range 2.2--2.8 except for one outlier, G292.2-0.5, for which $\alpha_2$= 4.1. The low energy part is instead
much harder, in the range 1.0--1.6. These results are consistent with previous studies of part of the sample we have treated, see, for instance,
\citet{bucciantini11}. The breaks, on the other hand, appear at a Lorentz factor in the range $10^5-10^{6.7}$, and for most of the models are
actually concentrated in a narrower range around $5 \times 10^5$. These very small ranges of values of the slopes and break energies for modeling
sources that appear so different at first sight suggests that the processes at the pulsar wind termination shock are common. The only models that
are exceptional to these trends are G292.2-0.5, and the Model 2 of HESS J1356–645. For the PWN likely associated with HESS J1356-645, a broken
power law with parameters in agreement with the previous trends produces a good fit to the data; and the single power law was explored only as an
alternative to give account of ignorance or degeneracies in parameters such as age and pulsar braking index. G292.2-0.5 is also outlier to other
phenomenology discussed in this section. The spectral break of the injected electron needed in G292.2-0.5 is the highest of all PWNe studied.
Despite the obvious caveats in trying to model a spatially complex region with a one zone radiative model, we note that we are also uncomfortable
with the large ejected mass that would be needed in our model to have a good fit to G292.2-0.5 radiative data. It may well be the case that this
PWN is just different in their acceleration properties (the pulsar has one of the largest magnetic field in our sample, in excess of 10$^{13}$,
the other one being Kes 75), or that the model fails due to a large influence of more advanced dynamical states. In fact, the PWN is offset with
respect to the position of the pulsar, what could be originated if the nebula has been displaced after being crushed by an asymmetric reverse
shock caused by the presence of the dark cloud in the vicinity. Finally, it may also be that the steepness of the G292.2-0.5 spectrum points
towards an alternative origin, related to the SNR, a possibility discussed, but not favored, by \citet{kumar12}. All in all, due to the more
uncertain origin of the radiation at the highest energies, the case of G292.2-0.5 requires special attention when looking at the overall
properties of the population. We also note that G292.2-0.5 and the Model 2 of HESS J1356-645 are the only two cases in which we have braking
indices of 2 or lower.

\subsubsection{Pair multiplicity and bulk Lorentz factor}

We now consider the PWN injection rates resulting from our models. We will compare the injection rate with the electrodynamics minimum suggested
by \citet{goldreich69},
\begin{equation}
\dot{N}=\sqrt{\frac{c I \Omega \dot{\Omega}}{e^2}}=7.6 \times 10^{33} \sqrt{\frac{I_{45} \dot{P}}{P^3_{33} 4 \times 10^{-13}}}\ \text{s}^{-1},
\end{equation}
where $P$ and $\dot{P}$ of Crab have been used for normalizing ($P_{33}=P$/33 ms, $I_{45}=I/10^{45}$ g cm$^2$). We can directly compute the
injection rate by integrating $Q=\int Q(\gamma,t)\mathrm{d}\gamma$, from where the multiplicity follows
\begin{equation}
\label{multiplicity}
\kappa=\frac{Q}{\dot{N}}.
\end{equation}
The values of $\kappa$ for all the PWNe in our sample are shown in table \ref{mult}. Multiplicities are large in all cases, although they should
be taken as upper limits. We have found that at the level of the SED, the lower limit value of $\gamma_{min}$ (critical in defining the value of
$\kappa$) remains unconstrained in most cases. For instance, for the Crab nebula, $\gamma_{min}$ values larger than 10$^4$ would make very
difficult to realize a proper description of the synchrotron part of the SED, but instead, the SED is essentially unchanged for lower values. The
same happens in other cases, for instance, with a $\gamma_{min}=10^5$ it is already difficult to fit well the radio spectrum of G0.9+0.1 and
G21.5-0.9. The same happens with G54.1+0.3 for which $\gamma_{min}$ values up to 1000 would require no change in any of the parameters, and up to
$5 \times 10^4$, similarly good fits can be obtained with slight variations of the injection slopes. The only case in which we need a large value
of $\gamma_{min}$ is in fact in the Model 2 of HESS J1356-645, the particularities of which were discussed above.

\begin{table}
\scriptsize
\centering
\caption[\protect\citet{goldreich69} estimation and multiplicity computed from our models.]{\protect\citet{goldreich69} estimation and
multiplicity computed from our models (an upper limit). See the description in the text.}
  \begin{tabular}{  l l  lc cc c cc c   }
  \hline
    
 PWN  & $\dot N$ & $Q$ & $\kappa$ &$\gamma_w$\\
   & s$^{-1}$ & s$^{-1}$ &  \\      
    \hline
    
     Crab nebula       &  7.6$\times 10^{33}$      &    3.2$\times 10^{41}$ & $ 4.2\times 10^{7}  $ & $1.7 \times 10^3$  \\
     
     G54.1+0.3         & 1.2$\times 10^{33}$    &    7.4$\times 10^{38}$     & $ 6.2\times 10^{5}  $  & $2.0 \times 10^4$ \\
     
     G0.9+0.1 (M1)& 2.3$\times 10^{33}$     &    4.0$\times 10^{40}$ & $ 1.8\times 10^{7}  $ & $1.3 \times 10^3$ \\ 
     
     G0.9+0.1 (M2) & 2.3$\times 10^{33}$    &    1.3$\times 10^{40}$ & $ 5.6\times 10^{6}  $ & $4.0 \times 10^3$\\
     
     G21.5--0.9        & 2.1$\times 10^{33}$    &    1.7$\times 10^{39}$ & $ 8.0\times 10^{5}  $ & $2.4 \times 10^4$ \\
     
     MSH 15--52  (M1)      &   1.5$\times 10^{33}$    &    1.3$\times 10^{40}$ & $ 8.6\times 10^{6}  $ & $1.6 \times 10^3$\\
     
     MSH 15--52    (M2)    &   1.5$\times 10^{33}$    &    1.3$\times 10^{40}$ & $ 8.7\times 10^{6}  $ & $1.6 \times 10^3$\\
     
     G292.2--0.5       &  5.5$\times 10^{32}$    &    9.8$\times 10^{38}$ & $ 1.8\times 10^{6}  $ & $2.8 \times 10^3$\\
     
     Kes 75 (M1)  & 1.0$\times 10^{33}$    &    3.5$\times 10^{39}$ & $ 3.5\times 10^{6}  $ & $2.9 \times 10^3$\\
     
     Kes 75 (M2)  &1.0$\times 10^{33}$    &    1.4$\times 10^{40}$ & $ 1.4\times 10^{7}  $ & $7.2 \times 10^2$\\        
     
     HESS J1356--645 (M1)   & 6.4$\times 10^{32}$    &    2.2$\times 10^{38}$ & $ 3.4\times 10^{5}  $ & $1.6 \times 10^4$  \\
     
     HESS J1356--645 (M2)   & 6.4$\times 10^{32}$    &    1.3$\times 10^{37}$ & $ 2.1\times 10^{4}  $ & $2.7 \times 10^5$\\
     
     CTA~1             & 2.4$\times 10^{32}$    &    3.8$\times 10^{38}$    & $1.6 \times 10^6$ & $8.8 \times 10^2$\\                        
 \hline
 \hline
\end{tabular}
\label{mult}
\end{table}

If the wind is characterized by a single value of the Lorentz factor $\gamma_w$, we may write the average energy per particle in the spectrum as
\begin{equation}
<E>=\frac{(1-\eta) \dot{E}(t)}{\int Q(\gamma,t) \mathrm{d}\gamma} \equiv \gamma_w m_e c^2.
\end{equation}
The values of $\gamma_w$ are given in table \ref{mult}. To compute these values we have used the $\gamma_{min}$, $\gamma_{max}$, and $\gamma_b$
values, as well as the slopes $\alpha_1$ and $\alpha_2$ when broken power-laws are a good representation of the electron spectra, for each of the
nebulae. We see that in all cases, $\gamma_b$ is larger than $\gamma_w$ by up to several orders of magnitude. This can be understood from the
mean energy definition above, which can be analytically computed. This formula is time-independent and $\gamma_w$ is fully characterized by 5
parameters: $\gamma_{min}$, $\gamma_{max}$, and $\gamma_b$, $\alpha_1$ and $\alpha_2$. To get a better idea on the dependence of $\gamma_w$ on
each parameter, we can simplify the expression taking into account that normally $1<\alpha_1<2$, $\alpha_2>2$ and
$\gamma_{min}<\gamma_b<\gamma_{max}$. With this assumptions, we can simplify it to yield,
\begin{equation}
\gamma_w \simeq \left[\frac{\frac{1}{2-\alpha_2}-\frac{1}{2-\alpha_1}}{\frac{1}{1-\alpha_1}\left(\frac{\gamma_b}{\gamma_{min}} \right)^{\alpha_1-1}+\frac{1}{1-\alpha_2}} \right]\gamma_b,
\end{equation}
with the order of magnitude being $\gamma_w \sim \gamma_b (\gamma_b/\gamma_{min} )^{(1-\alpha_1)}$. Physically, the population of low energy
electrons is more numerous, and it is responsible for the radio to IR emission of the nebulae. 

\subsubsection{ISRF values compared with a Galactic model}

\begin{figure}[t!]
\centering
\includegraphics[width=1.0\textwidth]{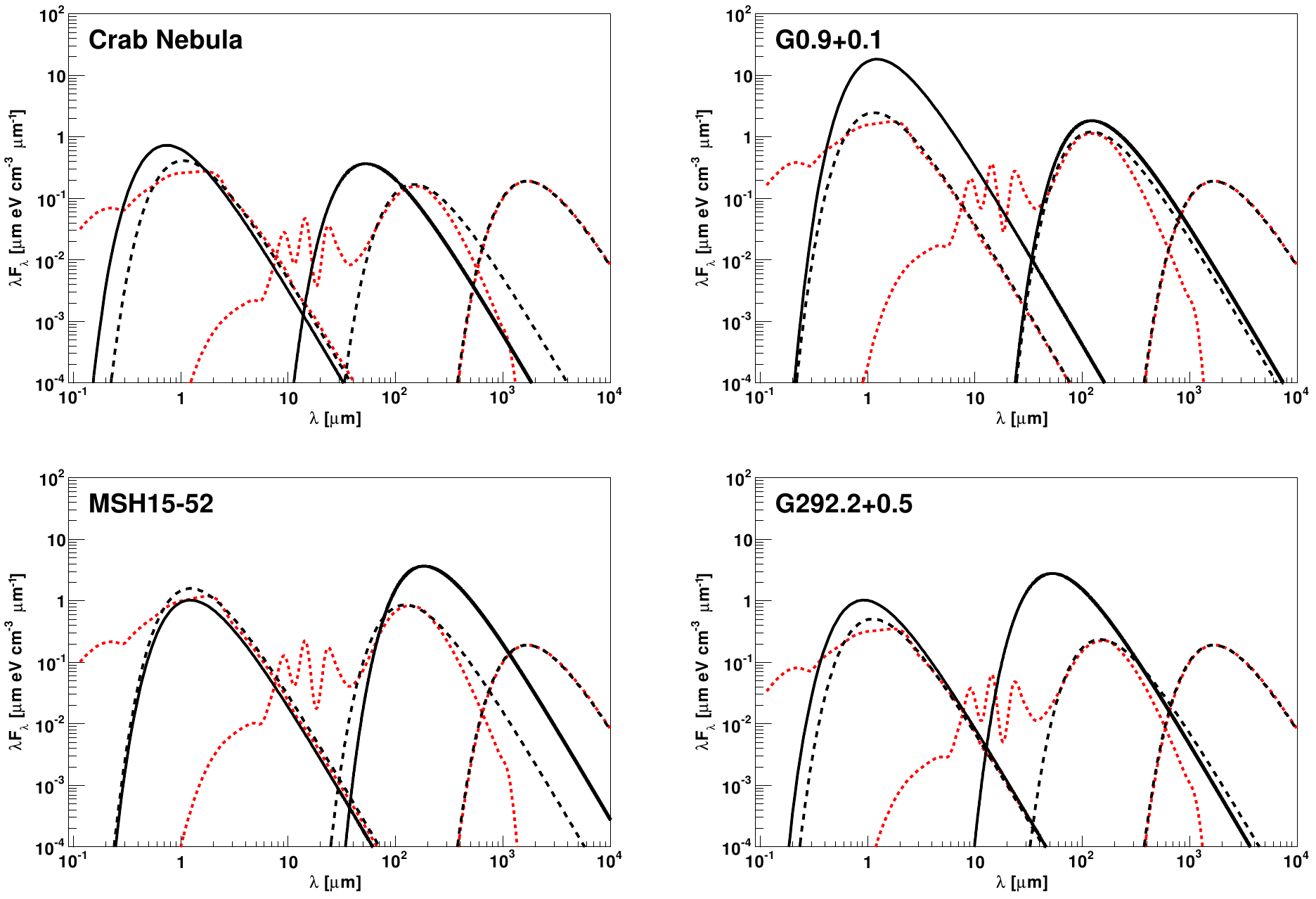}
\caption[Example of the comparison between the ISRF obtained from the GALPROP code and the assumptions made to fit the PWNe models]{Example of
the comparison between the ISRF obtained from the GALPROP code \protect\citep{porter06} and the assumptions made to fit the PWNe models. We show
the FIR and NIR diluted blackbodies (with the parameters of table \ref{param} in bold black curves), in comparison with the GALPROP raw results
(in red) and fits to these results using diluted blackbodies (in black thin lines, and as given in table \ref{isrftab}.) The rightmost component
stands for the CMB in all panels.}
\label{isrf}
\end{figure}

Table \ref{isrftab} compares the energy densities used to fit the observational data of each of the PWNe studied with those obtained from the
GALPROP code \citep{porter06}. In order to do this, we have obtained the ISRF from GALPROP and fitted three diluted blackbodies, for which the
energy densities and temperatures are referred to as $w^G$ and $T^G$, respectively. As shown in table \ref{isrftab}, the values of the FIR energy
densities obtained from GALPROP are generally lower (by up to a factor of a few) than what we found is needed to fit the PWN high-energy
emission. Figure \ref{isrf} shows four examples.

\begin{table}
\scriptsize
\centering
\caption[Comparison between modeled ($w$,$T$) and  GALPROP ($w^G$,$T^G$) energy densities and temperatures]{Comparison between modeled ($w$,$T$)
and  GALPROP ($w^G$,$T^G$) energy densities and temperatures. When the parameters ($w$,$T$) in the model are the same as the extracted from
GALPROP we quote \ldots}
  \begin{tabular}{lcccccccc}
  \hline    
   PWN  & $w_{FIR}$ & $T_{FIR}$ & $w_{NIR}$ & $T_{NIR}$ & $w^G_{FIR}$ & $T^G_{FIR}$ & $w^G_{NIR}$ & $T^G_{NIR}$ \\
    & (eV cm$^{-3}$) & (K) & (eV cm$^{-3}$) & (K) & (eV cm$^{-3}$) & (K) & (eV cm$^{-3}$) & (K)  \\
    \hline
     Crab nebula & 0.5 & 70 & 1.0 & 5000 & 0.2 & 25 & 0.6 & 3500\\
     G54.1+0.3 & 2.0 & 20 & 1.1 & 3000 & 0.8 & 25 & 1.1 & 3000\\
     G0.9+0.1 (M1) & 2.5 & 30 & 25 & 3000 & 1.4 & 35 & 10.5 & 3500\\
     G0.9+0.1 (M2) & 3.8 & 30 & 25 & 3000 & 1.7 & 30 & 3.4 & 3200\\
     G21.5--0.9 & \ldots & \ldots & \ldots & \ldots & 1.4 & 35 & 5.0 & 3500\\
     MSH 15--52 (M1)  & 5 & 20 & 1.4 & 3000 & 1.2 & 30 & 2.2 & 3000\\
     MSH 15--52 (M2) Ê& 4 & 20 & 20 & 400 & 1.2 & 30 & 2.2 & 3000\\
     G292.2--0.5 & 3.8 & 70 & 1.4 & 4000 & 0.3 & 25 & 0.7 & 3300\\
     Kes 75 (M1) & 2.5 & 25 & 1.4 & 5000 & 1.5 & 30 & 4.4 & 3500\\
     Kes 75 (M2) & 5.0 & 25 & 1.4 & 5000 & 1.6 & 30 & 2.2 & 3000\\
   HESS J1356--645 (M1) & 0.4 & 25 & 0.5 & 5000 & 0.6 & 25 & 1.2 & 3100\\
   HESS J1356--645 (M2) & 0.4 & 25 & 0.5 & 5000 & 0.6 & 25 & 1.2 & 3100\\
      CTA~1 & 0.1 & 70 & 0.1 & 5000 & 0.3 & 25 & 0.6 & 3000\\
 \hline
 \hline
\end{tabular}
\label{isrftab}
\end{table}

The use of GALPROP ISRFs all along the Galaxy is known to be subject to local uncertainties. Galactic locations in which freshly accelerated
electrons target overdensities of FIR photons contributed by nearby stars, star-forming regions, or the supernova remnants themselves, could
produce these local variations. As mentioned above in some of the individual PWNe studied, the need of larger energy densities than those found
in GALPROP when time-dependent models have been used has been spotted in the past, but for scattered PWN. The possibility of finding relatively
high energy densities in the background photon fields nearby PWNe is interesting from a couple of perspectives: On the one hand, it would imply
that CTA could be mapping PWNe also at averaged (and thus lower) Galactic photon backgrounds, ultimately helping determine the latter. On the
other hand, detailed studies of the IR emission around PWNe should reveal significant sources. This is in general true, as examples, one could
quote the case of G54.1+0.3 in which \citet{temim10} proposed that the SN dust is being heated by early-type stars belonging to a cluster in
which the SN exploded; or MSH 15-52 where there is an O star 13 arcsec away from the corresponding pulsar \citep{arendt91,koo11}. A statistical
study of the correlation between mass (traced by CO and dust) and TeV sources has been recently performed by \citet{pedaletti14}, finding that
there are hints of a positive correlation with IR excess at the level of 2–3$\sigma$, which still needs to be confirmed.

\subsubsection{Magnetization of the nebulae}

\begin{figure}
\centering
\includegraphics[width=1.0\textwidth]{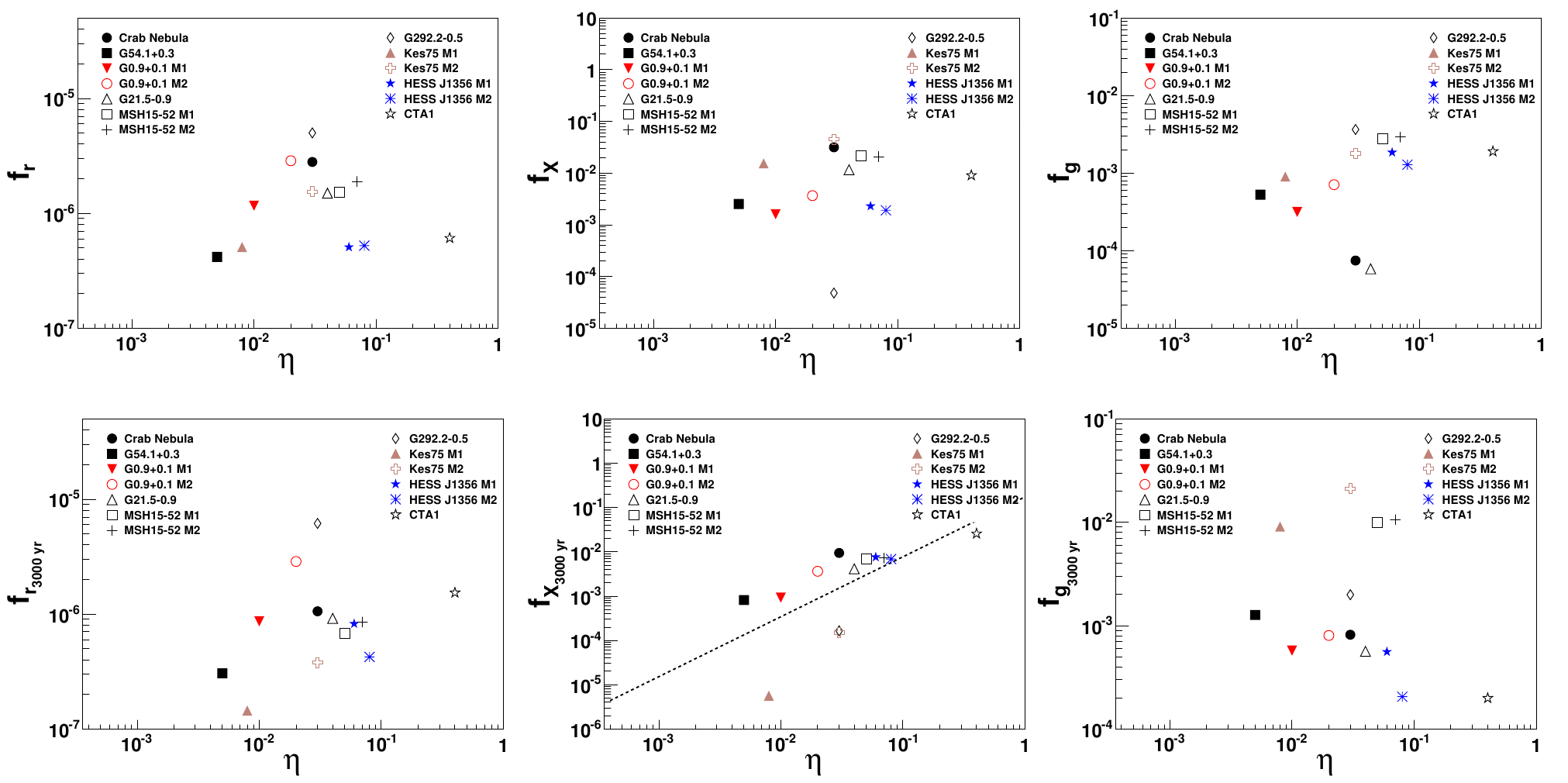}
\caption[Magnetization of PWNe as a function of the radio, X-ray, and $\gamma$-ray efficiency]{Magnetization of PWNe as a function of the radio,
X-ray, and $\gamma$-ray efficiency. In the first row, all luminosity fraction values correspond to those today; in the second row, to the values
they have from the evolution of each of the PWN when considered at 3000 years.}
\label{eta_eff}
\end{figure}

From table \ref{param} we see that all young nebulae detected at TeV are particle dominated, with magnetic fractions that in all cases except CTA
1, never exceed a few percent. Figure \ref{eta_eff} shows the values of the obtained radio, X-ray, and $\gamma$-ray efficiencies as a function of
the magnetic fraction of the nebulae (which in our model is constant along the evolution). The two sets of panels distinguish the values of the
efficiencies obtained today (at different ages for each of the nebulae considered) from those obtained at the same age, fixed at 3000 years.

To consider whether there is a correlation in any of these (and subsequent) magnitudes we use a Pearson test. The Pearson $r$ estimator is
computed using 9 PWN models (unless otherwise clarified). When more than one model was considered plausible for a given PWN we use M1, although
we have verified that considering the alternatives would not introduce a significant change to the results. We do not emphasize here the search
for precise fit parameters (unless an obvious connection would appear), but of plausible correlations. The latter will be hinted in those cases
in which the Pearson coefficient for the pair of magnitudes considered yields to a non-directional probability of incorrectly rejecting the null
hypothesis (i.e., no correlation) smaller than 0.05. In these cases, we quote the fit parameters in table \ref{fitspar}, as well as we show the
fit in the corresponding figure.

There is no apparent correlation of the efficiencies with the magnetization except when we consider the X-ray efficiency $f_X$ of the nebulae
normalized at the same age. In that case, the Pearson coefficient yields to a probability of 0.043 of incorrectly rejecting the null hypothesis,
but the coefficients of a linear fit are poorly determined because of the dispersion of the data. The significance of the correlation barely
meets our cut. The radio and $\gamma$-ray efficiencies computed at the same age present significances of the order of 10\%. The fact that we do
not see a correlation of the gamma-ray efficiency with the magnetization implies that $\eta$ is neither the only nor the dominant order parameter
to impact the luminosities. The fact that we see essentially very similarly magnetized PWNe from a magnetic point of view reduces the
$\eta$-distinguishing power further.

\begin{table}
\scriptsize
\centering
\caption[Correlation fits shown in the figures]{Correlation fits shown in the figures. We use $y=p_1 x+p_0$, where variables can be in
logarithmic scale, as shown in the corresponding figures. Numbering of panels goes alphabetically, from left to right and top to bottom. Unless
otherwise clarified we used all PWNe for fitting (in cases where we have two models, we use Model 1). We show the Pearson's correlation
coefficient $r$ and the non-directional significance implied by it.}
\begin{tabular}{lllllcccc}
\hline
$x$-Magnitude &  $y$-Magnitude & Fig. & $p_0$ & $p_1$ & Pearson's $r$ & $P$ \\
\hline
\hline
$\eta$ & $f_{X}$             & Fig. \ref{eta_eff} - panel e & $   -0.75 \pm 0.89$ & $   1.35 \pm  0.55 $  & 0.68  & $4.3 \times 10^{-2}$&\\
\hline
$\dot{E}$ & $L_{r}$       & Fig. \ref{corr1} - panel a & $  -11.50\pm  5.67  $ & $  1.15  \pm  0.15 $ & 0.94 &  $1.4 \times 10^{-4}$ \\
$\dot{E}$ & $L_{X}$       & Fig. \ref{corr1} - panel b & $ -17.67 \pm   12.78 $ & $ 1.41 \pm 0.34   $ & 0.84 & $4.5 \times 10^{-3}$ \\
$\tau_c$  & $L_{\gamma}$  & Fig. \ref{corr1} - panel f & $   36.95 \pm 1.31   $ & $  -0.88  \pm 0.38 $ & -0.67 & $4.8 \times 10^{-2}$\\
$\dot{E}$ & $L_{\gamma}/L_{r}$  & Fig. \ref{corr1} - panel h & $  30.20 \pm 7.69 $ & $ -0.74 \pm  0.21 $ & -0.80 &$9.6 \times 10^{-3}$\\
$\dot{E}$ & $L_{\gamma}/L_{X}$  & Fig. \ref{corr1} - panel i & $   36.38  \pm 15.76   $ & $ -1.00 \pm 0.42 $ & -0.67 & $4.8 \times 10^{-2}$\\
\hline 
\hline
$\dot{E}$ & $\gamma_{max}$ & Fig. \ref{corr2}  - panel a & $ -2.85 \pm 3.53 $ & $ 0.32 \pm 0.10 $ & 0.79 &  $1.0  \times 10^{-2}$ &\\
$B_{LC}$ & $\gamma_{max}$ &  Fig. \ref{corr2} - panel c & $  7.31 \pm 0.47 $ & $ 0.37 \pm 0.10 $ & 0.82 & $6.8  \times 10^{-3}$&\\
$\Delta V$ & $\gamma_{max}$ &  Fig. \ref{corr2} - panel d & $ -1.16 \pm 3.03 $ & $ 0.65 \pm 0.19 $ & 0.78 & $1.3  \times 10^{-2}$ &\\
$\dot{E}$ & $B$ & Fig. \ref{corr2} - panel p & $  -18.13 \pm 4.28 $ & $ 0.52 \pm 0.11 $  & 0.86 & $2.9  \times 10^{-3}$ &\\
$B_{LC}$ & $B$  & Fig. \ref{corr2} - panel r & $ -1.51  \pm 0.70 $ & $ 0.55 \pm 0.14 $  & 0.83 & $6.0  \times 10^{-3}$ &\\
$\Delta V$ & $B$  & Fig. \ref{corr2} - panel s & $ -15.40  \pm 3.68 $ & $ 1.04 \pm 0.23 $  & 0.86 &   $2.9  \times 10^{-3}$& \\
\hline
\hline
\end{tabular}
\label{fitspar}
\end{table}

\subsubsection{Is there a low-magnetization observational bias?}

The only high-magnetization nebula we found in the sample we study is CTA 1, for which $\eta=0.4$, is close to equipartition. Should $\eta$ be
much lower than this value we would find TeV fluxes in excess of what has been detected. The possibility that CTA 1 is beyond free expansion
could play a role here; a compression of the nebula due to reverberation could lead to an increase of the magnetization. Note that in the model
of CTA 1 by \citet{aliu13}, where a reverberation has been taken into account, the magnetization was also found to be in the high end, more than
an order of magnitude larger than in Crab nebula. It is to note that the highest magnetized nebula in the sample is showing one of the lowest
magnetic fields (see table \ref{param}), something which has also been found with other models (e.g., \citealt{aliu13}). However, the conclusion
that all the other nebulae are heavily particle-dominated is not affected by uncertainties in the modeling. To prove this we have tried to fit
these nebulae data with an ad-hoc increase of $\eta$ up to 0.5 (equal distribution of the power between particles and field) and explored the
range of parameters, if any, which would allow for a good fit. Models with larger $\eta$ allow us to investigate whether we would have detected
the nebulae should they have an increased magnetic fraction. Earlier, we have concluded that if the injection and environment of PWNe were as
those of Crab, only in the case of a large, Crab-like, spin-down power feeding into a nebula located at 2 kpc or less, a H.E.S.S.-like telescope
would detect magnetically-dominated nebula beyond $\eta \sim 0.5$ \citep{torres13b}. Different to our earlier study, we here consider the
injection and environmental properties specifically derived for each nebulae.

\begin{figure}
\centering
\includegraphics[width=1.0\textwidth]{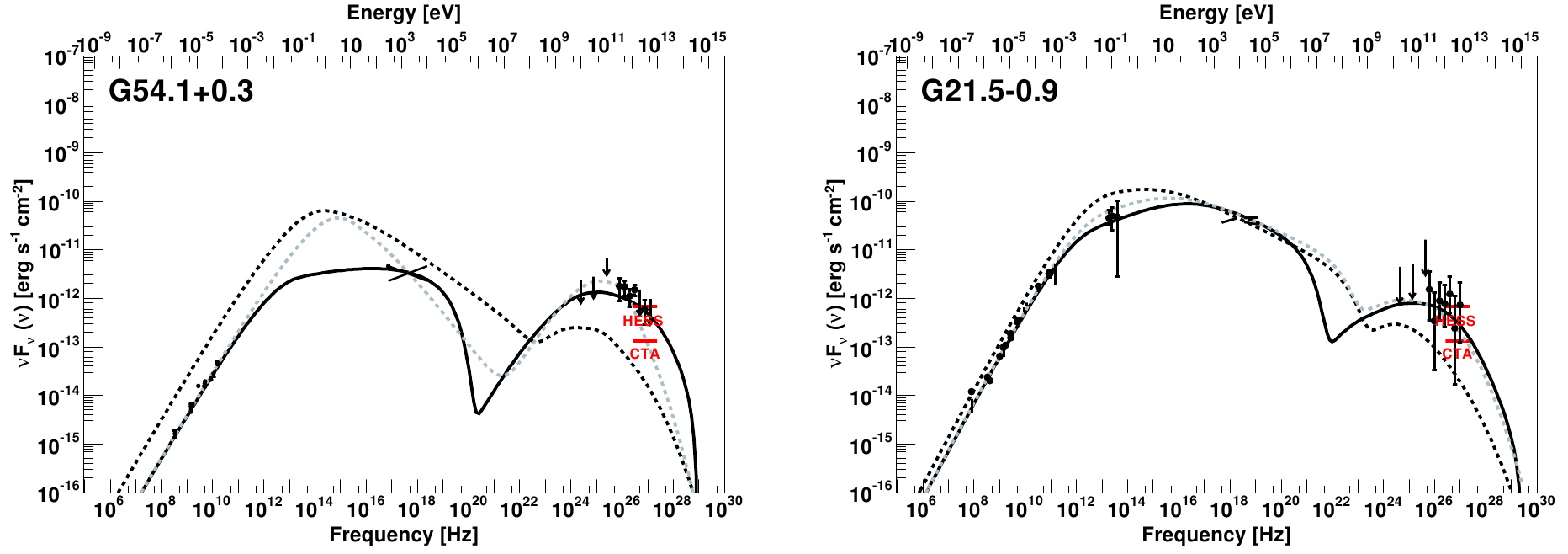}
\caption[G54.1+0.3 (left) and G21.5-0.9 (right) modeled with an imposed equipartition of the energetics as compared with the adopted models]
{G54.1+0.3 (left) and G21.5-0.9 (right) modeled with an imposed equipartition of the energetics ($\eta=0.5$) as compared with the adopted
(particle dominated) models. The solid line represents the fitted model of table \ref{param}, the dashed black line represents a model with
$\eta=0.5$ and no changes in other parameters with respect to the fitted model of table \ref{param}, and the dashed grey line stands for an
equipartition model where other parameters are adjusted ad-hoc so that a relatively good fit is attained. For a discussion of the caveats of
latter models see the text. The sensitivity of a H.E.S.S.-like telescope and of CTA are marked by the horizontal lines.}
\label{equipartition}
\end{figure}

Figure \ref{equipartition} shows two examples, for PWN G54.1+0.3 and G21.5-0.9, when modeled with imposed equipartition of the energetics keeping
other parameters the same (e.g., with the same FIR/NIR densities). The increase in $\eta$ implies enlarging it by a factor of $\sim$100 and
$\sim$10 in the fitted $\eta$-value, respectively. The predicted TeV emission fits the data badly, and the TeV fluxes are below the sensitivity
of CTA.

We have also searched for a fit in case the PWNe are in equipartition but all other parameters are allowed to vary. The solutions we found
require extreme values of other parameters and are thus not preferred. For instance, in the case of G21.5-0.9, a relatively good fit (albeit of
poorer quality than the one we show in figure \ref{g21}) can be found by increasing the FIR density to 6 eV cm$^{-3}$ (a factor of 6 larger than
the GALPROP outcome at the position) and reducing the ejected mass by a factor of 2 (what enlarges the nebula size in our model and contributes
to dilute the magnetic field energy). It is clear that there is no preference for these stretched parameters over the ones shown in our earlier
fit. The case of G54.1+0.3 is similar, although requires even larger changes in the FIR and NIR densities, and the ejected mass in order to yield
to a fit which is not even close to all data points, particularly those at high energies. In particular, figure \ref{equipartition} shows a model
with $\eta=0.5$ a FIR (NIR) density of 4 (40) eV cm$^{-3}$, and an ejected mass more than a factor of 3 smaller implying a factor of $\sim$2
larger nebula. It is clear that no equipartition model can be sustained in this case either. These conclusions are similarly obtained in the
analysis of other PWNe. The finding of CTA 1, however, shows that the fact that most of the PWNe we see are particle dominated cannot be fully
ascribed to an observational bias; at least in some cases (but not in the majority) we would be able to detect them with the current generation
of telescopes.

\subsubsection{Searching for a more meaningful SEDs and electron population comparison}

\begin{figure}
\centering
\includegraphics[width=1.0\textwidth]{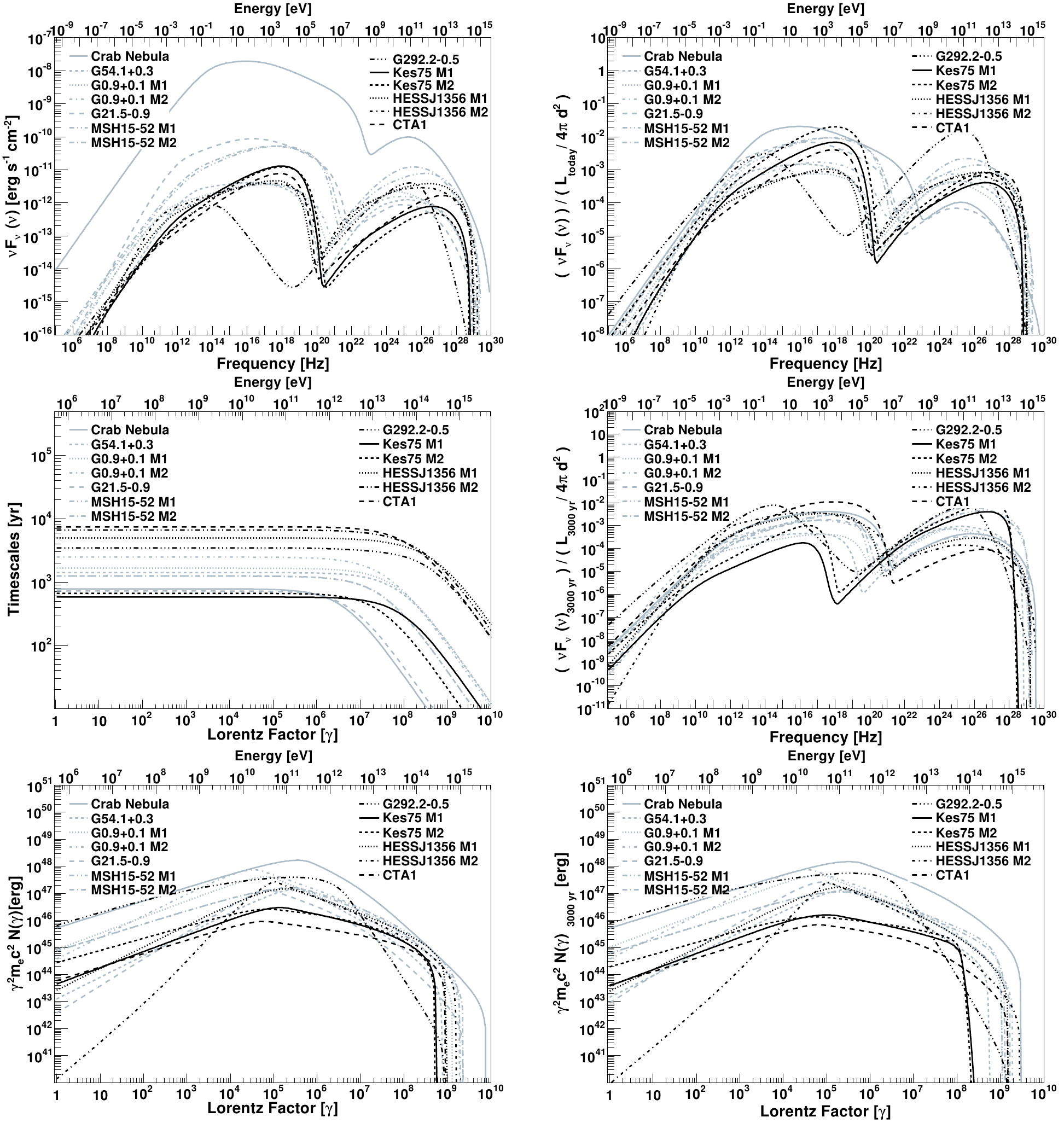}
\caption[Comparison of PWNe results]{Comparison of PWNe results. Left panels: from top to bottom, SEDs, electron losses, and electron
distributions today. Right panels: from top to bottom, SEDs normalized by the corresponding spin-down flux ($\dot{E}/4\pi D^2$, as obtained from
table \ref{param}), SEDs at 3000 yr normalized with the spin-down flux that each pulsar would have at that age, and electron populations at 3000
yr.}
\label{norm_mod}
\end{figure}

Figure \ref{norm_mod} put together the currently observed SEDs, the corresponding electron losses, and the electron populations. Whereas this is
an interesting figure to gather the variety of the sources detected, a direct comparison of the multi-frequency emission (as it is usually done)
has to be taken with care: we are looking at objects at different ages and powered by pulsars of different spin-down. The variety we found at the
SED level (top left panel) contrasts with the little dispersion (one order of magnitude) in the timescales for the losses that are operative in
all the PWN. From the SED results today, the two outliers from the bulk of models are the Crab nebula and G292.2-0.5. Whereas the former can be
understood due to the large difference in spin-down power, the reason for the latter discrepancy is less clear (see the discussion above).

In order to search for a more meaningful comparison we explore two normalizations of the SEDs. On the one hand, we normalize the SED of each PWN
by its corresponding spin-down flux ($F_{sd}=\dot{E}/4 \pi D^2$, as obtained from table \ref{param}) each pulsar has at its current age. On the
other hand, we compute the SEDs at the same age (arbitrarily chosen to be 3000 yr) for all pulsars, and normalize them with the spin-down flux
that each pulsar would have at that age ($\dot{E}$(3000 yr)/$4 \pi D^2$). These normalized SEDs are shown in the right panels of figure
\ref{norm_mod}. The bottom-right panel of figure \ref{norm_mod} shows the electron populations of all PWNe at the same age (3000 yr).

It is interesting to compare the Crab nebula's SED with respect to the others when one normalize it with the corresponding spin-down power and/or
look at all PWNe at the same age: the Crab nebula becomes an unnoticeable member of the same population of sources. It is also interesting to
notice that the other outlier, G292.2-0.5, is now also in the bulk of models (see second panel, right column). The population is only
distinguished by differences in the electron content, where slight variations in the position of the breaks and cutoffs is retained even when
looked at the same age.

\subsubsection{PWN versus PSR properties: $\dot{E}$ and $\tau_c$}

\begin{figure}[t!]
\centering
\includegraphics[width=1.0\textwidth]{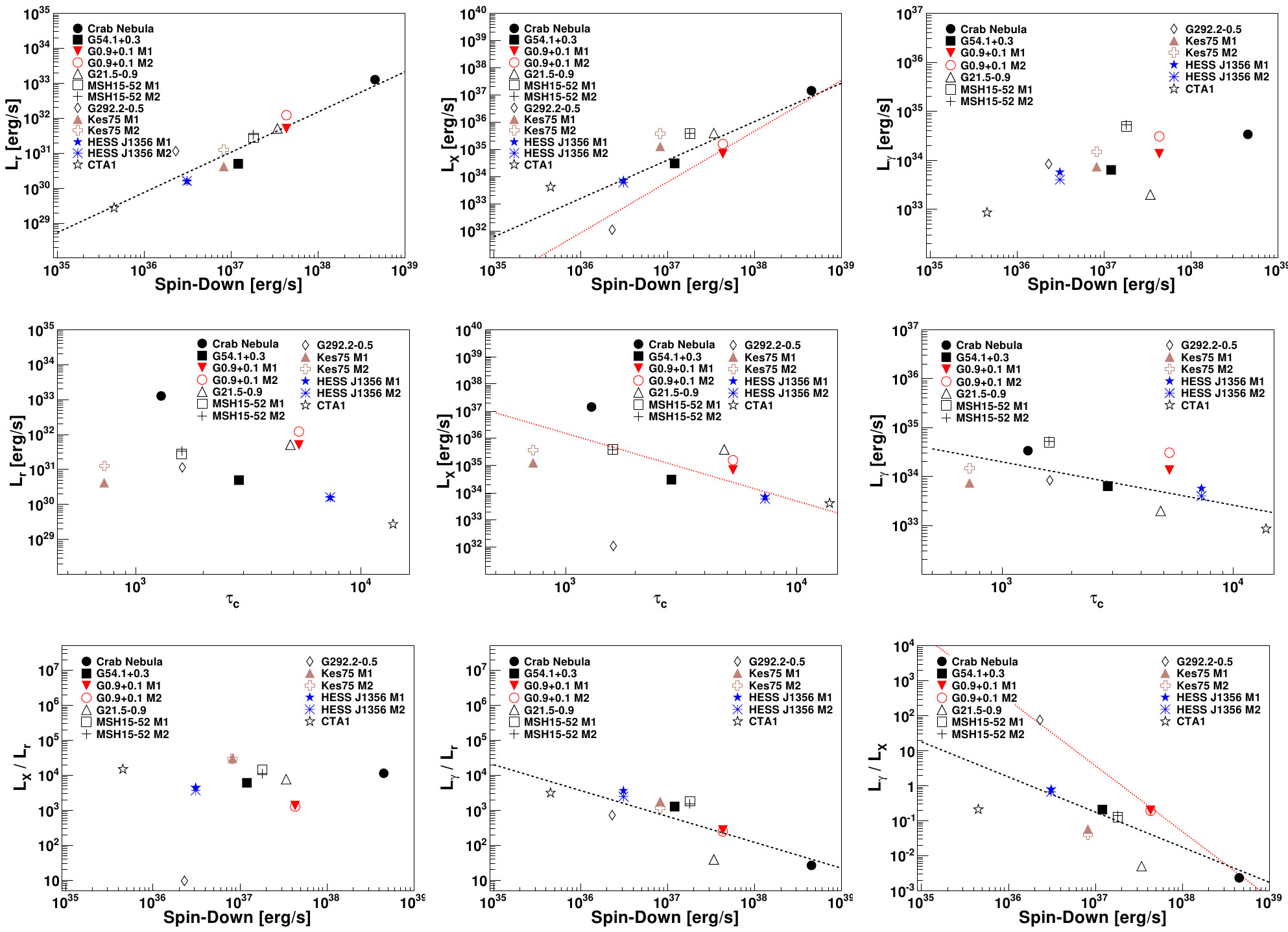}
\caption[Radio, X-ray, and $\gamma$-ray luminosities of young, TeV-detected PWNe as a function of $\dot{E}$ and $\tau_c$ of their pulsars]{Radio,
X-ray, and $\gamma$-ray luminosities of young, TeV-detected PWNe as a function of spin-down power and characteristic ages of their pulsars.
Linear fits to the data (black dashed lines) are also shown for magnitudes with a high Pearson coefficient (see text for details). Red dashed
lines stand for fits presented in \protect\citet{mattana09} using observational data on pulsars of up to 10$^5$ years of age. The bottom row
shows the ratios between the X-ray and radio, gamma-ray and radio, and gamma-ray and X-ray luminosities.}
\label{corr1}
\end{figure}

Possible correlations between the luminosities obtained from our models and two of the main features of the central pulsars, their spin-down
power and characteristic age, are explored in figure \ref{corr1}. It shows the distribution of radio, X-ray, and gamma-ray luminosities, and
their ratios (see table \ref{modprop}) as a function of spin-down power and characteristic ages. A line is added (and parameters are shown in
table \ref{fitspar}) when the Pearson coefficient is such that the correlation is significant to better than 95\% of confidence, as above. A red
line is added to those panels for which \citet{mattana09} provided a fit when considering observational values of TeV-detected PWNe up to 10$^5$
yr of age.

The possible correlation of the PWN luminosities with the PSR characteristic ages (second row in figure \ref{corr1}) is not clear for young PWNe;
for $L_r$ and $L_X$ we actually do not find them at the confidence cut imposed. At the latter case, however, the fit by \citet{mattana09} is in
agreement with the overall (visual) trend of our sample. The only correlation barely surviving our 95\% confidence cut is the one between
$\tau_c$ and $L_\gamma$ (see table \ref{fitspar}), which \citet{mattana09} did not find. We see that the larger the characteristic age the lower
the $\gamma$-ray luminosity. This trend is opposite to the example made in the introduction, where we find more $\gamma$-ray luminosity for
pulsars with larger $\tau_c$ when all other parameters were the same, and thus requires a careful look. On the one hand, we have in our sample
cases of similar spin-down power and $\tau_c$, for G21.5-0.9 and G0.9+0.1; but different real (or assumed real) age (the age assumed for G0.9+0.1
is a factor of 2 to 4 larger than that of G21.5-0.9). In this case, one should also expect variance in $L_\gamma$ (being smaller for the
youngest, as found) even if all other parameters influencing the gamma-ray production are the same (which usually are not). On the other hand,
CTA 1 (at the extreme of the distribution) has the largest magnetization and lowest spin-down power of the sample, what reduces its $\gamma$-ray
luminosity despite its larger $\tau_c$.

The possible correlation of the luminosities with the spin-down power is visually apparent for all three luminosities considered (see top row of
figure \ref{corr1} and table \ref{fitspar}), although in the case of the $\gamma$-ray luminosity the confidence cut is not met (the resulting
probability for no correlation is $6.2 \times 10^{-2}$). This is compatible with \citet{mattana09} results. The scaling between X-ray
luminosities and spin-down power was also noted by \citet{seward88} and \citet{becker97}; in the form $L_X \sim 10^{-3}\dot{E}$, see also
\citet{kargaltsev09}. The radio luminosity/spin-down power correlation is the best in the sample we study.

We have also found correlations in two of the ratios of luminosities explored, $L_\gamma/L_r$ and $L_\gamma/L_X$. That is, when we compare the IC
$\gamma$-ray luminosity with the synchrotron generated ones, we find that the larger the spin-down, the smaller their ratio. We have seen above
that all three luminosities apparently increase with the spin-down, with the luminosity of the synchrotron components increasing faster. The
larger the spin-down power, the more particles are in the nebulae and the larger is the maximum energy they attain. However, the timescale for
cooling of electrons via radiating synchrotron emission is faster than for IC, and whereas the radio emission is greatly enhanced, the
$\gamma$-ray emission grows at slower rate.

We have considered what happens to these correlations when all the systems are evolved to the same pulsar age, at 3000 yr. We see that the
correlations between the luminosities and the spin-down power (both at 3000 yr of age) still appear at our confidence cut level, but their
significances worsen with respect to the one pointed out above. This worsening makes for the correlation of the ratio of the luminosities to
disappear in this case.

\subsubsection{PWN versus PSR properties: other parameters}

We now consider possible correlations between other PWN properties resulting from our fits and those of the central pulsar. We compute for each
pulsar the surface magnetic field, the potential difference at the polar cap, the light cylinder, and the magnetic field at the light cylinder
(assuming the neutron star is a dipole). The definitions used for these quantities are summarized in table \ref{psrpar}, as well as the values
obtained for all pulsars in our study. These quantities relate to each other and to the spin-down power, all being functions of P and $\dot{P}$;
thus, it is to expect that if we find a correlation of any magnitude with the spin-down power, we would also find it with the potential
difference at the polar cap, and the magnetic field at the light cylinder. The spin-down--surface magnetic field dispersion can introduce
different correlations, depending on the values of P and $\dot{P}$.

\begin{table}
\scriptsize
\centering
\caption[Parameters used in search of correlations, as a function of $P$ and $\dot P$]{Parameters used in search of correlations, as a function
of $P$ and $\dot P$ and values for the pulsars associate with the PWNe considered in the study.}
\begin{tabular}{lcccccccccc}
\hline
  PSR associated with &  Surface Magnetic field & Light Cylinder Radius & Magnetic field at & Electric Potential\\
 & & & Light Cylinder & \\
\hline 
  &  Equation (\ref{bdip}) & $R_{LC}=4.77 \times 10^9 (P/{\rm s})$ cm & $B_{LC}=5.9 \times 10^8 (P/{\rm s})^{-5/2}$ & Equation (\ref{elecpot})\\
  & & & $\sqrt{(\dot P/{\rm s \, s^{-1}})}$ G & \\
\hline
     Crab                & $7.58\times 10^{12} $& $1.59\times 10^8  $&$ 1.88\times 10^6 $&$ 4.46\times 10^{16}$ \\
     G54.1+0.3           & $2.04\times 10^{13} $& $6.49\times 10^8  $&$ 7.49\times 10^4 $&$ 7.25\times 10^{15}$\\
     G0.9+0.1 (M1/M2)  & $5.66\times 10^{12} $& $2.49\times 10^8  $&$ 3.67\times 10^5 $&$ 1.36\times 10^{16}$ \\
     G21.5--0.9          & $7.12\times 10^{12} $&$ 2.95\times 10^8  $&$ 2.77\times 10^5 $&$ 1.22\times 10^{16}$ \\
     MSH 15--52 (M1/M2)          & $3.04\times 10^{13} $&$ 7.16\times 10^8  $&$ 8.29\times 10^4 $&$ 8.85\times 10^{15}$\\
     G292.2--0.5         & $8.18\times 10^{13} $&$ 1.95\times 10^9  $&$ 1.11\times 10^4 $&$ 3.22\times 10^{15}$ \\
     Kes 75 (M1/M2)    & $9.71\times 10^{13} $&$ 1.55\times 10^9  $&$ 2.63\times 10^4 $&$ 6.07\times 10^{15}$\\
      HESS J1356--645    & $1.56\times 10^{13} $&$ 7.92\times 10^8  $&$ 3.15\times 10^4 $&$ 3.73\times 10^{15}$\\
      CTA 1              & $2.16\times 10^{13} $&$ 1.51\times 10^9  $&$ 6.27\times 10^3 $&$ 1.41\times 10^{15}$ \\                       
 \hline
 \hline
\end{tabular}
\label{psrpar}
\end{table}

\begin{figure}
\centering
\includegraphics[width=1.0\textwidth]{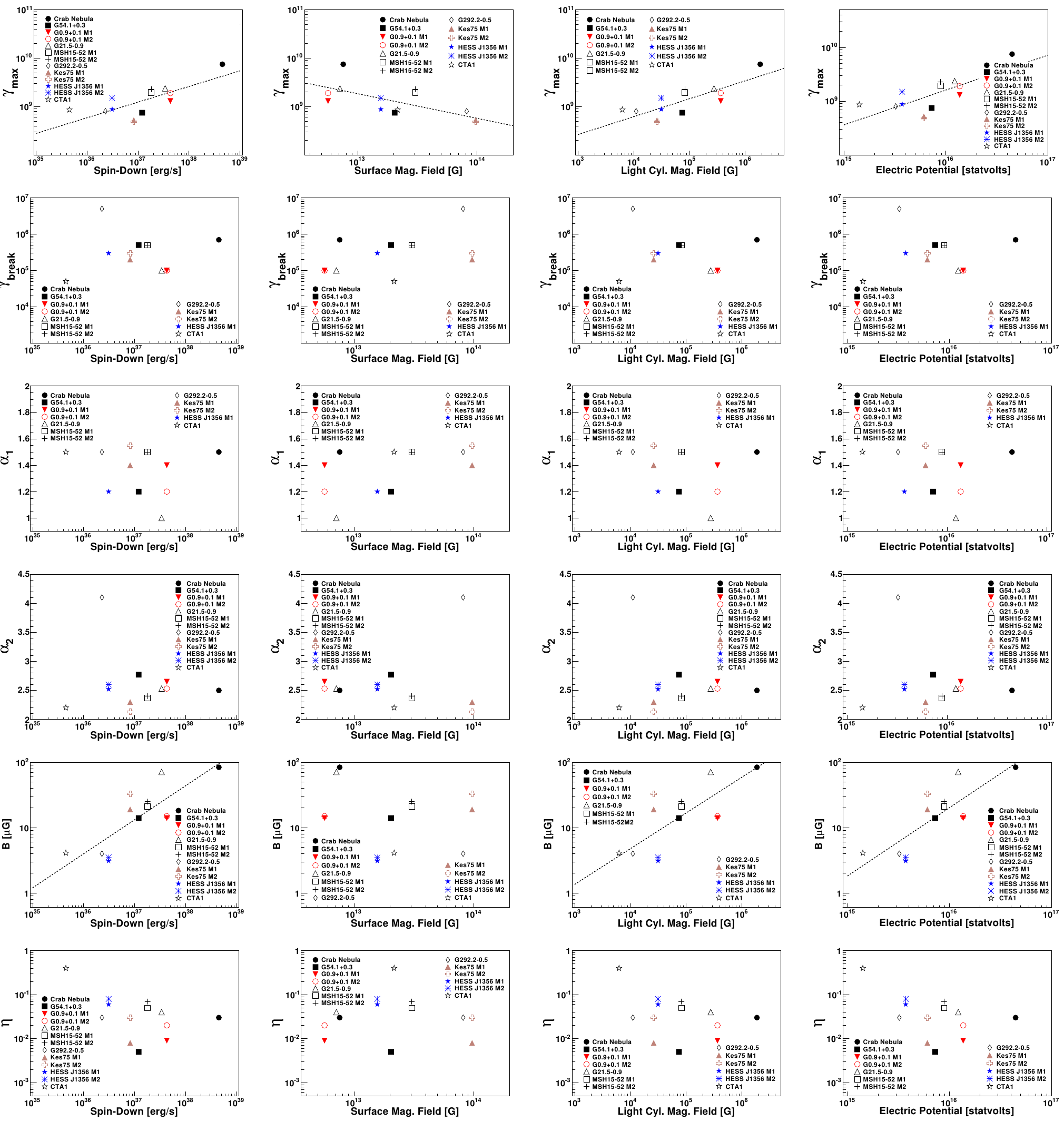}
\caption[PWNe properties as a function of pulsar properties]{PWNe properties in the y-axis of all plots as a function of pulsar properties in the
x-axis of all plots. The values of all magnitudes refer to the current time. From left to right, we plot the obtained values of $\gamma_{max}$,
$\gamma_b$, $\alpha_1$, $\alpha_2$, $B(t_{age})$, and $\eta$, as a function of (from top to bottom) spin-down, surface magnetic field, light
cylinder magnetic field, and potential.}
\label{corr2}
\end{figure}

The first four rows of figure \ref{corr2} plot the spectral parameters of the injected electrons as a function of the pulsar properties. We find
no correlation of the slopes $\alpha_1$ and $\alpha_2$, or $\gamma_b$ with the pulsar properties. In the case of $\alpha_2$, this is true even
disregarding the outlier, G292.2-0.5.

We do find a correlation of the the maximum Lorentz factor with the spin-down power (and thus the magnetic field at the light cylinder, and the
pulsar electric potential). The significance of the correlation surpasses 95\% CL. For the surface magnetic field, the significance we obtain is
the level of 94\%, and this is why we do not quote this fit in table \ref{fitspar} although we show it in the corresponding plot for visual
inspection. If this trend is considered, the $\gamma_{max}$ value is anti-correlated with the surface B field of the pulsar. On the contrary, the
larger is the spin-down power (or the magnetic field at the light cylinder or the electric potential), the larger is the Lorentz factor of
electrons in the nebulae. The maximum energy to which electrons are accelerated in the nebulae depends on the injected electrons at the bottom of
the wind zone. This correlation is to be expected via equation \ref{gmax2} and the fact that the dispersion that we find in the two other free
parameters appearing in there, $\varepsilon$ and $\eta$, is relatively not large for most of the sample.

Looked at the same age (at 3000 years) the $\gamma_{max}$--surface magnetic field anti-correlation is confirmed better than the 95\% level,
whereas the results for the other parameters are very similar.

The magnetic field in the nebulae is also correlated with the pulsar properties. Also here, the larger the spin-down power (or the magnetic field
at the light cylinder or the electric potential) the larger the nebula magnetic field, but this too can be ascribed to the way we define the
magnetic field in the model. The magnetization, however, is a free parameter in the fit, and with the confidence cut imposed, we see no relation
between $\eta$ and any of the pulsar characteristics. Take as an example the Crab Nebula: it is the pulsar with the largest spin-down power and
nebular B (today magnitudes) but its magnetization is similar to that of the remaining PWNe.

Taking the PWNe at the same age of 3000 years, we find that the PWN magnetic field correlation with the spin-down power (or the magnetic field at
the light cylinder or the electric potential) is lost. The nebular magnetic field and the spin-down power are both decreasing with the age of the
system, thus looking for its relationship at the same age increases the dispersion.

The multiplicity of the models studied is correlated (but only better than 94\% of CL) with the pulsar parameters, presenting positive
correlations with the spin-down power (or the magnetic field at the light cylinder or the electric potential) and negative correlation with the
surface magnetic field (albeit the scatter of the data points in this latter case seems to be worse). A caveat in this case is that the $\kappa$
parameter is already making use of the $P$ and $\dot{P}$ values to normalize the injected electrons (see equation \ref{multiplicity}), and in
fact, because of its definition $Q$ itself is obviously correlated with the spin-down.

\section{Concluding remarks}
\label{sec4.5}

The aim of this study was to present numerical models of the TeV-detected, young PWNe along more than 20 decades of frequencies; using a
radiatively complete, time-dependent numerical approach. For the first time, we have a coverage of many such PWNe analyzed under the same
framework, adopting similar assumptions, which allows for a more meaningful parameter comparison. Despite the caveats of the model used, we find
that one-zone, leptonic-only generated radiation provides a reasonably good fit to the multifrequency data for PWNe detected at TeV. Here we
summarize our findings.

\begin{itemize}
\item We favor a non-PWN origin for the radiation detected from HESS J1813--178. For the remaining 9 TeV sources studied, we find a plausible
PWN origin of the mutiwavelength emission. 
\item For all the TeV sources plausibly related with a PWN, only the Crab nebula is SSC dominated. All the remaining PWNe except for HESS
J1356--645 and CTA 1 are IC-FIR dominated. The dominance of the FIR contribution to IC is always significant.
\item The FIR energy densities that we found is needed to fit the PWN high-energy emission are generally larger than what is obtained from
GALPROP (usually by up to a factor of a few).
\item The efficiencies of emission are $\sim 10^{-6 \div 7}$ in radio, $\sim 10^{-2 \div 3}$ in X-rays, and $\sim 10^{-3 \div 4}$ in
$\gamma$-rays, with only one outlier in the sample presenting very low X-ray fluxes (G292.2--0.5). 
\item The electron population can be described by a broken power law in all cases. The parameters of the injection cluster in relatively narrow
ranges, especially, the break Lorenz factor, which is around $5 \times 10^5$. The high energy spectral slope is found to be in the range 2.2--2.8
(except for the steeper case of G292.2--0.5, which also present a higher energy break). The low energy part is instead much harder, with the low
energy index in the range 1.0--1.6.
\item All PWNe have large multiplicities, in general in excess of $10^5$. The population of low-energy electrons is large by number, and generate
a low medium energy per particle in the spectrum in all cases.
\item All the nebulae except CTA~1 have low values of magnetization, of only a few percent. CTA~1 presents the largest magnetization of our
sample, and reaches almost to equipartition. All the other PWNe are heavily particle dominated. This result is found to be stable against
uncertainties.
\item We do not find significant correlations between the efficiencies of emission at different frequencies and the magnetization, implying that
the specific environment and the injection effects play a dominant role in determining, e.g., the $\gamma$-ray luminosity. 
\item Comparing SEDs of the PWNe as observed today mixes pulsars of different spin-down power and age, and generates a variety of distributions. 
A normalized comparison of the SEDs (e.g., with the corresponding spin-down flux) at the same age significantly reduces the dispersion.
\item We do not find clear correlations between the pulsar's characteristic ages and the radio and X-ray luminosities. The gamma-ray luminosity
seems to be anti-correlated with the characteristic age. On the other hand, we do find correlations of the radio and X-ray (and at a slightly
lower confidence also $\gamma$-ray luminosities) with the spin-down, and an anti-correlation of the ratios of IC to synchrotron luminosities with
the spin-down.
\item The injection parameters do not appear to be correlated with the pulsar properties, except for the maximum Lorentz factor and the magnetic
field in the nebula  which are correlated with the spin-down power (or the magnetic field at the light cylinder or the electric potential), but
these cases can be ascribed to the model properties. 
\item We do not find a significant correlation of any PWN parameter with the surface magnetic field of the pulsars.
\end{itemize}

\chapter[Is there room for ultra-magnetized PWNe?]{Is there room for ultra-magnetized PWNe among those non-detected at TeV?}
\label{chap5}

\ifpdf
    \graphicspath{{Chapter5/Figs/Raster/}{Chapter5/Figs/PDF/}{Chapter5/Figs/}}
\else
    \graphicspath{{Chapter5/Figs/Vector/}{Chapter5/Figs/}}
\fi

The spectral energy distribution (SED) of the pulsar wind nebulae (PWNe) of the highest spin-down powered pulsars are varied. In particular,
luminous pulsars such as Crab ($\dot{E}=4.5 \times 10^{38}$ erg s$^{-1}$) and the Large Magellanic Cloud (LMC) J0537-6910 in N157B
($\dot{E}=4.9 \times 10^{38}$ erg s$^{-1}$) are TeV detected, as are others with spin-down power in the order of several 10$^{37}$ erg s$^{-1}$.
However, several PWNe with pulsars similarly luminous, are not. Why? Do they have significantly different interstellar environment, injection, or
nebular magnetization?

The X-ray luminosity efficiency of these high-spin down pulsars also presents a large range. A notable case is G76.9+1.0 for which the X-ray
efficiency is $L_X/\dot{E} \sim 2.4 \times 10^{-4}D^2_{10}$, where $D^2_{10}$ is the distance in units of 10 kpc
\citep{arzoumanian11}\footnote{The spin-down of the pulsar in G76.9+1.0 has been recently re-assessed due to a new measurement of the period (see
section \ref{sec5.1}), and while it is now lower than earlier claimed, it still qualifies as one the most energetic pulsars we know.}. This and
similar cases are challenging for PWNe spectral models since they imply an inefficient acceleration of high energy electrons in order to fit the
X-ray luminosity. For these cases, \citet{arzoumanian08} suggested that the pulsar wind has a high magnetization factor, speculating that because
particle-dominated winds are necessary for efficient conversion of wind to synchrotron power, PWNe with high magnetization would lead to dim
X-ray PWNe. This was confirmed by the phase space exploration of Crab-like nebulae done by \citet{torres13b}: the magnetization of the nebulae,
all other parameters being the same, can decide on TeV detectability. Thus, high-$\eta$ models point to an interesting alternative for the
interpretation of PWNe, which, despite their high spin-down, lack TeV emission and have weak X-ray counterparts. These PWNe would be different to
TeV detected ones. Except CTA 1, for which the magnetization reaches almost to equipartition, all TeV-PWNe with characteristic ages of 10 kyr or
less can be described with an spectral model with low $\eta$, and are thus strongly particle dominated \citep{torres14}.

An interesting case is that of G292.0+0.18, for which the central powering pulsar, J1124-5916, has essentially the same $P$, $\dot{P}$ (up to
three signicant decimal places) than J1930+1852, which powers G54.1+0.3. The distance for both nebulae is also similar ($\sim$6 kpc). Whereas the
latter is a TeV source, and modeled as particle dominated PWN (e.g., \citealt{tanaka11,torres14}), the former is not (at least at the level in
which it has been coveredin the Galactic Plane observations by H.E.S.S. \citep{carrigan13}. With the same spin-down power and located at a
similar Galactic distance, it seems that the flux at TeV energies depends on other factors such the environment (the FIR density, for instance)
or the nebula magnetization. Is then G292.0+0.18 simply like G54.1+0.3 but subject to a stronger magnetization?

\citet{tanaka13} have also analyzed several PWNe which have been undetected at TeV\footnote{For differences between their model and ours see the
discussion in \citet{martin12,torres14}. Their magnetic field evolution does not consider losses in magnetic energy due to expansion, and thus
their magnetization values are lower than ours typically by a factor 2-3.}. However, they assumed a fixed low magnetization ($3 \times 10^{-3}$)
compatible with usual particle dominated nebulae that have been detected at TeV to describe them.

In this chapter, we propose models for the non-detected PWNe at TeV with $\dot{E}>10^{37}$ erg s$^{-1}$ and we explore the phase space of PWNe
models also in magnetization, in order to distinguish whether there is preference for the existence of highly magnetized nebulae (or at least,
for nebula with magnetization close to equipartition) among those not yet seen at TeV.

This chapter is based on the work done in \citet{martin14b}.

\newpage

\section[Non-detected PWNe at TeV energies with high $\dot{E}$ PSRs]{Non-detected PWNe at TeV energies with high spin-down pulsars}
\label{sec5.1}

\begin{table}
\vspace{4cm}
\centering
\scriptsize
\vspace{0.2cm}
\caption[Model parameters for 3C 58, N157B, N158A, G76.9+1.0, G310--1.6 \& G292.0+1.8.]{Model parameters for 3C 58, N157B, N158A, G76.9+1.0,
G310--1.6 \& G292.0+1.8.}
\label{tabres}
\begin{tabular}{llllllll}
\hline
Magnitude & 3C 58 & N157B & N158A$^{\dag}$ & G76.9+1.0$^{\ddag}$ & G310.6--1.6 & G292.0+1.8\\
\hline
\hline
Pulsar magnitudes\\
\hline
$P$ (ms) & 65.7 & 16.12 & 50.50 & 48 & 31.18 & 135.48\\
$\dot{P}$ (s s$^{-1}$) & 1.93 $\times 10^{-13}$ & 5.18 $\times 10^{-14}$ & 4.79 $\times 10^{-13}$ & 8.64 $\times 10^{-14}$ & 3.89 $\times 10^{-14}$ & 7.53 $\times 10^{-13}$\\
$\tau_c$ (yr) & 5397 & 4936 & 1670 & 8970 & 12709 & 2854\\
$t_{age}$ (yr) & 2500 & 4600 & 760 & 5000 & 1100 & 2500\\
$L(t_{age})$ (erg s$^{-1}$) & 2.7 $\times 10^{37}$ & 4.9 $\times 10^{38}$ & 1.5 $\times 10^{38}$ & 2.96 $\times 10^{37}$ & 5.1 $\times 10^{37}$ & 1.2 $\times 10^{37}$\\
$L_0$ (erg s$^{-1}$) & 9.3 $\times 10^{37}$ & 1.1 $\times 10^{41}$ & 3.3 $\times 10^{38}$ & 1.5 $\times 10^{38}$ & 6.1 $\times 10^{37}$ & 7.8 $\times 10^{38}$\\
$n$ & 3 & 3 & 2.08 & 3 & 3 & 3\\
$\tau_0$ (yr) & 2897 & 336 & 2340 & 3970 & 11609 & 354\\
$d$ (kpc) & 3.2 & 48 & 49 & 10 & 7 & 6\\
$M_{ej}$ (M$_{\odot}$) & 8 & 20 & 25 & 20 & 9 & 9\\
$R_{PWN}$ (pc) & 3.7 & 10.6 & 0.7 & 4.7 & 1.3 & 3.5\\
\hline
Photon environment\\
\hline
$T_{FIR}^{(1)}$ (K) & 20 & 80 & 80 & 25 & 25 & 25\\
$w_{FIR}^{(1)}$ (eV cm$^{-3}$) & 5 (0.75) & 0.7 & 5 (0.2) & 0.13 & 0.62 & 0.42\\
$T_{FIR}^{(2)}$ (K) & - & 88 & - & - & - & -\\
$w_{FIR}^{(2)}$ (eV cm$^{-3}$) & - & 0.3 & - & - & - & -\\
$T_{NIR}$ (K) & - & - & - & 3200 & 3300 & 2800\\
$w_{NIR}$ (eV cm$^{-3}$) & - & - & - & 0.33 & 1.62 & 0.70\\
\hline
Injection parameters\\
\hline
$\gamma_{max}(t_{age})$ & 7.3 $\times 10^{9}$ & 3.8 $\times 10^{8}$ & 9.8 $\times 10^{8}$ & 5.7 $\times 10^{8}$ & 5.7 $\times 10^{8}$ & 2.4 $\times 10^{9}$\\
$\gamma_{min}$ & 1 & 1 & 1 & 1 & 1 & 1\\
$\gamma_b$ & 7.8 $\times 10^{4}$ & $10^{6}$ & 3 $\times 10^{7}$ & $10^3$ & 2 $\times 10^{6}$ & $10^{5}$\\
$\alpha_l$ & 1.05 & 1.5 & 1.8 & 1.5 & 1.5 & 1.5\\
$\alpha_h$ & 2.91 & 2.75 & 2.6 & 2.7 & 2.5 & 2.55\\
$\epsilon$ & 0.3 & 0.02 & 0.3 & 0.25 & 0.3 & 0.3\\
\hline
Magnetic field\\
\hline
$B(t_{age}) ({\mu}G)$ & 35 & 13 & 32 & 3.5 & 8.2 & 21\\
$\eta$ & 0.21 & 0.006 & 0.0007 & 0.0017 & 0.0007 & 0.05\\
\hline
\hline
\multicolumn{7}{l}{
\begin{minipage}{12cm}
Some alternative models are commented in the text.
\end{minipage}
}\\
\multicolumn{7}{l}{
\begin{minipage}{12cm}
$^{\dag}$The FIR energy density in the table is the one required for the PWN to be detected by H.E.S.S. (CTA) in 50 hours.
\end{minipage}
}\\
\multicolumn{7}{l}{
\begin{minipage}{12cm}
$^{\ddag}$These parameters correspond to model 1 in figure \ref{g76}, other models are described in the text.
\end{minipage}
}
\end{tabular}
\end{table}

\subsubsection{G310.6+1.6}

G310.6-1.6 (IGR J14003-6326) was discovered as a soft $\gamma$-ray source in a deep mosaic of the Circinus region done by INTEGRAL
\citep{keek06}. It was also observed in the Swift survey of INTEGRAL sources, but without conclusions about its origin \citep{malizia07}. With
Chandra observations, \citet{tomsick09} fitted the spectrum (0.3 and 10 keV) of the source with a power-law with a photon index of
$\Gamma=1.82 \pm 0.13$. \citet{renaud10} discovered 31.18 ms pulsations using RXTE, as well as reported the radio detection of PSR J1400-6325 and
its nebula. From the RXTE timing analysis, they obtained a period derivative for PSR J1400-6325 of $3.89 \times 10^{-14}$ s s$^{-1}$, which
implies an spin-down luminosity of $5.1 \times 10^{37}$ erg s$^{-1}$ and a characteristic age of 12.7 kyr. There are several estimations of the
PWN distance, covering a range between 6 and 10 kpc. We adopt the value of 7 kpc given in \citet{renaud10}.

\citet{renaud10} have studied the spectrum of G310.6-1.6, PSR J1400-6325 and its PWN from 0.8 to 100 keV. The spectrum is highly dominated by the
PWN and it is fitted with a broken power-law. The energy break is located at 6 keV and it is probably produced by the synchrotron cooling of the
particles. The spectral index for energies lower (higher) than the energy break is $1.90 \pm 0.10$ ($2.59 \pm 0.11$). The total flux for the PWN
at 20-100 keV is $5.3 \times 10^{-12}$ erg s$^{-1}$ cm$^{-2}$. The PWN flux in radio frequencies has also been measured, using data from the
Molonglo Galactic Plane Survey \citep{murphy07} at 843 MHz, as $217 \pm 9.4$ mJy, as well as from the Parkes-MIT-NRAO (PMN) survey
\citep{griffith93,condon93} at 4.85 GHz, as $113 \pm 10$ mJy. An upper limit of 0.6 mJy at 2.4 GHz is also established by the Parkes telescope
\citep{duncan95}. At TeV energies, G310.6-1.6 was observed by H.E.S.S. \citep{chaves08}, but only an upper limit of 4\% of the Crab Nebula was
established \citep{khelifi08}.

The spectrum of G310.6-1.6 PWN has been previously studied by \citet{tanaka13}. They assumed a magnetic fraction of 0.003, an age of 600 yr and a
distance of 7 kpc. For these parameters, they obtained an injection with a low (high) energy spectral index of 1.4 (3.0) with an energy break of
$\gamma_b=3 \times 10^6$ and a magnetic field of 17 $\mu$G. They assumed a 0.3 eV cm$^{-3}$ energy density for the FIR and NIR target photon
fields.

In our case, we firstly propose a low magnetized model (model 1), where we assume that the age of the PWN is 1.1 kyr, which is consistent with
the upper limit of 1.9 kyr established by \citet{renaud10}, but older than the one considered in \citet{tanaka13}. This assumption has been done
also taking into account the size of the nebula and a reasonable ejected mass of 9M$_\odot$ with a SN energy of 10$^{51}$ erg. \citet{renaud10}
proposed a subenergetic SN of $5 \times 10^{48}$ erg setting an ISM density of 0.01 cm$^{-3}$. This implies an ejected mass of 3M$_\odot$ to
explain the size of the nebula. This mass is very low for the ejecta of a star that explodes as a SN. We also prefer to consider the canonical
value for the SN explosion energy.

The target photon fields are obtained from those computed by GALPROP. The fitted black bodies of these photon fields have a temperature of 25 K
and 3300 K and an energy density of 0.63 eV cm$^{-3}$ and 1.62 eV cm$^{-3}$ for FIR and NIR, respectively. The obtained magnetic field is 8.2
$\mu$G and $\eta=0.0007$. The latter is the same value we find for the particle dominated models of N158A. The value of the magnetic field agrees
with the lower limit of 6 $\mu$G given by \citet{renaud10}. The intrinsic energy break of the injection in this model is located at
$\gamma=2 \times 10^6$ ($\sim$1 TeV). The injection index at low (high) energies is 1.5 (2.5).

\begin{figure}
\centering
\includegraphics[width=1.0\textwidth]{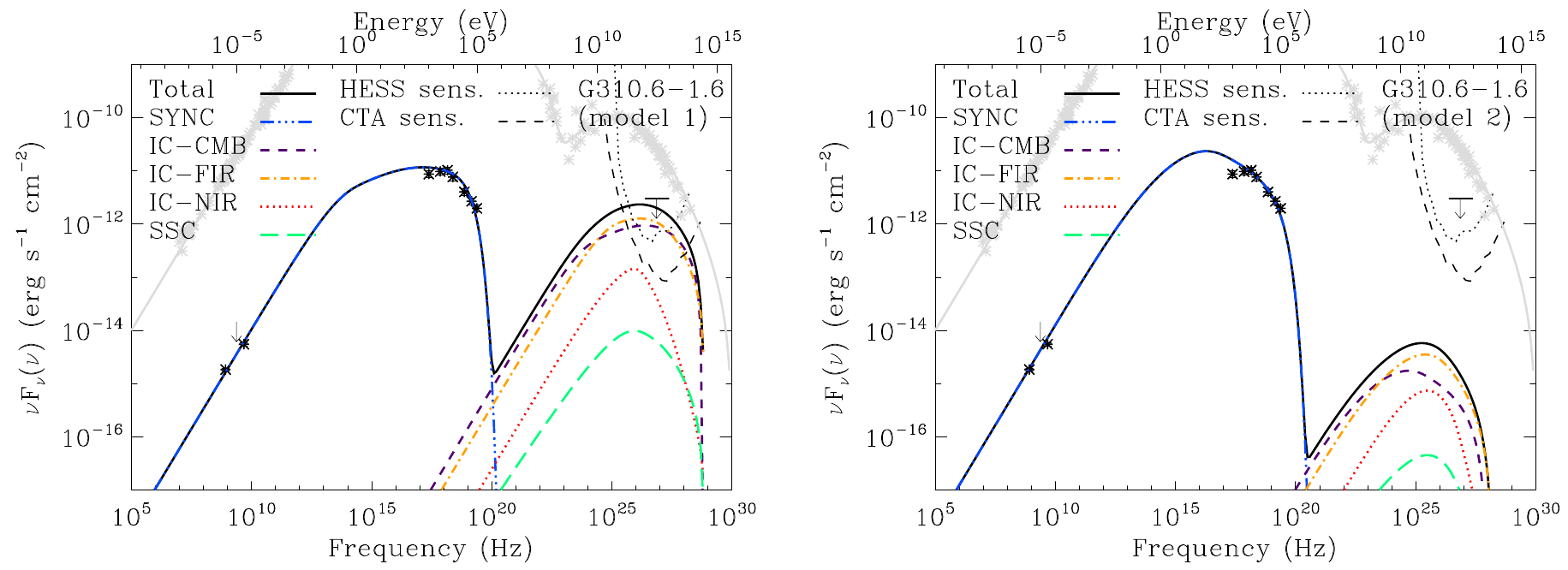}
\caption[Spectral fits for G310.6-1.6 PWN]{Spectral fits for G310.6-1.6 PWN. Data points are extracted from \protect\citet{renaud09}.}
\label{g310}
\end{figure}

The flux of G310.6-1.6 is a factor $\sim$4 over the H.E.S.S. sensitivity flux at 50 h of exposure time. Even with only the CMB contribution, this
sensitivity is surpassed by a factor $\sim$2. If this low-$\eta$ model is right, its detection is expected in a moderate exposure time with the
current Cherenkov telescopes.

\subsubsection{G76.9+1.0}

G76.9+1.0 hosts the pulsar PSR J2022+3842. The period and the period derivative of this pulsar was firstly determined by \citet{arzoumanian11}.
They obtained a period of 24 ms and a period derivative of $4.3 \times 10^{-14}$ s s$^{-1}$, which implies a spin-down luminosity of
$1.2 \times 10^{38}$ erg s$^{-1}$. This made PSR J2022+3842 the third pulsar with the highest spin-down known. In later observations with
XMM-Newton, \citet{arumugasamy13} discovered a factor 2 error in the determination of the pulsar period and period derivative. The new period is
then 48 ms and the spin-down luminosity reduces to $2.96 \times 10^{37}$ erg s$^{-1}$.

\begin{figure}
\centering
\includegraphics[width=0.6\textwidth]{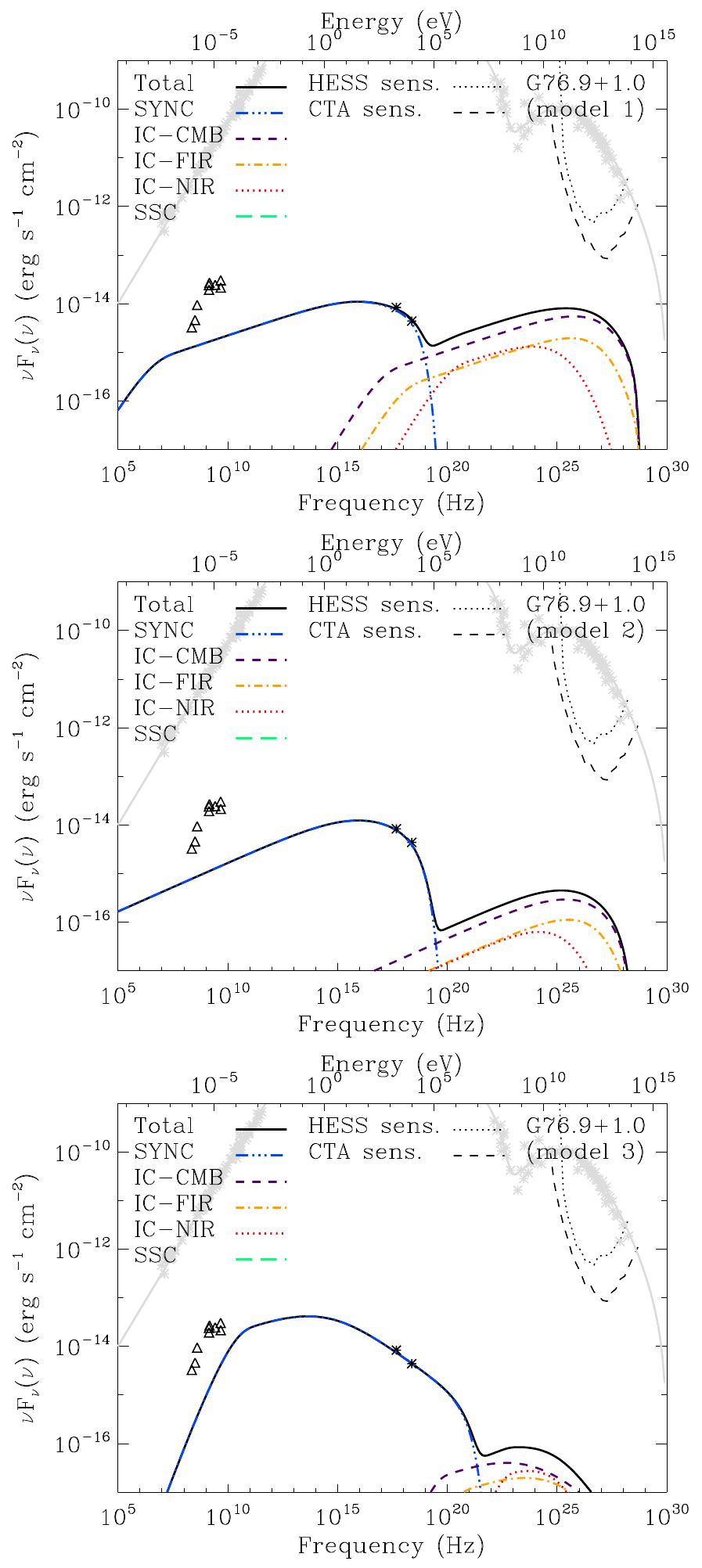}
\caption[Spectral fits for G76.9+1.0 PWN]{Spectral fits for G76.9+1.0 PWN (models 1 to 3, top to bottom). The triangle data points correspond to
the radio flux of the radio shell given in \protect\citet{landecker93}, here used as upper limits. The X-ray data is obtained from
\citet{arzoumanian11}.}
\label{g76}
\end{figure}

The remnant was observed in radio using the Very Large Array telescope (VLA) \citep{landecker93}. These authors assume a distance of 7 kpc, which
implies a size of $18 \times 24$ pc. The structure of the SNR is dominated by two lobes oriented in the north-south direction separated by 3
arcmin. The spectral index is $0.62 \pm 0.04$. They looked for an infrared counterpart using IRAS data but none was found. \citet{arzoumanian11}
observed PSR J2022+3842 in X-rays using Chandra, obtaining an absorbed X-ray flux (2–10 keV) of $5.3 \times 10^{-13}$ erg s$^{-1}$ cm$^{-2}$ and
detecting a very weak PWN with an absorbed flux of $4 \times 10^{-14}$ erg s$^{-1}$ cm$^{-2}$. In this case, there is no TeV detection either,
and we only have information about the spectrum in X-rays and upper limits in radio using the flux observed for the SNR radio shell.

We adopted an age of 5 kyr, which implies a reasonable ejected mass of 20M$_\odot$, also proposed by \citet{arzoumanian11}. There are no
estimations of the age of the remnant and of the ejected mass. \citet{arzoumanian11} has established an upper limit on the true age of the pulsar
depending on the braking index of $\sim$40 kyr, which is unconstraining.

We use the data simulated by GALPROP \citep{porter06} for the energy densities and temperatures for the FIR and NIR photon fields, essentially,
diluted black bodies with a temperature of 25 K and an energy density of 0.13 eV cm$^{-3}$ for the FIR field, and a temperature of 3200 K and an
energy density of 0.33 eV cm$^{-3}$ for the NIR field. As the PWN in X-rays is very diluted, its shell cannot be distinguished. For this reason,
to simulate the expansion of the nebula, we assumed a ballistic expansion of the SNR radio shell ($R_{SN}=V_0 t$) and compute the necessary
ejected mass. In this case, we obtain a value of $\sim$20M$_\odot$, which implies a radius of $\sim$6.3 pc. We assume a braking index of 3, which
implies a reasonable value for the initial period for PSR J2022+3842 of 32 ms.

In model 1, see table \ref{tabres}, we assume a broken power-law injection with a low-energy (high-energy) spectral index of 1.5 (2.7). The
resulting energy break is at $\gamma_b=10^3$. The magnetic field (3.5 $\mu$G) is close to the average ISM value. The magnetic fraction is 0.0017.
The low value of the injection energy break in this model argues for a possible simple power-law injection. This is assumed in model 2. In this
case, the spectral index is 2.6 and the magnetic field is 16.6 $\mu$G, with a magnetic fraction of 0.038. The resulting fits of these models are
shown in figure \ref{g76}.

The lack of observational data to put sufficient constrains to differenciate the models proposed. In any case, its detection at TeV energies
seems unexpected.

\subsubsection{3C 58}

Our study of 3C 58 is based on the work done in \citet{torres13a} where we have analyzed the detectability at TeV energies of this source. It was
suggested to be plausibly associated with the 831-year-old supernova SN 1181 (see, e.g. \citet{stephenson71,stephenson02}). However, recent
investigations of dynamical models for the PWN \citep{chevalier05} and the velocities of both the expansion rate of the radio nebula
\citep{bietenholz06} and the optical knots \citep{fesen08} imply an age of several thousand years. This is closer to the characteristic age of
the pulsar in the nebula, PSR J0205+6449 \citet{murray02}. A recent rekindling of the low age has been put forward by \citet{kothes10}, based on
a new estimation of the nebula distance.

3C 58 has a flat-radio spectrum with a spectral break between the radio and IR bands \citep{green92}. X-ray observations reveal a non-thermal
spectrum that varies with radius, becoming steeper toward the outer regions \citep{slane04b}. PSR J0205+6449 is one of the most energetic pulsars
known in the Galaxy. The pulsar powers a faint jet and is surrounded by a toroidal structure apparently associated with flows downstream of the
pulsar wind termination shock \citep{slane04b}. The shell of the thermal X-ray emission that was seen in 3C 58, e.g., by \citet{gotthelf07}, is
smaller than the maximum extent of the PWN. Therefore, this emission is likely associated with supernova ejecta swept up by the expanding PWN
rather than the original forward shock from the supernova. The pulsar has been recently detected at high-energy gamma-rays by Fermi, but only
upper limits were imposed for the nebula emission \citep{abdo09a}. Similarly, Whipple \citep{hall01}, and both MAGIC \citep{anderhub10} and
VERITAS \citep{konopelko08} observed the nebula, but only upper limits were imposed at TeV energies.

3C 58 and the Crab Nebula differ significantly both in luminosity and size. 3C 58 is larger, but less luminous, e.g., its TeV luminosity is at
least $\sim$100 \citep{anderhub10}, its X-ray luminosity is $\sim$2000 \citep{torii00}, and its radio luminosity is $\sim$10 times smaller than
Crab. The similarity in fact comes from morphology (e.g., \citealt{slane04b}).

PWN models for 3C 58 have been presented before by a few authors, e.g., \citet{bednarek03,bednarek05,bucciantini11}, with some disparity in the
results, particularly at the high-energy end of the spectrum. These studies use different assumptions for the primary particles assumed to
populate the wind, and differ also in the treatment of the radiative physics. With 3C 58 being a candidate for observations in the current or
forthcoming generation of Cherenkov telescopes, it is interesting to study under what conditions 3C 58 is observable at high energies.

We adopt a mass of the ejecta comparable to that of Crab, motivated by estimates of the total mass of the precursor \citep{rudie07}. Not all
estimates for the age and distance of 3C 58 (see \citet{fesen08} for a summary) can be consistently encompassed within the expansion model of the
nebula. Consider first an age of $\sim$5000 years or more (as in \citet{murray02,bietenholz06,slane02b}) and a distance of 3.2 kpc (as in
\citet{roberts93}). Following \citet{fesen08} for the angular size of the nebula, its physical size at that distance is $6 \times 9.5$ pc. We can
extrapolate these magnitudes to the spherical case by matching the projected area of the nebula to that of a circle, and so we obtain a radius of
3.7 pc. Using the observational parameters in table \ref{tabres} and using equation \ref{pwnrad}, we would, however, obtain a physical size of
about 18 pc, a result that worsens for larger ages. On the contrary, if we assume the observed size and compute the ejected mass needed to have a
radius of $\sim$3.7 pc, we would find an inconsistently large value. The scenario where we change the initial spin-down power is similarly
problematic, since it would be impossible to reach the current $\dot{E}$, being the initial one smaller than the current power. Such innuendos
are not solved by assuming a different value of braking index and are also stable (producing sizes in excess of 10 pc) for up to one order of
magnitude variations in the moment of inertia $I$. If 3C 58 is closer to Earth, the mismatch would be larger, given that the physical size of the
nebula would be smaller than at 3.2 kpc.

Consider next an 830-year-old nebula (as in \citealt{stephenson71,stephenson02}) and a distance of 2 kpc (as in \citealt{kothes10}). Geometry
implies that the physical size of the nebula should be around 2.3 pc, but using the observed, derived, and assumed parameters of table
\ref{tabres} we obtain a size of 0.8 pc, a factor of three smaller. This result, similar to the larger age case above, is stable against changes
in $n$, $I$, or other parameters. The only way to recover a larger nebula would be to assume a mass of the ejecta of the order of 1 M$_\odot$,
but this would be inconsistent with estimates based on the observed filamentary knots (which already account for a large fraction of 1 M$_\odot$)
or with evolutionary models (e.g., \citet{rudie07,fesen08,bocchino01,slane04b}). A larger distance to 3C 58 would imply a larger physical size of
the nebula, making the mismatch more severe.

We are a priori favorable to the case of an age of 2500 years and a distance of 3.2 kpc. For this set of parameters, the size of the nebula can
be easily accommodated within the model described in the previous section. Variations of the parameters in a reasonable manner maintain this
conclusion stable. In addition, the shock velocity agrees with estimates coming from the thermal X-ray emission (e.g., \citealt{bocchino01}) and
at the same time the swept-up mass resulting from these model parameters $M_{sw}=M_{ej}(R_{PWN}/V_0 t)^3 \sim 0.26$M$_\odot$ is in line with the
measurements of the mass contained in filaments \citep{bocchino01,slane04b}; i.e., we assume that the filamentary structure roughly corresponds
to the swept-up shell of the ejecta. Similar conclusions have been reached by \citet{chevalier04,chevalier05,bucciantini11} with others
arguments.

\begin{figure}
\centering
\includegraphics[width=0.6\textwidth]{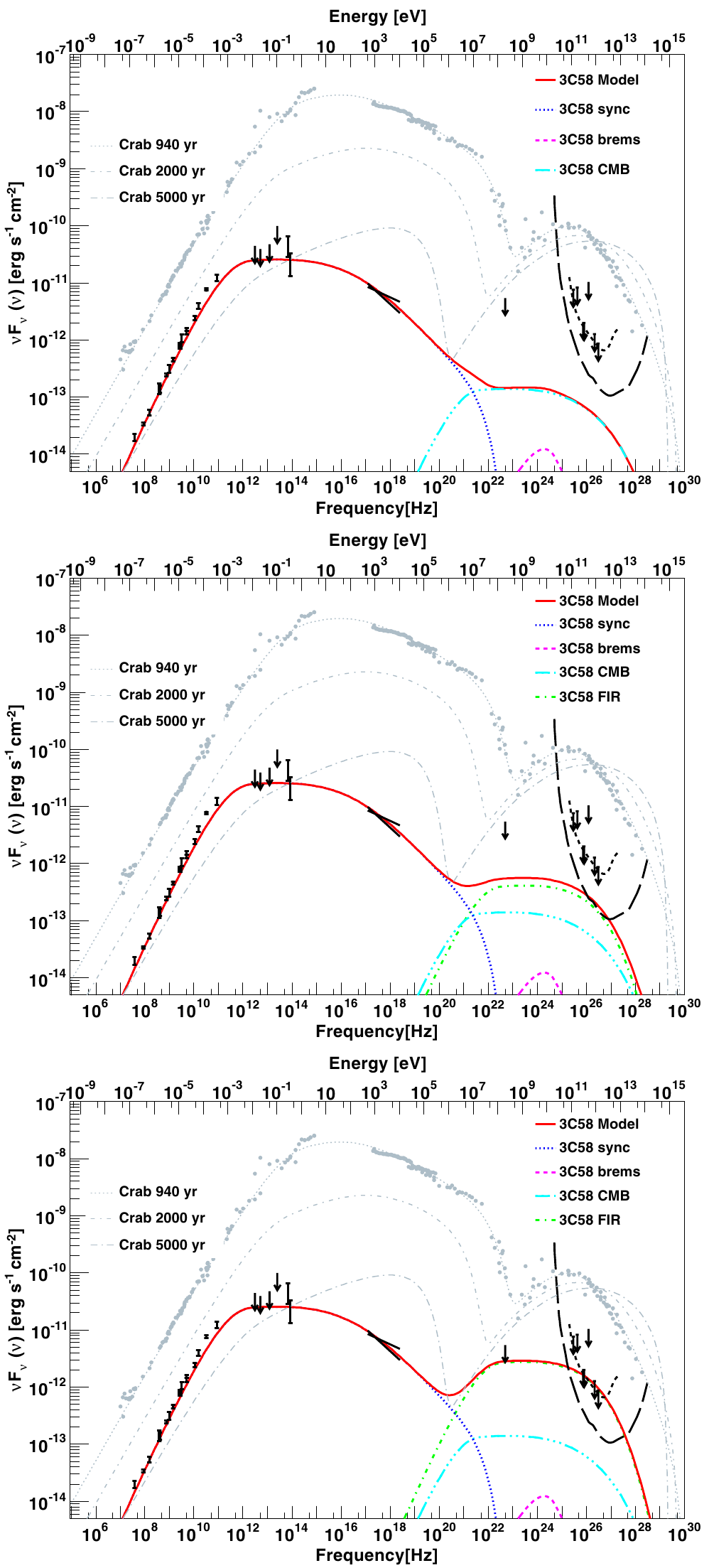}
\caption[Multi-frequency models of the PWN 3C 58 under different assumptions for the background photon fields]{Multi-frequency models of the PWN
3C 58 under different assumptions for the background photon fields. Top: CMB only. Middle: IR energy density up to the level where the emission
of 3C 58 reaches the CTA sensitivity. Bottom: same as the middle panel, for the MAGIC sensitivity. The SSC contribution is not visible in this
scale. Observational data comse from \protect\citet{green86,morsi87,salter89} (radio); \protect\citet{green94,slane08} (infrared);
\protect\citet{torii00} (X-rays); \protect\citet{abdo09a} (GeV); and \protect\citet{hall01,konopelko08,anderhub10} (TeV).}
\label{3c58}
\end{figure}

Multi-frequency results of our model for a 3C 58 age of 2500 years located at 3.2 kpc are shown in figure \ref{3c58}. We also show for comparison
the results corresponding to the Crab Nebula at different ages, i.e., 940 (where the data points are fit with a corresponding model; see e.g.
\citealt{martin12}), 2000, and 5000 years. The differences in the data of Crab and that of 3C 58 are evident. Figure \ref{3c58} shows results for
three different assumptions regarding the dominance of the IC contribution. In the first panel, only the CMB is assumed as background for IC. The
resulting parameters are given in table \ref{tabres}, and they show that a broken power law fits the current radio to X-ray data. The magnetic
field of the nebula results in 35 $\mu$G. The contribution of SSC to IC is sub-dominant to Bremsstrahlung under the assumption of a low medium
density of 0.1 cm$^{-3}$. As there is no clearly detected supernova remnant shell, we cannot reliably estimate the interstellar medium density.
It is impressive how low the prediction in the GeV and TeV regimes is, far beyond the reach of Fermi-LAT and Cherenkov telescopes.

Taking into account that the IR/FIR background is uncertain, we consider two limiting situations by exploring how large the energy density at 20
K should be for its corresponding IC contribution to reach the sensitivity of CTA; \citep{actis11} and of the currently operating MAGIC II
\citep{aleksic12}. The energy densities used in each panel of figure \ref{3c58} are shown in table \ref{tabres}. The FIR energy densities
range from 3 to 20 times that of the CMB. These can be compared with the typical background at that frequency from, e.g., GALPROP models to see
that the case where 3C 58 appears as an in-principle-observable nebula in MAGIC II or VERITAS would be highly unexpected. In both of these cases,
still, the GeV emission would be far below the Fermi-LAT upper limit.

We have explored models having different braking indices (keeping age and distance fixed at 2500 years and 3.2 kpc, respectively, as the test
bed). For these models, the overall quality of the fit is unchanged; some of the fit parameters are slightly modified, however. Lower values of
$n$ imply changes in the initial spin-down power (from 9.3, to 7.3, to $5.9 \times 10^{37}$ erg s$^{-1}$ for $n$=3, 2.5, and 2, respectively),
initial spin-down age (from 2897, to 4696 to 8294 years), and the PWN radius today (from 3.7 to 3.5 to 3.4 pc). The magnetic fraction changes
from 0.23, for $n$=2, to 0.22, for $n$=2.5, to 0.21, for $n$=3. The magnetic field today has a value between 35 and 37 $\mu$G in all these cases.
Letting the parameter vary allows us to explore the ability of the fit to adapt to the X-ray-measured spectra better. But again we find a very
small parameter dependence.

The total energetics is conserved in our model, since particles have a fraction (1-$\eta$), and the magnetic field a fraction $\eta$, of the
total power. 3C 58 features a 21\% magnetic fraction in our model, significantly higher than the one we obtain for Crab ($\sim$3\%). It is still
a particle-dominated nebula. These results differ from those in \citet{bucciantini11} work, where the total energy was not conserved by
$\sim$30\%, leading to a nebula in equipartition, and where it was said that models with energy conservation would always underpredict the radio
flux.

Note the two contiguous IR measurements around 10$^{14}$ GHz. Because of their location, it is nearly impossible to fit them both at once.
Indeed, there appears to be two subsequent steepening of the spectrum, one just beyond the radio band and one additional in the IR band (see
\citealt{slane08}). We have explored an injection containing another break, but results do not improve the fit significantly. We also tried to
improve the fit using an injection model based on the particle in cell (PIC) simulations done by \citet{spitkovsky08} keeping the ratios of the
additional parameters as in \citet{holler12}. These are not devoid of significant extrapolations and include a number of additional parameters
for which we have no constraints. We used a value for the energy break of $\gamma_b=2 \times 10^4$ to have an acceptable fit of the radio points,
but it is not possible to fit the IR and X-ray points correctly, even changing the ratios of the parameters. In all these cases for the
injection, the fits are of similar (two breaks power law) or lower (PIC motivated) quality than the ones presented above; and in none, the
high-energy yield is significantly affected.

\begin{table}
\scriptsize
\centering
\caption[Parameters proposed for 3C58 with the TeV flux given by MAGIC]{New parameters proposed for 3C58 with the TeV flux given by MAGIC.}
\vspace{0.2cm}
\label{new3c58}
\begin{tabular}{lll}
\hline
Magnitude & Model 1 & Model 2\\
\hline
$t_{age}$ (yr) & 2500 & 2000\\
$P(t_{age})$ (ms) & 65.7 & $\cdots$\\
$\dot{P}(t_{age})$ (s s$^{-1}$) & 1.94 $\times 10^{-13}$ & $\cdots$\\
$\tau_c$ (yr) & 5378 & $\cdots$\\
$L(t_{age})$ (erg s$^{-1}$) & $2.7 \times 10^{37}$ & $\cdots$\\
$n$ & 3 & $\cdots$\\
$d$ (kpc) & 3.2 & 2\\
\hline
$\gamma_{min}$ & 1 & $\cdots$\\
$\gamma_{max}$ & $7.3 \times 10^9$ & $2.5 \times 10^9$\\
$\gamma_b$ & $8 \times 10^4$ & $\cdots$\\
$\alpha_l$ & 1.05 & $\cdots$\\
$\alpha_h$ & 2.91 & 3\\
\hline
$L_0$ (erg s$^{-1}$) & $9.4 \times 10^{37}$ & $6.8 \times 10^{37}$\\
$\tau_0$ (yr) & 2878 & 3378\\
$B(t_{age})$ ($\mu$G) & 36 & 22\\
$\eta$ & 0.21 & 0.025\\
$\varepsilon$ & 0.3 & $\cdots$\\
$R_{PWN}(t_{age})$ (pc) & 3.7 & 2.3\\
\hline
$T_{CMB}$ (K) & 2.73 & $\cdots$\\
$w_{CMB}$ (eV/cm$^3$) & 0.26 & $\cdots$\\
$T_{FIR}$ (K) & 25 & $\cdots$\\
$w_{FIR}$ (eV/cm$^3$) & 0.8 & 0.2\\
$T_{NIR}$ (K) & 2800 & $\cdots$\\
$w_{NIR}$ (eV/cm$^3$) & 4 & 1.8\\
$n_H$ (cm$^{-3}$) & 0.1 & $\cdots$\\
$E_0$ (erg) & 10$^{51}$ & $\cdots$\\
$M_{ej}$ (M$_\odot$) & 6 & 9\\
\hline
\hline
\end{tabular}
\end{table}

\begin{figure}
\centering
\includegraphics[width=1.0\textwidth]{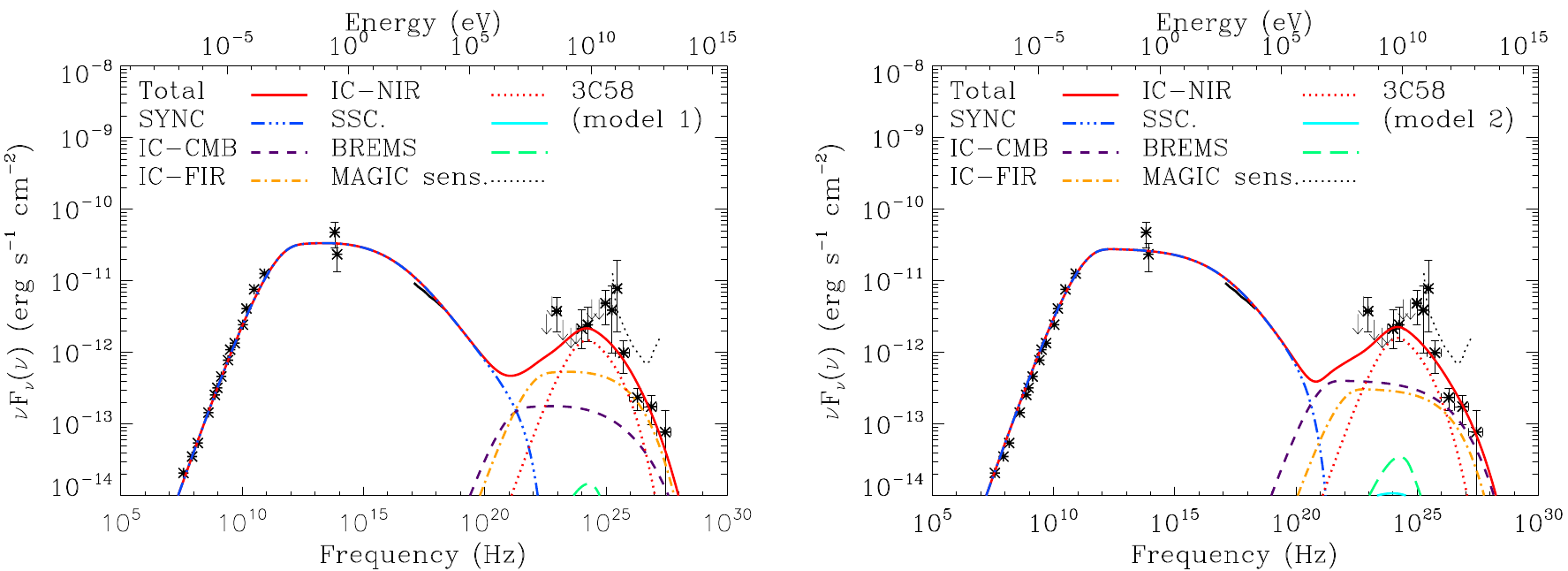}
\caption[New models for 3C 58 using the TeV flux detection by MAGIC]{New models for 3C 58 using the TeV flux detection by MAGIC. We include also
the flux points and upper limits given by Fermi in \protect\citet{abdo13}.}
\label{3c58_new}
\end{figure}

Recently, \citet{aleksic14} have claimed the detection at TeV of 3C 58 by the MAGIC telescopes. The detection has a significance of 5.7$\sigma$
and the flux between 400 GeV and 10 TeV is well described by a power law such that $F(E)=F_0 (E/1$ TeV$)^{-\Gamma}$ with $F_0=2 \times 10^{-13}$
TeV$^{-1}$ cm$^{-2}$ s$^{-1}$ and $\Gamma=2.4$. Considering this result, we propose two new models for 3C 58 in table \ref{new3c58} and show the
fits in figure \ref{3c58_new}. In model 1, we assume the same values for distance and age as in table \ref{tabres}. Note that the ejected mass
changes, because in these new fits we are taking into account spin-down luminosity variation during the expansion. We get similar values as those
obtained in \citet{torres13a}, but the target photon field energy densities are readjusted to the new TeV flux. We have included a NIR
contribution with an energy density of 4 eV cm$^{-3}$ and the FIR energy density is 0.8 eV cm$^{-3}$, close to the energy density limit to be
detected with CTA in \citet{torres13a} (0.75 eV cm$^{-3}$). In model 2, we use the distance of 2 kpc given by \citet{kothes10}, but assuming an
age of 2 kyr. In this case, the physical radius of the nebula is 2.3 pc. The magnetic field obtained to fit the synchrotron spectrum is 22 $\mu$G
and the magnetic fraction is 0.025. This latter value for the magnetic fraction fits better with the trend of low-magnetized nebula observed in
\cite{torres14}. The energy densities for the FIR and NIR photon fields are lower than in model 1 (0.2 and 1.8 eV cm$^{-3}$, respectively). We
have assumed the fiducial value of 0.1 cm$^{-3}$ for the ISM density to compute the Bremsstrahlung contribution, but it is neglectable in both
models.

\subsubsection{G292.2+1.8}

As stated in the introduction, the pulsars related with G54.1+0.3 and G292.0+0.18 both have a period of $\sim$135 ms, period derivative of
$\sim 7.5 \times 10^{-13}$, a spin-down power of $1.2 \times 10^{37}$ erg s$^{-1}$, a characteristic age of $\sim$2900 years and a distance of
$\sim$6 kpc. For both pulsars, the braking index is unknown.

The radius of G292.0+0.18 is based on the SNR size of 8' diameter \citep{gaensler03}, which means a physical radius of 3.5 pc. The distance
estimate is based on the HI absorption profile given by \citet{winkler09}. Based on measurement of the transverse motions of the filaments of the
SNR and assuming that the shell is expanding with transverse expansion velocity, \citet{winkler09} estimated an age between 3000 and 3400 years,
concurring with \citet{gaensler03}. The ejected mass of the SN explosion was estimated as $\sim$6 M$_\odot$ \citep{gaensler03}.

Radio observations for the nebula were obtained from the work of \citet{gaensler03}. The flux of the nebula in X-rays was measured by Chandra
\citep{hughes01}. The photon index of the X-ray spectra, as it is suggested in \citet{hughes01}, is considered the same as that of the pulsar. At
GeV energies, we only have upper limits from Fermi-LAT \citep{ackermann11}. Optical and near infrared observations were obtained for the torus of
the nebula, by \citet{zharikov08} and \citet{zharikov13}, respectively, but these are not considered in our fits, since do not include the entire
system. The background energy densities are unknown. We assume those given by GALPROP, for which the equivalent temperatures and densities of the
representing blackbodies are $T_{FIR}=25$ K, $w_{FIR}=0.42$ eV cm$^{-3}$, and $T_{NIR}=2800$ K, $w_{NIR}=0.70$ eV cm$^{-3}$.

\begin{figure}
\centering
\includegraphics[width=1.0\textwidth]{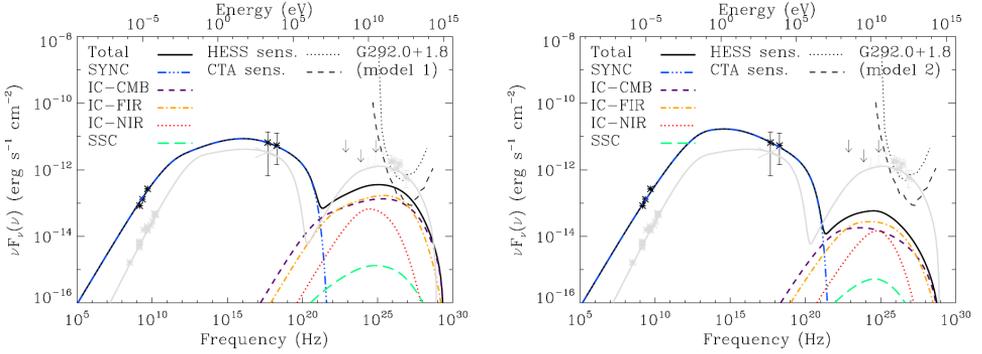}
\caption[Spectral fits for G292.0+1.8 PWN]{Spectral fits for G292.0+1.8 PWN. Data points for G292.0+1.8 are obtained from
\protect\citet{gaensler03} (radio), \protect\citet{hughes01} (X-rays) and \protect\citet{ackermann11} (Fermi upper limits). In grey, we show the
model and data for G54.1+0.3 extracted from \citet{torres14} (also shown in chapter \ref{chap4}).}
\label{g2920}
\end{figure}

Figure \ref{g2920} shows two models that fit the radio and the X-ray data for this nebula. In both cases the age of the system is 2500 years, and
the ejected mass is 9 M$_\odot$. In model 1 (see figure \ref{g2920}), we consider a low magnetic fraction model with $\eta$=0.05, which is 10
times larger than the magnetic fraction obtained in our model for G54.1+0.3 in \citet{torres14}. This model predicts that the nebula will be seen
by CTA, and it would reach H.E.S.S. sensitivity if the FIR energy density reaches 2 eV cm$^{-3}$. The TeV flux would be only a factor $\sim$2
lower than the H.E.S.S. sensitivity limit in 50 h exposure time. We obtain a magnetic field of 21 $\mu$G with an injection intrinsic break of
$\gamma_b=10^5$ ($\sim$51 GeV), with a low (high) energy index of 1.5 (2.55). These parameters differ from the ones obtained for G54.1+0.3 in
\citet{torres14} ($B$=14$\mu$G, $\gamma_b=5 \times 10^5$, $\alpha_1=1.2$, $\alpha_2=2.8$). With this model, the difference in the magnetic
fraction, the energy densities of the IC target photon fields and the age of the system explain why we observe G54.1+0.3 and not G292.0+1.8, even
when both are particle dominated.

\section[Non-Galactic PWNe at TeV energies]{Non-Galactic PWNe at TeV energies}
\label{sec5.2}

\subsubsection{N157B}

N157B is located in the LMC and it was the first extragalactic PWN detected in gamma rays \citep{abramowski12}. Its pulsar, PSR J0537-6910, has a
spin-down power of $4.9 \times 10^{38}$ erg s$^{-1}$ \citep{manchester05}. \citet{lazendic00} did radio observations of this PWN using the
Australia Telescope Compact Array (ATCA), obtaining a spectral index of $\sim$0.19. \citet{micelotta09} did infrared observations using the
Spitzer telescope but reported no infrared counterpart (no bright SNR). Studying the gas and dust properties of the vicinity, they deduced that
the mass of the progenitor star should not be higher than 25M$_\odot$. In X-rays, N157B was observed with ASCA and ROSAT \citep{wang98}, and4
\citep{wang01} detected the PWN with Chandra. \citet{chen06} analyzed the spectrum of N157B and the pulsar PSR J0537-6910 in X-rays. The spectrum
of the entire remnant is fitted with a dominant non-thermal component (a power-law with a spectral index of 2.29 and an unabsorbed flux of
$1.4 \times 10^{-11}$ erg s$^{-1}$ cm$^{-2}$) and a thermal component (a NEI model with a temperature of 0.72 keV and an unabsorbed flux of
$7 \times 10^{-12}$ erg s$^{-1}$ cm$^{-2}$).

In our study, we use the estimated distance of 48 kpc, see \citet{abramowski12}. There are two gas bubbles in the vicinity of N157B which
contribute to the far-infrared (FIR) photon fields: 30 Doradus complex and the OB association LH99. From the infrared observations done by
\citet{indebetouw09}, \citet{abramowski12} modelled the infrared emission as a black body with energy density of 8.9 eV cm$^{-3}$ and a
temperature of 88 K for the LH99 association, and 2.7 eV cm$^{-3}$ and 80 K for 30 Doradus. They consider these values as an upper limit since
the unprojected distance between these objects is unknown.

\begin{figure}
\centering
\includegraphics[width=1.0\textwidth]{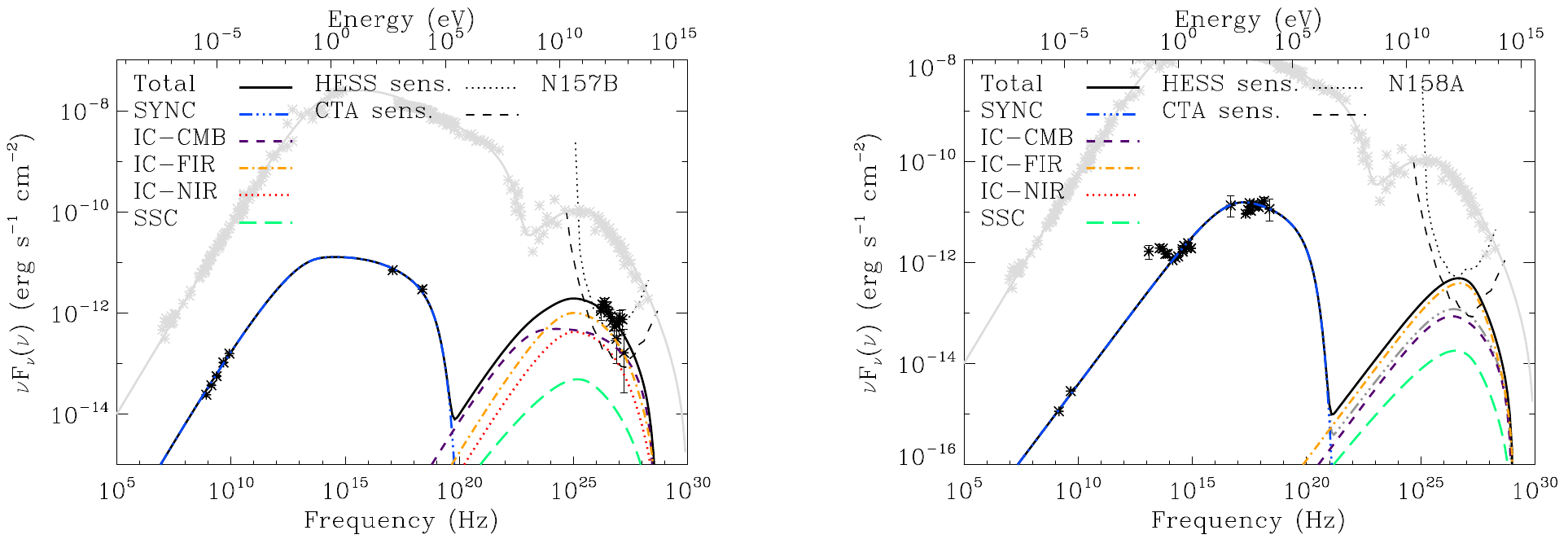}
\caption[PWNe in the LMC]{PWNe in the LMC. Right panel: Spectral fit for the N157B PWN. The fluxes and the fit of the Crab Nebula are overplotted
in grey for comparison. We plot also the sensitivity curves of H.E.S.S. and CTA for an exposure time of 50 hours. The data points are obtained
from: \protect\citet{lazendic00} (radio), \protect\citet{chen06} (X-rays), \protect\citet{abramowski12} (VHE). Left panel: Spectral fit for the
N158A PWN to reach H.E.S.S. (in solid black) and CTA (in triple-dot dashed grey) sensitivities. The data points are obtained from:
\protect\citet{manchester93b} (radio), \protect\citet{mignani12} (infrared \& optical), \protect\citet{kaaret01,campana08} (X-rays).}
\label{lmcpwn}
\end{figure}

Figure \ref{lmcpwn} (left panel) shows the fit we obtain for N157B. We assume the radius for the PWN given by \citet{lazendic00}, i.e., 10.6 pc
for a distance of 48 kpc. The PWN shell is not very well defined and some small contribution of the SNR could be included. In this first model,
we assumed an age of 4600 yr, which is consistent with the Sedov age of the SNR given by \citet{wang98} ($\sim$5 kyr) and an ejected mass of
20M$_\odot$, corresponding to the lower limit in the ejected mass given by \citet{chen06}. The electron injection has a low (high) energy index
of 1.5 (2.75) and the energy break is located at $\gamma=10^6$ ($\sim$511 GeV). From the synchrotron part of the spectrum, we inferred a magnetic
field of 13 $\mu$G and a magnetic fraction of 0.006. The energy density of the target photon fields, enhanced due to the near presence of LH99
and 30 Doradus, results in our fits much below the upper limits given by \citet{abramowski12}, i.e., 0.7 and 0.3 eV cm$^{-3}$, respectively.

If instead we assume the energy densities given by \citet{abramowski12}, we need to consider a lower age of 2.5 kyr to fit the TeV data.
Considering the lower limit on the ejected given by \citet{chen06}, then the radius decreases until 3.7 pc. Regarding the synchrotron spectrum,
the magnetic field reaches 35$\mu$G and $\mu$=0.01. The intrinsic energy break changes to $\gamma_b=2 \times 10^5$ ($\sim$102 GeV) and the
injection slopes change slightly ($\alpha_1$=1.5,$\alpha_2$=2.6). The value obtained for the radius in the latter model is only $\sim$50\% higher
than the radius observed in X-rays. This difference is small in comparison with other cases. For example, for the Crab nebula, we see that the
radius in the radio band is $\sim$2 pc and in X-rays $\sim$0.6 pc. As the shell is not well defined, the radius measured by \citet{lazendic00}
could include parts of the remnant, but the relation between the PWN radius in X-rays and the radius in the radio band seems to be more similar
to the Crab nebula case. \citet{vanderswaluw04a} suggested that N157B PWN could be interacting with the reverse shock of the SNR in a very
initial phase, explaining its elongated morphology. In any case, we find that N157B is a luminous particle dominated nebula.

\subsubsection{N158A}

N158A, known as the Crab twin, is also located in the LMC but has not been detected at TeV yet. This PWN is powered by the pulsar PSR B0540-69,
which has been observed in radio, infrared, optical and X-ray bands. The period of this pulsar is 50.5 ms \citep{seward84b} and the period
derivative is $4.7 \times 10^{-13}$s s$^{-1}$ \citep{livingstone05b}. The resulting spin-down luminosity is then $1.5 \times 10^{38}$ erg
s$^{-1}$. The diameter of N158A is 1.4 pc, as obtained from radio observations \citep{manchester93b}. The distance to PSR B0540-69 has been
estimated as $\sim$49 kpc \citep{seward84b,taylor93a,slowikowska07}. An age of 760 yr is deduced through measurements of the expansion velocity
of the SNR shell in the optical spectral range \citep{reynolds85,kirshner89}. There is no observational measurement of the ejected mass in N158A,
and we have left this parameter free in our model. The resulting ejected mass in our fits is 25M$_\odot$. According to \citet{heger03}, this mass
is at the limit for neutron star creation, which can grow with the quantity of helium in the core of the star and the energy of the supernova
explosion. In the infrared, \citet{caraveo92} did a high-resolution observation of N158A using the European Southern Observatory New Technology
Telescope (ESO-NTT) and concluded that the progenitor of the SNR could have belonged to the same generation of young stars in 30 Doradus
\citep{caraveo92,kirshner89}. \citet{williams08} did not find evidence of infrared emission from the SNR, but they inferred a mass of
20–25M$_\odot$ for the progenitor star. PSR B0540-69 is one of the few pulsars with optical pulsations and polarized emission. Its optical
spectrum is well fitted by a power-law, but joining it with the X-ray spectrum, a double break is required \citep{mignani12}. The braking index
for PSR B0540-69 is 2.08 \citep{kaaret01}. A high-resolution X-ray observation was done with Chandra \citep{gotthelf00a,kaaret01} and there is
also a compilation of the observations done with RXTE, Swift and INTEGRAL in the work by \citet{campana08}. The flux obtained for the PWN is
$\sim 8 \times 10^{-11}$ erg s$^{-1}$ cm$^{-2}$. There is no detection of the PWN at VHE.

For N158A, the injection spectrum resulting from our fit is a broken power-law with break at a large energy $\gamma=3 \times 10^7$ ($\sim$15.3
TeV) and a low (high) energy spectral index of 1.8 (2.6). The synchrotron component is fitted with a magnetic field of 32 $\mu$G. The magnetic
fraction in this case is low ($\eta=0.0007$). Due to the lack of information on the FIR and NIR fields, we assume a FIR field with a temperature
of 80 K and compute the energy density needed for the PWN to be detected by H.E.S.S. or CTA. For H.E.S.S., a minimum energy density of 5 eV
cm$^{-3}$ is required to be detected in a 50 hours observation, according to the sensitivity curve used here. For CTA, an energy density of 0.2
eV cm$^{-3}$ would be enough to allow detection, which foresees its identification in case our model is correct. Both models are shown in figure
\ref{lmcpwn} (right panel) and their parameters are given in table \ref{tabres}.

The NIR photon field could also be important depending on the density of nearby stars in the N158A field and could enhance the TeV yield, at the
same time reducing the required FIR densities for detection.

We conclude that N158A is a particle dominated nebulae that has been undetected because of sensitivity limitations.

\section[High-$\eta$ models for non-detected at TeV PWNe]{High magnetization models for non-detected at TeV PWNe}
\label{sec5.3}

According to the phase space exploration done in \citet{torres13b}, PWNe with high magnetic fractions would lead to dim X-ray PWNe and they could
explain also very low fluxes at VHE for PWNe which are bright from radio to X-rays. The lack of detection of some of the cases explained in the
previous sections have motivated considering alternative models in this direction \citep{martin14b}.

For G310.6-1.6, the lack of observational constraints allows considering an alternative model in which the nebula has a magnetic fraction of
$\eta=0.98$, well beyond equipartition. In this case (model 2, see figure \ref{g310}), the energy break moves to higher energies
($\gamma=6 \times 10^6$ or $\sim$3 TeV) and the magnetic field increases up to 306 $\mu$G. Model 2 explains also well the overall X-ray flux, but
fails in reproducing the break at 6 keV.

In the case of G76.9+1.0, model 3 explores whether G76.9+1.0 could be a highly magnetic PWN as speculated previously by
\citet{arzoumanian08,arzoumanian11}. We show an example with a magnetic field of 85.2 $\mu$G and a magnetic fraction of $\eta=0.998$. The
injection function in this case is a simple power-law with an spectral index of 2.65. In order to respect the upper limits in radio, we need to
impose a minimum energy at injection for particles of $\gamma=10^4$ ($\sim$5.1 GeV). The IC contribution decreases with respect the other models,
as expected due to the lower contribution of spin-down energy to particles and the larger synchrotron field, which maximizes their losses.

The radio and X-ray data are also compatible with a high-$\eta$ model for G292.0+1.8 (see figure \ref{g2920}, model 2) with $\eta=0.77$ and a
resulting magnetic field of 81 $\mu$G (similar to the Crab Nebula). The injection in this case has an energy break of $\gamma_b=2.5 \times 10^5$
($\sim$130 GeV) and the high energy spectral index changes slightly (2.5). In this case, G292.0+1.8 would not be detected even with CTA also
explaining the difference with G54.1+0.3. A deep TeV observation will distinguish between these two models.

We have also investigated highly magnetized models in which the detection of N158A is not possible even with CTA, unless the energy density of
the FIR increases up to $\sim$500 eV cm$^{-3}$ (assuming that there is no NIR contribution). The injection function in such models has an energy
break of $\gamma_b=6 \times 10^7$, and a low (high) spectral index is 1.45 (2.4). Taking into account that the maximum energy at injection is
$\gamma_{max}=1.2 \times 10^8$, a simple power-law model with an index of 1.45 could also be compatible with this fit. Here, we obtain a highly
magnetized nebula with a magnetic fraction of 0.9 and an extreme magnetic field of 1.15 mG. But whereas the radio and the infrared data are
fitted similarly well to particle dominated models, the predicted X-ray flux of these models is not quite in agreement with data. This fact and
the extreme values of the parameters we have just quoted make a high $\eta$ model unlikely. In equipartition (i.e., $\eta=0.5$), the radio and
X-ray flux surpasses the data flux in a factor $\sim$4. In this latter case, the magnetic field is lower ($B$=858 $\mu$G ), but the number of
particles is still high to fit the flux.

\section{Conclusions}
\label{sec5.4}

Despite having similar spin down, the value of the magnetic field differs from one to another PWN not only because of the value of $\eta$ 
($B\sim$ $\eta^{1/2}$) differs, but also because their size does ($B\sim R_{PWN}^{-3/2}$). Models with high values of $\eta$ would explain the
low efficiency of some PWNe at X-rays and make them undetectable at VHE. However, we here found that models with high magnetic field and fraction
can be constructed only for some of the  nebulae that are non-detected at TeV, at the price of stretching other parameters. They seem to work
worse than particle dominated models in general, and remain viable only for G76.9+1.0 (for which there are significantly less observational
constraints) and G310.6--1.6 (pending the scrutiny of deeper TeV observations). These are the specific conclusions.

\begin{itemize}
\item We propose a model for 3C 58 considering a distance of 3.2 kpc, which implies a physical radius of 3.7 pc. The magnetic fraction necessary
to fit the radio and X-ray data is 0.21, which is higher than the general trend seen in chapter \ref{chap4}, with the exception of CTA-1. The
energy density of the FIR target photon field should be $\sim$5 eV cm$^{-3}$ to be detected by MAGIC in a 50 h observation. In the case of CTA,
this reduces to $\sim$0.75 eV cm$^{-3}$. which makes 3C 58 PWN a good candidate to be detected with the current Cherenkov telescopes. These
estimations are pesimistic since they do not consider NIR contribution, which can be also important.
\item We propose a low magnetization model for N157B with an age of 4.6 kyr and a magnetic field of 13 $\mu$G. The size of the nebula is
compatible with the one given by \citet{lazendic00}, the age with the Sedov age of the remnant \citep{wang98} and the ejected mass with the lower
limit given by \citet{chen06}. A high magnetization model ($\eta>0.5$) does not agree with the detection of N157B at TeV energies, which would
imply FIR and NIR energy densities much higher than the upper limits obtained by \citet{abramowski12}.
\item N158A non-detection seems to happen because of its smaller age (perhaps also because of a lower photon background?) rather than by having a
large magnetization. If this is the case, it will certainly be detected with CTA and likely also by the current generation of instruments.
Indeed, just the CMB inverse Compton contribution would produce a CTA source. Without taking into account a possible significant NIR contribution
which would ease the required observation time, we find that if N158A is subject to a FIR energy density of 5 eV cm$^{-3}$, it can already lead
to a detection by H.E.S.S. in 50 hours (lower IC target fields leads to larger integration times, but still within plausible limits). The
high-$\eta$ model(s) explored for N158A has been disregarded as unlikely due to inability to produce a good match to the X-ray data.
\item G76.9+1.0 is subject to a large uncertainty given the lack of sufficient observational constraints (only X-ray data are available). This
leads to the possibility of accommodating both extremes in the magnetic fraction phase space. In none of the cases, a TeV detection is expected
and it will be difficult to differentiate among models. The injection of particles is inefficient at high energies or the energy that goes to
particles is so low that the energy density for the FIR and NIR target fields necessary to reach the CTA sensitivity would be more than a factor
100 in comparison with those obtained by GALPROP for model 1, and more than a factor 1000 for model 2. In such case, the inverse Compton
contribution at X-ray energies would make impossible to fit the spectral slope. Other important parameters as age or the radius of the nebula are
not well determined and are necessary to make a solid conclusion.
\item The low-$\eta$ models for G310.6--1.6 and G292.0+1.8 predict their detection with H.E.S.S. given sufficient integration time. The CMB
inverse Compton contribution reaches the sensitivity curve of a 50 hrs observation in the case of G310.6--1.6. The magnetic fraction for
G292.0+1.8 is one order of magnitude higher than the one obtained for G54.1+0.3 in \citet{torres14}. This fact and the slight difference in the
FIR and NIR energy densities considered in both cases, could explain the lack of detection of G292.0+1.8 at TeV. In both cases, radio and X-ray
data are also explained with a high-$\eta$ model with a magnetic field of 306 $\mu$G for G310.6--1.6 and 81 $\mu$G for G292.0+1.8. However, the
high-$\eta$ model for G310.6--1.6 is not preferred due to its inability to correctly reproduce the spectral break at 6 keV. For G292.0+1.8
instead, a high-$\eta$ model remains viable and TeV observations would solve the degeneracy.
\end{itemize}

\chapter[Comparing SNRs around magnetars and canonical PSRs]{Comparing supernova remnants around strongly magnetized and canonical pulsars}
\label{chap6}

\ifpdf
    \graphicspath{{Chapter6/Figs/Raster/}{Chapter6/Figs/PDF/}{Chapter6/Figs/}}
\else
    \graphicspath{{Chapter6/Figs/Vector/}{Chapter6/Figs/}}
\fi

As we already commented in chapter \ref{chap1}, the formation mechanism of the high magnetic fields found in magnetars is still not clear. The
alpha-dynamo model suggested by \citet{duncan92} would imply an excess of rotational energy release (with respect to normal pulsars) the effects
of which could in principle be observed in the SNRs. \citet{vink06} started the idea of studying the energetic of supernova remnants surrounding
magnetar with the aim of disentangling a possible energetic difference between SNRs associated with magnetars and others surrounding normal
pulsars. Their work did not find any clear evidence i.e. of an additional energy released in the remnant possibly due to an excess of rotational
energy at birth.

Following this study we decided to extend their work re-analysing all available {\it XMM-Newton} or {\it Chandra} data of all confirmed and
bright SNRs associated with magnetars, and to to a high-$B$ pulsar that showed magnetar-like activity, and comparing in a coherent and
comprehensive way all the extracted properties of these SNRs with other remnants: in particular line ionization and X-ray luminosity.

This chapter is based on the work done in \citet{martin14a}.

\section{Data analysis and reduction}
\label{sec6.1}

In this work, our approach has tried to be as conservative and model independent as possible. In particular, our target sample has been chosen
such to include all confirmed associations (see the McGill catalog\footnote{http://www.physics.mcgill.ca$/~$pulsar/magnetar/main.html} for all
proposed associations), and among those, we chose only those supernova remnants bright enough, and with sufficiently good spectra, to perform a
detailed analysis and classification of their spectral lines. We analyze the X-ray spectral lines of four SNRs hosting a neutron star that
showed magnetar-like activity in its center: Kes 73, CTB 109, N 49 and Kes 75. We use for all targets the best available archival data: from
the {\em XMM-Newton} telescope in the case of Kes 73, CTB 109 and N 49, and {\em Chandra} for Kes 75. The observations used are summarized in
table \ref{tab:obs}. To compare coherently all the spectral lines and fluxes we observed for these remnants we have chosen to use an empirical
spectral fitting for all SNRs. We have modeled all spectra using one or two Bremsstrahlung models for the spectral continuum, plus Gaussian
functions for each detected spectral line. We added spectral lines one by one until the addition of a further line did not significantly improve
the fit (by using the F-test). This approach is totally empirical, with respect of using more detailed ionized plasma models, but ensures a
coherent comparison between different remnants. In table \ref{plasmamodels}, we report also the results of our spectra modeled with ionized
plasma models, for a comparison with the literature.

\begin{table}[t!]
\centering
\scriptsize
\caption[Observations used in this work]{Observations used in this work.}
\label{tab:obs}
\begin{tabular}{@{}cccccc}
\hline
\bf{SNR} & \bf{Instrument} & \bf{ObsID} & \bf{Date} & \bf{Detector} & \bf{Exp. (s)}\\
\hline
Kes73 & XMM & 0013340101 & 2002-10-05 & PN & 6017\\
 & & & & MOS1 & 5773\\
 & & & & MOS2 & 5771\\
 & & 0013340201 & 2002-10-07 & PN & 6613\\
 & & & & MOS1 & 6372\\
 & & & & MOS2 & 6372\\
CTB 109 & XMM & 0057540101 & 2002-01-22 & PN & 12237\\
 & & & & MOS1 & 19027\\
 & & & & MOS2 & 19026\\
 & & 0057540201 & 2002-07-09 & PN & 14298\\
 & & & & MOS1 & 17679\\
 & & & & MOS2 & 17679\\
 & & 0057540301 & 2002-07-09 & PN & 14011\\
 & & & & MOS1 & 17379\\
 & & & & MOS2 & 17379\\
N49 & XMM & 0505310101 & 2007-11-10 & PN & 72172\\
Kes75 & Chandra & 748 & 2000-10-15 & ACIS-S & 37280\\
 & & 6686 & 2006-06-07 & ACIS-S & 54070\\
 & & 7337 & 2006-06-05 & ACIS-S & 17360\\
 & & 7338 & 2006-06-09 & ACIS-S & 39250\\
 & & 7339 & 2006-06-12 & ACIS-S & 44110\\
 \hline
 \hline
\end{tabular}
\end{table}

\subsubsection{XMM-Newton data}

We use images in full-frame mode obtained from the European Photon Imaging Camera (EPIC) PN \citep{struder01} and MOS \citep{turner01}. The
spectra of these images are fitted simultaneously in order to obtain the spectrum with the maximum possible number of counts. We used the
specific software for {\it XMM-Newton} data, Science Analysis System (SAS) v13.5.0 with the latest calibration files. To clean images of solar
flares, we used the SAS tool {\it tabtigen} to choose the good time intervals and extract them and the spectra with {\it evselect}. Source and
background spectra were extracted from each single image with pattern $\le$ 4 for PN images and pattern $\le$ 12 for MOS. The spectra and the
backgrounds corresponding to the same regions and the same detector were merged using the FTOOLS routine {\it mathpha} and we compute the
mean of the response matrices (RMF) and the ancillary files (ARF) weighted by the exposure time using the tools {\it addrmf} and {\it addarf}
(this means, that we keep PN, MOS1 and MOS2 data separately and we merge the spectra when they come from the same detector). Finally, we
binned the spectra demanding a minimum of 25 counts per bin to allow the use of $\chi^2$-statistics.

\begin{figure}
\centering
\includegraphics[width=0.75\textwidth]{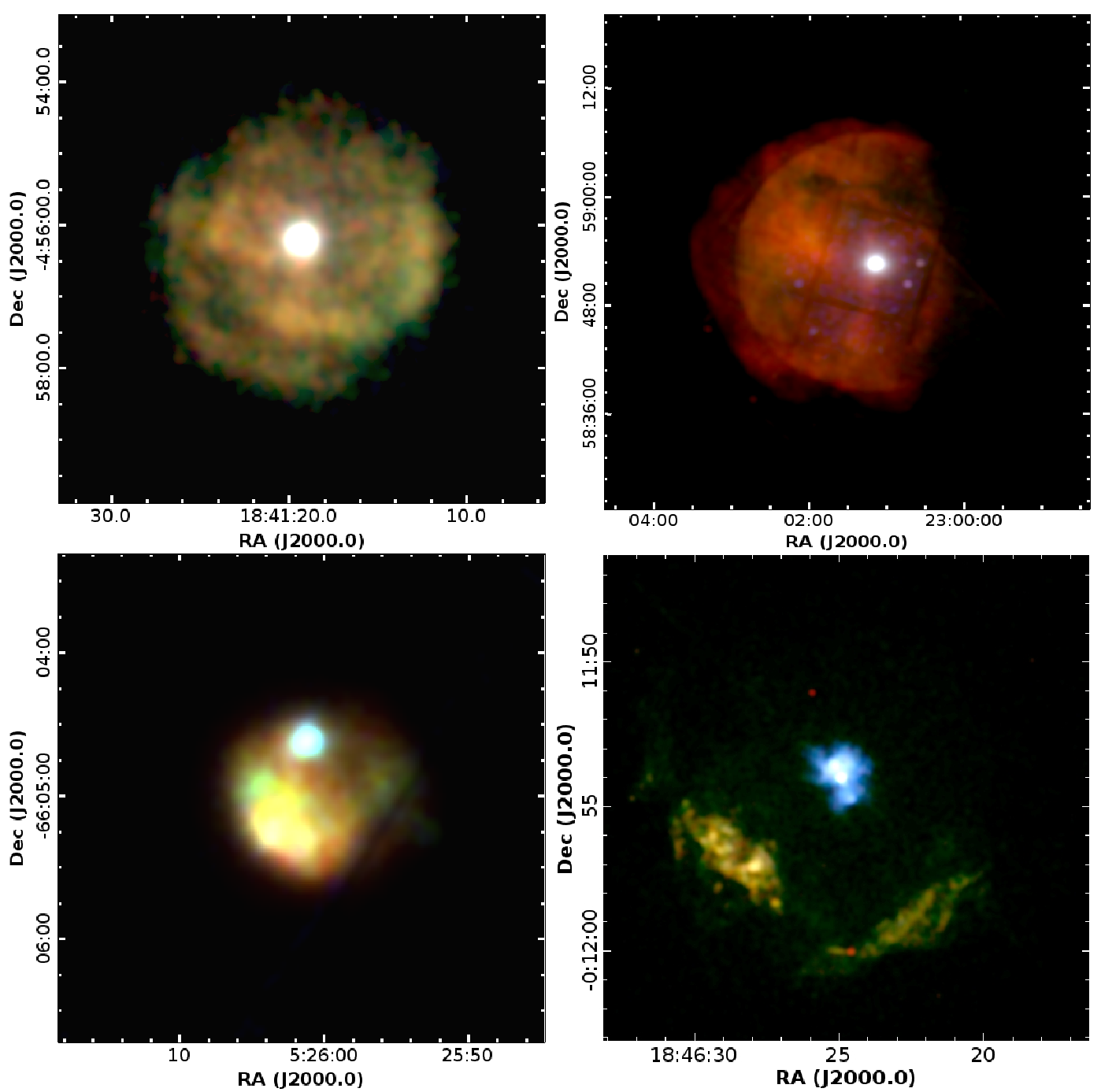}
\caption[Color images of Kes 73, CTB 109, N 49 \& Kes 75]{Combined color images of Kes 73 (top-left), CTB 109 (top-right), N49 (bottom-left) and
Kes 75 (bottom-right).}
\label{fig:neb}
\end{figure}

We analyze the spectrum of each nebula considering its entire extension. For Kes 73, the nebula is completely covered in the EPIC PN, MOS 1 and
MOS 2 detectors and we consider all of them in the analysis. In the case of CTB 109, the SNR is too large to be included entirely in a single
pointing. The images with the {\em XMM-Newton} data ID: 0057540101, 0057540201 and 0057540301 correspond to south, north and east pointings of
the remnant. We computed the spectra of each pointing, also considered the EPIC PN, MOS 1 and MOS 2 cameras. For N 49, the exposure time of the
MOS detectors is very low in comparison with PN. For this reason, we did not use the MOS data to avoid statistical noise in the data.

\begin{figure}
\centering
\includegraphics[width=0.75\textwidth]{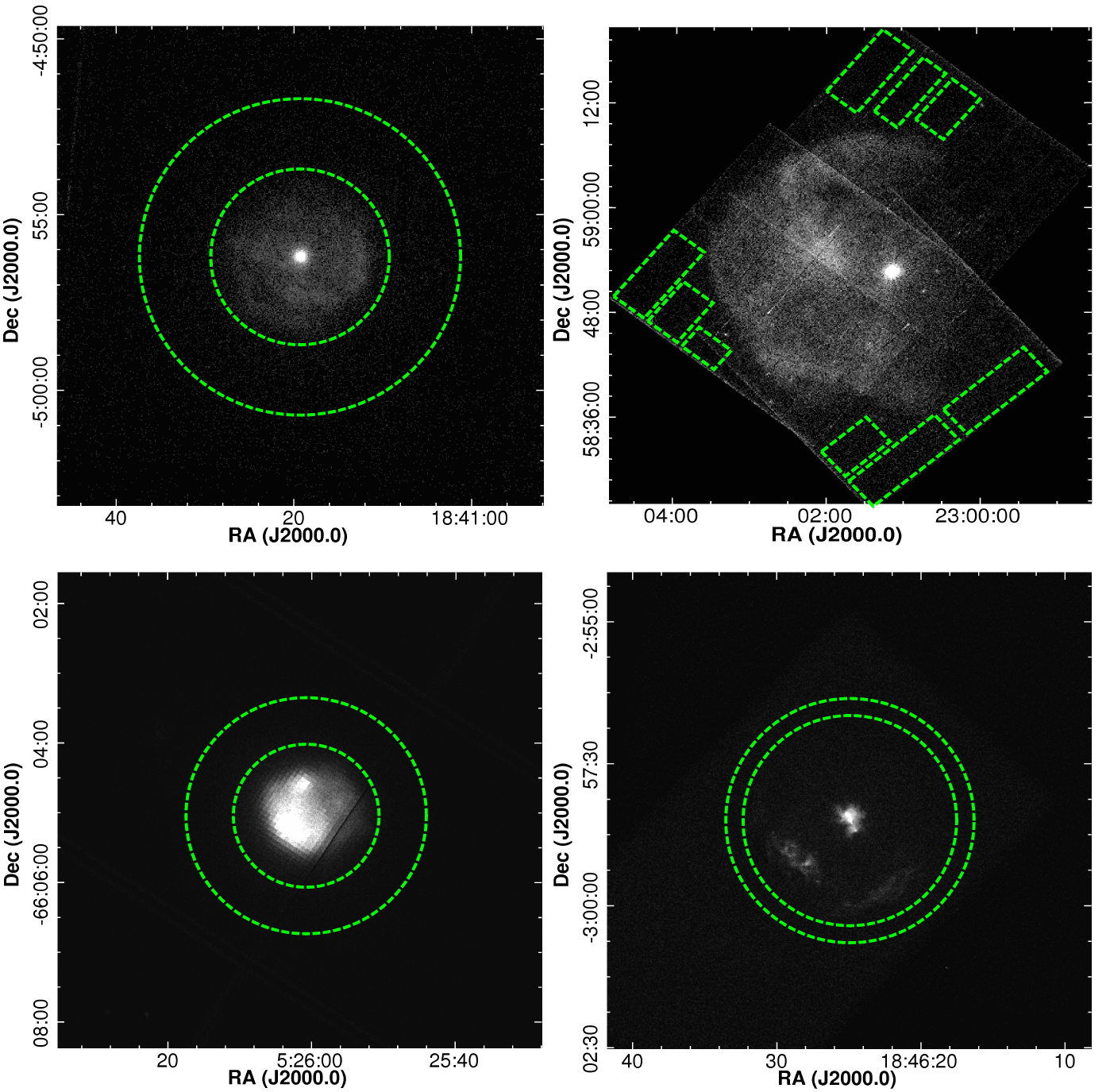}[t!]
\caption[Map of the backgrounds used in the spectrum analysis]{Map of the backgrounds used in the spectrum analysis. The order of the images is
the same as in figure \ref{fig:neb}.}
\label{fig:bg}
\end{figure}

\subsubsection{Chandra data}

In the case of Kes 75, the best available observations were performed with {\em Chandra} using the Advanced CCD Imaging Spectrometer (ACIS). The
ID numbers of the data used are in Table \ref{tab:obs}. We used the standard reduction software for {\em Chandra}, the Chandra Interactive
Analysis of Observations (CIAO) v4.5. The spectra and the backgrounds were extracted using the routine {\it specextract} and the RMFs and ARFs
using {\it mkacisrmf} and {\it mkwarf} respectively. Finally, we combine the spectra demanding a minimum of 25 counts per energy bin using
{\it combine\_spectra}.

\section{Spectral analysis and results}
\label{sec6.2}

\begin{figure}
\centering
\includegraphics[width=1.0\textwidth]{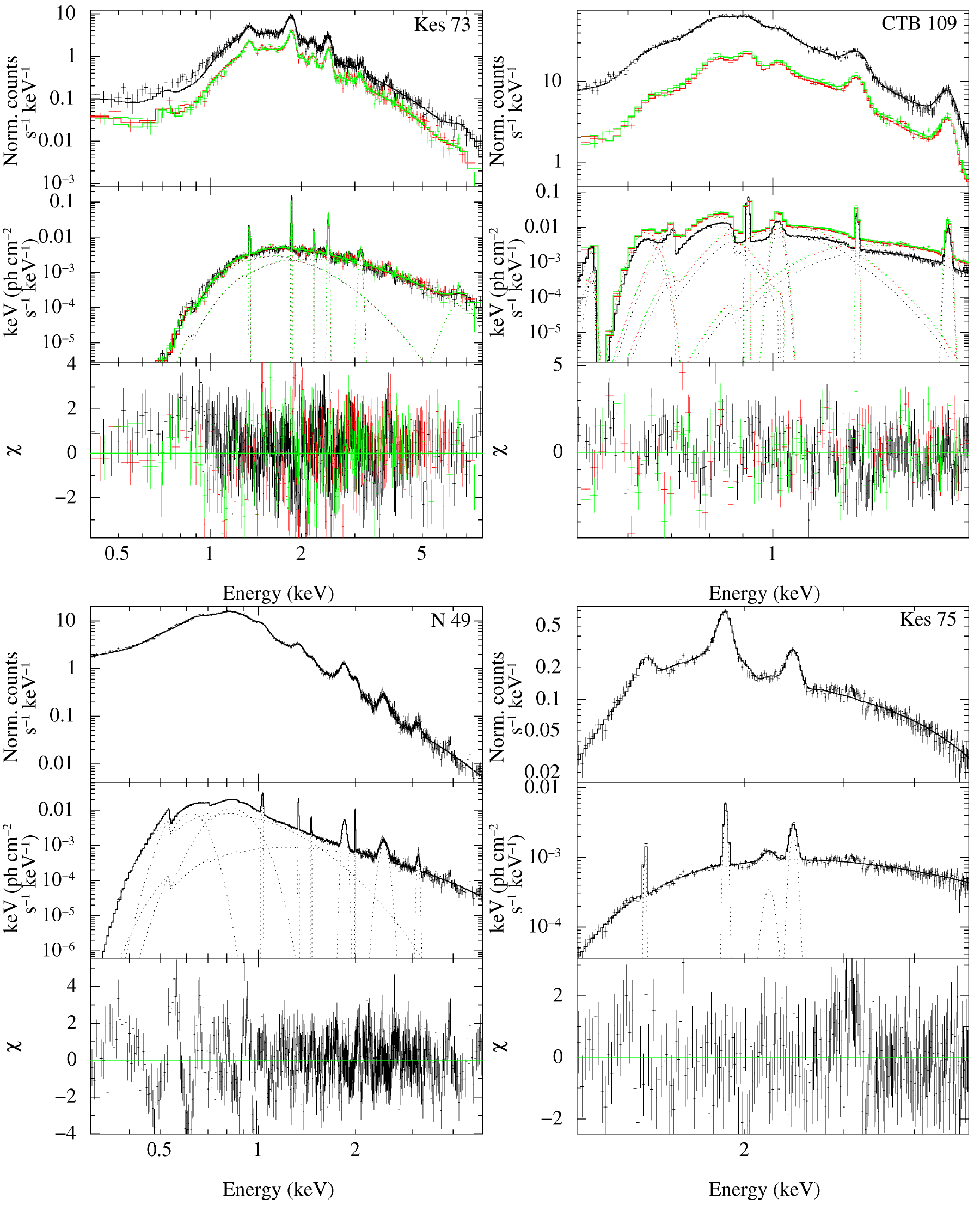}
\caption[Spectra obtained for the Kes 73, CTB 109, N 49 \& Kes 75]{Spectra obtained for the Kes 73, CTB 109, N 49 \& Kes 75. We used the EPIC PN
(in black), MOS 1 (in red) and MOS 2 (in green) data simultaneously to fit the models.}
\label{fig:spec}
\end{figure}

We report the fitted spectra in figure \ref{fig:spec}, while reporting the best fitting models and relative parameters in Tables \ref{tab1_snr}
and \ref{tab2_snr}. For the spectral analysis, we used the program {\em XSPEC} \citep{arnaud96} v12.8.1 from the package HEASOFT v6.15. As
anticipated above, we have used for all SNRs a spectral model comprised of photo-electric absorption ({\tt phabs}), one or two Bremsstrahlung
models ({\tt brems}), plus a series of Gaussian functions to model the emission lines. Even if more physical ionized plasma models such a
{\tt vnei}, {\tt vshock} or {\tt vpshock} could be used to fit those SNRs: e. g., \citet{kumar14} for Kes 73, \citet{sasaki04,sasaki13} for CTB
109, \citet{park12} for N 49 and \citet{temim12} for Kes 75; we prefer to use a more empirical approach to compare coherently the emission lines
and luminosities of those objects, which is the aim of our work. Below we summarize for each studied remnant our results in the context of the
general properties of the SNR.

In figure \ref{fig:bg} we show the background regions we have chosen for this analysis. We have tried several different regions finding
consistent results. During the spectral analysis we checked that subtracting the background spectra or fitting it separately from the remnant
spectra and subtracting its best fitting model, gave consistent results.

\subsubsection{Kes 73}

Kes 73 (also known as G27.4+0.0) is a shell-type SNR. Its dimensions are about $4.7' \times 4.5'$ and it is located between 7.5 and 9.8 kpc
\citep{tian08b}. The central source is the magnetar 1E 1841$-$045 discovered as a compact X-ray source with the Einstein Observatory
\citep{kriss85}, and confirmed as a magnetar in \citet{vasisht97,gotthelf99b}. The period of the magnetar is 11.78 s and its period derivative is
4.47 $\times 10^{-11}$ s s$^{-1}$. The resulting dipolar magnetic field is 7.3 $\times 10^{14}$ G, the spin-down luminosity is 1.1
$\times 10^{33}$ erg s$^{-1}$ and the characteristic age is 4180 yr. The age of the SNR shell is estimated around 1300 yr \citep{vink06}, which
is consistent with the age between 750 and 2100 yr estimated by \citet{kumar14}. Kes 73 has been also observed by {\em ROSAT} \citep{helfand94},
{\em ASCA} \citep{gotthelf97}, {\em Chandra} \citep{lopez11} and {\em Suzaku} \citep{sezer10}.

Kes 73 shows a quite spherical structure with 1E 1841-045 in the center of the remnant (see figure \ref{fig:neb}). In the western part of the
nebula (right-hand side of the images), we distinguish a shock ring which encloses the central source from west to east of the image passing
below the central source. Most of the flux is emitted between 1 and 3 keV. Finally, we analyzed the total spectrum of the nebula excluding a
circle of 40$"$ around the central source to exclude possible contamination from the central object. The background spectrum has been extracted
from a surrounding annular region shown in figure \ref{fig:bg}, avoiding gaps between the CCDs to ensure good convergence of the response
matrices. The continuum spectrum has been fitted with two plasmas with temperatures of 0.43 keV and 1.34 keV. The absorption column density
obtained is $N_{\rm H}=2 \times 10^{22}$ cm$^{-2}$. We detected 6 lines. The most prominent is the Fe XXV at 6.7 keV with an equivalent width
(EW) of 1.89 keV. Other lines are Mg XI at 1.35 keV (EW=95 eV), Si XIII at 1.85 keV (EW=0.37 keV), Si XIII at 2.19 keV (EW=46 eV), S XV at 2.45
keV (EW=0.38 keV) and Ar XVII at 3.13 keV (EW=0.12 keV).

\begin{table}[t!]
\scriptsize
\centering
\caption[Summary of the fitted models for Kes 73, CTB 109, N 49 \& Kes 75]{Summary of the fitted models for Kes 73, CTB 109, N 49 and Kes 75.}
\label{tab1_snr}
\begin{tabular}{lllll}
\hline
\bf{Parameter} & {\bf Kes 73} & {\bf CTB 109} & {\bf N 49}$^{\dagger}$ & {\bf Kes 75}\\
\hline
$N_H$ ($10^{22} cm^{-2}$) & $2.00^{-0.02}_{+0.01}$ & $2.83^{-0.06}_{+0.10}$ & $0.698^{-0.024}_{+0.006}$ & $1.79^{-0.05}_{+0.06}$\\
$kT_1$ (keV) & $0.43^{-0.05}_{+0.02}$ & $0.065^{-0.002}_{+0.001}$ & $0.230^{-0.003}_{+0.004}$ & $2.8^{-0.1}_{+0.2}$\\
$N_1^{brems}$ (Norm. counts s$^{-1}$) & $0.36^{-0.02}_{+0.15}$ & $9^{-1}_{+14} \times 10^6$ & $0.512^{-0.007}_{+0.067}$ & $(4.5^{-0.3}_{+0.2}) \times 10^{-3}$\\
$kT_2$ (keV) & $1.34^{-0.01}_{+0.01}$ & $0.20^{-0.02}_{+0.03}$ & $1.14^{-0.01}_{+0.04}$ & -\\
$N_2^{brems}$ (Norm. counts s$^{-1}$) & $(2.47^{-0.06}_{+0.41}) \times 10^{-2}$ & $18^{-4}_{+9}$ & $(3.5^{-0.15}_{+0.08}) \times 10^{-3}$ & -\\
\hline
{\bf N VII (3,4 $\rightarrow$ 1)} & & & & \\
$E$ (keV) & - & $0.515^{-0.008}_{+0.016}$ & - & -\\
$\sigma$ (keV) & - & $9.2^{-0.3}_{+0.1}) \times 10^{-2}$ & - & -\\
$N$ (Norm. counts s$^{-1}$) & - & $(4^{-1}_{+4}) \times 10^4$ & - & -\\
$EW^\ddagger$ (eV) & - & 737 & - & -\\
\hline
{\bf O VII (2,5 $\rightarrow$ 1)} & & & & \\
$E$ (keV) & - & - & $0.568^{-0.004}_{+0.004}$ & -\\
$\sigma$ (keV) & - & - & $(6.1^{-0.3}_{+0.1}) \times 10^{-2}$ & -\\
$N$ (Norm. counts s$^{-1}$) & - & - & $(4.7^{-0.3}_{+0.7}) \times 10^{-2}$ & -\\
$EW^\ddagger$ (eV) & - & - & 198 & -\\
\hline
{\bf N VII (6,7 $\rightarrow$ 1)/O VII (2,5,6 $\rightarrow$ 1)} & & & & \\
$E$ (keV) & - & $0.597^{-0.002}_{+0.003}$ & - & -\\
$\sigma$ (keV) & - & $<0.06$ & - & -\\
$N$ (Norm. counts s$^{-1}$) & - & $(2.4^{-0.4}_{+1.5}) \times 10^5$ & - & -\\
$EW^\ddagger$ (eV) & - & 472 & - & -\\
\hline
{\bf O VIII (6,7 $\rightarrow$ 1)/Fe XVIII (4,5 $\rightarrow$ 1)} & & & & \\
$E$ (keV) & - & - & $0.769^{-0.001}_{+0.001}$ & -\\
$\sigma$ (keV) & - & - & $0.112^{-0.003}_{+0.002}$ & -\\
$N$ (Norm. counts s$^{-1}$) & - & - & $(1.78^{-0.06}_{+0.11}) \times 10^{-2}$ & -\\
$EW^\ddagger$ (eV) & - & - & 338 & -\\
\hline
{\bf Ne IX (2,5 $\rightarrow$ 1)} & & & & \\
$E$ (keV) & - & $0.91^{-0.01}_{+0.01}$ & - & -\\
$\sigma$ (keV) & - & $<0.07$ & - & -\\
$N$ (Norm. counts s$^{-1}$) & - & $7.2^{-0.6}_{+0.2}$ & - & -\\
$EW^\ddagger$ (eV) & - & 147 & - & -\\
\hline
{\bf Ne X (3,4 $\rightarrow$ 1)} & & & & \\
$E$ (keV) & - & $1.014^{-0.003}_{+0.002}$ & $1.028^{-0.001}_{+0.004}$ & -\\
$\sigma$ (keV) & - & $<0.07$ & $<0.07$ & -\\
$N$ (Norm. counts s$^{-1}$) & - & $0.37^{-0.04}_{+0.03}$ & $(5.9^{-0.3}_{+0.3}) \times 10^{-4}$ & -\\
$EW^\ddagger$ (eV) & - & 68 & 33 & -\\
\hline
{\bf Mg XI (2 $\rightarrow$ 1)} & & & & \\
$E$ (keV) & $1.346^{-0.002}_{+0.001}$ & $1.347^{-0.004}_{+0.003}$ & $1.332^{-0.002}_{+0.006}$ & $1.33^{-0.02}_{+0.02}$\\
$\sigma$ (keV) & $<0.08$ & $<0.08$ & $<0.08$ & $<0.08$\\
$N$ (Norm. counts s$^{-1}$) & $2.6^{-0.1}_{+0.1} \times 10^{-3}$ & $(2.0^{-0.1}_{+0.3}) \times 10^{-3}$ & $(2.03^{-0.08}_{+0.08}) \times 10^{-4}$ & $(1.8^{-0.3}_{+0.3}) \times 10^{-4}$\\
$EW^\ddagger$ (eV) & 95 & 337 & 62 & 84\\
\hline
\hline
\multicolumn{5}{l}{
\begin{minipage}{12cm}
$^\dagger$ The absorption column density of N49 is fitted using the LMC abundances: He=0.89, C=0.30, N=0.12, O=0.26, Ne=0.33, Na=0.30, Mg=0.32, Al=0.30, Si=0.30, S=0.31, Cl=0.31, Ar=0.54, Ca=0.34, Cr=0.61, Fe=0.36, Co=0.30 \& Ni=0.62. We have added also the galactic absorption $N_H=6 \times 10^{20} cm^{-2}$.
\end{minipage}}\\
\multicolumn{5}{l}{
\begin{minipage}{12cm}
$^\ddagger$ Equivalent Width.
\end{minipage}}
\end{tabular}
\end{table}

\subsubsection{CTB 109}

CTB 109 (also G109.2-1.0) was discovered in X-rays with the Einstein Observatory by \citet{gregory80}, it is $30' \times 45'$ wide and the
estimated distance is about 3 kpc \citep{kothes02}. The central source is the magnetar 1E 2259+586 with a spin period of 6.98 s \citep{fahlman83}
and a period derivative of 4.83 $\times 10^{-13}$ \citep{iwasawa92}. The dipolar magnetic field is about 5.9 $\times 10^{13}$ G, the spin down
power is 5.6 $\times 10^{31}$ erg s$^{-1}$ and the characteristic age is 229 kyr. Despite the large characteristic age of the pulsar, the
estimated true age of the remnant is about 14 kyr \citep{sasaki13}. CTB 109 has been observed also in X-rays with {\em ASCA} \citep{rho98},
{\em BeppoSAX} \citep{parmar98} and {\em ROSAT} \citep{hurford95,rho97}.

The spectrum covers the entire shell and combines the three observations detailed in Table \ref{tab:obs}. The background regions used are shown
in Figure \ref{fig:bg}. We observe that the main contribution to the flux is in the energy range between 0.5 and 2 keV. Some known X-ray sources
in the field of view have been excluded in our analysis. 

In this case we used two Bremsstrahlung models to fit the continuum, with temperatures of 0.07 keV and 0.20 keV. The measured absorption density
is $N_{\rm H}=$2.83 $\times 10^{22}$ cm$^{-2}$, and we detected 6 lines: N VII at 0.52 keV (EW=0.74 keV) and at 0.60 keV (EW=0.47 keV), Ne IX at
0.91 keV (EW=0.15 keV), Ne X at 1.01 keV (EW=68 eV), Mg XI at 1.35 keV (EW=0.34 keV) and Si XIII at 1.86 keV (0.28 keV).

\subsubsection{N 49}

N49 (also SNR B0525-66.1) is a SNR located in the Large Magellanic Cloud (LMC). The associated central source is SGR 0526-66 with a period of
8.047 s \citep{mazets79} and a period derivative of 6.6 $\times 10^{-11}$ s s$^{-1}$ \citep{kulkarni03}. There is some uncertainty in the
association of SGR 0526-66 with N49 (see \citealt{gaensler01b}). The inferred dipolar magnetic field is 7.3 $\times 10^{14}$ G, the spin-down
luminosity is 4.9 $\times 10^{33}$ erg/s and the characteristic age is $\sim$2 kyr. The nebula is $1.5' \times 1.5'$, this means that assuming a
distance of 50 kpc the diameter of N49 is $\sim$22 pc. \citet{park12} establish a Sedov age for the nebula of $\sim$4.8 kyr and a SN explosion
energy of 1.8 $\times 10^{51}$ erg.

SGR 0526-66 is located in the north of the remnant. The brightest part of the nebula is in the southeast, coinciding with dense interstellar
clouds \citep{vancura92,banas97,park12}. This part of the remnant also has contributions between 3 and 10 keV, while the contribution of the rest
of the nebula is clearly negligible at this range. In Figure \ref{fig:neb}, we show a color image of N49. We analyze the total spectrum of the
nebula excluding a circle of 20$"$ around the central source to avoid its contribution to the spectrum.

The absorption of N49 has two components: one is related with the Galactic absorption and the other is the absorption produced by LMC. The Milky
Way photoelectric absorption towards N49 is fixed as $N_{\rm H}=6 \times 10^{20}$ cm$^{-2}$ \citep{dickey90,park12}. We include a second
absorption component to take into account the absorption column density for LMC, where we use the abundances given by
\citet{russell92,hughes98,park12}. We obtain an absorption column density of $N_{\rm H}=0.7 \times 10^{22}$ cm$^{-2}$ for the LMC contribution.
The continuum is represented by two Bremsstrahlung models with temperatures of 0.23 keV and 1.14 keV. In this case, we have detected 9 lines: O
VII at 0.57 keV (EW=0.20 keV), O VIII/Fe XVIII at 0.77 keV (EW=0.34 keV), Ne X at 1.03 keV (EW=33 eV), Mg XI at 1.33 keV (EW=62 eV), Mg XII at
1.46 keV (EW=20 eV), Si XIII at 1.85 keV (EW=0.30 keV), Si XIV at 2.00 keV (EW=0.13 keV), S XV at 2.44 keV (EW=0.30 keV) and Ar XVII at 3.12 keV
(EW=0.11 keV).

\begin{table}[t!]
\scriptsize
\centering
\caption[Table \ref{tab1_snr} continued.]{Continued.}
\label{tab2_snr}
\begin{tabular}{lllll}
\hline
\bf{Parameter} & {\bf Kes 73} & {\bf CTB 109} & {\bf N 49}$^{\dagger}$ & {\bf Kes 75}\\
\hline
{\bf Mg XII (3,4 $\rightarrow$ 1)} & & & & \\
$E$ (keV) & - & - & $1.459^{-0.005}_{+0.006}$ & -\\
$\sigma$ (keV) & - & - & $<0.08$ & -\\
$N$ (Norm. counts s$^{-1}$) & - & - & $(3.9^{-0.5}_{+0.6}) \times 10^{-5}$ & -\\
$EW^\ddagger$ (eV) & - & - & 20 & -\\
\hline
{\bf Si XIII (2,5,6,7 $\rightarrow$ 1)} & & & & \\
$E$ (keV) & $1.8521^{-0.0001}_{+0.0001}$ & $1.856^{-0.001}_{+0.006}$ & $1.848^{-0.003}_{+0.002}$ & $1.851^{-0.003}_{+0.012}$\\
$\sigma$ (keV) & $<0.02$ & $<0.02$ & $(2.3^{-0.6}_{+0.6}) \times 10^{-2}$ & $<0.02$\\
$N$ (Norm. counts s$^{-1}$) & $2.76^{-0.06}_{+0.06} \times 10^{-3}$ & $(7.0^{-0.2}_{+0.3}) \times 10^{-4}$ & $(1.68^{-0.04}_{+0.06}) \times 10^{-4}$ & $(2.6^{-0.1}_{+0.2}) \times 10^{-4}$\\
$EW^\ddagger$ (eV) & 368 & 278 & 299 & 232\\
\hline
{\bf Si XIV (3,4 $\rightarrow$ 1)} & & & & \\
$E$ (keV) & - & - & $1.998^{-0.002}_{+0.007}$ & -\\
$\sigma$ (keV) & - & - & $<0.09$ & -\\
$N$ (Norm. counts s$^{-1}$) & - & - & $(5.2^{-0.4}_{+0.3}) \times 10^{-5}$ & -\\
$EW^\ddagger$ (eV) & - & - & 132 & -\\
\hline
{\bf Si XIII (13 $\rightarrow$ 1)} & & & & \\
$E$ (keV) & $2.201^{-0.010}_{+0.009}$ & - & - & $2.21^{-0.02}_{+0.04}$\\
$\sigma$ (keV) & $<0.09$ & - & - & $<0.09$\\
$N$ (Norm. counts s$^{-1}$) & $(1.6^{-0.2}_{+0.2}) \times 10^{-4}$ & - & - & $(3.4^{-0.9}_{+1.1}) \times 10^{-5}$\\
$EW^\ddagger$ (eV) & 46 & - & - & 45\\
\hline
{\bf S XV (2,5,6,7 $\rightarrow$ 1)} & & & & \\
$E$ (keV) & $2.452^{-0.002}_{+0.002}$ & - & $2.444^{-0.005}_{+0.005}$ & $2.437^{-0.005}_{+0.007}$\\
$\sigma$ (keV) & $<0.09$ & - & $<0.09$ & $<0.09$\\
$N$ (Norm. counts s$^{-1}$) & $(8.0^{-0.3}_{+0.2}) \times 10^{-4}$ & - & $(6.8^{-0.4}_{+0.4}) \times 10^{-5}$ & $(1.09^{-0.12}_{+0.08}) \times 10^{-4}$\\
$EW^\ddagger$ (eV) & 375 & - & 299 & 178\\
\hline
{\bf S XV (13 $\rightarrow$ 1)} & & & & \\
$E$ (keV) & - & - & - & -\\
$\sigma$ (keV) & - & - & - & -\\
$N$ (Norm. counts s$^{-1}$) & - & - & - & -\\
$EW^\ddagger$ (eV) & - & - & - & -\\
\hline
{\bf Ar XVII (2,5,6,7 $\rightarrow$ 1)} & & & & \\
$E$ (keV) & $3.13^{-0.01}_{+0.01}$ & - & $3.12^{-0.02}_{+0.02}$ & -\\
$\sigma$ (keV) & $<0.1$ & - & $<0.1$ & -\\
$N$ (Norm. counts s$^{-1}$) & $(9^{-1}_{+1}) \times 10^{-5}$ & - & $(7^{-1}_{+1}) \times 10^{-6}$ & -\\
$EW^\ddagger$ (eV) & 120 & - & 110 & -\\
\hline
{\bf Fe XXV (7 $\rightarrow$ 1)} & & & & \\
$E$ (keV) & $6.7^{-0.2}_{+0.2}$ & - & - & -\\
$\sigma$ (keV) & $0.5^{-0.1}_{+0.2}$ & - & - & -\\
$N$ (Norm. counts s$^{-1}$) & $2.9^{-0.6}_{+0.6} \times 10^{-5}$ & - & - & -\\
$EW^\ddagger$ (eV) & 1890 & - & - & -\\
\hline
$\chi^2_r$ & 1.57 (985) & 2.05 (477) & 1.84 (578) & 1.12 (258)\\
\hline
\hline
\multicolumn{5}{l}{
\begin{minipage}{11cm}
$^\dagger$ The absorption column density of N49 is fitted using the LMC abundances: He=0.89, C=0.30, N=0.12, O=0.26, Ne=0.33, Na=0.30, Mg=0.32, Al=0.30, Si=0.30, S=0.31, Cl=0.31, Ar=0.54, Ca=0.34, Cr=0.61, Fe=0.36, Co=0.30 \& Ni=0.62. We have added also the galactic absorption $N_H=6 \times 10^{20} cm^{-2}$.
\end{minipage}}\\
\multicolumn{5}{l}{
\begin{minipage}{11cm}
$^\ddagger$ Equivalent Width.
\end{minipage}}
\end{tabular}
\end{table}

\subsubsection{Kes 75}

Kes 75 (G29.7-0.3) is a composite SNR. The X-ray emission of the partial shell is extended in two clouds in the southwest and southeast part of
the image (see figure \ref{fig:neb}). It was observed firstly in X-rays by {\em Einstein} \citep{becker83} showing an incomplete shell of 3' in
extent. In the center of the nebula, there is a bright pulsar wind nebula (PWN), which was spatially resolved by the {\em Chandra} observation
\citep{helfand03,ng08}, and PSR J1846-0258 powers it. This pulsar was discovered using the {\it RXTE} telescope and localized within an arc
minute of the remnant using {\it ASCA} \citep{gotthelf00b}. The period of the pulsar is $\sim$326 ms and the period derivative
$7.11 \times 10^{-12}$ s s$^{-1}$ (e. g., \citealt{livingstone11a}). This leads to a spin-down energy loss of $8.1 \times 10^{36}$ erg s$^{-1}$,
a magnetic field of $4.9 \times 10^{13}$ G and a characteristic age of 728 yr. \citet{livingstone06} estimated a braking index of 2.65$\pm$0.01.
Despite its early classification as a typical rotational powered pulsar, PSR J1846-0258 showed magnetar-like activity via short bursts and the
outburst of its persistent emission \citep{gavriil08,kumar08} enabling its classification as (at least sporadically) a magnetically powered
pulsar. There is a big uncertainty in the distance of this SNR in the literature \citep{caswell75,milne79,mcbride08,becker84}. Most recent
estimates give a distance between $\sim$5.1-7.5\,kpc based on H I absorption observations \citep{leahy08b}, and 10.6\,kpc using millimeter
observations of CO lines from an adjacent molecular cloud \citep{su09}. In our work, we adopt this value in order to compute the X-ray luminosity
and the size of the SNR.

The spectrum of Kes 75 has been fitted using only one thermal Bremsstrahlung component with a temperature of 2.8 keV and an absorption column
density of $1.79 \times 10^{22}$ cm$^{-2}$. Four clear lines are resolved using Gaussians: Mg XI line at 1.33 keV (EW=84 eV), two Si XIII lines
at 1.85 (EW=0.23 keV) and 2.21 keV (EW=45 eV) and S XV at 2.44 keV (EW=0.18 keV).

\section{Spectral line and photometric comparison with other SNRs}
\label{sec6.3}

In this work we have re-analyzed in a coherent way the X-ray emission from SNRs around magnetars, and compared their emission lines and
luminosities. The aim of this study was to search for any possible trend or significant difference in SNRs associated with different types of
neutron stars. This work complements and extends the work by \citet{vink06}, providing a detailed description of the spectra for Kes 73, Kes 75,
N 49 and CTB 109, and compares them directly with other remnants with similar spectroscopic X-ray studies. We also looked for any possible trend
or significant difference in the ionization state and X-ray luminosity of SNRs associated with different types of neutron stars.

X-ray spectra of SNRs are usually fit with plasma models (see also table \ref{plasmamodels}). In this work we proceed to fit the spectra of Kes
73, CTB 109, N 49 and Kes 75 using a thermal Bremsstrahlung model for the continuum emission and Gaussians for the lines. Our main aim is to have
an estimate of line centroid energy, to identify it properly. We have then used the simplest continuum model to reduce the free parameters of the
fit\footnote{Note that in the 0.5-1\,keV the detection of spectral lines are dependent on absorption model we adopted.}. One could expect that
the excess of rotational energy released by the magnetar during the alpha-dynamo process could be stored in the ionization level of the lines
present in the spectrum. If the energy release is higher than in a normal SNR, heavy elements such as silicon (Si), sulfur (S), argon (Ar),
calcium (Ca) or iron (Fe) could be systematically at a higher state of ionization. In table \ref{tab1_snr}, we collected all SNRs with detailed
spectroscopic studies in the literature and we see that the typical elements detected are O VII, O VIII, Ne IX, Ne X, Mg XI, Mg XII, Si XIII, Si
XIV, S XV, S XVI, Ar XVII, Ca XIX and Fe XXV. The only lines detected in all four of the spectra are the Mg XI line at 1.33 keV and Si XIII at
1.85 keV. For comparison, we also fitted the spectra of the SNRs using a {\tt vnei} model (e. g., \citealt{borkowski01}). The results are
summarized in the table \ref{plasmamodels}. We have added a thermal Bremsstrahlung component in some cases. The temperature of the {\tt vnei}
plasma is always higher than for the thermal Bremsstrahlung, with the exception of N49 in which the temperature for {\tt vnei} is 0.17 keV (0.99
keV for Bremsstrahlung). The abundances obtained in both models show similar tendencies. For Kes 73 and N 49, the abundances of Si and S are
quite above the solar ones. CTB 109 shows low abundances with respect to the solar ones for O, Ne, Mg, Si and Fe. Due to the complexity of the N
49 spectrum, some lines have not been reproduced well by the plasma models and we have added them using gaussian profiles to improve the fit. In
summary, our spectroscopic X-ray analysis of these sources shows compatible results with other non-magnetar SNRs already reported in literature.

\begin{table}[t!]
\tiny
\centering
\caption[Line detections for some important SNRs in comparison with our analysis]{Summary of the line detections in X-ray for some important SNRs
compared with lines detected in our analysis.}
\label{tab:snr_list}
\begin{tabular}{lcccccccc}
\hline
{\bf SNR} & {\bf Galaxy} & {\bf Age (yr)} & \multicolumn{6}{c}{{\bf Element}} \\
 & & & {\it O VII} & {\it O VIII} & {\it O VIII} & {\it Ne IX} & {\it Ne X} & {\it Ne X}\\
 & & & {\it (2,5,7 $\to$ 1)} & {\it (3,4 $\to$ 1)} & {\it (6,7 $\to$ 1)} & {\it (2,5 $\to$ 1)} & {\it (3,4 $\to$ 1)} & {\it (6,7 $\to$ 1)}\\
 & & & {\it (0.574 KeV)} & {\it (0.653 KeV)} & {\it (0.774 KeV)} & {\it (0.915 KeV)} & {\it (1.022 KeV)}& {\it (1.21 KeV)} \\
\hline
Kes73 & MW & 1100-1500 & & & & & &\\
CTB109 & MW & 7900-9700 & & & & X & X &\\
Kes75 & MW & 900-4300 & & & & & &\\
N49 & LMC & 5000 & X & & X & X & X &\\
\hline
G1.9+1.3 [2] & MW & 110-170 & & & & & &\\
Kepler [3],[8],[12] & MW & 408 & & & & X & &\\
Tycho [4],[5],[6],[13] & MW & 440 & & & & X & &\\
SN1006 [10],[19] & MW & 1006 & X & & X & & & X\\
Cas A [1],[9],[16] & MW & 316-352 & X & X & X & X & X &\\
MSH11-54 [11],[14] & MW & 2930-3050 & X & X & X & X & X &\\
Puppis A [7],[17],[18] & MW & 3700-5500 & X & X & X & X & X & X\\
B0509-67.5 [15] & LMC & 400 & X & X & & X & &\\
\hline
\hline
 & & & {\it Mg XI} & {\it Mg XII} & {\it Si XIII} & {\it Si XIV} & {\it Si XIII} & {\it S XV}\\
 & & & {\it (2,5,6,7 $\to$ 1)} & {\it (3,4 $\to$ 1)} & {\it (2,5,6,7 $\to$ 1)} & {\it (3,4 $\to$ 1)} & {\it (13 $\to$ 1)} & {\it (2,5,6,7 $\to$ 1)}\\
 & & & {\it (1.35 KeV)} & {\it (1.47 KeV)} & {\it (1.86 KeV)} & {\it (2.00 KeV)} & {\it (2.18 KeV)} & {\it (2.46 KeV)}\\
\hline
Kes73 & MW & 1100-1500 & X & & X & & X & X\\
CTB109 & MW & 7900-9700 & X & & X & & &\\
Kes75 & MW & 900-4300 & X & & X & & X & X\\
N49 & LMC & 5000 & X & X & X & X & & X\\
\hline
G1.9+1.3 & MW & 110-170 & X & & X & & & X\\
Kepler & MW & 408 & X & & X & X & X & X\\
Tycho & MW & 440 & & & X & X & X & X\\
SN1006 & MW & 1006 & X & & X & & &\\
Cas A & MW & 316-352 & X & X & X & X & X & X\\
MSH11-54 & MW & 2930-3050 & X & X & X & & & X\\
Puppis A & MW & 3700-5500 & X & & X & & X & X\\
B0509-67.5 & LMC & 400 & X & & X & & X & X\\
\hline
\hline
 & & & {\it S XV} & {\it Ar XVII} & {\it Ca XIX} & {\it Fe XXV}\\
 & & & {\it (13 $\to$ 1)} & {\it (2,5,6,7 $\to$ 1)} & {\it (2,5,6,7 $\to$ 1)} & {\it K-shell}\\
 & & & {\it (2.88 KeV)} & {\it (3.13 KeV)} & {\it (3.89 KeV)} & {\it (6.65 KeV)}\\
\hline
Kes73 & MW & 1100-1500 & & X & & X\\
CTB109 & MW & 7900-9700 & & & & \\
Kes75 & MW & 900-4300 & & & & \\
N49 & LMC & 5000 & & X & & \\
\hline
G1.9+1.3 & MW & 110-170 & & X & X & X\\
Kepler & MW & 408 & X & X & X & X\\
Tycho & MW & 440 & X & X & X & X\\
SN1006 & MW & 1006 & & & & \\
Cas A & MW & 316-352 & X & X & X & X\\
MSH11-54 & MW & 2930-3050 & & & & \\
Puppis A & MW & 3700-5500 & & & & \\
B0509-67.5 & LMC & 400 & & X & X & X\\
\hline
\hline
\multicolumn{9}{l}{
\begin{minipage}{12cm}
The references are: $^{[1]}$\citet{bleeker01}, $^{[2]}$\citet{borkowski10}, $^{[3]}$\citet{cassamchenai04}, $^{[4]}$\citet{decourchelle01},
$^{[5]}$\citet{hayato10}, $^{[6]}$\citet{hwang97}, $^{[7]}$\citet{hwang08}, $^{[8]}$\citet{kinugasa99}, $^{[9]}$\citet{maeda09},
$^{[10]}$\citet{miceli09}, $^{[11]}$\citet{park07}, $^{[12]}$\citet{reynolds07}, $^{[13]}$\citet{tamagawa09}, $^{[14]}$\citet{vink04},
$^{[15]}$\citet{warren04}, $^{[16]}$\citet{willingale02}, $^{[17]}$\citet{winkler81a}, $^{[18]}$\citet{winkler81b}, $^{[19]}$\citet{yamaguchi08}.
\end{minipage}}
\end{tabular}
\end{table}

\begin{table}
\scriptsize
\centering
\caption[Fits for Kes 73, CTB 109, N 49 \& Kes 75 using a {\tt vnei} plasma model.]{Fits for Kes 73, CTB 109, N 49 \& Kes 75 using a {\tt vnei}
plasma model. A second thermal Bremsstrahlung component is included in some cases.$^\dagger$ The absorption column density of N49 is
fitted using the LMC abundances: He=0.89, C=0.30, N=0.12, O=0.26, Ne=0.33, Na=0.30, Mg=0.32, Al=0.30, Si=0.30, S=0.31, Cl=0.31, Ar=0.54,
Ca=0.34, Cr=0.61, Fe=0.36, Co=0.30 \& Ni=0.62. We have added also the galactic absorption $N_H=6 \times 10^{20} cm^{-2}$.}
\begin{tabular}{lllll}
\hline
 \multicolumn{5}{c}{{\tt VNEI}}\\
\hline
\bf{Parameter} & \bf{Kes 73} & \bf{CTB 109} & \bf{N 49$^\dagger$} & \bf{Kes 75}\\
\hline
$N_H$ (cm$^{-2}$) & $2.51^{-0.08}_{+0.06}$ & $0.695^{-0.018}_{+0.005}$ & $1.03^{-0.02}_{+0.02}$ & $3.71^{-0.06}_{+0.07}$\\
kT$_{brems}$ (keV) & $0.41^{-0.03}_{+0.05}$ & - & $0.99^{-0.01}_{+0.02}$ & $0.31^{-0.04}_{+0.05}$\\
N$_{brems}$ (Norm. counts s$^{-1}$) & $0.5^{-0.2}_{+0.2}$ & - & $(5.4^{-0.3}_{+0.3}) \times 10^{-3}$ & $0.4^{-0.2}_{+0.5}$\\
$kT$ (keV) & $1.51^{-0.08}_{+0.15}$ & $0.297^{-0.004}_{+0.007}$ & $0.1650^{-0.0003}_{+0.0011}$ & $2.0^{-0.1}_{+0.2}$\\
$O$ & 1 (fixed) & $0.16^{-0.02}_{+0.01}$ & $0.137^{-0.003}_{+0.002}$ & 1 (fixed)\\
$Ne$ & 1 (fixed) & $0.27^{-0.01}_{+0.01}$ & $0.175^{-0.004}_{+0.004}$ & 1 (fixed)\\
$Mg$ & $1.30^{-0.11}_{+0.09}$ & $0.23^{-0.02}_{+0.01}$ & $0.36^{-0.01}_{+0.01}$ & $0.51^{-0.08}_{+0.09}$\\
$Si$ & $1.6^{-0.1}_{0.2}$ & $0.49^{-0.05}_{+0.03}$ & 1 (fixed) & $0.56^{-0.04}_{+0.05}$\\
$S$ & $2.1^{-0.2}_{+0.4}$ & 1 (fixed) & 1 (fixed) & $0.9^{-0.1}_{+0.2}$\\
$Ar$ & $3.1^{-0.6}_{+0.9}$ & 1 (fixed) & 1 (fixed) & $1.2^{-0.6}_{+0.8}$\\
$Ca$ & $6^{-2}_{+4}$ & 1 (fixed) & 1 (fixed) & 1 (fixed)\\
$Fe$ & 1 (fixed) & $0.226^{-0.024}_{+0.008}$ & 1 (fixed) & 1 (fixed)\\
$E_1$ (keV) & - & - & $0.729^{-0.002}_{+0.005}$ & -\\
$\sigma_1$ (keV) & - & - & $<0.07$ & -\\
$N_1$ (Norm. counts s$^{-1}$) & - & - & $(5.4^{-0.3}_{+0.3}) \times 10^{-3}$ & -\\
$E_2$ (keV) & - & - & $1.018^{-0.001}_{+0.001}$ & -\\
$\sigma_2$ (keV) & - & - & $<0.07$ & -\\
$N_2$ (Norm. counts s$^{-1}$) & - & - & $(1.20^{0.04}_{0.04}) \times 10^{-3}$ & -\\
$E_3$ (keV) & - & - & $1.467^{-0.008}_{+0.004}$ & -\\
$\sigma_3$ (keV) & - & - & $<0.08$ & -\\
$N_3$ (Norm. counts s$^{-1}$) & - & - & $(4.9^{0.6}_{0.6}) \times 10^{-5}$ & -\\
$E_4$ (keV) & - & - & $1.846^{-0.003}_{+0.003}$ & -\\
$\sigma_4$ (keV) & - & - & $<0.09$ & -\\
$N_4$ (Norm. counts s$^{-1}$) & - & - & $(1.56^{0.07}_{0.07}) \times 10^{-4}$ & -\\
$E_5$ (keV) & - & - & $1.998^{-0.003}_{+0.028}$ & -\\
$\sigma_5$ (keV) & - & - & $<0.09$ & -\\
$N_5$ (Norm. counts s$^{-1}$) & - & - & $(5.3^{0.5}_{0.5}) \times 10^{-5}$ & -\\
$E_6$ (keV) & - & - & $2.445^{-0.005}_{+0.005}$ & -\\
$\sigma_6$ (keV) & - & - & $<0.1$ & -\\
$N_6$ (Norm. counts s$^{-1}$) & - & - & $(6.4^{-0.4}_{+0.3}) \times 10^{-5}$ & -\\
$E_7$ (keV) & - & - & $3.12^{-0.02}_{+0.02}$ & -\\
$\sigma_7$ (keV) & - & - & $<0.1$ & -\\
$N_7$ (Norm. counts s$^{-1}$) & - & - & $(7^{-1}_{+1}) \times 10^{-6}$ & -\\
$\tau$ (s cm$^{-3}$) & $(5.1^{-0.8}_{+0.6}) \times 10^{10}$ & $(6.7^{-1.0}_{+0.8}) \times 10^{11}$ & $(1.3^{-0.2}_{+0.1}) \times 10^{12}$ & $(2.4^{-0.3}_{+0.3}) \times 10^{10}$\\
$N$ (Norm. counts s$^{-1}$) & $(3.9^{-0.9}_{+0.6}) \times 10^{-2}$ & $0.35^{-0.04}_{+0.02}$ & $1.69^{-0.02}_{+0.03}$ & $0.021^{-0.003}_{+0.003}$\\
$\chi^2_r$ & 1.56 (997) & 2.60 (491) & 1.87 (569) & 1.19 (236)\\
\hline
\hline
\end{tabular}
\label{plasmamodels}
\end{table}

\begin{figure}
\centering
\includegraphics[width=0.9\textwidth]{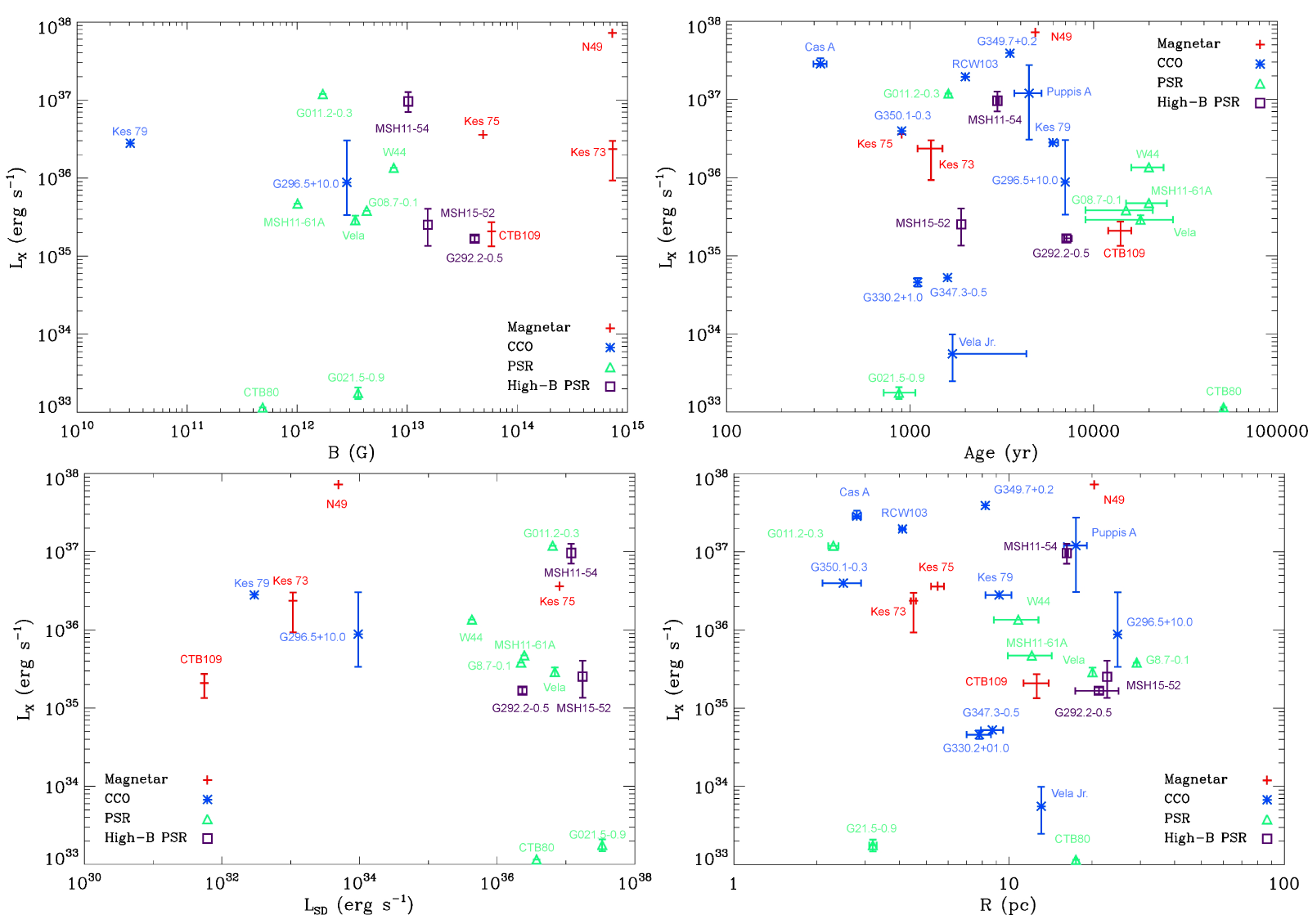}
\caption[X-ray luminosity of SNR versus PSR properties]{X-ray luminosity of the all observed, and securely associated, X-ray emitting SNRs
containing a magnetar, a CCO, a high-B pulsar or a normal pulsar, plotted versus magnetic-field (top-left), age (top-right), spin-down luminosity
(bottom-left) and radius (bottom-right).}
\label{snr_lum}
\end{figure}

In figure \ref{snr_lum} we have collected from the literature the X-ray luminosities from 0.5 to 10 keV of all observed SNRs brighter than
$\sim10^{33}$ erg s$^{-1}$, with an age lower than 100 kyr and having a confirmed association with a central source. For these remnants, we
obtain the age, distance, approximate radius, magnetic field and spin-down luminosity of the central source (whenever possible) from the
literature. All this information is summarized in table \ref{snrdata}. We have plotted the SNRs luminosities (excluding the contribution of the
central neutron star luminosity) as a function of the SNR age and dimension (although note that the latter parameter is highly dependent on the
environment of each remnant). For those remnants having a central neutron star with measured rotational properties, we plot the SNR luminosity as
a function of the pulsar surface dipolar magnetic field at the equator ($B=3.2\times10^{19}\sqrt{P\dot{P}}$ G), and the pulsar spin down
luminosity ($L_{sd}= 3.9\times10^{46}\dot{P}/P^{3}$ erg/s; always assuming the neutron star moment of inertia $I=10^{45}$ g cm$^2$), and where
$P$ is the pulsar rotation period in seconds and $\dot{P}$ its first derivative.

\begin{table}
\tiny
\centering
\caption[SNRs considered in our X-ray luminosity analysis]{SNRs considered in our X-ray luminosity analysis. The data without references is
extracted from this work or deduced from the data obtained in the literature.}
\label{snrdata}
\begin{tabular}{l@{\ \ }l@{\ \ }l@{\ \ }l@{\ \ }l@{\ \ }l@{\ \ }l@{\ \ }l@{\ \ }l}
\hline
 \multicolumn{9}{c}{SNRs with magnetars}\\
\hline
\bf{Name} & \bf{Central source} & \bf{Distance} &  \bf{Radius} & \bf{Age} & \bf{$\dot{E}$} & \bf{B$_s$} & \bf{F$_X$} & \bf{L$_X$}\\
 & & \bf{(kpc)} & \bf{(pc)} & \bf{(kyr)} & \bf{(erg s$^{-1}$)} & \bf{(G)} & \bf{(erg cm$^{-2}$ s$^{-1}$)} & \bf{(erg s$^{-1}$)}\\
\hline
Kes 75 & J1846-0258 [26] & 10.6 [66] & 5.5$^{-0.3}_{+0.3}$ [10] & 0.9 [6] & $8.06 \times 10^{36}$ [40] & $4.88 \times 10^{13}$ [40] & $2.69 \times 10^{-10}$ & $3.61 \times 10^{36}$\\
Kes 73 & 1E 1841-045 [72] & 6.7$^{-1.0}_{+1.8}$ [61] &  4.5$^{-0.1}_{+0.1}$ [10] & 1.3$^{-0.2}_{+0.2}$ [73] & $1.08 \times 10^{33}$ [31] & $7.34 \times 10^{14}$ [31] & $4.39 \times 10^{-10}$ & $2.36^{-0.65}_{+1.43} \times 10^{36}$\\
N 49 & RX J0526-6604 [36] & 50 [36] & 20.4 [10] & 4.8 [48] & $4.92 \times 10^{33}$ [36] & $7.32 \times 10^{14}$ [36] & $2.41 \times 10^{-10}$ & $7.21 \times 10^{37}$\\
CTB 109 & 1E 2259+586 [2] & 3$^{-0.5}_{+0.5}$ [33] & 12.6$^{-1.3}_{+1.3}$ [10] & 14$^{-2}_{+2}$ [62] & $5.54 \times 10^{31}$ [2] & $5.84 \times 10^{13}$ [2] & $1.94 \times 10^{-10}$ & $2.09^{-0.64}_{+0.75} \times 10^{35}$\\
\hline
\hline
 \multicolumn{9}{c}{SNRs with CCOs}\\
\hline
Cas A & CXO J2323+5848 [45] & 3.4$^{-0.1}_{+0.3}$ [53] & 2.8$^{-0.1}_{+0.1}$ [10] & 0.326$^{-27}_{+27}$ [17] & - & - & $2.06 \times 10^{-8}$ [10] & $2.85^{-0.20}_{+0.50} \times 10^{37}$ [10]\\
G350.1-0.3 & XMMU J1720-3726 [23] & 4.5 [23] & 2.5$^{-0.4}_{+0.4}$ [10] & 0.9 [23] & - & - & $1.64 \times 10^{-9}$ [10] & $3.97 \times 10^{36}$ [10]\\
G330.2+1.0 & CXOU J1601-5133 [47] & 4.9$^{-0.3}_{+0.3}$ [53] & 7.8$^{-0.8}_{+0.8}$ [10] & 1.1 [47] & - & - & $1.60 \times 10^{-11}$ [71] & $4.60^{-0.55}_{+0.57} \times 10^{34}$ [71]\\
G347.3-0.5 & 1 WGA J1713-3949 [38] & 1 [34] & 8.7$^{-0.8}_{+0.8}$ [18] & 1.6 [18] & - & - & $4.40 \times 10^{-10}$ [51] & $5.26 \times 10^{34}$\\
Vela Jr. & CXOU J0852-4617 [49] & 0.75$^{-0.55}_{+0.25}$ [31] &  13.1 [10] & 1.7$^{-0}_{+2.6}$ [31] & - & - & $8.30 \times 10^{-11}$ [1] & $5.58^{-3.10}_{+4.34} \times 10^{33}$\\
RCW 103 & 1E 1613-5055 [25] & 3.1 [55] &  4.1$^{-0.1}_{+0.1}$ [10] & 2 [7] & - & - & $1.70 \times 10^{-8}$ [10] & $1.95 \times 10^{37}$\\
G349.7+0.2 & CXOU J1718-3726 [39] & 22.4 [20] &  8.2 [39] & 3.5 [39] & - & - & $6.50 \times 10^{-10}$ [39] & $3.90 \times 10^{37}$\\
Puppis A & RX J0822-4300 [4] & 2.2$^{-0.3}_{+0.3}$ [54] &  17.5$^{-1.7}_{+1.7}$ [16] & 4.45$^{-0.75}_{+0.75}$ [4] & - & - & $2.16 \times 10^{-8}$ & $1.20^{0.90}_{+1.55} \times 10^{37}$ [14]\\
Kes 79 & J1852+0040 [63] & 7.1 [8] &  9.2$^{-1.0}_{+1.0}$ [10] & 6.0$^{-0.2}_{+0.4}$ [67] & $2.96 \times 10^{32}$ [27] & $3.05 \times 10^{10}$ [27] & $4.64 \times 10^{-10}$ [67] & $2.80 \times 10^{36}$ [67]\\
G296.5+10.0 & 1E 1207-5209 [24] & 2.1$^{-0.8}_{+1.8}$ [24] &  24.8 [32] & 7 [57] & $9.58 \times 10^{33}$ [50] & $2.83 \times 10^{12}$ [50] & $1.67 \times 10^{-9}$ [44] & $8.81^{-5.40}_{+21.60} \times 10^{34}$\\
\hline
\hline
 \multicolumn{9}{c}{SNRs with high-B PSRs}\\
\hline
MSH 15-52 & J1513-5908 [21] & 5.2$^{-1.4}_{+1.4}$ [15] & 22.7 [46] & 1.9 [15] & $1.75 \times 10^{37}$ [41] & $1.54 \times 10^{13}$ [41] & $7.80 \times 10^{-11}$ [46] & $2.52^{-1.17}_{+1.54} \times 10^{35}$\\
MSH 11-54 & J1124-5916 [29] & 6.2$^{-0.9}_{+0.9}$ [22] & 16.2$^{-0.2}_{+0.2}$ [10] & 2.99$^{-0.06}_{+0.06}$ [76] & $1.19 \times 10^{37}$ [52] & $1.02 \times 10^{13}$ [52] & $2.09 \times 10^{-9}$ [10] & $9.61^{-2.59}_{+2.99} \times 10^{36}$\\
G292.2-0.5 & J1119-6127 [37] & 8.4$^{-0.4}_{+0.4}$ [9] &  21.1$^{-3.8}_{+3.8}$ [10] & 7.1$^{-0.2}_{+0.5}$ [37] & $2.34 \times 10^{36}$ [75] & $4.10 \times 10^{13}$ [75] & $1.98 \times 10^{-11}$ [37] & $1.67^{-0.15}_{+0.16} \times 10^{35}$\\
\hline
\hline
 \multicolumn{9}{c}{SNRs with normal PSRs}\\
\hline
G21.5-0.9 & J1833-1034 [43] & 4.7$^{-0.4}_{+0.4}$ [69] & 3.2$^{-0.1}_{+0.1}$ [10] & 0.87$^{-1.5}_{+2.0}$ [5] & $3.37 \times 10^{37}$ [58] & $3.58 \times 10^{12}$ [58] & $6.69 \times 10^{-13}$ & $1.77^{-0.31}_{0.29} \times 10^{33}$ [43]\\
G11.2-0.3 & J1811-1925 [70] & 5 [30] & 2.3$^{-0.1}_{+0.1}$ [10] & 1.616 [68] & $6.42 \times 10^{36}$ [70] & $1.71 \times 10^{12}$ [70] & $3.98 \times 10^{-9}$ [10] & $1.19 \times 10^{37}$ [10]\\
G8.7-0.1 & J1803-2137 [19] & 4 [19] & 29.1 [19] & 15$^{-6}_{+6}$ [19] & $2.22 \times 10^{36}$ [77] & $4.92 \times 10^{12}$ [77] & $2.00 \times 10^{-10}$ [19] & $3.83 \times 10^{35}$\\
Vela & J0835-4510 [3] & 0.287$^{-0.017}_{+0.019}$ [13] & 20.1 [42] & 18$^{-9}_{+9}$ [3] & $6.92 \times 10^{36}$ [12] & $3.38 \times 10^{12}$ [12] & $2.94 \times 10^{-8}$ & $2.90^{-0.34}_{+0.39} \times 10^{35}$ [42]\\
MSH 11-61A & J1105-6107 [64] & 7 [64] & 12.1$^{-2.2}_{+2.2}$ [64] & 20$^{-5}_{+5}$ [64] & $2.48 \times 10^{36}$ [74] & $1.01 \times 10^{12}$ [74] & $8.06 \times 10^{-11}$ [10] & $4.71 \times 10^{35}$ [10]\\
W 44 & J1856+0113 [11] & 2.5 [11] & 10.8$^{-2.0}_{+2.0}$ [11] & 20$^{-4}_{+4}$ [11] & $4.30 \times 10^{35}$ [28] & $7.55 \times 10^{12}$ [28] & $1.80 \times 10^{-9}$ [56] & $1.35 \times 10^{36}$\\
CTB 80 & J1952+3252 [60] & 2 [65] & 1.5 [60] & 51 [78] & $3.74 \times 10^{36}$ [28] & $4.86 \times 10^{11}$ [28] & $2.40 \times 10^{-12}$ & $1.15 \times 10^{33}$ [59]\\
\hline
\hline
\multicolumn{9}{l}{
\begin{minipage}{12cm}
The references are: $^{[1]}$\citet{aharonian07}, $^{[2]}$\citet{archibald13}, $^{[3]}$\citet{aschenbach95}, $^{[4]}$\citet{becker12},
$^{[5]}$\citet{bietenholz08}, $^{[6]}$\citet{blanton96}, $^{[7]}$\citet{carter97}, $^{[8]}$\citet{case98}, $^{[9]}$\citet{caswell04},
$^{[10]}${\it Chandra} SNR catalog\footnote{\tt{http://hea-www.cfa.harvard.edu/ChandraSNR/}}, $^{[11]}$\citet{cox99}, $^{[12]}$\citet{dodson02},
$^{[13]}$\citet{dodson03}, $^{[14]}$\citet{dubner13}, $^{[15]}$\citet{fang10a}, $^{[16]}$\citet{ferrand12}, $^{[17]}$\citet{fesen06},
$^{[18]}$\citet{fesen12}, $^{[19]}$\citet{finley94}, $^{[20]}$\citet{frail96a}, $^{[21]}$\citet{gaensler99}, $^{[22]}$\citet{gaensler03},
$^{[23]}$\citet{gaensler08}, $^{[24]}$\citet{giacani00}, $^{[25]}$\citet{gotthelf99a}, $^{[26]}$\citet{gotthelf00b}, $^{[27]}$\citet{halpern10},
$^{[28]}$\citet{hobbs04}, $^{[29]}$\citet{hughes03}, $^{[30]}$\citet{kaspi01}, $^{[31]}$\citet{katsuda08}, $^{[32]}$\citet{kellett87},
$^{[33]}$\citet{kothes02}, $^{[34]}$\citet{koyama97}, $^{[31]}$\citet{kuiper06}, $^{[36]}$\citet{kulkarni03}, $^{[37]}$\citet{kumar12},
$^{[38]}$\citet{lazendic03}, $^{[39]}$\citet{lazendic05}, $^{[40]}$\citet{livingstone11a}, $^{[41]}$\citet{livingstone11b},
$^{[42]}$\citet{lu00}, $^{[43]}$\citet{matheson10}, $^{[44]}$\citet{matsui88}, $^{[45]}$\citet{mereghetti02}, $^{[46]}$\citet{mineo01},
$^{[47]}$\citet{park09}, $^{[48]}$\citet{park12}, $^{[49]}$\citet{pavlov01}, $^{[50]}$\citet{pavlov02}, $^{[51]}$\citet{pfeffermann96},
$^{[52]}$\citet{ray11}, $^{[53]}$\citet{reed95}, $^{[54]}$\citet{reynoso95}, $^{[55]}$\citet{reynoso04}, $^{[56]}$\citet{rho94},
$^{[57]}$\citet{roger88}, $^{[58]}$\citet{roy12}, $^{[59]}$\citet{safiharb94}, $^{[60]}$\citet{safiharb95}, $^{[61]}$\citet{sanbonmatsu92},
$^{[62]}$\citet{sasaki13}, $^{[63]}$\citet{seward03}, $^{[64]}$\citet{slane02a}, $^{[65]}$\citet{strom00}, $^{[66]}$\citet{su09},
$^{[67]}$\citet{sun04}, $^{[68]}$\citet{tam03}, $^{[69]}$\citet{tian08a}, $^{[70]}$\citet{torii99}, $^{[71]}$\citet{torii06},
$^{[72]}$\citet{vasisht97}, $^{[73]}$\citet{vink06}, $^{[74]}$\citet{wang00}, $^{[75]}$\citet{weltevrede11}, $^{[76]}$\citet{winkler09},
$^{[77]}$\citet{yuan10}, $^{[78]}$\citet{zeiger08}.
\end{minipage}}
\end{tabular}
\end{table}

In order to search for any correlations in the SNRs and pulsars characteristics (see figure \ref{snr_lum}), we run a Spearman test. We searched
for correlations between the X-ray luminosity and other features of the sources of our sample, such as dimension of the remnant, age, surface
magnetic field strength and spin down power of the associated pulsar. To this end, we employed a Spearman rank correlation test, and evaluated
the significance of the value of the coefficient of correlation $r$ obtained, by computing $t=r\sqrt{(N-2)/(1-r^2)}$, which is distributed
approximately as Student's distribution with $N-2$ degrees of freedom, where $N$ is the number of couples considered. The results we obtained are
listed in table \ref{spearman}; no correlation is found at a significance level larger than 99\% , or any significant difference in luminosity
between SNRs surrounding magnetars and those around other classes of isolated neutron stars.

\begin{table}[t!]
\centering
\caption[Spearman coefficient comparing the X-ray luminosity with the PSR parameters.]{Spearman correlation coefficient (r), number of couples
considered (N) and probability that the two samples are not correlated (p) evaluated by comparing the X-ray luminosity of the sources of our
sample with the age, radius, surface magnetic field strength and spin-down luminosity.}
\label{spearman}
\begin{tabular}{lccc}
\hline
Parameters & \bf{r} & \bf{N} & \bf{p}\\
\hline
$L_X$ vs. age & -0.158 & 24 & 0.46\\
$L_X$ vs. radius & -0.245 & 24 & 0.25\\
$L_X$ vs. $B$ & 0.271 & 16 & 0.31\\
$L_X$ vs. $L_{sd}$ & -0.309 & 16 & 0.25\\
\hline
\end{tabular}
\end{table}

We have also been looking at the number of pulsars having detected SNRs as a function of age, and compared it to the magnetar case. We caution,
however, that there are several systematic effects in this comparison (different detection wavebands, distance, low number of magnetars in
comparison with pulsars, etc.), but we were mostly interested in looking for a general trend. In figure \ref{fig:percentage} we plot the result
of this comparison, where we can see how on average (with all the due caveats) for a similar age, pulsars and magnetars seem to show a similar
probability to have a detected SNR.

\section{Conclusions}
\label{sec6.4}

We have reported on the re-analysis of the X-ray emission of SNRs surrounding magnetars, using an empirical modeling of their spectrum with a
Bremsstrahlung continuum plus several emission lines modeled by Gaussian functions. Our analysis, and the comparison of the emission of those
remnants with other bright SNR surrounding normal pulsars suggest the following conclusions:

\begin{figure}
\centering
\includegraphics[width=0.4\textwidth]{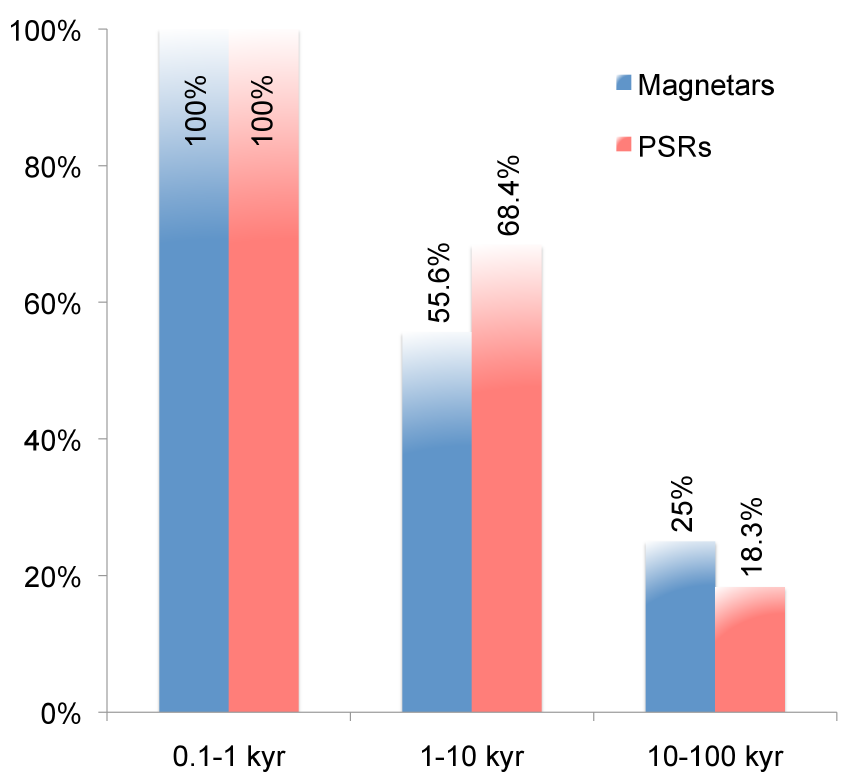}
\caption[Percentage of PSRs and magnetars having a detected SNR]{Percentage of pulsars and magnetars having a detected SNR as a function of the
age.}
\label{fig:percentage}
\end{figure}

\begin{itemize}
\item We find no evidence of generally enhanced ionization states in the elements observed in magnetars' SNRs compared to remnants observed
around lower magnetic pulsars.
\item No significant correlation is observed between the SNRs X-ray luminosities and the pulsar magnetic fields.
\item We show evidence that the percentage of magnetars and pulsars hosted in a detectable SNR are very similar, at a similar age.
\end{itemize}

Our findings do not support the claim of magnetars being formed via more energetic supernovae, or having a large rotational energy budget at
birth that is released in the surrounding medium in the first phases of the magnetar  formation. However, we note that although we do not
find any hint in the SNRs to support such an idea, we cannot exclude that: 1) most of the rotational energy has been emitted via neutrinos or
gravitational waves, hence with no interaction with the remnant; or 2) we are restricted to a very small sample, and with larger statistics
some correlation might be observed in the future.

\chapter{Concluding remarks}
\label{chap7}

\ifpdf
    \graphicspath{{Chapter7/Figs/Raster/}{Chapter7/Figs/PDF/}{Chapter7/Figs/}}
\else
    \graphicspath{{Chapter7/Figs/Vector/}{Chapter7/Figs/}}
\fi

\section{On this thesis}
\label{sec7.1}

In this thesis, we have studied a few of the current open problems on pulsars, pulsar wind nebulae and supernova remnants, from a theoretical
and/or observational point of view. We have made special emphasis in spectral modeling of PWNe, and the comparison of the X-ray emission of SNRs
hosting magnetars and canonical pulsars.

Spectral modeling of PWNe is a useful tool to characterize the pulsar wind and obtain information about the distribution of accelerated particles
as a function of the energy, degree of magnetization, energy densities of the background photon fields and the dynamical evolution of the PWN-SNR
complex. The luminosity of these objects is strongly related with the characteristics of the central pulsar and the interaction of the pulsar
wind with the SNR ejecta. We have developed a detailed spectral code which reproduces the electromagnetic spectrum of PWNe in free expansion, and
we have succeeded in fitting the spectra of many PWNe. We have studied approximations found in other models in the literature, and tested how
these approximations affect the resulting PWN predicted spectra, and its time evolution. We have seen that the differences between models can be
as large as 100\% in flux .This introduces large uncertainties in the parameters when we apply these approximations in a PWN population study at
different ages.

We have also shed light in other long standing PWN issues as: i) the synchrotron self Compton dominance in the Crab Nebula (why only in Crab?
which are the configuration parameters to have a SSC dominated nebula?); ii) the particle dominance for the PWNe detected at TeV (why are they
particle dominated? Is there any observational bias? can we detect magnetic dominated nebulae?); and iii) we have put some constrains the
detectability at TeV of Crab-like PWNe by doing a parameter phase space exploration. In particular, we have observed that no Crab-like PWNe is
SSC dominated if the spin-down luminosity is less than $\sim$70\% of $\dot{E}_{Crab}$ and low ages (less than a few thousand years), and only
particle dominated PWNe can be detected by the current Cherenkov telescopes (i.e. H.E.S.S.-like telescopes) above $\sim$1\% of $\dot{E}_{Crab}$.
No magnetic dominated PWNe could be detected for PSRs younger than $\sim$10 kyr.

We have made a systematic study of all the young TeV-detected PWNe to evaluate the possible existence of common evolutive trends. We confirmed
the unique SSC dominance in the Crab Nebula and for the rest of nebulae, the IC contribution is generally dominated by the FIR photon field.
TeV-detected PWNe show similar radio, X-ray and $\gamma$-ray flux efficiencies with the exception of G292.2$-$0.5, for which the X-ray efficiency
is very low. Generally, the electromagnetic spectrum is well described by a broken power law spectrum, which falls in a mean Lorentz factor of
$10^5$. Multiplicities are large $>10^5$ and they are generally particle dominated.  We do not find significant correlations between the
efficiencies of emission at different frequencies and the magnetization. The same happens with the pulsar's characteristic age and the radio and
X-ray luminosities. On the other hand, we find correlations of the radio and X-ray with the spin-down luminosity. Some anti-correlations have
been found between the ratios of IC to synchrotron luminosities and the spin-down luminosity, and also between the $\gamma$-ray luminosity and
the characteristic age.

The existence of PWNe with relatively high spin-down luminosity PSRs non-detected at TeV made us wondering if this could be explained in the
classical context of low magnetized nebulae, or otherwise these cases could be candidates of PWNe with high magnetization. We have modeled the
non-detected at TeV PWNe with PSRs with an spin-down luminosity $>10^{37}$ erg s$^{-1}$ and we have noticed that the non-detection of these
nebulae can be explained with low magnetization models and predict the detection of G292.0+1.8 and G310.6$-$1.6 in a reasonable exposure time
with H.E.S.S. The synchrotron spectrum of these nebulae have been fitted assuming high magnetization with some caveats, which predict no
detection of these sources neither H.E.S.S.-like or CTA-like telescopes. Future observations will discern between both models. 

Regarding the X-ray analysis of supernova remnants, X-ray spectroscopy gives us a powerful tool to know about the energetics, kinetics, chemical
composition, abundances, level of ionization and interaction with the ambient medium of these objects. Here, we applied this technique to
investigate further on the formation mechanism of magnetars. The alpha-dynamo mechanism proposed in the 1990's is believed to release a large
rotational energy that might be observed in the SNRs characteristics. \citet{vink06} studied the energetics of the explosion observing SNRs in
X-rays with an associated magnetar, but they did not find any clear evidence for this additional energy. We have extended this work and looked
for the element ionization and the X-ray luminosity in comparison with other SNRs with an associated neutron star. Our main conclusions have been
that we have not found enhanced ionization states in the elements observed in SNRs with magnetars compared to other X-ray bright SNRs observed
around lower magnetic pulsars, and no significant correlation is observed between the X-ray luminosities of the SNRs and the PSR magnetic fields.

\section{Ongoing work}
\label{sec7.2}

As we said in the introduction, the current number of PWNe detected at TeV is $\sim$30, but this number will increase severely with the
forthcoming generation of Cherenkov radiation telescopes, as CTA. The new level of sensitivity, spatial resolution and number of new sources will
be challenging for these models, which will have to include high precision at radiative and magnetohydrodynamic level. It is important to be
prepared for that moment and continue the improvement of the current radiative codes for PWNe to have enough capacity at physical and programming
level to analyse the data. We are currently working on the implementation of new subroutines in TIDE-PWN to reproduce the interaction of the
reverse shock of the SNR with the PWN. When the reverse shock collides with the PWN shell, the PWN contracts itself, increasing the magnetic
field and burning off the high energy electrons. The introduction of the dynamics after the reverberation phase increase the number of PWNe
suitable for modeling.


\begin{spacing}{0.9}


\bibliographystyle{mn2e}
\cleardoublepage
\bibliography{References/references} 

\begin{thebibliography}{533}
\expandafter\ifx\csname natexlab\endcsname\relax\def\natexlab#1{#1}\fi

\bibitem[{{Abdo} {et~al}\mbox{.}(2009{\natexlab{a}}){Abdo}, {Ackermann},
  {Ajello}, {Atwood}, {Axelsson}, {Baldini}, {Ballet}, {Barbiellini}, {Baring},
  {Bastieri}, {Baughman}, {Bechtol}, {Bellazzini}, {Berenji}, {Bloom},
  {Bonamente}, {Borgland}, {Bregeon}, {Brez}, {Brigida}, {Bruel}, {Caliandro},
  {Cameron}, {Camilo}, {Caraveo}, {Casandjian}, {Cecchi}, {Chekhtman},
  {Cheung}, {Chiang}, {Ciprini}, {Claus}, {Cognard}, {Cohen-Tanugi}, {Conrad},
  {de Angelis}, {de Palma}, {Dormody}, {Silva}, {Drell}, {Dubois}, {Dumora},
  {Farnier}, {Favuzzi}, {Frailis}, {Freire}, {Fukazawa}, {Funk}, {Fusco},
  {Gargano}, {Gehrels}, {Germani}, {Giebels}, {Giglietto}, {Giordano},
  {Glanzman}, {Godfrey}, {Grenier}, {Grondin}, {Grove}, {Guillemot}, {Guiriec},
  {Halpern}, {Hanabata}, {Harding}, {Hayashida}, {Hays}, {Hobbs}, {Hughes},
  {J{\'o}hannesson}, {Johnson}, {Johnson}, {Johnson}, {Johnson}, {Johnston},
  {Kamae}, {Katagiri}, {Kataoka}, {Kawai}, {Kerr}, {Kn{\"o}dlseder}, {Kocian},
  {Kramer}, {Kuehn}, {Kuss}, {Lande}, {Latronico}, {Lemoine-Goumard}, {Longo},
  {Loparco}, {Lott}, {Lovellette}, {Lubrano}, {Lyne}, {Makeev}, {Manchester},
  {Marelli}, {Mazziotta}, {McEnery}, {Meurer}, {Michelson}, {Mitthumsiri},
  {Mizuno}, {Moiseev}, {Monte}, {Monzani}, {Morselli}, {Moskalenko}, {Murgia},
  {Nolan}, {Norris}, {Noutsos}, {Nuss}, {Ohsugi}, {Omodei}, {Orlando}, {Ormes},
  {Ozaki}, {Paneque}, {Panetta}, {Parent}, {Pepe}, {Pesce-Rollins}, {Piron},
  {Porter}, {Rain{\`o}}, {Rando}, {Ransom}, {Razzano}, {Reimer}, {Reimer},
  {Reposeur}, {Rochester}, {Rodriguez}, {Romani}, {Roth}, {Ryde},
  {Sadrozinski}, {Sanchez}, {Sander}, {Saz Parkinson}, {Scargle}, {Sgr{\`o}},
  {Siskind}, {Smith}, {Smith}, {Spandre}, {Spinelli}, {Stappers}, {Strickman},
  {Suson}, {Tajima}, {Takahashi}, {Tanaka}, {Thayer}, {Thayer}, {Theureau},
  {Thompson}, {Thorsett}, {Tibaldo}, {Torres}, {Tosti}, {Uchiyama}, {Usher},
  {Van Etten}, {Vilchez}, {Vitale}, {Waite}, {Wang}, {Wang}, {Watters},
  {Weltevrede}, {Winer}, {Wood}, {Ylinen}, \& {Ziegler}}]{abdo09b}
{Abdo} A.~A. {et~al.}, 2009{\natexlab{a}}, ApJ, 706, 1331

\bibitem[{{Abdo} {et~al}\mbox{.}(2010{\natexlab{a}}){Abdo}, {Ackermann},
  {Ajello}, {Atwood}, {Axelsson}, {Baldini}, {Ballet}, {Barbiellini}, {Baring},
  {Bastieri}, {Bechtol}, {Bellazzini}, {Berenji}, {Blandford}, {Bloom},
  {Bonamente}, {Borgland}, {Bregeon}, {Brez}, {Brigida}, {Bruel}, {Burnett},
  {Caliandro}, {Cameron}, {Camilo}, {Caraveo}, {Casandjian}, {Cecchi}, {{\c
  C}elik}, {Chekhtman}, {Cheung}, {Chiang}, {Ciprini}, {Claus}, {Cognard},
  {Cohen-Tanugi}, {Cominsky}, {Conrad}, {Dermer}, {de Angelis}, {de Luca}, {de
  Palma}, {Digel}, {Silva}, {Drell}, {Dubois}, {Dumora}, {Espinoza}, {Farnier},
  {Favuzzi}, {Fegan}, {Ferrara}, {Focke}, {Frailis}, {Freire}, {Fukazawa},
  {Funk}, {Fusco}, {Gargano}, {Gasparrini}, {Gehrels}, {Germani}, {Giavitto},
  {Giebels}, {Giglietto}, {Giordano}, {Glanzman}, {Godfrey}, {Grenier},
  {Grondin}, {Grove}, {Guillemot}, {Guiriec}, {Hanabata}, {Harding},
  {Hayashida}, {Hays}, {Hughes}, {J{\'o}hannesson}, {Johnson}, {Johnson},
  {Johnson}, {Johnson}, {Johnston}, {Kamae}, {Katagiri}, {Kataoka}, {Kawai},
  {Kerr}, {Kn{\"o}dlseder}, {Kocian}, {Kramer}, {Kuehn}, {Kuss}, {Lande},
  {Latronico}, {Lee}, {Lemoine-Goumard}, {Longo}, {Loparco}, {Lott},
  {Lovellette}, {Lubrano}, {Lyne}, {Makeev}, {Marelli}, {Mazziotta}, {McEnery},
  {Meurer}, {Michelson}, {Mitthumsiri}, {Mizuno}, {Moiseev}, {Monte},
  {Monzani}, {Moretti}, {Morselli}, {Moskalenko}, {Murgia}, {Nakamori},
  {Nolan}, {Norris}, {Noutsos}, {Nuss}, {Ohsugi}, {Omodei}, {Orlando}, {Ormes},
  {Ozaki}, {Paneque}, {Panetta}, {Parent}, {Pelassa}, {Pepe}, {Pesce-Rollins},
  {Pierbattista}, {Piron}, {Porter}, {Rain{\`o}}, {Rando}, {Ray}, {Razzano},
  {Reimer}, {Reimer}, {Reposeur}, {Ritz}, {Rochester}, {Rodriguez}, {Romani},
  {Roth}, {Ryde}, {Sadrozinski}, {Sanchez}, {Sander}, {Saz Parkinson},
  {Scargle}, {Sgr{\`o}}, {Siskind}, {Smith}, {Smith}, {Spandre}, {Spinelli},
  {Stappers}, {Strickman}, {Suson}, {Tajima}, {Takahashi}, {Tanaka}, {Thayer},
  {Thayer}, {Theureau}, {Thompson}, {Thorsett}, {Tibaldo}, {Torres}, {Tosti},
  {Tramacere}, {Uchiyama}, {Usher}, {Van Etten}, {Vasileiou}, {Vilchez},
  {Vitale}, {Waite}, {Wallace}, {Wang}, {Watters}, {Weltevrede}, {Winer},
  {Wood}, {Ylinen}, \& {Ziegler}}]{abdo10a}
{Abdo} A.~A. {et~al.}, 2010{\natexlab{a}}, ApJ, 708, 1254

\bibitem[{{Abdo} {et~al}\mbox{.}(2010{\natexlab{b}}){Abdo}, {Ackermann},
  {Ajello}, {Atwood}, {Axelsson}, {Baldini}, {Ballet}, {Barbiellini}, {Baring},
  {Bastieri}, \& et~al.}]{abdo10b}
{Abdo} A.~A. {et~al.}, 2010{\natexlab{b}}, ApJS, 187, 460

\bibitem[{{Abdo} {et~al}\mbox{.}(2009{\natexlab{b}}){Abdo}, {Ackermann},
  {Ajello}, {Atwood}, {Axelsson}, {Baldini}, {Ballet}, {Barbiellini},
  {Bastieri}, {Baughman}, {Bechtol}, {Bellazzini}, {Berenji}, {Blandford},
  {Bloom}, {Bonamente}, {Borgland}, {Bouvier}, {Bregeon}, {Brez}, {Brigida},
  {Bruel}, {Burnett}, {Caliandro}, {Cameron}, {Camilo}, {Caraveo},
  {Casandjian}, {Cecchi}, {{\c C}elik}, {Chekhtman}, {Cheung}, {Chiang},
  {Ciprini}, {Claus}, {Cognard}, {Cohen-Tanugi}, {Conrad}, {Dermer}, {de
  Angelis}, {de Palma}, {Digel}, {Dormody}, {do Couto e Silva}, {Drell},
  {Dubois}, {Dumora}, {Edmonds}, {Espinoza}, {Farnier}, {Favuzzi}, {Focke},
  {Frailis}, {Freire}, {Fukazawa}, {Fusco}, {Gargano}, {Gehrels}, {Germani},
  {Giebels}, {Giglietto}, {Giordano}, {Glanzman}, {Godfrey}, {Grenier},
  {Grondin}, {Grove}, {Guillemot}, {Guiriec}, {Hanabata}, {Harding},
  {Hayashida}, {Hays}, {Hobbs}, {Hughes}, {J{\'o}hannesson}, {Johnson},
  {Johnson}, {Johnson}, {Johnson}, {Johnston}, {Kamae}, {Kaspi}, {Katagiri},
  {Kataoka}, {Kawai}, {Keith}, {Kerr}, {Kn{\"o}dlseder}, {Kramer}, {Kuehn},
  {Kuss}, {Lande}, {Latronico}, {Lemoine-Goumard}, {Livingstone}, {Longo},
  {Loparco}, {Lott}, {Lovellette}, {Lubrano}, {Lyne}, {Makeev}, {Manchester},
  {Marelli}, {Mazziotta}, {McEnery}, {Meurer}, {Michelson}, {Mitthumsiri},
  {Mizuno}, {Moiseev}, {Monte}, {Monzani}, {Morselli}, {Moskalenko}, {Murgia},
  {Nolan}, {Nuss}, {Ohsugi}, {Omodei}, {Orlando}, {Ormes}, {Paneque},
  {Panetta}, {Parent}, {Pelassa}, {Pepe}, {Pesce-Rollins}, {Pierbattista},
  {Piron}, {Porter}, {Rain{\`o}}, {Rando}, {Ransom}, {Razzano}, {Reimer},
  {Reimer}, {Reposeur}, {Ritz}, {Rochester}, {Rodriguez}, {Romani}, {Ryde},
  {Sadrozinski}, {Sanchez}, {Sander}, {Saz Parkinson}, {Sgr{\`o}}, {Siskind},
  {Smith}, {Smith}, {Spandre}, {Spinelli}, {Stappers}, {Striani}, {Strickman},
  {Suson}, {Tajima}, {Takahashi}, {Tanaka}, {Thayer}, {Thayer}, {Theureau},
  {Thompson}, {Thorsett}, {Tibaldo}, {Torres}, {Tosti}, {Tramacere},
  {Uchiyama}, {Usher}, {Van Etten}, {Vasileiou}, {Vilchez}, {Vitale}, {Waite},
  {Wang}, {Watters}, {Weltevrede}, {Winer}, {Wood}, {Ylinen}, \&
  {Ziegler}}]{abdo09a}
{Abdo} A.~A. {et~al.}, 2009{\natexlab{b}}, ApJL, 699, L102

\bibitem[{{Abdo} {et~al}\mbox{.}(2008){Abdo}, {Ackermann}, {Atwood}, {Baldini},
  {Ballet}, {Barbiellini}, {Baring}, {Bastieri}, {Baughman}, {Bechtol},
  {Bellazzini}, {Berenji}, {Blandford}, {Bloom}, {Bogaert}, {Bonamente},
  {Borgland}, {Bregeon}, {Brez}, {Brigida}, {Bruel}, {Burnett}, {Caliandro},
  {Cameron}, {Caraveo}, {Carlson}, {Casandjian}, {Cecchi}, {Charles},
  {Chekhtman}, {Cheung}, {Chiang}, {Ciprini}, {Claus}, {Cohen-Tanugi},
  {Cominsky}, {Conrad}, {Cutini}, {Davis}, {Dermer}, {de Angelis}, {de Palma},
  {Digel}, {Dormody}, {do Couto e Silva}, {Drell}, {Dubois}, {Dumora},
  {Edmonds}, {Farnier}, {Focke}, {Fukazawa}, {Funk}, {Fusco}, {Gargano},
  {Gasparrini}, {Gehrels}, {Germani}, {Giebels}, {Giglietto}, {Giordano},
  {Glanzman}, {Godfrey}, {Grenier}, {Grondin}, {Grove}, {Guillemot}, {Guiriec},
  {Harding}, {Hartman}, {Hays}, {Hughes}, {J{\'o}hannesson}, {Johnson},
  {Johnson}, {Johnson}, {Johnson}, {Kamae}, {Kanai}, {Kanbach}, {Katagiri},
  {Kawai}, {Kerr}, {Kishishita}, {Kiziltan}, {Kn{\"o}dlseder}, {Kocian},
  {Komin}, {Kuehn}, {Kuss}, {Latronico}, {Lemoine-Goumard}, {Longo}, {Lonjou},
  {Loparco}, {Lott}, {Lovellette}, {Lubrano}, {Makeev}, {Marelli}, {Mazziotta},
  {McEnery}, {McGlynn}, {Meurer}, {Michelson}, {Mineo}, {Mitthumsiri},
  {Mizuno}, {Moiseev}, {Monte}, {Monzani}, {Morselli}, {Moskalenko}, {Murgia},
  {Nakamori}, {Nolan}, {Nuss}, {Ohno}, {Ohsugi}, {Okumura}, {Omodei},
  {Orlando}, {Ormes}, {Ozaki}, {Paneque}, {Panetta}, {Parent}, {Pelassa},
  {Pepe}, {Pesce-Rollins}, {Piano}, {Pieri}, {Piron}, {Porter}, {Rain{\`o}},
  {Rando}, {Ray}, {Razzano}, {Reimer}, {Reimer}, {Reposeur}, {Ritz},
  {Rochester}, {Rodriguez}, {Romani}, {Roth}, {Ryde}, {Sadrozinski}, {Sanchez},
  {Sander}, {Parkinson}, {Schalk}, {Sellerholm}, {Sgr{\`o}}, {Siskind},
  {Smith}, {Smith}, {Spandre}, {Spinelli}, {Starck}, {Strickman}, {Suson},
  {Tajima}, {Takahashi}, {Takahashi}, {Tanaka}, {Thayer}, {Thayer}, {Thompson},
  {Thorsett}, {Tibaldo}, {Torres}, {Tosti}, {Tramacere}, {Usher}, {Van Etten},
  {Vilchez}, {Vitale}, {Wang}, {Watters}, {Winer}, {Wood}, {Yasuda}, {Ylinen},
  \& {Ziegler}}]{abdo08}
{Abdo} A.~A. {et~al.}, 2008, Science, 322, 1218

\bibitem[{{Abdo} {et~al}\mbox{.}(2013){Abdo}, {Ajello}, {Allafort}, {Baldini},
  {Ballet}, {Barbiellini}, {Baring}, {Bastieri}, {Belfiore}, {Bellazzini}, \&
  et~al.}]{abdo13}
{Abdo} A.~A. {et~al.}, 2013, ApJS, 208, 17

\bibitem[{{Abdo} {et~al}\mbox{.}(2012){Abdo}, {Wood}, {DeCesar}, {Gargano},
  {Giordano}, {Ray}, {Parent}, {Harding}, {Miller}, {Wood}, \&
  {Wolff}}]{abdo12}
{Abdo} A.~A. {et~al.}, 2012, ApJ, 744, 146

\bibitem[{{Acciari} {et~al}\mbox{.}(2010){Acciari}, {Aliu}, {Arlen}, {Aune},
  {Bautista}, {Beilicke}, {Benbow}, {Boltuch}, {Bradbury}, {Buckley}, {Bugaev},
  {Butt}, {Byrum}, {Cesarini}, {Ciupik}, {Cui}, {Dickherber}, {Duke}, {Finley},
  {Finnegan}, {Fortson}, {Furniss}, {Galante}, {Gall}, {Gillanders}, {Godambe},
  {Gotthelf}, {Grube}, {Guenette}, {Gyuk}, {Hanna}, {Holder}, {Hui},
  {Humensky}, {Imran}, {Kaaret}, {Karlsson}, {Kertzman}, {Kieda}, {Konopelko},
  {Krawczynski}, {Krennrich}, {Lang}, {LeBohec}, {Maier}, {McArthur}, {McCann},
  {McCutcheon}, {Moriarty}, {Muhkerjee}, {Ong}, {Otte}, {Pandel}, {Perkins},
  {Pohl}, {Quinn}, {Ragan}, {Reyes}, {Reynolds}, {Roache}, {Rose},
  {Schroedter}, {Sembroski}, {Senturk}, {Slane}, {Smith}, {Steele}, {Swordy},
  {T{\v e}si{\'c}}, {Theiling}, {Thibadeau}, {Vassiliev}, {Vincent}, {Wakely},
  {Ward}, {Weekes}, {Weinstein}, {Weisgarber}, {Williams}, {Wissel}, {Wood}, \&
  {Zitzer}}]{acciari10}
{Acciari} V.~A. {et~al.}, 2010, ApJL, 719, L69

\bibitem[{{Acero} {et~al}\mbox{.}(2013){Acero}, {Ackermann}, {Ajello},
  {Allafort}, {Baldini}, {Ballet}, {Barbiellini}, {Bastieri}, {Bechtol},
  {Bellazzini}, {Blandford}, {Bloom}, {Bonamente}, {Bottacini}, {Brandt},
  {Bregeon}, {Brigida}, {Bruel}, {Buehler}, {Buson}, {Caliandro}, {Cameron},
  {Caraveo}, {Cecchi}, {Charles}, {Chaves}, {Chekhtman}, {Chiang}, {Chiaro},
  {Ciprini}, {Claus}, {Cohen-Tanugi}, {Conrad}, {Cutini}, {Dalton},
  {D'Ammando}, {de Palma}, {Dermer}, {Di Venere}, {Silva}, {Drell},
  {Drlica-Wagner}, {Falletti}, {Favuzzi}, {Fegan}, {Ferrara}, {Focke},
  {Franckowiak}, {Fukazawa}, {Funk}, {Fusco}, {Gargano}, {Gasparrini},
  {Giglietto}, {Giordano}, {Giroletti}, {Glanzman}, {Godfrey}, {Gr{\'e}goire},
  {Grenier}, {Grondin}, {Grove}, {Guiriec}, {Hadasch}, {Hanabata}, {Harding},
  {Hayashida}, {Hayashi}, {Hays}, {Hewitt}, {Hill}, {Horan}, {Hou}, {Hughes},
  {Inoue}, {Jackson}, {Jogler}, {J{\'o}hannesson}, {Johnson}, {Kamae},
  {Kawano}, {Kerr}, {Kn{\"o}dlseder}, {Kuss}, {Lande}, {Larsson}, {Latronico},
  {Lemoine-Goumard}, {Longo}, {Loparco}, {Lovellette}, {Lubrano}, {Marelli},
  {Massaro}, {Mayer}, {Mazziotta}, {McEnery}, {Mehault}, {Michelson},
  {Mitthumsiri}, {Mizuno}, {Monte}, {Monzani}, {Morselli}, {Moskalenko},
  {Murgia}, {Nakamori}, {Nemmen}, {Nuss}, {Ohsugi}, {Okumura}, {Orienti},
  {Orlando}, {Ormes}, {Paneque}, {Panetta}, {Perkins}, {Pesce-Rollins},
  {Piron}, {Pivato}, {Porter}, {Rain{\`o}}, {Rando}, {Razzano}, {Reimer},
  {Reimer}, {Reposeur}, {Ritz}, {Roth}, {Rousseau}, {Saz Parkinson}, {Schulz},
  {Sgr{\`o}}, {Siskind}, {Smith}, {Spandre}, {Spinelli}, {Suson}, {Takahashi},
  {Takeuchi}, {Thayer}, {Thayer}, {Thompson}, {Tibaldo}, {Tibolla},
  {Tinivella}, {Torres}, {Tosti}, {Troja}, {Uchiyama}, {Vandenbroucke},
  {Vasileiou}, {Vianello}, {Vitale}, {Werner}, {Winer}, {Wood}, \&
  {Yang}}]{acero13}
{Acero} F. {et~al.}, 2013, ApJ, 773, 77

\bibitem[{{Ackermann} {et~al}\mbox{.}(2011){Ackermann}, {Ajello}, {Baldini},
  {Ballet}, {Barbiellini}, {Bastieri}, {Bechtol}, {Bellazzini}, {Berenji},
  {Bloom}, {Bonamente}, {Borgland}, {Bouvier}, {Bregeon}, {Brez}, {Brigida},
  {Bruel}, {Buehler}, {Buson}, {Caliandro}, {Cameron}, {Camilo}, {Caraveo},
  {Casandjian}, {Cecchi}, {{\c C}elik}, {Charles}, {Chekhtman}, {Cheung},
  {Chiang}, {Ciprini}, {Claus}, {Cognard}, {Cohen-Tanugi}, {Conrad}, {Dermer},
  {de Angelis}, {de Luca}, {de Palma}, {Digel}, {Silva}, {Drell}, {Dubois},
  {Dumora}, {Favuzzi}, {Focke}, {Frailis}, {Fukazawa}, {Funk}, {Fusco},
  {Gargano}, {Germani}, {Giglietto}, {Giommi}, {Giordano}, {Giroletti},
  {Glanzman}, {Godfrey}, {Grenier}, {Grondin}, {Grove}, {Guillemot}, {Guiriec},
  {Hadasch}, {Hanabata}, {Harding}, {Hayashi}, {Hays}, {Hobbs}, {Hughes},
  {J{\'o}hannesson}, {Johnson}, {Johnson}, {Johnston}, {Kamae}, {Katagiri},
  {Kataoka}, {Keith}, {Kerr}, {Kn{\"o}dlseder}, {Kramer}, {Kuss}, {Lande},
  {Latronico}, {Lee}, {Lemoine-Goumard}, {Longo}, {Loparco}, {Lovellette},
  {Lubrano}, {Lyne}, {Makeev}, {Marelli}, {Mazziotta}, {McEnery}, {Mehault},
  {Michelson}, {Mizuno}, {Moiseev}, {Monte}, {Monzani}, {Morselli},
  {Moskalenko}, {Murgia}, {Nakamori}, {Naumann-Godo}, {Nolan}, {Noutsos},
  {Nuss}, {Ohsugi}, {Okumura}, {Ormes}, {Paneque}, {Panetta}, {Parent},
  {Pelassa}, {Pepe}, {Pesce-Rollins}, {Piron}, {Porter}, {Rain{\`o}}, {Rando},
  {Ransom}, {Ray}, {Razzano}, {Rea}, {Reimer}, {Reimer}, {Reposeur}, {Ripken},
  {Ritz}, {Romani}, {Sadrozinski}, {Sander}, {Saz Parkinson}, {Sgr{\`o}},
  {Siskind}, {Smith}, {Smith}, {Spandre}, {Spinelli}, {Strickman}, {Suson},
  {Takahashi}, {Takahashi}, {Tanaka}, {Thayer}, {Thayer}, {Theureau},
  {Thompson}, {Thorsett}, {Tibaldo}, {Torres}, {Tosti}, {Tramacere},
  {Uchiyama}, {Uehara}, {Usher}, {Vandenbroucke}, {Van Etten}, {Vasileiou},
  {Vilchez}, {Vitale}, {Waite}, {Wang}, {Weltevrede}, {Winer}, {Wood}, {Yang},
  {Ylinen}, \& {Ziegler}}]{ackermann11}
{Ackermann} M. {et~al.}, 2011, ApJ, 726, 35

\bibitem[{{Actis} {et~al}\mbox{.}(2011){Actis}, {Agnetta}, {Aharonian},
  {Akhperjanian}, {Aleksi{\'c}}, {Aliu}, {Allan}, {Allekotte}, {Antico},
  {Antonelli}, \& et~al.}]{actis11}
{Actis} M. {et~al.}, 2011, Experimental Astronomy, 32, 193

\bibitem[{{Aharonian} {et~al}\mbox{.}(2002){Aharonian}, {Akhperjanian},
  {Beilicke}, {Bernl{\"o}hr}, {B{\"o}rst}, {Bojahr}, {Bolz}, {Coarasa},
  {Contreras}, {Cortina}, {Denninghoff}, {Fonseca}, {Girma}, {G{\"o}tting},
  {Heinzelmann}, {Hermann}, {Heusler}, {Hofmann}, {Horns}, {Jung}, {Kankanyan},
  {Kestel}, {Kettler}, {Kohnle}, {Konopelko}, {Kornmeyer}, {Kranich},
  {Krawczynski}, {Lampeitl}, {Lopez}, {Lorenz}, {Lucarelli}, {Magnussen},
  {Mang}, {Meyer}, {Milite}, {Mirzoyan}, {Moralejo}, {Ona}, {Panter},
  {Plyasheshnikov}, {Prahl}, {P{\"u}hlhofer}, {Rauterberg}, {Reyes}, {Rhode},
  {Ripken}, {R{\"o}hring}, {Rowell}, {Sahakian}, {Samorski}, {Schilling},
  {Schr{\"o}der}, {Siems}, {Sobzynska}, {Stamm}, {Tluczykont}, {V{\"o}lk},
  {Wiedner}, {Wittek}, {Uchiyama}, {Takahashi}, \& {HEGRA
  Collaboration}}]{aharonian02}
{Aharonian} F. {et~al.}, 2002, A\&A, 393, L37

\bibitem[{{Aharonian} {et~al}\mbox{.}(2004){Aharonian}, {Akhperjanian},
  {Beilicke}, {Bernl{\"o}hr}, {B{\"o}rst}, {Bojahr}, {Bolz}, {Coarasa},
  {Contreras}, {Cortina}, {Denninghoff}, {Fonseca}, {Girma}, {G{\"o}tting},
  {Heinzelmann}, {Hermann}, {Heusler}, {Hofmann}, {Horns}, {Jung}, {Kankanyan},
  {Kestel}, {Kohnle}, {Konopelko}, {Kranich}, {Lampeitl}, {Lopez}, {Lorenz},
  {Lucarelli}, {Mang}, {Mazin}, {Meyer}, {Mirzoyan}, {Moralejo},
  {O{\~n}a-Wilhelmi}, {Panter}, {Plyasheshnikov}, {P{\"u}hlhofer}, {de los
  Reyes}, {Rhode}, {Ripken}, {Rowell}, {Sahakian}, {Samorski}, {Schilling},
  {Siems}, {Sobzynska}, {Stamm}, {Tluczykont}, {Vitale}, {V{\"o}lk}, {Wiedner},
  \& {Wittek}}]{aharonian04}
{Aharonian} F. {et~al.}, 2004, ApJ, 614, 897

\bibitem[{{Aharonian} {et~al}\mbox{.}(2005{\natexlab{a}}){Aharonian},
  {Akhperjanian}, {Aye}, {Bazer-Bachi}, {Beilicke}, {Benbow}, {Berge},
  {Berghaus}, {Bernl{\"o}hr}, {Boisson}, {Bolz}, {Borgmeier}, {Braun},
  {Breitling}, {Brown}, {Bussons Gordo}, {Chadwick}, {Chounet}, {Cornils},
  {Costamante}, {Degrange}, {Djannati-Ata{\"i}}, {O'C.~Drury}, {Dubus},
  {Ergin}, {Espigat}, {Feinstein}, {Fleury}, {Fontaine}, {Funk}, {Gallant},
  {Giebels}, {Gillessen}, {Goret}, {Hadjichristidis}, {Hauser}, {Heinzelmann},
  {Henri}, {Hermann}, {Hinton}, {Hofmann}, {Holleran}, {Horns}, {de Jager},
  {Jung}, {Kh{\'e}lifi}, {Komin}, {Konopelko}, {Latham}, {Le Gallou},
  {Lemi{\`e}re}, {Lemoine}, {Leroy}, {Lohse}, {Marcowith}, {Masterson},
  {McComb}, {de Naurois}, {Nolan}, {Noutsos}, {Orford}, {Osborne}, {Ouchrif},
  {Panter}, {Pelletier}, {Pita}, {P{\"u}hlhofer}, {Punch}, {Raubenheimer},
  {Raue}, {Raux}, {Rayner}, {Redondo}, {Reimer}, {Reimer}, {Ripken}, {Rob},
  {Rolland}, {Rowell}, {Sahakian}, {Saug{\'e}}, {Schlenker}, {Schlickeiser},
  {Schuster}, {Schwanke}, {Siewert}, {Sol}, {Steenkamp}, {Stegmann},
  {Tavernet}, {Terrier}, {Th{\'e}oret}, {Tluczykont}, {Vasileiadis}, {Venter},
  {Vincent}, {Visser}, {V{\"o}lk}, \& {Wagner}}]{aharonian05a}
{Aharonian} F. {et~al.}, 2005{\natexlab{a}}, A\&A, 432, L25

\bibitem[{{Aharonian} {et~al}\mbox{.}(2005{\natexlab{b}}){Aharonian},
  {Akhperjanian}, {Aye}, {Bazer-Bachi}, {Beilicke}, {Benbow}, {Berge},
  {Berghaus}, {Bernl{\"o}hr}, {Boisson}, {Bolz}, {Braun}, {Breitling}, {Brown},
  {Bussons Gordo}, {Chadwick}, {Chounet}, {Cornils}, {Costamante}, {Degrange},
  {Djannati-Ata{\"i}}, {O'C.~Drury}, {Dubus}, {Emmanoulopoulos}, {Espigat},
  {Feinstein}, {Fleury}, {Fontaine}, {Fuchs}, {Funk}, {Gallant}, {Giebels},
  {Gillessen}, {Glicenstein}, {Goret}, {Hadjichristidis}, {Hauser},
  {Heinzelmann}, {Henri}, {Hermann}, {Hinton}, {Hofmann}, {Holleran}, {Horns},
  {de Jager}, {Kh{\'e}lifi}, {Komin}, {Konopelko}, {Latham}, {Le Gallou},
  {Lemi{\`e}re}, {Lemoine-Goumard}, {Leroy}, {Lohse}, {Martineau-Huynh},
  {Marcowith}, {Masterson}, {McComb}, {de Naurois}, {Nolan}, {Noutsos},
  {Orford}, {Osborne}, {Ouchrif}, {Panter}, {Pelletier}, {Pita},
  {P{\"u}hlhofer}, {Punch}, {Raubenheimer}, {Raue}, {Raux}, {Rayner},
  {Redondo}, {Reimer}, {Reimer}, {Ripken}, {Rob}, {Rolland}, {Rowell},
  {Sahakian}, {Saug{\'e}}, {Schlenker}, {Schlickeiser}, {Schuster}, {Schwanke},
  {Siewert}, {Sol}, {Steenkamp}, {Stegmann}, {Tavernet}, {Terrier},
  {Th{\'e}oret}, {Tluczykont}, {Vasileiadis}, {Venter}, {Vincent}, {V{\"o}lk},
  \& {Wagner}}]{aharonian05b}
{Aharonian} F. {et~al.}, 2005{\natexlab{b}}, A\&A, 435, L17

\bibitem[{{Aharonian} {et~al}\mbox{.}(2008){Aharonian}, {Akhperjanian}, {Barres
  de Almeida}, {Bazer-Bachi}, {Behera}, {Beilicke}, {Benbow}, {Bernl{\"o}hr},
  {Boisson}, {Bolz}, {Borrel}, {Braun}, {Brion}, {Brown}, {B{\"u}hler},
  {Bulik}, {B{\"u}sching}, {Boutelier}, {Carrigan}, {Chadwick}, {Chounet},
  {Clapson}, {Coignet}, {Cornils}, {Costamante}, {Dalton}, {Degrange},
  {Dickinson}, {Djannati-Ata{\"i}}, {Domainko}, {Drury}, {Dubois}, {Dubus},
  {Dyks}, {Egberts}, {Emmanoulopoulos}, {Espigat}, {Farnier}, {Feinstein},
  {Fiasson}, {F{\"o}rster}, {Fontaine}, {Funk}, {F{\"u}{\ss}ling}, {Gallant},
  {Giebels}, {Glicenstein}, {Gl{\"u}ck}, {Goret}, {Hadjichristidis}, {Hauser},
  {Hauser}, {Heinzelmann}, {Henri}, {Hermann}, {Hinton}, {Hoffmann}, {Hofmann},
  {Holleran}, {Hoppe}, {Horns}, {Jacholkowska}, {de Jager}, {Jung},
  {Katarzy{\'n}ski}, {Kendziorra}, {Kerschhaggl}, {Kh{\'e}lifi}, {Keogh},
  {Komin}, {Kosack}, {Lamanna}, {Latham}, {Lemi{\`e}re}, {Lemoine-Goumard},
  {Lenain}, {Lohse}, {Martin}, {Martineau-Huynh}, {Marcowith}, {Masterson},
  {Maurin}, {Maurin}, {McComb}, {Moderski}, {Moulin}, {de Naurois}, {Nedbal},
  {Nolan}, {Ohm}, {Olive}, {de O{\~n}a Wilhelmi}, {Orford}, {Osborne},
  {Ostrowski}, {Panter}, {Pedaletti}, {Pelletier}, {Petrucci}, {Pita},
  {P{\"u}hlhofer}, {Punch}, {Ranchon}, {Raubenheimer}, {Raue}, {Rayner},
  {Renaud}, {Ripken}, {Rob}, {Rolland}, {Rosier-Lees}, {Rowell}, {Rudak},
  {Ruppel}, {Sahakian}, {Santangelo}, {Schlickeiser}, {Sch{\"o}ck},
  {Schr{\"o}der}, {Schwanke}, {Schwarzburg}, {Schwemmer}, {Shalchi}, {Sol},
  {Spangler}, {Stawarz}, {Steenkamp}, {Stegmann}, {Superina}, {Tam},
  {Tavernet}, {Terrier}, {van Eldik}, {Vasileiadis}, {Venter}, {Vialle},
  {Vincent}, {Vivier}, {V{\"o}lk}, {Volpe}, {Wagner}, {Ward}, {Zdziarski}, \&
  {Zech}}]{aharonian08}
{Aharonian} F. {et~al.}, 2008, A\&A, 477, 353

\bibitem[{{Aharonian} {et~al}\mbox{.}(2006{\natexlab{a}}){Aharonian},
  {Akhperjanian}, {Bazer-Bachi}, {Beilicke}, {Benbow}, {Berge}, {Bernl{\"o}hr},
  {Boisson}, {Bolz}, {Borrel}, {Braun}, {Breitling}, {Brown}, {B{\"u}hler},
  {B{\"u}sching}, {Carrigan}, {Chadwick}, {Chounet}, {Cornils}, {Costamante},
  {Degrange}, {Dickinson}, {Djannati-Ata{\"i}}, {O'C.~Drury}, {Dubus},
  {Egberts}, {Emmanoulopoulos}, {Espigat}, {Feinstein}, {Ferrero}, {Fiasson},
  {Fontaine}, {Funk}, {Funk}, {Gallant}, {Giebels}, {Glicenstein}, {Goret},
  {Hadjichristidis}, {Hauser}, {Hauser}, {Heinzelmann}, {Henri}, {Hermann},
  {Hinton}, {Hofmann}, {Holleran}, {Horns}, {Jacholkowska}, {de Jager},
  {Kh{\'e}lifi}, {Komin}, {Konopelko}, {Kosack}, {Latham}, {Le Gallou},
  {Lemi{\`e}re}, {Lemoine-Goumard}, {Lohse}, {Martin}, {Martineau-Huynh},
  {Marcowith}, {Masterson}, {McComb}, {de Naurois}, {Nedbal}, {Nolan},
  {Noutsos}, {Orford}, {Osborne}, {Ouchrif}, {Panter}, {Pelletier}, {Pita},
  {P{\"u}hlhofer}, {Punch}, {Raubenheimer}, {Raue}, {Rayner}, {Reimer},
  {Reimer}, {Ripken}, {Rob}, {Rolland}, {Rowell}, {Sahakian}, {Saug{\'e}},
  {Schlenker}, {Schlickeiser}, {Schwanke}, {Sol}, {Spangler}, {Spanier},
  {Steenkamp}, {Stegmann}, {Superina}, {Tavernet}, {Terrier}, {Th{\'e}oret},
  {Tluczykont}, {van Eldik}, {Vasileiadis}, {Venter}, {Vincent}, {V{\"o}lk},
  {Wagner}, \& {Ward}}]{aharonian06a}
{Aharonian} F. {et~al.}, 2006{\natexlab{a}}, A\&A, 457, 899

\bibitem[{{Aharonian} {et~al}\mbox{.}(2006{\natexlab{b}}){Aharonian},
  {Akhperjanian}, {Bazer-Bachi}, {Beilicke}, {Benbow}, {Berge}, {Bernl{\"o}hr},
  {Boisson}, {Bolz}, {Borrel}, {Braun}, {Breitling}, {Brown}, {Chadwick},
  {Chounet}, {Cornils}, {Costamante}, {Degrange}, {Dickinson},
  {Djannati-Ata{\"i}}, {Drury}, {Dubus}, {Emmanoulopoulos}, {Espigat},
  {Feinstein}, {Fontaine}, {Fuchs}, {Funk}, {Gallant}, {Giebels}, {Gillessen},
  {Glicenstein}, {Goret}, {Hadjichristidis}, {Hauser}, {Heinzelmann}, {Henri},
  {Hermann}, {Hinton}, {Hofmann}, {Holleran}, {Horns}, {Jacholkowska}, {de
  Jager}, {Kh{\'e}lifi}, {Komin}, {Konopelko}, {Latham}, {Le Gallou},
  {Lemi{\`e}re}, {Lemoine-Goumard}, {Leroy}, {Lohse}, {Martin},
  {Martineau-Huynh}, {Marcowith}, {Masterson}, {McComb}, {de Naurois}, {Nolan},
  {Noutsos}, {Orford}, {Osborne}, {Ouchrif}, {Panter}, {Pelletier}, {Pita},
  {P{\"u}hlhofer}, {Punch}, {Raubenheimer}, {Raue}, {Raux}, {Rayner}, {Reimer},
  {Reimer}, {Ripken}, {Rob}, {Rolland}, {Rowell}, {Sahakian}, {Saug{\'e}},
  {Schlenker}, {Schlickeiser}, {Schuster}, {Schwanke}, {Siewert}, {Sol},
  {Spangler}, {Steenkamp}, {Stegmann}, {Tavernet}, {Terrier}, {Th{\'e}oret},
  {Tluczykont}, {Vasileiadis}, {Venter}, {Vincent}, {V{\"o}lk}, \&
  {Wagner}}]{aharonian06b}
{Aharonian} F. {et~al.}, 2006{\natexlab{b}}, ApJ, 636, 777

\bibitem[{{Aharonian} {et~al}\mbox{.}(2007){Aharonian}, {Akhperjanian},
  {Bazer-Bachi}, {Beilicke}, {Benbow}, {Berge}, {Bernl{\"o}hr}, {Boisson},
  {Bolz}, {Borrel}, {Braun}, {Brown}, {B{\"u}hler}, {B{\"u}sching}, {Carrigan},
  {Chadwick}, {Chounet}, {Coignet}, {Cornils}, {Costamante}, {Degrange},
  {Dickinson}, {Djannati-Ata{\"i}}, {Drury}, {Dubus}, {Egberts},
  {Emmanoulopoulos}, {Espigat}, {Feinstein}, {Ferrero}, {Fiasson}, {Filipovic},
  {Fontaine}, {Fukui}, {Funk}, {Funk}, {F{\"u}{\ss}ling}, {Gallant}, {Giebels},
  {Glicenstein}, {Goret}, {Hadjichristidis}, {Hauser}, {Hauser}, {Heinzelmann},
  {Henri}, {Hermann}, {Hinton}, {Hiraga}, {Hoffmann}, {Hofmann}, {Holleran},
  {Hoppe}, {Horns}, {Ishisaki}, {Jacholkowska}, {de Jager}, {Kendziorra},
  {Kerschhaggl}, {Kh{\'e}lifi}, {Komin}, {Konopelko}, {Kosack}, {Lamanna},
  {Latham}, {Le Gallou}, {Lemi{\`e}re}, {Lemoine-Goumard}, {Lohse}, {Martin},
  {Martineau-Huynh}, {Marcowith}, {Masterson}, {Maurin}, {McComb}, {Moulin},
  {Moriguchi}, {de Naurois}, {Nedbal}, {Nolan}, {Noutsos}, {Orford}, {Osborne},
  {Ouchrif}, {Panter}, {Pelletier}, {Pita}, {P{\"u}hlhofer}, {Punch},
  {Ranchon}, {Raubenheimer}, {Raue}, {Rayner}, {Reimer}, {Ripken}, {Rob},
  {Rolland}, {Rosier-Lees}, {Rowell}, {Sahakian}, {Santangelo}, {Saug{\'e}},
  {Schlenker}, {Schlickeiser}, {Schr{\"o}der}, {Schwanke}, {Schwarzburg},
  {Schwemmer}, {Shalchi}, {Sol}, {Spangler}, {Spanier}, {Steenkamp},
  {Stegmann}, {Superina}, {Tam}, {Tavernet}, {Terrier}, {Tluczykont}, {van
  Eldik}, {Vasileiadis}, {Venter}, {Vialle}, {Vincent}, {V{\"o}lk}, {Wagner},
  \& {Ward}}]{aharonian07}
{Aharonian} F. {et~al.}, 2007, ApJ, 661, 236

\bibitem[{{Aharonian} \& {Atoyan}(1999)}]{aharonian99}
{Aharonian} F.~A., {Atoyan} A.~M., 1999, A\&A, 351, 330

\bibitem[{{Aharonian}, {Atoyan} \& {Kifune}(1997){Aharonian}, {Atoyan}, \&
  {Kifune}}]{aharonian97}
{Aharonian} F.~A., {Atoyan} A.~M., {Kifune} T., 1997, MNRAS, 291, 162

\bibitem[{{Ahmad} {et~al}\mbox{.}(2006){Ahmad}, {Greene}, {Moore}, {Ghelberg},
  {Ofan}, {Paul}, \& {Kutschera}}]{ahmad06}
{Ahmad} I., {Greene} J.~P., {Moore} E.~F., {Ghelberg} S., {Ofan} A., {Paul} M.,
  {Kutschera} W., 2006, Phys. Rev. C., 74, 065803

\bibitem[{{Albert} {et~al}\mbox{.}(2006{\natexlab{a}}){Albert}, {Aliu},
  {Anderhub}, {Antoranz}, {Armada}, {Asensio}, {Baixeras}, {Barrio}, {Bartel},
  {Bartko}, {Bastieri}, {Bavikadi}, {Bednarek}, {Berger}, {Bigongiari},
  {Biland}, {Bisesi}, {Blanch}, {Bock}, {Bretz}, {Britvitch}, {Camara},
  {Chilingarian}, {Ciprini}, {Coarasa}, {Commichau}, {Contreras}, {Cortina},
  {Curtev}, {Danielyan}, {Dazzi}, {De Angelis}, {de los Reyes}, {De Lotto},
  {Domingo-Santamaria}, {Dorner}, {Doro}, {Errando}, {Fagiolini}, {Ferenc},
  {Fern{\'a}ndez}, {Firpo}, {Flix}, {Fonseca}, {Font}, {Galante},
  {Garczarczyk}, {Gaug}, {Gebauer}, {Giller}, {Goebel}, {Hakobyan},
  {Hayashida}, {Hengstebeck}, {H{\"o}hne}, {Hose}, {Jacon}, {Kalekin},
  {Kranich}, {Laille}, {Lenisa}, {Liebing}, {Lindfors}, {Longo}, {L{\'o}pez},
  {L{\'o}pez}, {Lorenz}, {Lucarelli}, {Majumdar}, {Maneva}, {Mannheim},
  {Mariotti}, {Mart{\'{\i}}nez}, {Mase}, {Mazin}, {Merck}, {Merck}, {Meucci},
  {Meyer}, {Miranda}, {Mirzoyan}, {Mizobuchi}, {Moralejo}, {Nilsson},
  {O{\~n}a-Wilhelmi}, {Ordu{\~n}a}, {Otte}, {Oya}, {Paneque}, {Paoletti},
  {Pasanen}, {Pascoli}, {Pauss}, {Pavel}, {Pegna}, {Peruzzo}, {Piccioli},
  {Prandini}, {Rico}, {Rhode}, {Riegel}, {Rissi}, {Robert}, {Rossato},
  {R{\"u}gamer}, {Saggion}, {Sanchez}, {Sartori}, {Scalzotto}, {Schmitt},
  {Schweizer}, {Shayduk}, {Shinozaki}, {Shore}, {Sidro}, {Sillanp{\"a}{\"a}},
  {Sobczynska}, {Stamerra}, {Stark}, {Takalo}, {Temnikov}, {Tescaro},
  {Teshima}, {Tonello}, {Torres}, {Torres}, {Turini}, {Vankov}, {Vitale},
  {Wagner}, {Wibig}, {Wittek}, \& {Zapatero}}]{albert06a}
{Albert} J. {et~al.}, 2006{\natexlab{a}}, ApJL, 637, L41

\bibitem[{{Albert} {et~al}\mbox{.}(2006{\natexlab{b}}){Albert}, {Aliu},
  {Anderhub}, {Antoranz}, {Armada}, {Asensio}, {Baixeras}, {Barrio}, {Bartelt},
  {Bartko}, {Bastieri}, {Bavikadi}, {Bednarek}, {Berger}, {Bigongiari},
  {Biland}, {Bisesi}, {Bock}, {Bordas}, {Bosch-Ramon}, {Bretz}, {Britvitch},
  {Camara}, {Carmona}, {Chilingarian}, {Ciprini}, {Coarasa}, {Commichau},
  {Contreras}, {Cortina}, {Curtef}, {Dame}, {Danielyan}, {Dazzi}, {De Angelis},
  {de los Reyes}, {De Lotto}, {Domingo-Santamar{\'{\i}}}, {Dorner}, {Doro},
  {Errando}, {Fagiolini}, {Ferenc}, {Fern{\'a}ndez}, {Firpo}, {Flix},
  {Fonseca}, {Font}, {Fuchs}, {Galante}, {Garczarczyk}, {Gaug}, {Giller},
  {Goebel}, {Hakobyan}, {Hayashida}, {Hengstebeck}, {H{\"o}hne}, {Hose}, {Hsu},
  {Isar}, {Jacon}, {Kalekin}, {Kasyra}, {Kranich}, {Laatiaoui}, {Laille},
  {Lenisa}, {Liebing}, {Lindfors}, {Lombardi}, {Longo}, {L{\'o}pez},
  {L{\'o}pez}, {Lorenz}, {Lucarelli}, {Majumdar}, {Maneva}, {Mannheim},
  {Mansutti}, {Mariotti}, {Mart{\'{\i}}nez}, {Mase}, {Mazin}, {Merck},
  {Meucci}, {Meyer}, {Miranda}, {Mirzoyan}, {Mizobuchi}, {Moralejo}, {Nilsson},
  {O{\~n}a-Wilhelmi}, {Ordu{\~n}a}, {Otte}, {Oya}, {Paneque}, {Paoletti},
  {Paredes}, {Pasanen}, {Pascoli}, {Pauss}, {Pavel}, {Pegna}, {Persic},
  {Peruzzo}, {Piccioli}, {Poller}, {Prandini}, {Raymers}, {Rico}, {Rhode},
  {Rib{\'o}}, {Riegel}, {Rissi}, {Robert}, {R{\"u}gamer}, {Saggion},
  {S{\'a}nchez}, {Sartori}, {Scalzotto}, {Scapin}, {Schmitt}, {Schweizer},
  {Shayduk}, {Shinozaki}, {Shore}, {Sidro}, {Sillanp{\"a}{\"a}}, {Sobczynska},
  {Stamerra}, {Stark}, {Takalo}, {Temnikov}, {Tescaro}, {Teshima}, {Tonello},
  {Torres}, {Torres}, {Turini}, {Vankov}, {Vitale}, {Wagner}, {Wibig},
  {Wittek}, {Zanin}, \& {Zapatero}}]{albert06b}
{Albert} J. {et~al.}, 2006{\natexlab{b}}, ApJL, 643, L53

\bibitem[{{Albert} {et~al}\mbox{.}(2008{\natexlab{a}}){Albert}, {Aliu},
  {Anderhub}, {Antoranz}, {Armada}, {Baixeras}, {Barrio}, {Bartko}, {Bastieri},
  {Becker}, {Bednarek}, {Berger}, {Bigongiari}, {Biland}, {Bock}, {Bordas},
  {Bosch-Ramon}, {Bretz}, {Britvitch}, {Camara}, {Carmona}, {Chilingarian},
  {Coarasa}, {Commichau}, {Contreras}, {Cortina}, {Costado}, {Curtef},
  {Danielyan}, {Dazzi}, {De Angelis}, {Delgado}, {de los Reyes}, {De Lotto},
  {Domingo-Santamar{\'{\i}}a}, {Dorner}, {Doro}, {Errando}, {Fagiolini},
  {Ferenc}, {Fern{\'a}ndez}, {Firpo}, {Flix}, {Fonseca}, {Font}, {Fuchs},
  {Galante}, {Garc{\'{\i}}a-L{\'o}pez}, {Garczarczyk}, {Gaug}, {Giller},
  {Goebel}, {Hakobyan}, {Hayashida}, {Hengstebeck}, {Herrero}, {H{\"o}hne},
  {Hose}, {Hsu}, {Jacon}, {Jogler}, {Kosyra}, {Kranich}, {Kritzer}, {Laille},
  {Lindfors}, {Lombardi}, {Longo}, {L{\'o}pez}, {L{\'o}pez}, {Lorenz},
  {Majumdar}, {Maneva}, {Mannheim}, {Mansutti}, {Mariotti}, {Mart{\'{\i}}nez},
  {Mazin}, {Merck}, {Meucci}, {Meyer}, {Miranda}, {Mirzoyan}, {Mizobuchi},
  {Moralejo}, {Nieto}, {Nilsson}, {Ninkovic}, {O{\~n}a-Wilhelmi}, {Otte},
  {Oya}, {Paneque}, {Panniello}, {Paoletti}, {Paredes}, {Pasanen}, {Pascoli},
  {Pauss}, {Pegna}, {Persic}, {Peruzzo}, {Piccioli}, {Poller}, {Prandini},
  {Puchades}, {Raymers}, {Rhode}, {Rib{\'o}}, {Rico}, {Rissi}, {Robert},
  {R{\"u}gamer}, {Saggion}, {S{\'a}nchez}, {Sartori}, {Scalzotto}, {Scapin},
  {Schmitt}, {Schweizer}, {Shayduk}, {Shinozaki}, {Shore}, {Sidro},
  {Sillanp{\"a}{\"a}}, {Sobczynska}, {Stamerra}, {Stark}, {Takalo}, {Temnikov},
  {Tescaro}, {Teshima}, {Tonello}, {Torres}, {Turini}, {Vankov}, {Vitale},
  {Wagner}, {Wibig}, {Wittek}, {Zandanel}, {Zanin}, \& {Zapatero}}]{albert08a}
{Albert} J. {et~al.}, 2008{\natexlab{a}}, ApJ, 674, 1037

\bibitem[{{Albert} {et~al}\mbox{.}(2008{\natexlab{b}}){Albert}, {Aliu},
  {Anderhub}, {Antoranz}, {Baixeras}, {Barrio}, {Bartko}, {Bastieri}, {Becker},
  {Bednarek}, {Berger}, {Bigongiari}, {Biland}, {Bock}, {Bonnoli}, {Bordas},
  {Bosch-Ramon}, {Bretz}, {Britvitch}, {Camara}, {Carmona}, {Chilingarian},
  {Commichau}, {Contreras}, {Cortina}, {Costado}, {Curtef}, {Dazzi}, {De
  Angelis}, {Delgado}, {de los Reyes}, {Domingo-Santamar{\'{\i}}a}, {De Lotto},
  {De Maria}, {De Sabata}, {Dorner}, {Doro}, {Errando}, {Fagiolini}, {Ferenc},
  {Fern{\'a}ndez}, {Firpo}, {Fonseca}, {Font}, {Galante},
  {Garc{\'{\i}}a-L{\'o}pez}, {Garczarczyk}, {Gaug}, {Goebel}, {Hayashida},
  {Herrero}, {H{\"o}hne}, {Hose}, {Hsu}, {Huber}, {Jogler}, {Kosyra},
  {Kranich}, {Laille}, {Leonardo}, {Lindfors}, {Lombardi}, {Longo},
  {L{\'o}pez}, {Lorenz}, {Majumdar}, {Maneva}, {Mankuzhiyil}, {Mannheim},
  {Mariotti}, {Mart{\'{\i}}nez}, {Mazin}, {Merck}, {Meucci}, {Meyer},
  {Miranda}, {Mirzoyan}, {Mizobuchi}, {Moralejo}, {Nieto}, {Nilsson},
  {Ninkovic}, {O{\~n}a-Wilhelmi}, {Otte}, {Oya}, {Panniello}, {Paoletti},
  {Paredes}, {Pasanen}, {Pascoli}, {Pauss}, {Pegna}, {Persic}, {Peruzzo},
  {Piccioli}, {Prandini}, {Puchades}, {Raymers}, {Rhode}, {Rib{\'o}}, {Rico},
  {Rissi}, {Robert}, {R{\"u}gamer}, {Saggion}, {Saito}, {S{\'a}nchez},
  {Sartori}, {Scalzotto}, {Scapin}, {Schmitt}, {Schweizer}, {Shayduk},
  {Shinozaki}, {Shore}, {Sidro}, {Sillanp{\"a}{\"a}}, {Sobczynska}, {Spanier},
  {Stamerra}, {Stark}, {Takalo}, {Temnikov}, {Tescaro}, {Teshima}, {Torres},
  {Turini}, {Vankov}, {Venturini}, {Vitale}, {Wagner}, {Wittek}, {Zandanel},
  {Zanin}, \& {Zapatero}}]{albert08b}
{Albert} J. {et~al.}, 2008{\natexlab{b}}, ApJL, 675, L25

\bibitem[{{Aleksi{\'c}} {et~al}\mbox{.}(2012){Aleksi{\'c}}, {Alvarez},
  {Antonelli}, {Antoranz}, {Asensio}, {Backes}, {Barrio}, {Bastieri}, {Becerra
  Gonz{\'a}lez}, {Bednarek}, {Berdyugin}, {Berger}, {Bernardini}, {Biland},
  {Blanch}, {Bock}, {Boller}, {Bonnoli}, {Borla Tridon}, {Braun}, {Bretz},
  {Ca{\~n}ellas}, {Carmona}, {Carosi}, {Colin}, {Colombo}, {Contreras},
  {Cortina}, {Cossio}, {Covino}, {Dazzi}, {de Angelis}, {de Caneva}, {de Cea
  Del Pozo}, {de Lotto}, {Delgado Mendez}, {Diago Ortega}, {Doert},
  {Dom{\'{\i}}nguez}, {Dominis Prester}, {Dorner}, {Doro}, {Elsaesser},
  {Ferenc}, {Fonseca}, {Font}, {Fruck}, {Garc{\'{\i}}a L{\'o}pez},
  {Garczarczyk}, {Garrido}, {Giavitto}, {Godinovi{\'c}}, {Hadasch},
  {H{\"a}fner}, {Herrero}, {Hildebrand}, {H{\"o}hne-M{\"o}nch}, {Hose},
  {Hrupec}, {Huber}, {Jogler}, {Kellermann}, {Klepser}, {Kr{\"a}henb{\"u}hl},
  {Krause}, {La Barbera}, {Lelas}, {Leonardo}, {Lindfors}, {Lombardi},
  {L{\'o}pez}, {L{\'o}pez-Oramas}, {Lorenz}, {Makariev}, {Maneva},
  {Mankuzhiyil}, {Mannheim}, {Maraschi}, {Mariotti}, {Mart{\'{\i}}nez},
  {Mazin}, {Meucci}, {Miranda}, {Mirzoyan}, {Miyamoto}, {Mold{\'o}n},
  {Moralejo}, {Munar-Adrover}, {Nieto}, {Nilsson}, {Orito}, {Oya}, {Paneque},
  {Paoletti}, {Pardo}, {Paredes}, {Partini}, {Pasanen}, {Pauss},
  {Perez-Torres}, {Persic}, {Peruzzo}, {Pilia}, {Pochon}, {Prada}, {Prada
  Moroni}, {Prandini}, {Puljak}, {Reichardt}, {Reinthal}, {Rhode}, {Rib{\'o}},
  {Rico}, {R{\"u}gamer}, {Saggion}, {Saito}, {Saito}, {Salvati}, {Satalecka},
  {Scalzotto}, {Scapin}, {Schultz}, {Schweizer}, {Shayduk}, {Shore},
  {Sillanp{\"a}{\"a}}, {Sitarek}, {Snidaric}, {Sobczynska}, {Spanier}, {Spiro},
  {Stamatescu}, {Stamerra}, {Steinke}, {Storz}, {Strah}, {Suri{\'c}}, {Takalo},
  {Takami}, {Tavecchio}, {Temnikov}, {Terzi{\'c}}, {Tescaro}, {Teshima},
  {Tibolla}, {Torres}, {Treves}, {Uellenbeck}, {Vankov}, {Vogler}, {Wagner},
  {Weitzel}, {Zabalza}, {Zandanel}, \& {Zanin}}]{aleksic12}
{Aleksi{\'c}} J. {et~al.}, 2012, Astroparticle Physics, 35, 435

\bibitem[{{Aleksi{\'c}} {et~al}\mbox{.}(2014){Aleksi{\'c}}, {Ansoldi},
  {Antonelli}, {Antoranz}, {Babic}, {Bangale}, {de Almeida}, {Barrio},
  {Gonz{\'a}lez}, {Bednarek}, {Berger}, {Bernardini}, {Biland}, {Blanch},
  {Bock}, {Bonnefoy}, {Bonnoli}, {Borracci}, {Bretz}, {Carmona}, {Carosi},
  {Fidalgo}, {Colin}, {Colombo}, {Contreras}, {Cortina}, {Covino}, {Da Vela},
  {Dazzi}, {De Angelis}, {De Caneva}, {De Lotto}, {Mendez}, {Doert},
  {Dom{\'{\i}}nguez}, {Prester}, {Dorner}, {Doro}, {Einecke}, {Eisenacher},
  {Elsaesser}, {Farina}, {Ferenc}, {Fonseca}, {Font}, {Frantzen}, {Fruck},
  {L{\'o}pez}, {Garczarczyk}, {Terrats}, {Gaug}, {Giavitto}, {Godinovi{\'c}},
  {Mu{\~n}oz}, {Gozzini}, {Hadasch}, {Hayashida}, {Herrero}, {Hildebrand},
  {Hose}, {Hrupec}, {Idec}, {Kadenius}, {Kellermann}, {Knoetig}, {Kodani},
  {Konno}, {Krause}, {Kubo}, {Kushida}, {Barbera}, {Lelas}, {Lewandowska},
  {Lindfors}, {Lombardi}, {L{\'o}pez}, {L{\'o}pez-Coto}, {L{\'o}pez-Oramas},
  {Lorenz}, {Lozano}, {Makariev}, {Mallot}, {Maneva}, {Mankuzhiyil},
  {Mannheim}, {Maraschi}, {Marcote}, {Mariotti}, {Mart{\'{\i}}nez}, {Mazin},
  {Menzel}, {Meucci}, {Miranda}, {Mirzoyan}, {Moralejo}, {Munar-Adrover},
  {Nakajima}, {Niedzwiecki}, {Nilsson}, {Nishijima}, {Nowak}, {Orito},
  {Overkemping}, {Paiano}, {Palatiello}, {Paneque}, {Paoletti}, {Paredes},
  {Paredes-Fortuny}, {Partini}, {Persic}, {Prada}, {Moroni}, {Prandini},
  {Preziuso}, {Puljak}, {Reinthal}, {Rhode}, {Rib{\'o}}, {Rico}, {Garcia},
  {R{\"u}gamer}, {Saggion}, {Saito}, {Saito}, {Salvati}, {Satalecka},
  {Scalzotto}, {Scapin}, {Schultz}, {Schweizer}, {Shore}, {Sillanp{\"a}{\"a}},
  {Sitarek}, {Snidaric}, {Sobczynska}, {Spanier}, {Stamatescu}, {Stamerra},
  {Steinbring}, {Storz}, {Sun}, {Suri{\'c}}, {Takalo}, {Takami}, {Tavecchio},
  {Temnikov}, {Terzi{\'c}}, {Tescaro}, {Teshima}, {Thaele}, {Tibolla},
  {Torres}, {Toyama}, {Treves}, {Vogler}, {Wagner}, {Zandanel}, \&
  {Zanin}}]{aleksic14}
{Aleksi{\'c}} J. {et~al.}, 2014, ArXiv e-prints

\bibitem[{{Aliu} {et~al}\mbox{.}(2013){Aliu}, {Archambault}, {Arlen}, {Aune},
  {Beilicke}, {Benbow}, {Bouvier}, {Buckley}, {Bugaev}, {Cesarini}, {Ciupik},
  {Collins-Hughes}, {Connolly}, {Cui}, {Dickherber}, {Duke}, {Dumm},
  {Dwarkadas}, {Errando}, {Falcone}, {Federici}, {Feng}, {Finley}, {Finnegan},
  {Fortson}, {Furniss}, {Galante}, {Gall}, {Gillanders}, {Godambe}, {Gotthelf},
  {Griffin}, {Grube}, {Gyuk}, {Hanna}, {Holder}, {Hughes}, {Humensky},
  {Kaaret}, {Kargaltsev}, {Karlsson}, {Khassen}, {Kieda}, {Krawczynski},
  {Krennrich}, {Lang}, {Lee}, {Madhavan}, {Maier}, {Majumdar}, {McArthur},
  {McCann}, {Moriarty}, {Mukherjee}, {Nelson}, {O'Faol{\'a}in de Bhr{\'o}ithe},
  {Ong}, {Orr}, {Otte}, {Park}, {Perkins}, {Pohl}, {Prokoph}, {Quinn}, {Ragan},
  {Reyes}, {Reynolds}, {Roache}, {Roberts}, {Saxon}, {Schroedter}, {Sembroski},
  {Slane}, {Smith}, {Staszak}, {Telezhinsky}, {Te{\v s}i{\'c}}, {Theiling},
  {Thibadeau}, {Tsurusaki}, {Tyler}, {Varlotta}, {Vassiliev}, {Vincent},
  {Vivier}, {Wakely}, {Weekes}, {Weinstein}, {Welsing}, {Williams}, \&
  {Zitzer}}]{aliu13}
{Aliu} E. {et~al.}, 2013, ApJ, 764, 38

\bibitem[{{Altenhoff} {et~al}\mbox{.}(1979){Altenhoff}, {Downes}, {Pauls}, \&
  {Schraml}}]{altenhoff79}
{Altenhoff} W.~J., {Downes} D., {Pauls} T., {Schraml} J., 1979, A\&AS, 35, 23

\bibitem[{{Anderhub} {et~al}\mbox{.}(2010){Anderhub}, {Antonelli}, {Antoranz},
  {Backes}, {Baixeras}, {Balestra}, {Barrio}, {Bastieri}, {Becerra
  Gonz{\'a}lez}, {Becker}, {Bednarek}, {Berger}, {Bernardini}, {Biland},
  {Bock}, {Bonnoli}, {Bordas}, {Borla Tridon}, {Bosch-Ramon}, {Bose}, {Braun},
  {Bretz}, {Britzger}, {Camara}, {Carmona}, {Carosi}, {Colin}, {Commichau},
  {Contreras}, {Cortina}, {Costado}, {Covino}, {Dazzi}, {De Angelis}, {de Cea
  del Pozo}, {De los Reyes}, {De Lotto}, {De Maria}, {De Sabata}, {Delgado
  Mendez}, {Dom{\'{\i}}nguez}, {Dominis Prester}, {Dorner}, {Doro},
  {Elsaesser}, {Errando}, {Ferenc}, {Fern{\'a}ndez}, {Firpo}, {Fonseca},
  {Font}, {Galante}, {Garc{\'{\i}}a L{\'o}pez}, {Garczarczyk}, {Gaug},
  {Godinovic}, {Goebel}, {Hadasch}, {Herrero}, {Hildebrand},
  {H{\"o}hne-M{\"o}nch}, {Hose}, {Hrupec}, {Hsu}, {Jogler}, {Klepser},
  {Kranich}, {La Barbera}, {Laille}, {Leonardo}, {Lindfors}, {Lombardi},
  {Longo}, {L{\'o}pez}, {Lorenz}, {Majumdar}, {Maneva}, {Mankuzhiyil},
  {Mannheim}, {Maraschi}, {Mariotti}, {Mart{\'{\i}}nez}, {Mazin}, {Meucci},
  {Miranda}, {Mirzoyan}, {Miyamoto}, {Mold{\'o}n}, {Moles}, {Moralejo},
  {Nieto}, {Nilsson}, {Ninkovic}, {Orito}, {Oya}, {Paoletti}, {Paredes},
  {Pasanen}, {Pascoli}, {Pauss}, {Pegna}, {Perez-Torres}, {Persic}, {Peruzzo},
  {Prada}, {Prandini}, {Puchades}, {Puljak}, {Reichardt}, {Rhode}, {Rib{\'o}},
  {Rico}, {Rissi}, {Robert}, {R{\"u}gamer}, {Saggion}, {Saito}, {Salvati},
  {S{\'a}nchez-Conde}, {Satalecka}, {Scalzotto}, {Scapin}, {Schweizer},
  {Shayduk}, {Shore}, {Sierpowska-Bartosik}, {Sillanp{\"a}{\"a}}, {Sitarek},
  {Sobczynska}, {Spanier}, {Spiro}, {Stamerra}, {Stark}, {Suric}, {Takalo},
  {Tavecchio}, {Temnikov}, {Tescaro}, {Teshima}, {Torres}, {Turini}, {Vankov},
  {Wagner}, {Zabalza}, {Zandanel}, {Zanin}, {Zapatero}, \&
  {Cognard}}]{anderhub10}
{Anderhub} H. {et~al.}, 2010, ApJ, 710, 828

\bibitem[{{Anderson} \& {Itoh}(1975)}]{anderson75}
{Anderson} P.~W., {Itoh} N., 1975, Nature, 256, 25

\bibitem[{{Antoniadis} {et~al}\mbox{.}(2013){Antoniadis}, {Freire}, {Wex},
  {Tauris}, {Lynch}, {van Kerkwijk}, {Kramer}, {Bassa}, {Dhillon}, {Driebe},
  {Hessels}, {Kaspi}, {Kondratiev}, {Langer}, {Marsh}, {McLaughlin},
  {Pennucci}, {Ransom}, {Stairs}, {van Leeuwen}, {Verbiest}, \&
  {Whelan}}]{antoniadis13}
{Antoniadis} J. {et~al.}, 2013, Science, 340, 448

\bibitem[{{Archibald} {et~al}\mbox{.}(2013){Archibald}, {Kaspi}, {Ng},
  {Gourgouliatos}, {Tsang}, {Scholz}, {Beardmore}, {Gehrels}, \&
  {Kennea}}]{archibald13}
{Archibald} R.~F. {et~al.}, 2013, Nature, 497, 591

\bibitem[{{Arendt}(1991)}]{arendt91}
{Arendt} R.~G., 1991, AJ, 101, 2160

\bibitem[{{Arnaud}(1996)}]{arnaud96}
{Arnaud} K.~A., 1996, in Astronomical Society of the Pacific Conference Series,
  Vol. 101, Astronomical Data Analysis Software and Systems V, {Jacoby} G.~H.,
  {Barnes} J., eds., p.~17

\bibitem[{{Arons}(2002)}]{arons02}
{Arons} J., 2002, in Astronomical Society of the Pacific Conference Series,
  Vol. 271, Neutron Stars in Supernova Remnants, {Slane} P.~O., {Gaensler}
  B.~M., eds., p.~71

\bibitem[{{Arumugasamy}, {Pavlov} \& {Kargaltsev}(2013){Arumugasamy}, {Pavlov},
  \& {Kargaltsev}}]{arumugasamy13}
{Arumugasamy} P., {Pavlov} G., {Kargaltsev} O., 2013, in The Fast and the
  Furious: Energetic Phenomena in Isolated Neutron Stars, Pulsar Wind Nebulae
  and Supernova Remnants, {Ness} J.-U., ed.

\bibitem[{{Arzoumanian} {et~al}\mbox{.}(2011){Arzoumanian}, {Gotthelf},
  {Ransom}, {Safi-Harb}, {Kothes}, \& {Landecker}}]{arzoumanian11}
{Arzoumanian} Z., {Gotthelf} E.~V., {Ransom} S.~M., {Safi-Harb} S., {Kothes}
  R., {Landecker} T.~L., 2011, ApJ, 739, 39

\bibitem[{{Arzoumanian} {et~al}\mbox{.}(2008){Arzoumanian}, {Safi-Harb},
  {Landecker}, {Kothes}, \& {Camilo}}]{arzoumanian08}
{Arzoumanian} Z., {Safi-Harb} S., {Landecker} T.~L., {Kothes} R., {Camilo} F.,
  2008, ApJ, 687, 505

\bibitem[{{Aschenbach}, {Egger} \& {Tr{\"u}mper}(1995){Aschenbach}, {Egger}, \&
  {Tr{\"u}mper}}]{aschenbach95}
{Aschenbach} B., {Egger} R., {Tr{\"u}mper} J., 1995, Nature, 373, 587

\bibitem[{{Asvarov} {et~al}\mbox{.}(1990){Asvarov}, {Guseinov}, {Kasumov}, \&
  {Dogel'}}]{asvarov90}
{Asvarov} A.~I., {Guseinov} O.~H., {Kasumov} F.~K., {Dogel'} V.~A., 1990, A\&A,
  229, 196

\bibitem[{{Atoyan} \& {Aharonian}(1996)}]{atoyan96}
{Atoyan} A.~M., {Aharonian} F.~A., 1996, MNRAS, 278, 525

\bibitem[{{Baade} \& {Zwicky}(1934)}]{baade34}
{Baade} W., {Zwicky} F., 1934, Physical Review, 46, 76

\bibitem[{{Backer} {et~al}\mbox{.}(1982){Backer}, {Kulkarni}, {Heiles},
  {Davis}, \& {Goss}}]{backer82}
{Backer} D., {Kulkarni} S., {Heiles} C., {Davis} M., {Goss} M., 1982, IAU
  Circ., 3743, 2

\bibitem[{{Baldwin}(1971)}]{baldwin71}
{Baldwin} J.~E., 1971, in IAU Symposium, Vol.~46, The Crab Nebula, {Davies}
  R.~D., {Graham-Smith} F., eds., p.~22

\bibitem[{{Banas} {et~al}\mbox{.}(1997){Banas}, {Hughes}, {Bronfman}, \&
  {Nyman}}]{banas97}
{Banas} K.~R., {Hughes} J.~P., {Bronfman} L., {Nyman} L.-A., 1997, ApJ, 480,
  607

\bibitem[{{Becker} \& {Helfand}(1984)}]{becker84}
{Becker} R.~H., {Helfand} D.~J., 1984, ApJ, 283, 154

\bibitem[{{Becker}, {Helfand} \& {Szymkowiak}(1983){Becker}, {Helfand}, \&
  {Szymkowiak}}]{becker83}
{Becker} R.~H., {Helfand} D.~J., {Szymkowiak} A.~E., 1983, ApJL, 268, L93

\bibitem[{{Becker} \& {Kundu}(1976)}]{becker76}
{Becker} R.~H., {Kundu} M.~R., 1976, ApJ, 204, 427

\bibitem[{{Becker} {et~al}\mbox{.}(2012){Becker}, {Prinz}, {Winkler}, \&
  {Petre}}]{becker12}
{Becker} W., {Prinz} T., {Winkler} P.~F., {Petre} R., 2012, ApJ, 755, 141

\bibitem[{{Becker} \& {Truemper}(1997)}]{becker97}
{Becker} W., {Truemper} J., 1997, A\&A, 326, 682

\bibitem[{{Bednarek} \& {Bartosik}(2003)}]{bednarek03}
{Bednarek} W., {Bartosik} M., 2003, A\&A, 405, 689

\bibitem[{{Bednarek} \& {Bartosik}(2005)}]{bednarek05}
{Bednarek} W., {Bartosik} M., 2005, Journal of Physics G Nuclear Physics, 31,
  1465

\bibitem[{{Beiersdorfer} {et~al}\mbox{.}(1993){Beiersdorfer}, {Phillips},
  {Jacobs}, {Hill}, {Bitter}, {von Goeler}, \& {Kahn}}]{beiersdorfer93}
{Beiersdorfer} P., {Phillips} T., {Jacobs} V.~L., {Hill} K.~W., {Bitter} M.,
  {von Goeler} S., {Kahn} S.~M., 1993, ApJ, 409, 846

\bibitem[{{Bietenholz}(2006)}]{bietenholz06}
{Bietenholz} M.~F., 2006, ApJ, 645, 1180

\bibitem[{{Bietenholz} \& {Bartel}(2008)}]{bietenholz08}
{Bietenholz} M.~F., {Bartel} N., 2008, MNRAS, 386, 1411

\bibitem[{{Bietenholz} {et~al}\mbox{.}(2004){Bietenholz}, {Hester}, {Frail}, \&
  {Bartel}}]{bietenholz04}
{Bietenholz} M.~F., {Hester} J.~J., {Frail} D.~A., {Bartel} N., 2004, ApJ, 615,
  794

\bibitem[{{Blanton} \& {Helfand}(1996)}]{blanton96}
{Blanton} E.~L., {Helfand} D.~J., 1996, ApJ, 470, 961

\bibitem[{{Bleeker} {et~al}\mbox{.}(2001){Bleeker}, {Willingale}, {van der
  Heyden}, {Dennerl}, {Kaastra}, {Aschenbach}, \& {Vink}}]{bleeker01}
{Bleeker} J.~A.~M., {Willingale} R., {van der Heyden} K., {Dennerl} K.,
  {Kaastra} J.~S., {Aschenbach} B., {Vink} J., 2001, A\&A, 365, L225

\bibitem[{{Blondin}, {Chevalier} \& {Frierson}(2001){Blondin}, {Chevalier}, \&
  {Frierson}}]{blondin01}
{Blondin} J.~M., {Chevalier} R.~A., {Frierson} D.~M., 2001, ApJ, 563, 806

\bibitem[{{Blumenthal}(1971)}]{blumenthal71}
{Blumenthal} G.~R., 1971, PhRvD, 3, 2308

\bibitem[{{Blumenthal} \& {Gould}(1970)}]{blumenthal70}
{Blumenthal} G.~R., {Gould} R.~J., 1970, Reviews of Modern Physics, 42, 237

\bibitem[{{Bocchino} {et~al}\mbox{.}(2005){Bocchino}, {van der Swaluw},
  {Chevalier}, \& {Bandiera}}]{bocchino05}
{Bocchino} F., {van der Swaluw} E., {Chevalier} R., {Bandiera} R., 2005, A\&A,
  442, 539

\bibitem[{{Bocchino} {et~al}\mbox{.}(2001){Bocchino}, {Warwick}, {Marty},
  {Lumb}, {Becker}, \& {Pigot}}]{bocchino01}
{Bocchino} F., {Warwick} R.~S., {Marty} P., {Lumb} D., {Becker} W., {Pigot} C.,
  2001, A\&A, 369, 1078

\bibitem[{{Bock} \& {Gaensler}(2005)}]{bock05}
{Bock} D.~C.-J., {Gaensler} B.~M., 2005, ApJ, 626, 343

\bibitem[{{Bogovalov} {et~al}\mbox{.}(2005){Bogovalov}, {Chechetkin},
  {Koldoba}, \& {Ustyugova}}]{bogovalov05}
{Bogovalov} S.~V., {Chechetkin} V.~M., {Koldoba} A.~V., {Ustyugova} G.~V.,
  2005, MNRAS, 358, 705

\bibitem[{{Bogovalov} \& {Khangoulyan}(2002)}]{bogovalov02}
{Bogovalov} S.~V., {Khangoulyan} D.~V., 2002, Astronomy Letters, 28, 373

\bibitem[{{Borkowski}, {Lyerly} \& {Reynolds}(2001){Borkowski}, {Lyerly}, \&
  {Reynolds}}]{borkowski01}
{Borkowski} K.~J., {Lyerly} W.~J., {Reynolds} S.~P., 2001, ApJ, 548, 820

\bibitem[{{Borkowski} {et~al}\mbox{.}(2010){Borkowski}, {Reynolds}, {Green},
  {Hwang}, {Petre}, {Krishnamurthy}, \& {Willett}}]{borkowski10}
{Borkowski} K.~J., {Reynolds} S.~P., {Green} D.~A., {Hwang} U., {Petre} R.,
  {Krishnamurthy} K., {Willett} R., 2010, ApJL, 724, L161

\bibitem[{{Brazier} {et~al}\mbox{.}(1998){Brazier}, {Reimer}, {Kanbach}, \&
  {Carraminana}}]{brazier98}
{Brazier} K.~T.~S., {Reimer} O., {Kanbach} G., {Carraminana} A., 1998, MNRAS,
  295, 819

\bibitem[{{Brogan} {et~al}\mbox{.}(2005){Brogan}, {Gaensler}, {Gelfand},
  {Lazendic}, {Lazio}, {Kassim}, \& {McClure-Griffiths}}]{brogan05}
{Brogan} C.~L., {Gaensler} B.~M., {Gelfand} J.~D., {Lazendic} J.~S., {Lazio}
  T.~J.~W., {Kassim} N.~E., {McClure-Griffiths} N.~M., 2005, ApJL, 629, L105

\bibitem[{{Bucciantini} {et~al}\mbox{.}(2004){Bucciantini}, {Amato},
  {Bandiera}, {Blondin}, \& {Del Zanna}}]{bucciantini04b}
{Bucciantini} N., {Amato} E., {Bandiera} R., {Blondin} J.~M., {Del Zanna} L.,
  2004, A\&A, 423, 253

\bibitem[{{Bucciantini}, {Arons} \& {Amato}(2011){Bucciantini}, {Arons}, \&
  {Amato}}]{bucciantini11}
{Bucciantini} N., {Arons} J., {Amato} E., 2011, MNRAS, 410, 381

\bibitem[{{Bucciantini} {et~al}\mbox{.}(2003){Bucciantini}, {Blondin}, {Del
  Zanna}, \& {Amato}}]{bucciantini03}
{Bucciantini} N., {Blondin} J.~M., {Del Zanna} L., {Amato} E., 2003, A\&A, 405,
  617

\bibitem[{{Bucciantini} {et~al}\mbox{.}(2012){Bucciantini}, {Metzger},
  {Thompson}, \& {Quataert}}]{bucciantini12}
{Bucciantini} N., {Metzger} B.~D., {Thompson} T.~A., {Quataert} E., 2012,
  MNRAS, 419, 1537

\bibitem[{{Burgay} {et~al}\mbox{.}(2003){Burgay}, {D'Amico}, {Possenti},
  {Manchester}, {Lyne}, {Joshi}, {McLaughlin}, {Kramer}, {Sarkissian},
  {Camilo}, {Kalogera}, {Kim}, \& {Lorimer}}]{burgay03}
{Burgay} M. {et~al.}, 2003, Nature, 426, 531

\bibitem[{{B{\"u}sching} {et~al}\mbox{.}(2008){B{\"u}sching}, {de Jager},
  {Potgieter}, \& {Venter}}]{busching08}
{B{\"u}sching} I., {de Jager} O.~C., {Potgieter} M.~S., {Venter} C., 2008,
  ApJL, 678, L39

\bibitem[{{Camilo} {et~al}\mbox{.}(2000){Camilo}, {Kaspi}, {Lyne},
  {Manchester}, {Bell}, {D'Amico}, {McKay}, \& {Crawford}}]{camilo00}
{Camilo} F., {Kaspi} V.~M., {Lyne} A.~G., {Manchester} R.~N., {Bell} J.~F.,
  {D'Amico} N., {McKay} N.~P.~F., {Crawford} F., 2000, ApJ, 541, 367

\bibitem[{{Camilo} {et~al}\mbox{.}(2002){Camilo}, {Lorimer}, {Bhat},
  {Gotthelf}, {Halpern}, {Wang}, {Lu}, \& {Mirabal}}]{camilo02}
{Camilo} F., {Lorimer} D.~R., {Bhat} N.~D.~R., {Gotthelf} E.~V., {Halpern}
  J.~P., {Wang} Q.~D., {Lu} F.~J., {Mirabal} N., 2002, ApJL, 574, L71

\bibitem[{{Camilo} {et~al}\mbox{.}(2004){Camilo}, {Manchester}, {Lyne},
  {Gaensler}, {Possenti}, {D'Amico}, {Stairs}, {Faulkner}, {Kramer}, {Lorimer},
  {McLaughlin}, \& {Hobbs}}]{camilo04}
{Camilo} F. {et~al.}, 2004, ApJL, 611, L25

\bibitem[{{Camilo} {et~al}\mbox{.}(2009){Camilo}, {Ransom}, {Gaensler}, \&
  {Lorimer}}]{camilo09}
{Camilo} F., {Ransom} S.~M., {Gaensler} B.~M., {Lorimer} D.~R., 2009, ApJL,
  700, L34

\bibitem[{{Camilo} {et~al}\mbox{.}(2006){Camilo}, {Ransom}, {Gaensler},
  {Slane}, {Lorimer}, {Reynolds}, {Manchester}, \& {Murray}}]{camilo06}
{Camilo} F., {Ransom} S.~M., {Gaensler} B.~M., {Slane} P.~O., {Lorimer} D.~R.,
  {Reynolds} J., {Manchester} R.~N., {Murray} S.~S., 2006, ApJ, 637, 456

\bibitem[{{Campana} {et~al}\mbox{.}(2008){Campana}, {Mineo}, {de Rosa},
  {Massaro}, {Dean}, \& {Bassani}}]{campana08}
{Campana} R., {Mineo} T., {de Rosa} A., {Massaro} E., {Dean} A.~J., {Bassani}
  L., 2008, MNRAS, 389, 691

\bibitem[{{Caraveo} {et~al}\mbox{.}(1992){Caraveo}, {Bignami}, {Mereghetti}, \&
  {Mombelli}}]{caraveo92}
{Caraveo} P.~A., {Bignami} G.~F., {Mereghetti} S., {Mombelli} M., 1992, ApJL,
  395, L103

\bibitem[{{Caraveo} {et~al}\mbox{.}(2010){Caraveo}, {De Luca}, {Marelli},
  {Bignami}, {Ray}, {Saz Parkinson}, \& {Kanbach}}]{caraveo10}
{Caraveo} P.~A., {De Luca} A., {Marelli} M., {Bignami} G.~F., {Ray} P.~S., {Saz
  Parkinson} P.~M., {Kanbach} G., 2010, ApJL, 725, L6

\bibitem[{{Carrigan} {et~al}\mbox{.}(2013){Carrigan}, {Brun}, {Chaves}, {Deil},
  {Donath}, {Gast}, {Marandon}, {Renaud}, \& {for the
  H.~E.~S.~S.~collaboration}}]{carrigan13}
{Carrigan} S. {et~al.}, 2013, ArXiv e-prints

\bibitem[{{Carter}, {Dickel} \& {Bomans}(1997){Carter}, {Dickel}, \&
  {Bomans}}]{carter97}
{Carter} L.~M., {Dickel} J.~R., {Bomans} D.~J., 1997, PASP, 109, 990

\bibitem[{{Case} \& {Bhattacharya}(1998)}]{case98}
{Case} G.~L., {Bhattacharya} D., 1998, ApJ, 504, 761

\bibitem[{{Cassam-Chena{\"i}} {et~al}\mbox{.}(2004){Cassam-Chena{\"i}},
  {Decourchelle}, {Ballet}, {Hwang}, {Hughes}, {Petre}, \& {et
  al.}}]{cassamchenai04}
{Cassam-Chena{\"i}} G., {Decourchelle} A., {Ballet} J., {Hwang} U., {Hughes}
  J.~P., {Petre} R., {et al.}, 2004, A\&A, 414, 545

\bibitem[{{Castelletti} {et~al}\mbox{.}(2011){Castelletti}, {Giacani},
  {Dubner}, {Joshi}, {Rao}, \& {Terrier}}]{castelletti11}
{Castelletti} G., {Giacani} E., {Dubner} G., {Joshi} B.~C., {Rao} A.~P.,
  {Terrier} R., 2011, A\&A, 536, A98

\bibitem[{{Castro} {et~al}\mbox{.}(2013){Castro}, {Slane}, {Carlton}, \&
  {Figueroa-Feliciano}}]{castro13}
{Castro} D., {Slane} P., {Carlton} A., {Figueroa-Feliciano} E., 2013, ApJ, 774,
  36

\bibitem[{{Caswell} \& {Haynes}(1987)}]{caswell87}
{Caswell} J.~L., {Haynes} R.~F., 1987, A\&A, 171, 261

\bibitem[{{Caswell}, {McClure-Griffiths} \& {Cheung}(2004){Caswell},
  {McClure-Griffiths}, \& {Cheung}}]{caswell04}
{Caswell} J.~L., {McClure-Griffiths} N.~M., {Cheung} M.~C.~M., 2004, MNRAS,
  352, 1405

\bibitem[{{Caswell}, {Milne} \& {Wellington}(1981){Caswell}, {Milne}, \&
  {Wellington}}]{caswell81}
{Caswell} J.~L., {Milne} D.~K., {Wellington} K.~J., 1981, MNRAS, 195, 89

\bibitem[{{Caswell} {et~al}\mbox{.}(1975){Caswell}, {Murray}, {Roger}, {Cole},
  \& {Cooke}}]{caswell75}
{Caswell} J.~L., {Murray} J.~D., {Roger} R.~S., {Cole} D.~J., {Cooke} D.~J.,
  1975, A\&A, 45, 239

\bibitem[{{Chadwick}(1932)}]{chadwick32}
{Chadwick} J., 1932, Nature, 129, 312

\bibitem[{{Chandrasekhar}(1935)}]{chandrasekhar35}
{Chandrasekhar} S., 1935, MNRAS, 95, 226

\bibitem[{{Chang} {et~al}\mbox{.}(2012){Chang}, {Pavlov}, {Kargaltsev}, \&
  {Shibanov}}]{chang12}
{Chang} C., {Pavlov} G.~G., {Kargaltsev} O., {Shibanov} Y.~A., 2012, ApJ, 744,
  81

\bibitem[{{Chaves}, {de O{\~n}a Wilhemi} \& {Hoppe}(2008){Chaves}, {de O{\~n}a
  Wilhemi}, \& {Hoppe}}]{chaves08}
{Chaves} R.~C.~G., {de O{\~n}a Wilhemi} E., {Hoppe} S., 2008, in American
  Institute of Physics Conference Series, Vol. 1085, American Institute of
  Physics Conference Series, {Aharonian} F.~A., {Hofmann} W., {Rieger} F.,
  eds., pp. 219--222

\bibitem[{{Chen} {et~al}\mbox{.}(2006){Chen}, {Wang}, {Gotthelf}, {Jiang},
  {Chu}, \& {Gruendl}}]{chen06}
{Chen} Y., {Wang} Q.~D., {Gotthelf} E.~V., {Jiang} B., {Chu} Y.-H., {Gruendl}
  R., 2006, ApJ, 651, 237

\bibitem[{{Chevalier}(1982)}]{chevalier82}
{Chevalier} R.~A., 1982, ApJ, 258, 790

\bibitem[{{Chevalier}(1998)}]{chevalier98}
{Chevalier} R.~A., 1998, MmSAI, 69, 977

\bibitem[{{Chevalier}(2004)}]{chevalier04}
{Chevalier} R.~A., 2004, Advances in Space Research, 33, 456

\bibitem[{{Chevalier}(2005)}]{chevalier05}
{Chevalier} R.~A., 2005, ApJ, 619, 839

\bibitem[{{Clark} \& {Stephenson}(1982)}]{clark1982}
{Clark} D.~H., {Stephenson} F.~R., 1982, in NATO ASIC Proc. 90: Supernovae: A
  Survey of Current Research, {Rees} M.~J., {Stoneham} R.~J., eds., pp.
  355--370

\bibitem[{{Clark} {et~al}\mbox{.}(2014){Clark}, {Ritchie}, {Najarro}, {Langer},
  \& {Negueruela}}]{clark14}
{Clark} J.~S., {Ritchie} B.~W., {Najarro} F., {Langer} N., {Negueruela} I.,
  2014, A\&A, 565, A90

\bibitem[{{Condon}, {Broderick} \& {Seielstad}(1989){Condon}, {Broderick}, \&
  {Seielstad}}]{condon89}
{Condon} J.~J., {Broderick} J.~J., {Seielstad} G.~A., 1989, AJ, 97, 1064

\bibitem[{{Condon}, {Griffith} \& {Wright}(1993){Condon}, {Griffith}, \&
  {Wright}}]{condon93}
{Condon} J.~J., {Griffith} M.~R., {Wright} A.~E., 1993, AJ, 106, 1095

\bibitem[{{Cordes} \& {Lazio}(2003)}]{cordes03}
{Cordes} J.~M., {Lazio} T.~J.~W., 2003, ArXiv Astrophysics e-prints

\bibitem[{{Cox}(1972)}]{cox72}
{Cox} D.~P., 1972, ApJ, 178, 159

\bibitem[{{Cox} {et~al}\mbox{.}(1999){Cox}, {Shelton}, {Maciejewski}, {Smith},
  {Plewa}, {Pawl}, \& {R{\'o}{\.z}yczka}}]{cox99}
{Cox} D.~P., {Shelton} R.~L., {Maciejewski} W., {Smith} R.~K., {Plewa} T.,
  {Pawl} A., {R{\'o}{\.z}yczka} M., 1999, ApJ, 524, 179

\bibitem[{{Crawford} {et~al}\mbox{.}(2001){Crawford}, {Gaensler}, {Kaspi},
  {Manchester}, {Camilo}, {Lyne}, \& {Pivovaroff}}]{crawford01}
{Crawford} F., {Gaensler} B.~M., {Kaspi} V.~M., {Manchester} R.~N., {Camilo}
  F., {Lyne} A.~G., {Pivovaroff} M.~J., 2001, ApJ, 554, 152

\bibitem[{{Danilenko} {et~al}\mbox{.}(2012){Danilenko}, {Kirichenko},
  {Mennickent}, {Pavlov}, {Shibanov}, {Zharikov}, \& {Zyuzin}}]{danilenko12}
{Danilenko} A., {Kirichenko} A., {Mennickent} R.~E., {Pavlov} G., {Shibanov}
  Y., {Zharikov} S., {Zyuzin} D., 2012, A\&A, 540, A28

\bibitem[{{Davies} {et~al}\mbox{.}(2009){Davies}, {Figer}, {Kudritzki},
  {Trombley}, {Kouveliotou}, \& {Wachter}}]{davies09}
{Davies} B., {Figer} D.~F., {Kudritzki} R.-P., {Trombley} C., {Kouveliotou} C.,
  {Wachter} S., 2009, ApJ, 707, 844

\bibitem[{{de Jager} \& {Djannati-Ata{\"i}}(2009)}]{dejager09b}
{de Jager} O.~C., {Djannati-Ata{\"i}} A., 2009, in Astrophysics and Space
  Science Library, Vol. 357, Astrophysics and Space Science Library, {Becker}
  W., ed., p. 451

\bibitem[{{de O{\~n}a-Wilhelmi} {et~al}\mbox{.}(2013){de O{\~n}a-Wilhelmi},
  {Rudak}, {Barrio}, {Contreras}, {Gallant}, {Hadasch}, {Hassan}, {Lopez},
  {Mazin}, {Mirabal}, {Pedaletti}, {Renaud}, {de los Reyes}, {Torres}, \& {CTA
  Consortium}}]{deonawilhelmi13}
{de O{\~n}a-Wilhelmi} E. {et~al.}, 2013, Astroparticle Physics, 43, 287

\bibitem[{{de Rosa} {et~al}\mbox{.}(2009){de Rosa}, {Ubertini}, {Campana},
  {Bazzano}, {Dean}, \& {Bassani}}]{derosa09}
{de Rosa} A., {Ubertini} P., {Campana} R., {Bazzano} A., {Dean} A.~J.,
  {Bassani} L., 2009, MNRAS, 393, 527

\bibitem[{{Decourchelle} {et~al}\mbox{.}(2001){Decourchelle}, {Sauvageot},
  {Audard}, {Aschenbach}, {Sembay}, {Rothenflug}, {Ballet}, {Stadlbauer}, \&
  {West}}]{decourchelle01}
{Decourchelle} A. {et~al.}, 2001, A\&A, 365, L218

\bibitem[{{Del Zanna}, {Amato} \& {Bucciantini}(2004){Del Zanna}, {Amato}, \&
  {Bucciantini}}]{delzanna04}
{Del Zanna} L., {Amato} E., {Bucciantini} N., 2004, A\&A, 421, 1063

\bibitem[{{Del Zanna} {et~al}\mbox{.}(2006){Del Zanna}, {Volpi}, {Amato}, \&
  {Bucciantini}}]{delzanna06}
{Del Zanna} L., {Volpi} D., {Amato} E., {Bucciantini} N., 2006, A\&A, 453, 621

\bibitem[{{DeLaney} \& {Rudnick}(2003)}]{delaney03}
{DeLaney} T., {Rudnick} L., 2003, ApJ, 589, 818

\bibitem[{{Demorest} {et~al}\mbox{.}(2010){Demorest}, {Pennucci}, {Ransom},
  {Roberts}, \& {Hessels}}]{demorest10}
{Demorest} P.~B., {Pennucci} T., {Ransom} S.~M., {Roberts} M.~S.~E., {Hessels}
  J.~W.~T., 2010, Nature, 467, 1081

\bibitem[{{Dickey} \& {Lockman}(1990)}]{dickey90}
{Dickey} J.~M., {Lockman} F.~J., 1990, ARA\&A, 28, 215

\bibitem[{{Djannati-Ata\"i}(2009)}]{djannatiatai09}
{Djannati-Ata\"i} A., 2009, in MA

\bibitem[{{Dodson} {et~al}\mbox{.}(2003){Dodson}, {Legge}, {Reynolds}, \&
  {McCulloch}}]{dodson03}
{Dodson} R., {Legge} D., {Reynolds} J.~E., {McCulloch} P.~M., 2003, ApJ, 596,
  1137

\bibitem[{{Dodson}, {McCulloch} \& {Lewis}(2002){Dodson}, {McCulloch}, \&
  {Lewis}}]{dodson02}
{Dodson} R.~G., {McCulloch} P.~M., {Lewis} D.~R., 2002, ApJL, 564, L85

\bibitem[{{Donati} {et~al}\mbox{.}(2002){Donati}, {Babel}, {Harries},
  {Howarth}, {Petit}, \& {Semel}}]{donati02}
{Donati} J.-F., {Babel} J., {Harries} T.~J., {Howarth} I.~D., {Petit} P.,
  {Semel} M., 2002, MNRAS, 333, 55

\bibitem[{{Donati} {et~al}\mbox{.}(2006){Donati}, {Howarth}, {Bouret}, {Petit},
  {Catala}, \& {Landstreet}}]{donati06}
{Donati} J.-F., {Howarth} I.~D., {Bouret} J.-C., {Petit} P., {Catala} C.,
  {Landstreet} J., 2006, MNRAS, 365, L6

\bibitem[{{Du Plessis} {et~al}\mbox{.}(1995){Du Plessis}, {de Jager},
  {Buchner}, {Nel}, {North}, {Raubenheimer}, \& {van der Walt}}]{duplessis95}
{Du Plessis} I., {de Jager} O.~C., {Buchner} S., {Nel} H.~I., {North} A.~R.,
  {Raubenheimer} B.~C., {van der Walt} D.~J., 1995, ApJ, 453, 746

\bibitem[{{Dubner}, {Giacani} \& {Decourchelle}(2008){Dubner}, {Giacani}, \&
  {Decourchelle}}]{dubner08}
{Dubner} G., {Giacani} E., {Decourchelle} A., 2008, A\&A, 487, 1033

\bibitem[{{Dubner} {et~al}\mbox{.}(2013){Dubner}, {Loiseau},
  {Rodr{\'{\i}}guez-Pascual}, {Smith}, {Giacani}, \& {Castelletti}}]{dubner13}
{Dubner} G., {Loiseau} N., {Rodr{\'{\i}}guez-Pascual} P., {Smith} M.~J.~S.,
  {Giacani} E., {Castelletti} G., 2013, A\&A, 555, A9

\bibitem[{{Dubner} {et~al}\mbox{.}(2002){Dubner}, {Gaensler}, {Giacani},
  {Goss}, \& {Green}}]{dubner02}
{Dubner} G.~M., {Gaensler} B.~M., {Giacani} E.~B., {Goss} W.~M., {Green} A.~J.,
  2002, AJ, 123, 337

\bibitem[{{Duncan} {et~al}\mbox{.}(1995){Duncan}, {Haynes}, {Stewart}, \&
  {Jones}}]{duncan95}
{Duncan} A.~R., {Haynes} R.~F., {Stewart} R.~T., {Jones} K.~L., 1995, MNRAS,
  277, 319

\bibitem[{{Duncan} \& {Thompson}(1992)}]{duncan92}
{Duncan} R.~C., {Thompson} C., 1992, ApJL, 392, L9

\bibitem[{{Duncan} \& {Thompson}(1996)}]{duncan96}
{Duncan} R.~C., {Thompson} C., 1996, in American Institute of Physics
  Conference Series, Vol. 366, High Velocity Neutron Stars, {Rothschild} R.~E.,
  {Lingenfelter} R.~E., eds., pp. 111--117

\bibitem[{{Espinoza} {et~al}\mbox{.}(2011){Espinoza}, {Lyne}, {Stappers}, \&
  {Kramer}}]{espinoza11}
{Espinoza} C.~M., {Lyne} A.~G., {Stappers} B.~W., {Kramer} M., 2011, MNRAS,
  414, 1679

\bibitem[{{Esposito} {et~al}\mbox{.}(2007){Esposito}, {Mereghetti}, {Tiengo},
  {Zane}, {Turolla}, {G{\"o}tz}, {Rea}, {Kawai}, {Ueno}, {Israel}, {Stella}, \&
  {Feroci}}]{esposito07}
{Esposito} P. {et~al.}, 2007, A\&A, 476, 321

\bibitem[{{Fahlman} \& {Gregory}(1981)}]{fahlman81}
{Fahlman} G.~G., {Gregory} P.~C., 1981, Nature, 293, 202

\bibitem[{{Fahlman} \& {Gregory}(1983)}]{fahlman83}
{Fahlman} G.~G., {Gregory} P.~C., 1983, in IAU Symposium, Vol. 101, Supernova
  Remnants and their X-ray Emission, {Danziger} J., {Gorenstein} P., eds., pp.
  445--453

\bibitem[{{Fang} \& {Zhang}(2010{\natexlab{a}})}]{fang10a}
{Fang} J., {Zhang} L., 2010{\natexlab{a}}, A\&A, 515, A20

\bibitem[{{Fang} \& {Zhang}(2010{\natexlab{b}})}]{fang10b}
{Fang} J., {Zhang} L., 2010{\natexlab{b}}, ApJ, 718, 467

\bibitem[{{Ferrand} \& {Safi-Harb}(2012)}]{ferrand12}
{Ferrand} G., {Safi-Harb} S., 2012, Advances in Space Research, 49, 1313

\bibitem[{{Ferrario} \& {Wickramasinghe}(2006)}]{ferrario06}
{Ferrario} L., {Wickramasinghe} D., 2006, MNRAS, 367, 1323

\bibitem[{{Fesen} {et~al}\mbox{.}(2008){Fesen}, {Rudie}, {Hurford}, \&
  {Soto}}]{fesen08}
{Fesen} R., {Rudie} G., {Hurford} A., {Soto} A., 2008, ApJS, 174, 379

\bibitem[{{Fesen} {et~al}\mbox{.}(2006){Fesen}, {Hammell}, {Morse},
  {Chevalier}, {Borkowski}, {Dopita}, {Gerardy}, {Lawrence}, {Raymond}, \& {van
  den Bergh}}]{fesen06}
{Fesen} R.~A. {et~al.}, 2006, ApJ, 645, 283

\bibitem[{{Fesen} {et~al}\mbox{.}(2012){Fesen}, {Kremer}, {Patnaude}, \&
  {Milisavljevic}}]{fesen12}
{Fesen} R.~A., {Kremer} R., {Patnaude} D., {Milisavljevic} D., 2012, AJ, 143,
  27

\bibitem[{{Figer} {et~al}\mbox{.}(2005){Figer}, {Najarro}, {Geballe}, {Blum},
  \& {Kudritzki}}]{figer05}
{Figer} D.~F., {Najarro} F., {Geballe} T.~R., {Blum} R.~D., {Kudritzki} R.~P.,
  2005, ApJL, 622, L49

\bibitem[{{Finley} \& {Oegelman}(1994)}]{finley94}
{Finley} J.~P., {Oegelman} H., 1994, ApJL, 434, L25

\bibitem[{{Forot} {et~al}\mbox{.}(2006){Forot}, {Hermsen}, {Renaud}, {Laurent},
  {Grenier}, {Goret}, {Khelifi}, \& {Kuiper}}]{forot06}
{Forot} M., {Hermsen} W., {Renaud} M., {Laurent} P., {Grenier} I., {Goret} P.,
  {Khelifi} B., {Kuiper} L., 2006, ApJL, 651, L45

\bibitem[{{Frail} {et~al}\mbox{.}(1996{\natexlab{a}}){Frail}, {Giacani},
  {Goss}, \& {Dubner}}]{frail96b}
{Frail} D.~A., {Giacani} E.~B., {Goss} W.~M., {Dubner} G., 1996{\natexlab{a}},
  ApJL, 464, L165

\bibitem[{{Frail} {et~al}\mbox{.}(1996{\natexlab{b}}){Frail}, {Goss},
  {Reynoso}, {Giacani}, {Green}, \& {Otrupcek}}]{frail96a}
{Frail} D.~A., {Goss} W.~M., {Reynoso} E.~M., {Giacani} E.~B., {Green} A.~J.,
  {Otrupcek} R., 1996{\natexlab{b}}, AJ, 111, 1651

\bibitem[{{Fuchs} {et~al}\mbox{.}(1999){Fuchs}, {Mirabel}, {Chaty}, {Claret},
  {Cesarsky}, \& {Cesarsky}}]{fuchs99}
{Fuchs} Y., {Mirabel} F., {Chaty} S., {Claret} A., {Cesarsky} C.~J., {Cesarsky}
  D.~A., 1999, A\&A, 350, 891

\bibitem[{{Funk} {et~al}\mbox{.}(2007{\natexlab{a}}){Funk}, {Hinton},
  {Moriguchi}, {Aharonian}, {Fukui}, {Hofmann}, {Horns}, {P{\"u}hlhofer},
  {Reimer}, {Rowell}, {Terrier}, {Vink}, \& {Wagner}}]{funk07a}
{Funk} S. {et~al.}, 2007{\natexlab{a}}, A\&A, 470, 249

\bibitem[{{Funk} {et~al}\mbox{.}(2007{\natexlab{b}}){Funk}, {Hinton},
  {P{\"u}hlhofer}, {Aharonian}, {Hofmann}, {Reimer}, \& {Wagner}}]{funk07b}
{Funk} S., {Hinton} J.~A., {P{\"u}hlhofer} G., {Aharonian} F.~A., {Hofmann} W.,
  {Reimer} O., {Wagner} S., 2007{\natexlab{b}}, ApJ, 662, 517

\bibitem[{{Gabici}, {Aharonian} \& {Casanova}(2009){Gabici}, {Aharonian}, \&
  {Casanova}}]{gabici09}
{Gabici} S., {Aharonian} F.~A., {Casanova} S., 2009, MNRAS, 396, 1629

\bibitem[{{Gaensler} {et~al}\mbox{.}(2002){Gaensler}, {Arons}, {Kaspi},
  {Pivovaroff}, {Kawai}, \& {Tamura}}]{gaensler02}
{Gaensler} B.~M., {Arons} J., {Kaspi} V.~M., {Pivovaroff} M.~J., {Kawai} N.,
  {Tamura} K., 2002, ApJ, 569, 878

\bibitem[{{Gaensler} {et~al}\mbox{.}(1999){Gaensler}, {Brazier}, {Manchester},
  {Johnston}, \& {Green}}]{gaensler99}
{Gaensler} B.~M., {Brazier} K.~T.~S., {Manchester} R.~N., {Johnston} S.,
  {Green} A.~J., 1999, MNRAS, 305, 724

\bibitem[{{Gaensler} {et~al}\mbox{.}(2005){Gaensler}, {McClure-Griffiths},
  {Oey}, {Haverkorn}, {Dickey}, \& {Green}}]{gaensler05}
{Gaensler} B.~M., {McClure-Griffiths} N.~M., {Oey} M.~S., {Haverkorn} M.,
  {Dickey} J.~M., {Green} A.~J., 2005, ApJL, 620, L95

\bibitem[{{Gaensler}, {Pivovaroff} \& {Garmire}(2001){Gaensler}, {Pivovaroff},
  \& {Garmire}}]{gaensler01a}
{Gaensler} B.~M., {Pivovaroff} M.~J., {Garmire} G.~P., 2001, ApJL, 556, L107

\bibitem[{{Gaensler} \& {Slane}(2006)}]{gaensler06}
{Gaensler} B.~M., {Slane} P.~O., 2006, ARA\&A, 44, 17

\bibitem[{{Gaensler} {et~al}\mbox{.}(2001){Gaensler}, {Slane}, {Gotthelf}, \&
  {Vasisht}}]{gaensler01b}
{Gaensler} B.~M., {Slane} P.~O., {Gotthelf} E.~V., {Vasisht} G., 2001, ApJ,
  559, 963

\bibitem[{{Gaensler} {et~al}\mbox{.}(2008){Gaensler}, {Tanna}, {Slane},
  {Brogan}, {Gelfand}, {McClure-Griffiths}, {Camilo}, {Ng}, \&
  {Miller}}]{gaensler08}
{Gaensler} B.~M. {et~al.}, 2008, ApJL, 680, L37

\bibitem[{{Gaensler} \& {Wallace}(2003)}]{gaensler03}
{Gaensler} B.~M., {Wallace} B.~J., 2003, ApJ, 594, 326

\bibitem[{{Gallant} {et~al}\mbox{.}(2008){Gallant}, {Carrigan},
  {Djannati-Ata{\"i}}, {Funk}, {Hinton}, {Hoppe}, {de Jager}, {Kh{\'e}lifi},
  {Komin}, {Kosack}, {Lemi{\`e}re}, \& {Masterson}}]{gallant08}
{Gallant} Y.~A. {et~al.}, 2008, in American Institute of Physics Conference
  Series, Vol. 983, 40 Years of Pulsars: Millisecond Pulsars, Magnetars and
  More, {Bassa} C., {Wang} Z., {Cumming} A., {Kaspi} V.~M., eds., pp. 195--199

\bibitem[{{Gallant} \& {Tuffs}(1998)}]{gallant98}
{Gallant} Y.~A., {Tuffs} R.~J., 1998, MmSAI, 69, 963

\bibitem[{{Gallant} \& {Tuffs}(1999)}]{gallant99}
{Gallant} Y.~A., {Tuffs} R.~J., 1999, in ESA Special Publication, Vol. 427, The
  Universe as Seen by ISO, {Cox} P., {Kessler} M., eds., p. 313

\bibitem[{{Gast} {et~al}\mbox{.}(2011){Gast}, {Brun}, {Carrigan}, {Chaves},
  {Deil}, {Djannati-Ata{\"i}}, {Gallant}, {Marandon}, {de Naurois}, \& {de los
  Reyes}}]{gast11}
{Gast} H. {et~al.}, 2011, International Cosmic Ray Conference, 7, 157

\bibitem[{{Gavriil} {et~al}\mbox{.}(2008){Gavriil}, {Gonzalez}, {Gotthelf},
  {Kaspi}, {Livingstone}, \& {Woods}}]{gavriil08}
{Gavriil} F.~P., {Gonzalez} M.~E., {Gotthelf} E.~V., {Kaspi} V.~M.,
  {Livingstone} M.~A., {Woods} P.~M., 2008, Science, 319, 1802

\bibitem[{{Gelfand}, {Slane} \& {Zhang}(2009){Gelfand}, {Slane}, \&
  {Zhang}}]{gelfand09}
{Gelfand} J.~D., {Slane} P.~O., {Zhang} W., 2009, ApJ, 703, 2051

\bibitem[{{Giacani} {et~al}\mbox{.}(2013){Giacani}, {Rovero}, {Cillis},
  {Pichel}, \& {Dubner}}]{giacani13}
{Giacani} E., {Rovero} A., {Cillis} A.~N., {Pichel} A., {Dubner} G., 2013

\bibitem[{{Giacani} {et~al}\mbox{.}(2000){Giacani}, {Dubner}, {Green}, {Goss},
  \& {Gaensler}}]{giacani00}
{Giacani} E.~B., {Dubner} G.~M., {Green} A.~J., {Goss} W.~M., {Gaensler} B.~M.,
  2000, AJ, 119, 281

\bibitem[{{Giacani} {et~al}\mbox{.}(1997){Giacani}, {Dubner}, {Kassim},
  {Frail}, {Goss}, {Winkler}, \& {Williams}}]{giacani97}
{Giacani} E.~B., {Dubner} G.~M., {Kassim} N.~E., {Frail} D.~A., {Goss} W.~M.,
  {Winkler} P.~F., {Williams} B.~F., 1997, AJ, 113, 1379

\bibitem[{{Giacconi} {et~al}\mbox{.}(1971){Giacconi}, {Gursky}, {Kellogg},
  {Schreier}, \& {Tananbaum}}]{giacconi71}
{Giacconi} R., {Gursky} H., {Kellogg} E., {Schreier} E., {Tananbaum} H., 1971,
  ApJL, 167, L67

\bibitem[{{Ginzburg}(1979)}]{ginzburg79}
{Ginzburg} V.~L., 1979, {Theoretical physics and astrophysics}. Pergamon Press

\bibitem[{{Ginzburg} \& {Syrovatskii}(1964)}]{ginzburg64}
{Ginzburg} V.~L., {Syrovatskii} S.~I., 1964, {The Origin of Cosmic Rays}.
  Springer

\bibitem[{{Ginzburg} \& {Syrovatskii}(1965)}]{ginzburg65}
{Ginzburg} V.~L., {Syrovatskii} S.~I., 1965, ARA\&A, 3, 297

\bibitem[{{Goldreich} \& {Julian}(1969)}]{goldreich69}
{Goldreich} P., {Julian} W.~H., 1969, ApJ, 157, 869

\bibitem[{{Gonzalez} \& {Safi-Harb}(2003)}]{gonzalez03}
{Gonzalez} M., {Safi-Harb} S., 2003, ApJL, 591, L143

\bibitem[{{Gonzalez} {et~al}\mbox{.}(2005){Gonzalez}, {Kaspi}, {Camilo},
  {Gaensler}, \& {Pivovaroff}}]{gonzalez05}
{Gonzalez} M.~E., {Kaspi} V.~M., {Camilo} F., {Gaensler} B.~M., {Pivovaroff}
  M.~J., 2005, ApJ, 630, 489

\bibitem[{{Gotthelf} \& {Halpern}(2009)}]{gotthelf09b}
{Gotthelf} E.~V., {Halpern} J.~P., 2009, ApJL, 700, L158

\bibitem[{{Gotthelf}, {Helfand} \& {Newburgh}(2007){Gotthelf}, {Helfand}, \&
  {Newburgh}}]{gotthelf07}
{Gotthelf} E.~V., {Helfand} D.~J., {Newburgh} L., 2007, ApJ, 654, 267

\bibitem[{{Gotthelf}, {Petre} \& {Vasisht}(1999){Gotthelf}, {Petre}, \&
  {Vasisht}}]{gotthelf99a}
{Gotthelf} E.~V., {Petre} R., {Vasisht} G., 1999, ApJL, 514, L107

\bibitem[{{Gotthelf} \& {Vasisht}(1997)}]{gotthelf97}
{Gotthelf} E.~V., {Vasisht} G., 1997, ApJL, 486, L133

\bibitem[{{Gotthelf} {et~al}\mbox{.}(2000){Gotthelf}, {Vasisht},
  {Boylan-Kolchin}, \& {Torii}}]{gotthelf00b}
{Gotthelf} E.~V., {Vasisht} G., {Boylan-Kolchin} M., {Torii} K., 2000, ApJL,
  542, L37

\bibitem[{{Gotthelf}, {Vasisht} \& {Dotani}(1999){Gotthelf}, {Vasisht}, \&
  {Dotani}}]{gotthelf99b}
{Gotthelf} E.~V., {Vasisht} G., {Dotani} T., 1999, ApJL, 522, L49

\bibitem[{{Gotthelf} \& {Wang}(2000)}]{gotthelf00a}
{Gotthelf} E.~V., {Wang} Q.~D., 2000, ApJL, 532, L117

\bibitem[{{Grasdalen}(1979)}]{grasdalen79}
{Grasdalen} G.~L., 1979, PASP, 91, 436

\bibitem[{{Green}(1986)}]{green86}
{Green} D.~A., 1986, MNRAS, 218, 533

\bibitem[{{Green}(1994)}]{green94}
{Green} D.~A., 1994, ApJS, 90, 817

\bibitem[{{Green}(2009)}]{green09}
{Green} D.~A., 2009, Bulletin of the Astronomical Society of India, 37, 45

\bibitem[{{Green} \& {Scheuer}(1992)}]{green92}
{Green} D.~A., {Scheuer} P.~A.~G., 1992, MNRAS, 258, 833

\bibitem[{{Green}, {Tuffs} \& {Popescu}(2004){Green}, {Tuffs}, \&
  {Popescu}}]{green04}
{Green} D.~A., {Tuffs} R.~J., {Popescu} C.~C., 2004, MNRAS, 355, 1315

\bibitem[{{Gregory} \& {Fahlman}(1980)}]{gregory80}
{Gregory} P.~C., {Fahlman} G.~G., 1980, Nature, 287, 805

\bibitem[{{Griffith} {et~al}\mbox{.}(1990){Griffith}, {Langston}, {Heflin},
  {Conner}, {Lehar}, \& {Burke}}]{griffith90}
{Griffith} M., {Langston} G., {Heflin} M., {Conner} S., {Lehar} J., {Burke} B.,
  1990, ApJS, 74, 129

\bibitem[{{Griffith} \& {Wright}(1993)}]{griffith93}
{Griffith} M.~R., {Wright} A.~E., 1993, AJ, 105, 1666

\bibitem[{{H.~E.~S.~S.~Collaboration}
  {et~al}\mbox{.}(2007){H.~E.~S.~S.~Collaboration}, {:}, {Djannati-Atai}, {De
  Jager}, {Terrier}, {Gallant}, \& {Hoppe}}]{djannatiatai07}
{H.~E.~S.~S.~Collaboration}, {:}, {Djannati-Atai} A., {De Jager} O.~C.,
  {Terrier} R., {Gallant} Y.~A., {Hoppe} S., 2007, ArXiv e-prints

\bibitem[{{Hall}, {Wakely} \& {VERITAS Collaboration}(2001){Hall}, {Wakely}, \&
  {VERITAS Collaboration}}]{hall01}
{Hall} T.~A., {Wakely} S.~P., {VERITAS Collaboration}, 2001, International
  Cosmic Ray Conference, 6, 2485

\bibitem[{{Halpern} {et~al}\mbox{.}(2001){Halpern}, {Camilo}, {Gotthelf},
  {Helfand}, {Kramer}, {Lyne}, {Leighly}, \& {Eracleous}}]{halpern01}
{Halpern} J.~P., {Camilo} F., {Gotthelf} E.~V., {Helfand} D.~J., {Kramer} M.,
  {Lyne} A.~G., {Leighly} K.~M., {Eracleous} M., 2001, ApJL, 552, L125

\bibitem[{{Halpern} \& {Gotthelf}(2010)}]{halpern10}
{Halpern} J.~P., {Gotthelf} E.~V., 2010, ApJ, 709, 436

\bibitem[{{Halpern}, {Gotthelf} \& {Camilo}(2012){Halpern}, {Gotthelf}, \&
  {Camilo}}]{halpern12}
{Halpern} J.~P., {Gotthelf} E.~V., {Camilo} F., 2012, ApJL, 753, L14

\bibitem[{{Halpern} {et~al}\mbox{.}(2002){Halpern}, {Gotthelf}, {Camilo},
  {Collins}, \& {Helfand}}]{halpern02}
{Halpern} J.~P., {Gotthelf} E.~V., {Camilo} F., {Collins} B., {Helfand} D.~J.,
  2002, in Astronomical Society of the Pacific Conference Series, Vol. 271,
  Neutron Stars in Supernova Remnants, {Slane} P.~O., {Gaensler} B.~M., eds.,
  p. 199

\bibitem[{{Halpern} {et~al}\mbox{.}(2004){Halpern}, {Gotthelf}, {Camilo},
  {Helfand}, \& {Ransom}}]{halpern04}
{Halpern} J.~P., {Gotthelf} E.~V., {Camilo} F., {Helfand} D.~J., {Ransom}
  S.~M., 2004, ApJ, 612, 398

\bibitem[{{Harding}(2013)}]{harding13}
{Harding} A.~K., 2013, Frontiers of Physics, 8, 679

\bibitem[{{Harris} \& {Roberts}(1960)}]{harris60}
{Harris} D.~E., {Roberts} J.~A., 1960, PASP, 72, 237

\bibitem[{{Haug}(1998)}]{haug98}
{Haug} E., 1998, Solar Physics, 178, 341

\bibitem[{{Haug}(2004)}]{haug04}
{Haug} E., 2004, A\&A, 423, 793

\bibitem[{{Hayato} {et~al}\mbox{.}(2010){Hayato}, {Yamaguchi}, {Tamagawa},
  {Katsuda}, {Hwang}, {Hughes}, {Ozawa}, {Bamba}, {Kinugasa}, {Terada},
  {Furuzawa}, {Kunieda}, \& {Makishima}}]{hayato10}
{Hayato} A. {et~al.}, 2010, ApJ, 725, 894

\bibitem[{{Heger} {et~al}\mbox{.}(2003){Heger}, {Fryer}, {Woosley}, {Langer},
  \& {Hartmann}}]{heger03}
{Heger} A., {Fryer} C.~L., {Woosley} S.~E., {Langer} N., {Hartmann} D.~H.,
  2003, ApJ, 591, 288

\bibitem[{{Heger}, {Woosley} \& {Spruit}(2005){Heger}, {Woosley}, \&
  {Spruit}}]{heger05}
{Heger} A., {Woosley} S.~E., {Spruit} H.~C., 2005, ApJ, 626, 350

\bibitem[{{Helfand} \& {Becker}(1987)}]{helfand87}
{Helfand} D.~J., {Becker} R.~H., 1987, ApJ, 314, 203

\bibitem[{{Helfand} {et~al}\mbox{.}(1994){Helfand}, {Becker}, {Hawkins}, \&
  {White}}]{helfand94}
{Helfand} D.~J., {Becker} R.~H., {Hawkins} G., {White} R.~L., 1994, ApJ, 434,
  627

\bibitem[{{Helfand}, {Collins} \& {Gotthelf}(2003){Helfand}, {Collins}, \&
  {Gotthelf}}]{helfand03}
{Helfand} D.~J., {Collins} B.~F., {Gotthelf} E.~V., 2003, ApJ, 582, 783

\bibitem[{{Helfand} {et~al}\mbox{.}(2007){Helfand}, {Gotthelf}, {Halpern},
  {Camilo}, {Semler}, {Becker}, \& {White}}]{helfand07}
{Helfand} D.~J., {Gotthelf} E.~V., {Halpern} J.~P., {Camilo} F., {Semler}
  D.~R., {Becker} R.~H., {White} R.~L., 2007, ApJ, 665, 1297

\bibitem[{{Hennessy} {et~al}\mbox{.}(1992){Hennessy}, {O'Connell}, {Cheng},
  {Bohlin}, {Collins}, {Gull}, {Hintzen}, {Isensee}, {Landsman}, {Roberts},
  {Smith}, {Smith}, \& {Stecher}}]{hennessy92}
{Hennessy} G.~S. {et~al.}, 1992, ApJL, 395, L13

\bibitem[{{H.E.S.S.~Collaboration}
  {et~al}\mbox{.}(2012){H.E.S.S.~Collaboration}, {Abramowski}, {Acero},
  {Aharonian}, {Akhperjanian}, {Anton}, {Balenderan}, {Balzer}, {Barnacka},
  {Becherini}, {Becker}, {Bernl{\"o}hr}, {Birsin}, {Biteau}, {Bochow},
  {Boisson}, {Bolmont}, {Bordas}, {Brucker}, {Brun}, {Brun}, {Bulik},
  {Carrigan}, {Casanova}, {Cerruti}, {Chadwick}, {Charbonnier}, {Chaves},
  {Cheesebrough}, {Cologna}, {Conrad}, {Couturier}, {Dalton}, {Daniel},
  {Davids}, {Degrange}, {Deil}, {Dickinson}, {Djannati-At{\"a}{\i}},
  {Domainko}, {Drury}, {Dubus}, {Dutson}, {Dyks}, {Dyrda}, {Egberts}, {Eger},
  {Espigat}, {Fallon}, {Farnier}, {Fegan}, {Feinstein}, {Fernandes},
  {Fernandez}, {Fiasson}, {Fontaine}, {F{\"o}rster}, {F{\"u}{\ss}ling},
  {Gajdus}, {Gallant}, {Garrigoux}, {Gast}, {G{\'e}rard}, {Giebels},
  {Glicenstein}, {Gl{\"u}ck}, {G{\"o}ring}, {Grondin}, {H{\"a}ffner}, {Hague},
  {Hahn}, {Hampf}, {Harris}, {Hauser}, {Heinz}, {Heinzelmann}, {Henri},
  {Hermann}, {Hillert}, {Hinton}, {Hofmann}, {Hofverberg}, {Holler}, {Horns},
  {Jacholkowska}, {de Jager}, {Jahn}, {Jamrozy}, {Jung}, {Kastendieck},
  {Katarzy{\'n}ski}, {Katz}, {Kaufmann}, {Kh{\'e}lifi}, {Klochkov},
  {Klu{\'z}niak}, {Kneiske}, {Komin}, {Kosack}, {Kossakowski}, {Krayzel},
  {Laffon}, {Lamanna}, {Lenain}, {Lennarz}, {Lohse}, {Lopatin}, {Lu},
  {Marandon}, {Marcowith}, {Masbou}, {Maurin}, {Maxted}, {Mayer}, {McComb},
  {Medina}, {M{\'e}hault}, {Menzler}, {Moderski}, {Mohamed}, {Moulin},
  {Naumann}, {Naumann-Godo}, {de Naurois}, {Nedbal}, {Nguyen}, {Nicholas},
  {Niemiec}, {Nolan}, {Ohm}, {de O{\~n}a Wilhelmi}, {Opitz}, {Ostrowski},
  {Oya}, {Panter}, {Paz Arribas}, {Pekeur}, {Pelletier}, {Perez}, {Petrucci},
  {Peyaud}, {Pita}, {P{\"u}hlhofer}, {Punch}, {Quirrenbach}, {Raue}, {Reimer},
  {Reimer}, {Renaud}, {de los Reyes}, {Rieger}, {Ripken}, {Rob}, {Rosier-Lees},
  {Rowell}, {Rudak}, {Rulten}, {Sahakian}, {Sanchez}, {Santangelo},
  {Schlickeiser}, {Schulz}, {Schwanke}, {Schwarzburg}, {Schwemmer}, {Sheidaei},
  {Skilton}, {Sol}, {Spengler}, {Stawarz}, {.}, {Steenkamp}, {Stegmann},
  {Stinzing}, {Stycz}, {Sushch}, {Szostek}, {Tavernet}, {Terrier},
  {Tluczykont}, {Valerius}, {van Eldik}, {Vasileiadis}, {Venter}, {Viana},
  {Vincent}, {V{\"o}lk}, {Volpe}, {Vorobiov}, {Vorster}, {Wagner}, {Ward},
  {White}, {Wierzcholska}, {Zacharias}, {Zajczyk}, {Zdziarski}, {Zech}, \&
  {Zechlin}}]{abramowski12}
{H.E.S.S.~Collaboration} {et~al.}, 2012, A\&A, 545, L2

\bibitem[{{H.E.S.S.~Collaboration}
  {et~al}\mbox{.}(2011{\natexlab{a}}){H.E.S.S.~Collaboration}, {Abramowski},
  {Acero}, {Aharonian}, {Akhperjanian}, {Anton}, {Balzer}, {Barnacka}, {Barres
  de Almeida}, {Becherini}, {Becker}, {Behera}, {Bernl{\"o}hr}, {Bochow},
  {Boisson}, {Bolmont}, {Bordas}, {Brucker}, {Brun}, {Brun}, {Bulik},
  {B{\"u}sching}, {Carrigan}, {Casanova}, {Cerruti}, {Chadwick}, {Charbonnier},
  {Chaves}, {Cheesebrough}, {Chounet}, {Clapson}, {Coignet}, {Cologna},
  {Conrad}, {Dalton}, {Daniel}, {Davids}, {Degrange}, {Deil}, {Dickinson},
  {Djannati-Ata{\"i}}, {Domainko}, {O'C.~Drury}, {Dubois}, {Dubus}, {Dutson},
  {Dyks}, {Dyrda}, {Egberts}, {Eger}, {Espigat}, {Fallon}, {Farnier}, {Fegan},
  {Feinstein}, {Fernandes}, {Fiasson}, {Fontaine}, {F{\"o}rster},
  {F{\"u}ssling}, {Gallant}, {Gast}, {G{\'e}rard}, {Gerbig}, {Giebels},
  {Glicenstein}, {Gl{\"u}ck}, {Goret}, {G{\"o}ring}, {H{\"a}ffner}, {Hague},
  {Hampf}, {Hauser}, {Heinz}, {Heinzelmann}, {Henri}, {Hermann}, {Hinton},
  {Hoffmann}, {Hofmann}, {Hofverberg}, {Holler}, {Horns}, {Jacholkowska}, {de
  Jager}, {Jahn}, {Jamrozy}, {Jung}, {Kastendieck}, {Katarzynski}, {Katz},
  {Kaufmann}, {Keogh}, {Khangulyan}, {Kh{\'e}lifi}, {Klochkov}, {Kluzniak},
  {Kneiske}, {Komin}, {Kosack}, {Kossakowski}, {Laffon}, {Lamanna}, {Lennarz},
  {Lohse}, {Lopatin}, {Lu}, {Marandon}, {Marcowith}, {Masbou}, {Maurin},
  {Maxted}, {Mayer}, {McComb}, {Medina}, {M{\'e}hault}, {Moderski}, {Moulin},
  {Naumann}, {Naumann-Godo}, {de Naurois}, {Nedbal}, {Nekrassov}, {Nguyen},
  {Nicholas}, {Niemiec}, {Nolan}, {Ohm}, {de Ona Wilhelmi}, {Opitz},
  {Ostrowski}, {Oya}, {Panter}, {Paz Arribas}, {Pedaletti}, {Pelletier},
  {Petrucci}, {Pita}, {P{\"u}hlhofer}, {Punch}, {Quirrenbach}, {Raue},
  {Rayner}, {Reimer}, {Reimer}, {Renaud}, {de Los Reyes}, {Rieger}, {Ripken},
  {Rob}, {Rosier-Lees}, {Rowell}, {Rudak}, {Rulten}, {Ruppel}, {Sahakian},
  {Sanchez}, {Santangelo}, {Schlickeiser}, {Sch{\"o}ck}, {Schulz}, {Schwanke},
  {Schwarzburg}, {Schwemmer}, {Sikora}, {Skilton}, {Sol}, {Spengler},
  {Stawarz}, {Steenkamp}, {Stegmann}, {Stinzing}, {Stycz}, {Sushch}, {Szostek},
  {Tavernet}, {Terrier}, {Tluczykont}, {Valerius}, {van Eldik}, {Vasileiadis},
  {Venter}, {Vialle}, {Viana}, {Vincent}, {V{\"o}lk}, {Volpe}, {Vorobiov},
  {Vorster}, {Wagner}, {Ward}, {White}, {Wierzcholska}, {Zacharias}, {Zajczyk},
  {Zdziarski}, {Zech}, \& {Zechlin}}]{abramowski11b}
{H.E.S.S.~Collaboration} {et~al.}, 2011{\natexlab{a}}, A\&A, 533, A103

\bibitem[{{H.E.S.S.~Collaboration}
  {et~al}\mbox{.}(2011{\natexlab{b}}){H.E.S.S.~Collaboration}, {Abramowski},
  {Acero}, {Aharonian}, {Akhperjanian}, {Anton}, {Barnacka}, {Barres de
  Almeida}, {Bazer-Bachi}, {Becherini}, {Becker}, {Behera}, {Bernl{\"o}hr},
  {Bochow}, {Boisson}, {Bolmont}, {Bordas}, {Borrel}, {Brucker}, {Brun},
  {Brun}, {Bulik}, {B{\"u}sching}, {Boutelier}, {Casanova}, {Cerruti},
  {Chadwick}, {Charbonnier}, {Chaves}, {Cheesebrough}, {Conrad}, {Chounet},
  {Clapson}, {Coignet}, {Dalton}, {Daniel}, {Davids}, {Degrange}, {Deil},
  {Dickinson}, {Djannati-Ata{\"i}}, {Domainko}, {Drury}, {Dubois}, {Dubus},
  {Dyks}, {Dyrda}, {Egberts}, {Eger}, {Espigat}, {Fallon}, {Farnier}, {Fegan},
  {Feinstein}, {Fernandes}, {Fiasson}, {F{\"o}rster}, {Fontaine},
  {F{\"u}{\ss}ling}, {Gabici}, {Gallant}, {G{\'e}rard}, {Gerbig}, {Giebels},
  {Glicenstein}, {Gl{\"u}ck}, {Goret}, {G{\"o}ring}, {Hague}, {Hampf},
  {Hauser}, {Heinz}, {Heinzelmann}, {Henri}, {Hermann}, {Hinton}, {Hoffmann},
  {Hofmann}, {Hofverberg}, {Holleran}, {Hoppe}, {Horns}, {Jacholkowska}, {de
  Jager}, {Jahn}, {Jung}, {Katarzy{\'n}ski}, {Katz}, {Kaufmann}, {Kerschhaggl},
  {Khangulyan}, {Kh{\'e}lifi}, {Keogh}, {Klochkov}, {Klu{\'z}niak}, {Kneiske},
  {Komin}, {Kosack}, {Kossakowski}, {Lamanna}, {Lenain}, {Lennarz}, {Lohse},
  {Lu}, {Marandon}, {Marcowith}, {Masbou}, {Maurin}, {McComb}, {Medina},
  {M{\'e}hault}, {Moderski}, {Moulin}, {Naumann-Godo}, {de Naurois}, {Nedbal},
  {Nekrassov}, {Nguyen}, {Nicholas}, {Niemiec}, {Nolan}, {Ohm}, {Olive}, {de
  O{\~n}a Wilhelmi}, {Opitz}, {Orford}, {Ostrowski}, {Panter}, {Paz Arribas},
  {Pedaletti}, {Pelletier}, {Petrucci}, {Pita}, {P{\"u}hlhofer}, {Punch},
  {Quirrenbach}, {Raubenheimer}, {Raue}, {Rayner}, {Reimer}, {Reimer},
  {Renaud}, {de los Reyes}, {Rieger}, {Ripken}, {Rob}, {Rosier-Lees}, {Rowell},
  {Rudak}, {Rulten}, {Ruppel}, {Ryde}, {Sahakian}, {Santangelo},
  {Schlickeiser}, {Sch{\"o}ck}, {Sch{\"o}nwald}, {Schwanke}, {Schwarzburg},
  {Schwemmer}, {Shalchi}, {Sushch}, {Sikora}, {Skilton}, {Sol}, {Spengler},
  {Stawarz}, {Steenkamp}, {Stegmann}, {Stinzing}, {Szostek}, {Tam}, {Tavernet},
  {Terrier}, {Tibolla}, {Tluczykont}, {Valerius}, {van Eldik}, {Vasileiadis},
  {Venter}, {Vialle}, {Vincent}, {Vivier}, {V{\"o}lk}, {Volpe}, {Wagner},
  {Ward}, {Zdziarski}, {Zech}, {Zechlin}, {Fukui}, {Furukawa}, {Ohama}, {Sano},
  {Dawson}, {Kawamura}, \& {H.E.S.S.~Collaboration}}]{abramowski11a}
{H.E.S.S.~Collaboration} {et~al.}, 2011{\natexlab{b}}, A\&A, 525, A46

\bibitem[{{Hester} {et~al}\mbox{.}(1996){Hester}, {Stone}, {Scowen}, {Jun},
  {Gallagher}, {Norman}, {Ballester}, {Burrows}, {Casertano}, {Clarke},
  {Crisp}, {Griffiths}, {Hoessel}, {Holtzman}, {Krist}, {Mould}, {Sankrit},
  {Stapelfeldt}, {Trauger}, {Watson}, \& {Westphal}}]{hester96}
{Hester} J.~J. {et~al.}, 1996, ApJ, 456, 225

\bibitem[{{Hewish} {et~al}\mbox{.}(1968){Hewish}, {Bell}, {Pilkington},
  {Scott}, \& {Collins}}]{hewish68}
{Hewish} A., {Bell} S.~J., {Pilkington} J.~D.~H., {Scott} P.~F., {Collins}
  R.~A., 1968, Nature, 217, 709

\bibitem[{{Hillas} {et~al}\mbox{.}(1998){Hillas}, {Akerlof}, {Biller},
  {Buckley}, {Carter-Lewis}, {Catanese}, {Cawley}, {Fegan}, {Finley}, {Gaidos},
  {Krennrich}, {Lamb}, {Lang}, {Mohanty}, {Punch}, {Reynolds}, {Rodgers},
  {Rose}, {Rovero}, {Schubnell}, {Sembroski}, {Vacanti}, {Weekes}, {West}, \&
  {Zweerink}}]{hillas98}
{Hillas} A.~M. {et~al.}, 1998, ApJ, 503, 744

\bibitem[{{Hinton} \& {Hofmann}(2009)}]{hinton09}
{Hinton} J.~A., {Hofmann} W., 2009, ARA\&A, 47, 523

\bibitem[{{Hobbs} {et~al}\mbox{.}(2004){Hobbs}, {Lyne}, {Kramer}, {Martin}, \&
  {Jordan}}]{hobbs04}
{Hobbs} G., {Lyne} A.~G., {Kramer} M., {Martin} C.~E., {Jordan} C., 2004,
  MNRAS, 353, 1311

\bibitem[{{Holler} {et~al}\mbox{.}(2012){Holler}, {Sch{\"o}ck}, {Eger},
  {Kie{\ss}ling}, {Valerius}, \& {Stegmann}}]{holler12}
{Holler} M., {Sch{\"o}ck} F.~M., {Eger} P., {Kie{\ss}ling} D., {Valerius} K.,
  {Stegmann} C., 2012, A\&A, 539, A24

\bibitem[{{Hughes}, {Hayashi} \& {Koyama}(1998){Hughes}, {Hayashi}, \&
  {Koyama}}]{hughes98}
{Hughes} J.~P., {Hayashi} I., {Koyama} K., 1998, ApJ, 505, 732

\bibitem[{{Hughes} \& {Helfand}(1985)}]{hughes85}
{Hughes} J.~P., {Helfand} D.~J., 1985, ApJ, 291, 544

\bibitem[{{Hughes} {et~al}\mbox{.}(2001){Hughes}, {Slane}, {Burrows},
  {Garmire}, {Nousek}, {Olbert}, \& {Keohane}}]{hughes01}
{Hughes} J.~P., {Slane} P.~O., {Burrows} D.~N., {Garmire} G., {Nousek} J.~A.,
  {Olbert} C.~M., {Keohane} J.~W., 2001, ApJL, 559, L153

\bibitem[{{Hughes} {et~al}\mbox{.}(2003){Hughes}, {Slane}, {Park}, {Roming}, \&
  {Burrows}}]{hughes03}
{Hughes} J.~P., {Slane} P.~O., {Park} S., {Roming} P.~W.~A., {Burrows} D.~N.,
  2003, ApJL, 591, L139

\bibitem[{{Hurford} \& {Fesen}(1995)}]{hurford95}
{Hurford} A.~P., {Fesen} R.~A., 1995, MNRAS, 277, 549

\bibitem[{{Hurley-Walker} {et~al}\mbox{.}(2009){Hurley-Walker}, {Scaife},
  {Green}, {Davies}, {Grainge}, {Hobson}, {Jones}, {Kaneko}, {Lasenby},
  {Pooley}, {Saunders}, {Scott}, {Titterington}, {Waldram}, \&
  {Zwart}}]{hurleywalker09}
{Hurley-Walker} N. {et~al.}, 2009, MNRAS, 396, 365

\bibitem[{{Hwang} \& {Gotthelf}(1997)}]{hwang97}
{Hwang} U., {Gotthelf} E.~V., 1997, ApJ, 475, 665

\bibitem[{{Hwang}, {Petre} \& {Flanagan}(2008){Hwang}, {Petre}, \&
  {Flanagan}}]{hwang08}
{Hwang} U., {Petre} R., {Flanagan} K.~A., 2008, ApJ, 676, 378

\bibitem[{{Indebetouw} {et~al}\mbox{.}(2009){Indebetouw}, {de Messi{\`e}res},
  {Madden}, {Engelbracht}, {Smith}, {Meixner}, {Brandl}, {Smith}, {Boulanger},
  {Galliano}, {Gordon}, {Hora}, {Sewilo}, {Tielens}, {Werner}, \&
  {Wolfire}}]{indebetouw09}
{Indebetouw} R. {et~al.}, 2009, ApJ, 694, 84

\bibitem[{{Itoh}(1977)}]{itoh77}
{Itoh} H., 1977, PASJ, 29, 813

\bibitem[{{Iwasawa}, {Koyama} \& {Halpern}(1992){Iwasawa}, {Koyama}, \&
  {Halpern}}]{iwasawa92}
{Iwasawa} K., {Koyama} K., {Halpern} J.~P., 1992, PASJ, 44, 9

\bibitem[{{Jackson}(1962)}]{jackson62}
{Jackson} J.~D., 1962, {Classical Electrodynamics}. Wiley

\bibitem[{{Jones}(1968)}]{jones68}
{Jones} F.~C., 1968, Physical Review, 167, 1159

\bibitem[{{Kaaret} {et~al}\mbox{.}(2001){Kaaret}, {Marshall}, {Aldcroft},
  {Graessle}, {Karovska}, {Murray}, {Rots}, {Schulz}, \& {Seward}}]{kaaret01}
{Kaaret} P. {et~al.}, 2001, ApJ, 546, 1159

\bibitem[{{Kaastra} \& {Jansen}(1993)}]{kaastra93}
{Kaastra} J.~S., {Jansen} F.~A., 1993, A\&AS, 97, 873

\bibitem[{{Kaastra} {et~al}\mbox{.}(2008){Kaastra}, {Paerels}, {Durret},
  {Schindler}, \& {Richter}}]{kaastra08}
{Kaastra} J.~S., {Paerels} F.~B.~S., {Durret} F., {Schindler} S., {Richter} P.,
  2008, SSRv, 134, 155

\bibitem[{{Kargaltsev} \& {Pavlov}(2008)}]{kargaltsev08}
{Kargaltsev} O., {Pavlov} G.~G., 2008, in American Institute of Physics
  Conference Series, Vol. 983, 40 Years of Pulsars: Millisecond Pulsars,
  Magnetars and More, {Bassa} C., {Wang} Z., {Cumming} A., {Kaspi} V.~M., eds.,
  pp. 171--185

\bibitem[{{Kargaltsev} \& {Pavlov}(2010)}]{kargaltsev10}
{Kargaltsev} O., {Pavlov} G.~G., 2010, X-ray Astronomy 2009; Present Status,
  Multi-Wavelength Approach and Future Perspectives, 1248, 25

\bibitem[{{Kargaltsev}, {Pavlov} \& {Wong}(2009){Kargaltsev}, {Pavlov}, \&
  {Wong}}]{kargaltsev09}
{Kargaltsev} O., {Pavlov} G.~G., {Wong} J.~A., 2009, ApJ, 690, 891

\bibitem[{{Kaspi} \& {Boydstun}(2010)}]{kaspi10}
{Kaspi} V.~M., {Boydstun} K., 2010, ApJL, 710, L115

\bibitem[{{Kaspi} {et~al}\mbox{.}(1998){Kaspi}, {Crawford}, {Manchester},
  {Lyne}, {Camilo}, {D'Amico}, \& {Gaensler}}]{kaspi98}
{Kaspi} V.~M., {Crawford} F., {Manchester} R.~N., {Lyne} A.~G., {Camilo} F.,
  {D'Amico} N., {Gaensler} B.~M., 1998, ApJL, 503, L161

\bibitem[{{Kaspi} {et~al}\mbox{.}(1994){Kaspi}, {Manchester}, {Siegman},
  {Johnston}, \& {Lyne}}]{kaspi94}
{Kaspi} V.~M., {Manchester} R.~N., {Siegman} B., {Johnston} S., {Lyne} A.~G.,
  1994, ApJL, 422, L83

\bibitem[{{Kaspi} {et~al}\mbox{.}(2001){Kaspi}, {Roberts}, {Vasisht},
  {Gotthelf}, {Pivovaroff}, \& {Kawai}}]{kaspi01}
{Kaspi} V.~M., {Roberts} M.~E., {Vasisht} G., {Gotthelf} E.~V., {Pivovaroff}
  M., {Kawai} N., 2001, ApJ, 560, 371

\bibitem[{{Katsuda}, {Tsunemi} \& {Mori}(2008){Katsuda}, {Tsunemi}, \&
  {Mori}}]{katsuda08}
{Katsuda} S., {Tsunemi} H., {Mori} K., 2008, ApJL, 678, L35

\bibitem[{{Keek}, {Kuiper} \& {Hermsen}(2006){Keek}, {Kuiper}, \&
  {Hermsen}}]{keek06}
{Keek} S., {Kuiper} L., {Hermsen} W., 2006, The Astronomer's Telegram, 810, 1

\bibitem[{{Kellett} {et~al}\mbox{.}(1987){Kellett}, {Branduardi-Raymont},
  {Culhane}, {Mason}, {Mason}, \& {Whitehouse}}]{kellett87}
{Kellett} B.~J., {Branduardi-Raymont} G., {Culhane} J.~L., {Mason} I.~M.,
  {Mason} K.~O., {Whitehouse} D.~R., 1987, MNRAS, 225, 199

\bibitem[{{Kennel} \& {Coroniti}(1984{\natexlab{a}})}]{kennel84a}
{Kennel} C.~F., {Coroniti} F.~V., 1984{\natexlab{a}}, ApJ, 283, 694

\bibitem[{{Kennel} \& {Coroniti}(1984{\natexlab{b}})}]{kennel84b}
{Kennel} C.~F., {Coroniti} F.~V., 1984{\natexlab{b}}, ApJ, 283, 710

\bibitem[{{Khelifi}(2002)}]{khelifi02}
{Khelifi} B., 2002, PhD thesis, University of Caen

\bibitem[{{Kh{\'e}lifi} {et~al}\mbox{.}(2008){Kh{\'e}lifi}, {Masterson},
  {Pita}, \& {et al.}}]{khelifi08}
{Kh{\'e}lifi} B., {Masterson} C., {Pita} S., {et al.}, 2008, International
  Cosmic Ray Conference, 2, 803

\bibitem[{{Kinugasa} \& {Tsunemi}(1999)}]{kinugasa99}
{Kinugasa} K., {Tsunemi} H., 1999, PASJ, 51, 239

\bibitem[{{Kirshner} {et~al}\mbox{.}(1989){Kirshner}, {Morse}, {Winkler}, \&
  {Blair}}]{kirshner89}
{Kirshner} R.~P., {Morse} J.~A., {Winkler} P.~F., {Blair} W.~P., 1989, ApJ,
  342, 260

\bibitem[{{Klein} \& {Nishina}(1929)}]{klein29}
{Klein} O., {Nishina} Y., 1929, Z. Phys., 52, 853

\bibitem[{{Komissarov} \& {Lyubarsky}(2004)}]{komissarov04}
{Komissarov} S.~S., {Lyubarsky} Y.~E., 2004, MNRAS, 349, 779

\bibitem[{{Konopelko}(2008)}]{konopelko08}
{Konopelko} A., 2008, International Cosmic Ray Conference, 2, 767

\bibitem[{{Koo} {et~al}\mbox{.}(2008){Koo}, {McKee}, {Lee}, {Lee}, {Lee},
  {Moon}, {Hong}, {Kaneda}, \& {Onaka}}]{koo08}
{Koo} B.-C. {et~al.}, 2008, ApJL, 673, L147

\bibitem[{{Koo} {et~al}\mbox{.}(2011){Koo}, {McKee}, {Suh}, {Moon}, {Onaka},
  {Burton}, {Hiramatsu}, {Bessell}, {Gaensler}, {Kim}, {Lee}, {Jeong}, {Lee},
  {Im}, {Tatematsu}, {Kohno}, {Kawabe}, {Ezawa}, {Wilson}, {Yun}, \&
  {Hughes}}]{koo11}
{Koo} B.-C. {et~al.}, 2011, ApJ, 732, 6

\bibitem[{{Kothes}(2010)}]{kothes10}
{Kothes} R., 2010, in Astronomical Society of the Pacific Conference Series,
  Vol. 438, Astronomical Society of the Pacific Conference Series, {Kothes} R.,
  {Landecker} T.~L., {Willis} A.~G., eds., p. 347

\bibitem[{{Kothes}, {Reich} \& {Uyan{\i}ker}(2006){Kothes}, {Reich}, \&
  {Uyan{\i}ker}}]{kothes06}
{Kothes} R., {Reich} W., {Uyan{\i}ker} B., 2006, ApJ, 638, 225

\bibitem[{{Kothes}, {Uyaniker} \& {Pineault}(2001){Kothes}, {Uyaniker}, \&
  {Pineault}}]{kothes01}
{Kothes} R., {Uyaniker} B., {Pineault} S., 2001, ApJ, 560, 236

\bibitem[{{Kothes}, {Uyaniker} \& {Yar}(2002){Kothes}, {Uyaniker}, \&
  {Yar}}]{kothes02}
{Kothes} R., {Uyaniker} B., {Yar} A., 2002, ApJ, 576, 169

\bibitem[{{Koyama} {et~al}\mbox{.}(1997){Koyama}, {Kinugasa}, {Matsuzaki},
  {Nishiuchi}, {Sugizaki}, {Torii}, {Yamauchi}, \& {Aschenbach}}]{koyama97}
{Koyama} K., {Kinugasa} K., {Matsuzaki} K., {Nishiuchi} M., {Sugizaki} M.,
  {Torii} K., {Yamauchi} S., {Aschenbach} B., 1997, PASJ, 49, L7

\bibitem[{{Kriss} {et~al}\mbox{.}(1985){Kriss}, {Becker}, {Helfand}, \&
  {Canizares}}]{kriss85}
{Kriss} G.~A., {Becker} R.~H., {Helfand} D.~J., {Canizares} C.~R., 1985, ApJ,
  288, 703

\bibitem[{{Kuiper} \& {Hermsen}(2009)}]{kuiper09}
{Kuiper} L., {Hermsen} W., 2009, ArXiv e-prints

\bibitem[{{Kuiper} {et~al}\mbox{.}(2001){Kuiper}, {Hermsen}, {Cusumano},
  {Diehl}, {Sch{\"o}nfelder}, {Strong}, {Bennett}, \& {McConnell}}]{kuiper01}
{Kuiper} L., {Hermsen} W., {Cusumano} G., {Diehl} R., {Sch{\"o}nfelder} V.,
  {Strong} A., {Bennett} K., {McConnell} M.~L., 2001, A\&A, 378, 918

\bibitem[{{Kuiper} {et~al}\mbox{.}(2006){Kuiper}, {Hermsen}, {den Hartog}, \&
  {Collmar}}]{kuiper06}
{Kuiper} L., {Hermsen} W., {den Hartog} P.~R., {Collmar} W., 2006, ApJ, 645,
  556

\bibitem[{{Kuiper} {et~al}\mbox{.}(1999){Kuiper}, {Hermsen}, {Krijger},
  {Bennett}, {Carrami{\~n}ana}, {Sch{\"o}nfelder}, {Bailes}, \&
  {Manchester}}]{kuiper99}
{Kuiper} L., {Hermsen} W., {Krijger} J.~M., {Bennett} K., {Carrami{\~n}ana} A.,
  {Sch{\"o}nfelder} V., {Bailes} M., {Manchester} R.~N., 1999, A\&A, 351, 119

\bibitem[{{Kulkarni} {et~al}\mbox{.}(2003){Kulkarni}, {Kaplan}, {Marshall},
  {Frail}, {Murakami}, \& {Yonetoku}}]{kulkarni03}
{Kulkarni} S.~R., {Kaplan} D.~L., {Marshall} H.~L., {Frail} D.~A., {Murakami}
  T., {Yonetoku} D., 2003, ApJ, 585, 948

\bibitem[{{Kumar} \& {Safi-Harb}(2008)}]{kumar08}
{Kumar} H.~S., {Safi-Harb} S., 2008, ApJL, 678, L43

\bibitem[{{Kumar}, {Safi-Harb} \& {Gonzalez}(2012){Kumar}, {Safi-Harb}, \&
  {Gonzalez}}]{kumar12}
{Kumar} H.~S., {Safi-Harb} S., {Gonzalez} M.~E., 2012, ApJ, 754, 96

\bibitem[{{Kumar} {et~al}\mbox{.}(2014){Kumar}, {Safi-Harb}, {Slane}, \&
  {Gotthelf}}]{kumar14}
{Kumar} H.~S., {Safi-Harb} S., {Slane} P.~O., {Gotthelf} E.~V., 2014, ApJ, 781,
  41

\bibitem[{{Laming}(2001)}]{laming01}
{Laming} J.~M., 2001, ApJ, 546, 1149

\bibitem[{{Landau}(1932)}]{landau32}
{Landau} L., 1932, Phy. Z. Sowejtunion, 1, 285

\bibitem[{{Landecker}, {Higgs} \& {Wendker}(1993){Landecker}, {Higgs}, \&
  {Wendker}}]{landecker93}
{Landecker} T.~L., {Higgs} L.~A., {Wendker} H.~J., 1993, A\&A, 276, 522

\bibitem[{{Landi} {et~al}\mbox{.}(2007){Landi}, {de Rosa}, {Dean}, {Bassani},
  {Ubertini}, \& {Bird}}]{landi07}
{Landi} R., {de Rosa} A., {Dean} A.~J., {Bassani} L., {Ubertini} P., {Bird}
  A.~J., 2007, MNRAS, 380, 926

\bibitem[{{Lang} {et~al}\mbox{.}(2010){Lang}, {Wang}, {Lu}, \&
  {Clubb}}]{lang10}
{Lang} C.~C., {Wang} Q.~D., {Lu} F., {Clubb} K.~I., 2010, ApJ, 709, 1125

\bibitem[{{Lazendic} {et~al}\mbox{.}(2000){Lazendic}, {Dickel}, {Haynes},
  {Jones}, \& {White}}]{lazendic00}
{Lazendic} J.~S., {Dickel} J.~R., {Haynes} R.~F., {Jones} P.~A., {White} G.~L.,
  2000, ApJ, 540, 808

\bibitem[{{Lazendic} {et~al}\mbox{.}(2003){Lazendic}, {Slane}, {Gaensler},
  {Plucinsky}, {Hughes}, {Galloway}, \& {Crawford}}]{lazendic03}
{Lazendic} J.~S., {Slane} P.~O., {Gaensler} B.~M., {Plucinsky} P.~P., {Hughes}
  J.~P., {Galloway} D.~K., {Crawford} F., 2003, ApJL, 593, L27

\bibitem[{{Lazendic} {et~al}\mbox{.}(2005){Lazendic}, {Slane}, {Hughes},
  {Chen}, \& {Dame}}]{lazendic05}
{Lazendic} J.~S., {Slane} P.~O., {Hughes} J.~P., {Chen} Y., {Dame} T.~M., 2005,
  ApJ, 618, 733

\bibitem[{{Leahy}, {Tian} \& {Wang}(2008){Leahy}, {Tian}, \& {Wang}}]{leahy08a}
{Leahy} D.~A., {Tian} W., {Wang} Q.~D., 2008, AJ, 136, 1477

\bibitem[{{Leahy} \& {Tian}(2008)}]{leahy08b}
{Leahy} D.~A., {Tian} W.~W., 2008, A\&A, 480, L25

\bibitem[{{Lemiere} {et~al}\mbox{.}(2009){Lemiere}, {Slane}, {Gaensler}, \&
  {Murray}}]{lemiere09}
{Lemiere} A., {Slane} P., {Gaensler} B.~M., {Murray} S., 2009, ApJ, 706, 1269

\bibitem[{{Lemoine-Goumard} {et~al}\mbox{.}(2011){Lemoine-Goumard}, {Zavlin},
  {Grondin}, {Shannon}, {Smith}, {Burgay}, {Camilo}, {Cohen-Tanugi}, {Freire},
  {Grove}, {Guillemot}, {Johnston}, {Keith}, {Kramer}, {Manchester},
  {Michelson}, {Parent}, {Possenti}, {Ray}, {Renaud}, {Thorsett}, {Weltevrede},
  \& {Wolff}}]{lemoinegoumard11}
{Lemoine-Goumard} M. {et~al.}, 2011, A\&A, 533, A102

\bibitem[{{Li}, {Chen} \& {Zhang}(2010){Li}, {Chen}, \& {Zhang}}]{li10}
{Li} H., {Chen} Y., {Zhang} L., 2010, MNRAS, 408, L80

\bibitem[{{Lin} {et~al}\mbox{.}(2010){Lin}, {Huang}, {Takata}, {Hwang}, {Kong},
  \& {Hui}}]{lin10}
{Lin} L.~C.~C., {Huang} R.~H.~H., {Takata} J., {Hwang} C.~Y., {Kong} A.~K.~H.,
  {Hui} C.~Y., 2010, ApJL, 725, L1

\bibitem[{{Livingstone} \& {Kaspi}(2011)}]{livingstone11b}
{Livingstone} M.~A., {Kaspi} V.~M., 2011, ApJ, 742, 31

\bibitem[{{Livingstone}, {Kaspi} \& {Gavriil}(2005){Livingstone}, {Kaspi}, \&
  {Gavriil}}]{livingstone05b}
{Livingstone} M.~A., {Kaspi} V.~M., {Gavriil} F.~P., 2005, ApJ, 633, 1095

\bibitem[{{Livingstone} {et~al}\mbox{.}(2005){Livingstone}, {Kaspi}, {Gavriil},
  \& {Manchester}}]{livingstone05a}
{Livingstone} M.~A., {Kaspi} V.~M., {Gavriil} F.~P., {Manchester} R.~N., 2005,
  ApJ, 619, 1046

\bibitem[{{Livingstone} {et~al}\mbox{.}(2006){Livingstone}, {Kaspi},
  {Gotthelf}, \& {Kuiper}}]{livingstone06}
{Livingstone} M.~A., {Kaspi} V.~M., {Gotthelf} E.~V., {Kuiper} L., 2006, ApJ,
  647, 1286

\bibitem[{{Livingstone} {et~al}\mbox{.}(2011){Livingstone}, {Ng}, {Kaspi},
  {Gavriil}, \& {Gotthelf}}]{livingstone11a}
{Livingstone} M.~A., {Ng} C.-Y., {Kaspi} V.~M., {Gavriil} F.~P., {Gotthelf}
  E.~V., 2011, ApJ, 730, 66

\bibitem[{{Longair}(1994)}]{longair94}
{Longair} M.~S., 1994, {High energy astrophysics. Volume 2. Stars, the Galaxy
  and the interstellar medium.} Cambridge University Press

\bibitem[{{Lopez} {et~al}\mbox{.}(2009){Lopez}, {Ramirez-Ruiz}, {Badenes},
  {Huppenkothen}, {Jeltema}, \& {Pooley}}]{lopez09}
{Lopez} L.~A., {Ramirez-Ruiz} E., {Badenes} C., {Huppenkothen} D., {Jeltema}
  T.~E., {Pooley} D.~A., 2009, ApJL, 706, L106

\bibitem[{{Lopez} {et~al}\mbox{.}(2011){Lopez}, {Ramirez-Ruiz}, {Huppenkothen},
  {Badenes}, \& {Pooley}}]{lopez11}
{Lopez} L.~A., {Ramirez-Ruiz} E., {Huppenkothen} D., {Badenes} C., {Pooley}
  D.~A., 2011, ApJ, 732, 114

\bibitem[{{Lu} \& {Aschenbach}(2000)}]{lu00}
{Lu} F.~J., {Aschenbach} B., 2000, A\&A, 362, 1083

\bibitem[{{Lu}, {Aschenbach} \& {Song}(2001){Lu}, {Aschenbach}, \&
  {Song}}]{lu01}
{Lu} F.~J., {Aschenbach} B., {Song} L.~M., 2001, A\&A, 370, 570

\bibitem[{{Lu} {et~al}\mbox{.}(2002){Lu}, {Wang}, {Aschenbach}, {Durouchoux},
  \& {Song}}]{lu02}
{Lu} F.~J., {Wang} Q.~D., {Aschenbach} B., {Durouchoux} P., {Song} L.~M., 2002,
  ApJL, 568, L49

\bibitem[{{Lyne}, {Pritchard} \& {Smith}(1988){Lyne}, {Pritchard}, \&
  {Smith}}]{lyne88}
{Lyne} A.~G., {Pritchard} R.~S., {Smith} F.~G., 1988, MNRAS, 233, 667

\bibitem[{{Mac{\'{\i}}as-P{\'e}rez}
  {et~al}\mbox{.}(2010){Mac{\'{\i}}as-P{\'e}rez}, {Mayet}, {Aumont}, \&
  {D{\'e}sert}}]{maciasperez10}
{Mac{\'{\i}}as-P{\'e}rez} J.~F., {Mayet} F., {Aumont} J., {D{\'e}sert} F.-X.,
  2010, ApJ, 711, 417

\bibitem[{{Maeda} {et~al}\mbox{.}(2009){Maeda}, {Uchiyama}, {Bamba}, {Kosugi},
  {Tsunemi}, {Helder}, {Vink}, {Kodaka}, {Terada}, {Fukazawa}, {Hiraga},
  {Hughes}, {Kokubun}, {Kouzu}, {Matsumoto}, {Miyata}, {Nakamura}, {Okada},
  {Someya}, {Tamagawa}, {Tamura}, {Totsuka}, {Tsuboi}, {Ezoe}, {Holt},
  {Ishida}, {Kamae}, {Petre}, \& {Takahashi}}]{maeda09}
{Maeda} Y. {et~al.}, 2009, PASJ, 61, 1217

\bibitem[{{Malizia} {et~al}\mbox{.}(2007){Malizia}, {Landi}, {Bassani}, {Bird},
  {Molina}, {De Rosa}, {Fiocchi}, {Gehrels}, {Kennea}, \& {Perri}}]{malizia07}
{Malizia} A. {et~al.}, 2007, ApJ, 668, 81

\bibitem[{{Manchester} {et~al}\mbox{.}(2005){Manchester}, {Hobbs}, {Teoh}, \&
  {Hobbs}}]{manchester05}
{Manchester} R.~N., {Hobbs} G.~B., {Teoh} A., {Hobbs} M., 2005, AJ, 129, 1993

\bibitem[{{Manchester}, {Staveley-Smith} \& {Kesteven}(1993){Manchester},
  {Staveley-Smith}, \& {Kesteven}}]{manchester93b}
{Manchester} R.~N., {Staveley-Smith} L., {Kesteven} M.~J., 1993, ApJ, 411, 756

\bibitem[{{Manchester}, {Tuohy} \& {Damico}(1982){Manchester}, {Tuohy}, \&
  {Damico}}]{manchester82}
{Manchester} R.~N., {Tuohy} I.~R., {Damico} N., 1982, ApJL, 262, L31

\bibitem[{{Marsden} {et~al}\mbox{.}(1984){Marsden}, {Gillett}, {Jennings},
  {Emerson}, {de Jong}, \& {Olnon}}]{marsden84}
{Marsden} P.~L., {Gillett} F.~C., {Jennings} R.~E., {Emerson} J.~P., {de Jong}
  T., {Olnon} F.~M., 1984, ApJL, 278, L29

\bibitem[{{Mart{\'i}n} {et~al}\mbox{.}(2014{\natexlab{a}}){Mart{\'i}n}, {Rea},
  {Torres}, \& {Papitto}}]{martin14a}
{Mart{\'i}n} J., {Rea} N., {Torres} D.~F., {Papitto} A., 2014{\natexlab{a}},
  ArXiv e-prints

\bibitem[{{Mart{\'i}n} {et~al}\mbox{.}(2014{\natexlab{b}}){Mart{\'i}n},
  {Torres}, {Cillis}, \& {de O\~na Wilhelmi}}]{martin14b}
{Mart{\'i}n} J., {Torres} D.~F., {Cillis} A., {de O\~na Wilhelmi} E.,
  2014{\natexlab{b}}, MNRAS, 443, 138

\bibitem[{{Mart{\'i}n}, {Torres} \& {Rea}(2012){Mart{\'i}n}, {Torres}, \&
  {Rea}}]{martin12}
{Mart{\'i}n} J., {Torres} D.~F., {Rea} N., 2012, MNRAS, 427, 415

\bibitem[{{Matheson} \& {Safi-Harb}(2005)}]{matheson05}
{Matheson} H., {Safi-Harb} S., 2005, Advances in Space Research, 35, 1099

\bibitem[{{Matheson} \& {Safi-Harb}(2010)}]{matheson10}
{Matheson} H., {Safi-Harb} S., 2010, ApJ, 724, 572

\bibitem[{{Matsui}, {Long} \& {Tuohy}(1988){Matsui}, {Long}, \&
  {Tuohy}}]{matsui88}
{Matsui} Y., {Long} K.~S., {Tuohy} I.~R., 1988, ApJ, 329, 838

\bibitem[{{Mattana} {et~al}\mbox{.}(2009){Mattana}, {Falanga}, {G{\"o}tz},
  {Terrier}, {Esposito}, {Pellizzoni}, {De Luca}, {Marandon}, {Goldwurm}, \&
  {Caraveo}}]{mattana09}
{Mattana} F. {et~al.}, 2009, ApJ, 694, 12

\bibitem[{{Mattox} {et~al}\mbox{.}(1996){Mattox}, {Koh}, {Lamb}, {Macomb},
  {Prince}, \& {Ray}}]{mattox96}
{Mattox} J.~R., {Koh} D.~T., {Lamb} R.~C., {Macomb} D.~J., {Prince} T.~A.,
  {Ray} P.~S., 1996, A\&AS, 120, C95

\bibitem[{{Mayer}(2010)}]{mayer10}
{Mayer} M., 2010, PhD thesis, Erlangen Center for Astroparticle physics

\bibitem[{{Mazets}, {Golenetskij} \& {Guryan}(1979){Mazets}, {Golenetskij}, \&
  {Guryan}}]{mazets79a}
{Mazets} E.~P., {Golenetskij} S.~V., {Guryan} Y.~A., 1979, Soviet Astronomy
  Letters, 5, 343

\bibitem[{{Mazets} {et~al}\mbox{.}(1979{\natexlab{a}}){Mazets}, {Golentskii},
  {Ilinskii}, {Aptekar}, \& {Guryan}}]{mazets79b}
{Mazets} E.~P., {Golentskii} S.~V., {Ilinskii} V.~N., {Aptekar} R.~L., {Guryan}
  I.~A., 1979{\natexlab{a}}, Nature, 282, 587

\bibitem[{{Mazets} {et~al}\mbox{.}(1979{\natexlab{b}}){Mazets}, {Golentskii},
  {Ilinskii}, {Aptekar}, \& {Guryan}}]{mazets79}
{Mazets} E.~P., {Golentskii} S.~V., {Ilinskii} V.~N., {Aptekar} R.~L., {Guryan}
  I.~A., 1979{\natexlab{b}}, Nature, 282, 587

\bibitem[{{McBride} {et~al}\mbox{.}(2008){McBride}, {Dean}, {Bazzano}, {Bird},
  {Hill}, {de Rosa}, {Landi}, {Sguera}, \& {Malizia}}]{mcbride08}
{McBride} V.~A. {et~al.}, 2008, A\&A, 477, 249

\bibitem[{{M{\'e}hault}(2011)}]{mehault11}
{M{\'e}hault} J., 2011, International Cosmic Ray Conference, 7, 181

\bibitem[{{Melatos}(1998)}]{melatos98}
{Melatos} A., 1998, MmSAI, 69, 1009

\bibitem[{{Mendoza} {et~al}\mbox{.}(2004){Mendoza}, {Kallman}, {Bautista}, \&
  {Palmeri}}]{mendoza04}
{Mendoza} C., {Kallman} T.~R., {Bautista} M.~A., {Palmeri} P., 2004, A\&A, 414,
  377

\bibitem[{{Mereghetti}(2008)}]{mereghetti08}
{Mereghetti} S., 2008, A\&ARv, 15, 225

\bibitem[{{Mereghetti}, {Sidoli} \& {Israel}(1998){Mereghetti}, {Sidoli}, \&
  {Israel}}]{mereghetti98}
{Mereghetti} S., {Sidoli} L., {Israel} G.~L., 1998, A\&A, 331, L77

\bibitem[{{Mereghetti}, {Tiengo} \& {Israel}(2002){Mereghetti}, {Tiengo}, \&
  {Israel}}]{mereghetti02}
{Mereghetti} S., {Tiengo} A., {Israel} G.~L., 2002, ApJ, 569, 275

\bibitem[{{Metzger} {et~al}\mbox{.}(2011){Metzger}, {Giannios}, {Thompson},
  {Bucciantini}, \& {Quataert}}]{metzger11}
{Metzger} B.~D., {Giannios} D., {Thompson} T.~A., {Bucciantini} N., {Quataert}
  E., 2011, MNRAS, 413, 2031

\bibitem[{{Mewe}(1999)}]{mewe99}
{Mewe} R., 1999, in Lecture Notes in Physics, Berlin Springer Verlag, Vol. 520,
  X-Ray Spectroscopy in Astrophysics, {van Paradijs} J., {Bleeker} J.~A.~M.,
  eds., p. 109

\bibitem[{{Miceli} {et~al}\mbox{.}(2009){Miceli}, {Bocchino}, {Iakubovskyi},
  {Orlando}, {Telezhinsky}, {Kirsch}, {Petruk}, {Dubner}, \&
  {Castelletti}}]{miceli09}
{Miceli} M. {et~al.}, 2009, A\&A, 501, 239

\bibitem[{{Micelotta}, {Brandl} \& {Israel}(2009){Micelotta}, {Brandl}, \&
  {Israel}}]{micelotta09}
{Micelotta} E.~R., {Brandl} B.~R., {Israel} F.~P., 2009, A\&A, 500, 807

\bibitem[{{Mignani}(2012)}]{mignani12b}
{Mignani} R.~P., 2012, in Astronomical Society of the Pacific Conference
  Series, Vol. 466, Electromagnetic Radiation from Pulsars and Magnetars,
  {Lewandowski} W., {Maron} O., {Kijak} J., eds., p.~3

\bibitem[{{Mignani} {et~al}\mbox{.}(2012){Mignani}, {De Luca}, {Hummel},
  {Zajczyk}, {Rudak}, {Kanbach}, \& {S{\l}owikowska}}]{mignani12}
{Mignani} R.~P., {De Luca} A., {Hummel} W., {Zajczyk} A., {Rudak} B., {Kanbach}
  G., {S{\l}owikowska} A., 2012, A\&A, 544, A100

\bibitem[{{Mignani} {et~al}\mbox{.}(2013){Mignani}, {de Luca}, {Rea},
  {Shearer}, {Collins}, {Torres}, {Hadasch}, \& {Caliandro}}]{mignani13}
{Mignani} R.~P., {de Luca} A., {Rea} N., {Shearer} A., {Collins} S., {Torres}
  D.~F., {Hadasch} D., {Caliandro} A., 2013, MNRAS, 430, 1354

\bibitem[{{Milne}(1979)}]{milne79}
{Milne} D.~K., 1979, Australian Journal of Physics, 32, 83

\bibitem[{{Mineo} {et~al}\mbox{.}(2001){Mineo}, {Cusumano}, {Maccarone},
  {Massaglia}, {Massaro}, \& {Trussoni}}]{mineo01}
{Mineo} T., {Cusumano} G., {Maccarone} M.~C., {Massaglia} S., {Massaro} E.,
  {Trussoni} E., 2001, A\&A, 380, 695

\bibitem[{{Misanovic}, {Kargaltsev} \& {Pavlov}(2011){Misanovic}, {Kargaltsev},
  \& {Pavlov}}]{misanovic11}
{Misanovic} Z., {Kargaltsev} O., {Pavlov} G.~G., 2011, ApJ, 735, 33

\bibitem[{{Mori} {et~al}\mbox{.}(2013){Mori}, {Gotthelf}, {Zhang}, {An},
  {Baganoff}, {Barri{\`e}re}, {Beloborodov}, {Boggs}, {Christensen}, {Craig},
  {Dufour}, {Grefenstette}, {Hailey}, {Harrison}, {Hong}, {Kaspi}, {Kennea},
  {Madsen}, {Markwardt}, {Nynka}, {Stern}, {Tomsick}, \& {Zhang}}]{mori13}
{Mori} K. {et~al.}, 2013, ApJL, 770, L23

\bibitem[{{Morsi} \& {Reich}(1987)}]{morsi87}
{Morsi} H.~W., {Reich} W., 1987, A\&As, 69, 533

\bibitem[{{Morton}(1964)}]{morton64}
{Morton} D.~C., 1964, Nature, 201, 1308

\bibitem[{{Morton} {et~al}\mbox{.}(2007){Morton}, {Slane}, {Borkowski},
  {Reynolds}, {Helfand}, {Gaensler}, \& {Hughes}}]{morton07}
{Morton} T.~D., {Slane} P., {Borkowski} K.~J., {Reynolds} S.~P., {Helfand}
  D.~J., {Gaensler} B.~M., {Hughes} J.~P., 2007, ApJ, 667, 219

\bibitem[{{Muno} {et~al}\mbox{.}(2006){Muno}, {Clark}, {Crowther}, {Dougherty},
  {de Grijs}, {Law}, {McMillan}, {Morris}, {Negueruela}, {Pooley}, {Portegies
  Zwart}, \& {Yusef-Zadeh}}]{muno06}
{Muno} M.~P. {et~al.}, 2006, ApJL, 636, L41

\bibitem[{{Murphy} {et~al}\mbox{.}(2007){Murphy}, {Mauch}, {Green}, {Hunstead},
  {Piestrzynska}, {Kels}, \& {Sztajer}}]{murphy07}
{Murphy} T., {Mauch} T., {Green} A., {Hunstead} R.~W., {Piestrzynska} B.,
  {Kels} A.~P., {Sztajer} P., 2007, MNRAS, 382, 382

\bibitem[{{Murray} {et~al}\mbox{.}(2002){Murray}, {Slane}, {Seward}, {Ransom},
  \& {Gaensler}}]{murray02}
{Murray} S.~S., {Slane} P.~O., {Seward} F.~D., {Ransom} S.~M., {Gaensler}
  B.~M., 2002, ApJ, 568, 226

\bibitem[{{Nagataki} {et~al}\mbox{.}(1998){Nagataki}, {Hashimoto}, {Sato},
  {Yamada}, \& {Mochizuki}}]{nagataki98}
{Nagataki} S., {Hashimoto} M.-A., {Sato} K., {Yamada} S., {Mochizuki} Y.~S.,
  1998, ApJL, 492, L45

\bibitem[{{Nakamori} {et~al}\mbox{.}(2008){Nakamori}, {Kubo}, {Yoshida},
  {Tanimori}, {Enomoto}, {Bicknell}, {Clay}, {Edwards}, {Gunji}, {Hara},
  {Hara}, {Hattori}, {Hayashi}, {Higashi}, {Hirai}, {Inoue}, {Kabuki},
  {Kajino}, {Katagiri}, {Kawachi}, {Kifune}, {Kiuchi}, {Kushida}, {Matsubara},
  {Mizukami}, {Mizumoto}, {Mizuniwa}, {Mori}, {Muraishi}, {Muraki}, {Naito},
  {Nakano}, {Nishida}, {Nishijima}, {Ohishi}, {Sakamoto}, {Seki}, {Stamatescu},
  {Suzuki}, {Swaby}, {Thornton}, {Tokanai}, {Tsuchiya}, {Watanabe}, {Yamada},
  {Yamazaki}, {Yanagita}, {Yoshikoshi}, \& {Yukawa}}]{nakamori08}
{Nakamori} T. {et~al.}, 2008, ApJ, 677, 297

\bibitem[{{Ney} \& {Stein}(1968)}]{ney68}
{Ney} E.~P., {Stein} W.~A., 1968, ApJL, 152, L21

\bibitem[{{Ng} {et~al}\mbox{.}(2008){Ng}, {Slane}, {Gaensler}, \&
  {Hughes}}]{ng08}
{Ng} C.-Y., {Slane} P.~O., {Gaensler} B.~M., {Hughes} J.~P., 2008, ApJ, 686,
  508

\bibitem[{{Olausen} \& {Kaspi}(2014)}]{olausen14}
{Olausen} S.~A., {Kaspi} V.~M., 2014, ApJS, 212, 6

\bibitem[{{Oppenheimer} \& {Volkoff}(1939)}]{oppenheimer39}
{Oppenheimer} J.~R., {Volkoff} G.~M., 1939, Physical Review, 55, 374

\bibitem[{{Oskinova} {et~al}\mbox{.}(2011){Oskinova}, {Todt}, {Ignace},
  {Brown}, {Cassinelli}, \& {Hamann}}]{oskinova11}
{Oskinova} L.~M., {Todt} H., {Ignace} R., {Brown} J.~C., {Cassinelli} J.~P.,
  {Hamann} W.-R., 2011, MNRAS, 416, 1456

\bibitem[{{Pacini} \& {Salvati}(1973)}]{pacini73}
{Pacini} F., {Salvati} M., 1973, ApJ, 186, 249

\bibitem[{{Palmeri} {et~al}\mbox{.}(2003){Palmeri}, {Mendoza}, {Kallman},
  {Bautista}, \& {Mel{\'e}ndez}}]{palmeri03}
{Palmeri} P., {Mendoza} C., {Kallman} T.~R., {Bautista} M.~A., {Mel{\'e}ndez}
  M., 2003, A\&A, 410, 359

\bibitem[{{Parent} {et~al}\mbox{.}(2011){Parent}, {Kerr}, {den Hartog},
  {Baring}, {DeCesar}, {Espinoza}, {Gotthelf}, {Harding}, {Johnston}, {Kaspi},
  {Livingstone}, {Romani}, {Stappers}, {Watters}, {Weltevrede}, {Abdo},
  {Burgay}, {Camilo}, {Craig}, {Freire}, {Giordano}, {Guillemot}, {Hobbs},
  {Keith}, {Kramer}, {Lyne}, {Manchester}, {Noutsos}, {Possenti}, \&
  {Smith}}]{parent11}
{Parent} D. {et~al.}, 2011, ApJ, 743, 170

\bibitem[{{Park} {et~al}\mbox{.}(2007){Park}, {Hughes}, {Slane}, {Burrows},
  {Gaensler}, \& {Ghavamian}}]{park07}
{Park} S., {Hughes} J.~P., {Slane} P.~O., {Burrows} D.~N., {Gaensler} B.~M.,
  {Ghavamian} P., 2007, ApJL, 670, L121

\bibitem[{{Park} {et~al}\mbox{.}(2012){Park}, {Hughes}, {Slane}, {Burrows},
  {Lee}, \& {Mori}}]{park12}
{Park} S., {Hughes} J.~P., {Slane} P.~O., {Burrows} D.~N., {Lee} J.-J., {Mori}
  K., 2012, ApJ, 748, 117

\bibitem[{{Park} {et~al}\mbox{.}(2009){Park}, {Kargaltsev}, {Pavlov}, {Mori},
  {Slane}, {Hughes}, {Burrows}, \& {Garmire}}]{park09}
{Park} S., {Kargaltsev} O., {Pavlov} G.~G., {Mori} K., {Slane} P.~O., {Hughes}
  J.~P., {Burrows} D.~N., {Garmire} G.~P., 2009, ApJ, 695, 431

\bibitem[{{Parmar} {et~al}\mbox{.}(1998){Parmar}, {Oosterbroek}, {Favata},
  {Pightling}, {Coe}, {Mereghetti}, \& {Israel}}]{parmar98}
{Parmar} A.~N., {Oosterbroek} T., {Favata} F., {Pightling} S., {Coe} M.~J.,
  {Mereghetti} S., {Israel} G.~L., 1998, A\&A, 330, 175

\bibitem[{{Patnaude} \& {Fesen}(2009)}]{patnaude09}
{Patnaude} D.~J., {Fesen} R.~A., 2009, ApJ, 697, 535

\bibitem[{{Pavlov} {et~al}\mbox{.}(2001){Pavlov}, {Sanwal}, {K{\i}z{\i}ltan},
  \& {Garmire}}]{pavlov01}
{Pavlov} G.~G., {Sanwal} D., {K{\i}z{\i}ltan} B., {Garmire} G.~P., 2001, ApJL,
  559, L131

\bibitem[{{Pavlov} {et~al}\mbox{.}(2002){Pavlov}, {Zavlin}, {Sanwal}, \&
  {Tr{\"u}mper}}]{pavlov02}
{Pavlov} G.~G., {Zavlin} V.~E., {Sanwal} D., {Tr{\"u}mper} J., 2002, ApJL, 569,
  L95

\bibitem[{{Pedaletti}, {de O{\~n}a Wilhelmi} \& {Torres}(2014){Pedaletti}, {de
  O{\~n}a Wilhelmi}, \& {Torres}}]{pedaletti14}
{Pedaletti} G., {de O{\~n}a Wilhelmi} E., {Torres} D.~F., 2014, ArXiv e-prints

\bibitem[{{Pfeffermann} \& {Aschenbach}(1996)}]{pfeffermann96}
{Pfeffermann} E., {Aschenbach} B., 1996, in Roentgenstrahlung from the
  Universe, {Zimmermann} H.~U., {Tr{\"u}mper} J., {Yorke} H., eds., pp.
  267--268

\bibitem[{{Pineault} \& {Joncas}(2000)}]{pineault00}
{Pineault} S., {Joncas} G., 2000, AJ, 120, 3218

\bibitem[{{Pineault} {et~al}\mbox{.}(1993){Pineault}, {Landecker}, {Madore}, \&
  {Gaumont-Guay}}]{pineault93}
{Pineault} S., {Landecker} T.~L., {Madore} B., {Gaumont-Guay} S., 1993, AJ,
  105, 1060

\bibitem[{{Pineault} {et~al}\mbox{.}(1997){Pineault}, {Landecker}, {Swerdlyk},
  \& {Reich}}]{pineault97}
{Pineault} S., {Landecker} T.~L., {Swerdlyk} C.~M., {Reich} W., 1997, A\&A,
  324, 1152

\bibitem[{{Pletsch} {et~al}\mbox{.}(2012){Pletsch}, {Guillemot}, {Allen},
  {Kramer}, {Aulbert}, {Fehrmann}, {Baring}, {Camilo}, {Caraveo}, {Grove},
  {Kerr}, {Marelli}, {Ransom}, {Ray}, \& {Saz Parkinson}}]{pletsch12}
{Pletsch} H.~J. {et~al.}, 2012, ApJL, 755, L20

\bibitem[{{Porquet}, {Decourchelle} \& {Warwick}(2003){Porquet},
  {Decourchelle}, \& {Warwick}}]{porquet03}
{Porquet} D., {Decourchelle} A., {Warwick} R.~S., 2003, A\&A, 401, 197

\bibitem[{{Porter}, {Moskalenko} \& {Strong}(2006){Porter}, {Moskalenko}, \&
  {Strong}}]{porter06}
{Porter} T.~A., {Moskalenko} I.~V., {Strong} A.~W., 2006, ApJL, 648, L29

\bibitem[{{Prantzos}(2011)}]{prantzos11}
{Prantzos} N., 2011, ArXiv e-prints

\bibitem[{{Qiao}, {Zhang} \& {Fang}(2009){Qiao}, {Zhang}, \& {Fang}}]{qiao09}
{Qiao} W.-F., {Zhang} L., {Fang} J., 2009, Research in Astronomy and
  Astrophysics, 9, 449

\bibitem[{{Ray} {et~al}\mbox{.}(2011){Ray}, {Kerr}, {Parent}, {Abdo},
  {Guillemot}, {Ransom}, {Rea}, {Wolff}, {Makeev}, {Roberts}, {Camilo},
  {Dormody}, {Freire}, {Grove}, {Gwon}, {Harding}, {Johnston}, {Keith},
  {Kramer}, {Michelson}, {Romani}, {Saz Parkinson}, {Thompson}, {Weltevrede},
  {Wood}, \& {Ziegler}}]{ray11}
{Ray} P.~S. {et~al.}, 2011, ApJS, 194, 17

\bibitem[{{Rea} \& {Esposito}(2011)}]{rea11}
{Rea} N., {Esposito} P., 2011, in High-Energy Emission from Pulsars and their
  Systems, {Torres} D.~F., {Rea} N., eds., p. 247

\bibitem[{{Rea} {et~al}\mbox{.}(2013){Rea}, {Esposito}, {Pons}, {Turolla},
  {Torres}, {Israel}, {Possenti}, {Burgay}, {Vigan{\`o}}, {Papitto}, {Perna},
  {Stella}, {Ponti}, {Baganoff}, {Haggard}, {Camero-Arranz}, {Zane}, {Minter},
  {Mereghetti}, {Tiengo}, {Sch{\"o}del}, {Feroci}, {Mignani}, \&
  {G{\"o}tz}}]{rea13}
{Rea} N. {et~al.}, 2013, ApJL, 775, L34

\bibitem[{{Rea} {et~al}\mbox{.}(2010){Rea}, {Esposito}, {Turolla}, {Israel},
  {Zane}, {Stella}, {Mereghetti}, {Tiengo}, {G{\"o}tz}, {G{\"o}{\u g}{\"u}{\c
  s}}, \& {Kouveliotou}}]{rea10}
{Rea} N. {et~al.}, 2010, Science, 330, 944

\bibitem[{{Rea} {et~al}\mbox{.}(2012){Rea}, {Israel}, {Esposito}, {Pons},
  {Camero-Arranz}, {Mignani}, {Turolla}, {Zane}, {Burgay}, {Possenti},
  {Campana}, {Enoto}, {Gehrels}, {G{\"o}{\v g}{\"u}{\c s}}, {G{\"o}tz},
  {Kouveliotou}, {Makishima}, {Mereghetti}, {Oates}, {Palmer}, {Perna},
  {Stella}, \& {Tiengo}}]{rea12}
{Rea} N. {et~al.}, 2012, ApJ, 754, 27

\bibitem[{{Rea} {et~al}\mbox{.}(2014){Rea}, {Vigan{\`o}}, {Israel}, {Pons}, \&
  {Torres}}]{rea14}
{Rea} N., {Vigan{\`o}} D., {Israel} G.~L., {Pons} J.~A., {Torres} D.~F., 2014,
  ApJL, 781, L17

\bibitem[{{Reed} {et~al}\mbox{.}(1995){Reed}, {Hester}, {Fabian}, \&
  {Winkler}}]{reed95}
{Reed} J.~E., {Hester} J.~J., {Fabian} A.~C., {Winkler} P.~F., 1995, ApJ, 440,
  706

\bibitem[{{Rees} \& {Gunn}(1974)}]{rees74}
{Rees} M.~J., {Gunn} J.~E., 1974, MNRAS, 167, 1

\bibitem[{{Reich} {et~al}\mbox{.}(1985){Reich}, {Fuerst}, {Altenhoff}, {Reich},
  \& {Junkes}}]{reich85}
{Reich} W., {Fuerst} E., {Altenhoff} W.~J., {Reich} P., {Junkes} N., 1985,
  A\&A, 151, L10

\bibitem[{{Reich} {et~al}\mbox{.}(1984){Reich}, {Fuerst}, {Haslam}, {Steffen},
  \& {Reif}}]{reich84}
{Reich} W., {Fuerst} E., {Haslam} C.~G.~T., {Steffen} P., {Reif} K., 1984,
  A\&AS, 58, 197

\bibitem[{{Reimer} {et~al}\mbox{.}(2008){Reimer}, {Hinton}, {Hofmann}, \& {et
  al.}}]{reimer08}
{Reimer} O., {Hinton} J., {Hofmann} W., {et al.}, 2008, International Cosmic
  Ray Conference, 2, 567

\bibitem[{{Renaud}(2009)}]{renaud09}
{Renaud} M., 2009, ArXiv e-prints

\bibitem[{{Renaud} {et~al}\mbox{.}(2010){Renaud}, {Marandon}, {Gotthelf},
  {Rodriguez}, {Terrier}, {Mattana}, {Lebrun}, {Tomsick}, \&
  {Manchester}}]{renaud10}
{Renaud} M. {et~al.}, 2010, ApJ, 716, 663

\bibitem[{{Reynolds}(1985)}]{reynolds85}
{Reynolds} S.~P., 1985, ApJ, 291, 152

\bibitem[{{Reynolds}(1988)}]{reynolds88a}
{Reynolds} S.~P., 1988, ApJ, 327, 853

\bibitem[{{Reynolds}(1998)}]{reynolds98}
{Reynolds} S.~P., 1998, ApJ, 493, 375

\bibitem[{{Reynolds}(2003)}]{reynolds03}
{Reynolds} S.~P., 2003, ArXiv Astrophysics e-prints

\bibitem[{{Reynolds} \& {Aller}(1988)}]{reynolds88b}
{Reynolds} S.~P., {Aller} H.~D., 1988, ApJ, 327, 845

\bibitem[{{Reynolds} {et~al}\mbox{.}(2007){Reynolds}, {Borkowski}, {Hwang},
  {Hughes}, {Badenes}, {Laming}, \& {Blondin}}]{reynolds07}
{Reynolds} S.~P., {Borkowski} K.~J., {Hwang} U., {Hughes} J.~P., {Badenes} C.,
  {Laming} J.~M., {Blondin} J.~M., 2007, ApJL, 668, L135

\bibitem[{{Reynolds} \& {Chevalier}(1984)}]{reynolds84}
{Reynolds} S.~P., {Chevalier} R.~A., 1984, ApJ, 278, 630

\bibitem[{{Reynoso} {et~al}\mbox{.}(1995){Reynoso}, {Dubner}, {Goss}, \&
  {Arnal}}]{reynoso95}
{Reynoso} E.~M., {Dubner} G.~M., {Goss} W.~M., {Arnal} E.~M., 1995, AJ, 110,
  318

\bibitem[{{Reynoso} {et~al}\mbox{.}(2004){Reynoso}, {Green}, {Johnston},
  {Goss}, {Dubner}, \& {Giacani}}]{reynoso04}
{Reynoso} E.~M., {Green} A.~J., {Johnston} S., {Goss} W.~M., {Dubner} G.~M.,
  {Giacani} E.~B., 2004, PASA, 21, 82

\bibitem[{{Rho} \& {Petre}(1997)}]{rho97}
{Rho} J., {Petre} R., 1997, ApJ, 484, 828

\bibitem[{{Rho} {et~al}\mbox{.}(1994){Rho}, {Petre}, {Schlegel}, \&
  {Hester}}]{rho94}
{Rho} J., {Petre} R., {Schlegel} E.~M., {Hester} J.~J., 1994, ApJ, 430, 757

\bibitem[{{Rho}, {Petre} \& {Ballet}(1998){Rho}, {Petre}, \& {Ballet}}]{rho98}
{Rho} J.-H., {Petre} R., {Ballet} J., 1998, Advances in Space Research, 22,
  1039

\bibitem[{{Rieger}, {de O{\~n}a-Wilhelmi} \& {Aharonian}(2013){Rieger}, {de
  O{\~n}a-Wilhelmi}, \& {Aharonian}}]{rieger13}
{Rieger} F.~M., {de O{\~n}a-Wilhelmi} E., {Aharonian} F.~A., 2013, Frontiers of
  Physics, 8, 714

\bibitem[{{Ritchie} {et~al}\mbox{.}(2010){Ritchie}, {Clark}, {Negueruela}, \&
  {Langer}}]{ritchie10}
{Ritchie} B.~W., {Clark} J.~S., {Negueruela} I., {Langer} N., 2010, A\&A, 520,
  A48

\bibitem[{{Roberts} {et~al}\mbox{.}(1993){Roberts}, {Goss}, {Kalberla},
  {Herbstmeier}, \& {Schwarz}}]{roberts93}
{Roberts} D.~A., {Goss} W.~M., {Kalberla} P.~M.~W., {Herbstmeier} U., {Schwarz}
  U.~J., 1993, A\&A, 274, 427

\bibitem[{{Roger} {et~al}\mbox{.}(1988){Roger}, {Milne}, {Kesteven},
  {Wellington}, \& {Haynes}}]{roger88}
{Roger} R.~S., {Milne} D.~K., {Kesteven} M.~J., {Wellington} K.~J., {Haynes}
  R.~F., 1988, ApJ, 332, 940

\bibitem[{{Roy}, {Gupta} \& {Lewandowski}(2012){Roy}, {Gupta}, \&
  {Lewandowski}}]{roy12}
{Roy} J., {Gupta} Y., {Lewandowski} W., 2012, MNRAS, 424, 2213

\bibitem[{{Rudie} \& {Fesen}(2007)}]{rudie07}
{Rudie} G.~C., {Fesen} R.~A., 2007, in Revista Mexicana de Astronomia y
  Astrofisica Conference Series, Vol.~30, Revista Mexicana de Astronomia y
  Astrofisica Conference Series, pp. 90--95

\bibitem[{{Russell} \& {Dopita}(1992)}]{russell92}
{Russell} S.~C., {Dopita} M.~A., 1992, ApJ, 384, 508

\bibitem[{{Rybicki} \& {Lightman}(1979)}]{rybicki79}
{Rybicki} G.~B., {Lightman} A.~P., 1979, {Radiative processes in astrophysics}.
  Wiley

\bibitem[{{Safi-Harb} {et~al}\mbox{.}(2001){Safi-Harb}, {Harrus}, {Petre},
  {Pavlov}, {Koptsevich}, \& {Sanwal}}]{safiharb01}
{Safi-Harb} S., {Harrus} I.~M., {Petre} R., {Pavlov} G.~G., {Koptsevich} A.~B.,
  {Sanwal} D., 2001, ApJ, 561, 308

\bibitem[{{Safi-Harb} \& {Kumar}(2008)}]{safiharb08}
{Safi-Harb} S., {Kumar} H.~S., 2008, ApJ, 684, 532

\bibitem[{{Safi-Harb} \& {Kumar}(2013)}]{safiharb13}
{Safi-Harb} S., {Kumar} H.~S., 2013, in IAU Symposium, Vol. 291, IAU Symposium,
  {van Leeuwen} J., ed., pp. 480--482

\bibitem[{{Safi-Harb}, {Ogelman} \& {Finley}(1994){Safi-Harb}, {Ogelman}, \&
  {Finley}}]{safiharb94}
{Safi-Harb} S., {Ogelman} H., {Finley} J.~P., 1994, in Bulletin of the American
  Astronomical Society, Vol.~26, American Astronomical Society Meeting
  Abstracts \#184, p. 950

\bibitem[{{Safi-Harb}, {Ogelman} \& {Finley}(1995){Safi-Harb}, {Ogelman}, \&
  {Finley}}]{safiharb95}
{Safi-Harb} S., {Ogelman} H., {Finley} J.~P., 1995, ApJ, 439, 722

\bibitem[{{Salter} {et~al}\mbox{.}(1989){Salter}, {Reynolds}, {Hogg}, {Payne},
  \& {Rhodes}}]{salter89}
{Salter} C.~J., {Reynolds} S.~P., {Hogg} D.~E., {Payne} J.~M., {Rhodes} P.~J.,
  1989, ApJ, 338, 171

\bibitem[{{Sanbonmatsu} \& {Helfand}(1992)}]{sanbonmatsu92}
{Sanbonmatsu} K.~Y., {Helfand} D.~J., 1992, AJ, 104, 2189

\bibitem[{{Sasaki} {et~al}\mbox{.}(2013){Sasaki}, {Plucinsky}, {Gaetz}, \&
  {Bocchino}}]{sasaki13}
{Sasaki} M., {Plucinsky} P.~P., {Gaetz} T.~J., {Bocchino} F., 2013, A\&A, 552,
  A45

\bibitem[{{Sasaki} {et~al}\mbox{.}(2004){Sasaki}, {Plucinsky}, {Gaetz},
  {Smith}, {Edgar}, \& {Slane}}]{sasaki04}
{Sasaki} M., {Plucinsky} P.~P., {Gaetz} T.~J., {Smith} R.~K., {Edgar} R.~J.,
  {Slane} P.~O., 2004, ApJ, 617, 322

\bibitem[{{Schlickeiser}(2002)}]{schlickeiser02}
{Schlickeiser} R., 2002, {Cosmic Ray Astrophysics}. Springer

\bibitem[{{Schure} {et~al}\mbox{.}(2009){Schure}, {Kosenko}, {Kaastra},
  {Keppens}, \& {Vink}}]{schure09}
{Schure} K.~M., {Kosenko} D., {Kaastra} J.~S., {Keppens} R., {Vink} J., 2009,
  A\&A, 508, 751

\bibitem[{{Sedov}(1959)}]{sedov59}
{Sedov} L.~I., 1959, {Similarity and Dimensional Methods in Mechanics}. CRC
  Press

\bibitem[{{Seward} \& {Harnden}(1982)}]{seward82}
{Seward} F.~D., {Harnden}, Jr. F.~R., 1982, ApJL, 256, L45

\bibitem[{{Seward}, {Harnden} \& {Helfand}(1984){Seward}, {Harnden}, \&
  {Helfand}}]{seward84b}
{Seward} F.~D., {Harnden}, Jr. F.~R., {Helfand} D.~J., 1984, ApJL, 287, L19

\bibitem[{{Seward} {et~al}\mbox{.}(1984){Seward}, {Harnden}, {Szymkowiak}, \&
  {Swank}}]{seward84a}
{Seward} F.~D., {Harnden}, Jr. F.~R., {Szymkowiak} A., {Swank} J., 1984, ApJ,
  281, 650

\bibitem[{{Seward}, {Schmidt} \& {Slane}(1995){Seward}, {Schmidt}, \&
  {Slane}}]{seward95}
{Seward} F.~D., {Schmidt} B., {Slane} P., 1995, ApJ, 453, 284

\bibitem[{{Seward} {et~al}\mbox{.}(2003){Seward}, {Slane}, {Smith}, \&
  {Sun}}]{seward03}
{Seward} F.~D., {Slane} P.~O., {Smith} R.~K., {Sun} M., 2003, ApJ, 584, 414

\bibitem[{{Seward} \& {Wang}(1988)}]{seward88}
{Seward} F.~D., {Wang} Z.-R., 1988, ApJ, 332, 199

\bibitem[{{Sezer} {et~al}\mbox{.}(2010){Sezer}, {G{\"u}k}, {Hudaverdi},
  {Aktekin}, \& {Ercan}}]{sezer10}
{Sezer} A., {G{\"u}k} F., {Hudaverdi} M., {Aktekin} E., {Ercan} E.~N., 2010, in
  Astronomical Society of the Pacific Conference Series, Vol. 424, 9th
  International Conference of the Hellenic Astronomical Society, {Tsinganos}
  K., {Hatzidimitriou} D., {Matsakos} T., eds., p. 171

\bibitem[{{Sguera} {et~al}\mbox{.}(2009){Sguera}, {Romero}, {Bazzano},
  {Masetti}, {Bird}, \& {Bassani}}]{sguera09}
{Sguera} V., {Romero} G.~E., {Bazzano} A., {Masetti} N., {Bird} A.~J.,
  {Bassani} L., 2009, ApJ, 697, 1194

\bibitem[{{Shapiro} \& {Teukolsky}(2004)}]{shapiro04}
{Shapiro} S.~L., {Teukolsky} S.~A., 2004, Black Holes, White Dwarfs and Neutron
  Stars, Verlag W.-V., ed. Cornell University

\bibitem[{{Sidoli} {et~al}\mbox{.}(2000){Sidoli}, {Mereghetti}, {Israel}, \&
  {Bocchino}}]{sidoli00}
{Sidoli} L., {Mereghetti} S., {Israel} G.~L., {Bocchino} F., 2000, A\&A, 361,
  719

\bibitem[{{Slane} {et~al}\mbox{.}(2010){Slane}, {Castro}, {Funk}, {Uchiyama},
  {Lemiere}, {Gelfand}, \& {Lemoine-Goumard}}]{slane10}
{Slane} P., {Castro} D., {Funk} S., {Uchiyama} Y., {Lemiere} A., {Gelfand}
  J.~D., {Lemoine-Goumard} M., 2010, ApJ, 720, 266

\bibitem[{{Slane} {et~al}\mbox{.}(2000){Slane}, {Chen}, {Schulz}, {Seward},
  {Hughes}, \& {Gaensler}}]{slane00}
{Slane} P., {Chen} Y., {Schulz} N.~S., {Seward} F.~D., {Hughes} J.~P.,
  {Gaensler} B.~M., 2000, ApJL, 533, L29

\bibitem[{{Slane} {et~al}\mbox{.}(2008){Slane}, {Helfand}, {Reynolds},
  {Gaensler}, {Lemiere}, \& {Wang}}]{slane08}
{Slane} P., {Helfand} D.~J., {Reynolds} S.~P., {Gaensler} B.~M., {Lemiere} A.,
  {Wang} Z., 2008, ApJL, 676, L33

\bibitem[{{Slane} {et~al}\mbox{.}(2004{\natexlab{a}}){Slane}, {Helfand}, {van
  der Swaluw}, \& {Murray}}]{slane04b}
{Slane} P., {Helfand} D.~J., {van der Swaluw} E., {Murray} S.~S.,
  2004{\natexlab{a}}, ApJ, 616, 403

\bibitem[{{Slane} {et~al}\mbox{.}(1997){Slane}, {Seward}, {Bandiera}, {Torii},
  \& {Tsunemi}}]{slane97}
{Slane} P., {Seward} F.~D., {Bandiera} R., {Torii} K., {Tsunemi} H., 1997, ApJ,
  485, 221

\bibitem[{{Slane} {et~al}\mbox{.}(2002){Slane}, {Smith}, {Hughes}, \&
  {Petre}}]{slane02a}
{Slane} P., {Smith} R.~K., {Hughes} J.~P., {Petre} R., 2002, ApJ, 564, 284

\bibitem[{{Slane} {et~al}\mbox{.}(2004{\natexlab{b}}){Slane}, {Zimmerman},
  {Hughes}, {Seward}, {Gaensler}, \& {Clarke}}]{slane04a}
{Slane} P., {Zimmerman} E.~R., {Hughes} J.~P., {Seward} F.~D., {Gaensler}
  B.~M., {Clarke} M.~J., 2004{\natexlab{b}}, ApJ, 601, 1045

\bibitem[{{Slane}, {Helfand} \& {Murray}(2002){Slane}, {Helfand}, \&
  {Murray}}]{slane02b}
{Slane} P.~O., {Helfand} D.~J., {Murray} S.~S., 2002, ApJL, 571, L45

\bibitem[{{Slowikowska} {et~al}\mbox{.}(2007){Slowikowska}, {Kanbach},
  {Borkowski}, \& {Becker}}]{slowikowska07}
{Slowikowska} A., {Kanbach} G., {Borkowski} J., {Becker} W., 2007, in
  WE-Heraeus Seminar on Neutron Stars and Pulsars 40 years after the Discovery,
  {Becker} W., {Huang} H.~H., eds., p.~44

\bibitem[{{Smith}(2003)}]{smith03}
{Smith} N., 2003, MNRAS, 346, 885

\bibitem[{{Smith} \& {Hughes}(2010)}]{smith10}
{Smith} R.~K., {Hughes} J.~P., 2010, ApJ, 718, 583

\bibitem[{{Spitkovsky}(2008)}]{spitkovsky08}
{Spitkovsky} A., 2008, ApJL, 682, L5

\bibitem[{{Stephenson}(1971)}]{stephenson71}
{Stephenson} F.~R., 1971, Q. J. R. Astron. Soc., 12, 10

\bibitem[{{Stephenson} \& {Green}(2002)}]{stephenson02}
{Stephenson} F.~R., {Green} D.~A., 2002, Historical supernovae and their
  remnants, by F.~Richard Stephenson and David A.~Green.~International series
  in astronomy and astrophysics, vol.~5.~Oxford: Clarendon Press, 2002, ISBN
  0198507666, 5

\bibitem[{{Strom} \& {Stappers}(2000)}]{strom00}
{Strom} R.~G., {Stappers} B.~W., 2000, in Astronomical Society of the Pacific
  Conference Series, Vol. 202, IAU Colloq. 177: Pulsar Astronomy - 2000 and
  Beyond, {Kramer} M., {Wex} N., {Wielebinski} R., eds., p. 509

\bibitem[{{Str{\"u}der} {et~al}\mbox{.}(2001){Str{\"u}der}, {Briel}, {Dennerl},
  {Hartmann}, {Kendziorra}, {Meidinger}, {Pfeffermann}, {Reppin}, {Aschenbach},
  {Bornemann}, {Br{\"a}uninger}, {Burkert}, {Elender}, {Freyberg}, {Haberl},
  {Hartner}, {Heuschmann}, {Hippmann}, {Kastelic}, {Kemmer}, {Kettenring},
  {Kink}, {Krause}, {M{\"u}ller}, {Oppitz}, {Pietsch}, {Popp}, {Predehl},
  {Read}, {Stephan}, {St{\"o}tter}, {Tr{\"u}mper}, {Holl}, {Kemmer}, {Soltau},
  {St{\"o}tter}, {Weber}, {Weichert}, {von Zanthier}, {Carathanassis}, {Lutz},
  {Richter}, {Solc}, {B{\"o}ttcher}, {Kuster}, {Staubert}, {Abbey}, {Holland},
  {Turner}, {Balasini}, {Bignami}, {La Palombara}, {Villa}, {Buttler},
  {Gianini}, {Lain{\'e}}, {Lumb}, \& {Dhez}}]{struder01}
{Str{\"u}der} L. {et~al.}, 2001, A\&A, 365, L18

\bibitem[{{Su} {et~al}\mbox{.}(2009){Su}, {Chen}, {Yang}, {Koo}, {Zhou},
  {Jeong}, \& {Zhang}}]{su09}
{Su} Y., {Chen} Y., {Yang} J., {Koo} B.-C., {Zhou} X., {Jeong} I.-G., {Zhang}
  C.-G., 2009, ApJ, 694, 376

\bibitem[{{Sun} {et~al}\mbox{.}(2004){Sun}, {Seward}, {Smith}, \&
  {Slane}}]{sun04}
{Sun} M., {Seward} F.~D., {Smith} R.~K., {Slane} P.~O., 2004, ApJ, 605, 742

\bibitem[{{Syrovatskii}(1959)}]{syrovatskii59}
{Syrovatskii} S.~I., 1959, Soviet Astronomy, 3, 22

\bibitem[{{Tam} \& {Roberts}(2003)}]{tam03}
{Tam} C., {Roberts} M.~S.~E., 2003, ApJL, 598, L27

\bibitem[{{Tamagawa} {et~al}\mbox{.}(2009){Tamagawa}, {Hayato}, {Nakamura},
  {Terada}, {Bamba}, {Hiraga}, {Hughes}, {Hwang}, {Kataoka}, {Kinugasa},
  {Kunieda}, {Tanaka}, {Tsunemi}, {Ueno}, {Holt}, {Kokubun}, {Miyata},
  {Szymkowiak}, {Takahashi}, {Tamura}, {Ueno}, \& {Makishima}}]{tamagawa09}
{Tamagawa} T. {et~al.}, 2009, PASJ, 61, 167

\bibitem[{{Tanaka} \& {Takahara}(2010)}]{tanaka10}
{Tanaka} S.~J., {Takahara} F., 2010, ApJ, 715, 1248

\bibitem[{{Tanaka} \& {Takahara}(2011)}]{tanaka11}
{Tanaka} S.~J., {Takahara} F., 2011, ApJ, 741, 40

\bibitem[{{Tanaka} \& {Takahara}(2013)}]{tanaka13}
{Tanaka} S.~J., {Takahara} F., 2013, MNRAS, 429, 2945

\bibitem[{{Taylor}(1946)}]{taylor46}
{Taylor} G.~I., 1946, Royal Society of London Proceedings Series A, 186, 273

\bibitem[{{Taylor} \& {Cordes}(1993)}]{taylor93a}
{Taylor} J.~H., {Cordes} J.~M., 1993, ApJ, 411, 674

\bibitem[{{Taylor} {et~al}\mbox{.}(1974){Taylor}, {Hulse}, {Margon},
  {Davidsen}, {Mason}, {Sanford}, {Liller}, {Bernacca}, {Ciatti}, {John},
  {Regener}, {Papaliolios}, {Pennypacker}, {Canizares}, {McClintock}, {Jones},
  {Graham}, \& {Wielebinski}}]{taylor74}
{Taylor} J.~H. {et~al.}, 1974, IAU Circ., 2704, 1

\bibitem[{{Taylor}, {Manchester} \& {Lyne}(1993){Taylor}, {Manchester}, \&
  {Lyne}}]{taylor93b}
{Taylor} J.~H., {Manchester} R.~N., {Lyne} A.~G., 1993, ApJS, 88, 529

\bibitem[{{Temim} {et~al}\mbox{.}(2006){Temim}, {Gehrz}, {Woodward}, {Roellig},
  {Smith}, {Rudnick}, {Polomski}, {Davidson}, {Yuen}, \& {Onaka}}]{temim06}
{Temim} T. {et~al.}, 2006, AJ, 132, 1610

\bibitem[{{Temim} {et~al}\mbox{.}(2012){Temim}, {Slane}, {Arendt}, \&
  {Dwek}}]{temim12}
{Temim} T., {Slane} P., {Arendt} R.~G., {Dwek} E., 2012, ApJ, 745, 46

\bibitem[{{Temim} {et~al}\mbox{.}(2009){Temim}, {Slane}, {Gaensler}, {Hughes},
  \& {Van Der Swaluw}}]{temim09}
{Temim} T., {Slane} P., {Gaensler} B.~M., {Hughes} J.~P., {Van Der Swaluw} E.,
  2009, ApJ, 691, 895

\bibitem[{{Temim} {et~al}\mbox{.}(2010){Temim}, {Slane}, {Reynolds}, {Raymond},
  \& {Borkowski}}]{temim10}
{Temim} T., {Slane} P., {Reynolds} S.~P., {Raymond} J.~C., {Borkowski} K.~J.,
  2010, ApJ, 710, 309

\bibitem[{{Thompson} \& {Duncan}(1993)}]{thompson93}
{Thompson} C., {Duncan} R.~C., 1993, ApJ, 408, 194

\bibitem[{{Thompson} \& {Duncan}(1996)}]{thompson96}
{Thompson} C., {Duncan} R.~C., 1996, ApJ, 473, 322

\bibitem[{{Tian} \& {Leahy}(2008{\natexlab{a}})}]{tian08b}
{Tian} W.~W., {Leahy} D.~A., 2008{\natexlab{a}}, ApJ, 677, 292

\bibitem[{{Tian} \& {Leahy}(2008{\natexlab{b}})}]{tian08a}
{Tian} W.~W., {Leahy} D.~A., 2008{\natexlab{b}}, MNRAS, 391, L54

\bibitem[{{Tian} {et~al}\mbox{.}(2007){Tian}, {Li}, {Leahy}, \&
  {Wang}}]{tian07}
{Tian} W.~W., {Li} Z., {Leahy} D.~A., {Wang} Q.~D., 2007, ApJL, 657, L25

\bibitem[{{Toledo-Roy} {et~al}\mbox{.}(2009){Toledo-Roy}, {Vel{\'a}zquez}, {de
  Colle}, {Gonz{\'a}lez}, {Reynoso}, {Kurtz}, \&
  {Reyes-Iturbide}}]{toledoroy09}
{Toledo-Roy} J.~C., {Vel{\'a}zquez} P.~F., {de Colle} F., {Gonz{\'a}lez} R.~F.,
  {Reynoso} E.~M., {Kurtz} S.~E., {Reyes-Iturbide} J., 2009, MNRAS, 395, 351

\bibitem[{{Tomsick} {et~al}\mbox{.}(2009){Tomsick}, {Chaty}, {Rodriguez},
  {Walter}, \& {Kaaret}}]{tomsick09}
{Tomsick} J.~A., {Chaty} S., {Rodriguez} J., {Walter} R., {Kaaret} P., 2009,
  ApJ, 701, 811

\bibitem[{{Torii} {et~al}\mbox{.}(2000){Torii}, {Slane}, {Kinugasa},
  {Hashimotodani}, \& {Tsunemi}}]{torii00}
{Torii} K., {Slane} P.~O., {Kinugasa} K., {Hashimotodani} K., {Tsunemi} H.,
  2000, PASJ, 52, 875

\bibitem[{{Torii} {et~al}\mbox{.}(1999){Torii}, {Tsunemi}, {Dotani}, {Mitsuda},
  {Kawai}, {Kinugasa}, {Saito}, \& {Shibata}}]{torii99}
{Torii} K., {Tsunemi} H., {Dotani} T., {Mitsuda} K., {Kawai} N., {Kinugasa} K.,
  {Saito} Y., {Shibata} S., 1999, ApJL, 523, L69

\bibitem[{{Torii} {et~al}\mbox{.}(2006){Torii}, {Uchida}, {Hasuike}, {Tsunemi},
  {Yamaguchi}, \& {Shibata}}]{torii06}
{Torii} K., {Uchida} H., {Hasuike} K., {Tsunemi} H., {Yamaguchi} Y., {Shibata}
  S., 2006, PASJ, 58, L11

\bibitem[{{Torres} {et~al}\mbox{.}(2014){Torres}, {Cillis}, {Mart{\'{\i}}n}, \&
  {de O{\~n}a Wilhelmi}}]{torres14}
{Torres} D.~F., {Cillis} A., {Mart{\'{\i}}n} J., {de O{\~n}a Wilhelmi} E.,
  2014, Journal of High Energy and Astrophysics, 1, 31

\bibitem[{{Torres}, {Cillis} \& {Mart{\'{\i}}n Rodriguez}(2013){Torres},
  {Cillis}, \& {Mart{\'{\i}}n Rodriguez}}]{torres13a}
{Torres} D.~F., {Cillis} A.~N., {Mart{\'{\i}}n Rodriguez} J., 2013, ApJL, 763,
  L4

\bibitem[{{Torres}, {Marrero} \& {de Cea Del Pozo}(2010){Torres}, {Marrero}, \&
  {de Cea Del Pozo}}]{torres10}
{Torres} D.~F., {Marrero} A.~Y.~R., {de Cea Del Pozo} E., 2010, MNRAS, 408,
  1257

\bibitem[{{Torres} {et~al}\mbox{.}(2013){Torres}, {Mart{\'{\i}}n}, {de O{\~n}a
  Wilhelmi}, \& {Cillis}}]{torres13b}
{Torres} D.~F., {Mart{\'{\i}}n} J., {de O{\~n}a Wilhelmi} E., {Cillis} A.,
  2013, MNRAS, 436, 3112

\bibitem[{{Trussoni} {et~al}\mbox{.}(1996){Trussoni}, {Massaglia}, {Caucino},
  {Brinkmann}, \& {Aschenbach}}]{trussoni96}
{Trussoni} E., {Massaglia} S., {Caucino} S., {Brinkmann} W., {Aschenbach} B.,
  1996, A\&A, 306, 581

\bibitem[{{Tsujimoto} {et~al}\mbox{.}(2011){Tsujimoto}, {Guainazzi},
  {Plucinsky}, {Beardmore}, {Ishida}, {Natalucci}, {Posson-Brown}, {Read},
  {Saxton}, \& {Shaposhnikov}}]{tsujimoto11}
{Tsujimoto} M. {et~al.}, 2011, A\&A, 525, A25

\bibitem[{{Turner} {et~al}\mbox{.}(2001){Turner}, {Abbey}, {Arnaud},
  {Balasini}, {Barbera}, {Belsole}, {Bennie}, {Bernard}, {Bignami}, {Boer},
  {Briel}, {Butler}, {Cara}, {Chabaud}, {Cole}, {Collura}, {Conte}, {Cros},
  {Denby}, {Dhez}, {Di Coco}, {Dowson}, {Ferrando}, {Ghizzardi}, {Gianotti},
  {Goodall}, {Gretton}, {Griffiths}, {Hainaut}, {Hochedez}, {Holland},
  {Jourdain}, {Kendziorra}, {Lagostina}, {Laine}, {La Palombara}, {Lortholary},
  {Lumb}, {Marty}, {Molendi}, {Pigot}, {Poindron}, {Pounds}, {Reeves},
  {Reppin}, {Rothenflug}, {Salvetat}, {Sauvageot}, {Schmitt}, {Sembay},
  {Short}, {Spragg}, {Stephen}, {Str{\"u}der}, {Tiengo}, {Trifoglio},
  {Tr{\"u}mper}, {Vercellone}, {Vigroux}, {Villa}, {Ward}, {Whitehead}, \&
  {Zonca}}]{turner01}
{Turner} M.~J.~L. {et~al.}, 2001, A\&A, 365, L27

\bibitem[{{Tziamtzis}, {Lundqvist} \& {Djupvik}(2009){Tziamtzis}, {Lundqvist},
  \& {Djupvik}}]{tziamtzis09}
{Tziamtzis} A., {Lundqvist} P., {Djupvik} A.~A., 2009, A\&A, 508, 221

\bibitem[{{Ubertini} {et~al}\mbox{.}(2005){Ubertini}, {Bassani}, {Malizia},
  {Bazzano}, {Bird}, {Dean}, {De Rosa}, {Lebrun}, {Moran}, {Renaud}, {Stephen},
  {Terrier}, \& {Walter}}]{ubertini05}
{Ubertini} P. {et~al.}, 2005, ApJL, 629, L109

\bibitem[{{van den Bergh}(1978)}]{vandenbergh78}
{van den Bergh} S., 1978, ApJL, 220, L9

\bibitem[{{van der Swaluw}(2003)}]{vanderswaluw03}
{van der Swaluw} E., 2003, A\&A, 404, 939

\bibitem[{{van der Swaluw}(2004)}]{vanderswaluw04a}
{van der Swaluw} E., 2004, Advances in Space Research, 33, 475

\bibitem[{{Van Der Swaluw}, {Achterberg} \& {Gallant}(1998){Van Der Swaluw},
  {Achterberg}, \& {Gallant}}]{vanderswaluw98}
{Van Der Swaluw} E., {Achterberg} A., {Gallant} Y.~A., 1998, MmSAI, 69, 1017

\bibitem[{{van der Swaluw} {et~al}\mbox{.}(2001){van der Swaluw}, {Achterberg},
  {Gallant}, \& {T{\'o}th}}]{vanderswaluw01}
{van der Swaluw} E., {Achterberg} A., {Gallant} Y.~A., {T{\'o}th} G., 2001,
  A\&A, 380, 309

\bibitem[{{van der Swaluw}, {Downes} \& {Keegan}(2004){van der Swaluw},
  {Downes}, \& {Keegan}}]{vanderswaluw04b}
{van der Swaluw} E., {Downes} T.~P., {Keegan} R., 2004, A\&A, 420, 937

\bibitem[{{Van Etten} \& {Romani}(2011)}]{vanetten11}
{Van Etten} A., {Romani} R.~W., 2011, ApJ, 742, 62

\bibitem[{{van Veelen} {et~al}\mbox{.}(2009){van Veelen}, {Langer}, {Vink},
  {Garc{\'{\i}}a-Segura}, \& {van Marle}}]{vanveelen09}
{van Veelen} B., {Langer} N., {Vink} J., {Garc{\'{\i}}a-Segura} G., {van Marle}
  A.~J., 2009, A\&A, 503, 495

\bibitem[{{Vancura} {et~al}\mbox{.}(1992){Vancura}, {Blair}, {Long}, \&
  {Raymond}}]{vancura92}
{Vancura} O., {Blair} W.~P., {Long} K.~S., {Raymond} J.~C., 1992, ApJ, 394, 158

\bibitem[{{Vasisht} \& {Gotthelf}(1997)}]{vasisht97}
{Vasisht} G., {Gotthelf} E.~V., 1997, ApJL, 486, L129

\bibitem[{{Velusamy} \& {Becker}(1988)}]{velusamy88}
{Velusamy} T., {Becker} R.~H., 1988, AJ, 95, 1162

\bibitem[{{Venter} \& {de Jager}(2007)}]{venter07}
{Venter} C., {de Jager} O.~C., 2007, in WE-Heraeus Seminar on Neutron Stars and
  Pulsars 40 years after the Discovery, {Becker} W., {Huang} H.~H., eds., p.~40

\bibitem[{{Veron-Cetty} \& {Woltjer}(1993)}]{veroncetty93}
{Veron-Cetty} M.~P., {Woltjer} L., 1993, A\&A, 270, 370

\bibitem[{{Vink}(2012)}]{vink12}
{Vink} J., 2012, A\&ARv, 20, 49

\bibitem[{{Vink} {et~al}\mbox{.}(2004){Vink}, {Bleeker}, {Kaastra}, \&
  {Rasmussen}}]{vink04}
{Vink} J., {Bleeker} J., {Kaastra} J.~S., {Rasmussen} A., 2004, Nuclear Physics
  B Proceedings Supplements, 132, 62

\bibitem[{{Vink} {et~al}\mbox{.}(1998){Vink}, {Bloemen}, {Kaastra}, \&
  {Bleeker}}]{vink98}
{Vink} J., {Bloemen} H., {Kaastra} J.~S., {Bleeker} J.~A.~M., 1998, A\&A, 339,
  201

\bibitem[{{Vink}, {Kaastra} \& {Bleeker}(1996){Vink}, {Kaastra}, \&
  {Bleeker}}]{vink96}
{Vink} J., {Kaastra} J.~S., {Bleeker} J.~A.~M., 1996, A\&A, 307, L41

\bibitem[{{Vink}, {Kaastra} \& {Bleeker}(1997){Vink}, {Kaastra}, \&
  {Bleeker}}]{vink97}
{Vink} J., {Kaastra} J.~S., {Bleeker} J.~A.~M., 1997, A\&A, 328, 628

\bibitem[{{Vink} \& {Kuiper}(2006)}]{vink06}
{Vink} J., {Kuiper} L., 2006, MNRAS, 370, L14

\bibitem[{{Vinyaikin}(2007)}]{vinyaikin07}
{Vinyaikin} E.~N., 2007, Astronomy Reports, 51, 570

\bibitem[{{Volpi} {et~al}\mbox{.}(2008){Volpi}, {Del Zanna}, {Amato}, \&
  {Bucciantini}}]{volpi08}
{Volpi} D., {Del Zanna} L., {Amato} E., {Bucciantini} N., 2008, A\&A, 485, 337

\bibitem[{{Vorster} \& {Moraal}(2013)}]{vorster13b}
{Vorster} M.~J., {Moraal} H., 2013, ApJ, 765, 30

\bibitem[{{Vorster} {et~al}\mbox{.}(2013){Vorster}, {Tibolla}, {Ferreira}, \&
  {Kaufmann}}]{vorster13c}
{Vorster} M.~J., {Tibolla} O., {Ferreira} S.~E.~S., {Kaufmann} S., 2013, ApJ,
  773, 139

\bibitem[{{Vrba} {et~al}\mbox{.}(2000){Vrba}, {Henden}, {Luginbuhl}, {Guetter},
  {Hartmann}, \& {Klose}}]{vrba00}
{Vrba} F.~J., {Henden} A.~A., {Luginbuhl} C.~B., {Guetter} H.~H., {Hartmann}
  D.~H., {Klose} S., 2000, ApJL, 533, L17

\bibitem[{{Wang} {et~al}\mbox{.}(2000){Wang}, {Manchester}, {Pace}, {Bailes},
  {Kaspi}, {Stappers}, \& {Lyne}}]{wang00}
{Wang} N., {Manchester} R.~N., {Pace} R.~T., {Bailes} M., {Kaspi} V.~M.,
  {Stappers} B.~W., {Lyne} A.~G., 2000, MNRAS, 317, 843

\bibitem[{{Wang} \& {Gotthelf}(1998)}]{wang98}
{Wang} Q.~D., {Gotthelf} E.~V., 1998, ApJ, 494, 623

\bibitem[{{Wang} {et~al}\mbox{.}(2001){Wang}, {Gotthelf}, {Chu}, \&
  {Dickel}}]{wang01}
{Wang} Q.~D., {Gotthelf} E.~V., {Chu} Y.-H., {Dickel} J.~R., 2001, ApJ, 559,
  275

\bibitem[{{Wang} {et~al}\mbox{.}(1986){Wang}, {Liu}, {Gorenstein}, \&
  {Zombeck}}]{wang86}
{Wang} Z.~R., {Liu} J.~Y., {Gorenstein} P., {Zombeck} M.~V., 1986, Highlights
  of Astronomy, 7, 583

\bibitem[{{Warren} \& {Hughes}(2004)}]{warren04}
{Warren} J.~S., {Hughes} J.~P., 2004, ApJ, 608, 261

\bibitem[{{Warwick} {et~al}\mbox{.}(2001){Warwick}, {Bernard}, {Bocchino},
  {Decourchelle}, {Ferrando}, {Griffiths}, {Haberl}, {La Palombara}, {Lumb},
  {Mereghetti}, {Read}, {Schaudel}, {Schurch}, {Tiengo}, \&
  {Willingale}}]{warwick01}
{Warwick} R.~S. {et~al.}, 2001, A\&A, 365, L248

\bibitem[{{Weekes} {et~al}\mbox{.}(1989){Weekes}, {Cawley}, {Fegan}, {Gibbs},
  {Hillas}, {Kowk}, {Lamb}, {Lewis}, {Macomb}, {Porter}, {Reynolds}, \&
  {Vacanti}}]{weekes89}
{Weekes} T.~C. {et~al.}, 1989, ApJ, 342, 379

\bibitem[{{Weiler} \& {Panagia}(1978)}]{weiler78}
{Weiler} K.~W., {Panagia} N., 1978, A\&A, 70, 419

\bibitem[{{Weisberg} {et~al}\mbox{.}(2008){Weisberg}, {Stanimirovi{\'c}},
  {Xilouris}, {Hedden}, {de la Fuente}, {Anderson}, \& {Jenet}}]{weisberg08}
{Weisberg} J.~M., {Stanimirovi{\'c}} S., {Xilouris} K., {Hedden} A., {de la
  Fuente} A., {Anderson} S.~B., {Jenet} F.~A., 2008, ApJ, 674, 286

\bibitem[{{Weisskopf} {et~al}\mbox{.}(2000){Weisskopf}, {Hester}, {Tennant},
  {Elsner}, {Schulz}, {Marshall}, {Karovska}, {Nichols}, {Swartz},
  {Kolodziejczak}, \& {O'Dell}}]{weisskopf00}
{Weisskopf} M.~C. {et~al.}, 2000, ApJL, 536, L81

\bibitem[{{Weltevrede}, {Johnston} \& {Espinoza}(2011){Weltevrede}, {Johnston},
  \& {Espinoza}}]{weltevrede11}
{Weltevrede} P., {Johnston} S., {Espinoza} C.~M., 2011, MNRAS, 411, 1917

\bibitem[{{Whiteoak} \& {Green}(1996)}]{whiteoak96}
{Whiteoak} J.~B.~Z., {Green} A.~J., 1996, A\&AS, 118, 329

\bibitem[{{Wickramasinghe} \& {Ferrario}(2005)}]{wickramasinghe05}
{Wickramasinghe} D.~T., {Ferrario} L., 2005, MNRAS, 356, 1576

\bibitem[{{Williams} {et~al}\mbox{.}(2008){Williams}, {Borkowski}, {Reynolds},
  {Raymond}, {Long}, {Morse}, {Blair}, {Ghavamian}, {Sankrit}, {Hendrick},
  {Smith}, {Points}, \& {Winkler}}]{williams08}
{Williams} B.~J. {et~al.}, 2008, ApJ, 687, 1054

\bibitem[{{Willingale} {et~al}\mbox{.}(2001){Willingale}, {Aschenbach},
  {Griffiths}, {Sembay}, {Warwick}, {Becker}, {Abbey}, \&
  {Bonnet-Bidaud}}]{willingale01}
{Willingale} R., {Aschenbach} B., {Griffiths} R.~G., {Sembay} S., {Warwick}
  R.~S., {Becker} W., {Abbey} A.~F., {Bonnet-Bidaud} J.-M., 2001, A\&A, 365,
  L212

\bibitem[{{Willingale} {et~al}\mbox{.}(2002){Willingale}, {Bleeker}, {van der
  Heyden}, {Kaastra}, \& {Vink}}]{willingale02}
{Willingale} R., {Bleeker} J.~A.~M., {van der Heyden} K.~J., {Kaastra} J.~S.,
  {Vink} J., 2002, A\&A, 381, 1039

\bibitem[{{Wilson} \& {Weiler}(1976)}]{wilson76}
{Wilson} A.~S., {Weiler} K.~W., 1976, A\&A, 53, 89

\bibitem[{{Winkler} {et~al}\mbox{.}(1981{\natexlab{a}}){Winkler}, {Clark},
  {Markert}, {Kalata}, {Schnopper}, \& {Canizares}}]{winkler81b}
{Winkler} P.~F., {Clark} G.~W., {Markert} T.~H., {Kalata} K., {Schnopper}
  H.~W., {Canizares} C.~R., 1981{\natexlab{a}}, ApJL, 246, L27

\bibitem[{{Winkler} {et~al}\mbox{.}(1981{\natexlab{b}}){Winkler}, {Clark},
  {Markert}, {Petre}, \& {Canizares}}]{winkler81a}
{Winkler} P.~F., {Clark} G.~W., {Markert} T.~H., {Petre} R., {Canizares} C.~R.,
  1981{\natexlab{b}}, ApJ, 245, 574

\bibitem[{{Winkler} {et~al}\mbox{.}(2009){Winkler}, {Twelker}, {Reith}, \&
  {Long}}]{winkler09}
{Winkler} P.~F., {Twelker} K., {Reith} C.~N., {Long} K.~S., 2009, ApJ, 692,
  1489

\bibitem[{{Woltjer}(1964)}]{woltjer64}
{Woltjer} L., 1964, ApJ, 140, 1309

\bibitem[{{Woods} \& {Thompson}(2006)}]{woods06}
{Woods} P.~M., {Thompson} C., 2006, {Soft gamma repeaters and anomalous X-ray
  pulsars: magnetar candidates}, {Lewin} W.~H.~G., {van der Klis} M., eds.,
  Cambridge University Press, pp. 547--586

\bibitem[{{Yamaguchi} {et~al}\mbox{.}(2008){Yamaguchi}, {Koyama}, {Katsuda},
  {Nakajima}, {Hughes}, {Bamba}, {Hiraga}, {Mori}, {Ozaki}, \&
  {Tsuru}}]{yamaguchi08}
{Yamaguchi} H. {et~al.}, 2008, PASJ, 60, 141

\bibitem[{{Yuan} {et~al}\mbox{.}(2010){Yuan}, {Wang}, {Manchester}, \&
  {Liu}}]{yuan10}
{Yuan} J.~P., {Wang} N., {Manchester} R.~N., {Liu} Z.~Y., 2010, MNRAS, 404, 289

\bibitem[{{Zajczyk} {et~al}\mbox{.}(2012){Zajczyk}, {Gallant}, {Slane},
  {Reynolds}, {Bandiera}, {Gouiff{\`e}s}, {Le Floc'h}, {Comer{\'o}n}, \& {Koch
  Miramond}}]{zajczyk12}
{Zajczyk} A. {et~al.}, 2012, A\&A, 542, A12

\bibitem[{{Zeiger} {et~al}\mbox{.}(2008){Zeiger}, {Brisken}, {Chatterjee}, \&
  {Goss}}]{zeiger08}
{Zeiger} B.~R., {Brisken} W.~F., {Chatterjee} S., {Goss} W.~M., 2008, ApJ, 674,
  271

\bibitem[{{Zhang}, {Chen} \& {Fang}(2008){Zhang}, {Chen}, \& {Fang}}]{zhang08}
{Zhang} L., {Chen} S.~B., {Fang} J., 2008, ApJ, 676, 1210

\bibitem[{{Zhang}, {Jiang} \& {Lin}(2009){Zhang}, {Jiang}, \& {Lin}}]{zhang09}
{Zhang} L., {Jiang} Z.~J., {Lin} G.~F., 2009, ApJ, 699, 507

\bibitem[{{Zharikov} {et~al}\mbox{.}(2008){Zharikov}, {Shibanov}, {Zyuzin},
  {Mennickent}, \& {Komarova}}]{zharikov08}
{Zharikov} S.~V., {Shibanov} Y.~A., {Zyuzin} D.~A., {Mennickent} R.~E.,
  {Komarova} V.~N., 2008, A\&A, 492, 805

\bibitem[{{Zharikov} {et~al}\mbox{.}(2013){Zharikov}, {Zyuzin}, {Shibanov}, \&
  {Mennickent}}]{zharikov13}
{Zharikov} S.~V., {Zyuzin} D.~A., {Shibanov} Y.~A., {Mennickent} R.~E., 2013,
  A\&A, 554, A120

\bibitem[{{Ziegler} {et~al}\mbox{.}(2008){Ziegler}, {Baughman}, {Johnson}, \&
  {Atwood}}]{ziegler08}
{Ziegler} M., {Baughman} B.~M., {Johnson} R.~P., {Atwood} W.~B., 2008, ApJ,
  680, 620

\bibitem[{{Zirakashvili} \& {Aharonian}(2007)}]{zirakashvili07}
{Zirakashvili} V.~N., {Aharonian} F., 2007, A\&A, 465, 695

\end{thebibliography}



\end{spacing}


\begin{appendices} 


\chapter{Energy losses equations}
\label{appa}

\section*{Synchrotron energy losses}

To derive equation (\ref{synclosses}), we will consider the second Newton's law in its relativistic version
\begin{equation}
f^\mu=\frac{dp^\mu}{d\tau}
\end{equation}
where $f^\mu$ is the four-vector force, $p^\mu$ is the four-momentum and $\tau$ is the proper time. The electromagnetic force is given by
\begin{equation}
\label{force}
f^\mu=\frac{q}{c}F^\mu_\nu u^\nu
\end{equation}
where $q$ is the charge of the particle, $c$ is the speed of light and $u^\nu$ is the four-velocity defined as $u^\nu=(\gamma c,\gamma \vec{v})$.
$F^\mu_\nu$ is the so-called Faraday tensor. In the absence of electric field, equation (\ref{force}) yields
\begin{equation}
m \frac{d}{dt}(\gamma \vec{v})=\frac{q}{c} \vec{v} \times \vec{B},
\end{equation}
\begin{equation}
m \frac{d}{dt}(\gamma c)=0.
\end{equation}
The second equation implies that $\gamma$ is constant and it can be written out of the derivative. Now, we take the first equation and we
separate the velocity in components along the magnetic field ($\vec{v}_\parallel$) an the perpendicular one ($\vec{v}_\perp$)
\begin{equation}
\frac{d\vec{v_\parallel}}{dt}=0,
\end{equation}
\begin{equation}
\label{gyr}
\frac{d\vec{v_\perp}}{dt}=\frac{q}{\gamma m c} \vec{v}_\perp \times \vec{B}.
\end{equation}
The solution of equation (\ref{gyr}) is a circular motion projected on the normal plane to the magnetic field. The frequency of rotation is
called {\it gyration} and is given by
\begin{equation}
\omega_B=\frac{qB}{\gamma m c}.
\end{equation}
The relativistic form of Larmor's formula, decomposed by the parallel and perpendicular components of the acceleration, is given by
\begin{equation}
\left(\frac{dE}{dt} \right)_{L}=\frac{2q^2}{3c^3}\gamma^4 \left(a^2_\perp+\gamma^2 a^2_\parallel \right).
\end{equation}
Taking into account that the acceleration is perpendicular to the velocity with magnitude $a_\perp=\omega_B v_\perp$, substituing this in
equation \ref{larmor}, we find
\begin{equation}
\label{sync1}
\frac{dE}{dt}=\frac{2}{3}r^2_0 c \beta^2_\perp \gamma^2 B^2,
\end{equation}
where $r_0=q^4/(m c^2)^2$ is the classical radius of the electron and $\beta_\perp=v_\perp/c$. If we consider an isotropic distribution of
velocities, we must do the average over all possible pitch angles $\alpha$
\begin{equation}
<\beta^2_\perp>=\frac{\beta^2}{4\pi} \int \sin^2 \alpha \ \mathrm{d}\Omega=\frac{2}{3}\beta^2.
\end{equation}
Subtituting this result in equation (\ref{sync1}), expliciting the sign of the derivative and writing in terms of the Lorentz factor, we recover
equation (\ref{synclosses})
\begin{equation}
\dot{\gamma}_{sync}(\gamma,t)=-\frac{4}{3}\frac{\sigma_T}{m_e c}U_B(t)\gamma^2.
\end{equation}
Note that we assume $\beta \simeq 1$.

\section*{IC energy losses}

The IC energy losses are calculated with the formalism used in \citet{blumenthal70}. Consider a target photon gas with a differential number
density $\mathrm{d}n=n(\nu_i,\hat{i}) \mathrm{d}\nu_i \mathrm{d}\Omega$, such that $\mathrm{d}n$ is the number of photons per cm$^{-3}$ with a
frequency within $\nu_i$ and $\nu_i+\mathrm{d}\nu_i$ moving in the direction defined by the unitary vector $\hat{i}$. In the most general case,
the interacting electrons passing through the gas would be described by $\mathrm{d} n_e=n_e(\gamma,\hat{i}_e)$. The distribution of the
scattered photons with frequency $\nu_f$ per unit volume and unit time is this case is very complicated to treat, but for our purpose we can do
some approximations as considering all the electrons with $\gamma \gg 1$. Depending on the energy of the target photons in the electron rest
frame, we will consider two different regimes: the Thomson limit ($h\nu_i \ll m_e c^2$) or the extreme Klein-Nishina limit
($h\nu_i \gg m_e c^2$).

The IC energy losses of the electrons would be the same the the energy cointained in the emitted radiation, thus
\begin{equation}
\dot{\gamma}_{IC}=-\frac{1}{m_e c^2}\frac{dE_{rad}}{dt}.
\end{equation}
In the Thomson limit, the energy of the photons $\varepsilon'$ in the electron rest frame before and after scattering is
$\varepsilon'_f=\varepsilon'_i$. The quantity $dE_{rad}/dt$ is invariant since it is the ratio of the same components of two parallel
four-vectors. Then,
\begin{equation}
\dot{\gamma}_{IC}=-\frac{1}{m_e c^2}\frac{dE'_{rad}}{dt'}=-\frac{1}{m_e c^2} \int \sigma_T c \varepsilon' \mathrm{d}n'=-\frac{\sigma_T U'_\gamma}{m_e c},
\end{equation}
being $\sigma_T$ the Thomson cross section and $U'_\gamma$ the target photon energy density. In the laboratory frame,
\begin{equation}
U'_\gamma=\gamma^2 \int (1-\cos \theta)^2 \varepsilon \mathrm{d}n.
\end{equation}
Considering an isotropic distribution of photons, we average over angles, $<(1-\cos \theta)^2>=4/3$, and we get
\begin{equation}
U'_\gamma=\frac{4}{3} \gamma^2 \int \varepsilon \mathrm{d}n=\frac{4}{3} \gamma^2 U_\gamma,
\end{equation}
and finally,
\begin{equation}
\dot{\gamma}_{IC}=-\frac{4}{3}\frac{\sigma_T}{m_e c} U_\gamma \gamma^2,
\end{equation}
recovering equation (\ref{icthomson}). As we said before, equation (\ref{icthomson}) give the IC energy losses for electrons when the scattering
is produced in the Thomson regime. The general case is solved using the Klein-Nishina cross section \citep{blumenthal70}
\begin{equation}
\frac{d\sigma_{KN}}{d\Omega'_f d\nu'_f}=\frac{3}{16 \pi}h \sigma_T \left(\frac{\nu'_f}{\nu'_i} \right)^2 \left(\frac{\nu'_i}{\nu'_f}+\frac{\nu'_f}{\nu'_i}-\sin^2 \theta'_f \right) \delta \left(\nu'_f-\frac{\nu'_i}{1+(\nu'_i/m_e c^2)(1-\cos \theta'_f)} \right)
\end{equation}
\citet{jones68} computed the scattered photon spectrum per electron is this case giving
\begin{equation}
\label{icsp}
\frac{dN}{dt d\nu_f}=\frac{3}{4} \frac{c \sigma_T}{\gamma^2}\frac{n(\nu_i) \mathrm{d}\nu}{\nu_i}f(q,\Gamma_{\varepsilon}),
\end{equation}
where $f(q,\Gamma_{\varepsilon})$, $\Gamma_{\varepsilon}$ and $q$ are defined by equations (\ref{fqg}), (\ref{geps}) and (\ref{qkin}),
respectively. The total energy loss rate is given by
\begin{equation}
\dot{\gamma_{IC}}=-\frac{h}{m_e c^2} \int (\nu_f-\nu_i) \frac{dN}{dt d\nu_f} \mathrm{d}\nu_f. 
\end{equation}
The final frequency $\nu_f$ will be in general much greater than the initial one $\nu_i$. Then, we neglect $\nu_i$ and substituing equation
(\ref{icsp}) inside the integral, we get
\begin{equation}
\dot{\gamma}_{IC}(\gamma)=-\frac{3}{4}\frac{\sigma_T h}{m_e c}\frac{1}{\gamma^2}\int_0^\infty \nu_f \mathrm{d}\nu_f \int_0^\infty \frac{n(\nu_i)}{\nu_i} f(q,\Gamma_{\varepsilon}) H(1-q) H \left(q-\frac{1}{4\gamma^2} \right) \mathrm{d}\nu_i.
\end{equation}
The Heaviside functions make explicit the extrem values of the kinematic coefficient $q$ ($1/4\gamma^2 \le q \le 1$, see \citealt{jones68}).

\section*{Bremsstrahlung energy losses}

The average energy loss per unit time by Bremsstrahlung emission of an electron is \citep{haug04}
\begin{equation}
\label{eqbrem}
\dot{\gamma}_{Brems}=-N v \int_0^{\gamma-1} k \frac{\mathrm{d}\sigma}{\mathrm{d}k}\mathrm{d}k,
\end{equation}
where $N$ is the number density of particles in the medium, $v$ is the velocity of the electrons, $k=h\nu/mc^2$ is the photon energy in units of
the electron rest energy and $\mathrm{d}\sigma/\mathrm{d}k$ is the Bremsstrahlung differential cross section. The velocity $v$ can be expressed
in terms of the Lorentz factor as $v=c\sqrt{\gamma^2-1}/\gamma$. The exact Breemsstrahlung cross section is a very complicated expression (see
e.g., \citealt{haug98}), but we can obtain some useful formulae in the Born approximation. We distinguish two kinds of Bremsstrahlung
interaction: the electron-nuclei and the electron-electron Breemsstrahlung.

For the electron-nuclei Breemsstrahlung, the cross section in Born approximation is proportional to the sum $S$ given in equation (\ref{s}). The
integral of the cross section can be derived avoiding the calculation of the dilogarithm ocurring in the exact formula.
\begin{equation}
\int_0^{\gamma-1} k \frac{\mathrm{d}\sigma}{\mathrm{d}k}\mathrm{d}k \simeq \frac{3}{\pi} \alpha \sigma_T Z^2 \frac{\gamma^3}{\gamma^2+p^2} \left [\frac{\gamma}{p} \ln(\gamma+p)-\frac{1}{3}+\frac{p^2}{\gamma^6} \left(\frac{2}{9}\gamma^2-\frac{19}{675}\gamma p^2-0.06 \frac{p^4}{\gamma} \right) \right],
\end{equation}
where $p=\sqrt{\gamma^2-1}$ is the linear momentum of the electron. This expression has a relative error less than 0.54\% throughout
\citep{haug04}. For the electron-electron Bremsstrahlung, the integral has a defined upper limit $k_{max}=(\gamma-1)/(\gamma-p+1)$. The integral
of the cross section in this case is fitted using the function $\Phi_{rad}^{e-e}$ defined as
\begin{equation}
\Phi_{rad}^{e-e}=\frac{1}{\gamma-1} \int_0^{k_{max}} k \frac{\mathrm{d}\sigma}{\mathrm{d}k}\mathrm{d}k.
\end{equation}
The fits of this function are given in equation (\ref{bremsfit}). Summing both contributions and substituing in equation \ref{eqbrem}, we obtain
\begin{multline}
\gamma_{Brems}=\frac{3\alpha}{\pi} c S \frac{\gamma^2}{\gamma^2+p^2}\left[\gamma \ln(\gamma+p)-\frac{p}{3}+\frac{p^3}{\gamma^6} \left(\frac{2}{9}\gamma^2-\frac{19}{675}\gamma p^2-0.06 \frac{p^4}{\gamma} \right) \right]+\\
c \left(\sum_Z Z N_Z \right) \frac{p}{\gamma} (\gamma-1) \Phi_{rad}^{ee}(\gamma).
\end{multline}

\section*{Adiabatic energy losses}

To derive the expression of this energy loss, let us assume that electrons form a non-relativistic Maxwellian gas (afterwards, we will show the
relativistic expression). The loss of internal energy $\mathrm{d}U$ due to the work done by the expanding gas in a volume $\mathrm{d}V$ is
\begin{equation}
\label{intene1}
\mathrm{d}U=-P\mathrm{d}V,
\end{equation}
where $P$ is the pressure of the gas. Assuming a perfect gas, we can write its equation of state
\begin{equation}
U=\frac{3}{2} n k T V,
\end{equation}
\begin{equation}
P=n k T,
\end{equation}
where $n$ is the number density of particles. Thus, combining these two equations and taking into account that the average energy of each
particle is $E=\frac{3}{2} k T$, we can deduce
\begin{equation}
\label{intene2}
\mathrm{d}U=n V \mathrm{d}E.
\end{equation}
Therefore, diving by the unit time interval $\mathrm{d}t$, from equations (\ref{intene1}) and (\ref{intene2}), we get
\begin{equation}
\label{enerate}
\frac{dE}{dt}=-\frac{2}{3}\frac{E}{V} \frac{dV}{dt}.
\end{equation}
The term $dV/dt$ is the expansion rate of the volume $V$ and this depends on the expansion velocity of the gas $\vec{v}(\vec{r})$. Considering a
volume as a cube of sides $\mathrm{d}x$, $\mathrm{d}y$ and $\mathrm{d}z$ moving with the flow, we can write the expansion of the volume as
\begin{equation}
 \frac{\mathrm{d}V}{\mathrm{d}t}=\left(v_{x+\mathrm{d}x}-v_x \right)\mathrm{d}y \, \mathrm{d}z+\left(v_{y+\mathrm{d}y}-v_y \right)\mathrm{d}x \, \mathrm{d}z+\left(v_{z+\mathrm{d}z}-v_z \right)\mathrm{d}x \, \mathrm{d}y.
\end{equation}
Expanding this expression using Taylor series, we find
\begin{equation}
\frac{\mathrm{d}V}{\mathrm{d}t}=\left(\frac{\partial v_x}{\partial x}+\frac{\partial v_y}{\partial y}+\frac{\partial v_z}{\partial z} \right)\mathrm{d}x \, \mathrm{d}y \, \mathrm{d}z=\left(\vec{\nabla} \cdot \vec{v} \right)V.
\end{equation}
Substituting this result into equation (\ref{enerate}), we get in terms of the Lorentz factor
\begin{equation}
\dot{\gamma}_{ad}=-\frac{2}{3} \left(\vec{\nabla} \cdot \vec{v} \right) \gamma,
\end{equation}
which is the general expression for the adiabatic energy losses. If we want to obtain the expression for the relativistic case, we do the same
procedure but we have to consider the internal energy of the gas and its pressure as $U=3 n k T V$ and $P=\frac{1}{3}U$, respectively. Finally,
the relativistic form of the adiabatic losses is
\begin{equation}
\label{adgen}
\dot{\gamma}_{ad}=-\frac{1}{3} \left(\vec{\nabla} \cdot \vec{v} \right) \gamma,
\end{equation}
which differs only in a factor 2 with the non-relativistic expression. In our model, we consider the PWN as an uniform expanding sphere, so the
expansion velocity of the gas can be written as
\begin{equation}
\label{vel}
v(r)=v_{PWN}(t) \left[\frac{r}{R_{PWN}(t)} \right].
\end{equation}
The divergence operator for a vector $\vec{f}(r,\theta,\varphi)$ in spherical coordinates has the form
\begin{equation}
\vec{\nabla} \cdot \vec{f}(r,\theta,\varphi)=\frac{1}{r^2}\frac{\partial}{\partial r}\left[r^2 f_r(r,\theta,\varphi) \right]+\frac{1}{r \sin \theta}\frac{\partial}{\partial \theta}\left[\sin \theta f_\theta (r,\theta,\varphi) \right]+\frac{1}{r \sin \theta}\frac{\partial f_\varphi (r,\theta,\varphi)}{\partial \varphi},
\end{equation}
thus, applying this operator to equation (\ref{vel}), we obtain
\begin{equation}
\vec{\nabla} \cdot \vec{v}=\frac{1}{r^2}\frac{\partial v(r)}{\partial r}=3\frac{v_{PWN}(t)}{R_{PWN}(t)}.
\end{equation}
Finally, subtituting in equation (\ref{adgen}), we get the expression that we use in our model for the adiabatic energy losses
\begin{equation}
\dot{\gamma}_{ad}(\gamma,t)=-\frac{v_{PWN}(t)}{R_{PWN}(t)}\gamma.
\end{equation}

\ifpdf
    \graphicspath{{Appendix2/Figs/Raster/}{Appendix2/Figs/PDF/}{Appendix2/Figs/}}
\else
    \graphicspath{{Appendix2/Figs/Vector/}{Appendix2/Figs/}}
\fi

\chapter{Luminosity equations}
\label{appb}

\section*{Synchrotron luminosity}

\begin{figure}
\centering
\includegraphics[width=0.5\textwidth]{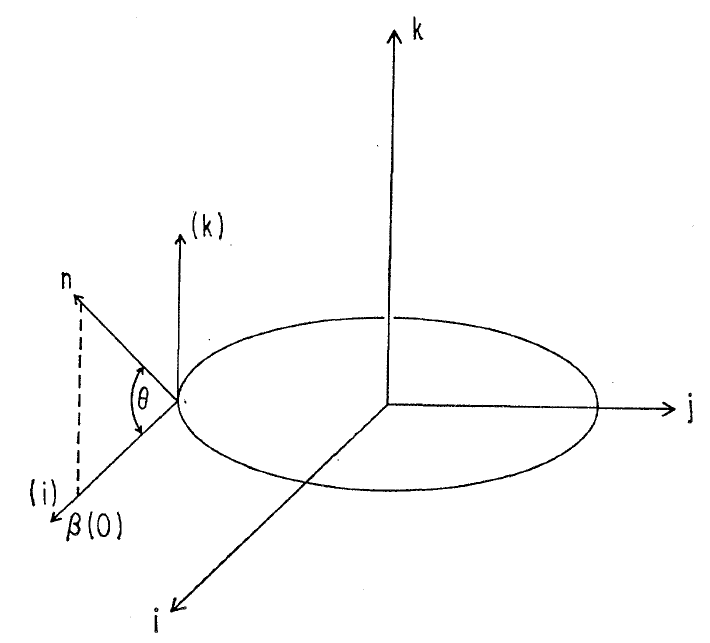}
\caption[Electron spiraling a magnetic field line]{Trajectory of an electron spiraling a magnetic field. The energy is emitted in the direction
$\vec{n}$. Image taken from \protect\citet{blumenthal70}.}
\label{elecspir}
\end{figure}

The formula for the synchrotron luminosity is explained in detail in many publications (e.g.,
\citealt{ginzburg64,ginzburg65,blumenthal70,rybicki79}). In this appendix, we will summarized the calculations following the prescription given
by \citet{blumenthal70}. For an electron with energy $\gamma \gg 1$ spiraling around a magnetic field line (see figure \ref{elecspir}), its
velocity vector is described as
\begin{equation}
\vec{\beta}=\beta(\hat{i} \cos \Omega t+\hat{j} \sin \Omega t)
\end{equation}
In the laboratory frame, the energy emitted by an electron per unit observer's time ($\tilde{t}$) per unit solid angle in the direction $\hat{n}$
is given by \citet{jackson62}:
\begin{equation}
\frac{dP(t)}{d\Omega_{\hat{n}}}=\frac{e^2}{4 \pi c} \frac{\vert \hat{n} \times [(\hat{n}-\vec{\beta}) \times \dot{\vec{\beta}}] \vert^2}{(1-\hat{n} \cdot \vec{\beta})^6}
\end{equation}
It can be demonstrated (see appendix of \citealt{blumenthal70}) that the spectrum of the radiation is written as
\begin{equation}
\frac{dI_\omega}{d\Omega_{\hat{n}}}=\frac{e^2}{4 \pi c} \vert f_\omega \vert^2
\end{equation}
with
\begin{equation}
\label{fom}
f_\omega=\int_{-\infty}^\infty \frac{\hat{n} \times [(\hat{n}-\vec{\beta}) \times \dot{\vec{\beta}}]}{(1-\hat{n} \cdot \vec{\beta})^3} \exp(i \omega \tilde{t}) \mathrm{d}\tilde{t}.
\end{equation}
The time $t$ at electron's rest frame is related with $\tilde{t}$ by
\begin{equation}
\frac{d\tilde{t}}{dt}=1-\hat{n} \cdot \vec{\beta}.
\end{equation}
Integrating this expression, for very distant observers, we have
\begin{equation}
\tilde{t}=t-\frac{\hat{n} \cdot \vec{r(t)}}{c},
\end{equation}
where a constant term has been ignored as it only contributes an over-all phase factor to $f_\omega$. Changing the variable $\tilde{t}$ by $t$ in
equation (\ref{fom}) and integrating, we obtain
\begin{equation}
\label{fom2}
f_\omega=\omega \int_{-\infty}^\infty \hat{n} \times (\hat{n} \times \vec{\beta}) \exp \left[i \omega \left(t-\frac{\hat{n} \cdot \vec{r}}{c} \right) \right] \mathrm{d}t.
\end{equation}
In the laboratory frame, the synchrotron radiation is emitted in an angle $\theta \sim \gamma^{-1}$. Thus, the electron radiates in a given
direction for a time $\Delta t \sim (\Omega \gamma)^{-1}$. For times greater than this, the exponential in $f_\omega$ oscillates rapidly and the
integral ends to zero. \citet{jackson62} gives an expression for $t-\hat{n} \cdot \vec{r}/c$ as a function of $\theta$, $\gamma$ and $\Omega$
\begin{equation}
t-\frac{\hat{n} \cdot \vec{r}}{c} \simeq \frac{1}{2} \left[\left(\theta^2+\frac{1}{\gamma^2} \right)t+\frac{\Omega^2 t^3}{3} \right].
\end{equation}
In order to solve the double cross product inside the integral in equation (\ref{fom2}), we change the coordinate system defining the unitary
vector $\hat{e}=\hat{n} \times \hat{j}$. In this system, the velocity has the form
\begin{equation}
\vec{\beta}=\beta(\hat{j} \sin \Omega t+\hat{e} \sin \theta \cos \Omega t+\hat{n} \cos \theta \cos \Omega t),
\end{equation}
and then, reducing to the lowest order in $\theta$ and $\Omega t$, we have
\begin{equation}
\hat{n} \times (\hat{n} \times \vec{\beta})=(\hat{e} \theta+\hat{j} \Omega t).
\end{equation}
Defining $\xi=\Omega t$ and substituting in equation (\ref{fom2}), we get
\begin{equation}
f_\omega=\frac{\omega}{\Omega} \int_{-\infty}^\infty (\hat{e} \theta+\hat{j} \xi) \exp \left \{ \frac{i \omega}{2 \Omega} \left[\left(\theta^2+\frac{1}{\gamma^2} \right) \xi+\frac{\xi^3}{3} \right] \right \} \mathrm{d} \xi.
\end{equation}
To compute the square of $f_\omega$, we define first the parameters $\mu=\omega/2\Omega$ and $\eta^2=\theta^2+1/\gamma^2$. We also make a change
of variables, defining $x=u+v$ and $y=u-v$ (note that the Jacobian is 2.). Thus,
\begin{equation}
\vert f_\omega \vert^2=\frac{2 \omega^2}{\Omega^2} \int_{-\infty}^\infty \exp \left[2 i \mu \left(\eta^2 v+\frac{1}{3} v^3 \right) \right] \mathrm{d}v \int_{-\infty}^\infty (u^2-v^2+\theta^2) e^{2 i \mu v u^2} \mathrm{d}u.
\end{equation}
We perform the integration in $u$ obtaining
\begin{equation}
\vert f_\omega \vert^2=\frac{2 \sqrt{\pi} \omega^2}{\Omega^2} \int_{-\infty}^\infty \left[\sqrt{2 \mu v} e^{i \pi (\theta^2-v^2)}-\frac{1}{2}(2 \mu v)^{-3/2} e^{-i \frac{\pi}{4}} \right] \exp \left[2 i \mu \left(\eta^2 v+\frac{1}{3} v^3 \right) \right] \mathrm{d}v.
\end{equation}

Now, we can substitute $\vert f_\omega \vert^2$ in $dI_\omega/d\Omega_{\hat{n}}$. To integrate the latter term over the solid angle, we can
approximate the differential section of solid angle as $d\Omega_{\hat{n}}=2\pi \sin (\pi/2-\theta) d\theta \simeq 2 \pi d\theta$. Applying this
approximation, one integrates $dI_\omega/d\Omega_{\hat{n}}$ and gets
\begin{equation}
I_\omega=-\frac{e^2 \omega i}{\Omega c} \int_{-\infty}^\infty \left[v-\frac{1}{2 \mu i v^2} \right] \exp \left[2 i \mu \left(\frac{v}{\gamma^2}+\frac{1}{3} v^3 \right) \right] \mathrm{d}v.
\end{equation}
The second term may be integrated by parts to yield,
\begin{equation}
\label{iom}
I_\omega=-\frac{e^2 \omega i}{\Omega c} \int_{-\infty}^\infty \left(2v-\frac{1}{\gamma^2 v} \right) \exp \left[2 i \mu \left(\frac{v}{\gamma^2}+\frac{1}{3} v^3 \right) \right] \mathrm{d}v.
\end{equation}
\\
\\
We can rewrite equation (\ref{iom}) such that
\begin{equation}
I_\omega=-\frac{e^2 \omega i}{\Omega c} (I_\omega^{(1)}+I_\omega^{(2)}).
\end{equation}
Setting $v=x/\gamma$ and $\xi=4\mu/3\gamma^3$, the function $I_\omega^{(1)}$ yields
\begin{equation}
I_\omega^{(1)}=\frac{2}{\gamma^2} \int_{-\infty}^\infty x \exp \left[i \frac{3}{2}\xi \left(x+\frac{1}{3}x^3 \right) \right] \mathrm{d}x.
\end{equation}
It is useful to write the integrand of the latter equation as an integral,
\begin{equation}
I_\omega^{(1)}=-\frac{i}{\gamma^2} \int_{\frac{2 \omega}{3 \Omega \gamma^3}}^\infty \mathrm{d}\xi \int_{-\infty}^\infty (3x^2+x^4) \left[i \frac{3}{2}\xi \left(x+\frac{1}{3}x^3 \right) \right] \mathrm{d}x.
\end{equation}
Regarding the term $I_\omega^{(2)}$, we have
\begin{equation}
I_\omega^{(2)}=\frac{1}{\gamma^2} \int_{-\infty}^\infty \frac{1}{v} \exp \left[2 i \mu \left(\frac{v}{\gamma^2}+\frac{1}{3}v^3 \right) \right] \mathrm{d}v.
\end{equation}
Differentiating with respect to $\gamma^{-2}$ and then integrating from $\gamma^{-2}$ to infinity, we get
\begin{equation}
I_\omega^{(2)}=-\frac{2 i \mu}{\gamma^2} \int_{\frac{1}{\gamma^2}}^\infty \mathrm{d}y \int_{-\infty}^\infty \exp \left[2 i \mu \left(v y+\frac{1}{3}v^3 \right) \right] \mathrm{d} v.
\end{equation}
Finally, substituting $\mu=\omega/2\Omega$ and changing the variables $x=v y^{-1/2}$ and $\xi=(4/3)\mu y^{3/2}$ yields
\begin{equation}
I_\omega^{(2)}=-\frac{i}{\gamma^2} \int_{\frac{2 \omega}{3 \Omega \gamma^3}}^\infty \mathrm{d}\xi \int_{-\infty}^\infty \exp \left[i \frac{3}{2} \xi \left(x+\frac{1}{3}x^3 \right) \right] \mathrm{d}x.
\end{equation}
Using the expressions obtained for $I_\omega^{(1)}$ and $I_\omega^{(2)}$, the formula for $I_\omega$ leads
\begin{equation}
I_\omega=-\frac{e^2 \omega}{\gamma^2 \Omega c} \int_{\frac{2 \omega}{3 \Omega \gamma^3}}^\infty \mathrm{d}\xi \int_{-\infty}^\infty (1+3x^2+x^4) \exp \left[i \frac{3}{2} \xi \left(x+\frac{1}{3}x^3 \right) \right] \mathrm{d}x.
\end{equation}
Integrating over $x$ gives
\begin{equation}
I_\omega=\frac{2 e^2 \omega}{\sqrt{3}\gamma^2 \Omega c} \int_{\frac{2 \omega}{3 \Omega \gamma^3}}^\infty K_{5/3}(\xi) \mathrm{d}\xi,
\end{equation}
where $K_{5/3}(\xi)$ is the modified Bessel function of 5/3 order. The latter equation gives the spectrum of the electron per revolution.
To obtain the total power, we must multiply by $\Omega/2\pi$ and considering that $\nu=\omega/2\pi$, we recover equation (\ref{psync})
\begin{equation}
P_{syn}(\nu,\gamma,B(t))=\frac{\sqrt{3}e^3 B}{m_e c^2} \frac{\nu}{\nu_c} \int_{\frac{\nu}{\nu_c}}^\infty K_{5/3}(\xi) \mathrm{d}\xi
\end{equation}
with $\nu_c$ defined as in equation (\ref{critfreq}). Finally, the total synchrotron luminosity for a distribution of electrons is given by
equation (\ref{synclum}).

\section*{IC luminosity}

As for the synchrotron luminosity, we also use the prescription given by \citet{blumenthal70}. The IC luminosity is deduced easily using equation
(\ref{icsp}), which is the scattered photon spectrum per electron used in the computation of the IC energy losses. Multiplying equation
(\ref{icsp}) by the energy of each photon, we recover equation (\ref{pic})
\begin{equation}
P_{IC}(\gamma,\nu,t)=\frac{3}{4} \frac{\sigma_T c h \nu}{\gamma^2} \int_0^\infty \frac{n(\nu_i)}{\nu_i}f(q,\Gamma_{\varepsilon}) H(1-q) H \left(q-\frac{1}{4\gamma^2} \right) \mathrm{d}\nu_i.
\end{equation}
The total IC luminosity is finally obtained by doing
\begin{equation}
L_{IC}(\nu,t)=\int_0^\infty N(\gamma,t) P_{IC}(\gamma,\nu,t) \mathrm{d}\nu,
\end{equation}
recovering equation (\ref{iclum}).

\section*{Bremsstrahlung luminosity}

According to \citet{blumenthal70}, the Bremsstrahlung spectrum per electron is given by
\begin{equation}
\label{dnbrems}
\frac{dN}{dt d\nu}=c \sum_s ns \frac{d \sigma_s}{d\nu},
\end{equation}
where $\nu$ is the radiated photon frequency. The quantity $d\sigma_s/d\nu$ is the differential Bremsstrahlung cross section, which is given by
\begin{equation}
\label{sigmas}
\frac{d\sigma_s}{d\nu}=\frac{3\alpha}{8\pi} \frac{d\nu}{\nu} \frac{1}{\gamma_i^2} \left[(\gamma_i^2+\gamma_f^2)\phi_1-\frac{2}{3}\gamma_i \gamma_f \phi_2 \right],
\end{equation}
where $\gamma_i$, $\gamma_f$ are the initial and final energies of the electron. The functions $\phi_1$ and $\phi_2$ depend on $\gamma_i$,
$\gamma_f$ and $\nu$. Assuming a complete ionized plasma, $\phi_1=\phi_2=Z^2\phi_u$ with $Z$ is the ion charge number. The function $\phi_u$ is
given by
\begin{equation}
\phi_u=4 \left[\ln \left(\frac{2 \gamma_i \gamma_f m_e c^2}{h \nu} \right)-\frac{1}{2} \right].
\end{equation}
Substituting equation (\ref{sigmas}) in equation (\ref{dnbrems}), we obtain
\begin{equation}
P_{Brems}(\gamma_i,\nu)=h \nu \frac{dN}{dt d\nu}=\frac{3}{2 \pi} \frac{\alpha \sigma_T h c S}{\gamma_i^2} \left(\gamma_i^2+\gamma_f^2-\frac{2}{3} \gamma_i \gamma_f \right) \left(\ln \frac{2 \gamma_i \gamma_f m c^2}{h \nu}-\frac{1}{2} \right),
\end{equation}
where $S$ is given by equation (\ref{s}). From the kinetic condition $h \nu/m_e c^2=\gamma_i-\gamma_f$ and the factor
$\ln[2 \gamma_i \gamma_f m_e c^2/(h \nu)]-1/2$, we get the minimum initial energy of the electron for a given frequency $\nu$ and final energy of
the electron $\gamma_f$
\begin{equation}
\gamma_i^{min}(\nu)=\frac{1}{2}\left[\frac{h \nu}{m_e c^2}+\sqrt{\left(\frac{h \nu}{m_e c^2} \right)^2+\frac{2h \nu}{m_e c^2} \exp \left(\frac{1}{2} \right)} \right].
\end{equation}
Thus, the total Bremsstrahlung luminosity from a distribution of electrons is
\begin{multline}
L_{Brems}(\nu,t)=h \nu \int N(\gamma,t) \frac{dN}{dt d\nu} \mathrm{d}\gamma=\frac{3}{2 \pi} \alpha \sigma_T h c S \int_{\gamma_i^{min}(\nu)}^{\gamma_{max}} \frac{N(\gamma_i)}{\gamma_i^2} \left(\gamma_i^2+\gamma_f^2-\frac{2}{3} \gamma_i \gamma_f \right)\\
\times \left(\ln \frac{2 \gamma_i \gamma_f m c^2}{h \nu}-\frac{1}{2} \right) \mathrm{d} \gamma_i.
\end{multline}
as is written in equation (\ref{bremslum}).

\end{appendices}

\printthesisindex 

\end{document}